# MODULAR GRAPH FORMS AND SCATTERING AMPLITUDES IN STRING THEORY

JAN E. GERKEN

Max Planck Institute For Gravitational Physics
Albert Einstein Institute

Humboldt-University Berlin
Faculty of Mathematics and Natural Sciences

# ABSTRACT

In this thesis, we investigate the low-energy expansion of scattering amplitudes of closed strings at one-loop level (i.e. at genus one) in a ten-dimensional Minkowski background using a special class of functions called modular graph forms. These allow for a systematic evaluation of the low-energy expansion and satisfy many non-trivial algebraic and differential relations. We study these relations in detail, leading to basis decompositions for a large number of modular graph forms which greatly reduce the complexity of the expansions of the integrals appearing in the amplitude. One of the results of this thesis is a `Mathematica` package which automatizes these simplifications.

We use these techniques to compute the leading low-energy orders of the scattering amplitude of four gluons in the heterotic string at one-loop level. Furthermore, we decompose the amplitude into building blocks of uniform transcendentality, a property known from field-theory amplitudes.

For tree-level string amplitudes, the single-valued map of multiple zeta values maps open-string amplitudes to closed-string amplitudes. The definition of a suitable one-loop generalization, a so-called elliptic single-valued map, is an active area of research and we show that a certain conjectural definition for this map, which was successfully applied to maximally supersymmetric amplitudes, cannot reproduce all terms in the heterotic string which has half-maximal supersymmetry.

In order to arrive at a more systematic treatment of modular graph forms and at a different perspective on the elliptic single-valued map, we then study a generating function which conjecturally contains the torus integrals of all perturbative closed-string theories. We determine a differential equation satisfied by this generating function and solve it in terms of path-ordered exponentials, leading to iterated integrals of holomorphic Eisenstein series. Since these are linearly independent, we can use this approach to arrive at a more rigorous characterization of the space of modular graph forms than was possible before. Moreover, since a similar construction is available for the open string, this opens a new perspective on the elliptic single-valued map.

The original version of this thesis, as submitted in June 2020 to the Humboldt University Berlin, is available under the DOI 10.18452/21829. The present text contains minor updates compared to this version reflecting further developments in the literature, in particular concerning the construction of an elliptic single-valued map.

*One thing I have learned in a long life: that all our science, measured against reality, is primitive and childlike — and yet it is the most precious thing we have.*

— Albert Einstein

# ACKNOWLEDGMENTS

Research is a group endeavor and it would have been impossible for me to complete the work presented in this thesis without the support of a large number of people. They provided ideas and encouragement and made the experience so much more enjoyable.

The most important ally on this journey was my supervisor Axel Kleinschmidt, who is not only a brilliant physicist but with his inexhaustible patience and clear explanations bridged many gaps on this path. He has moreover provided me with much encouragement and gave me the freedom to explore my own ideas. His breadth of knowledge about physics and mathematics allowed me to deepen my understanding of many areas of these fascinating fields, which was immensely gratifying. Thank you very much, Axel!

Early on in my PhD, Oliver Schlotterer became an invaluable support for my research. Bubbling with ideas, Oli was a driving force in our projects and he was always happy to patiently share his vast knowledge of (string) amplitudes. We have probably never been to a restaurant without leaving napkins filled with equations behind. By inviting me numerous times, he gave me the opportunity to travel and to meet many fascinating people. His encouragement was crucial and over time, he has become a good friend.

I am grateful to my collaborator Justin Kaidi for his quick and precise thinking that led to swift progress in our project.

Nils Matthes provided a mathematician's point of view on the research in this field which has greatly expanded my horizon. I am grateful for his invitation to Fukuoka and have fond memories of his great sense of humor.

One of the backbones for the research presented here was the stimulating environment provided by the Max Planck Institute for Gravitational Physics (the Albert Einstein Institute) in Potsdam and in particular by the quantum gravity division led by Hermann Nicolai. I am grateful to him for welcoming me into his group and feel privileged for having been able to do my PhD under such favorable conditions. I have benefited from the framework of the International Max Planck Research School (IMPRS) for Mathematical and Physical Aspects of Gravitation, Cosmology and Quantum Field Theory which allowed me to acquire many valuable skills, travel extensively and offered numerous fascinating lectures. All this was organized marvelously by Hadi Godazgar, Axel Kleinschmidt and Oliver Schlotterer.

What really brought the IMPRS to live, however, were my fellow PhD students. They made the excursions, lectures and seminars interesting and enjoyable and I have learned a lot from them. It was a great pleasure to share an office with Olof Ahlén, Matteo Broccoli and Lars Kreutzer. I had many interesting discussions about string theory, politics and culture in Germany and Korea with Seungjin Lee. My lunch breaks, excursions and our student seminars were made enjoyable and interesting by Sebastian Bramberger, Hugo Camargo, Lorenzo Casarin, Marco Finocchiaro, Caroline Jonas, Johannes Knaute, Isha Kotecha, Hannes Malcha, Tung Tran and Alice Di Tucci.

I want to thank Kevin Rieger and Konstantin Steinweg for providing a link to high school days as well as some counterbalance to physics.

Finally, I am deeply grateful to my parents for their unwavering support and encouragement. Their love and inspiration has been a strong basis for my endeavors and without it, I could have never reached this point.

# CONTENTS

# ACRONYMS

| | |
|---|---|
| BCFW | Britto–Cachazo–Feng–Witten |
| BCJ | Bern–Carrasco–Johansson |
| CFT | conformal field theory |
| CHY | Cachazo–He–Yuan |
| eMZV | elliptic multiple zeta value |
| GSO | Gliozzi–Scherk–Olive |
| HSR | holomorphic subgraph reduction |
| IR | infrared |
| K3 | Kummer–Kähler–Kodaira |
| KLT | Kawai–Lewellen–Tye |
| KZB | Knizhnik–Zamolodchikov–Bernard |
| LHS | left-hand side |
| MGF | modular graph function / form |
| MZV | multiple zeta value |
| NS | Neveu–Schwarz |
| QFT | quantum field theory |
| R | Ramond |
| RHS | right-hand side |
| RNS | Ramond–Neveu–Schwarz |
| teMZV | twisted elliptic multiple zeta value |
| UV | ultraviolet |

# PUBLICATIONS CONTAINED IN THIS THESIS

# 1

# INTRODUCTION

String theory is by far the most widely studied candidate theory of quantum gravity, unifying general relativity with quantum field theory (QFT). This is achieved by replacing zero-dimensional particles by one-dimensional objects, called *strings*. These can either form a closed loop, leading to *closed strings*, or have two endpoints, yielding *open strings*. Similarly to how particles interact in QFT, strings can scatter off of each other, in the quantum theory this is described by scattering amplitudes. These amplitudes give rise to the low-energy effective field theory description of string theory, therefore giving access to the physics which would first be observed at low energies. In particular, important insights into field-theory amplitudes were obtained by considering the low-energy limit of string amplitudes [1–6]. Higher-order terms in the low-energy expansion provide a testing ground for string dualities [7–13]. Aside from these practical motivations to study string amplitudes, also conceptually, they lie at the heart of the subject: String theory started in 1968 with the construction of a scattering amplitude [14], well before it was realized that this amplitude describes the scattering of strings. Since then, the study of scattering amplitudes in string theory has grown into a rich subject area of its own right which has benefited enormously from a fruitful interaction with pure mathematics.

Similarly to field theory amplitudes, string amplitudes admit a perturbative expansion in the coupling constant $g_s$. In this framework, the loop expansion known from QFT becomes an expansion in the genus of the worldsheet, the two-dimensional surface swept out by the string in spacetime, cf. Figure 1.1. In this thesis, we study the low-energy expansion of closed-string amplitudes in a ten-dimensional Minkowski spacetime at one-loop level, i.e. at genus one, using a class of functions called *modular graph functions / forms* (MGFs) [15, 16]. Our goal here is two-fold: On the one hand, we aim for a systematic way of computing this

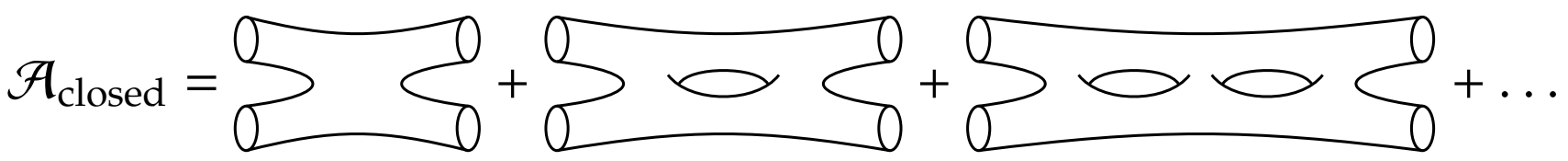

Figure 1.1: Expansion of four-point closed-string amplitude as a sum over genera.

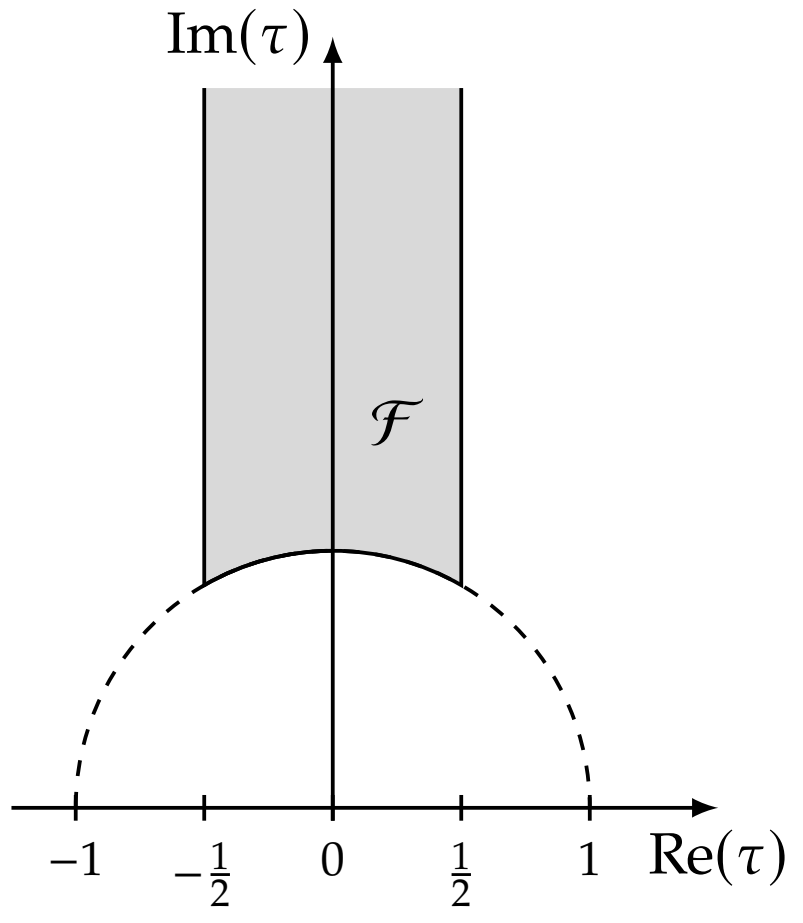

(a) Fundamental domain $\mathcal{F}$ of the torus, as defined in (3.14).

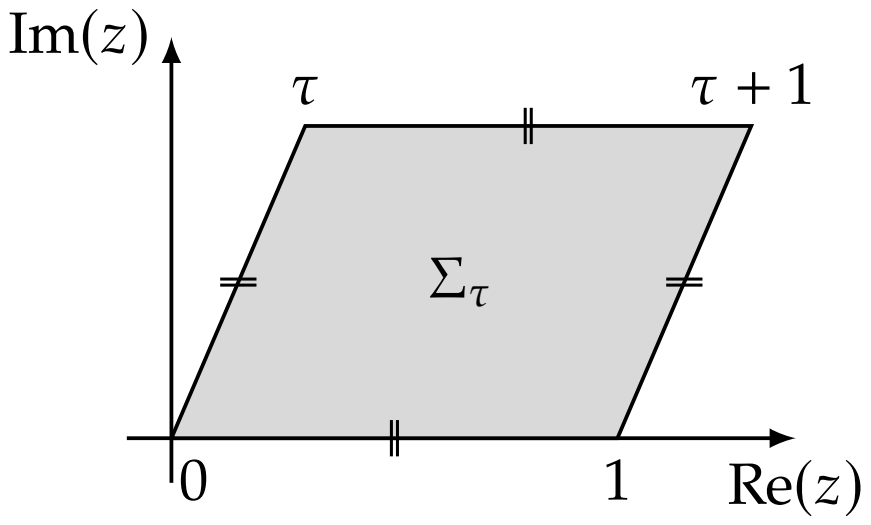

(b) Parametrization of the torus by the parallelogram $\Sigma_\tau$ with opposite edges identified.

Figure 1.2: Integration domains in (1.1).

low-energy expansion at genus one, making as much of the computation algorithmic as possible. On the other hand, we work towards extending a certain relation between open- and closed-strings known from tree-level string amplitudes to genus-one amplitudes, leading to a concrete proposal [17]. This relation, the so-called *single-valued map* [18, 19], is a formal operation on the number-theoretic ingredients of the amplitude.

At one-loop level, the worldsheet is a torus on which vertex operators, corresponding to the external string states, live. The amplitude is given as an integral of the correlator of the vertex operators in the conformal field theory (CFT) defining the string theory over their positions (the *punctures*) and the shape of the torus, encoded by the modular parameter $\tau$. Schematically, for $n$ external states,[1]

$$\mathcal{A}^{\text{closed}}_{\text{genus one}} = \int_{\mathcal{F}} d^2\tau \int_{\Sigma_\tau} \prod_{i=1}^{n} d^2 z_i \, \langle \mathcal{V}_1(z_1)\mathcal{V}_2(z_2)\cdots\mathcal{V}_n(z_n)\rangle \,, \tag{1.1}$$

where the fundamental domain $\mathcal{F}$ and the torus parametrization $\Sigma_\tau$ are depicted in Figure 1.2 and (1.1) contributes at the order $g_s^0$. In this thesis, we will mainly focus on the integral over the $z_i$ and mention the final integral over $\tau$ only briefly. The gauge invariance of string theory implies that the integral over the $z_i$ is invariant under modular transformations

$$\tau \to \frac{\alpha\tau+\beta}{\gamma\tau+\delta}\,, \qquad \begin{pmatrix}\alpha & \beta\\ \gamma & \delta\end{pmatrix} \in \mathrm{SL}(2,\mathbb{Z})\,. \tag{1.2}$$

1 In (1.1), we have absorbed the (super)-ghost operators in the CFT correlator into the vertex operators, cf. (2.21).

Functions of $\tau$ which are invariant under (1.2) (and satisfy a certain moderate growth condition, cf. (3.17)) are called *modular functions*. Non-holomorphic functions of $\tau$ with the transformation property

$$f\left(\frac{\alpha\tau+\beta}{\gamma\tau+\delta}\right) = (\gamma\tau+\delta)^a(\gamma\bar{\tau}+\delta)^b f(\tau) \tag{1.3}$$

(and which are of moderate growth) are referred to as *(non-holomorphic) modular forms* of *modular weight* $(a, b)$.

In order to obtain the low-energy expansion of the amplitude, we expand the CFT correlator in $\alpha'$, the inverse string tension, leading to a loop expansion in Feynman-like graphs in the worldsheet-CFT. These graphs are integrated over the puncture positions $z_i$ by performing a Fourier transformation, as is familiar from standard QFT, trivializing the integration over positions while yielding momentum conserving delta functions at the vertices and propagators of the form $1/|p|^2$. The resulting non-zero graphs are one-particle irreducible vacuum bubbles, for example,

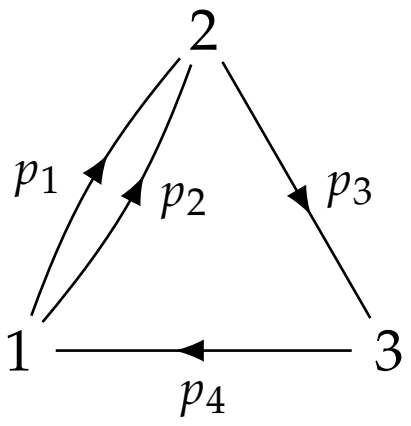

$$= \sum_{p_1,p_2,p_3,p_4}{}' \frac{\delta(p_1+p_2-p_3)\delta(p_3-p_4)}{|p_1|^2|p_2|^2|p_3|^2|p_4|^2}\,. \tag{1.4}$$

Since the torus is compact, the usual integrals over loop momenta are replaced by sums over discrete lattice points: In (1.4), $p = m\tau + n$ and the sum runs over $(m, n) \in \mathbb{Z}^2 \setminus \{(0, 0)\}$ to exclude the poles of the summand. The resulting objects are modular functions of $\tau$ associated to a graph and therefore called *modular graph functions* [15]. This construction can be generalized to functions with non-trivial modular transformation properties (the modular invariance of the integrand is then ensured by further $z$-independent contributions), leading to *modular graph forms* [16].

In this way, the low-energy expansion of closed-string one-loop amplitudes can be obtained systematically, up to the final integral over $\tau$. However, the resulting lattice sums are hard to evaluate and there are many non-trivial relations between sums associated to different graphs. In this thesis, we will study these relations systematically and find basis decompositions for a large class of MGFs [I, II] in Chapter 5. These do not only help to simplify the resulting expression for the low-energy expansion of the amplitude, they also facilitate the final integration over $\tau$, as we will demonstrate for the case of four-gluon scattering in heterotic string theory [III] in Chapter 6.

On top of their relevance for the explicit calculation of closed-string amplitudes, MGFs also aid in illuminating a deep relation between open-and closed-string amplitudes at one-loop, extending the single-valued map from tree-level.

At tree-level, the relevant closed-string worldsheet is a sphere. Hence, there is no analog of the modular parameter $\tau$ and after integrating over the puncture positions, the coefficients in the $\alpha'$ expansion are numbers, not functions. More specifically, only a certain class of numbers appears [20–24], so-called *multiple zeta values* (MZVs). These are obtained by evaluating a multi-variable generalization of the Riemann zeta function at arguments in $\mathbb{N}$ and exhibit a rich algebraic structure. A similar calculation can be done for tree-level open-string integrals, yielding also MZVs. Recently, it was proven by several groups [25–27] that, at tree-level, the closed-string coefficients can be obtained from the open-string coefficients via the single-valued map, a certain homomorphism on the algebra of MZVs. Here, we will work towards defining a suitable *elliptic single-valued map* which extends this construction to one-loop amplitudes, as detailed in [17] which appeared after the submission of this thesis. This map aims to ultimately reduce the problem of computing closed-string amplitudes at genus one to the simpler problem of computing open-string amplitudes at genus one. At the same time, the elliptic single-valued map supports the idea that the single-valued correspondence has a deeper relevance to string theory and is not just a coincidence at tree-level.

The coefficients in the low-energy expansion of open-string amplitudes at genus one were identified [28, 29] to be functions of the modular parameter called *elliptic multiple zeta values* (eMZVs) [30]. At this level, the expansion coefficients for open and closed strings are very different functions, making a comparison difficult. However, both eMZVs and MGFs can be written in terms of iterated integrals over holomorphic Eisenstein series, which are recursively defined by

$$\begin{aligned} \mathcal{E}(k_1,\dots,k_r;\tau) &= \frac{1}{(2\pi i)^{k_r-1}}\int_\tau^{i\infty} \mathrm{d}\tau_r\, \mathrm{G}_{k_r}(\tau_r)\,\mathcal{E}(k_1,\dots,k_{r-1};\tau_r) \\ \mathcal{E}(\emptyset;\tau) &= 1\,, \end{aligned} \tag{1.5}$$

together with a suitable regularization.[2] The labels $k_1,\dots,k_r$ take the values 0 and $4{+}2n$, $n\in\mathbb{N}_0$. The classic holomorphic Eisenstein series are given by the sums

$$\begin{aligned} \mathrm{G}_{2k}(\tau) &= \sum_{(m,n)\in\mathbb{Z}^2\setminus\{(0,0)\}} \frac{1}{(m\tau+n)^k}\,, \quad k\geq 2\in\mathbb{N} \\ \mathrm{G}_0 &= -1\,. \end{aligned} \tag{1.6}$$

Currently, finding a single-valued map for iterated Eisenstein integrals

2 Below, we will take this to be the tangential-base-point regularization [31]. However, in (4.15), we define the slightly modified integrals $\mathcal{E}_0$ which are convergent and span the same space as the $\mathcal{E}$.

is an active area of research with contributions from mathematicians and physicists alike [15, 17, 32–34].

On top of their importance for the comparison to open strings, the formulation of MGFs in terms of iterated Eisenstein integrals also facilitates the understanding of the space of MGFs itself, since iterated Eisenstein integrals with different labels are linearly independent [35].

In the literature, the translation of MGFs into iterated Eisenstein integrals has so far been done only on a case-by-case basis [15, 34, 36]. To achieve a more systematic treatment, we will study a generating function of closed-string integrals [IV] which captures all integrals appearing in string amplitudes and can in particular be expanded in terms of MGFs. We will investigate the differential equation w.r.t. $\tau$ of this function and in this way systematically obtain differential equations for MGFs in Chapter 7. Furthermore, a similar calculation was done in the open string [37, 38] and we will see that the closed-string differential equation is largely determined by the single-valued image of the open-string differential equation.

The differential equation for the closed-string generating function can be solved perturbatively in $\alpha'$ using so-called *Picard iteration*, yielding path-ordered exponentials and leading to an expression in terms of iterated Eisenstein integrals [V], as we will see in Chapter 8. Comparing this form of the generating function to the form obtained using the MGFs techniques from [I, II], leads to a systematic expression of MGFs in terms of iterated Eisenstein integrals. Furthermore, this correspondence allows for a basis-counting in the space of MGFs and proves that the decompositions obtained in [II] are in fact complete.

Similarly, the corresponding generating function in the open string can also be expressed in terms of iterated Eisenstein integrals and a comparison of the two expressions is a concrete step towards a single-valued prescription at genus one, as detailed in [17].

## 1.1 RESULTS OF THIS WORK

The work presented in this thesis was published in the five papers [I–V].

In [I], a certain simplification technique for MGFs, the so-called *holomorphic subgraph reduction* (HSR), introduced for two-point holomorphic subgraphs in [16], is extended to the $n$-point case. In particular, a central step in the procedure of HSR is the evaluation of certain conditionally convergent sums, which is put on firm mathematical grounds in [I] and extended to $n$ points, making HSR algorithmic for graphs with arbitrarily many points. Furthermore, a closed formula for three-point HSR is provided.

In [II], a systematic description of modular graph forms with up to four points was introduced and divergent MGFs studied for the first time. In particular, it was shown how their appearance is linked to poles in the kinematic variables of the associated string integral. Furthermore, it is shown that the Fay identities obeyed by Kronecker–Eisenstein series are equivalent to HSR and allow for a more efficient treatment of holomorphic subgraphs than the traditional techniques. The bases of all MGFs of total modular weight at most 12 are described and, by combining all the known properties of MGFs, basis decompositions for all two- and three-point MGFs with weight at most 12 are obtained. Finally, a `Mathematica` framework which implements these basis decompositions and the other simplification techniques is provided.

In [III], four-gluon scattering in heterotic string theory is studied, providing a practical application of the MGF techniques discussed here. This is in particular the first time that modular graph forms were used in the computation of a pure-gauge amplitude (as opposed to a pure-gravity amplitude). Furthermore, the final integral over $\tau$ is performed to second order (previously, only the zeroth order was known) and the amplitude is decomposed into building blocks of conjectured uniform transcendentality. In the literature, a conjectural prescription for the single-valued map acting on iterated Eisenstein integrals at maximal supersymmetry is available in [34] and it is shown that for the half-maximally supersymmetric heterotic amplitudes, this map cannot reproduce all necessary terms.

In [IV] a generating series for all closed-string integrals, and hence for all MGFs, is defined. Its Cauchy–Riemann and Laplace equation in $\tau$ are determined at $n$-point and closed expressions for the differential equations satisfied by the component integrals are obtained at two- and three-point, yielding Cauchy–Riemann and Laplace equations for infinite families of MGFs which were studied extensively in the literature [16, 39–41]. Using the MGF techniques discussed in [II], these differential equations are verified explicitly in a number of cases.

In [V], the Cauchy–Riemann equation of the generating function of closed-string integrals discussed in [IV] is solved perturbatively using path-ordered exponentials, yielding a series in iterated Eisenstein integrals. Since this solution generates closed-string integrals, it can also be expanded in terms of MGFs, making a translation between iterated Eisenstein integrals and MGFs possible. In this process, the techniques from [II] are instrumental in supplying the initial condition for the differential equation and simplified expressions for the comparison. Studying the solution obtained in this way leads to several results for iterated Eisenstein integrals and MGFs:

- The modular properties of the iterated Eisenstein integrals, which are generically hard to obtain, are fixed explicitly in a large class of cases.
- A dictionary between all basis-MGFs of total modular weight at most 12 and iterated Eisenstein integrals is established.
- The number of iterated Eisenstein integrals — and hence MGFs at a certain modular weight — are counted. In this way, basis dimen-

sions for MGFs are obtained explicitly for total modular weight at most 14, imaginary cusp forms can be counted independently. This confirms the bases found explicitly in [II] and ensures that they do not have to be extended for higher-point graphs.

- The structure of the solution shows a close similarity to the corresponding solution in the open string, paving the way towards an explicit understanding of the single-valued map acting on iterated Eisenstein integrals [17].

## 1.2 OUTLINE

This thesis is structured as follows: Chapter 2 contains a brief general introduction into string theory and in particular into string perturbation theory. The state of the art of this field is summarized in Section 2.3.2. The single-valued map at tree-level, introduced above, is presented in more detail in Section 2.4.

Chapter 3 specializes the discussion from the previous chapter to the case of closed-string one-loop amplitudes. How the modular group arises out of the symmetries of the string is explained in Section 3.1, Section 3.2 discusses the computation of the CFT correlator, focusing mostly on the case of four graviton scattering in type-IIB string theory. Modular graph forms are introduced in Section 3.3, together with a literature review on the topic. The integral over the modular parameter $\tau$ is briefly discussed in Section 3.4.

Chapter 4 gives a concise introduction to the objects appearing in the calculation of open-string amplitudes at genus one, to serve as a reference for the discussion of the genus-one single-valued map in the following chapters.

Chapter 5 discusses MGFs in great detail and focuses in particular on the derivation of identities between MGFs. Section 5.1 contains a brief overview of the `Modular Graph Forms` `Mathematica` package, which implements the techniques discussed in this chapter and contains in particular the basis decompositions for dihedral and trihedral graphs of total modular weight at most 12. A complete reference of all functions and symbols defined in the package is provided in Appendix A. Chapter 5 covers the material published in [II] as well as the material published in [I] in the Sections 5.4.2 and 5.4.3. These sections have extensive text overlap with the reference.

In Chapter 6 we will apply the techniques for simplifications of MGFs obtained in the previous chapter to the case of four-gluon scattering at genus one in the heterotic string. We will start in Section 6.1 by explaining how the kinds of integrals introduced in Chapter 3 appear in the evaluation of the CFT correlator of the vertex operators in the heterotic string. In Section 6.2 we will evaluate these integrals using the techniques discussed in Chapter 5, decompose them into building blocks of uniform transcendentality and perform the integral over $\tau$ at leading low-energy orders. We will then use these results in Section 6.3 to extend the proposal from [34] for an elliptic single-valued map to the heterotic string and show which contributions can and cannot be reproduced in this way. The material in this chapter was published in [III] and Chapter 6 has extensive text overlap with the reference.

Chapter 7 introduces the generating function for closed-string integrals and its differential equations. In Section 7.1, we will define the generating function and compute its expansion in terms of MGFs for some two- and three-point instances. In Section 7.2, we will derive some necessary identities and determine the Cauchy–Riemann and Laplace equations for the generating series for two points. We will also discuss the implications of these equations for MGFs. In Section 7.3 we will then derive the general $n$-point Cauchy–Riemann equation and specialize it to three and four points. In Section 7.4, we will then compute the Laplace equation at $n$-points and discuss some special cases at three- and $n$-points. This material was published in [IV] and Chapter 7 has extensive overlap with the reference.

In Chapter 8 we will discuss the solution of the differential equation introduced in the previous chapter in terms of iterated Eisenstein integrals. To this end, in Section 8.1, we will rewrite the generating function defined in Chapter 7 to obtain a differential equation which is amenable to Picard iteration. In Section 8.2 we will then solve this differential equation perturbatively, obtaining a solution in terms of iterated Eisenstein integrals. We will study the two- and three-point instances of this solution in the Sections 8.3 and 8.4. In the final Section 8.5, we will discuss the modular properties of the iterated Eisenstein integrals and count the number of basis elements for modular graph forms of total modular weight at most 14. Furthermore, we will show that the generating series satisfies uniform transcendentality if its initial value does. The material presented in this chapter was published in [V] and has extensive text overlap with the reference.

Some concluding remarks and an outlook are given in Chapter 9. Several appendices contain complementary material, an index can be found on page 358.

# 2

# BACKGROUND

This thesis is concerned with the calculation of one-loop amplitudes in string perturbation theory. To set the scene, we will review the most important concepts relevant to string amplitude calculations in this chapter.

We will start in Section 2.1 with a general overview of bosonic- and superstring theory which will in particular introduce the string theories for which we calculate amplitudes in later chapters. In Section 2.2 we continue with a short review of the structure and importance of scattering amplitudes in field theories.

Section 2.3 discusses the general structure of scattering amplitudes in string theory, their different contributions and relative weights as well as what has been achieved so far in their calculation. In Section 2.3.3 we will review the important relation between field- and string theory amplitudes via the low-energy expansion of the latter. Finally, Section 2.4 discusses the so-called *single-valued map*, an important number-theoretic relation which can be used to map open- to closed string amplitudes at tree level. One of the primary goals of the research presented in this thesis is to extend this map from tree-level- to one-loop amplitudes.

## 2.1 STRING THEORY

String theory is a theory of fundamental interactions that tries (in its modern understanding) to unify the standard model of particle physics with general relativity. We will give here a very brief introduction into the field which is geared very much towards perturbative string theory, the main subject of this work. Good introductory textbooks include the two-volume book by Polchinski [42, 43], with an emphasis on branes, the two-volume book by Green, Schwarz and Witten [44, 45] which uses the traditional Green–Schwarz formalism but puts more emphasis on perturbative string theory and the book by Blumenhagen, Lüst and Theisen [46] with a particularly thorough treatment of conformal field theory.

### 2.1.1 *Bosonic Strings*

In string theory, zero-dimensional fundamental particles (objects with a one-dimensional worldline) are replaced by one-dimensional strings (objects with a two-dimensional *worldsheet*). One distinguishes *open* strings with two endpoints and *closed* strings that form a loop. In open string theories, the worldsheet can have boundaries, in closed string theories, boundaries are forbidden. Hence, open theories always also contain closed strings. If we allow for worldsheets which are non-orientable as two-dimensional surfaces, the theory is called *unoriented*.

String theory is formulated in terms of a two-dimensional conformal field theory (CFT) on the worldsheet for the field $X : \mathbb{R}^2 \longrightarrow \mathbb{R}^D$ that describes the embedding of the worldsheet into the $D$-dimensional ambient spacetime. For the worldsheet, we will use coordinates $\sigma^0, \sigma^1$ with Latin indices and signature $(-,+)$ and for spacetime we use Greek indices and the mostly-plus signature $(-,+,\dots,+)$. The *Nambu-Goto action* for this field is proportional to the volume of the worldsheet,

$$S_{\mathrm{NG}}[X] = -T \int \mathrm{d}^2\sigma \sqrt{-\det h} \tag{2.1}$$

where $h$ is the pull-back under $X$ of the spacetime metric $G$,

$$h_{ab}(\sigma) = \frac{\partial X^\mu}{\partial \sigma^a}\frac{\partial X^\nu}{\partial \sigma^b} G_{\mu\nu}(X(\sigma)) , \quad a, b = 0, 1 , \tag{2.2}$$

and $T$ is the *string tension*. $T$ is the only free parameter in string theory and we will write it as

$$T = \frac{1}{2\pi\alpha'} , \tag{2.3}$$

where $\alpha' = l_s^2$ is the square of the natural length scale of string theory. We set $\hbar = c = 1$ in this thesis.

To quantize (2.1), one considers the equivalent *Polyakov action*,

$$S_{\mathrm{Poly}}[X, \gamma] = -\frac{1}{4\pi\alpha'} \int \mathrm{d}^2\sigma \sqrt{-\det\gamma}\, \gamma^{ab} \partial_a X^\mu \partial_b X^\nu G_{\mu\nu} \tag{2.4}$$

in which the worldsheet metric is promoted to an independent field $\gamma$, at the expense of introducing additional gauge symmetries on top of the diffeomorphism symmetries already present in the Nambu–Goto action. The Polyakov action takes the form of a non-linear sigma model with spacetime being the target space.

The gauge symmetries of the Polyakov action (2.4) are diffeomorphisms and Weyl rescalings of $\gamma$. The Weyl symmetry develops an anomaly at the quantum level and requiring this to vanish fixes the dimensions of spacetime to $D = 26$. Furthermore, we can Wick rotate to the coordinates $(\sigma^1, \sigma^2) = (\sigma^1, i\sigma^0)$ and use these gauge freedoms to locally make the metric flat, $\gamma_{ab} = \delta_{ab}$, leaving a residual conformal symmetry.[1] This is how conformal field theory enters into string theory. The action (2.4) then becomes

$$S_{\text{Poly}}[X] = \frac{1}{\pi\alpha'}\int \mathrm{d}^2 z\, \partial X^\mu \bar{\partial} X_\mu\,, \tag{2.5}$$

where we have introduced the complex coordinates $z = \sigma^1 - i\sigma^2$ and $\bar{z} = \sigma^1 + i\sigma^2$ and their derivatives $\partial = \frac{1}{2}(\partial_1 + i\partial_2)$ and $\bar{\partial} = \frac{1}{2}(\partial_1 - i\partial_2)$. The equation of motion of (2.5) is $\partial\bar{\partial}X = 0$ and implies that $X$ can be decomposed into holomorphic and antiholomorphic left- and right-mover $X_L(z)$ and $X_R(\bar{z})$,

$$X(z,\bar{z}) = X_L(z) + X_R(\bar{z})\,. \tag{2.6}$$

For open strings, we can impose Dirichlet or Neumann boundary conditions at the endpoints,[2] effectively removing either left- or right-movers. Since for Dirichlet boundary conditions momentum can flow off the end of the string, this endpoint needs to be attached to a dynamical higher-dimensional object, a *D-brane*. Therefore, string theory is actually a theory of strings and branes and since D-branes are non-perturbative objects, they can be thought of as the instantons or solitons of string theory.

Consider now the quantization of the Polyakov action (2.5). Fixing the gauge in the path integral requires a transformation of the field variables whose Jacobian is captured by introducing (Grassmannian) Faddeev–Popov ghost fields $b$ and $c$ with action

$$S_{\text{ghosts}}[b,c] = \frac{1}{\pi}\int \mathrm{d}^2 z \left(b\bar{\partial}c + \bar{b}\partial\bar{c}\right)\,. \tag{2.7}$$

In order to quantize the string, we expand $X$ into modes,

$$X^\mu = x^\mu - i\frac{\alpha'}{2}p^\mu \log|z|^2 + i\left(\frac{\alpha'}{2}\right)^{1/2} \sum_{m\in\mathbb{Z}\setminus\{0\}} \frac{1}{m}\left(\frac{\alpha_m^\mu}{z_m} + \frac{\tilde{\alpha}_m^\mu}{\bar{z}^m}\right)\,, \tag{2.8}$$

where $x^\mu$ and $p^\mu$ are the position and momentum of the center of mass of the string and $\alpha_m^\mu$ and $\tilde{\alpha}_m^\mu$ are raising and lowering operators for the left- and right-mover satisfying the commutation relations

$$[\alpha_m^\mu, \alpha_n^\nu] = [\tilde{\alpha}_m^\mu, \tilde{\alpha}_n^\nu] = m\delta_{m,-n}\eta^{\mu\nu}\,. \tag{2.9}$$

---

1 A conformal transformation is a diffeomorphism which locally rescales the metric, $\gamma \to \Omega(x)\gamma$. Since this can be compensated by a Weyl transformation $\gamma \to \Omega^{-1}(x)\gamma$, conformal transformations preserve the gauge condition $\gamma = \delta$ and hence are a residual gauge freedom.

2 Dirichlet boundary conditions fix the endpoint to a certain spacetime coordinate, $\delta X^\mu = 0$, Neumann boundary conditions set the derivative to zero, $\partial_1 X^\mu = 0$. They can be specified independently for both endpoints and for every direction.

The spacetime spectrum of the string is obtained by acting with the raising operators $\alpha^\mu_m$ and $\tilde{\alpha}^\mu_m$ with $m < 0$ on the ground state $|0; p\rangle$ defined by $\alpha^\mu_m|0; p\rangle = \tilde{\alpha}^\mu_m|0; p\rangle = 0$ for $m > 0$ (the lowering operators). Note that the ground state also carries the momentum $p^\mu$ of the string.

However, the Fock space generated in this way contains unphysical negative norm states related to the minus sign in the Minkowski metric in (2.9). Gauge invariance at the quantum level implies that the energy–momentum tensor acting on physical states must vanish, the so-called *Virasoro constraints*. In particular, this means that physical states satisfy

$$M^2 = \frac{4}{\alpha'}(N-1) = \frac{4}{\alpha'}(\tilde{N}-1)\,, \tag{2.10}$$

where $M$ is the mass of the state, $N$ is the number of left-movers (the sum $m_1 + \cdots + m_l$ for the state $\alpha^{\mu_1}_{-m_1} \ldots \alpha^{\mu_l}_{-m_l}|0; p\rangle$) and $\tilde{N}$ is the number of right-movers. The condition $N = \tilde{N}$ is called *level matching*. Furthermore, the Virasoro constraints imply that negative norm states vanish, however, the theory still contains null states. Modding out by them implies that we should remove two degrees of freedom from each oscillator, so $\alpha^\mu_m \to \alpha^i_m$ for $i = 2, \ldots, 25$ and similarly for $\tilde{\alpha}$. In order to find the spacetime spectrum, the representation of the little group, under which the state transforms, has to be decomposed into irreducible representations.

For the closed bosonic string, we find that the ground state is a scalar with negative $M^2$ (a *tachyon*). The first excited states are massless and form a $24 \times 24$ matrix which decomposes into a traceless, symmetric part (the graviton), an anti-symmetric part (the *B-field*) and a trace (the *dilaton*). Since the massless modes mediate long-range forces, the full Polyakov action should also contain couplings of the string to them.

For the open string, we have just one set of raising and lowering operators and physical states satisfy

$$M^2 = \frac{1}{\alpha'}(N-1)\,. \tag{2.11}$$

Also for the open string, the vacuum state is a tachyon and the first excited states form a massless gauge field. On top of the massless states, there is an infinite tower of massive higher-spin states in the spectrum of all open- and closed string theories.

The tachyonic states in the spectrum of the bosonic string are a serious problem and can be interpreted as an instability of the theory. This instability has to be studied in the context of *string field theory*, a second quantized formulation of string theory in which the string itself is quantized and not just the vibration modes (for a review, see [47]). In this context, the condensation of the tachyon in the bosonic string has been studied [48], but the the ultimate fate of the instability remains unknown.

### 2.1.2 *Superstrings*

One way to remove the tachyon from the spectrum and to include fermions into it, is to supersymmetrize the Polyakov action (2.5) to obtain a superconformal worldsheet action whose gauge fixed form is (in closed-string normalization)

$$S[X,\psi,\bar{\psi}] = \frac{1}{2\pi}\int \mathrm{d}^2 z\left(\frac{2}{\alpha'}\partial X^\mu \bar{\partial} X_\mu + \psi^\mu \bar{\partial}\psi_\mu + \bar{\psi}^\mu \partial \bar{\psi}_\mu\right), \tag{2.12}$$

where $\psi^\mu(z)$, $\bar{\psi}^\mu(\bar{z})$ are anticommuting worldsheet fields. Requiring (2.12) to be invariant under $z \longrightarrow e^{2\pi i}z$ leaves two possible transformation properties for the fields $\psi$, $\bar{\psi}$ and we have to distinguish between *Ramond* (R) fields satisfying $\psi^\mu(e^{2\pi i}z) = -\psi^\mu(z)$ and *Neveu–Schwarz* (NS) fields satisfying $\psi^\mu(e^{2\pi i}z) = \psi^\mu(z)$.[3] The modes of a Ramond field give rise to spacetime fermions, the modes of a Neveu–Schwarz field correspond to spacetime bosons.

In order to gauge fix the path integral, we have to include (commuting) Faddeev–Popov superghosts $\beta, \gamma$ for the fermionic fields. Their action is

$$S_{\text{superghosts}}[\beta,\gamma] = \frac{1}{\pi}\int \mathrm{d}^2 z\left(\beta\bar{\partial}\gamma + \bar{\beta}\partial\bar{\gamma}\right). \tag{2.13}$$

Upon quantizing this theory, one obtains a spacetime spectrum which includes a tachyon, bosonic and fermionic particles and is not supersymmetric. However, imposing gauge invariance of the one-loop amplitude introduces the *Gliozzi–Scherk–Olive* (GSO) *projection* which renders the spacetime spectrum supersymmetric and removes the tachyon. The vanishing of the Weyl anomaly requires $D = 10$ for all superstring theories.

The formulation of the superstring above, in which worldsheet supersymmetry is manifest but spacetime supersymmetry is not, is the so-called *Ramond–Neveu–Schwarz* (RNS) superstring. There are two other formulations however, the *Green–Schwarz* superstring [49] and the *pure-spinor* superstring [50] which both break manifest worldsheet supersymmetry but have manifest spacetime supersymmetry. Both the RNS and the pure-spinor superstring manifest spacetime Lorentz-symmetry, whereas the Green–Schwarz superstring was so far only quantized in light-cone gauge, which breaks spacetime Lorentz-symmetry. All formulations lead to the same physical results.

Using the actions (2.12) and (2.5), five inequivalent consistent superstring theories can be constructed. These are listed in the following:

3 Antiperiodicity of $X$ breaks spacetime Poincaré invariance and appears for twisted strings on an orbifold, but we will not consider it here.

**TYPE-IIA AND TYPE-IIB** These are theories of closed, oriented strings with worldsheet action (2.12). For left- and right-movers we have an R and NS sector each, leading to the four sectors NS-NS, R-R (both bosonic) and R-NS, NS-R (both fermionic). In the type-IIB theory the chiralities of left- and right moving R fields are aligned, in the type-IIA theory they are opposite. These theories have $\mathcal{N} = 2$ (maximal) spacetime supersymmetry in ten dimensions.

**HETEROTIC SO(32) AND HETEROTIC $E_8 \times E_8$** The heterotic string is a hybrid of a right-moving ten-dimensional superstring and a left-moving 26-dimensional bosonic string whose target space is the product of the ten-dimensional spacetime and a 16-dimensional internal torus (it is *compactified* on $T^{16}$). Since $T^{16}$ is a compact space, the momenta in these directions live on a lattice which is fixed by gauge invariance of the one-loop amplitude to be the root lattice of either $E_8 \times E_8$ or SO(32), giving rise to gauge fields with these gauge groups in spacetime. The heterotic theories have $\mathcal{N} = 1$ (half-maximal) spacetime supersymmetry in ten dimensions.

**TYPE-I SO(32)** The type-I theory is a theory of open and closed oriented and unoriented strings which has $\mathcal{N} = 1$ supersymmetry in ten dimensions. This theory also includes 32 spacetime-filling D-branes which implies an SO(32) gauge field in the spectrum of the open string. The gauge group is fixed by requiring the vanishing of gauge- and diffeomorphism anomalies [51].

According to (2.10), the masses of the massive string states are set by the *string scale* $\alpha'^{-1/2}$ which, for quantum gravity, is set to the Planck scale ($10^{19}$ GeV).[4] At energies much lower than this, only the massless modes of the string are relevant and an effective QFT for these modes can be derived. The low-energy effective actions of the five theories above are exactly the five consistent supergravity actions in ten dimensions.[5] More details on obtaining these effective actions from the calculation of scattering amplitudes in string theory can be found in Section 2.3.3.

The superstring theories are not completely independent however, but are conjectured to be connected by a web of (non-perturbative) dualities called S- and T-duality which also link them to the putative M-theory in eleven dimensions whose low-energy limit is eleven-dimensional supergravity.

---

4 Although the string scale is traditionally set to $10^{19}$ GeV, lower string scales in the TeV range were also considered [52, 53].

5 The anomaly cancellation for ten-dimensional supergravity theories with half-maximal supersymmetry also allows for the gauge groups $U(1)^{496}$ and $E_8 \times U(1)^{248}$ [51], but no string theories with these gauge groups are known. However, it was shown that these supergravity theories are inconsistent at the quantum level [54, 55].

### 2.1.3 *String theory as a unifying theory of physics*

Although string theory was historically invented to describe strong interactions (for a review, see [56]), it is nowadays primarily understood as a unifying theory of quantum gravity. This is because the spectrum of all string theories contains gravitons, quantized perturbations of spacetime. Since the spectra of the known superstring theories contain also scalars, gauge bosons and fermions, string theory has the potential to unify also the other forces in the standard model into one coherent picture.

One drawback of the string theory approach to quantum gravity is its background dependence: We have to couple the Polyakov action to a fixed background on top of which the graviton modes of the string propagate. This is remedied somewhat by the interpretation of the coupling to the spacetime metric as vertex operators for a coherent state of gravitons [57]. In string field theory, this problem is overcome [58–60].

The central difficulty in the quest for quantum gravity is that naive quantizations of general relativity are perturbatively non-renormalizable. String theory finds a beautiful cure for this problem: Intuitively, since the string has a non-zero extension, it cannot form an infinitely tight loop. Indeed, loop amplitudes in string theory are proven to be ultraviolet (UV) finite [61–63] and all known amplitudes are also infrared (IR) finite. For more details, cf. Section 2.3.1.

Unfortunately, the superstring theories listed above are only consistent in ten spacetime dimensions. A possible solution for this problem is given by *Kaluza–Klein compactification*, in which the spacetime manifold is assumed to be a product of a four-dimensional non-compact manifold and a six-dimensional compact manifold and the fields are Fourier expanded in the compact directions. If the radii of these compact directions are chosen small enough, the masses associated to the non-constant Fourier modes become large and at low energies, one obtains an effective four-dimensional theory. For the superstring, requiring the non-compact theory to retain quarter-maximal supersymmetry in four dimensions leads to the condition that the compact submanifold should be a *Calabi–Yau* manifold [64] of complex dimension three, a compact, Ricci-flat Kähler manifold with vanishing first Chern class. Since the effective theory in the non-compact directions depends on the shape of the compact submanifold, the possible four-dimensional theories correspond to the possible Calabi–Yau manifolds.

In one complex dimension, only tori are Calabi–Yau. In two complex dimensions, the only simply connected Calabi–Yau manifolds are *Kummer–Kähler–Kodaira* (K3) *manifolds*, (for a review, see [65]). In three complex dimensions, the case relevant to realistic string compactifications, the problem of classifying Calabi–Yau manifolds is unsolved and a very large number of infinite families is known. This is referred to as the *string landscape*, the study of possible four-dimensional effective theories coming out of string theory, for a recent review, see [66].

### 2.1.4 *String theory and mathematics*

Apart from its relevance to quantum gravity research, string theory has led to important insights in pure mathematics that were later proven rigorously.

The most well-known of these is concerned with the compactification of string theories on Calabi–Yau manifolds as discussed in the last section. Based on the observation that string theories compactified on different Calabi–Yau manifolds can lead to the same low-energy physics, string theorists found that Calabi–Yau manifolds can be organized in pairs, a property called *mirror symmetry*. This statement was linked to the counting of rational curves on a Calabi–Yau manifold [67], conjecturing many new results in enumerative geometry. Mirror symmetry can be formulated rigorously in the context of topological string theory [68], a simplified version of string theory and was proven in this context [69].

Another area, in which string theory has led to important insights is the interplay of modular functions and representations of finite groups: In the context of string theory, a surprising identification of expansion coefficients of the *Klein j-function* (a modular function) and the dimensions of the irreducible representations of the largest sporadic group, the *monster group* was observed [70] and is known as *monstrous moonshine*. This identification could be proven rigorously using techniques from string theory [71] and extended to the *Mathieu group*, another sporadic group [72].

Finally, a further area with fruitful interplay between string theory and mathematics is the study of automorphic forms, as summarized in the comprehensive textbook [73]. In the study of string amplitudes, requiring invariance of coefficients of operators in the low-energy effective action under the non-perturbative string dualities together with supersymmetry fix these coefficients to be automorphic forms of the background fields. In some cases, this is restrictive enough to fix them uniquely, yielding a fully non-perturbative expression [8]. Expanding the automorphic form in the parameters results in predictions for all perturbative- and instanton-contributions to the part of the amplitude in question.

Also the topic of this work, the study of modular graph forms, lies at the intersection of string theory and mathematics: On the one hand, modular graph forms are a tool to obtain closed-string one-loop amplitudes, as will be detailed in Chapter 3, on the other hand, they are also a fascinating class of (non-holomorphic) modular forms which has been studied by mathematicians purely for their interesting number-theoretic properties, cf. Section 3.3.3. In particular, the approach of studying modular graph forms from the perspective of iterated

Eisenstein integrals as detailed in Chapter 8, has also been followed in mathematics [32, 33, 74].

## 2.2 SCATTERING AMPLITUDES IN FIELD THEORY

Scattering amplitudes describe the interaction probability of particles and are the most important observables in quantum field theory. They are measured in collider experiments and have been studied since the birth of the field.

Quantum fields can be interpreted in a statistical way: The path integral integrates over all possible field configurations, weighted by the exponential of the action $S$ so that the partition function is given by

$$Z = \int \mathcal{D}\phi \, e^{iS[\phi]} \, . \tag{2.14}$$

In this framework, a two-point correlation function between the field $\phi$ at points $x$ and $y$ in spacetime is given by

$$\langle \phi(x)\phi(y) \rangle = \int \mathcal{D}\phi \, \phi(x)\phi(y) e^{iS[\phi]} \tag{2.15}$$

and computes the statistical correlation between the two points of the field.

A scattering amplitude $\mathcal{A}(k_1, \ldots, k_n)$ is a transition function $\langle i|f\rangle$ between an initial state $|i\rangle$ of a $r$ particles which at $t \to -\infty$ are localized at infinity with momenta $k_1, \ldots, k_r$ and a final state $|f\rangle$ of $n-r$ particles localized at infinity with momenta $k_{r+1}, \ldots, k_n$ at $t \to \infty$,

$$\mathcal{A}(k_1, \ldots, k_n) = \langle i|f\rangle \, . \tag{2.16}$$

Note that if the asymptotic particles carry spin, the amplitude also depends on the polarization tensors of those particles. An amplitude with $n$ external states is often referred to as an *$n$-point amplitude*. The initial and final states correspond to field configurations $\phi_i$ and $\phi_f$ at $t \to \mp\infty$ which solve the classical equations of motion. Hence, the transition function (2.16) is given by

$$\langle i|f\rangle = \frac{1}{Z} \int_{\phi_i}^{\phi_f} \mathcal{D}\phi \, e^{iS[\phi]} \, , \tag{2.17}$$

where the limits of the integral instruct to integrate only over those field configurations which are asymptotically $\phi_i$ and $\phi_f$ at $t \to \mp\infty$. Since it is hard to compute the path integral (2.17) with non-trivial boundary conditions in practice, one uses the LSZ reduction formula to express $\langle i|f\rangle$ as a Fourier transform of an $n$-point correlation function of the form (2.15).

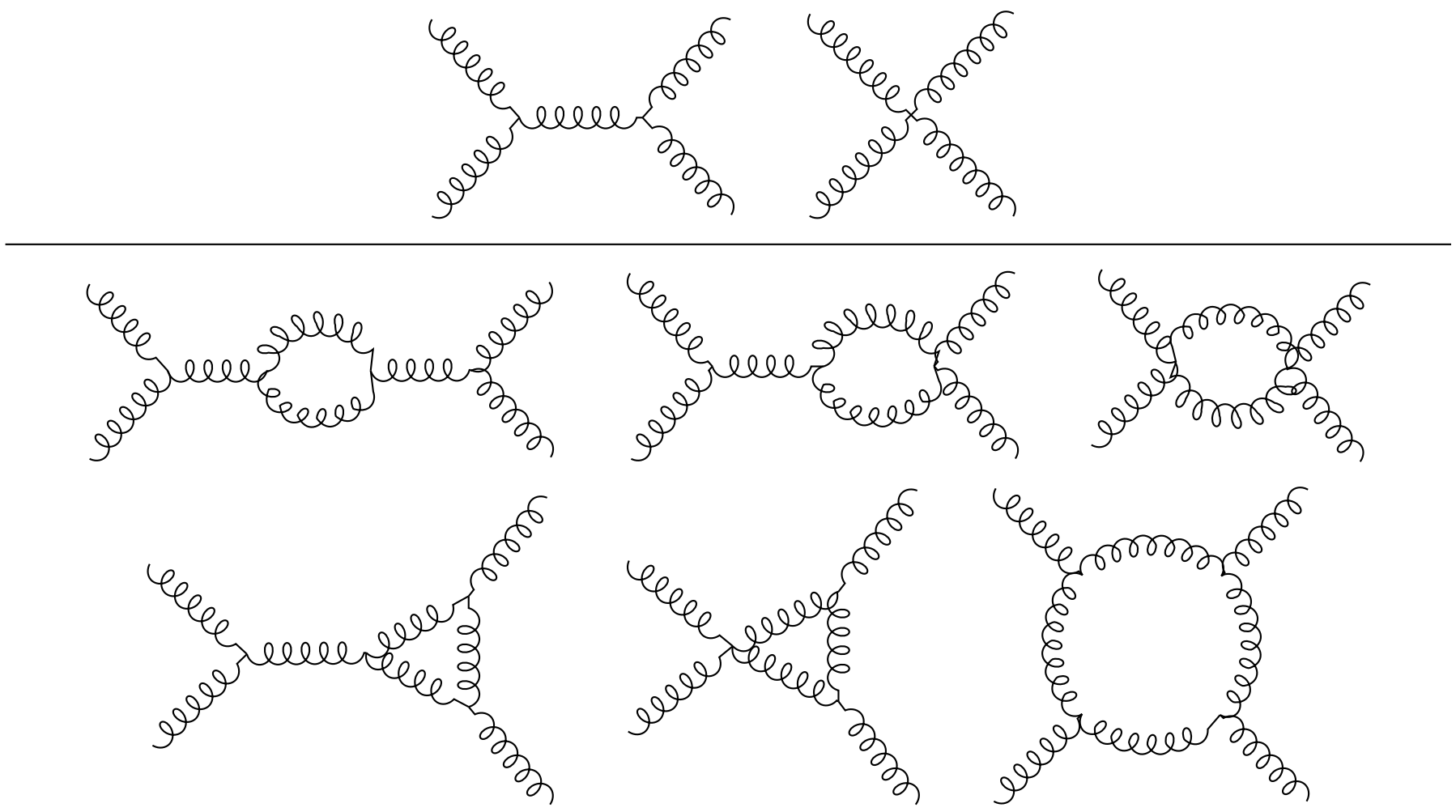

Figure 2.1: Tree-level (top) and one-loop (bottom) diagrams of the four-point Yang-Mills amplitude. For all diagrams, one has to sum over the ways to assign the external particles to the external edges.

In scattering experiments, typically the momenta of the incoming particles are fixed by the beam line, but particles coming out of the scattering process are measured in (almost) every direction. Therefore, to get the total interaction probability, or cross-section, we integrate the absolute value squared of (2.16) over the outgoing momenta $k_{r+1}, \dots, k_n$.

### 2.2.1 *Feynman diagrams*

Computing path integrals of the form (2.15) is very hard in practice and so in almost all cases, we rely on an expansion[6] of the amplitude around the non-interacting theory, treating the interactions of the field with itself and other fields as small perturbations whose strength is parametrized by the coupling constant $g \ll 1$. This expansion can be organized in terms of Feynman diagrams which are translated into contributions to the amplitude by the means of Feynman rules that are derived from the action. In order to obtain the contribution at a certain order in $g$, we draw all diagrams compatible with the asymptotic states which contribute at this power of $g$. The diagrams have a very natural interpretation in terms of particle interaction processes: The lines can be thought of as particles moving in spacetime and the vertices as particles interacting. Finally, all relevant diagrams have to be summed up to obtain (2.16). As an example, Figure 2.1 shows the diagrams contributing to the four-point amplitude in Yang-Mills theory at the first two orders.

6 Strictly speaking, the resulting series in the coupling constant is not convergent, but asymptotic instead (c.f. e.g. [75] for an early account of this in quantum electrodynamics). Since usually, only very few orders in $g$ are computed, this is not a problem in practice.

Since the external states are fixed, as the order of $g$ grows, more loops are added to the diagram, as can be seen in Figure 2.1. The number of loops is therefore used to classify the diagrams, with a diagram without loops referred to as *tree-level*. As is also exemplified in Figure 2.1, the number of diagrams grows factorially with the order of $g$. The calculation of field theory amplitudes from evaluating Feynman diagrams is a large research area on its own, two textbooks that summarize many of the techniques are [76] by Henn and Plefka and [77] by Elvang and Huang.

Each edge in the Feynman diagram is associated a momentum and momentum conservation is imposed at the vertices. Since this leaves the momenta running around in the loops undetermined, the Feynman rules instruct to integrate over them. These integrals can be very hard to evaluate and a lot of effort has gone in the last years into performing these. Recently, interesting classes of higher-loop Feynman integrals have been found to be expressible in terms of elliptic multiple polylogarithms [78, 79], exactly the objects which are used to expand open-string one-loop integrals, as reviewed in Chapter 4.

### 2.2.2 *Amplitudes without Feynman diagrams?*

As mentioned in the last section, the number of Feynman diagrams grows factorially with the loop order. At the same time, the only physically meaningful quantity is the sum of all diagrams. As can be seen in the following example, there are often cancellations in those sums that make the end result surprisingly simple.

Consider a tree-level $n$-gluon amplitude in four spacetime dimensions with the helicities of the external particles all $+$, apart from two particles with helicity $-$. Then, the amplitude is given by the Parke–Taylor formula [80]

$$\mathcal{A}^{\text{tree}}_{n\text{-point}} = \frac{\langle ij\rangle^4}{\langle 12\rangle\langle 23\rangle\dots\langle (n-1)n\rangle\langle n1\rangle}\,, \tag{2.18}$$

where $i$ and $j$ are the particles with helicity $-$ and the dependence on the external momenta and polarizations is expressed using the spinor-helicity formalism as reviewed e.g. in [81]. Results like this show that Feynman diagrams obscure a lot of the hidden simplicity in amplitudes and have motivated the search for ways to compute amplitudes without relying on Feynman diagrams. This has led to powerful methods for the computation of and elegant geometric descriptions for certain gauge-theory amplitudes, for instance Britto–Cachazo–Feng–Witten (BCFW) recursions [82, 83] and the amplituhedron [84]. In particular, the Cachazo–He–Yuan (CHY) formalism [85–87] and ambitwistor strings [88–90] elucidate the structure of $D$-dimensional field-theory amplitudes and make use of worldsheet methods similar to those in this thesis.

In theories of gravity, amplitudes are perturbative expansions around a fixed background geometry. For instance, for a Minkowski background, the metric tensor is written as $g_{\mu\nu} = \kappa h_{\mu\nu} + \eta_{\mu\nu}$. This is plugged into the action and the Feynman rules for $h_{\mu\nu}$ are computed. For the standard Einstein–Hilbert term, this leads to an infinite number of vertices with arbitrary multiplicities, implying a particularly rapid growth of the number of diagrams with the powers of the coupling constant. Furthermore, these vertices are much more complex than gauge theory vertices, e.g. the three-graviton vertex has in the standard de Donder gauge about 100 terms. These properties of the gravity Feynman rules mean that a direct computation via Feynman diagrams becomes quickly unfeasible.

A very successful line of research to circumvent this problem goes by the names *Bern–Carrasco–Johansson* (BCJ) *duality*, *color-kinematics duality* and *double copy* [91–93] . The key finding is that if the kinematical numerators of the integrand of a gauge theory amplitude are written in such a way that they satisfy the same relations as their color counterparts, then replacing the color factors by a second copy of kinematical numerators transforms the integrand into an integrand of a gravity theory. These results can be proven at tree-level by taking the $\alpha' \to 0$ limit of certain relations between scattering amplitudes in string theory: That kinematical numerators satisfy gauge-theory relations follows from studying string amplitudes in the pure-spinor formalism [5] and the double copy can be proven from the KLT formula in string theory [2] as explained in Section 2.3.3. At loop level, the BCJ duality is conjectural. Although the underlying reason for the BCJ duality is unknown, it is currently the main tool to calculate higher-loops amplitudes in gravity, as e.g. the calculation of a five-loop amplitude in supergravity [94] demonstrated, since the number of diagrams in gauge theory are considerably lower than the number of diagrams in gravity.

## 2.3 SCATTERING AMPLITUDES IN STRING THEORY

As in field theory, also in string theory scattering amplitudes can be computed as an expansion in the coupling constant. However, there are two conceptual differences between amplitudes in string theory and amplitudes in field theory: First, the string coupling $g_s$ is not an independent parameter, but related to the vacuum expectation value $\phi_0$ of the dilaton by $g_s = e^{\phi_0}$ and second, the standard formulation of string theory is only first quantized, i.e. there are no off-shell amplitudes and no correlation functions for the strings themselves, a problem that is addressed in string field theory. In this section, we will take a high-level view at the process of calculating string amplitudes without going into much detail.

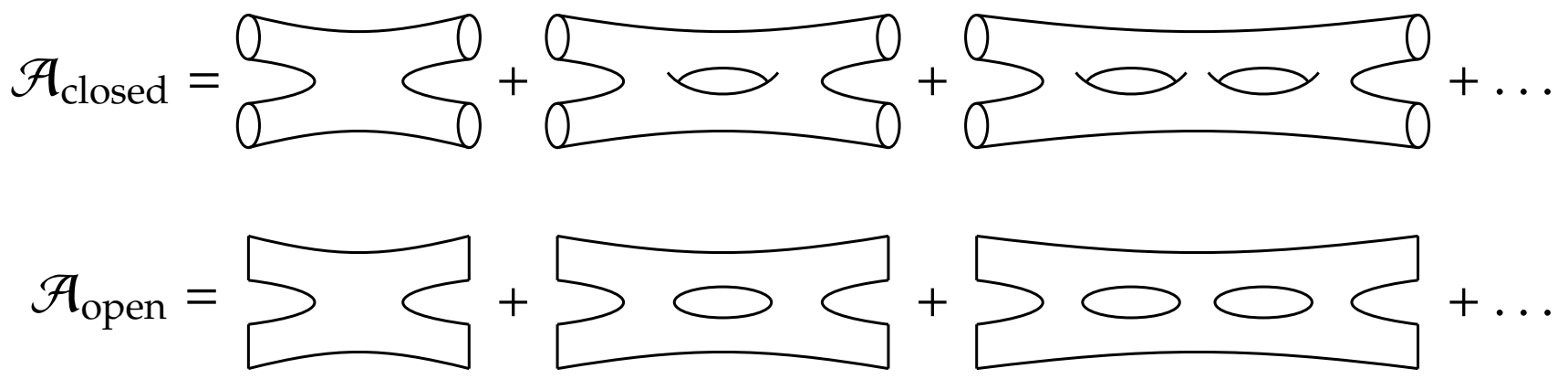

Figure 2.2: Sum over topologies for oriented closed (top) and open (bottom) string four-point amplitude.

### 2.3.1 *Structure of string amplitudes*

The asymptotic states for which we want to calculate a scattering amplitude live in string theory at the end of infinitely long worldsheets, namely strips for open strings and tubes for closed strings. In the spirit of "summing over possibilities", we have to sum over all worldsheets allowed in the considered theory, that connect these asymptotic pieces. Focusing for a moment on closed oriented strings, the relevant worldsheets are connected, compact, closed surfaces, which are classified by their genus (number of handles). Hence, we obtain a sum over genera, as depicted in Figure 2.2 . Since the dilaton contribution to the worldsheet action is just $\phi\chi$, where $\chi = 2(1-g)$ is the Euler characteristic of the genus $g$ worldsheet, in the path integral, this worldsheet is weighted by a factor $e^{-2\phi_0(1-g)} = g_s^{-2+2g}$. Hence, as long as $g_s \ll 1$, the genus expansion is a perturbative expansion in the string coupling where the genus corresponds to the loop order.

For each genus, the string amplitude is given as a path integral over the worldsheet metric $\gamma$ and the embedding field $X$ in the Polyakov action (2.4),

$$\frac{1}{\text{vol}(\text{diff} \times \text{Weyl})} \int \mathcal{D}X\mathcal{D}\gamma\, e^{-S_{\text{Poly}}[X,\gamma]}\,, \tag{2.19}$$

where the integral is taken over a worldsheet with the desired asymptotic states at the boundaries and we have divided by the volume of the gauge group to account for the gauge freedom. Note that this expression does not explicitly include fermionic fields, but this and the following arguments in this section apply equally to the superstring.

In order to compute (2.19), we use the conformal symmetry of the action to pull the infinite stretches of the worldsheet back to the compact surface in the center and replace the asymptotic states by insertions of appropriate operators on the worldsheet, which now becomes a punctured Riemann surface, as depicted in Figure 2.3. These operators are called *vertex operators* and are obtained from the CFT operator–state correspondence. Open-string vertex operators are inserted on the boundary of the worldsheet and closed-string operators in the bulk.

The path integral over $X$ in (2.19) then becomes a CFT correlator of the vertex operators and after fixing the remaining diffeomorphism ×

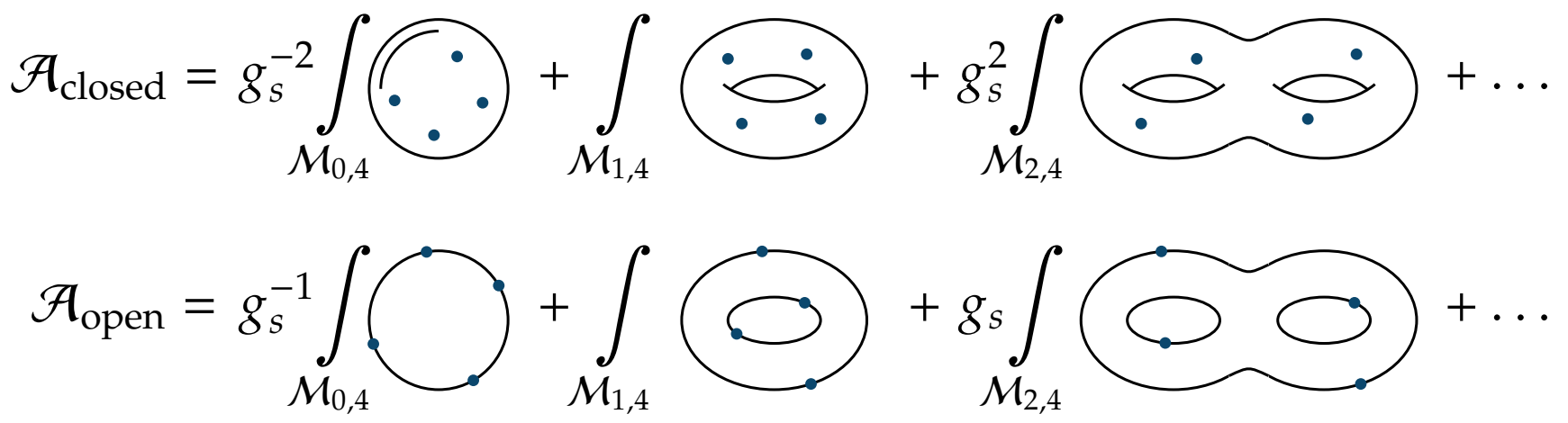

Figure 2.3: Expansion of oriented four-point open and closed string amplitude in terms of integrals over moduli spaces of punctured Riemann surfaces. In the open string, the integral over the moduli space implicitly contains a sum over all distributions of the punctures over the boundaries and their permutations. At each order in $g_s$, only one contribution is shown and we have for instance suppressed the Möbius strip at genus one. We have also absorbed the powers $g_s^{n_c}$ and $g_s^{n_o/2}$ into the vertex operators, cf. (2.21).

Weyl gauge freedom, the integral over the metric $\gamma$ becomes an integral over the moduli space $\mathcal{M}_{g,n}$ of the genus $g$ Riemann surface with $n$ marked points. Fixing the gauge cancels the volume of the gauge group in (2.19) and introduces ghosts, as mentioned in Sections 2.1.1 and 2.1.2.

When we are considering general string interactions involving open and closed strings on a possibly non-orientable worldsheet, we have to sum over compact, connected surfaces. These are characterized by their number of handles $g$, the number of boundaries $b$ and the number of cross-caps $c$ that one has to attach to the sphere to obtain them. The Euler characteristic is given by

$$\chi = 2 - 2g - b - c\,. \tag{2.20}$$

Adding a handle to the worldsheet decreases $\chi$ by 2 and corresponds to emission and absorption of a closed string, hence closed string vertex operators come with a factor of $g_s$. Adding a boundary decreases $\chi$ by one and corresponds to emission and absorption of an open string, hence, open string vertex operators come with a factor of $g_s^{1/2}$.

Putting everything together, the resulting expression has the form

$$\mathcal{A}_g(k_1,\dots,k_n) = g_s^{-\chi+n_c+\frac{1}{2}n_o}\int_{\mathcal{M}_{g,n}} d\mu_{g,n}\,\langle\prod_{i=1}^{n}\mathcal{V}_g^i(k_i,z_i)\text{ ghosts}\rangle\,, \tag{2.21}$$

where $n_o$ and $n_c$ are the number of open and closed vertex operators, $d\mu_{g,n}$ is the measure on $\mathcal{M}_{g,n}$, $k_1,\dots,k_n$ jointly denote momenta, polarizations and other data of the asymptotic states and $\mathcal{V}_g^i(k_i,z_i)$ is the genus $g$ vertex operator of the $i^{\text{th}}$ external state, inserted at position $z_i$. Note that the prefactors $g_s^{n_c}$ and $g_s^{n_o/2}$ are usually absorbed into the vertex operators, as in Figure 2.3. In (2.21), an integral over the (unfixed) insertion positions and a sum over the ways how to distribute the open string vertex operators over the boundaries and how to order them, is included in $d\mu_{g,n}$. The detailed structure of the ghost contribution depends on the external states and the genus of the worldsheet.

| closed/open | orientability | surface | $g$ | $b$ | $c$ | $\chi$ |
|---|---|---|---|---|---|---|
| closed | orientable | sphere | 0 | 0 | 0 | 2 |
| closed | orientable | torus | 1 | 0 | 0 | 0 |
| closed | non-orientable | real projective plane | 0 | 0 | 1 | 1 |
| closed | non-orientable | Klein bottle | 0 | 0 | 2 | 0 |
| open | orientable | disk | 0 | 1 | 0 | 1 |
| open | orientable | cylinder | 0 | 2 | 0 | 0 |
| open | non-orientable | Möbius strip | 0 | 1 | 1 | 0 |

Table 2.1: Tree-level and one-loop open and closed worldsheets of oriented and unoriented strings with $g$ handles, $b$ boundaries, $c$ cross-caps and Euler number $\chi$.

Note how peculiar (2.21) is: A spacetime amplitude in 10 or 26 dimensions is given in terms of a 2d CFT correlator. In particular, momentum conservation for the external momenta $k_i$ arises very indirectly from the zero mode integral of the worldsheet fields, as will be demonstrated in Section 3.2.1.

The tree-level and one-loop open and closed, orientable and non-orientable worldsheets are collected together with their number of handles $g$, boundaries $b$, cross-caps $c$ and Euler number $\chi$ in Table 2.1.

In the type-II and heterotic theories, only orientable worldsheets without boundaries are allowed and hence there are only closed, oriented strings, whose tree-level and one-loop contributions come from the sphere and torus, respectively.

In type-I theory, the worldsheets can be non-orientable and can have boundaries, hence all surfaces in Table 2.1 contribute. For closed string scattering, we have only the sphere at tree-level ($\chi = 2$), but at one-loop ($\chi = 0$) not only the torus, but also the Klein bottle, cylinder and Möbius strip contribute. Additionally, we have the real projective plane and the disk at "one-half-loop" with $\chi = 1$. For open strings in type-I, we have the disk at tree-level ($\chi = 1$) and the cylinder and Möbius strip at one-loop ($\chi = 0$).

As mentioned in Section 2.1.3, the UV finiteness of string perturbation theory is an important argument when considering string theory as a theory of quantum gravity and hence it is interesting to consider possible divergences in the expression (2.21). On the one hand, the integrand in (2.21) could have a singularity for some points of the moduli space. For unitary CFTs, correlation functions have no singularities and hence all possible singularities have to come from the non-unitary ghost CFT. Arguments for the absence of these singularities were given e.g. in [61–63] and a systematic procedure to avoid them is given in [95, 96]. On the other hand, the integral over $\mathcal{M}_{g,n}$ in (2.21) could diverge and although in all known examples the integral is finite, it is not known whether this is true in general. Divergences of the integral over $\mathcal{M}_{g,n}$ are due to degenerations of the Riemann surface and comparing the structure of (2.21) to field theory amplitudes shows that these degenerations correspond to IR divergences in the field theory language and similarly singularities of the integrand correspond to UV divergences [97]. In summary, string theory appears to be UV finite, while the question of IR finiteness is still open.

### 2.3.2 *State of the art in string amplitude calculations*

Although the structure of string amplitudes as written in (2.21) remains the same for all topologies and in particular the factorial growth of the number of Feynman diagrams known from field theory amplitudes is absent in string theory, in practice is often very hard to evaluate (2.21) explicitly. At the time of writing, only the tree-level case is fully under control, i.e. the $n$-point amplitude for arbitrary external states can be computed algorithmically in the type-I and II superstring [50, 98, 99] and in the bosonic and heterotic string [100].

At genus one, the CFT correlator of the vertex operators for massless external states has been computed until seven-points [51, 101–104]. Results for the CFT correlator of vertex operators for fermionic external states are available in [105]. The integral over the moduli space $\mathcal{M}_{1,n}$ for closed strings is the main focus of this work and is performed in terms of modular graph forms. The moduli space integral for open strings can be performed efficiently in terms of so-called *elliptic multiple zeta value* (eMZV), see Chapter 4 for a review.

At genus two, the CFT correlator was computed in terms of modular forms for up to four points in the RNS formalism [106–112] and the pure-spinor formalism [113, 114]. The low-energy limit of the genus-two five-point amplitude in type-II was calculated in [115] and checked against S-duality predictions. The techniques to calculate the moduli space integral for closed genus-one amplitudes in terms of modular graph forms were extended to genus two and some results obtained in certain limits [116–118].

At genus three, the low-energy limit of the four-point closed string amplitude was calculated in the pure-spinor formalism and shown to agree with predictions from S-duality [13]. Very little is known about amplitudes at even higher genera, but at genus $g \geq 5$, the moduli space of super Riemann surfaces without punctures is not split any more [119] (the bound on $g$ changes if punctures are introduced) and hence the fermions cannot be integrated out in the first step of the calculation as is usually done in the RNS formalism.

All the achievements mentioned above were performed in a flat ten-dimensional Minkowski background. For compactified backgrounds, much less is known, in particular genus-one CFT correlators for certain Calabi–Yau and K3 backgrounds were studied in [120–122] and an extension of the formalism of modular graph functions to plane wave backgrounds is considered in [123].

Instead of trying to calculate higher-loop amplitudes, some work has also been done on obtaining a deeper understanding of the structure of the tree-level amplitude. This is the direction of the classic result by Kawai–Lewellen–Tye (KLT) [2], which relates open and closed amplitudes at tree-level. The key idea of the KLT relations is that, since the closed string carries right- and left-moving modes, its vertex operators can be written as a product of vertex operators of open strings. Surprisingly, this factorization property can be lifted to the complete tree-level amplitude in the form of the KLT relations,

$$\mathcal{M}_n^{\text{tree}} = \sum_{\tau,\rho\in S_{n-3}} \mathcal{A}_n(1,\rho(2,\dots,n-2),n-1,n)\, S_{\alpha'}(\rho|\tau) \times \overline{\mathcal{A}_n(1,\tau(2,\dots,n-2),n,n-1)}\,. \tag{2.22}$$

Here, $\mathcal{M}_n^{\text{tree}}$ denotes an $n$-point closed-string tree-level amplitude, $\mathcal{A}_n(1,\dots,n)$ an $n$-point open-string tree-level amplitude with the order of the vertex operators on the disk boundary given in the argument. The sum runs over the two $(n-3)!$ independent open-string amplitudes[7] and the $\alpha'$-dependent $(n-3)! \times (n-3)!$ matrix $S_{\alpha'}(\rho|\tau)$ linking the two open-string amplitudes is known as the *KLT kernel* [124, 125]. Note that from the worldsheet perspective, (2.22) can be understood as making the intuition that a sphere cut in half are two half-spheres precise.

The KLT relations can be used to construct closed-string amplitudes with various amounts of supersymmetry, i.e. two bosonic open-string amplitudes combine to a bosonic closed string amplitude, two type-I open-string amplitudes combine to a type-II closed-string amplitude and a bosonic and a type-I open-string amplitude combine to a heterotic closed-string amplitude. Remarkably, the KLT kernel remains the same for all these combinations. It was first derived using deformations of the integration contours and monodromies but was recently shown to be computable in terms of associahedra and intersection numbers [126].

### 2.3.3 *Low-energy expansion and field theory*

The string amplitude (2.21) implicitly still depends on the parameter $\alpha'$ and a tractable way to approach the evaluation of (2.21) is to expand in $\alpha'$. This expansion is of great interest physically since it corresponds to calculating higher-order corrections to the supergravity action and therefore obtaining a low-energy effective field theory description for the string theory at hand. Note that on top of the analytic contributions discussed below, the string amplitude also contains non-analytic contributions corresponding to cuts of loop diagrams in this effective

7 There are a priori $(n-1)!$ independent orderings for the vertex operators on a circle. However, monodromy relations of disk integrals imply a smaller basis of $(n-3)!$ [3, 4].

field theory [127, 128]. As an example, consider the gravity sector of the low-energy effective action of type-IIB is in Einstein frame given by

$$S_{\text{eff}} = \frac{1}{\kappa^2} \int \mathrm{d}^{10}x \sqrt{-g} \sum_{m=1}^{\infty} \sum_{n=0}^{\infty} (\alpha')^{m+n-1} c_{m,n}(\eta) \nabla^{2n} R^m + \dots \,. \tag{2.23}$$

Here, $\nabla^{2n}R^m$ denotes a particular contraction of $2n$ covariant derivatives with $m$ powers of the Riemann tensor such that the $m = 1, n = 0$ contribution is just the Einstein–Hilbert term. The detailed structure of the $R^4$ and $\nabla^4 R^4$ contributions is spelled out in (3.77) and (3.78), respectively. In general, the form of the higher-order operators is constrained by supersymmetry and kinematics, this forces e.g. the coefficients of $R^2$, $R^3$ and $\nabla^2 R^4$ to vanish. The $c_{m,n}(\eta)$ are functions of the axio-dilaton $\eta = \chi + ie^{-\phi}$. The genus expansion of the amplitude corresponds to an expansion of the $c_{m,n}(\eta)$ in $(\operatorname{Im} \eta)^{-1} = e^{\phi}$ and the number of external gravitons in the amplitude corresponds to $m$. For example, by calculating the genus-one four-graviton scattering amplitude and expanding it in $\alpha'$, one obtains the genus-one contribution to the coefficients of $R^4$, $\nabla^4 R^4$, $\nabla^6 R^4$ etc.

The non-perturbative S-duality of type-IIB acts via a modular transformation of the form given in (1.2) on the axio-dilaton $\eta$ and the coefficients $c_{m,n}(\eta)$ are invariant under this transformation. Furthermore, supersymmetry implies that the $c_{m,n}(\eta)$ satisfy Laplace eigenvalue-equations w.r.t. $\eta$ [129]. For low orders, these properties are constraining enough to determine the functions $c_{m,n}(\eta)$ completely [8], leading to a full non-perturbative expression which can be shown to expand to the known results from string perturbation theory. For an extensive review of this area of research, see the textbook [73]. Recently, the coefficients on $\text{AdS}_5 \times \text{S}_5$ were connected via the AdS/CFT correspondence to correlators of a deformation of $\mathcal{N} = 2$ super Yang–Mills theory known non-perturbatively from constraints due to localization and the conformal bootstrap [130–135]. In the flat-space limit, these results confirm also the non-perturbative contributions to the $c_{m,n}(\eta)$ under consideration there and hence constitute a precision test of the AdS/CFT correspondence. Similarly, the interplay of correlation functions in $\mathcal{N} = 4$ super Yang–Mills with string-theory and supergravity amplitudes in $\text{AdS}_5 \times \text{S}_5$ and their flat space limit have for instance been discussed in [136–144].

Interestingly, aside from the calculation of string amplitudes, a different way to obtain the same low-energy effective theory is by computing the $\beta$ functions of the renormalization group flow of the worldsheet CFT. Requiring the gauge symmetries to also hold at the quantum level imposes constraints on the background fields which enter the $\beta$ functions and are the equations of motion of the low-energy effective field theory. In this context, higher orders in $\alpha'$ correspond to higher loop orders in the CFT.

The first order in the $\alpha'$ expansion, or the limit $\alpha' \to 0$, is particularly interesting, because it corresponds to taking the field theory limit. Since the low-energy effective theories of the five consistent superstring theories correspond to the three consistent supergravity theories in ten dimensions (the field-theory limits of the gravitational sectors of the heterotic and type-I theories are the same), this way we can obtain supergravity amplitudes. Noting that the calculation of amplitudes in gravity theories is notoriously hard due to the large number of Feynman diagrams, this can be a shortcut since in string theory only one (or in the unoriented case a small number) of worldsheets contribute at each loop order.

At tree-level, taking the limit $\alpha' \to 0$ is sufficient to obtain the field theory limit, however at one-loop and beyond also an appropriate limit has to be taken in the moduli space of the corresponding Riemann surface. At genus-one this limit is $\operatorname{Im}\tau \to \infty$, where $\tau$ is the complex structure modulus of the worldsheet torus, and leads (together with $\alpha' \to 0$) to one-loop amplitudes in supergravity.

Using this correspondence between field theory and string-theory amplitudes, properties of string amplitudes carry over to field theory amplitudes. E.g. taking the field theory limit of the KLT relations (2.22) leads directly to the double copy relations of tree-level amplitudes discussed in Section 2.2.2. This exemplifies the power of this approach: The KLT relations comprise as a concise proof of the field theory double copy at tree-level [5, 124]. Furthermore, the double copy construction can also be performed at one-loop level [92], where the kinematic numerators satisfying the BCJ duality can be obtained from genus-one string amplitudes in the pure-spinor formalism [6, 145].

## 2.4 SINGLE-VALUED MAP OF TREE-LEVEL STRING AMPLITUDES

After having reviewed the general structure of string perturbation theory in the previous section, we will review the *single-valued map*, an idea from analytic number theory, and its application to tree-level string amplitudes in this section. This is a more specialized and technical concept as compared to the previous sections, however, it is central for the research presented in this work.

### 2.4.1 *The Veneziano- and Virasoro–Shapiro amplitudes*

The calculation of the tree-level open (bosonic, oriented)-string amplitude for four tachyons in 1968 by Veneziano [14] was arguably the first calculation ever done in string theory, even before it was recognized as a theory of strings. The central disk integral of the Veneziano amplitude evaluates to

$$Z(s_{12}, s_{23}) = \frac{\Gamma(s_{12})\Gamma(1+s_{23})}{\Gamma(1+s_{12}+s_{23})}\,, \tag{2.24}$$

where the $s_{ij}$ are the dimensionless Mandelstam invariants defined by[8]

$$s_{ij} = -\frac{\alpha'}{4}\big(k_i + k_j\big)^2 \tag{2.25}$$

for the momenta $k_i$ of the external particles and $\Gamma$ is the Euler gamma function.

For later reference, note that for $n$ momenta $k_i$ of particles with masses $m_i$, the definition (2.25) and momentum conservation $\sum_{i=1}^{n} k_i = 0$ yield the relations

$$s_{ij} = s_{ji}\,, \qquad s_{ii} = \alpha' m_i^2\,, \qquad \sum_{j=1}^{n} s_{ij} = \frac{\alpha'}{4}\left(n m_i^2 + \sum_{j=1}^{n} m_j^2\right), \tag{2.26}$$

leaving in total $\frac{n(n-3)}{2}$ independent Mandelstam variables.[9] Hence, for three external particles, the kinematical space is trivial and for four particles, two independent Mandelstam variables are left, which we pick to be $s_{12}$ and $s_{23}$. To ease the notation, we will also use $s_{13} = -s_{12} - s_{23}$. Similarly to (2.25), we define multi-particle Mandelstam variables by

$$\underbrace{s_{i_1,\ldots,i_p}}_{I} = -\frac{\alpha'}{4}\Big(\sum_{j\in I} k_j\Big)^2 = \sum_{\substack{j<\ell\\ j,\ell\in I}} s_{j,l} - (p-2)\frac{\alpha'}{4}\sum_{j\in I} m_j^2\,. \tag{2.27}$$

Shortly after Veneziano, Virasoro and Shapiro calculated the tree-level closed (bosonic, oriented)-string amplitude for four tachyons whose central expression is a sphere integral which evaluates to [147, 148]

$$J(s_{12}, s_{23}) = \frac{1}{s_{12}}\frac{\Gamma(1+s_{12})\Gamma(1+s_{23})\Gamma(1+s_{13})}{\Gamma(1-s_{12})\Gamma(1-s_{23})\Gamma(1-s_{13})}\,. \tag{2.28}$$

Since the Mandelstam invariants are proportional to $\alpha'$, the low-energy expansion of (2.24) and (2.28) is an expansion in the Mandelstams. It can be performed using the identity

$$\log\Gamma(1+z) = -z\gamma_{\mathrm{E}} + \sum_{k=2}^{\infty}(-z)^k\frac{\zeta(k)}{k}\,, \qquad \text{for } |z|<1\,, \tag{2.29}$$

8 Strictly speaking, these are the conventions suitable for the closed string. The Mandelstam variables in the open string are normalized as $s_{ij} = \alpha'(k_i+k_j)^2$ and the actual expression appearing in the Veneziano amplitude is (2.24) with $s_{ij} \to -4s_{ij}$.

9 To be precise, there are more relations if $n > D+1$ in $D$ dimensions since then the $n-1$ independent momenta cannot be linearly independent, and hence the Gram determinants $\det s_{ij}$ vanish [146]. But since these are polynomial constraints, we will not use them here.

where $\gamma_{\mathrm{E}}$ is the Euler–Mascheroni constant and

$$\zeta(s) = \sum_{n=1}^{\infty} \frac{1}{n^s}, \qquad \mathrm{Re}(s) > 1 \tag{2.30}$$

the Riemann zeta function. Using (2.29), the expansions of (2.24) and (2.28) in powers of Mandelstam variables are given by

$$Z(s_{12}, s_{23}) = \frac{1}{s_{12}} \exp\left( \sum_{k=2}^{\infty} (-1)^k \frac{\zeta(k)}{k} [s_{12}^k + s_{23}^k - (s_{12} + s_{23})^k] \right) \tag{2.31}$$

$$J(s_{12}, s_{23}) = \frac{1}{s_{12}} \exp\left( -2 \sum_{k=2}^{\infty} \frac{\zeta(2k+1)}{2k+1} [s_{12}^{2k+1} + s_{23}^{2k+1} - (s_{12}+s_{23})^{2k+1}] \right). \tag{2.32}$$

The central observation is now that one can go from $Z$ to $J$ by replacing

$$\zeta(2k) \mapsto 0 \qquad\qquad \zeta(2k+1) \mapsto 2\zeta(2k+1). \tag{2.33}$$

This map is a special case of the *single-valued map* of multiple zeta values, as detailed in the next section.

### 2.4.2 *Multiple zeta values and the single-valued map*

The Riemann zeta function (2.30) defines the *single zeta values* $\zeta_k = \zeta(k)$, $k \geq 2 \in \mathbb{N}$. The even zeta values $\zeta_{2k}$ are rational multiples of $\pi^{2k}$,[10]

$$\zeta_{2k} = (-1)^{k+1} \frac{B_{2k}(2\pi)^{2k}}{2(2k)!}, \tag{2.34}$$

where the $B_{2k} \in \mathbb{Q}$ are the *Bernoulli numbers*

$$B_n = \sum_{k=0}^{n} \sum_{j=0}^{k} (-1)^j \binom{k}{j} \frac{j^n}{k+1}. \tag{2.35}$$

The single-valued map (2.33) for single zeta values can be motivated by considering the single zeta values as *polylogarithms* $\mathrm{Li}_k(z)$ evaluated at one, $\zeta_k = \mathrm{Li}_k(1)$. Here, $\mathrm{Li}_k(z)$ is defined by the sum

$$\mathrm{Li}_k(z) = \sum_{n=1}^{\infty} \frac{z^n}{n^k}, \qquad |z| < 1, \tag{2.36}$$

which can be extended to $|z| \geq 1$ by analytic continuation. Polylogarithms are a generalization of the logarithm since for $k = 1$ we recover

10 Interestingly, much less is known about odd zeta values: We know that $\zeta_3$ is irrational [149], that infinitely many odd zeta values are irrational [150] and e.g. that one of $\zeta_5$, $\zeta_7$, $\zeta_9$ and $\zeta_{11}$ is irrational [151].

the usual logarithm via $\mathrm{Li}_1(z) = -\log(1-z)$. As the logarithm, also polylogarithms are multi-valued functions on the complex plane. However, they can be made into single-valued functions by subtracting their monodromies in a suitable way, leading to *single-valued polylogarithms* $\mathrm{Li}_k^{\mathrm{sv}}(z)$ [152], defined by

$$\mathrm{Li}_k^{\mathrm{sv}}(z) = \mathrm{Li}_k(z) - \sum_{n=0}^{k-1} (-1)^{k-n} \frac{2^n}{n!} \log^n(|z|)\, \mathrm{Li}_{k-n}(\bar{z})\,. \tag{2.37}$$

E.g., $\mathrm{Li}_1^{\mathrm{sv}}(z) = -\log(|1-z|^2)$ is trivially single-valued. Using (2.37), we define *single-valued zeta values* $\zeta_k^{\mathrm{sv}}$ as single-valued polylogarithms evaluated at one,

$$\zeta_k^{\mathrm{sv}} = \mathrm{Li}_k^{\mathrm{sv}}(1)\,, \qquad k \geq 2 \in \mathbb{N}\,. \tag{2.38}$$

Remarkably, single-valued zeta values are a subset of ordinary zeta values and furthermore have exactly the property desired in (2.33),

$$\zeta_{2k}^{\mathrm{sv}} = 0 \qquad\qquad \zeta_{2k+1}^{\mathrm{sv}} = 2\zeta_{2k+1} \tag{2.39}$$

and therefore, order-by-order in $\alpha'$, $J$ is the *single-valued map* of $Z$,

$$J = \mathrm{sv}(Z)\,. \tag{2.40}$$

Of course, at the level of (2.33), this looks more like a coincidence rather than a deep fact about the structure of tree-level string amplitudes, but there is a much more intricate generalization to higher-point amplitudes as follows. The periods (integrals) of the moduli space $\mathcal{M}_{0,n}$ of genus zero Riemann surfaces with $n$ marked points are *multiple zeta values* (MZVs) [153], generalizations of single zeta values defined by

$$\zeta_{k_1,\ldots,k_r} = \sum_{0<n_1<\cdots<n_r}^{\infty} \frac{1}{n_1^{k_1}} \cdots \frac{1}{n_r^{k_r}}\,, \qquad k_1 \ldots k_r \in \mathbb{N}\,, \qquad k_r \geq 2\,, \tag{2.41}$$

where $r$ is the *depth* and $k_1 + \cdots + k_r$ the *weight* of the MZV. Therefore, the coefficients in the $\alpha'$ expansion of a general open or closed tree-level string amplitude are MZVs [20, 23, 24, 154, 155].

Also the polylogarithms (2.36) allow for a higher-depth generalization, namely *multiple polylogarithms*, defined by

$$\mathrm{Li}_{k_1,\ldots,k_r}(z_1,\ldots,z_r) = \sum_{0<n_1<\cdots<n_r}^{\infty} \frac{z_1^{n_1}}{n_1^{k_1}} \cdots \frac{z_r^{n_r}}{n_r^{k_r}}\,, \quad k_1 \ldots k_r \in \mathbb{N}\,, \quad |z_i| < 1\,. \tag{2.42}$$

Hence, MZVs are special values of multiple polylogarithms,

$$\zeta_{k_1,\ldots,k_r} = \mathrm{Li}_{k_1,\ldots,k_r}(1,\ldots,1)\,. \tag{2.43}$$

Multiple polylogarithms have a representation in terms of iterated integrals [156],

$$\begin{aligned} &\mathrm{Li}_{k_1,\ldots,k_r}(z_1,\ldots,z_r) \\ &= (-1)^r G((z_1\cdots z_r)^{-1},0^{k_1-1},(z_2\cdots z_r)^{-1},0^{k_2-1},\ldots,z_r^{-1},0^{k_r-1};1)\,, \end{aligned} \tag{2.44}$$

where $0^n$ is the row vector with $n$ entries of 0 and $G$ is the iterated integral defined recursively by

$$\begin{aligned} G(a_1,\ldots,a_n;x) &= \int_0^x \frac{\mathrm{d}t_1}{t_1-a_1} G(a_2,\ldots,a_n;t_1)\,, \\ G(\varnothing;x) &= 1\,. \end{aligned} \tag{2.45}$$

According to (2.43), MZVs are therefore given by

$$\zeta_{k_1,\ldots,k_r} = (-1)^r G(1,0^{k_1-1},\ldots,1,0^{k_r-1};1)\,. \tag{2.46}$$

Due to their series representation (2.42), multiple polylogarithms satisfy *stuffle relations*, which for depth one read

$$\mathrm{Li}_k(z)\,\mathrm{Li}_\ell(w) = \mathrm{Li}_{k+\ell}(zw) + \mathrm{Li}_{k,\ell}(z,w) + \mathrm{Li}_{\ell,k}(w,z)\,. \tag{2.47}$$

Furthermore, the integrals $G$ defined in (2.45) satisfy shuffle relations (as all iterated integrals do),

$$G(K;z)G(L;z) = G(K \sqcup\!\sqcup L;z)\,, \tag{2.48}$$

where $K, L$ are (non-commutative) words and $\sqcup\!\sqcup$ is the shuffle product,[11] implying further relations between multiple polylogarithms. For a detailed review of these relations, see e.g. [157].

The shuffle and stuffle relations satisfied by multiple polylogarithms imply similar identities for MZVs. E.g. the stuffle relation (2.47) becomes

$$\zeta_k\zeta_l = \zeta_{k,l} + \zeta_{l,k} + \zeta_{k+l}\,. \tag{2.50}$$

In fact, integer linear combinations of MZVs form a ring [158] and the graded $\mathbb{Q}$ algebra of MZVs is conjectured to be a Hopf algebra and proven to be one in the more abstract motivic setup [159, 160]. The dimension $d_w$ (over $\mathbb{Q}$) of the space of MZVs of weight $w$ is conjectured to be $d_w = d_{w-2} + d_{w-3}$, $w \geq 3$ and $d_0 = 1$, $d_1 = 0$, $d_2 = 1$ [158]. A

11 The shuffle product for two words $K = (k_1,\ldots,k_r)$ and $L = (l_1,\ldots,l_s)$ is defined as

$$G(K \sqcup\!\sqcup L;z) = \sum_{\sigma\in\Sigma(K,L)} G(\sigma(1),\ldots,\sigma(r+s);z)\,, \tag{2.49}$$

where $\Sigma(K,L) \subset S_{r+s}$ are the permutations of $(K,L)$ which are $(k_1,\ldots,k_r)$ if all $l_i$ are dropped and $(l_1,\ldots,l_s)$ if all $k_i$ are dropped. Hence, $K$ and $L$ are "shuffled" together like two decks of cards.

computer implementation for the decomposition of MZVs of weight at most 22 into this basis is available [161].

Also for multiple polylogarithms on can cancel the monodromies to obtain *single-valued multiple polylogarithms* $\mathrm{Li}^{\mathrm{sv}}_{k_1,\ldots,k_r}(z_1,\ldots,z_r)$ which define *single-valued MZVs* via [18, 19]

$$\zeta^{\mathrm{sv}}_{k_1,\ldots,k_r} = \mathrm{Li}^{\mathrm{sv}}_{k_1,\ldots,k_r}(1,\ldots,1)\,. \tag{2.51}$$

Remarkably, these are not only a subset of the ordinary MZVs, but in fact form a subalgebra w.r.t. the shuffle and stuffle products, hence (2.51) is an algebra homomorphism. The single-valued map for higher-depth MZVs is furthermore much more intricate than its single-zeta version (2.39), e.g.

$$\zeta^{\mathrm{sv}}_{5,3} = 14\zeta_3\zeta_5 \qquad \zeta^{\mathrm{sv}}_{3,5,3} = 2\zeta_{3,5,3} - 2\zeta_3\zeta_{3,5} - 10\zeta_3^2\zeta_5\,. \tag{2.52}$$

In 2013, it was observed that the MZVs in the $\alpha'$ expansion of closed-string tree-level integrals are the single-valued images of the MZVs in the $\alpha'$ expansion of the corresponding open-string amplitude [23, 162, 163]. In 2018, this result was proven by several groups [25–27].

# 3

# ONE-LOOP CLOSED-STRING AMPLITUDES

In this chapter, we discuss the calculation of one-loop closed-string amplitudes. As was mentioned in Section 2.3.1, the genus-one closed-string amplitude is an integral over the moduli space of a punctured torus,

$$\mathcal{A}^{\text{closed}}_{\text{genus one}}(k_1,\ldots,k_n) = g_s^n \int_{\mathcal{M}_{1,n}} \mathrm{d}\mu \,\langle \prod_{i=1}^{n} \mathcal{V}_1^i(k_i,z_i)\,\text{ghosts}\rangle\,. \tag{3.1}$$

In Section 3.1, we will review the parametrization of this worldsheet torus, its moduli space and and how the modular group enters the calculation. This discussion will closely follow the one in [46], Chapter 6. We will also introduce various differential operators that are useful when dealing with modular quantities.

The integrand on the worldsheet is a CFT-correlator of vertex operators and ghosts. In Section 2.3.1, only the basic idea was given, in Section 3.2, we will show the calculation for the genus-one case in more detail.

Finally, we will review how the integration over puncture positions can be performed using modular graph forms in Section 3.3 and also comment on the final integration over the moduli space of an unpunctured torus in Section 3.4.

## 3.1 TORI AND MODULARITY

In order to perform the integral over $\mathcal{M}_{1,n}$ in (3.1), we map the punctured torus to a parallelogram in the complex plane whose opposite edges are identified and which is spanned by the two complex numbers $\lambda_1$ and $\lambda_2$,

$$z \equiv z + n\lambda_1 + m\lambda_2\,, \quad m,n \in \mathbb{Z}\,. \tag{3.2}$$

The two cycles of the torus can be identified with $\lambda_1$ and $\lambda_2$ and we choose $\lambda_1$ to be along the $A$-cycle and $\lambda_2$ along the $B$-cycle, cf. the LHS of Figure 3.1. As mentioned in Section 2.3.1, the gauge-fixed Polyakov action has a residual conformal symmetry, hence the path integral over the worldsheet metric $\gamma$ is reduced to an integral over conformally inequivalent metrics. In the parametrization above, conformal transfor-

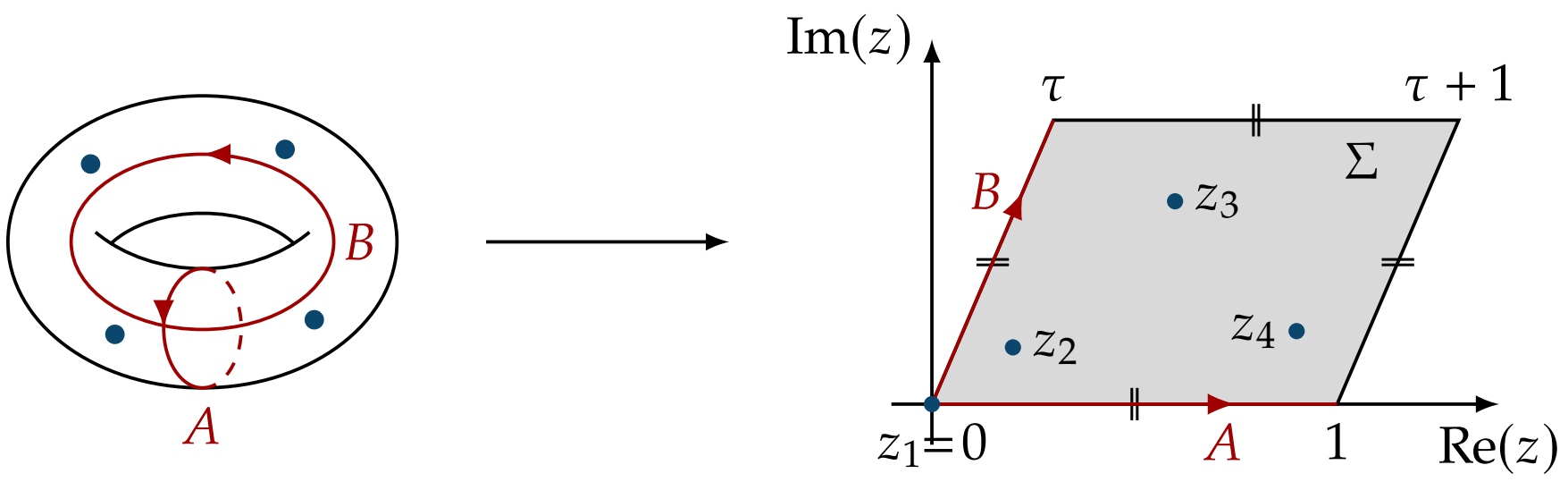

Figure 3.1: Parametrization of the worldsheet torus $\Sigma$ (here with four marked points $z_1, \ldots, z_4$) by a parallelogram spanned by $\tau$ and 1 whose opposite edges are identified. The two homology cycles $A$ and $B$ are identified with the lines $0 \to 1$ and $0 \to \tau$, respectively.

mations are rescalings and rotations of $\lambda_{1,2}$, so the *modular parameter*

$$\tau = \tau_1 + i\tau_2 = \frac{\lambda_2}{\lambda_1}, \qquad \tau_1, \tau_2 \in \mathbb{R} \tag{3.3}$$

is invariant under conformal transformations. Hence, we can set $\lambda_1 = 1$ and restrict to $\tau_2 > 0$, the resulting parallelogram is denoted by $\Sigma$ and illustrated in Figure 3.1. By choosing the origin of the coordinate system, we can furthermore fix one of the punctures to zero and we set w.l.o.g. $z_1 = 0$.

The identification of opposite edges in Figure 3.1 means that functions on $\Sigma$ have to be *doubly periodic*,

$$f(z+1) = f(z+\tau) = f(z)\,. \tag{3.4}$$

If $f$ is additionally meromorphic on $\mathbb{C}$, it is an *elliptic function*. Ellipticity is very constraining and a lot is known about elliptic functions, for a comprehensive textbook with proofs of all classic results, see [164]. E.g. every elliptic function can be written as a rational function in the Weierstraß function

$$\wp(z,\tau) = \frac{1}{z^2} + \sum_{(m,n)\in\mathbb{Z}^2\setminus\{(0,0)\}} \left[\frac{1}{(z+m\tau+n)^2} - \frac{1}{(m\tau+n)^2}\right] \tag{3.5}$$

and its derivative. Since the pole structure is sufficient to fix this rational function, every elliptic function is determined by its poles, zeros and constant term. Furthermore, the sum of the residues of the simple poles must vanish as follows directly from Stokes' theorem.

### 3.1.1 *Large diffeomorphisms and the modular group*

Following (3.3) and $\tau_2 > 0$, the path integral over $\gamma$ becomes an integral over $\tau$ over the upper half plane

$$\mathbb{H} = \{\tau \in \mathbb{C} | \operatorname{Im}(\tau) > 0\} , \tag{3.6}$$

the *Teichmüller space* of the torus.[1] However, in the construction above, we have only considered diffeomorphisms which are smoothly connected to the identity and discarded global diffeomorphisms. On the torus, there are two kinds of global diffeomorphisms, the *Dehn twists*. The first Dehn twist cuts the torus along the $A$ cycle, twists by $2\pi$ and glues it back together. This sends

$$\lambda_1 \to \lambda_1 , \quad \lambda_2 \to \lambda_1 + \lambda_2 \quad \Rightarrow \quad \tau \to \tau + 1 . \tag{3.8}$$

The second Dehn twist performs the same operation on the $B$ cycle and sends

$$\lambda_1 \to \lambda_1 + \lambda_2 , \quad \lambda_2 \to \lambda_2 \quad \Rightarrow \quad \tau \to \frac{\tau}{\tau + 1} . \tag{3.9}$$

Together, $\lambda_1$ and $\lambda_2$ generate the *modular group* $\mathrm{PSL}(2,\mathbb{Z})$ acting on $\tau$ via[2]

$$\tau \to \frac{\alpha\tau + \beta}{\gamma\tau + \delta} , \qquad \begin{pmatrix} \alpha & \beta \\ \gamma & \delta \end{pmatrix} \in \mathrm{SL}(2,\mathbb{Z}) . \tag{3.10}$$

Often, the modular T and S transformations, defined by

$$\mathrm{T}: \quad \tau \to \tau + 1 \qquad\qquad \mathrm{S}: \quad \tau \to -\frac{1}{\tau} \tag{3.11}$$

are used to generate the modular group, instead of the Dehn twists. Under a modular transformation (3.10), a point $z \in \Sigma$ on the torus transforms as

$$z \to \frac{z}{\gamma\tau + \delta} . \tag{3.12}$$

1 A different way to arrive at the same conclusion is the following: Although it is possible to make the worldsheet metric flat locally on a torus using diffeomorphisms and Weyl transformations, it is not possible to do this globally. Globally, we can only reach the form

$$\gamma = |\mathrm{d}\sigma^1 + \tau \mathrm{d}\sigma^2|^2 = (\mathrm{d}\sigma^1)^2 + |\tau|^2(\mathrm{d}\sigma^2)^2 + 2\tau_1 \mathrm{d}\sigma^1 \mathrm{d}\sigma^2 , \tag{3.7}$$

so the metric is only flat for $\tau = i$. Since (3.7) is invariant under complex conjugation of $\tau$ and degenerate for $\tau \in \mathbb{R}$, we can restrict to $\tau_2 > 0$.

2 Since sending $\left(\begin{smallmatrix} \alpha & \beta \\ \gamma & \delta \end{smallmatrix}\right) \to -\left(\begin{smallmatrix} \alpha & \beta \\ \gamma & \delta \end{smallmatrix}\right)$ leaves $\frac{\alpha\tau+\beta}{\gamma\tau+\delta}$ invariant, the modular group is $\mathrm{PSL}(2,\mathbb{Z})$, not $\mathrm{SL}(2,\mathbb{Z})$.

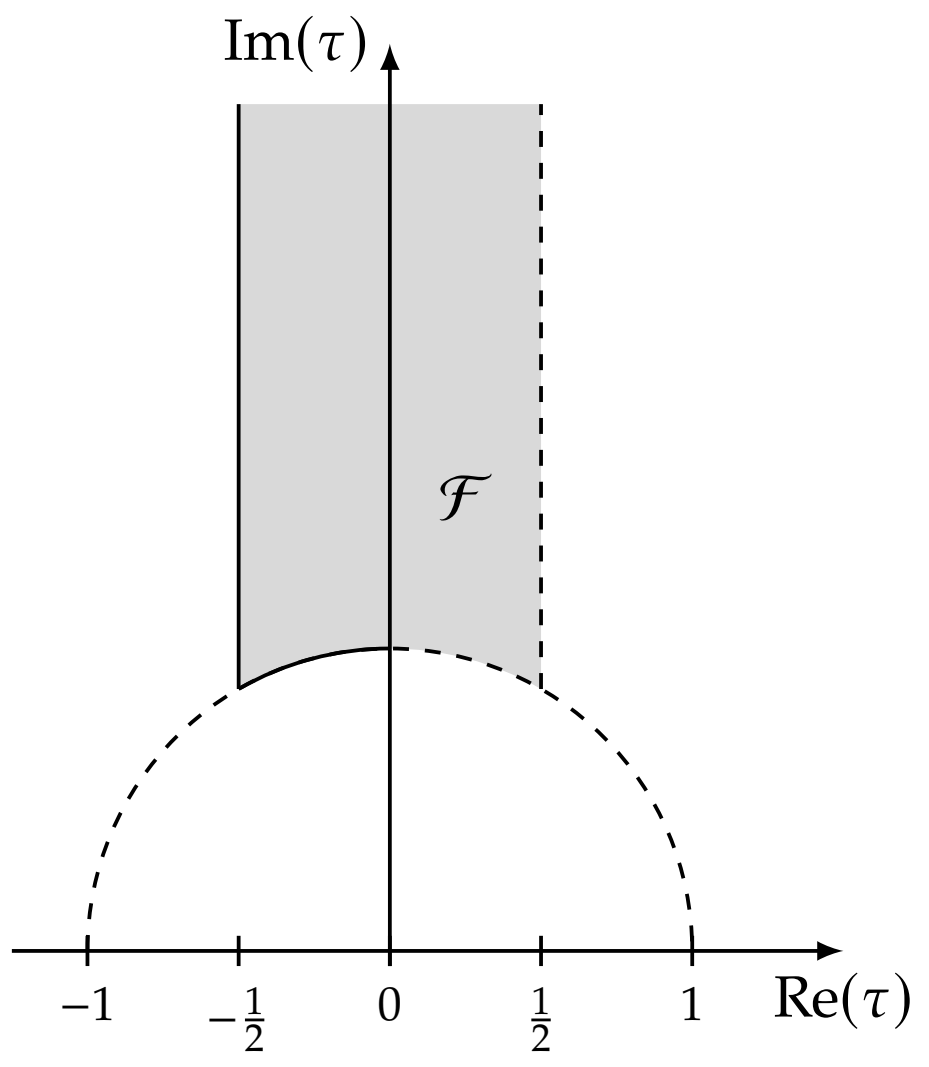

Figure 3.2: The fundamental domain $\mathcal{F}$ of the modular group of the torus as defined in (3.14). The solid part of the boundary belongs to $\mathcal{F}$, the dashed part does not.

In order to not overcount metrics which are related by large diffeomorphisms, we have to integrate over the quotient of the Teichmüller space and the modular group, the *moduli space* $\mathcal{M}_1$ of the torus,

$$\mathcal{M}_1 = \frac{\text{Teichmüller space}}{\text{modular group}} . \tag{3.13}$$

Hence, instead of integrating the modular parameter over $\mathbb{H}$, we integrate it over a *fundamental domain* $\mathcal{F}$ of the modular group, defined by

$$\begin{aligned} \mathcal{F} = &\left\{ \tau \in \mathbb{H} \,\middle|\, -\frac{1}{2} \leq \text{Re}(\tau) \leq 0, |\tau|^2 \geq 1 \right\} \\ &\cup \left\{ \tau \in \mathbb{H} \,\middle|\, 0 \leq \text{Re}(\tau) < \frac{1}{2}, |\tau|^2 > 1 \right\} \end{aligned} \tag{3.14}$$

and illustrated in Figure 3.2. The point $\tau = i\infty$ is often referred to as the *cusp*. Every fundamental domain has the following two properties,

$$\forall\, \tau \in \mathcal{F}, \quad \frac{\alpha\tau + \beta}{\gamma\tau + \delta} \notin \mathcal{F} \quad \forall \begin{pmatrix} \alpha & \beta \\ \gamma & \delta \end{pmatrix} \in \text{SL}(2, \mathbb{Z}) \setminus \{\mathbb{1}\} \tag{3.15a}$$

$$\forall\, \tau \in \mathbb{H} \quad \exists \begin{pmatrix} \alpha & \beta \\ \gamma & \delta \end{pmatrix} \in \text{SL}(2, \mathbb{Z}), \quad \text{s.t.} \quad \frac{\alpha\tau + \beta}{\gamma\tau + \delta} \in \mathcal{F}, \tag{3.15b}$$

which guarantee that an integral over $\mathcal{F}$ integrates over all conformally inequivalent tori once.

Of course the representation (3.14) of the fundamental domain is not unique and every image under modular transformations is an equivalent representation. In string theory this means that the integrand of the integral over $\tau$ should be modular invariant and this form of gauge invariance puts strong constraints on the structure of the theory, e.g. the gauge groups $E_8 \times E_8$ and SO(32) of the heterotic theories were fixed in this way.

### 3.1.2 *Modular functions and forms*

Having defined a group action of $SL(2,\mathbb{Z})$ on $\mathbb{H}$ via (3.10), it is natural to consider functions on $\mathbb{H}$ that transform nicely under modular transformations. We will define four such function classes and always set $\begin{pmatrix} \alpha & \beta \\ \gamma & \delta \end{pmatrix} \in SL(2,\mathbb{Z})$.

#### HOLOMORPHIC MODULAR FORMS

A holomorphic function $f : \mathbb{H} \to \mathbb{C}$ which transforms under modular transformations as

$$f\left(\frac{\alpha\tau + \beta}{\gamma\tau + \delta}\right) = (\gamma\tau + \delta)^a f(\tau) \tag{3.16}$$

and satisfies the growth condition

$$|f(\tau)| \leq C\tau_2^N \quad \text{as} \quad \tau_2 \to \infty \quad \forall\, \tau_1 \in \mathbb{R} \tag{3.17}$$

for constants $C$ and $N$ is a *holomorphic modular form* of *weight* $a$. In most cases considered in this work, we will have $a \in \mathbb{Z}$. Note that for $a$ odd, (3.16) implies that $f = 0$. If $a = 0$, $f$ is constant.

An important example of a class of holomorphic modular forms are the *holomorphic Eisenstein series* $G_k$, which carry weight $a = k$ and are defined by

$$G_k(\tau) = \sum_{(m,n)\in\mathbb{Z}^2\setminus\{(0,0)\}} \frac{1}{(m\tau+n)^k} = \sum_{(m,n)\in\mathbb{Z}^2}{}' \frac{1}{(m\tau+n)^k}\,, \qquad k \geq 3 \in \mathbb{N}. \tag{3.18}$$

Note that $G_{2n+1} = 0$ due to antisymmetry of the summand. We will often use the prime on the sum to indicate the omission of the origin from the lattice. In fact, all holomorphic modular forms of weight $k \geq 4 \in \mathbb{Z}$ can be written as sums of products of $G_4$ and $G_6$ with rational coefficients and hence form a ring over $\mathbb{Q}$. Furthermore, the Laurent expansion in $z$ of the Weierstraß function (3.5) has holomorphic Eisenstein series as coefficients,

$$\wp(z,\tau) = \frac{1}{z^2} + \sum_{k=4}^{\infty}(k-1)z^{k-2}G_k(\tau)\,. \tag{3.19}$$

Using the *Dedekind eta function*

$$\eta(\tau) = e^{\frac{\pi i \tau}{12}} \prod_{n=1}^{\infty} (1 - e^{2n\pi i \tau}) \,, \tag{3.20}$$

we obtain the weight 12 holomorphic modular from $\eta^{24}(\tau)$ which will be used later.[3]

The transformation property (3.16) implies in particular invariance under modular T transformations and hence a holomorphic modular form can be Fourier expanded in $\tau_1$,

$$f(\tau) = \sum_{n=0}^{\infty} a_n q^n \,. \tag{3.21}$$

This Fourier expansion is written as a series in

$$q = e^{2\pi i \tau} = e^{-2\pi \tau_2} e^{2\pi i \tau_1} \,, \tag{3.22}$$

where, since $\tau \in \mathbb{H}$, $|q| < 1$ and the cusp is at $q = 0$. Due to holomorphicity,[4] the prefactor cannot depend on $\tau_2$ and the sum starts at 0 to satisfy the growth condition (3.17). E.g. holomorphic Eisenstein series have the $q$ expansion

$$\mathrm{G}_{2k} = 2\zeta_{2k} - \frac{8k\zeta_{2k}}{B_{2k}} \sum_{n=1}^{\infty} \sigma_{2k-1}(n) q^n \,, \tag{3.24}$$

where $B_{2k}$ are the Bernoulli numbers (2.35) and $\sigma_k(n)$ is the *divisor sum*

$$\sigma_k(n) = \sum_{d|n} d^k \,. \tag{3.25}$$

## NON-HOLOMORPHIC MODULAR FORMS

A function $f : \mathbb{H} \to \mathbb{C}$ which transforms under modular transformations as

$$f\left(\frac{\alpha\tau + \beta}{\gamma\tau + \delta}\right) = (\gamma\tau + \delta)^a (\gamma\bar{\tau} + \delta)^b f(\tau) \tag{3.26}$$

3 Note that $\eta$ itself is not a modular form, only $\eta^{24}$.

4 Starting from a Fourier expansion of the form

$$f(\tau) = \sum_{n\in\mathbb{Z}} a_n(\tau_2) e^{2\pi i n \tau_1} \,, \tag{3.23}$$

requiring that $\partial_{\bar{\tau}} f(\tau) = 0$ imposes the differential equation $a_n'(\tau_2) = -2\pi n a_n(\tau_2)$ which forces the form (3.21).

and has $q$ expansion

$$f(\tau) = \sum_{m,n=0}^{\infty} a_{m,n}(\tau_2) q^m \bar{q}^n \,, \tag{3.27}$$

is a *non-holomorphic modular form* of *holomorphic weight $a$* and *antiholomorphic weight $b$*. We will write the weight often as $(a, b)$ and refer to the sum $a + b$ as the *total modular weight*. For $a + b$ odd, $f$ vanishes.

The behavior of $f$ at the cusp is determined by $a_{0,0}(\tau_2)$ which is often expanded in a Laurent polynomial in $\tau_2$ and hence we will refer to it as *the Laurent polynomial of $f$*. Note however, that the complete Fourier zero mode also contains infinitely many exponentially suppressed terms, since for $f(\tau) = \sum_{n\in\mathbb{Z}} b_n(\tau_2) e^{2\pi i n \tau_1}$, we obtain

$$b_0(\tau_2) = \sum_{k\geq 0} a_{k,k}(\tau_2) e^{-4\pi\tau_2 k} \,, \tag{3.28}$$

where $k = \frac{m+n}{2}$ in (3.27).

An important example of a non-holomorphic modular form is $\tau_2 = \operatorname{Im}\tau$. Since it transforms as

$$\operatorname{Im}\left(\frac{\alpha\tau + \beta}{\gamma\tau + \delta}\right) = \frac{\tau_2}{|\gamma\tau + \delta|^2} \,, \tag{3.29}$$

it is a modular form of weight $(-1, -1)$.

Another important non-holomorphic modular form arises when considering the weight-two case for holomorphic Eisenstein series. For $k = 2$ the sum in (3.18) is conditionally convergent and hence needs to be supplied with a summation prescription. We will use the *Eisenstein summation* and define

$$\mathrm{G}_2(\tau) = \sum_{n\neq 0} \frac{1}{n^2} + \sum_{m\neq 0} \sum_{n\in\mathbb{Z}} \frac{1}{(m\tau + n)^2} \,. \tag{3.30}$$

This function is modular T invariant and has Fourier expansion (3.24) with $k = 1$, but it is not modular. A way to obtain a modular version of $\mathrm{G}_2$ is by introducing a regulator into (3.18), leading to the definition

$$\widehat{\mathrm{G}}_2(\tau) = \lim_{s\to 0} \sideset{}{'}\sum_{(m,n)\in\mathbb{Z}^2} \frac{1}{(m\tau + n)^2 |m\tau + n|^s} \,. \tag{3.31}$$

The limit can be performed (see e.g. [165]) and we obtain

$$\widehat{\mathrm{G}}_2(\tau) = \mathrm{G}_2(\tau) - \frac{\pi}{\tau_2} \,, \tag{3.32}$$

which is modular of weight $(2, 0)$, but non-holomorphic and has Laurent polynomial $\frac{\pi^2}{3} - \frac{\pi}{\tau_2}$. We will denote the complex conjugate of $\widehat{\mathrm{G}}_2$ by $\overline{\widehat{\mathrm{G}}}_2$.

### MODULAR FUNCTIONS

A (non-holomorphic) modular form of weight $(0,0)$ (i.e. which is invariant under modular transformations) is a *modular function*.

An important example of a class of modular functions are the *non-holomorphic (real-analytic) Eisenstein series* $\mathrm{E}_s : \mathbb{H} \to \mathbb{R}$, defined by

$$\mathrm{E}_s(\tau) = \left(\frac{\tau_2}{\pi}\right)^s \sideset{}{'}\sum_{(m,n)\in\mathbb{Z}^2} \frac{1}{|m\tau+n|^{2s}}\,, \quad s\in\mathbb{C}\,, \quad \mathrm{Re}(s)>1\,. \tag{3.33}$$

Note that the sum transforms with modular weight $(s,s)$ and the prefactor with weight $(-s,-s)$, rendering the entire expression invariant.

Non-holomorphic Eisenstein series have $q,\bar{q}$ expansion (for a derivation, see e.g. [73])

$$\begin{aligned}\mathrm{E}_s =& \frac{2\zeta_{2s}}{\pi^s}\tau_2^s + \frac{2\Gamma(s-\frac{1}{2})\zeta_{2s-1}}{\Gamma(s)\pi^{s-\frac{1}{2}}}\tau_2^{1-s} \\ &+ \frac{4\sqrt{\tau_2}}{\Gamma(s)}\sum_{N\neq 0}|N|^{s-\frac{1}{2}}\sigma_{1-2s}(|N|)K_{s-\frac{1}{2}}(2\pi|N|\tau_2)e^{2i\pi N\tau_1}\,,\end{aligned} \tag{3.34}$$

where $K_\nu(z)$ is the modified Bessel function of order $\nu$. The Laurent polynomials of the non-holomorphic Eisenstein series are the two terms in the first line of (3.34).

Some modular functions can be represented as a sum over images of modular transformations, a *Poincaré sum*,

$$f(\tau) = \sum_{\gamma\in B(\mathbb{Z})\backslash\mathrm{SL}(2,\mathbb{Z})} \sigma(\gamma\cdot\tau)\,, \tag{3.35}$$

where $\sigma$ is the *seed function* satisfying $\sigma(\tau)=\sigma(\tau+m)$ for $m\in\mathbb{Z}$ and we mod out by the *Borel subgroup* $B(\mathbb{Z})$ of $\mathrm{SL}(2,\mathbb{Z})$,

$$B(\mathbb{Z}) = \left\{\pm\begin{pmatrix}1 & m\\ 0 & 1\end{pmatrix}\middle|\, m\in\mathbb{Z}\right\}\,, \tag{3.36}$$

to not overcount. The group action $\gamma\cdot\tau$ is the standard modular action (3.10). As a sum over images of the group action, the form (3.35) makes the modular invariance of $f$ manifest. E.g. the Poincaré series of the non-holomorphic Eisenstein series (3.33) is given by

$$\mathrm{E}_s(\tau) = \frac{2\zeta_{2s}}{\pi^s}\sum_{\gamma\in B(\mathbb{Z})\backslash\mathrm{SL}(2,\mathbb{Z})} (\mathrm{Im}(\gamma\cdot\tau))^s\,. \tag{3.37}$$

#### JACOBI FORMS

A weight $(a, b)$ (non-holomorphic) *Jacobi form* of index $k$ is a function $f : \mathbb{C} \times \mathbb{H} \to \mathbb{C}$ which transforms under elliptic transformations as

$$f(\tau, z + m\tau + n) = e^{-2\pi i k(m^2\tau + 2mz)} f(\tau, z) \quad \forall m, n \in \mathbb{Z} \tag{3.38}$$

and under modular transformations as

$$f\left(\frac{z}{\gamma\tau + \delta}, \frac{\alpha\tau + \beta}{\gamma\tau + \delta}\right) = e^{\frac{2\pi i k\gamma z^2}{\gamma\tau+\delta}} (\gamma\tau + \delta)^a (\gamma\bar{\tau} + \delta)^b f(z_1, \tau) \tag{3.39}$$

and has at most a pole at the cusp. We will often encounter Jacobi forms of several elliptic variables which transform as (3.38) for each elliptic variable and as (3.39) when all variables are modular transformed. Since we will only encounter Jacobi forms of vanishing index, we will not mention the index in the following. For more details on Jacobi forms, cf. [166]. We will often use the weight $(a, b)$ non-holomorphic Jacobi forms $C^{(a,b)} : \Sigma \times \mathbb{H} \to \mathbb{C}$ defined by

$$C^{(a,b)}(z, \tau) = \sum_{(m,n)\in\mathbb{Z}}' \frac{e^{2\pi i(mv - nu)}}{(m\tau + n)^a (m\bar{\tau} + n)^b}, \tag{3.40}$$

where $u, v \in [0, 1] \subset \mathbb{R}$ on the RHS are defined by

$$u = \frac{\mathrm{Im}(z)}{\tau_2}, \quad v = \mathrm{Re}(z) - \frac{\tau_1}{\tau_2}\mathrm{Im}(z) \quad \Rightarrow \quad z = u\tau + v\,. \tag{3.41}$$

If a modular form or function vanishes at the cusp, it is called a *cusp form*. Holomorphic cusp forms have vanishing Fourier zero mode ($a_0 = 0$ in (3.21)), non-holomorphic modular functions or forms have vanishing Laurent polynomial ($a_{0,0} = 0$ in (3.27)).

For holomorphic modular forms, the lowest weight cusp form is the *modular discriminant* $\Delta$ of weight 12, which can be found by subtracting the zero modes of $\mathrm{G}_4^3$ and $\mathrm{G}_6^2$,

$$\Delta(\tau) = (60\mathrm{G}_4)^3 - 27(140\mathrm{G}_6)^2\,, \tag{3.42}$$

and is proportional to $\eta^{24}$,

$$\Delta(\tau) = (2\pi)^{12}\eta^{24}(\tau)\,. \tag{3.43}$$

### 3.1.3 *Modular differential operators*

Having defined various functions on $\mathbb{H}$ with nice properties with respect to modular transformations, we will now introduce invariant and covariant differential operators on $\mathbb{H}$.

First, we introduce the modular invariant *Poincaré metric*

$$ds^2 = \frac{(d\tau_1)^2 + (d\tau_2)^2}{\tau_2^2} = \frac{d\tau d\bar{\tau}}{\tau_2^2}\,, \tag{3.44}$$

which turns $\mathbb{H}$ into a two-dimensional hyperbolic space. This metric induces a modular invariant integration measure on $\mathbb{H}$ given by

$$\frac{d\tau_1 \wedge d\tau_2}{\tau_2^2} = \frac{i d\tau \wedge d\bar{\tau}}{2\tau_2^2}\,. \tag{3.45}$$

Under this measure, the volume of the fundamental domain (3.14) is $\frac{\pi}{3}$.

Similarly, for points on the torus $z \in \Sigma$, the integration measure

$$\frac{d\,\mathrm{Re}(z) \wedge d\,\mathrm{Im}(z)}{\tau_2} \tag{3.46}$$

is invariant under (3.12). Often, we will choose coordinates $u, v \in [0, 1]$ which are aligned with $\tau$ by setting $z = u\tau + v$ as in (3.41). Then, (3.46) becomes

$$dv \wedge du\,. \tag{3.47}$$

With this, we have all necessary ingredients to define the integration domain and measure in (3.1),

$$\int_{\mathcal{M}_{1,n}} d\mu = \int_{\mathcal{F}} \frac{d\tau_1 \wedge d\tau_2}{\tau_2^2} \prod_{k=2}^{n} \int_{\Sigma} \frac{d\,\mathrm{Re}(z_k) \wedge d\,\mathrm{Im}(z_k)}{\tau_2} \tag{3.48}$$

$$= \int_{\mathcal{F}} \frac{d\tau_1 \wedge d\tau_2}{\tau_2^2} \prod_{k=2}^{n} \int_{[0,1]^2} dv_k \wedge du_k\,. \tag{3.49}$$

Note that the first puncture position is again fixed to zero, $z_1 = 0$, by translation invariance. For later convenience, we introduce the integration measure

$$d\mu_{n-1} = \prod_{k=2}^{n} \frac{d\,\mathrm{Re}(z_k) \wedge d\,\mathrm{Im}(z_k)}{\tau_2} = \prod_{k=2}^{n} dv_k \wedge du_k \tag{3.50}$$

for the integral over the $n-1$ unfixed puncture positions.

In order to take derivatives of modular functions and forms, we define the *Maaß operators*, also called the *Cauchy–Riemann operators* [167]

$$\nabla^{(a)} = 2i\tau_2\partial_\tau + a \qquad \overline{\nabla}^{(b)} = -2i\tau_2\partial_{\bar{\tau}} + b\,. \tag{3.51}$$

On modular forms of weight $(a, b)$ they act as

$$\nabla^{(a)} : (a, b) \to (a+1, b-1) \qquad \overline{\nabla}^{(b)} : (a, b) \to (a-1, b+1) \tag{3.52}$$

and are hence compatible with the modular properties. One should think about the Maaß operators as the raising and lowering operators of the action of $\mathrm{SL}(2,\mathbb{R})$ on modular forms. Since $\nabla^{(a)}$ and $\overline{\nabla}^{(b)}$ leave $a+b$ invariant, this induces a grading of constant $a+b$ on the space of modular forms. The operators furthermore obey the product rule

$$\nabla^{(a+a')}(fg) = (\nabla^{(a)}f)g + f(\nabla^{(a')}g) \qquad \text{and c.c.}\,. \tag{3.53}$$

For later convenience, we introduce the notation

$$\begin{aligned}\nabla^{(a)^n} &= \nabla^{(a+n)}\nabla^{(a+n-1)}\ldots\nabla^{(a)}\\ \overline{\nabla}^{(b)^n} &= \overline{\nabla}^{(b+n)}\overline{\nabla}^{(b+n-1)}\ldots\overline{\nabla}^{(b)}\end{aligned} \tag{3.54}$$

for higher derivatives. Another commonly used set of differential operators for $\tau$ are

$$\nabla_0 = \tau_2\nabla^{(0)} = 2i\tau_2^2\partial_\tau \qquad \overline{\nabla}_0 = \tau_2\overline{\nabla}^{(0)} = -2i\tau_2^2\partial_{\bar\tau}\,, \tag{3.55}$$

which act covariantly on modular forms of weight $(0,b)$ and $(a,0)$, respectively,

$$\nabla_0 : (0,b) \to (0,b-2) \qquad \overline{\nabla}_0 : (a,0) \to (a-2,0)\,. \tag{3.56}$$

The Maaß operators can be used to define a modular invariant Laplacian $\Delta^{(a,b)}$ mapping the space of modular forms of weight $(a,b)$ into itself via

$$\Delta^{(a,b)} = \nabla^{(a-1)}\overline{\nabla}^{(b)} - b(a-1) = \overline{\nabla}^{(b-1)}\nabla^{(a)} - a(b-1) \tag{3.57}$$

$$= 4\tau_2^2\partial_\tau\partial_{\bar\tau} + 2i\tau_2 b\partial_\tau - 2i\tau_2 a\partial_{\bar\tau}\,. \tag{3.58}$$

For modular functions $(a,b) = (0,0)$ (3.58) reduces to

$$\Delta^{(0,0)} = 4\tau_2^2\partial_\tau\partial_{\bar\tau} = \tau_2^2(\partial_{\tau_1}^2 + \partial_{\tau_2}^2)\,, \tag{3.59}$$

which is just the Laplace-Beltrami operator of the Poincaré metric (3.44). The study of modular forms which satisfy eigenvalue equations of $\Delta^{(0,0)}$, so-called *Maaß forms* is an important field in its own right, with cuspidal Maaß forms being particularly interesting. E.g., the non-holomorphic Eisenstein series (3.33) satisfy

$$\Delta^{(0,0)}\mathrm{E}_s = s(s-1)\mathrm{E}_s\,. \tag{3.60}$$

## 3.2 CFT CORRELATORS IN THE RNS FORMALISM

In the last section, we discussed the domain and integration measure of the integral (3.1) and the modular properties of the integrand. In this section we will discuss the integrand, which is a CFT correlator of the vertex operators of the external states and the (super) ghosts. We will first review the derivation of the universal Koba–Nielsen factor and then list the tensor structures that appear for four-gluon and graviton scattering in the heterotic and type-IIB theory, respectively.

### 3.2.1 *Green function and Koba–Nielsen factor*

As in any computation of correlation functions in quantum field theory, also for the computation of correlators in the worldsheet CFT, the *Green function* $G(z,\tau)$ plays an important role, since it becomes the propagator of the free theory. On the torus, it is defined by the differential equation[5]

$$\partial_z \partial_{\bar z} G(z,\tau) = -\pi \delta^{(2)}(z,\bar z) + \frac{\pi}{\tau_2}\,, \tag{3.61}$$

where we used the complex derivatives as introduced in (2.5). The additional term $+\frac{\pi}{\tau_2}$ on the RHS is necessary for consistency: $\int_\Sigma d^2z/\tau_2(\partial\bar\partial G) = 0$ by Stokes and the identification of the boundaries, so the integral over the RHS should also vanish. (3.61) is solved by

$$G(z,\tau) = -\log\left|\frac{\theta_1(z,\tau)}{\eta(\tau)}\right|^2 - \frac{\pi}{2\tau_2}(z-\bar z)^2\,. \tag{3.62}$$

where $\theta_1(z,\tau)$ is the first *Jacobi theta function* defined by

$$\theta_1(z,\tau) = 2q^{1/8}\sin(\pi z)\prod_{n=1}^{\infty}(1-q^n)(1-e^{2\pi i z}q^n)(1-e^{-2\pi i z}q^n)\,. \tag{3.63}$$

For later reference, we also introduce the higher theta functions

$$\begin{aligned}
\theta_2(z,\tau) &= 2q^{1/8}\cos(\pi z)\prod_{n=1}^{\infty}(1-q^n)(1+e^{2\pi i z}q^n)(1+e^{-2\pi i z}q^n)\\
\theta_3(z,\tau) &= \prod_{n=1}^{\infty}(1-q^n)(1+e^{2\pi i z}q^{n-1/2})(1+e^{-2\pi i z}q^{n-1/2})\\
\theta_4(z,\tau) &= \prod_{n=1}^{\infty}(1-q^n)(1-e^{2\pi i z}q^{n-1/2})(1-e^{-2\pi i z}q^{n-1/2})\,.
\end{aligned} \tag{3.64}$$

Since the Green function is doubly periodic in $z = u\tau + v$, it allows for a double Fourier expansion in $u$ and $v$,

$$G(z,\tau) = \frac{\tau_2}{\pi}\sum_{(m,n)\in\mathbb{Z}^2}{}' \frac{e^{2\pi i(mv-nu)}}{|m\tau+n|^2} = \frac{\tau_2}{\pi}\sum_{p}{}' \frac{e^{2\pi i\langle p,z\rangle}}{|p|^2}\,, \tag{3.65}$$

5 Note that we denote the Green function by an italic $G$ and the holomorphic Eisenstein series by an upright G.

where we have introduced

$$p = m\tau + n \qquad \langle p, z\rangle = mv - nu = \frac{(p\bar{z} - \bar{p}z)}{2i\tau_2}\,. \tag{3.66}$$

This notation is supposed to make the analogy to the usual field theory propagators manifest: $p$ can be thought of as the momentum of the propagator which is discrete due to the compactness of the torus and hence the Fourier integral becomes a sum. The representation (3.65) also manifests that $G$ is symmetric in $z$, $G(-z,\tau) = G(z,\tau)$, and transforms as a non-holomorphic Jacobi form of weight $(0,0)$, i.e. it is modular invariant,

$$G\left(\frac{z}{\gamma\tau+\delta}, \frac{\alpha\tau+\beta}{\gamma\tau+\delta}\right) = G(z,\tau)\,, \qquad \begin{pmatrix}\alpha & \beta\\ \gamma & \delta\end{pmatrix} \in \mathrm{SL}(2,\mathbb{Z})\,. \tag{3.67}$$

As was mentioned in Section 2.3.1, the vertex operators correspond to asymptotic string states. Since the plane-wave operator $e^{ik_\mu X^\mu(z)}$ produces a momentum eigenstate with eigenvalue $k_\mu$ when acting on the worldsheet vacuum, every vertex operator carries a plane-wave part in addition to raising operators for the bosonic and fermionic worldsheet fields as well as the ghosts. Therefore, let us consider the correlator

$$\left\langle \prod_{j=1}^{n} e^{ik_j\cdot X(z_j)} \right\rangle^{\tau} = \int \mathcal{D}X \prod_{j=1}^{n} e^{ik_j\cdot X(z_j)} e^{-S_{\mathrm{Poly}}[X]}\,, \tag{3.68}$$

where the $k_j$ are massless momenta, the superscript $\tau$ indicates that the correlator should be evaluated on a torus with modular parameter $\tau$ and $S_{\mathrm{Poly}}[X]$ is the gauge-fixed bosonic Polyakov action (2.5). The following argument closely follows Appendix B.4.2 of [168].

In order to evaluate (3.68), we rewrite the integrand by introducing the currents $J^\nu(z,\bar{z}) = \sum_{j=1}^{n} k_j^\nu \delta^{(2)}(z - z_j, \bar{z} - \bar{z}_j)$, obtaining

$$\int \mathcal{D}X \exp\left[\int \mathrm{d}^2 z \left(\frac{1}{\pi\alpha'} X^\mu \partial\bar{\partial} X_\mu + iJ^\nu X_\nu\right)\right]\,, \tag{3.69}$$

where we integrated by parts in the first term. This integral is of Gaussian type and can be evaluated by inverting the operator $\partial\bar{\partial}$ using the Green function. The result is

$$\begin{aligned}\left\langle \prod_{j=1}^{n} e^{ik_j\cdot X(z_j)} \right\rangle^{\tau} &\sim \exp\left(-\frac{\alpha'}{4} \int \mathrm{d}^2 z \int \mathrm{d}^2 w\, J^\nu(z) G(z-w,\tau) J_\nu(w)\right)\\ &= \exp\left(\sum_{1\le j<\ell}^{n} s_{j\ell}\, G(z_{j\ell},\tau)\right).\end{aligned} \tag{3.70}$$

Here, we have used the Mandelstam invariants (2.25) and introduced the shorthand notation $z_{ij} = z_i - z_j$, which will be used frequently in the following. We will also use abbreviation $G_{ij} = G(z_{ij}, \tau)$ for the Green function.

In evaluating (3.69), we were not quite precise: The zero mode $x^\mu$ of $X^\mu$ (cf. (2.8)) lies in the kernel of $\partial\bar{\partial}$, so its integral is not of Gaussian form and should be treated separately. For the zero mode, the path integral becomes an ordinary integral and we obtain (for $D$ spacetime dimensions) the additional contribution

$$\int \mathrm{d}^D x \, \exp\left(\int \mathrm{d}^2 z \, i J^\nu x_\nu\right) = \frac{1}{(2\pi)^D} \delta^{(D)}\left(\sum_{j=1}^{k} k_j\right) \tag{3.71}$$

to (3.69). This is just momentum conservation for the external string states. We will make use of this e.g. in the form of the Mandelstam identities (2.26) to reduce the number of parameters, but not write it explicitly. Note how indirectly momentum conservation arises in the worldsheet calculation from the zero-mode contribution of the path integral.

The expression (3.70) is the *Koba–Nielsen factor* and will be frequently used in the following, so we introduce the abbreviating notation

$$\mathrm{KN}_n = \exp\left(\sum_{1\le i<j}^{n} s_{ij} G_{ij}\right). \tag{3.72}$$

### 3.2.2 *Tensor structure of gluon and graviton scattering*

In the last section, we discussed the plane-wave contribution to the vertex operators. In general, the full expression also contains additional bosonic, fermionic and (super) ghost contributions. In the case of four-point genus-one scattering, these contribute only to the kinematical and color prefactors of the amplitude, the *tensor structure*. Since the tensor structure does not depend on the puncture positions $z_i$ and the modular parameter $\tau$, it can be pulled out of the integral (3.48).

For the case of four gluon scattering at genus one in the heterotic string, the vertex operators are of the form

$$\mathcal{V}^a(z, \epsilon, k) = J^a(z) V_{\mathrm{SUSY}}(\bar{z}, \epsilon, k) e^{ik\cdot X(z,\bar{z})}, \tag{3.73}$$

where $a$ is the color index of the gauge boson, $\epsilon$ its polarization tensor and $k$ its momentum. $V_{\mathrm{SUSY}}$ carries (super) ghost and fermionic contributions and the $J^a$ are Kac-Moody currents which will be discussed in Section 6.1. The Kac-Moody correlators factorize and the remaining correlator takes the form

$$\begin{aligned}&\left\langle \prod_{j=1}^{4} V_{\text{SUSY}}(\bar{z}_j, \epsilon_j, k_j) e^{ik_j\cdot X(z_j,\bar{z}_j)} \right\rangle \\ &= t^{\mu\nu\sigma\rho\alpha\beta\gamma\delta} k_{1\mu}\epsilon_{1\nu}k_{2\sigma}\epsilon_{2\rho}k_{3\alpha}\epsilon_{3\beta}k_{4\gamma}\epsilon_{4\delta}\,\text{KN}_4 \\ &= (k_1\cdot k_2)(k_2\cdot k_3) A^{\text{tree}}_{\text{SYM}}(1,2,3,4)\,\text{KN}_4\,,\end{aligned} \tag{3.74}$$

where $A^{\text{tree}}_{\text{SYM}}(1,2,3,4)$ is the four-point color-ordered tree-level amplitude of ten-dimensional maximally supersymmetric Yang–Mills theory and $t^{\mu\nu\sigma\rho\alpha\beta\gamma\delta}$ is the $t_8$ tensor, defined by

$$\begin{aligned}&t^{\mu\nu\sigma\rho\alpha\beta\gamma\delta} k_{1\mu}\epsilon_{1\nu}k_{2\sigma}\epsilon_{2\rho}k_{3\alpha}\epsilon_{3\beta}k_{4\gamma}\epsilon_{4\delta} \\ &= \frac{1}{8}\Big(4M^1_{\mu\nu}M^2_{\nu\sigma}M^3_{\sigma\rho}M^4_{\rho\mu} - M^1_{\mu\nu}M^2_{\nu\mu}M^3_{\sigma\rho}M^4_{\rho\sigma}\Big) + \text{cyc}(2,3,4)\end{aligned} \tag{3.75}$$

where $M^i_{\mu\nu} = k_{i\mu}\epsilon_{i\nu} - \epsilon_{i\mu}k_{i\nu}$ and cyc(2, 3, 4) instructs to add the same term with the permutations (1, 3, 4, 2) and (1, 4, 2, 3) of the upper indices on the $M^i_{\mu\nu}$.

Another process which we will encounter frequently is four graviton scattering at genus one in type-IIA and type-IIB. In this case, the CFT correlator evaluates to [1]

$$t^{\mu_1\nu_1\ldots\mu_4\nu_4} t^{\rho_1\sigma_1\ldots\rho_4\sigma_4} \prod_{j=1}^{4} \epsilon_{j\mu_j\rho_j} k_{j\nu_j} k_{j\sigma_j}\,\text{KN}_4\ . \tag{3.76}$$

Note that from five points and genus one onward, the type-IIA and type-IIB amplitudes for external gravitons differ [169, 170]. The tensor structure in (3.76) is the linear contribution to

$$R^4 = t^{\mu_1\nu_1\ldots\mu_4\nu_4} t^{\rho_1\sigma_1\ldots\rho_4\sigma_4} R_{\mu_1\nu_1\rho_1\sigma_1} R_{\mu_2\nu_2\rho_2\sigma_2} R_{\mu_3\nu_3\rho_3\sigma_3} R_{\mu_4\nu_4\rho_4\sigma_4}\,, \tag{3.77}$$

where $R_{\alpha\beta\gamma\delta}$ is the ten-dimensional Riemann tensor. As explained in Section 2.3.3, this corresponds to the first term in the $\alpha'$ expansion of the low-energy effective action (2.23) of the gravitational sector of type-IIB. Since $\alpha'$ only appears in the Mandelstams via (2.25), the $\alpha'$ expansion is an expansion in the Mandelstam invariants and higher orders come with additional momenta and hence translate into derivatives acting on (3.77). E.g. the first correction to four-graviton scattering appears at order $(\alpha')^2$ and hence carries four additional momenta as compared to (3.76). The corresponding term in the effective action (2.23) has the structure [171]

$$\begin{aligned}\nabla^4 R^4 = {}& t^{\mu_1\nu_1\ldots\mu_4\nu_4} t^{\rho_1\sigma_1\ldots\rho_4\sigma_4} (\nabla_\alpha \nabla_\beta R_{\mu_1\nu_1\rho_1\sigma_1}) \\ &\times (\nabla^\alpha \nabla^\beta R_{\mu_2\nu_2\rho_2\sigma_2}) R_{\mu_3\nu_3\rho_3\sigma_3} R_{\mu_4\nu_4\rho_4\sigma_4}\,.\end{aligned} \tag{3.78}$$

### 3.2.3 *Structure of the integrands and Kronecker–Eisenstein series*

As we saw in the last section, in the case of four graviton scattering in type-IIB, the only $z$-dependent part of the CFT correlator was the Koba–Nielsen factor. In general however, the correlator will produce certain elliptic functions multiplying the Koba–Nielsen factor. This happens e.g. for the scattering of more than four gravitons in type-IIB [101, 169, 170, 172] and, as we will see in Section 6.1, the Kac-Moody correlator of the currents $J^a(z)$ in (3.73) will evaluate to an elliptic function.

In order to characterize the functions to which the CFT correlator evaluates, we start by considering the *Kronecker–Eisenstein series* [173] $F : \mathbb{C} \times \mathbb{C} \times \mathbb{H} \to \mathbb{C}$, defined by

$$F(z, \eta, \tau) = \frac{\theta_1'(0, \tau)\theta_1(z + \eta, \tau)}{\theta_1(z, \tau)\theta_1(\eta, \tau)}\,. \tag{3.79}$$

We can expand $F$ in $\eta$ to define the functions $g^{(a)}(z, \tau)$,

$$F(z, \eta, \tau) = \sum_{a \geq 0} \eta^{a-1} g^{(a)}(z, \tau)\,. \tag{3.80}$$

The Kronecker–Eisenstein series is meromorphic and symmetric in $z$ and $\eta$, but not doubly-periodic in these variables,

$$F(z + 1, \eta, \tau) = F(z, \eta, \tau) \qquad F(z + \tau, \eta, \tau) = e^{-2\pi i \eta} F(z, \eta, \tau)\,. \tag{3.81}$$

However, by adding a suitable prefactor, it can be lifted to a doubly-periodic function in $z$, $\Omega : \Sigma \times \mathbb{C} \times \mathbb{H} \to \mathbb{C}$, as follows:

$$\Omega(z, \eta, \tau) = \exp\left(2\pi i \eta \frac{\operatorname{Im} z}{\tau_2}\right) F(z, \eta, \tau)\,. \tag{3.82}$$

This version of the Kronecker–Eisenstein series satisfies

$$\Omega(z + n + m\tau, \eta, \tau) = \Omega(z, \eta, \tau)\,, \qquad m, n \in \mathbb{Z}\,, \tag{3.83}$$

but is no longer symmetric in $z$ and $\eta$. The double periodicity of $\Omega$ in $z = u\tau + v$ allows for the same double Fourier expansion in $u$ and $v$ that we already saw for the Green function,

$$\Omega(z, \eta, \tau) = \sum_p \frac{e^{2\pi i \langle p, z \rangle}}{p + \eta}\,, \tag{3.84}$$

where we have used the notation (3.66). Note that the sum here includes the $p = 0$ term. In the form (3.84), the reflection property

$$F(-z, -\eta, \tau) = -F(z, \eta, \tau) \tag{3.85a}$$

$$\Omega(-z, -\eta, \tau) = -\Omega(z, \eta, \tau) \tag{3.85b}$$

of the Kronecker–Eisenstein series is easy to see and (3.84) also makes the modular transformation behavior of $\Omega$ manifest: since

$$\Omega\left(\frac{z}{\gamma\tau+\delta}, \frac{\eta}{\gamma\tau+\delta}, \frac{\alpha\tau+\beta}{\gamma\tau+\delta}\right) = (\gamma\tau+\delta)\Omega(z,\eta,\tau) \tag{3.86}$$

for $\left(\begin{smallmatrix}\alpha & \beta \\ \gamma & \delta\end{smallmatrix}\right) \in \mathrm{SL}(2,\mathbb{Z})$ it is a non-holomorphic Jacobi form of weight $(1,0)$. On top of the properties mentioned so far, the Kronecker–Eisenstein series satisfies further algebraic and differential equations like the Fay identity discussed in Section 5.4.4 and the differential equations discussed in Section 7.2.1.

In analogy to (3.80), expanding $\Omega$ in $\eta$ defines doubly-periodic but non-holomorphic functions $f^{(a)}$ via

$$\Omega(z,\eta,\tau) = \sum_{a\geq 0} \eta^{a-1} f^{(a)}(z,\tau)\,. \tag{3.87}$$

From the representation (3.82) we deduce for the first two instances

$$f^{(0)}(z,\tau) = 1 \tag{3.88a}$$

$$f^{(1)}(z,\tau) = \partial_z \log\theta(z,\tau) + \frac{\pi}{\tau_2}(z-\bar{z})\,. \tag{3.88b}$$

Comparing this to (3.62) manifests the important relation

$$\partial_z G(z,\tau) = -f^{(1)}(z,\tau)\,. \tag{3.89}$$

Together with the differential equation (3.61) of the Green function, this implies

$$\partial_{\bar{z}} f^{(1)}(z) = \pi\delta^{(2)}(z,\bar{z}) - \frac{\pi}{\tau_2}\,. \tag{3.90}$$

In analogy to the notation $G_{ij}$ for the Green function, we write $f^{(a)}_{ij} = f^{(a)}(z_i - z_j, \tau)$. We calculate the formal Fourier expansion of the $f^{(a)}$ by expanding (3.84) according to (3.87) and obtain

$$f^{(a)}(z,\tau) = (-1)^{a-1} \sum_p{}' \frac{2\pi i\langle p,z\rangle}{p^a}\,, \quad a>0\,. \tag{3.91}$$

In the form (3.91), it is manifest that the $f^{(a)}(z,\tau)$ satisfy $f^{(a)}(-z,\tau) = (-1)^a f^{(a)}(z,\tau)$. For the complex conjugate functions, we have

$$\overline{f^{(a)}(z,\tau)} = -\sum_p{}' \frac{2\pi i\langle p,z\rangle}{\bar{p}^a}\,, \quad a>0\,. \tag{3.92}$$

For $a > 2$, (3.91) converges absolutely and $f(0,\tau)$ is given in terms of holomorphic Eisenstein series,

$$f^{(a)}(0,\tau) = -\mathrm{G}_a(\tau),. \tag{3.93}$$

In the sum (3.91), we can take the derivative w.r.t. $\bar{z}$ term-by-term and obtain, together with (3.90),

$$\partial_{\bar{z}} f^{(a)}(z,\tau) = -\frac{\pi}{\tau_2} f^{(a-1)}(z,\tau) + \pi \delta_{a,1} \delta^{(2)}(z,\bar{z}), \quad a \geq 1. \tag{3.94}$$

The form (3.91) also shows that the $f^{(a)}$ transform as non-holomorphic Jacobi forms of weight $(a,0)$,

$$f^{(a)}\left(\frac{z}{\gamma\tau+\delta}, \frac{\alpha\tau+\beta}{\gamma\tau+\delta}\right) = (\gamma\tau+\delta)^a f^{(a)}(z,\tau). \tag{3.95}$$

An important class of elliptic functions $V_a$ can be generated by a cyclic product of Kronecker–Eisenstein series via [28, 174]

$$\begin{aligned} &F(z_{12},\eta,\tau)F(z_{23},\eta,\tau)\dots F(z_{n-1,n},\eta,\tau)F(z_{n,1},\eta,\tau) \\ &= \Omega(z_{12},\eta,\tau)\Omega(z_{23},\eta,\tau)\dots\Omega(z_{n-1,n},\eta,\tau)\Omega(z_{n,1},\eta,\tau) \\ &= \eta^{-n}\sum_{a=0}^{\infty}\eta^a V_a(1,2,\dots,n). \end{aligned} \tag{3.96}$$

Here, the argument of the $V_a$ refers to the order of the $z_{ij}$ on the LHS and the dependency on the modular parameter and the puncture positions is left implicit. Since (3.96) is an elliptic function in $\eta$, the coefficients $V_{a>n}$ of $\eta^{k>0}$ are determined by the coefficients $V_{a\leq n}$ of the poles and constant term in $\eta$. Hence, we will only consider $V_a$ with $a \leq n$ in the following. Furthermore, the reflection property

$$F(-z,-\eta,\tau) = -F(z,\eta,\tau), \qquad \Omega(-z,-\eta,\tau) = -\Omega(z,\eta,\tau) \tag{3.97}$$

of the Kronecker–Eisenstein series implies the symmetries

$$\begin{aligned} V_a(1,2,\dots,n) &= V_a(2,\dots,n,1) \\ V_a(n,n-1,\dots,2,1) &= (-1)^a V_a(1,2,\dots,n). \end{aligned} \tag{3.98}$$

Using (3.87), the $V_a$ can be expressed in terms of the $f^{(a)}$ as

$$\begin{aligned} V_0(1,2,\dots,n) &= 1 \\ V_1(1,2,\dots,n) &= \sum_{j=1}^{n} f^{(1)}_{j,j+1} \\ V_2(1,2,\dots,n) &= \sum_{j=1}^{n} f^{(2)}_{j,j+1} + \sum_{i=1}^{n}\sum_{j=i+1}^{n} f^{(1)}_{i,i+1} f^{(1)}_{j,j+1} \quad \text{etc,} \end{aligned} \tag{3.99}$$

where $z_{n+1} = z_1$. In this form, it is also clear that the $V_a$ are meromorphic Jacobi forms of weight $(a, 0)$, i.e.

$$V_a(1, 2, \ldots, n)\Big|^{z_j \to \frac{z_j}{\gamma\tau+\delta}}_{\tau \to \frac{\alpha\tau+\beta}{\gamma\tau+\delta}} = (\gamma\tau + \delta)^a V_a(1, 2, \ldots, n)\,. \qquad (3.100)$$

The reason for introducing the Jacobi forms above is that the CFT correlators of bosonic, type-II and heterotic closed-string genus-one integrals are argued in [III] (cf. Section 6.2.5) to be expressible in terms of the $f^{(a)}$. In particular, the $f^{(a)}$ were shown to arise from spin sums in the RNS superstring [28] and from Kac-Moody correlators in the heterotic string [174], as we will see in detail in Section 6.1. For the type-IIB graviton amplitude, we saw that only the Koba–Nielsen factor contributes at four points. The additional terms in the five-point correlator were shown to be expressible in terms of the $f^{(a)}$ in the RNS formalism in [101] and in the pure-spinor formalism in [169, 170]. This was extended to six points in [172] and to seven points in [102–104]. In general, the closed string correlator is comprised of two chiral halves of the open string, coupled via the zero modes of the worldsheet bosons. That open string amplitudes can be expressed in terms of the $f^{(a)}$ was shown in [28, 101]. The additional terms from the left-right interactions also do not break this pattern, supporting the claim in [III]. More details on this argument can be found in [IV], cf. Appendix D.1. The same general argument also applies to orbifold compactifications with reduced supersymmetry [122].

The pattern of $f^{(a)}$ to which the Kac-Moody correlator of heterotic gauge bosons and the correlators of the supersymmetric parts of the vertex operators of gauge bosons and gravitons evaluate, takes a particularly simple form. These correlators can be expressed in terms of holomorphic Eisenstein series and the $V_a$ from (3.96), as argued in [III].

## 3.3 INTEGRATION OVER PUNCTURE POSITIONS

After having fixed the domain and measure of the integral to be computed in Section 3.1 and the form of the integrand in Section 3.2, we will now perform the integral over the first set of moduli, namely the positions of the punctures on the torus.

Unfortunately, it is not known how to preform this integral in closed form, hence, as alluded to before in Section 2.3.3, we will expand the integrand in $\alpha'$ and perform the integral order-by-order using modular graph forms. Since $\alpha'$ enters the amplitude only through the Mandelstam invariants (2.25), we obtain a low-momentum expansion of the amplitude which can be used e.g. to construct the low-energy effective field theory description of the string theory at hand.

### 3.3.1 *Structure of the low-energy expansion*

In this section, we will exemplify the structure of the low-energy expansion of the genus-one amplitude by considering four-graviton scattering in type-IIB, following [127, 175] where the study of modular graph functions was initiated. As seen above, in this case, the amplitude is given by

$$\mathcal{A}(s_{12}, s_{23}) = \int_{\mathcal{F}} \frac{d^2\tau}{\tau_2^2} \mathcal{I}(s_{12}, s_{23}, \tau) \tag{3.101}$$

with

$$\mathcal{I}(s_{12}, s_{23}, \tau) = \int_{\Sigma^3} d\mu_3 \, \mathrm{KN}_4 \,, \tag{3.102}$$

where we have dropped the tensor structure and coupling-constant prefactor.

Although the complete amplitude (3.101) is finite, its low-momentum expansion contains analytic and non-analytic pieces in the Mandelstams,

$$\mathcal{A}(s_{12}, s_{23}) = \mathcal{A}_{\text{an}}(s_{12}, s_{23}) + \mathcal{A}_{\text{non-an}}(s_{12}, s_{23}) \,, \tag{3.103}$$

where the analytic part can be expanded as

$$\mathcal{A}_{\text{an}}(s_{12}, s_{23}) = \sum_{p=0}^{\infty} \sum_{q=0}^{\infty} \sigma_2^p \sigma_3^q J^{(p,q)} \tag{3.104}$$

with

$$\sigma_2 = (s_{12}^2 + s_{23}^2 + s_{13}^2) \qquad \sigma_3 = (s_{12}^3 + s_{23}^3 + s_{13}^3) \tag{3.105}$$

and constant $J^{(p,q)}$. The expression (3.104) is the most general symmetric power series in three Mandelstams with $s_{12} + s_{23} + s_{13} = 0$. The non-analytic terms have branch cuts and are due to infrared effects of massless states in the loop. They decompose into the one-loop contribution from supergravity and stringy effects from the expansion of the tree amplitude [127]. In particular, by studying the structure of the non-analytic terms, one can obtain insights into the UV divergences of loop-level amplitudes in supergravity [176].

The non-analytic pieces of the amplitude arise from the region $\tau_2 \to \infty$ in moduli space (the degeneration limit of the torus) and therefore we cut the fundamental domain into a region $\mathcal{F}_L$ with $\tau_2 < L$ ($L \gg 1$) and a rectangular semi-infinite region $\mathcal{R}_L$ with $\tau_2 > L$ and $\frac{1}{2} < \tau_1 < \frac{1}{2}$. The amplitude can then be written as

$$\mathcal{A}(s_{12}, s_{23}) = \mathcal{A}_{\mathcal{F}_L}(s_{12}, s_{23}) + \mathcal{A}_{\mathcal{R}_L}(s_{12}, s_{23}) \,, \tag{3.106}$$

where

$$\begin{aligned} \mathcal{A}_{\mathcal{F}_L}(s_{12}, s_{23}) &= \int_{\mathcal{F}_L} \frac{\mathrm{d}^2\tau}{\tau_2^2} \mathcal{I}(s_{12}, s_{23}, \tau) \\ &= \mathcal{A}_{\mathrm{an}}(s_{12}, s_{23}) + R(s_{12}, s_{23}, L) \end{aligned} \tag{3.107}$$

$$\begin{aligned} \mathcal{A}_{\mathcal{R}_L}(s_{12}, s_{23}) &= \int_{\mathcal{R}_L} \frac{\mathrm{d}^2\tau}{\tau_2^2} \mathcal{I}(s_{12}, s_{23}, \tau) \\ &= \mathcal{A}_{\text{non-an}}(s_{12}, s_{23}) - R(s_{12}, s_{23}, L)\,. \end{aligned} \tag{3.108}$$

The splitting of the fundamental domain leads to an additional cutoff-dependent term $R(s_{12}, s_{23}, L)$ in both $\mathcal{A}_{\mathcal{F}_L}$ and $\mathcal{A}_{\mathcal{R}_L}$ which is analytic in the Mandelstams and cancels in the sum. Hence, if we are only interested in the analytic contributions, it is sufficient to calculate $\mathcal{A}_{\mathcal{F}_L}$ and drop all $L$-dependent terms, as we will do in Section 6.2.4.

The $\mathcal{R}_L$ contribution can be obtained by taking the $\tau_2 \to \infty$ limit of $\mathcal{I}$ and integrating over the semi-infinite rectangle. For $\mathcal{A}_{\mathcal{F}_L}$, we expand the Koba–Nielsen factor in Mandelstams and integrate order-by-order. Expanding the integral $\mathcal{I}$ in (3.102) using multinomial coefficients yields

$$\mathcal{I}(s_{12}, s_{23}, \tau) = \sum_{\ell=0}^{\infty} \frac{1}{\ell!} \sum_{\sum \ell_{ij} = \ell} \binom{\ell}{\ell_{12}, \ell_{13}, \dots, \ell_{34}} \prod_{1 \le i < j}^{4} s_{ij}^{\ell_{ij}} D_{\ell_{12},\dots,\ell_{34}}(\tau)\,, \tag{3.109}$$

where we have introduced the integrals [127]

$$D_{\ell_{12},\dots,\ell_{34}}(\tau) = \int \mathrm{d}\mu_3 \prod_{1 \le i < j}^{4} G_{ij}^{\ell_{ij}}(\tau)\,. \tag{3.110}$$

The expression (3.109) can be brought into the form (3.104) by imposing the Mandelstam relations (2.26) and relabeling the integration variables. For the next section, we will focus on the integrals (3.110). Since the Green function and the integration measure are modular invariant and the $D_{\ell_{12},\dots,\ell_{34}}$ were shown to have a $q, \bar{q}$ expansion of the type (3.27) [177], these integrals are modular functions. They are in fact, as we will see soon, our first example of modular graph functions.

### 3.3.2 *Modular graph functions and -forms*

When considering an integral of the form (3.110) it is natural to represent it as a graph, in which the punctures $z_1, \dots, z_4$ become vertices and the $G_{ij}$ become edges between the vertices $i$ and $j$. We denote $\ell_{ij}$ parallel edges with a label $[\ell_{ij}]$, resulting in the graph

$$D_{\ell_{12},\dots,\ell_{34}} = \left(\frac{\tau_2}{\pi}\right)^{\sum_{i<j}\ell_{ij}} \begin{array}{ccccc} 2 & \text{—} & [\ell_{23}] & \text{—} & 3 \\ | & \diagdown & & \diagup & | \\ | & & [\ell_{24}]\ [\ell_{13}] & & | \\ [\ell_{12}] & & \times & & [\ell_{34}] \\ | & \diagup & & \diagdown & | \\ 1 & \text{—} & [\ell_{14}] & \text{—} & 4 \end{array} \,, \tag{3.111}$$

where we have pulled out the factors of $\frac{\tau_2}{\pi}$ from the Green function (3.65) for compatibility with later conventions. This representation is the reason for the name "modular graph function" [15]. The evaluation of (3.111) is done in full analogy to the evaluation of Feynman diagrams: We can think of the Green functions between the vertices as internal propagators carrying discrete momenta $p = m\tau + n$ and use their Fourier representation (3.65) to solve the integral over the puncture positions yielding momentum conserving delta functions at the vertices.

As an example, consider the simpler modular invariant two-point integral [39]

$$D_\ell(\tau) = \int \mathrm{d}\mu_1 G_{12}^\ell = \left(\frac{\tau_2}{\pi}\right)^\ell \underset{1}{\bullet}\text{———} [\ell] \text{———}\underset{2}{\bullet} \,, \tag{3.112}$$

consisting of $\ell$ parallel edges. In order to evaluate (3.112), we arbitrarily assign discrete momenta $p_1, \dots, p_\ell$ to the edges and pull the sum from the Fourier expansion (3.65) out of the integral,

$$D_\ell(\tau) = \left(\frac{\tau_2}{\pi}\right)^\ell \sum_{p_1,\dots,p_\ell}{}' \frac{1}{|p_1|^2 \cdots |p_\ell|^2} \int \mathrm{d}\mu_1 e^{2\pi i \langle p_1 + \cdots + p_\ell, z_2\rangle} \,, \tag{3.113}$$

where we used that $z_1 = 0$. Now we use (with $p = m\tau + n$)

$$\int_\Sigma \frac{\mathrm{d}^2 z}{\tau_2} e^{2\pi i \langle p, z\rangle} = \delta(p) = \delta_{m,0}\delta_{n,0} \tag{3.114}$$

to obtain

$$D_\ell(\tau) = \left(\frac{\tau_2}{\pi}\right)^\ell \sum_{p_1,\dots,p_\ell}{}' \frac{\delta(p_1 + \cdots + p_\ell)}{|p_1|^2 \cdots |p_\ell|^2} \,. \tag{3.115}$$

Of course, we can preform one of the sums by removing the delta.

In general, we get one sum over momentum $p$ per Green function $G_{ij}$, along with a factor of $\left(\frac{\tau_2}{\pi}\right)\frac{1}{|p|^2}$ and a momentum-conserving delta function for each vertex. Note that one of the momentum conservation constraints is implied by the others (consistent with the vanishing of one contribution in the exponential due to $z_1 = 0$). In particular, (3.111) can be written as

$$D_{\ell_{12},\dots,\ell_{34}} = \left(\frac{\tau_2}{\pi}\right)^{\sum_{i<j}\ell_{ij}} \sum_{\{p_{ij}^{(k)}\}}' \prod_{i<j} \prod_{k=1}^{\ell_{ij}} \frac{1}{|p_{ij}^{(k)}|^2} \times \delta(p_{12}-p_{23}-p_{24})\delta(p_{13}+p_{23}-p_{34})\delta(p_{14}+p_{24}+p_{34})\,, \tag{3.116}$$

where we have assigned momentum $p_{ij}^{(k)}$ to the $k^{\text{th}}$ propagator from $i$ to $j$ and we used the shorthand

$$p_{ij} = \sum_{k=1}^{\ell_{ij}} p_{ij}^{(k)}\,. \tag{3.117}$$

In this way, we can associate a modular function to any graph. Before we go into more details about some of the properties of modular graph functions, we extend the discussion to the case of modular graph forms [16] (MGF),[6] namely functions which transform like modular forms with non-trivial modular weight and can be associated to labeled graphs.

As discussed in Section 3.2.3, the Koba–Nielsen factor is just part of the CFT correlator and in general the Koba–Nielsen factor is multiplied by a polynomial in the non-holomorphic Jacobi forms $f_{ij}^{(a)}$ and $\overline{f_{ij}^{(b)}}$ of weight $(a,0)$ and $(0,b)$, respectively. Their Fourier expansions (3.91) and (3.92) can be used to integrate them in exactly the same way as the Green functions, yielding edges of non-trivial modular weight.

To illustrate the resulting structure in detail, we make one further generalization by considering integrands which carry factors of the Jacobi forms $C^{(a,b)}(z,\tau)$ of weight $(a,b)$ introduced in (3.40). Recall that these have the Fourier expansion

$$C^{(a,b)}(z,\tau) = \sum_p' \frac{e^{2\pi i\langle p,z\rangle}}{p^a \bar{p}^b} \tag{3.118}$$

and we will write $C_{ij}^{(a,b)}$ for $C^{(a,b)}(z_{ij},\tau)$. Comparing to the Fourier expansion (3.65) of the Green function and (3.91) and (3.92) of $f^{(a)}$ and $\overline{f^{(b)}}$, it is clear that these are just special cases of the $C^{(a,b)}$, namely

$$\begin{aligned} G(z,\tau) &= \frac{\tau_2}{\pi} C^{(1,1)}(z,\tau) \\ f^{(a)}(z,\tau) &= (-1)^{a-1} C^{(a,0)}(z,\tau) \quad && a>0 \\ \overline{f^{(b)}}(z,\tau) &= -C^{(0,b)}(z,\tau) && b>0\,. \end{aligned} \tag{3.119}$$

6 We will use the abbreviation MGF for both modular graph forms and modular graph functions.

And we have $C^{(a,b)}(-z,\tau) = (-1)^{a+b}C^{(a,b)}(z,\tau)$, compatible with the previously observed symmetry properties. Hence, in order to integrate general genus-one CFT correlators at an arbitrary order in $\alpha'$, it is sufficient to consider an integral of the form

$$\mathcal{C}_\Gamma(\tau) = \int \mathrm{d}\mu_{n-1} \prod_{e\in E_\Gamma} C^{(a_e,b_e)}(z_e,\tau)\,, \tag{3.120}$$

where the product runs over a set of weights $(a,b)$ and differences $z_{ij}$ between puncture positions $z_1,\dots,z_n$.

The notation in (3.120) is suggestive of the graphical representation of the integral in terms of a graph $\Gamma$ with edge set $E_\Gamma$. In this language, each $C^{(a,b)}_{ij}$ in the integrand corresponds to an edge from vertex $i$ to vertex $j$ with label (weight) $(a,b)$:

$$C^{(a,b)}_{ij} \quad \leftrightarrow$$

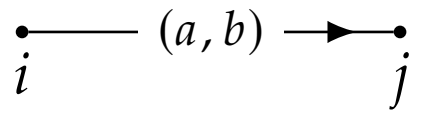

(3.121)

Note that this label $(a,b)$ assigns a weight to the edge and is very different from the labels $[\ell]$ used before which indicate $\ell$ parallel edges of weight $(1,1)$.

E.g. a *dihedral modular graph form* [16] with $R$ edges has the graph[7]

$$\mathcal{C}\left[\begin{smallmatrix} a_1 & \dots & a_R \\ b_1 & \dots & b_R\end{smallmatrix}\right] = \int \mathrm{d}\mu_1 \prod_{i=1}^{R} C^{(a_i,b_i)}_{12} =$$

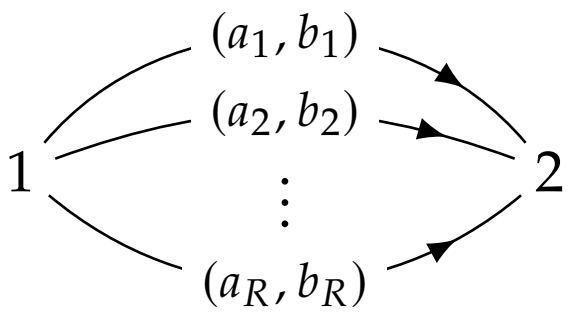

(3.122)

In order to evaluate a modular graph form (3.120), we assign momenta $p_e$ to the edges, aligned with their direction (in contrast to modular graph functions, where the direction was arbitrary) and impose momentum conservation at the vertices. Each edge then contributes a factor $\frac{1}{p^a \bar{p}^b}$ to the resulting sum and each vertex a delta function. For $\mathcal{C}_\Gamma$, we obtain

$$\mathcal{C}_\Gamma(\tau) = \sum_{\{p_e\}}{}' \prod_{e\in E_\Gamma} \frac{1}{p_e^{a_e}\bar{p}_e^{b_e}} \prod_{i\in V_\Gamma} \delta\left(\sum_{e'\in E_\Gamma} \Gamma_{ie'}p_{e'}\right)\,, \tag{3.123}$$

where $E_\Gamma$ is the set of edges of $\Gamma$, $V_\Gamma$ is the set of vertices and

$$\Gamma_{ie} = \begin{cases} 1 & \text{if } e \text{ is directed into } i \\ -1 & \text{if } e \text{ is directed out of } i \\ 0 & \text{if } e \text{ is not connected to } i \end{cases} \tag{3.124}$$

7 In the literature, various different conventions as to how many factors of $\pi$ and $\tau_2$ are included in the definition are being used. In our conventions, the modular weight is $\sum_i (a_i, b_i)$.

is the incidence matrix of vertex $i$. We will often use the notation

$$|A| = \sum_{e \in E_\Gamma} a_e \qquad |B| = \sum_{e \in E_\Gamma} b_e \,. \tag{3.125}$$

Since we now allow for arbitrary exponents $(a_e, b_e)$ of the momenta, it is not clear any more that the sum (3.120) converges and in fact for some choices of decorations $(a_e, b_e)$ it does not. This will become important in the following chapters and we will discuss details of the convergence properties in Section 5.6.

The sum representation (3.123) shows that modular graph forms transform as non-holomorphic modular forms[8] of weight $(|A|, |B|)$. Note that modular graph functions are modular graph forms with $a_e = b_e = k_e$, where $k_e$ are the labels of the modular graph function and a prefactor $(\frac{\tau_2}{\pi})^{|A|}$ to cancel the modular weight of the sum. This implies in particular that modular graph functions are real and non-holomorphic.[9] E.g. the dihedral modular graph functions [39],

$$C_{a,b,c} = \left(\frac{\tau_2}{\pi}\right)^{a+b+c} \mathcal{C}\left[\begin{smallmatrix} a & b & c \\ a & b & c \end{smallmatrix}\right] \tag{3.126}$$

$$C_{a,b,c,d} = \left(\frac{\tau_2}{\pi}\right)^{a+b+c+d} \mathcal{C}\left[\begin{smallmatrix} a & b & c & d \\ a & b & c & d \end{smallmatrix}\right], \tag{3.127}$$

have been studied extensively in the literature. Further special cases of dihedral MGF include

$$\mathcal{C}\left[\begin{smallmatrix} a & 0 \\ b & 0 \end{smallmatrix}\right] = \sum_p{}' \frac{1}{p^a \bar{p}^b} \tag{3.128}$$

$$\mathcal{C}\left[\begin{smallmatrix} k & 0 \\ 0 & 0 \end{smallmatrix}\right] = \mathrm{G}_k \,, \qquad k > 2 \tag{3.129}$$

$$\mathcal{C}\left[\begin{smallmatrix} s & 0 \\ s & 0 \end{smallmatrix}\right] = \left(\frac{\pi}{\tau_2}\right)^s \mathrm{E}_s \,, \qquad \mathrm{Re}(s) > 1 \,. \tag{3.130}$$

For lattice sums which are conditionally convergent, we assume a regularization of the form (3.31), so that we have in particular

$$\mathcal{C}\left[\begin{smallmatrix} 2 & 0 \\ 0 & 0 \end{smallmatrix}\right] = \widehat{\mathrm{G}}_2 \,. \tag{3.131}$$

We will now list a few important properties of MGFs:

- If $|A| + |B|$ is odd then $\mathcal{C}_\Gamma = 0$, this follows directly from the modular properties.
- Swapping the direction of an edge of weight $(a, b)$ produces a sign $(-1)^{a+b}$

8 It is not proven in general that the $q, \bar{q}$ expansion of modular graph forms has no negative powers, as required by (3.27), although a proof of this is given in [177] for special cases and it should be possible to generalize the argument in the reference.

9 In the literature, also the (weaker) condition $|A| = |B|$ is used to define modular graph functions which does not imply reality.

- If the graph of the MGF contains a one-valent vertex, the MGF vanishes, since momentum conservation forces the momentum of the corresponding edge to vanish, but $p = 0$ is excluded in the sum (3.123). In the language of Feynman diagrams, this means that all non-trivial MGF are vacuum bubbles.

- By the same argument, if the graph of the MGF can be disconnected by removing a single edge (for Feynman diagrams, this is called one-particle reducible), the momentum of the edge has to be zero, and hence the lattice sum vanishes. I.e. all non-trivial MGFs are one-particle irreducible.

- Two-valent vertices can be dropped by adding the weights of their edges:

$$\underset{i}{\bullet}\!\!-\!\!-(a_1,b_1)\!\!\longrightarrow\!\!\underset{j}{\bullet}\!\!-\!\!-(a_2,b_2)\!\!\longrightarrow\!\!\underset{k}{\bullet} \;=\; \underset{i}{\bullet}\!\!-\!\!-(a_1{+}a_2,b_1{+}b_2)\!\!\longrightarrow\!\!\underset{k}{\bullet} \tag{3.132}$$

  This shows in particular that the $C_{a,b,c}$ defined in (3.126) can be thought of as consisting of three chains of $a$, $b$ and $c$ Green functions, respectively,

$$C_{a,b,c} = \left(\frac{\tau_2}{\pi}\right)^{a+b+c}$$ 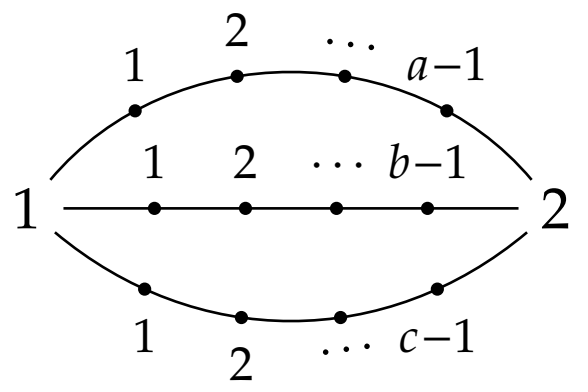

, (3.133)

  where each edge carries a label $(1,1)$ to indicate one Green function. Furthermore, this shows that

$$C^{(a,b)}_{ij} = (-1)^a \int \frac{\mathrm{d}^2 z_0}{\tau_2} f^{(a)}_{i0} \overline{f^{(b)}_{0j}} \tag{3.134}$$

  and hence every modular graph form can be written as a Koba–Nielsen integral with a monomial in $f^{(a)}$ and $\overline{f^{(b)}}$ in front of the Koba–Nielsen factor.

- If $\Gamma$ is disconnected, the MGFs for the disconnected components multiply

- If $\Gamma$ has connectivity one (i.e. by removing one vertex, it can be disconnected) then $C_\Gamma$ is the product of the MGFs of the components which become disconnected if the vertex is removed,

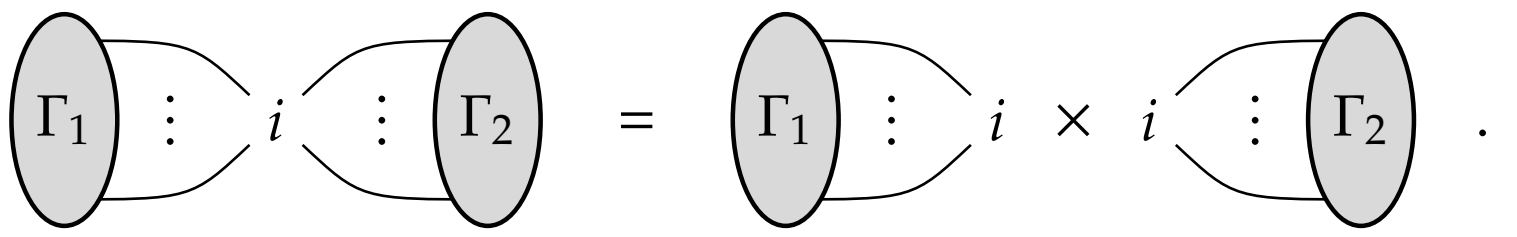

(3.135)

- Upon complex conjugating the MGF, the labels of the vertices are swapped,

$$\overline{C_\Gamma} = C_\Gamma \Big|_{a_e \leftrightarrow b_e} . \tag{3.136}$$

This implies e.g. for dihedral graphs

$$\overline{C\left[\begin{smallmatrix} a_1 & \dots & a_R \\ b_1 & \dots & b_R \end{smallmatrix}\right]} = C\left[\begin{smallmatrix} b_1 & \dots & b_R \\ a_1 & \dots & a_R \end{smallmatrix}\right] . \tag{3.137}$$

This shows that there are numerous non-trivial relations between MGFs and using these and more properties to be discussed in detail in Chapter 5 allows to simplify the final result dramatically. In this way, the integral (3.104) can be cast into the form [39]

$$\begin{aligned} \mathcal{I}(s_{12}, s_{23}, \tau) = 1 &+ \mathrm{E}_2\sigma_2 + \frac{1}{3}(5\mathrm{E}_3 + \zeta_3)\sigma_3 + \frac{1}{2}(\mathrm{E}_2^2 - \mathrm{E}_4 + 2C_{1,1,2})\sigma_2^2 \\ &+ \frac{1}{15}(35C_{1,1,3} + 25\mathrm{E}_2\mathrm{E}_3 - 34\mathrm{E}_5 + 5\mathrm{E}_2\zeta_3 + 3\zeta_5)\sigma_2\sigma_3 \\ &+ O(\alpha'^6) , \end{aligned} \tag{3.138}$$

which contains only modular graph functions with at most two loops. The terms shown here correspond to the terms $R^4$, $\nabla^4 R^4$, $\nabla^6 R^4$, $\nabla^8 R^4$ and $\nabla^{10} R^4$ in the effective action (2.23). In [40], the necessary identities between modular graph functions to simplify the order $\alpha'^6$, corresponding to $\nabla^{12} R^4$, were derived.

### 3.3.3 *Previous literature on modular graph functions and -forms*

Since the first papers on the subject appeared, the literature on modular graph functions and -forms has grown into a considerable body of work with contributions both from the physics- and mathematics community. In this section, we will give an (incomplete) overview over what has been achieved so far.

Graphical organizations of the $\alpha'$ expansion of the integral over puncture positions have been used for a long time and appear e.g. in [178] in the context of the heterotic string. The first papers [127, 175] in which modular graph functions appeared in a modern language studied the four-graviton amplitude in type-IIB. The $D_{\ell_{12},\dots,\ell_{34}}$ from (3.111) appeared first in [127]. In [39], modular graph functions were defined for general graphs and studied systematically for the first time. In this paper the $C_{a,b,c}$ from (3.126) were introduced, their Laplace eigenvalue equations investigated and their Poincaré series expansion introduced. The study of Laplace eigenvalue equations was continued for three-loop graphs with four vertices for the special case $D_{1,1,1,1,1,1}$ of (3.111) in [179] and for more general cases in [41]. The Laplace equation of general three-loop graphs with two vertices was discussed in [180].

In [181], a decomposition of the three-loop modular graph function $C_{1,1,1,1}$ into lower-loop functions was proven. Various identities between modular graph functions with four, five and six edges were proven in [182]. The relations between modular graph functions and elliptic polylogarithms (to be defined in Section 4.2) was investigated in [15].

The $q\bar{q}$ Fourier expansion of the form (3.27) and in particular the Laurent polynomial for modular graph functions has attracted a lot of attention in both physics and mathematics. Already in [127], Laurent polynomials of some modular graph functions were computed and used to derive identities between the lattice sums. In [183], an algorithm was given to compute the Laurent polynomial of the $D_{\ell_{12},\dots,\ell_{34}}$ in terms of conical sums. In this paper, it was also conjectured that the Laurent polynomials would only contain single-valued MZVs and rational numbers as coefficients. In [184], a formula for the Laurent polynomial of the $C_{a,b,c}$ functions was derived, which was extended in [185] to the complete space of weight $(a, a)$ two-loop modular graph functions. Also the Poincaré representation of the $C_{a,b,c}$ functions was worked out in this paper and it was proven that the one-loop graphs $C\left[\begin{smallmatrix} a & 0 \\ b & 0 \end{smallmatrix}\right]$ are linearly independent. A prescription for how to calculate the complete Fourier zero mode, including exponentially suppressed terms (cf. (3.28)), from the Poincaré series representation of modular graph functions using techniques from resurgence was worked out in [186]. In [187], the Laurent polynomials of certain linear combinations of $C_{a,b,c}$ functions were computed using their Poincaré representation. Finally, it was proven in several ways that the Laurent polynomials of modular graph functions of the form $C\left[\begin{smallmatrix} 1_n \\ 1_n \end{smallmatrix}\right]$, where $1_n$ is the vector $(1, \dots, 1)$ with $n$ entries, contain only odd single-zeta values, either by relating them to genus-zero amplitudes at four points [188, 189] or by recovering single-valued polylogarithms at genus zero [190].

Modular graph functions were derived for plane-wave backgrounds in [123] and appeared in little string theory [191]. Furthermore, they were generalized to genus-two surfaces for four graviton scattering [116–118, 192–194].

The concept of modular graph forms together with many important identities was introduced in [16]. In [195] modular graph forms were used to calculate the five-graviton amplitude. As we will see in Chapter 5, one prime interest in the study of modular graph forms are their numerous non-trivial relations. This was started in [16] and continued and extended to the study of Laplace equations in [40]. In [185] the space of two-loop cusp forms was investigated. From a mathematical perspective, in particular concerning their relation to iterated Eisenstein integrals (cf. Section 4.3.1 and Chapter 8), modular graph forms were studied in [32, 33, 74].

## 3.4 INTEGRATION OVER THE MODULAR PARAMETER

After having expressed the integral over the puncture positions in terms of lattice sums $C_\Gamma(\tau)$ of the form (3.123), the final step to obtain the complete one-loop amplitude is the integration over the modular parameter (cf. (3.101) for the case of four-graviton scattering). This integral is divergent since we ignored the non-analytic terms in the expansion in the Mandelstam variables and should be regulated with a cutoff in $\tau_2$. Integrals of modular functions over the fundamental domain can be computed e.g. by the technique of Rankin–Selberg unfolding which uses the Poincaré sum representation (3.35) of the integrand [196, 197]. In order to apply this technique to the divergent integrals mentioned above, it has to be extended to functions which are not of rapid decay. This can be done in a mathematically rigorous way [198], for an early application of these techniques to string amplitude calculations, see [199, 200].

In the cutoff-regularization introduced in Section 3.3.1, the $J^{(p,q)}$ from (3.104) will be of the form

$$J^{(p,q)} = \Xi^{(p,q)} + \sum_{i>1} a_i L^i + \sum_{i<1} a_i L^i + \log(L/\mu)\,. \tag{3.139}$$

The analytic contribution to the amplitude is, according to (3.107), given by the terms independent of $L$, $\Xi^{(p,q)}$ and $\log(\mu)$. Note however that the assignment of the $\log(\mu)$ term to the analytic part of the amplitude is ambiguous since there are corresponding threshold terms in the non-analytic part of the amplitude [127].

Unfortunately, there is no algorithmic way to perform this last integral over $\tau$, but on a case-by-case basis, various techniques exist to integrate large classes of modular graph functions. As an example, consider the important case that the function $f(\tau)$ to be integrated satisfies a Laplace eigenvalue equation of the form $\Delta f(\tau) = \omega f(\tau)$ (as is the case e.g. for the non-holomorphic Eisenstein series, cf. (3.60), or for certain linear combinations of $C_{a,b,c}$ functions [39] and four-point modular graph functions [41]). Then, the integrand is a total derivative and is reduced to a boundary integral by Stokes' theorem,

$$\int_{\mathcal{F}_L} \frac{d^2\tau}{\tau_2^2} f(\tau) = \int_{\mathcal{F}_L} \frac{d^2\tau}{\tau_2^2} \frac{\Delta f(\tau)}{\omega} = \frac{1}{\omega} \int_{-\frac{1}{2}}^{\frac{1}{2}} d\tau_1 \left(\partial_{\tau_2} f(\tau)\right)_{\tau_2 = L}, \tag{3.140}$$

where the other contributions of the boundary cancel due to modular invariance of $f(\tau)$. Since the cutoff is taken to be large ($L \gg 1$), we can replace $f$ by its Laurent polynomial, which trivializes the $\tau_1$ integral. This shows e.g. that $\int_{\mathcal{F}_L} \frac{d^2\tau}{\tau_2^2} \mathrm{E}_s$ has no $L$-independent contributions for $s > 1$.

In [39], the analytic contribution (3.104) to the genus-one four graviton amplitude was calculated by integrating (3.138) order-by-order in the Mandelstams and found to be[10]

$$\begin{aligned}\mathcal{A}_{\text{an}}(s_{12}, s_{23}) = \frac{\pi}{3} + \frac{\pi}{9}\zeta_3\sigma_3 + \frac{2\pi\zeta_3}{45}\left(\log(2) + \frac{\zeta_4'}{\zeta_4} - \frac{\zeta_3'}{\zeta_3} - \frac{1}{4}\right)\sigma_2^2 \\ + \frac{\pi}{3}\frac{29}{180}\zeta_5\sigma_2\sigma_3 + O(\alpha'^6)\,,\end{aligned} \tag{3.141}$$

where $\zeta_k'$ is the derivative of the Riemann zeta function (2.30) evaluated at $k$. In [128], this was extended to the order $\alpha'^6$ and the all-order version of (3.141) was proven to be free of irreducible MZVs (i.e. MZVs that cannot be written as products of single zeta values with rational coefficients) and the full amplitude was conjectured to be of maximal transcendentality, cf. Section 6.2.3 for a review of transcendentality.

Using the Poincaré series representation of the two-loop modular graph functions obtained in [185], a closed formula for the integrals over general two-loop modular graph functions was obtained in [201].

10 In [39], the authors chose to move the term proportional to $\sigma_2^2$ into the non-analytic part of the amplitude.

# 4

## ONE-LOOP OPEN-STRING AMPLITUDES

One of the main focus points of this work is the relation between open- and closed-string amplitudes at one-loop, in the spirit of the tree-level single-valued relation reviewed in Section 2.4. For this reason, we will give a brief overview over the structures appearing in the calculation of one-loop amplitudes in the open string in this chapter. In Sections 4.1 and 4.2, we review the general setup of one-loop open-string amplitudes and in Section 4.3, we will outline previous steps that were taken in the literature to generalize the single-valued relation from tree-level to one-loop.

### 4.1 STRUCTURE OF OPEN STRING AMPLITUDES

In the unoriented type-I string, the worldsheets of open strings are non-orientable Riemann surfaces with boundaries. The one-loop contributions ($\chi = 0$) are then, according to Table 2.1, the cylinder and the Möbius strip. The cylinder has two boundaries and the Möbius strip one and the vertex operators are inserted on those boundaries. The integral over moduli space then contains a sum over the distribution of the order of the vertex operators on these boundaries and (for the cylinder) also for the ways to distribute them on the different boundaries.

Amplitudes involving open and closed strings are also possible and have additional closed-string vertex operators inserted into the volume of the worldsheet of an open-string amplitude. For a discussion of these mixed amplitudes at tree-level, cf. [3, 202].

#### 4.1.1 *Open string one-loop worldsheets*

In order to parametrize the worldsheet cylinder (or annulus) it is treated as a special case of the torus with purely imaginary modular parameter $\tau = it$, $t \in \mathbb{R}_+$, as depicted in Figure 4.1. The two boundaries of the cylinder are realized by imposing the involution $z \equiv \bar{z}$, which has fixed points $\mathrm{Im}(z) = 0$ and $\mathrm{Im}(z) = \frac{\tau_2}{2}$ (when using the elliptic identification $z \equiv z + \tau$) and therefore generates boundaries on these lines. The identification $z \equiv z + 1$ from the torus is unaffected by this and again, we can fix $z_1 = 0$. If all vertex operators are inserted on one boundary, the amplitude is called *planar*, if they are inserted on several boundaries, the amplitude is *non-planar*.

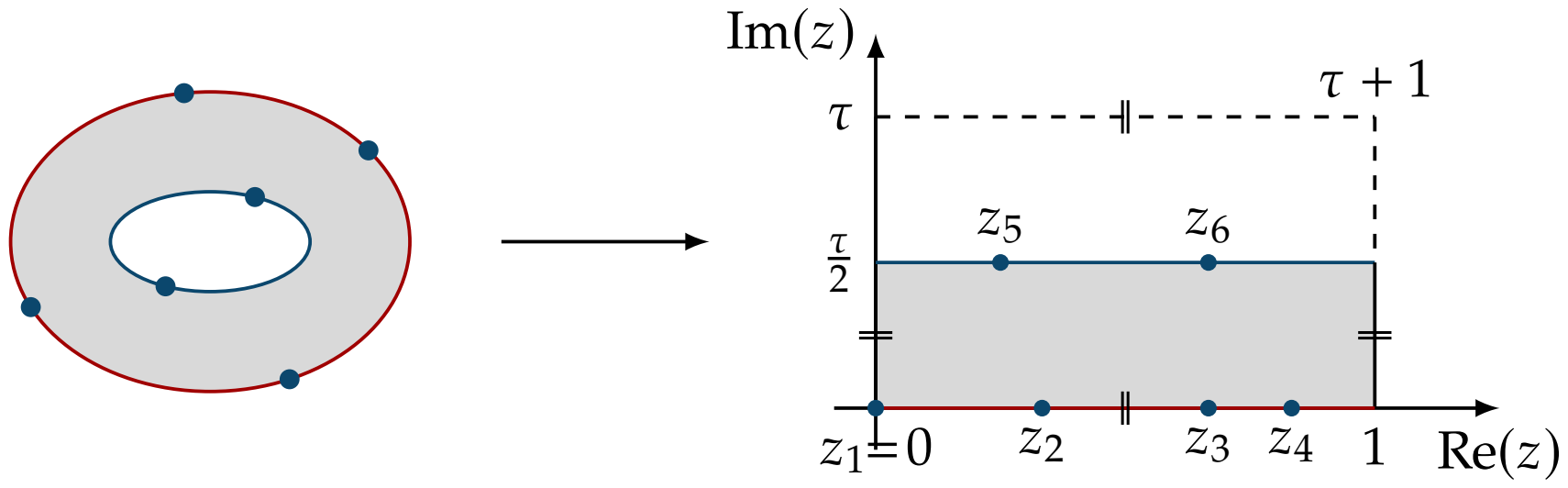

Figure 4.1: Parametrization of the worldsheet cylinder as a torus with modular parameter $\tau = it$, $t \in \mathbb{R}_+$, subject to the involution $z \equiv \bar{z}$. The lines $\text{Im}(z) = \frac{\tau}{2}$ and $\text{Im}(z) = 0$ are fixed points of the involution and become the cylinder boundaries.

The Möbius strip can be obtained by imposing the additional involution $z \equiv z + \frac{\tau}{2} + \frac{1}{2}$ on the cylinder which identifies the two boundaries and introduces a twist so that the resulting surface has only one boundary and is non-orientable. Hence, the contributions from the Möbius strip can be calculated in exactly the same way as the planar contributions to the cylinder [51], we just have to integrate $\tau$ over $\frac{1}{2} + i\mathbb{R}_+$, instead of $i\mathbb{R}_+$. For this reason, we will restrict the remaining discussion to cylinder amplitudes. For a more general recent discussion on how to obtain various open-string worldsheets from those of closed strings, see [190].

Planar amplitudes are labeled by the (cyclic) ordering of the vertex operators on the boundary and integrate over their positions while preserving this ordering. The integral over the modular parameter becomes an integral over the imaginary axis and the full expression is

$$\mathcal{A}^{\text{open}}_{\text{planar}}(1, 2, \ldots, n) = g_s^{\frac{n}{2}} \int_0^\infty \mathrm{d}t \int_0^1 \mathrm{d}z_n \int_0^{z_n} \mathrm{d}z_{n-1} \cdots \int_0^{z_3} \mathrm{d}z_2 \langle \prod_{i=1}^n \mathcal{V}_1^i(k_i, z_i) \text{ ghosts} \rangle \,. \tag{4.1}$$

For non-planar amplitudes, the two integration cycles are independent.

### 4.1.2 *Open string CFT correlators*

The vertex operators in the open string consist of one chiral half of the vertex operators of the closed string. The calculations are therefore very similar and in particular, we obtain the same Koba–Nielsen factor (3.72) for the correlator of plane-wave operators. The only difference to the closed string in the scaling of the Green function and the Mandelstam variables. To achieve comparability to the closed string, we use the definition

$$\mathrm{KN}_n^{\mathrm{open}} = \exp\left(\sum_{1\leq i<j}^{n} \frac{1}{2} s_{ij} G_{ij}\right) = (\mathrm{KN}_n)^{\frac{1}{2}} \tag{4.2}$$

for the Koba–Nielsen factor in the open string.[1]

As in the closed string, the additional contributions to the vertex operators yield the tensor structure of the amplitude. As mentioned in Section 2.1.2, the endpoints of the open strings in type-I theory are attached to 32 spacetime-filling D-branes which give rise to gauge fields with gauge group SO(32). Since the same gauge fields also appear in the heterotic string, the amplitudes carry the same tensor structures, namely (3.74). The color factor to the tensor structure, which for heterotic strings arises from the Kac–Moody correlator and will be discussed in Section 6.1.1, is for open strings due to the distribution of the vertex operators on the boundary. For each boundary with vertex operators $1, \ldots, r$ which carry color indices $a_1, \ldots, a_r$, we get a factor $\mathrm{Tr}(t^{a_1} \cdots t^{a_r})$, where the $t^{a_i}$ are the generators of SO(32). Hence, the full planar four-point amplitude is given by

$$\begin{aligned} \mathcal{A}^{\mathrm{open}}_{\mathrm{planar}} = \; & g_s^2 (k_1 \cdot k_2)(k_2 \cdot k_3) \\ & \times \sum_{\rho \in S_3} \mathrm{Tr}(t^{a_1} t^{a_{\rho(2)}} t^{a_{\rho(3)}} t^{a_{\rho(4)}}) A^{\mathrm{tree}}_{\mathrm{SYM}}(1, \rho(2), \rho(3), \rho(4)) \\ & \times \int_0^\infty \mathrm{d}t \, \mathcal{I}^{\mathrm{open}}_{1\rho(2)\rho(3)\rho(4)}(s_{12}, s_{23}, \tau) \, , \end{aligned} \tag{4.3}$$

where

$$\mathcal{I}^{\mathrm{open}}_{1234}(s_{12}, s_{23}, \tau) = \int_{0 \leq z_2 \leq z_3 \leq z_4 \leq 1} \mathrm{d}z_2 \mathrm{d}z_3 \mathrm{d}z_4 \, \mathrm{KN}_4^{\mathrm{open}} \, . \tag{4.4}$$

For amplitudes with more than four external states, there are additional contributions to the CFT correlator besides the Koba–Nielsen factor and the tensor structure, as was already the case for the closed string. These can be expressed as homogeneous polynomials in the $f^{(a)}$ functions introduced in (3.87). The same structure appeared also in the closed string. However, in the maximally supersymmetric type-II theories, the polynomial in $f^{(a)}$ is balanced by an equal-weight polynomial in $f^{(b)}$. This balancing is broken in the half maximally supersymmetric heterotic theories, in the half maximally supersymmetric type-I open string the $f^{(b)}$ are absent entirely.

1 This is the same convention which was also used in [34] and the definition in [28, 203] can be obtained by rescaling $s_{ij} \to 4 s_{ij}$ in (4.2).

## 4.2 INTEGRATION OVER PUNCTURE POSITIONS

The structure of the CFT correlator, together with the integration domain (4.1) manifests that open string one-loop amplitudes are naturally given as iterated integrals with integration kernels $f^{(a)}$. Iterated integrals of this type are known as *elliptic multiple zeta values* (eMZVs) [30] and in this section, we will review how the integration over puncture positions in the open string can be performed in terms of these objects. For the comparison to the closed string, we will also discuss a reformulation of the eMZVs in terms of iterated integrals over holomorphic Eisenstein series, since MGFs can be expressed in this language as well.

### 4.2.1 *Elliptic multiple zeta values*

Elliptic multiple zeta values are defined in terms of elliptic iterated integrals [28]

$$\begin{aligned}\Gamma\left(\begin{smallmatrix} n_1 & n_2 & \cdots & n_r \\ a_1 & a_2 & \cdots & a_r \end{smallmatrix} ; z, \tau\right) &= \int_0^z \mathrm{d}t\, f^{(n_1)}(t,\tau)\Gamma\left(\begin{smallmatrix} n_2 & \cdots & n_r \\ a_2 & \cdots & a_r \end{smallmatrix} ; t, \tau\right) \\ \Gamma(\emptyset; z, \tau) &= 1\,,\end{aligned} \tag{4.5}$$

where $z \in \mathbb{R}$ and we integrate along the real line. For this restriction, (4.5) coincide with the *multiple elliptic polylogarithms* defined in [204]. Using (4.5), eMZVs are given by the special values

$$\omega(n_1, n_2, \ldots, n_r; \tau) = \Gamma\left(\begin{smallmatrix} n_r & n_{r-1} & \cdots & n_1 \\ 0 & 0 & \cdots & 0 \end{smallmatrix} ; 1, \tau\right). \tag{4.6}$$

We will refer to the sum of the $n_i$ as the *weight* of the eMZV. The name of these objects refers to the similar construction for MZVs as special values of iterated integrals, cf. (2.46), however, despite their name, eMZVs are not numbers (as MZVs), but still functions of $\tau$.

In order to perform the integral (4.4) in terms of eMZVs, we note that according to (3.89), we have for $z \in \mathbb{R}$

$$\partial_z G(z,\tau) = -2 f^{(1)}(z,\tau) \quad \Rightarrow \quad G(z,\tau) = -2\Gamma\left(\begin{smallmatrix} 1 \\ 0 \end{smallmatrix}; z, \tau\right) + k(\tau). \tag{4.7}$$

The function $k(\tau)$ is required to regularize the elliptic integrals [203] and depends on the regularization scheme chosen, but it cancels from the Koba–Nielsen factor (4.2) upon momentum conservation $\sum_{i<j} s_{ij} = 0$. With (4.7), the integral (4.4) can be evaluated order-by-order in Mandelstams by using the definition (4.5), resulting in iterated elliptic integrals evaluated at one. However, some of those elliptic integrals will have $z$ dependent entries in the second row. As detailed in [28], these can be removed by exploiting differential equations in the $a_i$ labels. The end result can be written entirely in terms of eMZVs,

$$\begin{aligned}\mathcal{I}^{\text{open}}_{1234}(s_{12},s_{23}) = \frac{1}{6} &- 2s_{13}\omega(0,1,0,0) + 2(s_{12}^2+s_{23}^2)\omega(0,1,1,0,0) \\ &- 2s_{12}s_{23}\omega(0,1,0,1,0) + s_{13}(s_{13}^2-s_{12}s_{23})\beta_5 + s_{12}s_{23}s_{13}\beta_{2,3} + O(\alpha'^4)\,,\end{aligned} \tag{4.8}$$

where we have introduced the shorthand notation

$$\begin{aligned}\beta_5 = \frac{4}{3}\big[&\omega(0,0,1,0,0,2) + \omega(0,1,1,0,1,0) \\ &- \omega(2,0,1,0,0,0) - \zeta_2\omega(0,1,0,0)\big]\end{aligned} \tag{4.9}$$

$$\beta_{2,3} = \frac{\zeta_3}{12} + \frac{8}{3}\zeta_2\omega(0,1,0,0) - \frac{5}{18}\omega(0,3,0,0)\,. \tag{4.10}$$

Note that a database of eMZV relations to high weight is available on the website [205].

For non-planar amplitudes, the notion of eMZVs is generalized to *twisted elliptic multiple zeta values* (teMZVs), defined by [203]

$$\omega\left(\begin{smallmatrix} n_1 & n_2 & \cdots & n_r \\ a_1 & a_2 & \cdots & a_r \end{smallmatrix};\tau\right) = \Gamma\left(\begin{smallmatrix} n_r & n_{r-1} & \cdots & n_1 \\ a_r & a_{r-1} & \cdots & a_1 \end{smallmatrix};1,\tau\right), \tag{4.11}$$

where in string calculations $a_i \in \{0, \frac{\tau}{2}\}$. When the complete non-planar amplitude is assembled however, the teMZVs cancel and the result can be expressed purely in terms of ordinary eMZVs. E.g. the contribution to the four-point amplitude with two vertex operators on each boundary is given by

$$\mathcal{I}^{\text{open}}_{12|34}(s_{12},s_{23}) = \int_{\frac{\tau}{2}}^{\frac{\tau}{2}+1} \mathrm{d}z_4 \int_{\frac{\tau}{2}}^{z_4} \mathrm{d}z_3 \int_0^1 \mathrm{d}z_2\, \mathrm{KN}_4^{\text{open}} \tag{4.12}$$

$$\begin{aligned}&= \frac{q^{s_{12}/4}}{2}\Bigg[1 + s_{12}^2\left(\frac{7\zeta_2}{6} + 2\omega(0,0,2)\right) \\ &\qquad -2s_{13}s_{23}\left(\frac{\zeta_2}{3} + \omega(0,0,2)\right) + O(\alpha'^3)\Bigg]\,,\end{aligned} \tag{4.13}$$

where $z_1 = 0$, $z_2$ is integrated along the real line and $z_3 < z_4$ are integrated along the second boundary $\text{Im}(z) = \frac{\tau_2}{2}$. The cancellation of teMZVs was proven recently in [37, 38] for general $n$-point integrals to all orders in $\alpha'$ using a differential equation for a non-planar version of the generating function of Koba–Nielsen integrals defined in (7.3) below.

The conventions above, in which the cylinder boundaries are identified with the $A$ cycle $\text{Im}(z) = 0$ of the torus (cf. Figure 3.1) and the parallel $\text{Im}(z) = \frac{\tau_2}{2}$ to it, give rise to the *A-cycle eMZVs* of (4.5) and (4.6). An equivalent parametrization can be obtained by identifying the cylinder boundaries with the $B$ cycle $\text{Re}(z) = 0$ and its parallel $\text{Re}(z) = \frac{1}{2}$, giving rise to *B-cycle eMZVs*. These are special values of elliptic iterated integrals with integration path from zero to $\tau$. $A$- and $B$-cycle eMZVs are related by the modular S-transformation according to [30, 34]

$$\omega_A\left(n_1,\dots,n_r;-\frac{1}{\tau}\right) = \tau^{n_1+\dots+n_r-r}\omega_B(n_1,\dots,n_r;\tau)\,. \tag{4.14}$$

As we will see in Section 4.3, the $B$-cycle eMZVs are the starting point for the correspondence with the closed string.

### 4.2.2 *Iterated Eisenstein integrals*

Although the evaluation of open-string Koba–Nielsen integrals in terms of eMZVs is completely algorithmic, for the comparison with the closed string expansion, we need a different language. This is the language of *iterated Eisenstein integrals*, through which both eMZVs and MGFs can be expressed. As the name suggests, iterated Eisenstein integrals are iterated integrals over $\tau$ whose integration kernels are holomorphic Eisenstein series. They are recursively defined by

$$\begin{aligned}\mathcal{E}_0(k_1,\dots,k_r;\tau) &= 2\pi i\int_\tau^{i\infty}\mathrm{d}\tau_r\frac{\mathrm{G}^0_{k_r}(\tau_r)}{(2\pi i)^{k_r}}\mathcal{E}_0(k_1,\dots,k_{r-1};\tau_r)\\ \mathcal{E}_0(\emptyset;\tau) &= 1\,,\end{aligned} \tag{4.15}$$

where we subtracted the Fourier zero mode from the holomorphic Eisenstein series (cf. (3.24)), since the integral over the infinite region would otherwise diverge, and defined

$$\mathrm{G}^0_k = \begin{cases}-1 & \text{for } k=0\\ \mathrm{G}_k-2\zeta_k & \text{for } k=4,6,8\dots\end{cases}\,. \tag{4.16}$$

Therefore, the labels $k_1,\dots,k_r$ in the argument of $\mathcal{E}_0$ take on the values 0 and $4+2n$ for $n\in\mathbb{N}_0$ and for convergence, we require $k_1\geq 4$. We will refer to the number of nonzero $k_i$ in the argument as the *depth* of the iterated Eisenstein integral. The definition (4.15) implies immediately that the $\mathcal{E}_0$ satisfy the differential equation

$$\pi\nabla_0\,\mathcal{E}_0(k_1,\dots,k_r) = \frac{4y^2}{(2\pi i)^{k_r}}\mathrm{G}^0_{k_r}\,\mathcal{E}_0(k_1,\dots,k_{r-1})\,, \tag{4.17}$$

where $y=\pi\tau_2$ and $\nabla_0$ was defined in (3.55). Furthermore, integrating the $q$-series (3.24) term-by-term yields a closed expression for the $q$-series of the $\mathcal{E}_0$,

$$\begin{aligned}\mathcal{E}_0(k_1,0^{p_1-1},k_2,0^{p_2-1},\dots,k_r,0^{p_r-1};\tau) &= (-2)^r\left(\prod_{j=1}^r\frac{1}{(k_j-1)!}\right)\\ \times\sum_{m_i,n_i=1}^{\infty}&\frac{m_1^{k_1-1}m_2^{k_2-1}\cdots m_r^{k_r-1}\,q^{m_1n_1+m_2n_2+\dots+m_rn_r}}{(m_1n_1)^{p_1}\,(m_1n_1+m_2n_2)^{p_2}\cdots(m_1n_1+m_2n_2+\dots+m_rn_r)^{p_r}}\,.\end{aligned} \tag{4.18}$$

As all suitably regularized iterated integrals, also iterated Eisenstein integrals satisfy shuffle relations [206]. We therefore have

$$\mathcal{E}_0(A)\,\mathcal{E}_0(B) = \mathcal{E}_0(A \sqcup\!\sqcup B)\,, \tag{4.19}$$

where $A$ and $B$ are words in the entries $k_i$ and $\sqcup\!\sqcup$ is the same shuffle product as the one defined for MZVs in (2.49). Furthermore, $\mathcal{E}_0$ with different labels are linearly independent [35].

Since the $\mathrm{G}_k^0$ have a $q$ expansion with vanishing zero mode, so do the $\mathcal{E}_0$, hence they are in particular invariant under modular T-transformations. A closed formula for the entire $q$-series is available [29]. The modular S-transformation of iterated Eisenstein integrals gives rise to *multiple modular values* [31] which have been calculated for some cases in [34]. In particular, the $\mathcal{E}_0$ do not transform as modular forms as the following example illustrates:

$$\begin{aligned}\mathcal{E}_0\left(4,0;-\frac{1}{\tau}\right) = \frac{T^2}{1080} + \frac{\pi^2}{216} - \frac{i\zeta_3}{6T} - \frac{\pi^4}{360T^2}\\ + \mathcal{E}_0(4,0;\tau) + \frac{i}{T}\,\mathcal{E}_0(4,0,0;\tau)\,,\end{aligned} \tag{4.20}$$

where $T = \pi\tau$.

The significance of the integrals (4.15) lies in the fact that both eMZVs and MGFs can be written in terms of them and hence they are the right language to compare open- and closed-string $\alpha'$ expansions at one-loop. In particular, it is possible to obtain an expression for eMZVs in terms of iterated Eisenstein integrals because a closed formula is known for the derivative w.r.t. $\tau$ of eMZVs and it is given by eMZVs of lower length and holomorphic Eisenstein series [30], e.g. (with $\mathrm{G}_0 = -1$)

$$2\pi i\frac{\mathrm{d}}{\mathrm{d}\tau}\omega(0,n) = -2n\zeta_{n+1}\mathrm{G}_0 - n\mathrm{G}_{n+1}\,, \qquad n \text{ odd} \tag{4.21a}$$

$$2\pi i\frac{\mathrm{d}}{\mathrm{d}\tau}\omega(0,0,n) = n\omega(0,n+1)\mathrm{G}_0\,, \qquad n \text{ even}\,. \tag{4.21b}$$

Applying these kinds of equations successively and then integrating back builds up linear combinations of products of iterated Eisenstein integrals [29]. The integration constants generated in this procedure can be obtained from the elliptic *Knizhnik–Zamolodchikov–Bernard* (KZB) *associator*, the generating series of eMZVs, which, in the limit $\tau \longrightarrow i\infty$ is given in terms of the Drinfeld associator [30], the generating series of MZVs. In this way, one can obtain e.g. [34]

$$\omega_A(2,0,0) = -6\,\mathcal{E}_0(4,0) - \frac{1}{3}\zeta_2\,, \tag{4.22}$$

and in particular, the entire $\alpha'$ expansion of the open-string amplitude can be expressed in terms of iterated Eisenstein integrals. One further important property of the $\mathcal{E}_0$ is that they are linearly independent, making relations between eMZVs and MGFs apparent.

When translating eMZVs into iterated Eisenstein integrals, only certain combinations of iterated Eisenstein integrals appear, which are selected by the *derivation algebra* [207, 208]. The derivation algebra is the algebra satisfied by derivations $\epsilon_{2n}$ ($n \in \mathbb{N}_0$) appearing in the differential equation of the elliptic KZB associator [30] which act on non-commutative variables $x$ and $y$ via

$$\epsilon_0(x) = y \tag{4.23a}$$

$$\epsilon_0(y) = 0 \tag{4.23b}$$

$$\epsilon_{2n}(x) = (\mathrm{ad}_x)^{2n}(y)\,, \qquad n > 0 \tag{4.23c}$$

$$\begin{aligned}\epsilon_{2n}(y) = {}&[y, (\mathrm{ad}_x)^{2n-1}(y)]\\ &+ \sum_{1\leq j<n} (-1)^j [(\mathrm{ad}_x)^j(j), (\mathrm{ad}_x)^{2n-1-j}(y)]\,, \quad n > 0\,,\end{aligned} \tag{4.23d}$$

where $\mathrm{ad}_x(y) = [x, y] = xy - yx$. The definitions (4.23) imply infinitely many relations between the derivations $\epsilon_{2n}$, e.g.

$$0 = [\epsilon_{2n}, \epsilon_2]\,, \qquad n \geq 0 \tag{4.24a}$$

$$0 = (\mathrm{ad}_{\epsilon_0})^{2n-1}(\epsilon_{2n})\,, \qquad n > 0 \tag{4.24b}$$

$$0 = [\epsilon_{10}, \epsilon_4] - 3[\epsilon_8, \epsilon_6] \tag{4.24c}$$

$$0 = 2[\epsilon_{14}, \epsilon_4] - 7[\epsilon_{12}, \epsilon_6] + 11[\epsilon_{10}, \epsilon_8] \tag{4.24d}$$

$$\begin{aligned}0 = {}&80[\epsilon_{12},[\epsilon_4,\epsilon_0]]+16[\epsilon_4,[\epsilon_{12},\epsilon_0]]-250[\epsilon_{10},[\epsilon_6,\epsilon_0]]-125[\epsilon_6,[\epsilon_{10},\epsilon_0]]\\ &+280[\epsilon_8,[\epsilon_8,\epsilon_0]]-462[\epsilon_4,[\epsilon_4,\epsilon_8]]-1725[\epsilon_6,[\epsilon_6,\epsilon_4]]\,.\end{aligned} \tag{4.24e}$$

These relations can be associated to cusp forms [209] and have been made available to high weight in [205]. To obtain an action of the derivations on iterated Eisenstein integrals, the latter are represented in terms of formal words in the non-commutative letters $g_0, g_2, g_4, \ldots$ [29]. The derivations act on these by replacing the leftmost letter with a Kronecker delta,

$$\epsilon_{2n} g_{k_1} \cdots g_{k_n} = \delta_{2n,k_1} g_{k_2} \cdots g_{k_n}\,. \tag{4.25}$$

In this way, the relations (4.24) in the derivation algebra imply relations between iterated Eisenstein integrals which have to be satisfied by the particular combinations appearing in eMZVs.

## 4.3 COMPARING OPEN- AND CLOSED-STRING AMPLITUDES

Having reviewed the calculation of open-string amplitudes in the previous section, we are now in a position to discuss the results of [34], where first steps were undertaken to generalize the single-valued map from tree-level amplitudes (as discussed in Section 2.4) to a putative *elliptic single-valued map* which relates one-loop amplitudes. To this end, the authors write open-string $B$-cycle integrals from the four-gluon open-string amplitude in terms of iterated Eisenstein integrals and compare the resulting expression to iterated Eisenstein integrals they obtain from the expansion of the four-graviton closed-string amplitude in terms of modular graph functions. This was the first concrete realization of the elliptic single-valued map but that the closed-string amplitude is maximally supersymmetric is an important restriction on this result.

### 4.3.1 *Iterated Eisenstein integrals and modular graph functions*

After discussing how to express the open-string $\alpha'$ expansion in terms of iterated Eisenstein integrals, the last step for the comparison to the closed-string $\alpha'$ expansion is to express also modular graph forms in terms of iterated Eisenstein integrals.

The idea for this is analogous to the conversion from eMZVs to iterated Eisenstein integrals: We take derivatives in $\tau$ and integrate back using the definition of the $\mathcal{E}_0$ (4.15). The sieve algorithm [16], to be discussed in detail in Section 5.5, is a systematic procedure to identify a power of the Cauchy–Riemann derivative of an MGF which can be written as a linear combination of products of MGFs of lower weight and holomorphic Eisenstein series. Some examples of the resulting Cauchy–Riemann equations are

$$\nabla_0^2 \mathrm{E}_2 = 6\frac{\tau_2^4}{\pi^2}\mathrm{G}_4 \tag{4.26a}$$

$$\nabla_0^3 \mathrm{C}_{1,1,1} = 60\frac{\tau_2^6}{\pi^3}\mathrm{G}_6 \tag{4.26b}$$

$$\nabla_0^3 \mathrm{C}_{1,1,2} = \frac{9}{10}\nabla_0^3 \mathrm{E}_4 - 6\frac{\tau_2^4}{\pi^2}\mathrm{G}_4 \nabla_0 \mathrm{E}_2 \tag{4.26c}$$

$$\nabla_0^3\left(\left(\frac{\tau_2}{\pi}\right)^3 \mathcal{C}\left[\begin{smallmatrix} 0 & 1 & 2 \\ 1 & 2 & 2 \end{smallmatrix}\right]\right) = -6\frac{\tau_2^4}{\pi}\mathrm{E}_2\mathrm{G}_4 - \frac{3}{2}\pi(\nabla_0 \mathrm{E}_2)^2 - \frac{3}{5}\pi\nabla_0^2\mathrm{E}_4\,, \tag{4.26d}$$

where the factors of $\tau_2$ were added to make the argument of $\nabla_0$ modular invariant as required, cf. (3.56).[2] Integrating these Cauchy–Riemann equations over $\tau$ repeatedly results in a representation of the MGF in terms of iterated Eisenstein integrals and integration constants. These can be fixed by requiring the right modular behavior under S transformations, which is very non-trivial for the representation in terms of iterated Eisenstein integrals, cf. (4.20), and reality in case the MGF under consideration is real. A final integration constant can be

2 In general, the Cauchy–Riemann equations will also contain MGFs which cannot immediately be written in terms of Cauchy–Riemann derivatives of non-holomorphic Eisenstein series as in (4.26), cf. e.g. the second derivative of $\mathcal{C}\left[\begin{smallmatrix} 1 & 1 & 2 \\ 1 & 2 & 1 \end{smallmatrix}\right]$ in (5.142). These cases are discussed in Section 5.5.2.

fixed if the Laurent polynomial of the MGF is known. In this way, one obtains e.g. [34]

$$E_2 = \frac{y^2}{45} + \frac{\zeta_3}{y} - 12\,\mathrm{Re}[\mathcal{E}_0(4,0)] - \frac{6}{y}\,\mathrm{Re}[\mathcal{E}_0(4,0,0)] \tag{4.27a}$$

$$C_{1,1,1} = \frac{3y^3}{945} + \zeta_3 + \frac{3\zeta_5}{4y^2} - 120\,\mathrm{Re}[\mathcal{E}_0(6,0,0)] - \frac{180}{y}\,\mathrm{Re}[\mathcal{E}_0(6,0,0,0)] - \frac{90}{y^2}\,\mathrm{Re}[\mathcal{E}_0(6,0,0,0,0)]\,, \tag{4.27b}$$

where again $y = \pi\tau_2$. In practice, since the calculation of the modular S-transformation of iterated Eisenstein integrals is hard, it is easier to decompose all MGFs into a basis of simple MGFs and their derivatives and only compute the representation in terms of iterated Eisenstein integrals for the basis elements. We will discuss this basis in detail in Section 5.7.

It is convenient to choose basis elements for MGFs with a particularly simple representation in terms of iterated Eisenstein integrals by considering combinations of modular graph forms with simple Cauchy–Riemann equations (4.26). To this end, we define [34]

$$E_{2,2} = C_{1,1,2} - \frac{9}{10}E_4 \tag{4.28a}$$

$$E_{2,3} = C_{1,1,3} - \frac{43}{35}E_5 \tag{4.28b}$$

$$E_{3,3} = 3C_{1,2,3} + C_{2,2,2} - \frac{15}{14}E_6 \tag{4.28c}$$

$$E'_{3,3} = C_{1,2,3} + \frac{17}{60}C_{2,2,2} - \frac{59}{140}E_6 \tag{4.28d}$$

$$E_{2,4} = 9C_{1,1,4} + 3C_{1,2,3} + C_{2,2,2} - 13E_6 \tag{4.28e}$$

$$E_{2,2,2} = -C_{1,1,2,2} + \frac{232}{45}C_{2,2,2} + \frac{292}{15}C_{1,2,3} + \frac{2}{5}C_{1,1,4} + 2E_3^2 + E_2E_4 - \frac{466}{45}E_6\,, \tag{4.28f}$$

so that e.g.

$$\nabla_0^3 E_{2,2} = -6\frac{\tau_2^6}{\pi^2}G_4\nabla_0 E_2\,, \tag{4.29}$$

with one term less as compared to (4.26c). The Cauchy–Riemann equations of the remaining objects in (4.28) are listed in (5.211) below. Furthermore, the MGFs in (4.28) separate the different sectors of iterated Eisenstein integrals: E.g. $E_{2,2}$ contains only integrals over two instances of $G_4$, cf. (4.29), whereas $C_{1,1,2}$ contains also integrals over $G_8$, cf. (4.26c). For all the modular graph forms in (4.28), explicit representations in terms of iterated Eisenstein integrals are listed in [34], together with a closed expression for $\mathrm{E}_k$. Hence, given a decomposition of an MGF in terms of the objects in (4.28) and their Cauchy–Riemann derivatives, a representation in terms of iterated Eisenstein integrals can be computed using the known action (4.17) of $\nabla_0$ on $\mathcal{E}_0$.

### 4.3.2 *The elliptic single-valued map for maximal supersymmetry*

Having expressed both the open- and closed-string quantities in terms of iterated Eisenstein integrals, we can now start to compare. This can most easily be done at the level of individual contributions to the $\alpha'$ expansions. On the closed-string side, these are modular graph functions for the four-graviton amplitude we want to consider. On the open-string side, we consider the planar contribution to four-gluon scattering, but in order to obtain a similar-looking expansion, we replace the nested integration domain in (4.4) by an integration over the full range of all punctures,

$$\int_{0\leq z_2\leq z_3\leq z_4\leq 1} \mathrm{d}^3 z_i \longrightarrow \int_0^1 \mathrm{d}z_2 \int_0^1 \mathrm{d}z_3 \int_0^1 \mathrm{d}z_4 \,. \tag{4.30}$$

This can be thought of as considering auxiliary abelian external states (which do not exist in the spectrum of the type-I string), for which the color factors in (4.3) are trivial and the sum over permutations of vertex operator orderings assembles the integration domain (4.30). For this new integral, one can organize contributions of the corresponding four-gluon amplitude into similar graphs by expanding the open-string Koba–Nielsen factor (4.2), yielding *open string graph functions* $\mathcal{O}_\Gamma(\tau)$. It was found empirically in [34] that

$$\mathcal{C}_\Gamma(\tau) = \mathrm{esv}\left[\mathcal{O}_\Gamma\left(-\frac{1}{\tau}\right)\right] , \tag{4.31}$$

where the elliptic single-valued map esv acts on the representation in terms of iterated Eisenstein integrals via

$$\mathrm{esv} : \begin{cases} T \longrightarrow 2iy \\ \mathcal{E}_0(k_1,\dots,k_r) \longrightarrow 2\,\mathrm{Re}[\mathcal{E}_0(k_1,\dots,k_r)]\,, \quad k_1 \neq 0\,. \\ \zeta_{n_1,\dots,n_r} \longrightarrow \zeta^{\mathrm{sv}}_{n_1,\dots,n_r} \end{cases} \tag{4.32}$$

Note that $T \longrightarrow 2iy$ does not act on the $q$-series of the $\mathcal{E}_0$ and is in fact a special case of the second line since $\mathcal{E}_0(0) = 2\pi i\tau$. As an example, consider the graph $\Gamma =$ 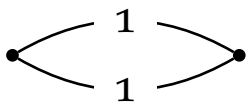

with two Green functions for which $\mathcal{C}_\Gamma(\tau) = \mathrm{E}_2(\tau)$ with its representation in terms of iterated Eisenstein integrals given in (4.27). The modular S-transformation of $O_\Gamma$ is given by [34]

$$O_\Gamma\Big(-\frac{1}{\tau}\Big) = -\frac{T^2}{180} + \frac{i\zeta_3}{T} - 6\,\mathcal{E}_0(4,0) - \frac{6i\,\mathcal{E}_0(4,0,0)}{T} \quad \mathrm{mod}\ \zeta_2\,, \tag{4.33}$$

where $\mathrm{mod}\,\zeta_2$ indicates that we suppressed terms proportional to $\zeta_2$, since $\zeta_2^{\mathrm{sv}} = 0$. It is easy to verify that in this case, (4.32) holds. The reason for introducing the modular S-transformation in (4.31) is that the $O_\Gamma$ do not produce the Laurent polynomials required by the modular graph functions. Although (4.31) is conjectural at this point, it was proven in [189] for two-point integrals at the level of the Laurent polynomials.

The conjecture (4.31) can easily be extended to the entire Koba–Nielsen integral. Taking the new integration domain (4.30) in the definition of the open-string graph functions into account yields the conjecture for the single-valued relation between the open-string Koba–Nielsen integral (4.4) and the closed-string Koba–Nielsen integral (3.109),

$$\mathcal{I}^{\mathrm{closed}}(\tau) = \mathrm{esv}\Bigg[\sum_{\rho\in S_3} \mathcal{I}^{\mathrm{open}}_{1\rho(2)\rho(3)\rho(4)}\Big(-\frac{1}{\tau}\Big)\Bigg] \tag{4.34}$$

order-by-order in $\alpha'$ [34].

Although the conjecture (4.34) has been tested until weight six and is an encouraging step towards an elliptic version of the tree-level single-valued relation, the Koba–Nielsen integrals (3.109) and (4.4) are both maximally supersymmetric and therefore very constrained. On the closed string side, this implies in particular that only modular graph functions appear, and no modular graph forms. We will go beyond this limitation by studying half-maximally supersymmetric heterotic amplitudes in Chapter 6 which in turn had key input on the general proposal for an elliptic single-valued map presented in [17].

Unfortunately, as discussed in [34], even for the maximally supersymmetric case, the conjecture (4.32) for the esv-map has a major conceptual problem: it is not an algebra homomorphism for the iterated Eisenstein integrals, i.e. the second line of (4.32) does not preserve the shuffle property. This raises the question on which shuffle-representation of the open-string $B$-cycle graph functions esv should be applied. Although no principle is known to select the right representation in general, empirically, it was always possible to find a representation in which (4.31) was correct. This shortcoming is expected to be absent in the construction of [17].

# 5

## PROPERTIES OF MODULAR GRAPH FORMS

As explained in the previous chapter, the modular graph forms introduced in Section 3.3.2 are the central objects in the integration over puncture positions of closed-string one-loop amplitudes. In fact, performing this integral is trivialized by the definition of MGFs, but the results are complicated nested lattice sums, cf. e.g. (3.116), which can be simplified dramatically, cf. (3.138). In order to compare the expansion of Koba–Nielsen integrals to open-string integrals, this simplification is unavoidable, as discussed in Section 4.3.2 and the same is true if one wants to integrate the MGFs over the modular parameter to obtain the full amplitude as in Section 3.4. This motivates a systematic study of the properties of modular graph forms and their relations.

Although many non-trivial relations between modular graph forms exist, these are often not obvious from the lattice sum representation. Indeed, arguably the simplest identity found for MGFs is

$$\begin{aligned} C_{1,1,1} &= \left(\frac{\tau_2}{\pi}\right)^3 \sum_{p_1,p_2}{}' \frac{1}{|p_1|^2|p_2|^2|p_1+p_2|^2} \\ &= \left(\frac{\tau_2}{\pi}\right)^3 \sum_{p}{}' \frac{1}{|p|^6} + \sum_{n=1}^{\infty} \frac{1}{n^3} \\ &= \mathrm{E}_3 + \zeta_3 \end{aligned} \tag{5.1}$$

which was obtained by direct evaluation of the sums [210]. Since finding identities in this way is tedious, we will discuss in this chapter more systematic ways of obtaining relations of the form (5.1). Using those techniques, we will be able to decompose all convergent two- and three-point MGFs with non-negative edge labels of weight $|A| + |B| \leq 12$ into linear combinations of the simple lattice sums listed in Table 5.3. The results discussed in Sections 5.4.2 and 5.4.3 were published in [I]. The remaining results were published in [II] and the present text has extensive overlap with these references.

This Chapter is structured as follows: We begin in Section 5.1 with a short overview of the `Mathematica` package `Modular Graph Forms` which implements many of the simplifications described in this chapter so that we can introduce the implementation of the various objects and formulae as we go along in the following sections. The notation for MGFs with up to four vertices used throughout is set in Section 5.2. In Section 5.3 we review several ways to obtain relations without performing any sums and in Section 5.4 we discuss a technique to perform certain sums in special MGFs systematically, going by the name of *holomorphic subgraph reduction* (HSR). Section 5.5 reviews an algorithm to combine HSR with the techniques of Section 5.3 to obtain relations for large classes of MGFs, up to an overall constant. Since in the derivation of identities between modular graph forms also divergent MGFs appear, we will discuss their properties in Section 5.6. Finally, we exhibit the structure of the basis of MGFs of weight $|A| + |B| \leq 12$ in Section 5.7. Further details about various points can be found in Appendix B

## 5.1 THE `Modular Graph Forms` MATHEMATICA PACKAGE

As alluded to in the last paragraphs, we will present a number of simplification techniques for MGFs in this chapter, which will allow us to derive basis decompositions for a large number of MGFs, as discussed in Section 5.7. To make the resulting decompositions accessible, it is convenient to have a computer database of them, together with an implementation for the various techniques to be discussed. This is realized in the `Mathematica` package `Modular Graph Forms` which is included in the arXiv submission of [II]. It contains 11079 identities to decompose all two- and three-point MGFs of total modular weight $a + b \leq 12$ into the basis given in Section 5.7 and functions for basic manipulations of four-point graphs. The package furthermore contains routines to automatically expand Koba–Nielsen integrals in terms of MGFs. In this section, we will outline the basic usage of the package and, as we discuss the manipulations for MGFs in the following sections, we will also describe how they are implemented in the `Modular Graph Forms` package. A complete reference of all defined symbols as well as all functions and their options is given in Appendix A.

### 5.1.1 *Basics*

The `Modular Graph Forms` package consists of the package itself in the `ModularGraphForms.m` file and two files containing identities between two- and three-point MGFs which were generated using the techniques presented in this chapter. To load the package, copy all files in a directory in `Mathematica`'s search path (e.g. into the directory in which the current notebook is saved) and run

```
In[1]:= Get["ModularGraphForms.m"]

        Dihedral identity file found at /home/user/DiIds.txt

        Trihedral identity file found at /home/user/TriIds.txt
```

```
Loaded 1559 identities for dihedral convergent MGF.

Loaded 9520 identities for trihedral convergent MGF.

Successfully loaded the ModularGraphForms package. Have
fun!
```

The notation used for $\tau$ is **tau**, $\bar{\tau}$ is **tauBar** and $\tau_2$ is **tau[2]**. The (multiple) zeta values (2.41) are written as **zeta[3,3,5]** and the holomorphic Eisenstein series (3.18) and their complex conjugates are **g[4]** and **gBar[4]**, respectively. For the modular version $\widehat{\mathrm{G}}_2$ of $\mathrm{G}_2$, we use **gHat[2]** and **gBarHat[2]**. The non-holomorphic Eisenstein series (3.33) and their higher-depth generalizations (4.28) are denoted for instance by **e[2,2]**. The normalizations are as described in Section 3.1.2.

The modular weight of an expression is determined by the function **CModWeight**, e.g.

In[2]:= $\texttt{CModWeight}\left[\texttt{g[4]} + \texttt{gHat[2]}^2 + \left(\frac{\texttt{tau[2]}}{\pi}\right)^4 \texttt{e[2,2] gBar[4] g[8]}\right]$

Out[2]= {4,0} .

Complex conjugation is performed by the function **CComplexConj**, e.g.

In[3]:= $\texttt{CComplexConj}\left[\texttt{g[4]} + \texttt{gHat[2]}^2 + \left(\frac{\texttt{tau[2]}}{\pi}\right)^4 \texttt{e[2,2] gBar[4] g[8]}\right]$

Out[3]= $\bar{\mathrm{G}}_4 + \hat{\bar{\mathrm{G}}}_2^2 + \frac{\mathrm{E}_{2,2}\,\mathrm{G}_4\,\bar{\mathrm{G}}_8\,\tau_2^4}{\pi^4}$ .

The most important function of the Modular Graph Forms package is the function **CSimplify**, which performs all known simplifications for MGFs on the expression in the argument, e.g. the identity (5.1) is hard-coded into the package and can be used as follows,

In[4]:= $\texttt{CSimplify}\left[\texttt{c}\left[\begin{smallmatrix}1 & 1 & 1\\ 1 & 1 & 1\end{smallmatrix}\right]\right]$

Out[4]= $\frac{\pi^3\,\mathrm{E}_3}{\tau_2^3} + \frac{\pi^3\,\zeta_3}{\tau_2^3}$ ,

where the notation for modular graph forms will be explained in Section 5.2 below. The function **CSimplify** calls the functions **DiCSimplify**, **TriCSimplify** and **TetCSimplify**, which perform simplifications on two- three- and four-point graphs, respectively.

The remaining functions in the Modular Graph Forms package will be discussed in the following sections, along with the manipulations for MGFs they implement. A complete reference for all functions and their options is given in Appendix A. Within Mathematica, short explanations for the various objects can be obtained using the **Information** function, e.g. by running **?CModWeight**. A complete list of all available objects can be printed by running **?ModularGraphForms`***.

| function | definition | Mathematica representation |
|---|---|---|
| $f_{ij}^{(a)}$ | (3.91) | `fz[a,i,j]` |
| $\overline{f_{ij}^{(b)}}$ | (3.92) | `fBarz[b,i,j]` |
| $G_{ij}$ | (3.65) | `gz[i,j]` |
| $C_{ij}^{(a,b)}$ | (3.118) | `cz[a,b,i,j]` |
| $V_a(k_1,\ldots,k_r)$ | (3.96) | `vz[k1,…,kr]` |
| $\overline{V_b(k_1,\ldots,k_r)}$ | (3.96) | `vBarz[k1,…,kr]` |

Table 5.1: Various $z$-dependent functions defined and used in Section 3.3 and their representation in Mathematica.

### 5.1.2 *Expanding Koba–Nielsen integrals*

As explained in Section 3.3, in string theory, modular graph forms arise as coefficients in the expansion of Koba–Nielsen integrals. The Modular Graph Forms package also contains the function **zIntegrate** which performs this expansion automatically. The syntax is as follows: **zIntegrate** has three arguments, the first one is the prefactor in front of the Koba–Nielsen factor, the second one is the number of points in the Koba–Nielsen factor (3.72) and the last one is the order in Mandelstam variables which is written in terms of MGFs. E.g. the second order in Mandelstams of the three-point integral

$$\int \mathrm{d}\mu_2 \, \mathrm{KN}_3 \tag{5.2}$$

is computed by

```
In[5]:= zIntegrate[1,3,2]//Factor
```

$$\text{Out[5]=} \ \frac{1}{2} \mathrm{E}_2 \left(s_{1,2}^2 + s_{1,3}^2 + s_{2,3}^2\right)$$

Note that all Mandelstam variables are treated as independent, no momentum conservation is imposed. A Koba–Nielsen factor which does not contain all Green functions and Mandelstam variables of (3.72) can be represented by replacing the second argument with a list of point pairs, corresponding to the Green functions appearing in the Koba–Nielsen factor. E.g. $\exp(s_{12}G_{12} + s_{13}G_{13})$ is represented by **{{1,2},{1,3}}**. For an integral without Koba–Nielsen factor, we can set the second argument of **zIntegrate** to an arbitrary number and the third argument to zero.

For the integrand in front of the Koba–Nielsen factor, the functions listed in Table 5.1 are available. To indicate their $z$ dependence, they all carry a suffix **z**. An arbitrary polynomial in these functions can be given as the first argument to **zIntegrate**. E.g. the first order in Mandelstams of the integral

$$\int \mathrm{d}\mu_3 V_2(1,2,3,4)\,\mathrm{KN}_4 \tag{5.3}$$

is computed by

```
In[6]:= zIntegrate[vz[2, {1, 2, 3, 4}], 4, 1] // Factor
```

$$\text{Out[6]=} \ -\frac{\mathrm{C}\left[\begin{smallmatrix}3 & 0\\ 1 & 0\end{smallmatrix}\right](s_{1,2} - 2\,s_{1,3} + s_{1,4} + s_{2,3} - 2\,s_{2,4} + s_{3,4})\,\tau_2}{\pi} \,.$$

The function **zIntegrate** returns MGFs in the notation introduced in Section 5.2 below for MGFs with up to four points, while exploiting the basic properties of MGFs listed in Section 3.3.2. If MGFs with more than four points appear in the expansion and they cannot be reduced by using these properties, they are printed as a graph, e.g.

```
In[7]:= zIntegrate[gz[1,2]^2 gz[2,3]^2 gz[3,4]^2 gz[4,5]^2 gz[5,1],5,0]
```

Out[7]=

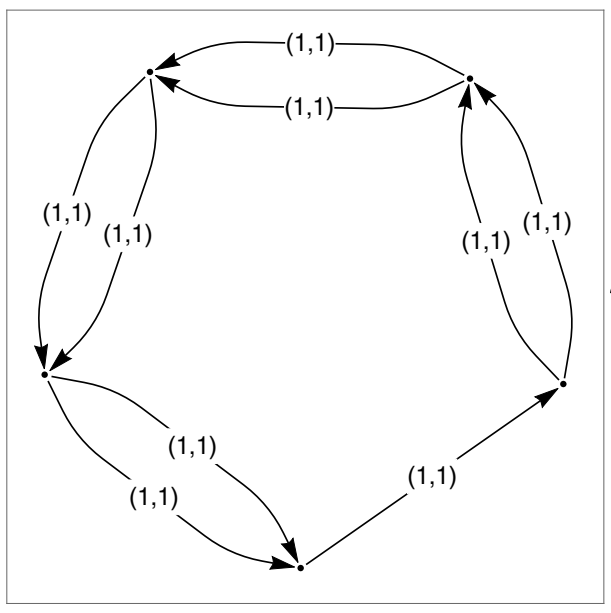

$\tau_2^9$

$\overline{\qquad\qquad}$

$\pi^9$ .

Note that if the Koba–Nielsen integral expanded using **zIntegrate** contains kinematic poles due to $f^{(1)}_{ij}\overline{f^{(1)}_{ij}}$ terms in the integrand, **zIntegrate** will contain divergent MGFs, as will be discussed in detail in Section 5.6.2.

Using the function **zIntegrate** and the decompositions discussed in Section 5.7 below, the two- and three-point generating functions for Koba–Nielsen integrals were evaluated in terms a few basis-MGFs up to total modular weight 12, as discussed in Chapter 8.

## 5.2 GRAPH TOPOLOGIES AND NOTATION

The general definition (3.123) for modular graph forms depends on a graph Γ with decorated and directed edges, where the decoration has the form $(a, b)$ with $a, b \in \mathbb{Z}$. Since it is inconvenient to specify the entire graph for every MGF, we introduce commonly used notations for graphs with up to four vertices, the only ones considered in this chapter.

### 5.2.1 *Two-point modular graph forms*

As introduced in (3.122), dihedral graphs have two vertices and all edges directed in the same way. They are denoted by

$$C\left[\begin{smallmatrix} a_1 & \cdots & a_R \\ b_1 & \cdots & b_R \end{smallmatrix}\right] = $$

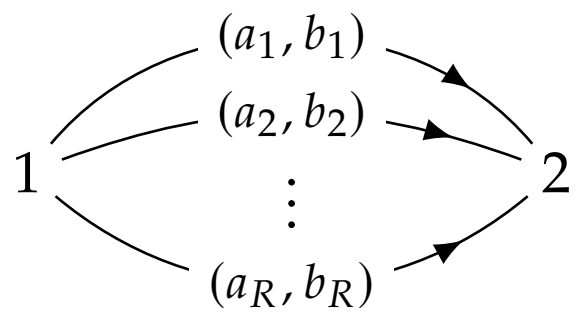

$$= \sum_{p_1,\dots,p_R}' \frac{\delta(p_1+\dots+p_R)}{p_1^{a_1}\bar{p}_1^{b_1}\cdots p_R^{a_R}\bar{p}_R^{b_R}} \,. \tag{5.4}$$

Since we will frequently encounter a bundle of parallel edges, we write

$$C\left[\begin{smallmatrix} A \\ B \end{smallmatrix}\right] = C\left[\begin{smallmatrix} a_1 & \cdots & a_R \\ b_1 & \cdots & b_R \end{smallmatrix}\right] \tag{5.5}$$

and call $\left[\begin{smallmatrix} A \\ B \end{smallmatrix}\right]$ a *block*. In graphs, we draw

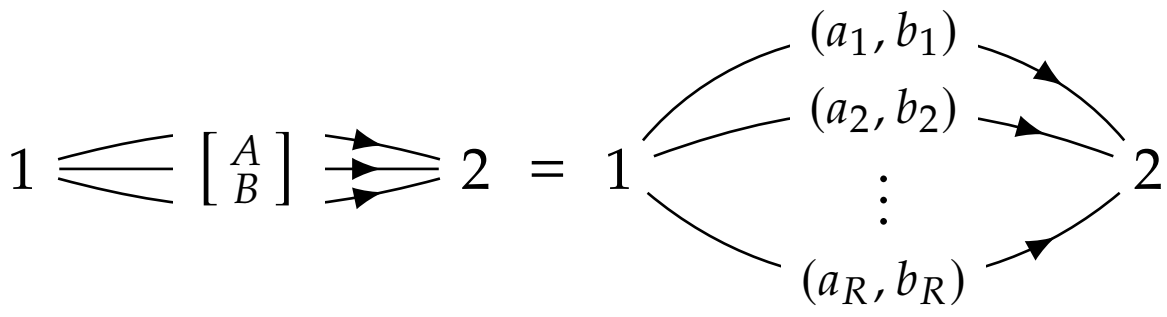

$$. \tag{5.6}$$

For $|A| = |B|$, (cf. (3.125)) we also introduce the antisymmetric version

$$\mathcal{A}\left[\begin{smallmatrix} A \\ B \end{smallmatrix}\right] = C\left[\begin{smallmatrix} A \\ B \end{smallmatrix}\right] - C\left[\begin{smallmatrix} B \\ A \end{smallmatrix}\right], \tag{5.7}$$

which is purely imaginary and was first studied in [185]. Under the transformation $\tau \to -\bar{\tau}$, any MGF satisfies $\mathcal{C}_\Gamma(-\bar{\tau}) = \overline{\mathcal{C}_\Gamma(\tau)}$ and hence we have that $\mathcal{A}\left[\begin{smallmatrix} A \\ B \end{smallmatrix}\right](-\bar{\tau}) = -\mathcal{A}\left[\begin{smallmatrix} A \\ B \end{smallmatrix}\right](\tau)$. Since $\tau_2$ is invariant under this transformation and the Laurent polynomial is mapped to its negative, the Laurent polynomial of $\mathcal{A}\left[\begin{smallmatrix} A \\ B \end{smallmatrix}\right]$ has to vanish, i.e. $\mathcal{A}\left[\begin{smallmatrix} A \\ B \end{smallmatrix}\right]$ is a cusp form.

In the `ModularGraphForms` package, MGFs have head **c**, i.e. they are formally given by the function **c** applied to various arguments. Dihedral MGFs have one argument which is a $2 \times R$ matrix which can, as any other matrix, be inserted in two-dimensional form or as a nested list,

```
In[8]:= c[1 2 3 / 1 1 1] + c[{{1,2,3},{1,1,1}}]

Out[8]= 2 C[1 2 3 / 1 1 1] .
```

Imaginary cusp forms of the form (5.7) have head **a**,

```
In[9]:= a[0 2 3 / 3 0 2]

Out[9]= A[0 2 3 / 3 0 2] .
```

### 5.2.2 *Three-point modular graph forms*

*Trihedral* graphs have three vertices. The notation we use is

$$C\left[\begin{smallmatrix}A_1\\B_1\end{smallmatrix}\middle|\begin{smallmatrix}A_2\\B_2\end{smallmatrix}\middle|\begin{smallmatrix}A_3\\B_3\end{smallmatrix}\right] = \tag{5.8}$$

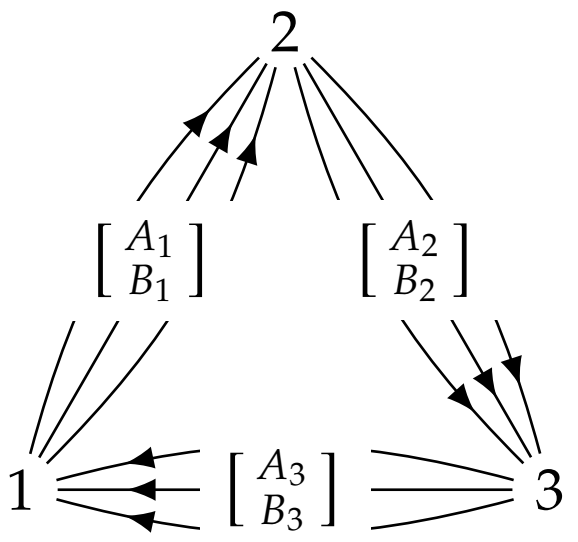

and hence

$$\begin{aligned} C\left[\begin{smallmatrix}A_1\\B_1\end{smallmatrix}\middle|\begin{smallmatrix}A_2\\B_2\end{smallmatrix}\middle|\begin{smallmatrix}A_3\\B_3\end{smallmatrix}\right] &= \sum_{\{p_i^{(j)}\}}{}' \left( \prod_{j=1}^{3} \prod_{i=1}^{R_j} \frac{1}{(p_i^{(j)})^{a_i^{(j)}} (\bar{p}_i^{(j)})^{b_i^{(j)}}} \right) \\ &\quad \delta\left( \sum_{i=1}^{R_1} p_i^{(1)} - \sum_{i=1}^{R_2} p_i^{(2)} \right) \delta\left( \sum_{i=1}^{R_2} p_i^{(2)} - \sum_{i=1}^{R_3} p_i^{(3)} \right), \end{aligned} \tag{5.9}$$

where the block $\left[\begin{smallmatrix}A_j\\B_j\end{smallmatrix}\right]$ has $R_j$ columns. We will use this notation henceforth. If two vertices are not connected by any edges, we write[1]

$$C\left[\begin{smallmatrix}A_1\\B_1\end{smallmatrix}\middle|\begin{smallmatrix}A_2\\B_2\end{smallmatrix}\middle|\varnothing\right] = \tag{5.10}$$

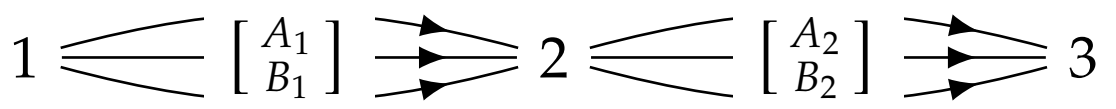

In the `Modular Graph Forms` package, the function **c** with three matrix-arguments is used,

```
In[10]:= c[(1 1 / 1 1), (2 3 / 1 1), (4 5 / 1 1)]

Out[10]= C[1 1 | 2 3 | 4 5 / 1 1 | 1 1 | 1 1] .
```

The edge directions and normalization are as in (5.8) and (5.9), respectively. For empty blocks, we use empty lists,

```
In[11]:= c[{}, (1 2 / 1 1), (3 4 / 1 1)]

Out[11]= C[{} | 1 2 / 1 1 | 3 4 / 1 1] .
```

### 5.2.3 *Four-point modular graph forms*

Due to their different symmetry properties, it is convenient to distinguish the following three topologies among four-point graphs.

---

1 This graph factorizes according to (5.25) into $C\left[\begin{smallmatrix}A_1\\B_1\end{smallmatrix}\right] C\left[\begin{smallmatrix}A_2\\B_2\end{smallmatrix}\right]$.

*Box graphs* have four edges in one cycle and are denoted by

$$C\left[\begin{smallmatrix}A_1\\B_1\end{smallmatrix}\middle|\begin{smallmatrix}A_2\\B_2\end{smallmatrix}\middle|\begin{smallmatrix}A_3\\B_3\end{smallmatrix}\middle|\begin{smallmatrix}A_4\\B_4\end{smallmatrix}\right] =$$

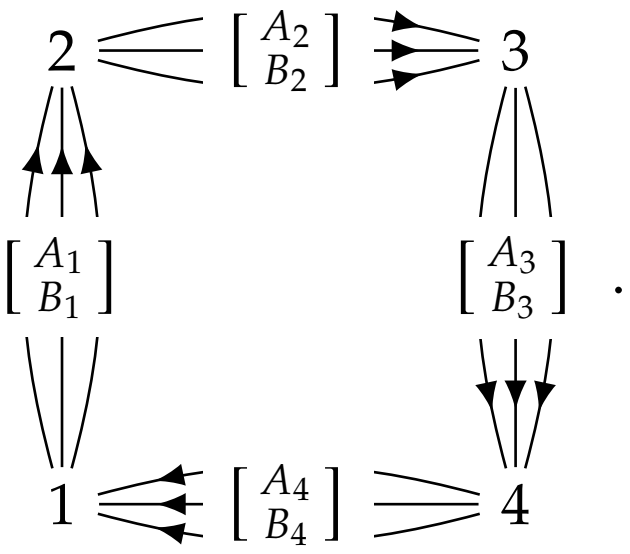

(5.11)

The lattice sum representation similarly to (5.9) can be read off straightforwardly from the graph. In `Mathematica`, we use **c** with four arguments,

In[12]:= $\mathbf{c}\left[\begin{smallmatrix}\mathbf{1}&\mathbf{2}\\\mathbf{1}&\mathbf{1}\end{smallmatrix},\begin{smallmatrix}\mathbf{3}&\mathbf{4}\\\mathbf{1}&\mathbf{1}\end{smallmatrix},\begin{smallmatrix}\mathbf{5}&\mathbf{6}\\\mathbf{1}&\mathbf{1}\end{smallmatrix},\begin{smallmatrix}\mathbf{7}&\mathbf{8}\\\mathbf{1}&\mathbf{1}\end{smallmatrix}\right]$

Out[12]= $\mathrm{C}\left[\begin{smallmatrix}1&2\\1&1\end{smallmatrix}\middle|\begin{smallmatrix}3&4\\1&1\end{smallmatrix}\middle|\begin{smallmatrix}5&6\\1&1\end{smallmatrix}\middle|\begin{smallmatrix}7&8\\1&1\end{smallmatrix}\right]$ .

*Kite graphs* have five edges: The cyclic ones from the box plus one diagonal. We write:

$$C\left[\begin{smallmatrix}A_1\\B_1\end{smallmatrix}\middle|\begin{smallmatrix}A_2\\B_2\end{smallmatrix}\middle\|\begin{smallmatrix}A_3\\B_3\end{smallmatrix}\middle|\begin{smallmatrix}A_4\\B_4\end{smallmatrix}\middle\|\begin{smallmatrix}A_5\\B_5\end{smallmatrix}\right] =$$

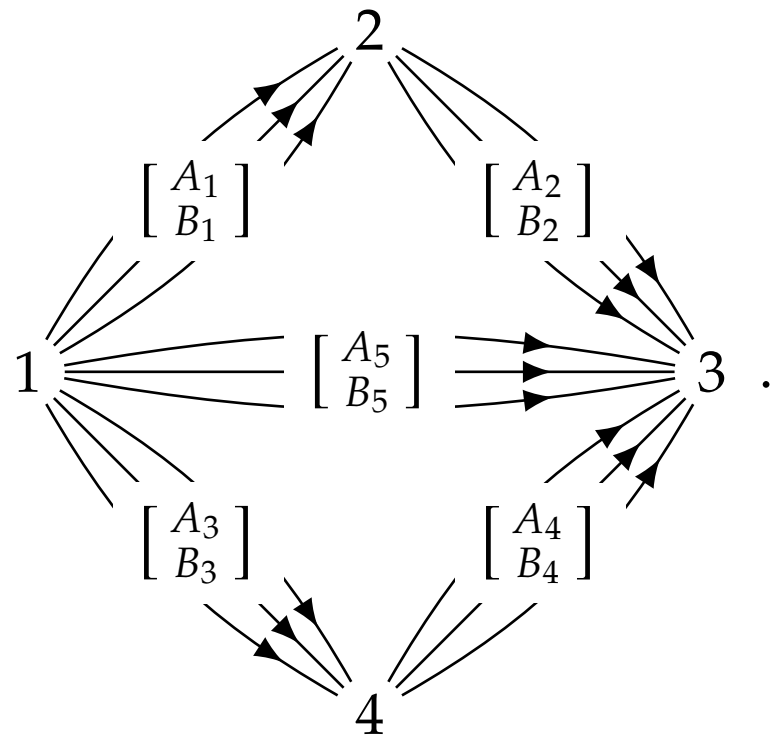

(5.12)

Note that the direction of the four outer edges is different from the box graph.

For kite graphs, **c** has five arguments,

In[13]:= $\mathbf{c}\left[\begin{smallmatrix}\mathbf{1}&\mathbf{2}\\\mathbf{1}&\mathbf{1}\end{smallmatrix},\begin{smallmatrix}\mathbf{1}&\mathbf{3}\\\mathbf{1}&\mathbf{1}\end{smallmatrix},\begin{smallmatrix}\mathbf{1}&\mathbf{4}\\\mathbf{1}&\mathbf{1}\end{smallmatrix},\begin{smallmatrix}\mathbf{1}&\mathbf{5}\\\mathbf{1}&\mathbf{1}\end{smallmatrix},\begin{smallmatrix}\mathbf{1}&\mathbf{6}\\\mathbf{1}&\mathbf{1}\end{smallmatrix}\right]$

Out[13]= $\mathrm{C}\left[\begin{smallmatrix}1&2\\1&1\end{smallmatrix}\middle|\begin{smallmatrix}1&3\\1&1\end{smallmatrix}\middle\|\begin{smallmatrix}1&4\\1&1\end{smallmatrix}\middle|\begin{smallmatrix}1&5\\1&1\end{smallmatrix}\middle\|\begin{smallmatrix}1&6\\1&1\end{smallmatrix}\right]$ .

Finally, the full *tetrahedral graph* (also known as *Mercedes graph*) has six edges connecting all pairs of points. As will become clear in the next section, due to its symmetry properties, it is convenient to arrange the six blocks in three columns as follows:[2]

$$C\left[\begin{array}{c||c||c} A_1 & A_2 & A_3 \\ B_1 & B_2 & B_3 \\ \hline A_4 & A_5 & A_6 \\ B_4 & B_5 & B_6 \end{array}\right] = $$

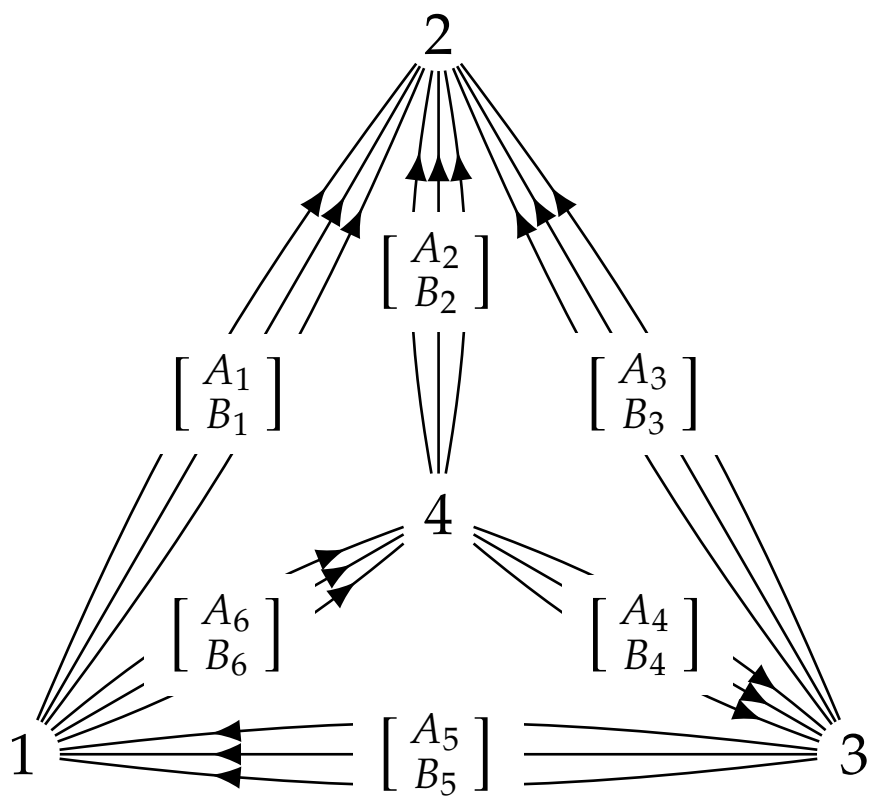

$$. \qquad (5.13)$$

Note that in this notation, edge bundles which do not share a common vertex correspond to blocks written in one column.

Tetrahedral graphs are written in the `Modular Graph Forms` package as **c** with six arguments,

In[14]:= $\mathbf{c}\left[\begin{smallmatrix}1&2\\1&1\end{smallmatrix}, \begin{smallmatrix}1&3\\1&1\end{smallmatrix}, \begin{smallmatrix}1&4\\1&1\end{smallmatrix}, \begin{smallmatrix}1&5\\1&1\end{smallmatrix}, \begin{smallmatrix}1&6\\1&1\end{smallmatrix}, \begin{smallmatrix}1&7\\1&1\end{smallmatrix}\right]$

Out[14]= $\mathrm{C}\left[\begin{array}{cc|cc|cc} 1 & 2 & 1 & 3 & 1 & 4 \\ 1 & 1 & 1 & 1 & 1 & 1 \\ \hline 1 & 5 & 1 & 6 & 1 & 7 \\ 1 & 1 & 1 & 1 & 1 & 1 \end{array}\right]$ .

For all four-point graphs, we will again use the symbol $\varnothing$ to denote blocks without any edges. In `Mathematica`, we again use empty lists.

## 5.3 SIMPLE RELATIONS

There are a number of relations between modular graph forms that follow directly from their definition in terms of graphs and lattice sums. These are easy to see, yet very powerful and already generate a lot of identities.

### 5.3.1 *Symmetries*

Given the graph of a modular graph form, the associated $C$-function as defined in the previous section is ambiguous and this generates relations between $C$-functions with different labels. In the simplest instance, permutations of the columns of a dihedral graph leave the MGF invariant. The same is true for permutations of columns in any block of the higher-point graphs.

If a vertex is connected to only two edge bundles, their total momenta have to agree and hence the two bundles can be swapped without

2 In the conventions of [41], the direction of the edges in third block is reversed.

changing the lattice sum associated to the graph. For trihedral- and box graphs this implies invariance under permutations of the blocks.

For the same reason, kite graphs are invariant under swapping blocks 1 and 2 as well as 3 and 4. Furthermore, swapping the vertices 2 and 4 leaves the graph invariant, so in total the symmetries are

$$\begin{aligned} &C\left[\begin{smallmatrix}A_1\\B_1\end{smallmatrix}\middle|\begin{smallmatrix}A_2\\B_2\end{smallmatrix}\middle\|\begin{smallmatrix}A_3\\B_3\end{smallmatrix}\middle|\begin{smallmatrix}A_4\\B_4\end{smallmatrix}\middle\|\begin{smallmatrix}A_5\\B_5\end{smallmatrix}\right] \\ =\;&C\left[\begin{smallmatrix}A_2\\B_2\end{smallmatrix}\middle|\begin{smallmatrix}A_1\\B_1\end{smallmatrix}\middle\|\begin{smallmatrix}A_3\\B_3\end{smallmatrix}\middle|\begin{smallmatrix}A_4\\B_4\end{smallmatrix}\middle\|\begin{smallmatrix}A_5\\B_5\end{smallmatrix}\right] \\ =\;&C\left[\begin{smallmatrix}A_1\\B_1\end{smallmatrix}\middle|\begin{smallmatrix}A_2\\B_2\end{smallmatrix}\middle\|\begin{smallmatrix}A_4\\B_4\end{smallmatrix}\middle|\begin{smallmatrix}A_3\\B_3\end{smallmatrix}\middle\|\begin{smallmatrix}A_5\\B_5\end{smallmatrix}\right] \\ =\;&C\left[\begin{smallmatrix}A_3\\B_3\end{smallmatrix}\middle|\begin{smallmatrix}A_4\\B_4\end{smallmatrix}\middle\|\begin{smallmatrix}A_1\\B_1\end{smallmatrix}\middle|\begin{smallmatrix}A_2\\B_2\end{smallmatrix}\middle\|\begin{smallmatrix}A_5\\B_5\end{smallmatrix}\right]. \end{aligned} \tag{5.14}$$

The double-line notation was chosen to make this intuitive. Note that the vertices in kite graphs are not all equivalent and this gives rise to the more complex symmetry properties (5.14).

Tetrahedral graphs have an $S_4$ permutation symmetry from relabeling the four equivalent vertices. These 24 permutations are generated by six permutations:

- three permutations of columns: Flipping a column comprised of two $(A_i, B_i)$-blocks in (5.13) with any other column produces a sign $(-1)^{|1|+|2|+|3|}$ where $|1|+|2|+|3| = |A_1|+|B_1|+|A_2|+|B_2|+|A_3|+|B_3|$ is a shorthand for the combined modular weight of the top row.[3] Explicitly:

$$\begin{aligned} C\left[\begin{array}{c\|c\|c} A_1 & A_2 & A_3 \\ B_1 & B_2 & B_3 \\ \hline A_4 & A_5 & A_6 \\ B_4 & B_5 & B_6 \end{array}\right] &= (-1)^{|1|+|2|+|3|}\, C\left[\begin{array}{c\|c\|c} A_2 & A_1 & A_3 \\ B_2 & B_1 & B_3 \\ \hline A_5 & A_4 & A_6 \\ B_5 & B_4 & B_6 \end{array}\right] \\ &= (-1)^{|1|+|2|+|3|}\, C\left[\begin{array}{c\|c\|c} A_3 & A_2 & A_1 \\ B_3 & B_2 & B_1 \\ \hline A_6 & A_5 & A_4 \\ B_6 & B_5 & B_4 \end{array}\right] \\ &= (-1)^{|1|+|2|+|3|}\, C\left[\begin{array}{c\|c\|c} A_1 & A_3 & A_2 \\ B_1 & B_3 & B_2 \\ \hline A_4 & A_6 & A_5 \\ B_4 & B_6 & B_5 \end{array}\right]. \end{aligned} \tag{5.15}$$

- three flips of two top/bottom pairs: Flipping the top/bottom blocks in any two columns changes the tetrahedral graph by a sign $(-1)^{|k|+|l|}$, where $k$ and $l$ in $|k|+|l| = |A_k|+|B_k|+|A_l|+|B_l|$ are given by the following prescription: Permute the three columns cyclically until the two columns in which top and bottom blocks

3 The sign does not depend on if we take the modular weight of the top- or bottom row since the total modular weight is even for non-vanishing MGFs.

are swapped are next to each other. The blocks in the left one of these has indices $k$ and $l$. Explicitly:

$$\begin{aligned} C\left[\begin{array}{c\|c\|c} A_1 & A_2 & A_3 \\ B_1 & B_2 & B_3 \\ \hline A_4 & A_5 & A_6 \\ B_4 & B_5 & B_6 \end{array}\right] &= (-1)^{|1|+|4|}\, C\left[\begin{array}{c\|c\|c} A_4 & A_5 & A_3 \\ B_4 & B_5 & B_3 \\ \hline A_1 & A_2 & A_6 \\ B_1 & B_2 & B_6 \end{array}\right] \\ &= (-1)^{|2|+|5|}\, C\left[\begin{array}{c\|c\|c} A_1 & A_5 & A_6 \\ B_1 & B_5 & B_6 \\ \hline A_4 & A_2 & A_3 \\ B_4 & B_2 & B_3 \end{array}\right] \\ &= (-1)^{|3|+|6|}\, C\left[\begin{array}{c\|c\|c} A_4 & A_2 & A_6 \\ B_4 & B_2 & B_6 \\ \hline A_1 & A_5 & A_3 \\ B_1 & B_5 & B_3 \end{array}\right] . \end{aligned} \tag{5.16}$$

The arrangement of the blocks in two rows of three columns was chosen to make these symmetries intuitive. For tetrahedral graphs, although all vertices are equivalent, the symmetry of the graph is broken by the direction of the edges, i.e. it is not possible to assign the directions in such a way that every vertex has the same number of ingoing and outgoing edges. Adjusting the edge direction when relabeling vertices leads to the signs in (5.15) and (5.16). These signs also mean that tetrahedral graphs can vanish by symmetry although they their sum of holomorphic and antiholomorphic labels is even. E.g., according to (5.15),

$$C\left[\begin{array}{c\|c\|c} A & A & A \\ B & B & B \\ \hline A & A & A \\ B & B & B \end{array}\right] = 0\,, \tag{5.17}$$

if $|A| + |B|$ odd, although $6(|A| + |B|)$ is even. This form of vanishing by symmetry does not exist for any of the other discussed graphs since no signs appear in their symmetry transformations.

In light of the above symmetry properties of it is convenient to define a *canonical representation* for the graph topologies discussed so far such that graphs related by a symmetry transformation are represented by the same arguments of the $C$-function. To this end, we define an ordering on the set of two-row columns and on the set of $2 \times R$ matrices. This will allow us to define an ordering on the MGFs of a certain topology and the smallest element in the symmetry orbit of an MGF will be the canonical representation of that graph.

The columns within an $\left[\begin{smallmatrix} A \\ B \end{smallmatrix}\right]$-block can be permuted arbitrarily for all graphs introduced above. The canonical representation of the MGFs therefore has the columns in each block in lexicographic order[4] w.r.t. the ordering defined by

- If $a_1 < a_2$ then $\left[\begin{smallmatrix} a_1 \\ b_1 \end{smallmatrix}\right] < \left[\begin{smallmatrix} a_2 \\ b_2 \end{smallmatrix}\right]$.
- If $a_1 = a_2$ then $\left[\begin{smallmatrix} a_1 \\ b_1 \end{smallmatrix}\right] < \left[\begin{smallmatrix} a_2 \\ b_2 \end{smallmatrix}\right]$ if $b_1 < b_2$.

4 In lexicographic order, the sequence $a_1, a_2, \ldots, a_n$ is smaller than $b_1, b_2, \ldots, b_n$ if $a_i < b_i$ for the first $i$ for which $a_i \neq b_i$.

Given two blocks $\left[\begin{smallmatrix}A_1\\B_1\end{smallmatrix}\right]$ and $\left[\begin{smallmatrix}A_2\\B_2\end{smallmatrix}\right]$ with canonical column order and $R_1$ and $R_2$ columns, respectively, we can define a canonical ordering of the two blocks by

- If $R_1 < R_2$ then $\left[\begin{smallmatrix}A_1\\B_1\end{smallmatrix}\right] < \left[\begin{smallmatrix}A_2\\B_2\end{smallmatrix}\right]$.
- If $R_1 = R_2$ then $\left[\begin{smallmatrix}A_1\\B_1\end{smallmatrix}\right] < \left[\begin{smallmatrix}A_2\\B_2\end{smallmatrix}\right]$ if $A_1 < A_2$ in lexicographic order.
- If $A_1 = A_2$ then $\left[\begin{smallmatrix}A_1\\B_1\end{smallmatrix}\right] < \left[\begin{smallmatrix}A_2\\B_2\end{smallmatrix}\right]$ if $B_1 < B_2$ in lexicographic order.

Using this ordering, we define

$$\mathcal{C}\left[\begin{smallmatrix}A_1\\B_1\end{smallmatrix}\right] < \mathcal{C}\left[\begin{smallmatrix}A_2\\B_2\end{smallmatrix}\right] \qquad \text{if} \qquad \left[\begin{smallmatrix}A_1\\B_1\end{smallmatrix}\right] < \left[\begin{smallmatrix}A_2\\B_2\end{smallmatrix}\right] \tag{5.18}$$

for dihedral graphs, unless the graph at hand is a one-loop graph. In this case, we write $\mathcal{C}\left[\begin{smallmatrix}a&0\\b&0\end{smallmatrix}\right]$ instead of $\mathcal{C}\left[\begin{smallmatrix}0&a\\0&b\end{smallmatrix}\right]$, to be consistent with the previous literature. For graphs with several blocks, we use lexicographic ordering on the set of blocks, hence

$$\begin{gathered}\mathcal{C}\left[\begin{smallmatrix}A_1\\B_1\end{smallmatrix}\middle|\begin{smallmatrix}A_2\\B_2\end{smallmatrix}\middle|\begin{smallmatrix}A_3\\B_3\end{smallmatrix}\right] < \mathcal{C}\left[\begin{smallmatrix}C_1\\D_1\end{smallmatrix}\middle|\begin{smallmatrix}C_2\\D_2\end{smallmatrix}\middle|\begin{smallmatrix}C_3\\D_3\end{smallmatrix}\right]\\ \text{if} \qquad \left(\left[\begin{smallmatrix}A_1\\B_1\end{smallmatrix}\right],\left[\begin{smallmatrix}A_2\\B_2\end{smallmatrix}\right],\left[\begin{smallmatrix}A_3\\B_3\end{smallmatrix}\right]\right) < \left(\left[\begin{smallmatrix}C_1\\D_1\end{smallmatrix}\right],\left[\begin{smallmatrix}C_2\\D_2\end{smallmatrix}\right],\left[\begin{smallmatrix}C_3\\D_3\end{smallmatrix}\right]\right)\end{gathered} \tag{5.19}$$

in lexicographic order and similarly for all four-point graphs with the numbering of the blocks as in Section 5.2.3.

For trihedral and box graphs, this just means that the canonical representation has the blocks (and in each block the columns) in lexicographic ordering. For kite graphs, the fifth block cannot be moved by the symmetries (5.14) and hence in the canonical representation, the smallest block out of the remaining four comes first, fixing the second one. The third block is the smaller one out of the remaining two, fixing the last block. Canonically represented tetrahedral graphs have the smallest block in the upper left slot, fixing the lower left block. The smallest block out of the remaining four blocks sits in the upper middle slot, fixing all remaining entries. The following examples are all in their canonical representation

$$\mathcal{C}\left[\begin{smallmatrix}3&0\\1&0\end{smallmatrix}\right] \tag{5.20a}$$

$$\mathcal{C}\left[\begin{smallmatrix}1&2&3\\7&5&4\end{smallmatrix}\right] \tag{5.20b}$$

$$\mathcal{C}\left[\begin{smallmatrix}2\\2\end{smallmatrix}\middle|\begin{smallmatrix}1&1\\1&1\end{smallmatrix}\middle|\begin{smallmatrix}0&0&1\\2&4&1\end{smallmatrix}\right] \tag{5.20c}$$

$$\mathcal{C}\left[\begin{smallmatrix}2\\2\end{smallmatrix}\middle|\begin{smallmatrix}1&2&3\\7&5&4\end{smallmatrix}\middle\|\begin{smallmatrix}1&1\\1&1\end{smallmatrix}\middle|\begin{smallmatrix}0&0&1\\2&4&1\end{smallmatrix}\middle\|\begin{smallmatrix}1\\1\end{smallmatrix}\right] \tag{5.20d}$$

$$\mathcal{C}\left[\begin{array}{c}\begin{smallmatrix}1\\1\end{smallmatrix}\\\hline\begin{smallmatrix}1&2&3\\7&5&4\end{smallmatrix}\end{array}\middle\|\begin{array}{c}\begin{smallmatrix}2\\2\end{smallmatrix}\\\hline\begin{smallmatrix}0&0&1\\2&4&1\end{smallmatrix}\end{array}\middle\|\begin{array}{c}\begin{smallmatrix}2&3\\1&2\end{smallmatrix}\\\hline\begin{smallmatrix}1&1\\1&1\end{smallmatrix}\end{array}\right]. \tag{5.20e}$$

In the `Modular Graph Forms` package, the function **`CSort`** brings MGFs into their canonical form, using the symmetries discussed above. For the MGFs in (5.20), we have e.g.

$$\text{In[15]:=}\ \mathtt{CSort}\Big[\Big\{\mathtt{c}\Big[\begin{smallmatrix}0&3\\0&1\end{smallmatrix}\Big],\mathtt{c}\Big[\begin{smallmatrix}2&1&3\\5&7&4\end{smallmatrix}\Big],\mathtt{c}\Big[\begin{smallmatrix}1&0&0\\1&4&2\end{smallmatrix},\begin{smallmatrix}2\\2\end{smallmatrix},\begin{smallmatrix}1&1\\1&1\end{smallmatrix}\Big],$$
$$\mathtt{c}\Big[\begin{smallmatrix}0&0&1\\4&2&1\end{smallmatrix},\begin{smallmatrix}1&1\\1&1\end{smallmatrix},\begin{smallmatrix}3&2&1\\4&5&7\end{smallmatrix},\begin{smallmatrix}2\\2\end{smallmatrix},\begin{smallmatrix}1\\1\end{smallmatrix}\Big],\mathtt{c}\Big[\begin{smallmatrix}2&3\\1&2\end{smallmatrix},\begin{smallmatrix}1&0&0\\1&4&2\end{smallmatrix},\begin{smallmatrix}3&2&1\\4&5&7\end{smallmatrix},\begin{smallmatrix}1&1\\1&1\end{smallmatrix},\begin{smallmatrix}2\\2\end{smallmatrix},\begin{smallmatrix}1\\1\end{smallmatrix}\Big]\Big\}\Big]$$

$$\text{Out[15]=}\ \Big\{\mathtt{C}\Big[\begin{smallmatrix}3&0\\1&0\end{smallmatrix}\Big],\mathtt{C}\Big[\begin{smallmatrix}1&2&3\\7&5&4\end{smallmatrix}\Big],\mathtt{C}\Big[\begin{smallmatrix}2\\2\end{smallmatrix}\Big|\begin{smallmatrix}1&1\\1&1\end{smallmatrix}\Big|\begin{smallmatrix}0&0&1\\2&4&1\end{smallmatrix}\Big],$$
$$\mathtt{C}\Big[\begin{smallmatrix}2\\2\end{smallmatrix}\Big|\begin{smallmatrix}1&2&3\\7&5&4\end{smallmatrix}\Big\|\begin{smallmatrix}1&1\\1&1\end{smallmatrix}\Big|\begin{smallmatrix}0&0&1\\2&4&1\end{smallmatrix}\Big\|\begin{smallmatrix}1\\1\end{smallmatrix}\Big],\mathtt{C}\left[\begin{array}{c|c|c}\begin{smallmatrix}1\\1\end{smallmatrix}&\begin{smallmatrix}2\\2\end{smallmatrix}&\begin{smallmatrix}2&3\\1&2\end{smallmatrix}\\\hline\begin{smallmatrix}1&2&3\\7&5&4\end{smallmatrix}&\begin{smallmatrix}0&0&1\\2&4&1\end{smallmatrix}&\begin{smallmatrix}1&1\\1&1\end{smallmatrix}\end{array}\right]\Big\}\ .$$

The output of the function **`CSimplify`** is always in canonical form. The property, that tetrahedral graphs can vanish by symmetry, as in the example (5.17), is implemented in the function **`TetCSimplify`**. E.g., we have

$$\text{In[16]:=}\ \mathtt{TetCSimplify}\Big[\mathtt{c}\Big[\begin{smallmatrix}1&2\\2&2\end{smallmatrix},\begin{smallmatrix}1&2\\2&2\end{smallmatrix},\begin{smallmatrix}1&2\\2&2\end{smallmatrix},\begin{smallmatrix}1&2\\2&2\end{smallmatrix},\begin{smallmatrix}1&2\\2&2\end{smallmatrix},\begin{smallmatrix}1&2\\2&2\end{smallmatrix}\Big]\Big]$$

$$\text{Out[16]=}\ 0\ .$$

### 5.3.2 *Topological simplifications*

For certain special cases of the graphs defined in Section 5.2, the MGF simplifies.

For the dihedral case, the fact that one-valent vertices lead to vanishing MGFs can be expressed as

$$\mathcal{C}\Big[\begin{smallmatrix}a\\b\end{smallmatrix}\Big]=0\,. \tag{5.21}$$

It is furthermore convenient to define

$$\mathcal{C}\big[\,\varnothing\,\big]=1\,. \tag{5.22}$$

The property (3.132) that two-valent vertices can be dropped translates for one-loop dihedral graphs into

$$\mathcal{C}\Big[\begin{smallmatrix}a_1&a_2\\b_1&b_2\end{smallmatrix}\Big]=(-1)^{a_2+b_2}\,\mathcal{C}\Big[\begin{smallmatrix}a_1+a_2&0\\b_1+b_2&0\end{smallmatrix}\Big]. \tag{5.23}$$

For trihedral graphs, (3.132) implies

$$\mathcal{C}\Big[\begin{smallmatrix}a_1\\b_1\end{smallmatrix}\Big|\begin{smallmatrix}a_2\\b_2\end{smallmatrix}\Big|\begin{smallmatrix}A_3\\B_3\end{smallmatrix}\Big]=(-1)^{a_1+b_1+a_2+b_2}\,\mathcal{C}\Big[\begin{smallmatrix}a_1+a_2&A_3\\b_1+b_2&B_3\end{smallmatrix}\Big] \tag{5.24}$$

and the factorization of one-particle reducible graphs (3.135) means that trihedral graphs with one empty block factorize into dihedral graphs,

$$\mathcal{C}\Big[\begin{smallmatrix}A_1\\B_1\end{smallmatrix}\Big|\begin{smallmatrix}A_2\\B_2\end{smallmatrix}\Big|\varnothing\Big]=\mathcal{C}\Big[\begin{smallmatrix}A_1\\B_1\end{smallmatrix}\Big]\mathcal{C}\Big[\begin{smallmatrix}A_2\\B_2\end{smallmatrix}\Big]. \tag{5.25}$$

Via (5.22), this also captures the case of two empty blocks.

Since two- and three-point graphs are special cases of four-point graphs, topological simplifications of four-point graphs should allow for simplifications down to dihedral graphs. We will provide a hierarchy of simplifications from tetrahedral graphs to box graphs which, if applied repeatedly together with (5.21) to (5.25), allow to identify any lower-point graph which is given as a tetrahedral MGF.

Tetrahedral graphs with one empty block are kite graphs,

$$\mathcal{C}\left[\begin{array}{c|c|c} \emptyset & \begin{smallmatrix}A_2\\B_2\end{smallmatrix} & \begin{smallmatrix}A_3\\B_3\end{smallmatrix} \\ \hline \begin{smallmatrix}A_4\\B_4\end{smallmatrix} & \begin{smallmatrix}A_5\\B_5\end{smallmatrix} & \begin{smallmatrix}A_6\\B_6\end{smallmatrix}\end{array}\right] = (-1)^{|A_2|+|B_2|+|A_4|+|B_4|}\,\mathcal{C}\left[\begin{smallmatrix}A_2\\B_2\end{smallmatrix}\middle|\begin{smallmatrix}A_3\\B_3\end{smallmatrix}\middle\|\begin{smallmatrix}A_5\\B_5\end{smallmatrix}\middle|\begin{smallmatrix}A_6\\B_6\end{smallmatrix}\middle\|\begin{smallmatrix}A_4\\B_4\end{smallmatrix}\right]. \tag{5.26}$$

A kite graph with one empty block is either a box graph or factorizes,

$$\mathcal{C}\left[\emptyset\middle|\begin{smallmatrix}A_2\\B_2\end{smallmatrix}\middle\|\begin{smallmatrix}A_3\\B_3\end{smallmatrix}\middle|\begin{smallmatrix}A_4\\B_4\end{smallmatrix}\middle\|\begin{smallmatrix}A_5\\B_5\end{smallmatrix}\right] = (-1)^{|A_5|+|B_5|}\,\mathcal{C}\left[\begin{smallmatrix}A_2\\B_2\end{smallmatrix}\right]\mathcal{C}\left[\begin{smallmatrix}A_3\\B_3\end{smallmatrix}\middle|\begin{smallmatrix}A_4\\B_4\end{smallmatrix}\middle|\begin{smallmatrix}A_5\\B_5\end{smallmatrix}\right] \tag{5.27a}$$

$$\mathcal{C}\left[\begin{smallmatrix}A_1\\B_1\end{smallmatrix}\middle|\begin{smallmatrix}A_2\\B_2\end{smallmatrix}\middle\|\begin{smallmatrix}A_3\\B_3\end{smallmatrix}\middle|\begin{smallmatrix}A_4\\B_4\end{smallmatrix}\middle\|\emptyset\right] = (-1)^{|A_3|+|B_3|+|A_4|+|B_4|}\,\mathcal{C}\left[\begin{smallmatrix}A_1\\B_1\end{smallmatrix}\middle|\begin{smallmatrix}A_2\\B_2\end{smallmatrix}\middle|\begin{smallmatrix}A_3\\B_3\end{smallmatrix}\middle|\begin{smallmatrix}A_4\\B_4\end{smallmatrix}\right]. \tag{5.27b}$$

If the two blocks in the first (or second) pair of blocks have only one column each, the vertex 2 (or 4) becomes two-valent end the kite graph simplifies into a box graph,

$$\mathcal{C}\left[\begin{smallmatrix}a_1\\b_1\end{smallmatrix}\middle|\begin{smallmatrix}a_2\\b_2\end{smallmatrix}\middle\|\begin{smallmatrix}A_3\\B_3\end{smallmatrix}\middle|\begin{smallmatrix}A_4\\B_4\end{smallmatrix}\middle\|\begin{smallmatrix}A_5\\B_5\end{smallmatrix}\right] = (-1)^{|A_3|+|B_3|+|A_4|+|B_4|}\,\mathcal{C}\left[\begin{smallmatrix}a_1+a_2 & A_5\\b_1+b_2 & B_5\end{smallmatrix}\middle|\begin{smallmatrix}A_3\\B_3\end{smallmatrix}\middle|\begin{smallmatrix}A_4\\B_4\end{smallmatrix}\right]. \tag{5.28}$$

A box graph with one (or more) empty blocks factorizes into dihedral graphs,

$$\mathcal{C}\left[\emptyset\middle|\begin{smallmatrix}A_2\\B_2\end{smallmatrix}\middle|\begin{smallmatrix}A_3\\B_3\end{smallmatrix}\middle|\begin{smallmatrix}A_4\\B_4\end{smallmatrix}\right] = \mathcal{C}\left[\begin{smallmatrix}A_2\\B_2\end{smallmatrix}\right]\mathcal{C}\left[\begin{smallmatrix}A_3\\B_3\end{smallmatrix}\right]\mathcal{C}\left[\begin{smallmatrix}A_4\\B_4\end{smallmatrix}\right] \tag{5.29}$$

and a box graph with two blocks of only one column each has a two-valent vertex and simplifies is a trihedral graph,

$$\mathcal{C}\left[\begin{smallmatrix}a_1\\b_1\end{smallmatrix}\middle|\begin{smallmatrix}a_2\\b_2\end{smallmatrix}\middle|\begin{smallmatrix}A_3\\B_3\end{smallmatrix}\middle|\begin{smallmatrix}A_4\\B_4\end{smallmatrix}\right] = \mathcal{C}\left[\begin{smallmatrix}a_1+a_2\\b_1+b_2\end{smallmatrix}\middle|\begin{smallmatrix}A_3\\B_3\end{smallmatrix}\middle|\begin{smallmatrix}A_4\\B_4\end{smallmatrix}\right]. \tag{5.30}$$

Combined, the relations above show e.g. that

$$\mathcal{C}\left[\begin{array}{c|c|c} \emptyset & \begin{smallmatrix}1\\1\end{smallmatrix} & \begin{smallmatrix}1\\1\end{smallmatrix} \\ \hline \begin{smallmatrix}1\\1\end{smallmatrix} & \begin{smallmatrix}1\\1\end{smallmatrix} & \begin{smallmatrix}1\\1\end{smallmatrix}\end{array}\right] = \mathcal{C}\left[\begin{smallmatrix}1&2&2\\1&2&2\end{smallmatrix}\right]. \tag{5.31}$$

In the `Mathematica` package `Modular Graph Forms`, the dihedral relations (5.21)–(5.23) are implement in the function **`DiCSimplify`**,

```
In[17]:= DiCSimplify[c[{}]c[0 3; 1 0] + c[3; 1]]

Out[17]= -c[3 0; 1 0] .
```

**`DiCSimplify`** also rewrites the special cases $\widehat{\mathrm{G}}_2$, $\mathrm{G}_k$ and $\mathrm{E}_k$ of one-loop graphs according to (3.129)–(3.131), whereas the one-loop simplification (5.23) is also performed by **`CSort`**. The function **`DiCSimplify`** has a Boolean option **`basisExpandG`** which, if set to **`True`**, causes **`DiCSimplify`** to expand all holomorphic Eisenstein series in the ring of $\mathrm{G}_4$ and $\mathrm{G}_6$, e.g.

```
In[18]:= DiCSimplify[g[24], basisExpandG → True]
```

Out[18]= $\dfrac{270\,G_4^6}{66079} + \dfrac{5400000\,G_4^3\,G_6^2}{151915621} + \dfrac{375\,G_6^4}{73853}$ .

The default value of **basisExpandG** is **False**.

The trihedral simplifications (5.24) and (5.25) are performed by **TriCSimplify**,

In[19]:= $\texttt{TriCSimplify}\left[\texttt{c}\left[\{\},\begin{matrix}1&2\\1&1\end{matrix},\begin{matrix}1&4\\1&1\end{matrix}\right]+\texttt{c}\left[\begin{matrix}1\\1\end{matrix},\begin{matrix}2\\2\end{matrix},\begin{matrix}1&3\\1&1\end{matrix}\right]\right]$

Out[19]= $\mathrm{C}\left[\begin{matrix}1&2\\1&1\end{matrix}\right]\mathrm{C}\left[\begin{matrix}1&4\\1&1\end{matrix}\right]+\mathrm{C}\left[\begin{matrix}3&1&3\\3&1&1\end{matrix}\right]$ .

Note that the dihedral graphs in Out[19] are not simplified or canonically represented, since **TriCSimplify** only acts on trihedral graphs. To simplify Out[19] further, we can apply **DiCSimplify**,

In[20]:= **DiCSimplify[Out[19], useIds → False]**

Out[20]= $\mathrm{C}\left[\begin{matrix}1&3&3\\1&1&3\end{matrix}\right]$ ,

where the Boolean option **useIds** was set to suppress the expansion using the result in the basis decompositions to be discussed in Section 5.7. The hierarchy of topological four-point simplifications (5.26)–(5.30) is implemented in the function **TetCSimplify**. Combining these functions, one can reproduce the example (5.31),

In[21]:= $\texttt{TetCSimplify}\left[\texttt{c}\left[\{\},\begin{matrix}1\\1\end{matrix},\begin{matrix}1\\1\end{matrix},\begin{matrix}1\\1\end{matrix},\begin{matrix}1\\1\end{matrix},\begin{matrix}1\\1\end{matrix}\right]\right]$
**TriCSimplify[%]**
**CSort[%]**

Out[21]= $\mathrm{C}\left[\begin{matrix}2&1\\2&1\end{matrix}\middle|\begin{matrix}1\\1\end{matrix}\middle|\begin{matrix}1\\1\end{matrix}\right]$

Out[22]= $\mathrm{C}\left[\begin{matrix}2&1&2\\2&1&2\end{matrix}\right]$

Out[23]= $\mathrm{C}\left[\begin{matrix}1&2&2\\1&2&2\end{matrix}\right]$ .

The function **CSimplify** acts on MGFs of all topologies and calls **DiCSimplify**, **TriCSimplify** and **TetCSimplify**. It also inherits the option **basisExpandG** from **DiCSimplify**. We have e.g.

In[24]:= $\texttt{CSimplify}\left[\texttt{c}\left[\{\},\begin{matrix}1&2\\1&2\end{matrix},\{\},\begin{matrix}1\\2\end{matrix},\begin{matrix}2\\1\end{matrix},\begin{matrix}2\\2\end{matrix}\right]\right]$

Out[24]= $\dfrac{\pi^8\,E_3\,E_5}{\tau_2^8}$ .

### 5.3.3 *Momentum conservation*

Momentum conservation [16] will be the central tool in our derivation of identities between modular graph forms and can be derived in the lattice sum representation (3.123) as well as the integral representation (3.120) of the MGF. As long as all graphs involved are convergent, as we will assume in this section, both approaches result in the same expression. If divergent graphs are involved, the integral representation allows one to use the tools of complex analysis to derive meaningful results, cf. Section 5.6.6.

Starting from the lattice sum representation (3.123) of an MGF with $|A| + |B|$ odd (hence, a vanishing MGF), which we will refer to as the *seed*, we have for each $j \in V_\Gamma$ the *momentum conservation identities*

$$0 = \sum_{e' \in E_\Gamma} \Gamma_{je'} \sum_{\{p_e\}}' \prod_{e \in E_\Gamma} \frac{p_{e'}}{p_e^{a_e} \bar{p}_e^{b_e}} \prod_{i \in V_\Gamma} \delta\Bigg( \sum_{e'' \in E_\Gamma} \Gamma_{ie''} p_{e''} \Bigg) \tag{5.32a}$$

$$0 = \sum_{e' \in E_\Gamma} \Gamma_{je'} \sum_{\{p_e\}}' \prod_{e \in E_\Gamma} \frac{\bar{p}_{e'}}{p_e^{a_e} \bar{p}_e^{b_e}} \prod_{i \in V_\Gamma} \delta\Bigg( \sum_{e'' \in E_\Gamma} \Gamma_{ie''} p_{e''} \Bigg) \tag{5.32b}$$

due to the momentum conserving delta functions. We will refer to (5.32a) as the *holomorphic-* and to (5.32b) as the *antiholomorphic momentum conservation identity*. By canceling the momenta from the numerators, (5.32) can be expressed entirely as a manipulation of the decorations of the graph and are therefore identities between MGFs,

$$0 = \sum_{e \in E_\Gamma} \Gamma_{je} C_{\Gamma_{a_e \to a_e - 1}}\,, \qquad 0 = \sum_{e \in E_\Gamma} \Gamma_{je} C_{\Gamma_{b_e \to b_e - 1}}\,, \qquad \forall\, j \in V_\Gamma\,. \tag{5.33}$$

If we had chosen a seed with $|A| + |B|$ even, the resulting MGFs would have all vanished trivially. Note that exchanging the sums over $e'$ and the $p_e$ in (5.32) required all sums to be convergent.

In the integral representation (3.120), the momentum conservation identities (5.32) correspond to integration-by-parts identities w.r.t. the puncture positions. To see this, note that due to (3.118),

$$\partial_z C^{(a,b)}(z) = -\frac{\pi}{\tau_2} C^{(a,b-1)}(z) \qquad \partial_{\bar{z}} C^{(a,b)}(z) = \frac{\pi}{\tau_2} C^{(a-1,b)}(z)\,. \tag{5.34}$$

If the integrand in (3.120) has no poles, the integral over the total derivative w.r.t. $z_j$ for each $j \in V_\Gamma$ vanishes and we have

$$0 = \sum_{e' \in E_\Gamma} \Gamma_{je'} \int d\mu_{n-1}\, C^{(a_{e'}, b_{e'}-1)}(z_{e'}) \prod_{\substack{e \in E_\Gamma \\ e \neq e'}} C^{(a_e, b_e)}(z_e) \tag{5.35a}$$

$$0 = \sum_{e' \in E_\Gamma} \Gamma_{je'} \int d\mu_{n-1}\, C^{(a_{e'}-1, b_{e'})}(z_{e'}) \prod_{\substack{e \in E_\Gamma \\ e \neq e'}} C^{(a_e, b_e)}(z_e)\,, \tag{5.35b}$$

agreeing with (5.33).

For dihedral graphs, the identities (5.33) for both vertices are identical and can be written as

$$0 = \sum_{i=1}^{R} C\Big[\begin{smallmatrix} A - S_i \\ B \end{smallmatrix}\Big] = \sum_{i=1}^{R} C\Big[\begin{smallmatrix} A \\ B - S_i \end{smallmatrix}\Big]\,, \tag{5.36}$$

the $j^{\text{th}}$ component of the row vector $S_i$ is $\delta_{ij}$. For trihedral MGFs, the momentum conservation identities involve two out of the three blocks and are given by

$$0 = \sum_{i=1}^{R_1} \mathcal{C}\left[\begin{array}{c|c|c} A_1-S_i & A_2 & A_3 \\ B_1 & B_2 & B_3 \end{array}\right] - \sum_{i=1}^{R_2} \mathcal{C}\left[\begin{array}{c|c|c} A_1 & A_2-S_i & A_3 \\ B_1 & B_2 & B_3 \end{array}\right] \tag{5.37}$$

and similarly for the complex conjugated identities. For box graphs, we have

$$0 = \sum_{i=1}^{R_1} \mathcal{C}\left[\begin{array}{c|c|c|c} A_1-S_i & A_2 & A_3 & A_4 \\ B_1 & B_2 & B_3 & B_4 \end{array}\right] - \sum_{i=1}^{R_2} \mathcal{C}\left[\begin{array}{c|c|c|c} A_1 & A_2-S_i & A_3 & A_4 \\ B_1 & B_2 & B_3 & B_4 \end{array}\right] \quad \text{and c.c.} \tag{5.38}$$

For kite graphs, we have to distinguish the cases in which the momentum conservation of vertex 2 or 4 is used, yielding

$$\begin{aligned} 0 = &\sum_{i=1}^{R_1} \mathcal{C}\left[\begin{array}{c|c||c|c||c} A_1-S_i & A_2 & A_3 & A_4 & A_5 \\ B_1 & B_2 & B_3 & B_4 & B_5 \end{array}\right] \\ &- \sum_{i=1}^{R_2} \mathcal{C}\left[\begin{array}{c|c||c|c||c} A_1 & A_2-S_i & A_3 & A_4 & A_5 \\ B_1 & B_2 & B_3 & B_4 & B_5 \end{array}\right] \quad \text{and c.c.} \end{aligned} \tag{5.39}$$

and the case in which the momentum conservation of vertex 1 or 3 is used, resulting in the identity

$$\begin{aligned} 0 = &\sum_{i=1}^{R_1} \mathcal{C}\left[\begin{array}{c|c||c|c||c} A_1-S_i & A_2 & A_3 & A_4 & A_5 \\ B_1 & B_2 & B_3 & B_4 & B_5 \end{array}\right] + \sum_{i=1}^{R_4} \mathcal{C}\left[\begin{array}{c|c||c|c||c} A_1 & A_2 & A_3-S_i & A_4 & A_5 \\ B_1 & B_2 & B_3 & B_4 & B_5 \end{array}\right] \\ &+ \sum_{i=1}^{R_5} \mathcal{C}\left[\begin{array}{c|c||c|c||c} A_1 & A_2 & A_3 & A_4 & A_5-S_i \\ B_1 & B_2 & B_3 & B_4 & B_5 \end{array}\right] \quad \text{and c.c.} \end{aligned} \tag{5.40}$$

The topology of tetrahedral graphs is completely symmetric, hence the momentum conservation identity for vertex 2,

$$\begin{aligned} 0 = &\sum_{i=1}^{R_1} \mathcal{C}\left[\begin{array}{c||c||c} A_1-S_i & A_2 & A_3 \\ B_1 & B_2 & B_3 \\ \hline A_4 & A_5 & A_6 \\ B_4 & B_5 & B_6 \end{array}\right] + \sum_{i=1}^{R_2} \mathcal{C}\left[\begin{array}{c||c||c} A_1 & A_2-S_i & A_3 \\ B_1 & B_2 & B_3 \\ \hline A_4 & A_5 & A_6 \\ B_4 & B_5 & B_6 \end{array}\right] \\ &+ \sum_{i=1}^{R_3} \mathcal{C}\left[\begin{array}{c||c||c} A_1 & A_2 & A_3-S_i \\ B_1 & B_2 & B_3 \\ \hline A_4 & A_5 & A_6 \\ B_4 & B_5 & B_6 \end{array}\right] \quad \text{and c.c.}\,, \end{aligned} \tag{5.41}$$

is related to those of all other vertices by the transformations (5.15) and (5.16).

In the `Modular Graph Forms` package, momentum conservation for dihedral and trihedral graphs is implemented in the functions **`DiHolMomConsId`** and **`TriHolMomConsId`** and their antiholomorphic versions **`DiAHolMomConsId`** and **`TriAHolMomConsId`**. In the dihedral case (5.36), the function **`DiHolMomConsId`** takes the seed as its only argument and we have e.g.

In[25]:= $\texttt{DiHolMomConsId}\left[\mathrm{c}\left[\begin{smallmatrix}1 & 1 & 2\\ 1 & 1 & 1\end{smallmatrix}\right]\right]$

Out[25]= $\mathrm{C}\left[\begin{smallmatrix}0 & 1 & 2\\ 1 & 1 & 1\end{smallmatrix}\right] + \mathrm{C}\left[\begin{smallmatrix}1 & 0 & 2\\ 1 & 1 & 1\end{smallmatrix}\right] + \mathrm{C}\left[\begin{smallmatrix}1 & 1 & 1\\ 1 & 1 & 1\end{smallmatrix}\right] == 0$ .

For trihedral momentum conservation (5.37), we have to specify which of the three vertices we use and hence which pair of blocks has its labels changed. The list of these blocks is passed as a second argument to **TriHolMomConsId**, e.g.

In[26]:= $\texttt{TriHolMomConsId}\left[\mathrm{c}\left[\begin{smallmatrix}1 & 2\\ 1 & 1\end{smallmatrix}, \begin{smallmatrix}1 & 3\\ 1 & 1\end{smallmatrix}, \begin{smallmatrix}1 & 4\\ 1 & 1\end{smallmatrix}\right], \{2,3\}\right]$

Out[26]= $\mathrm{C}\left[\begin{smallmatrix}1 & 2\\ 1 & 1\end{smallmatrix}\middle|\begin{smallmatrix}0 & 3\\ 1 & 1\end{smallmatrix}\middle|\begin{smallmatrix}1 & 4\\ 1 & 1\end{smallmatrix}\right] + \mathrm{C}\left[\begin{smallmatrix}1 & 2\\ 1 & 1\end{smallmatrix}\middle|\begin{smallmatrix}1 & 2\\ 1 & 1\end{smallmatrix}\middle|\begin{smallmatrix}1 & 4\\ 1 & 1\end{smallmatrix}\right] -$
$\mathrm{C}\left[\begin{smallmatrix}1 & 2\\ 1 & 1\end{smallmatrix}\middle|\begin{smallmatrix}1 & 3\\ 1 & 1\end{smallmatrix}\middle|\begin{smallmatrix}0 & 4\\ 1 & 1\end{smallmatrix}\right] - \mathrm{C}\left[\begin{smallmatrix}1 & 2\\ 1 & 1\end{smallmatrix}\middle|\begin{smallmatrix}1 & 3\\ 1 & 1\end{smallmatrix}\middle|\begin{smallmatrix}1 & 3\\ 1 & 1\end{smallmatrix}\right] == 0$ .

Note that the functions discussed here do not apply **CSort** to the resulting equation, so that it is more transparent which exponents were lowered. E.g. Out[25] simplifies to

In[27]:= $\texttt{CSort}[\texttt{Out[25]}]$

Out[27]= $2\,\mathrm{C}\left[\begin{smallmatrix}0 & 1 & 2\\ 1 & 1 & 1\end{smallmatrix}\right] + \mathrm{C}\left[\begin{smallmatrix}1 & 1 & 1\\ 1 & 1 & 1\end{smallmatrix}\right] == 0$ .

### 5.3.4 *Factorization*

Consider a modular graph form with a $(0,0)$-edge. In this case, the graph factorizes [16]. To see this, consider two vertices $x$ and $y$ and an edge from $x$ to $y$ with momentum $p$ and decoration $(0,0)$. Furthermore assume that all other edges connected to $x$ are directed away from $x$ and have momentum sum $p_x$ and all other edges connected to $y$ are directed away from $y$ and have momentum sum $p_y$ ,

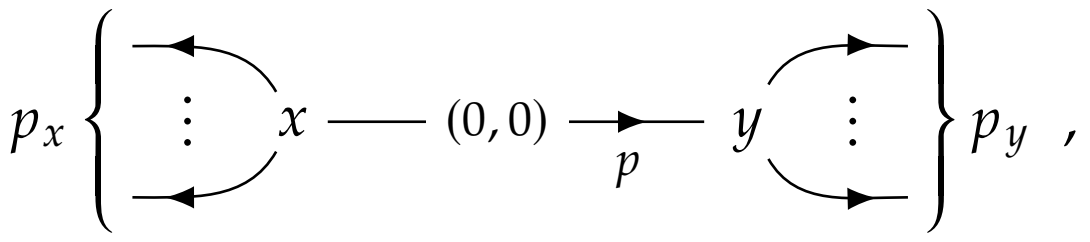

, (5.42)

where the $(0,0)$-edge is not necessarily the only edge between $x$ and $y$. In the sum representation, the momentum $p$ only appears in the momentum-conserving delta functions for the vertices $x$ and $y$. Isolating this contribution, we get

$$\begin{aligned}\sum_p{}' \delta(p_x + p)\delta(p_y - p) &= \sum_p \delta(p_x + p)\delta(p_y - p) - \delta(p_x)\delta(p_y)\\ &= \delta(p_x + p_y) - \delta(p_x)\delta(p_y)\,,\end{aligned} \tag{5.43}$$

where we added $p = 0$ to the sum to evaluate the deltas. When (5.43) appears in the nested lattice sum of an MGF, the first term gives rise to the original MGF with the vertices $x$ and $y$ identified, whereas the second term can be associated to the original MGF with the $(0,0)$-edge removed. Schematically, if the edge $e$ between vertices $x$ and $y$ carries decoration $(0,0)$, we have

$$\mathcal{C}_{\Gamma_{a_e=b_e=0}} = \mathcal{C}_{\Gamma_{x=y}} - \mathcal{C}_{\Gamma\setminus e}\,. \tag{5.44}$$

If the vertices $x$ and $y$ are connected by more edges than just $e$, these will factorize as one-loop graphs in the first term of (5.44).

In the integral representation, a $(0,0)$-edge is represented by a factor $C^{(0,0)}(z)$ in the integrand, which as special case of (3.118) can be simplified to

$$\begin{aligned} C^{(0,0)}(z) &= \sum_{m,n\in\mathbb{Z}} e^{2\pi i(mv-nu)} - 1 \\ &= \delta(v)\delta(u) - 1 = \tau_2\delta^{(2)}(z,\bar{z}) - 1\,, \end{aligned} \tag{5.45}$$

where we used (3.41). Note that (5.45) is not the $a=0$ case of (3.119), since $f^{(0)}(z)=1$, but is implied by the $a=1, b=0$ case of (5.34) and (3.90). The interpretation of (5.45) is exactly as in the sum representation: The delta identifies the two vertices connected by the $(0,0)$ edge and in the second term the $(0,0)$ edge is removed. In this way, we get again (5.44).

For dihedral MGFs, (5.44) implies[5]

$$\mathcal{C}\left[\begin{smallmatrix} 0 & A \\ 0 & B \end{smallmatrix}\right] = \prod_{j=1}^{R} \mathcal{C}\left[\begin{smallmatrix} a_j & 0 \\ b_j & 0 \end{smallmatrix}\right] - \mathcal{C}\left[\begin{smallmatrix} A \\ B \end{smallmatrix}\right] \tag{5.46}$$

for higher-point graphs we have

$$\begin{aligned} \mathcal{C}\left[\begin{smallmatrix} 0 & A_1 \\ 0 & B_1 \end{smallmatrix}\middle|\begin{smallmatrix} A_2 \\ B_2 \end{smallmatrix}\middle|\begin{smallmatrix} A_3 \\ B_3 \end{smallmatrix}\right] &= (-1)^{|2|}\, \mathcal{C}\left[\begin{smallmatrix} A_2 & A_3 \\ B_2 & B_3 \end{smallmatrix}\right] \prod_{i=1}^{R_1} \mathcal{C}\left[\begin{smallmatrix} a_1^{(i)} & 0 \\ b_1^{(i)} & 0 \end{smallmatrix}\right] \\ &\quad - \mathcal{C}\left[\begin{smallmatrix} A_1 \\ B_1 \end{smallmatrix}\middle|\begin{smallmatrix} A_2 \\ B_2 \end{smallmatrix}\middle|\begin{smallmatrix} A_3 \\ B_3 \end{smallmatrix}\right] \end{aligned} \tag{5.47}$$

$$\begin{aligned} \mathcal{C}\left[\begin{smallmatrix} 0 & A_1 \\ 0 & B_1 \end{smallmatrix}\middle|\begin{smallmatrix} A_2 \\ B_2 \end{smallmatrix}\middle|\begin{smallmatrix} A_3 \\ B_3 \end{smallmatrix}\middle|\begin{smallmatrix} A_4 \\ B_4 \end{smallmatrix}\right] &= \mathcal{C}\left[\begin{smallmatrix} A_2 \\ B_2 \end{smallmatrix}\middle|\begin{smallmatrix} A_3 \\ B_3 \end{smallmatrix}\middle|\begin{smallmatrix} A_4 \\ B_4 \end{smallmatrix}\right] \prod_{i=1}^{R_1} \mathcal{C}\left[\begin{smallmatrix} a_1^{(i)} & 0 \\ b_1^{(i)} & 0 \end{smallmatrix}\right] \\ &\quad - \mathcal{C}\left[\begin{smallmatrix} A_1 \\ B_1 \end{smallmatrix}\middle|\begin{smallmatrix} A_2 \\ B_2 \end{smallmatrix}\middle|\begin{smallmatrix} A_3 \\ B_3 \end{smallmatrix}\middle|\begin{smallmatrix} A_4 \\ B_4 \end{smallmatrix}\right] \end{aligned} \tag{5.48}$$

$$\begin{aligned} \mathcal{C}\left[\begin{smallmatrix} 0 & A_1 \\ 0 & B_1 \end{smallmatrix}\middle|\begin{smallmatrix} A_2 \\ B_2 \end{smallmatrix}\middle\|\begin{smallmatrix} A_3 \\ B_3 \end{smallmatrix}\middle|\begin{smallmatrix} A_4 \\ B_4 \end{smallmatrix}\middle\|\begin{smallmatrix} A_5 \\ B_5 \end{smallmatrix}\right] &= (-1)^{|2|+|5|}\, \mathcal{C}\left[\begin{smallmatrix} A_2 & A_5 \\ B_2 & B_5 \end{smallmatrix}\middle|\begin{smallmatrix} A_3 \\ B_3 \end{smallmatrix}\middle|\begin{smallmatrix} A_4 \\ B_4 \end{smallmatrix}\right] \prod_{i=1}^{R_1} \mathcal{C}\left[\begin{smallmatrix} a_1^{(i)} & 0 \\ b_1^{(i)} & 0 \end{smallmatrix}\right] \\ &\quad - \mathcal{C}\left[\begin{smallmatrix} A_1 \\ B_1 \end{smallmatrix}\middle|\begin{smallmatrix} A_2 \\ B_2 \end{smallmatrix}\middle\|\begin{smallmatrix} A_3 \\ B_3 \end{smallmatrix}\middle|\begin{smallmatrix} A_4 \\ B_4 \end{smallmatrix}\middle\|\begin{smallmatrix} A_5 \\ B_5 \end{smallmatrix}\right] \end{aligned} \tag{5.49}$$

5 For one-loop graphs $\mathcal{C}\left[\begin{smallmatrix} a & 0 \\ b & 0 \end{smallmatrix}\right]$, (5.46) is trivial upon using (5.21).

$$\mathcal{C}\left[\begin{smallmatrix}A_1\\B_1\end{smallmatrix}\middle|\begin{smallmatrix}A_2\\B_2\end{smallmatrix}\middle\|\begin{smallmatrix}A_3\\B_3\end{smallmatrix}\middle|\begin{smallmatrix}A_4\\B_4\end{smallmatrix}\middle\|\begin{smallmatrix}0&A_5\\0&B_5\end{smallmatrix}\right] = (-1)^{|1|+|3|}\,\mathcal{C}\left[\begin{smallmatrix}A_1&A_2\\B_1&B_2\end{smallmatrix}\right]\mathcal{C}\left[\begin{smallmatrix}A_3&A_4\\B_3&B_4\end{smallmatrix}\right]\prod_{i=1}^{R_5}\mathcal{C}\left[\begin{smallmatrix}a_5^{(i)}&0\\b_5^{(i)}&0\end{smallmatrix}\right] \\ - \mathcal{C}\left[\begin{smallmatrix}A_1\\B_1\end{smallmatrix}\middle|\begin{smallmatrix}A_2\\B_2\end{smallmatrix}\middle\|\begin{smallmatrix}A_3\\B_3\end{smallmatrix}\middle|\begin{smallmatrix}A_4\\B_4\end{smallmatrix}\middle\|\begin{smallmatrix}A_5\\B_5\end{smallmatrix}\right] \tag{5.50}$$

$$\mathcal{C}\left[\begin{smallmatrix}0&A_1\\0&B_1\\\hline A_4\\B_4\end{smallmatrix}\middle\|\begin{smallmatrix}A_2\\B_2\\\hline A_5\\B_5\end{smallmatrix}\middle\|\begin{smallmatrix}A_3\\B_3\\\hline A_6\\B_6\end{smallmatrix}\right] = (-1)^{|2|}\,\mathcal{C}\left[\begin{smallmatrix}A_2&A_6\\B_2&B_6\end{smallmatrix}\middle|\begin{smallmatrix}A_3&A_5\\B_3&B_5\end{smallmatrix}\middle|\begin{smallmatrix}A_4\\B_4\end{smallmatrix}\right]\prod_{i=1}^{R_1}\mathcal{C}\left[\begin{smallmatrix}a_1^{(i)}&0\\b_1^{(i)}&0\end{smallmatrix}\right] \\ - \mathcal{C}\left[\begin{smallmatrix}A_1\\B_1\\\hline A_4\\B_4\end{smallmatrix}\middle\|\begin{smallmatrix}A_2\\B_2\\\hline A_5\\B_5\end{smallmatrix}\middle\|\begin{smallmatrix}A_3\\B_3\\\hline A_6\\B_6\end{smallmatrix}\right], \tag{5.51}$$

where we used the abbreviation $|i| = |A_i| + |B_i|$ as above. Note that the RHSs have one vertex less in the first term and one loop order less in the second term and hence (5.46) to (5.51) are powerful identities to simplify MGFs. Together with the momentum conservation identities from Section 5.3.3, these identities form the backbone of all the simplifications we will carry out in the following.

In the `Modular Graph Forms` package, factorization of $(0,0)$-edges for dihedral and trihedral graph is also done by the functions **`DiCSimplify`** and **`TriCSimplify`**. E.g. in the trihedral case (5.47), we have

In[28]:= $\mathtt{TriCSimplify}\left[\mathtt{c}\left[\begin{smallmatrix}0&2&1\\0&1&2\end{smallmatrix},\begin{smallmatrix}1&2\\1&1\end{smallmatrix},\begin{smallmatrix}1&4\\1&1\end{smallmatrix}\right]\right]$

Out[28]= $-\,\mathtt{c}\left[\begin{smallmatrix}1&0\\2&0\end{smallmatrix}\right]\mathtt{c}\left[\begin{smallmatrix}2&0\\1&0\end{smallmatrix}\right]\mathtt{c}\left[\begin{smallmatrix}2&2&1&4\\1&2&1&1\end{smallmatrix}\right]-\mathtt{c}\left[\begin{smallmatrix}1&2\\2&1\end{smallmatrix}\middle|\begin{smallmatrix}1&4\\1&1\end{smallmatrix}\middle|\begin{smallmatrix}2&2\\1&2\end{smallmatrix}\right]$ .

If several $(0,0)$-edges are present, the factorization is repeated until no more $(0,0)$-edges in the respective topology appear. E.g. we have

In[29]:= $\mathtt{TriCSimplify}\left[\mathtt{c}\left[\begin{smallmatrix}0&2&1\\0&1&2\end{smallmatrix},\begin{smallmatrix}0&2&2\\0&1&2\end{smallmatrix},\begin{smallmatrix}1&4\\1&1\end{smallmatrix}\right]\right]$

Out[29]= $\mathtt{c}\left[\begin{smallmatrix}2&0\\1&0\end{smallmatrix}\right]\mathtt{c}\left[\begin{smallmatrix}2&0\\2&0\end{smallmatrix}\right]\mathtt{c}\left[\begin{smallmatrix}1&2&1&4\\2&1&1&1\end{smallmatrix}\right]-\mathtt{c}\left[\begin{smallmatrix}1&0\\2&0\end{smallmatrix}\right]\mathtt{c}\left[\begin{smallmatrix}2&0\\1&0\end{smallmatrix}\right]\mathtt{c}\left[\begin{smallmatrix}0&2&2&1&4\\0&1&2&1&1\end{smallmatrix}\right]+$
$\mathtt{c}\left[\begin{smallmatrix}1&2\\2&1\end{smallmatrix}\middle|\begin{smallmatrix}1&4\\1&1\end{smallmatrix}\middle|\begin{smallmatrix}2&2\\1&2\end{smallmatrix}\right]$ ,

where the remaining dihedral factorization can be preformed by applying **`DiCSimplify`**.

### 5.3.5 *Taking derivatives*

On top of momentum conservation and factorization, another way to obtain new identities for MGFs is by taking derivatives of known identities using the modular differential operators $\nabla^{(a)}$ and $\overline{\nabla}^{(b)}$ defined in (3.51). Since these change the modular weight according to (3.52), one obtains an identity between MGFs of different weights.

Consider the action of $\nabla^{(|A|)}$ and $\overline{\nabla}^{(|B|)}$ and on an MGF of weight $(|A|, |B|)$ in its lattice sum representation (3.123). Using the product rule (3.53) and [16]

$$\nabla^{(a)}\left(\frac{1}{p^a}\right) = a\,\frac{1}{p^{a+1}\bar{p}^{-1}} \qquad \overline{\nabla}^{(b)}\left(\frac{1}{\bar{p}^b}\right) = b\,\frac{1}{p^{-1}\bar{p}^{b+1}}\,, \tag{5.52}$$

the derivatives are given by

$$\nabla^{(|A|)} \mathcal{C}_\Gamma = \sum_{e \in E_\Gamma} a_e \mathcal{C}_{\Gamma_{(a_e,b_e) \to (a_e+1,b_e-1)}} \tag{5.53a}$$

$$\overline{\nabla}^{(|B|)} \mathcal{C}_\Gamma = \sum_{e \in E_\Gamma} b_e \mathcal{C}_{\Gamma_{(a_e,b_e) \to (a_e-1,b_e+1)}} . \tag{5.53b}$$

In the integral representation, $\nabla^{(|A|)}$ and $\overline{\nabla}^{(|B|)}$ act on the Jacobi forms $C^{(a,b)}(z,\tau)$ given in (3.118). According to (5.52), we have

$$\nabla^{(a)} C^{(a,b)}(z,\tau) = a\, C^{(a+1,b-1)}(z,\tau) \tag{5.54a}$$

$$\overline{\nabla}^{(b)} C^{(a,b)}(z,\tau) = b\, C^{(a-1,b+1)}(z,\tau) \tag{5.54b}$$

and using this together with the product rule (3.53), we obtain again (5.53).

For a dihedral MGF, (5.53) implies [16]

$$\nabla^{(|A|)} \mathcal{C}\big[\begin{smallmatrix} A \\ B \end{smallmatrix}\big] = \sum_{i=1}^{R} a_i\, \mathcal{C}\big[\begin{smallmatrix} A+S_i \\ B-S_i \end{smallmatrix}\big] \tag{5.55a}$$

$$\nabla^{(|B|)} \mathcal{C}\big[\begin{smallmatrix} A \\ B \end{smallmatrix}\big] = \sum_{i=1}^{R} b_i\, \mathcal{C}\big[\begin{smallmatrix} A-S_i \\ B+S_i \end{smallmatrix}\big] , \tag{5.55b}$$

where the $j^{\text{th}}$ component of $S_i$ is $\delta_{ij}$ as above. A special case of (5.55) is the important relation

$$\nabla_0^n \mathrm{E}_k = \frac{\tau_2^{k+n}}{\pi^k} \frac{(k+n-1)!}{(k-1)!} \mathcal{C}\big[\begin{smallmatrix} k+n & 0 \\ k-n & 0 \end{smallmatrix}\big] , \tag{5.56}$$

where $\nabla_0$ is defined in (3.55). Since (5.53) does not depend on the topology of the graph, the higher-point versions of (5.55) are completely analogous, so for trihedral graphs, we have e.g. [16]

$$\begin{aligned} \nabla^{(|A|)} \mathcal{C}\big[\begin{smallmatrix} A_1 \\ B_1 \end{smallmatrix}\big|\begin{smallmatrix} A_2 \\ B_2 \end{smallmatrix}\big|\begin{smallmatrix} A_3 \\ B_3 \end{smallmatrix}\big] = \sum_{i=1}^{R_1} a_1^{(i)} \mathcal{C}\big[\begin{smallmatrix} A_1+S_i \\ B_1-S_i \end{smallmatrix}\big|\begin{smallmatrix} A_2 \\ B_2 \end{smallmatrix}\big|\begin{smallmatrix} A_3 \\ B_3 \end{smallmatrix}\big] + \sum_{i=1}^{R_2} a_2^{(i)} \mathcal{C}\big[\begin{smallmatrix} A_1 \\ B_1 \end{smallmatrix}\big|\begin{smallmatrix} A_2+S_i \\ B_2-S_i \end{smallmatrix}\big|\begin{smallmatrix} A_3 \\ B_3 \end{smallmatrix}\big] \\ + \sum_{i=1}^{R_3} a_3^{(i)} \mathcal{C}\big[\begin{smallmatrix} A_1 \\ B_1 \end{smallmatrix}\big|\begin{smallmatrix} A_2 \\ B_2 \end{smallmatrix}\big|\begin{smallmatrix} A_3+S_i \\ B_3-S_i \end{smallmatrix}\big] \quad \text{and c.c.} , \end{aligned} \tag{5.57}$$

where in the complex conjugation, we swap all $a$ and $b$ labels everywhere and replace $S_i \longrightarrow -S_i$. Similar identities hold for all four-point graphs.

When taking the Cauchy–Riemann derivative of a holomorphic Eisenstein series, one obtains

$$\nabla^{(2k)} \mathrm{G}_{2k} = 2k\, \mathcal{C}\big[\begin{smallmatrix} 2k+1 & 0 \\ -1 & 0 \end{smallmatrix}\big] , \quad k \geq 2 , \tag{5.58}$$

which cannot be simplified further with the methods presented so far. However, the $\bar{\tau}$-derivative of the weight $(2k+2,0)$ modular form

$$\frac{\pi}{\tau_2}\mathcal{C}\left[\begin{smallmatrix}2k+1 & 0\\ -1 & 0\end{smallmatrix}\right]+\mathrm{G}_{2k}\widehat{\mathrm{G}}_2\,,\quad k\geq 2 \tag{5.59}$$

vanishes, and hence it can be expanded in the ring of holomorphic Eisenstein series. To this end, we calculate the $q$ expansion

$$\frac{1}{2k}\nabla^{(2k)}\mathrm{G}_{2k}=2\zeta_{2k}-\frac{4\zeta_{2k}}{B_{2k}}\sum_{n=1}^{\infty}\sigma_{2k-1}(n)(2k-4\pi n\tau_2)q^n\,,\quad k\geq 1\,, \tag{5.60}$$

by taking the Cauchy–Riemann derivative of (3.24). Now, by comparing a finite number of terms, we can expand (5.58) in holomorphic Eisenstein series. Since for low weights this ring is one-dimensional, we can give a closed formula in these cases,

$$\mathcal{C}\left[\begin{smallmatrix}2k+1 & 0\\ -1 & 0\end{smallmatrix}\right]=\frac{\tau_2}{\pi}\left(\frac{2\zeta_2\zeta_{2k}}{\zeta_{2k+2}}\mathrm{G}_{2k+2}-\mathrm{G}_{2k}\widehat{\mathrm{G}}_2\right),\quad k=2,3,4\,. \tag{5.61}$$

For the non-holomorphic but modular version $\widehat{\mathrm{G}}_2=\mathrm{G}_2-\frac{\pi}{\tau_2}$ of $\mathrm{G}_2$, we obtain

$$\nabla^{(2)}\widehat{\mathrm{G}}_2=2\,\mathcal{C}\left[\begin{smallmatrix}3 & 0\\ -1 & 0\end{smallmatrix}\right]=\frac{\tau_2}{\pi}\left(5\mathrm{G}_4-\widehat{\mathrm{G}}_2^2\right), \tag{5.62}$$

as can be verified by explicitly comparing the $q$ expansions term by term. Note that (5.62) and (5.61) for $k=2,3$ are equivalent to the classic Ramanujan identities

$$q\frac{\mathrm{d}\mathrm{G}_2}{\mathrm{d}q}=\frac{\mathrm{G}_2^2-5\mathrm{G}_4}{4\pi^2} \tag{5.63a}$$

$$q\frac{\mathrm{d}\mathrm{G}_4}{\mathrm{d}q}=\frac{2\mathrm{G}_2\mathrm{G}_4-7\mathrm{G}_6}{2\pi^2} \tag{5.63b}$$

$$q\frac{\mathrm{d}\mathrm{G}_6}{\mathrm{d}q}=\frac{21\mathrm{G}_2\mathrm{G}_6-30\mathrm{G}_4^2}{14\pi^2}\,. \tag{5.63c}$$

Since the expressions above allow to write the derivative of any holomorphic Eisenstein series back into a polynomial in holomorphic Eisenstein series, we can iterate these expressions and simplify arbitrarily high derivatives of holomorphic Eisenstein series. E.g. we have

$$\mathcal{C}\left[\begin{smallmatrix}4 & 0\\ -2 & 0\end{smallmatrix}\right]=\frac{1}{6}\nabla^{(2)^2}\widehat{\mathrm{G}}_2=\left(\frac{\tau_2}{\pi}\right)^2\left(\frac{35}{3}\mathrm{G}_6-5\mathrm{G}_4\widehat{\mathrm{G}}_2+\frac{1}{3}\widehat{\mathrm{G}}_2^3\right) \tag{5.64a}$$

$$\mathcal{C}\left[\begin{smallmatrix}6 & 0\\ -2 & 0\end{smallmatrix}\right]=\frac{1}{20}\nabla^{(4)^2}\mathrm{G}_4=\left(\frac{\tau_2}{\pi}\right)^2\left(\mathrm{G}_4\widehat{\mathrm{G}}_2^2-7\mathrm{G}_6\widehat{\mathrm{G}}_2+5\mathrm{G}_4^2\right), \tag{5.64b}$$

where we used the notation (3.54) for the second Cauchy–Riemann derivatives.

In the `Modular Graph Forms` package, the Cauchy–Riemann derivatives (5.53) are implemented in the function **`CHolCR`** for the holomorphic case and **`CAHolCR`** for the antiholomorphic case. For clarity, the result is returned as it comes out of the action (5.53) of $\nabla^{(a)}$ hence, to obtain the derivative in canonical representation, we have to apply **`CSort`**, e.g.

In[30]:= $\texttt{CHolCR}\Big[\texttt{c}\Big[\begin{smallmatrix}1&1\\1&1\end{smallmatrix},\begin{smallmatrix}1&1\\1&1\end{smallmatrix},\begin{smallmatrix}1&1\\1&1\end{smallmatrix},\begin{smallmatrix}1&1\\1&1\end{smallmatrix},\begin{smallmatrix}1&1\\1&1\end{smallmatrix}\Big]\Big]\texttt{//CSort}$

Out[30]= $8\,\mathrm{C}\Big[\begin{smallmatrix}1&1\\1&1\end{smallmatrix}\Big|\begin{smallmatrix}1&1\\1&1\end{smallmatrix}\Big\|\begin{smallmatrix}1&1\\1&1\end{smallmatrix}\Big|\begin{smallmatrix}1&2\\1&0\end{smallmatrix}\Big\|\begin{smallmatrix}1&1\\1&1\end{smallmatrix}\Big] + 2\,\mathrm{C}\Big[\begin{smallmatrix}1&1\\1&1\end{smallmatrix}\Big|\begin{smallmatrix}1&1\\1&1\end{smallmatrix}\Big\|\begin{smallmatrix}1&1\\1&1\end{smallmatrix}\Big|\begin{smallmatrix}1&1\\1&1\end{smallmatrix}\Big\|\begin{smallmatrix}1&2\\1&0\end{smallmatrix}\Big]\,.$

The functions **`CHolCR`** and **`CAHolCR`** can also be used to calculate derivatives of holomorphic Eisenstein series,

In[31]:=
```
CHolCR[g[4]]
CHolCR[%]
CHolCR[gHat[2]]
```

Out[31]= $4\,\mathrm{C}\Big[\begin{smallmatrix}5&0\\-1&0\end{smallmatrix}\Big]$

Out[32]= $20\,\mathrm{C}\Big[\begin{smallmatrix}6&0\\-2&0\end{smallmatrix}\Big]$

Out[33]= $2\,\mathrm{C}\Big[\begin{smallmatrix}3&0\\-1&0\end{smallmatrix}\Big]\,.$

The simplifications of these expressions by means of the Ramanujan identities (5.61), (5.62) and (5.64) and higher-weight generalizations is performed by the function **`DiCSimplify`**, if the option **`basisExpandG`** is set to **`True`**, e.g.

In[34]:=
```
DiCSimplify[Out[31], basisExpandG → True]
DiCSimplify[Out[32], basisExpandG → True]
DiCSimplify[Out[33], basisExpandG → True]
```

Out[34]= $\dfrac{14\,G_6\,\tau_2}{\pi} - \dfrac{4\,G_4\,\hat{G}_2\,\tau_2}{\pi}$

Out[35]= $\dfrac{100\,G_4^2\,\tau_2^2}{\pi^2} - \dfrac{140\,G_6\,\hat{G}_2\,\tau_2^2}{\pi^2} + \dfrac{20\,G_4\,\hat{G}_2^2\,\tau_2^2}{\pi^2}$

Out[36]= $\dfrac{5\,G_4\,\tau_2}{\pi} - \dfrac{\hat{G}_2^2\,\tau_2}{\pi}\,.$

Using the techniques outlined above, **`DiCSimplify`** can decompose any MGF of the form $\mathcal{C}\Big[\begin{smallmatrix}k&0\\-n&0\end{smallmatrix}\Big]$ or $\mathcal{C}\Big[\begin{smallmatrix}-n&0\\k&0\end{smallmatrix}\Big]$ with $k, n \in \mathbb{N}_0$ and $k > n$ into the ring of holomorphic Eisenstein series and powers of $\widehat{\mathrm{G}}_2$ and $\frac{\tau_2}{\pi}$ (or c.c.).

## 5.4 HOLOMORPHIC SUBGRAPH REDUCTION

Using the relatively straightforward techniques discussed in the previous section, many identities between MGFs can be derived. However, an important class of identities is still missing to decompose all relevant MGFs into the basis to be presented in Section 5.7, namely holomorphic subgraph reduction. In this section, we will review HSR as it was introduced first for dihedral graphs [16] and the extension of this technique to higher-point graphs [I].[6]

The basic idea behind HSR is the following: If an MGF has a closed subgraph (i.e. a subgraph which forms a loop) in which all edges have only holomorphic momenta (i.e. the decorations are all of the form $(a, 0)$), then one can apply the partial-fraction decomposition

$$\begin{aligned}\frac{1}{p^a(q-p)^b} &= \sum_{k=1}^{a} \binom{a+b-k-1}{a-k} \frac{1}{p^k q^{a+b-k}} \\ &\quad + \sum_{k=1}^{b} \binom{a+b-k-1}{b-k} \frac{1}{q^{a+b-k}(q-p)^k}\end{aligned} \tag{5.65}$$

to the summand and perform the sum over the loop momentum explicitly. Since this sum is for certain values of $a$, $b$ only conditionally convergent, it has to be supplied with a summation prescription, which we will choose to be *Eisenstein summation*, to be defined below in (5.72) and discussed in more detail in Appendix B.1. This procedure however breaks the modular transformation properties at the level of the individual contributions. As shown in [16] for two-point graphs and in [I] for general graphs, the terms with incorrect modular properties cancel out in the final expression and one obtains a decomposition of the original MGF into terms which all have at least one loop order less. If a graph has several (possibly overlapping) closed holomorphic subgraphs, we can apply HSR iteratively, as discussed in Section 5.4.5.

As an example, consider the trihedral graph $C\left[\begin{smallmatrix}1&2\\0&1\end{smallmatrix}\middle|\begin{smallmatrix}1&2\\0&1\end{smallmatrix}\middle|\begin{smallmatrix}1&2\\0&1\end{smallmatrix}\right]$, which has a closed three-point holomorphic subgraph. Using the techniques discussed in this section, it can be decomposed into

$$\begin{aligned}C\left[\begin{smallmatrix}1&2\\0&1\end{smallmatrix}\middle|\begin{smallmatrix}1&2\\0&1\end{smallmatrix}\middle|\begin{smallmatrix}1&2\\0&1\end{smallmatrix}\right] &= 6\,C\left[\begin{smallmatrix}2\\1\end{smallmatrix}\middle|\begin{smallmatrix}1&2\\0&1\end{smallmatrix}\middle|\begin{smallmatrix}2&2\\0&1\end{smallmatrix}\right] - 3\,C\left[\begin{smallmatrix}2&3&4\\1&0&2\end{smallmatrix}\right] + 3\,C\left[\begin{smallmatrix}1&2&4\\0&1&2\end{smallmatrix}\right]\widehat{\mathrm{G}}_2 \\ &\quad + \frac{\pi}{\tau_2}\,C\left[\begin{smallmatrix}2&2&4\\-1&1&2\end{smallmatrix}\right] - 2\frac{\pi}{\tau_2}\,C\left[\begin{smallmatrix}2\\1\end{smallmatrix}\middle|\begin{smallmatrix}1&2\\-1&1\end{smallmatrix}\middle|\begin{smallmatrix}1&2\\0&1\end{smallmatrix}\right].\end{aligned} \tag{5.66}$$

6 In the references, a different convention for MGFs was used, which differs from the one used here by factors of $\tau_2$ and $\pi$.

### 5.4.1 *Dihedral holomorphic subgraph reduction*

Holomorphic subgraph reduction was first worked out for dihedral graphs in [16]. The following review is largely identical to Section 3 in [I].

A generic dihedral modular graph form with a holomorphic subgraph may be represented by

$$\mathcal{C}\left[\begin{smallmatrix} a_+ & a_- & A \\ 0 & 0 & B \end{smallmatrix}\right] = \qquad (5.67)$$

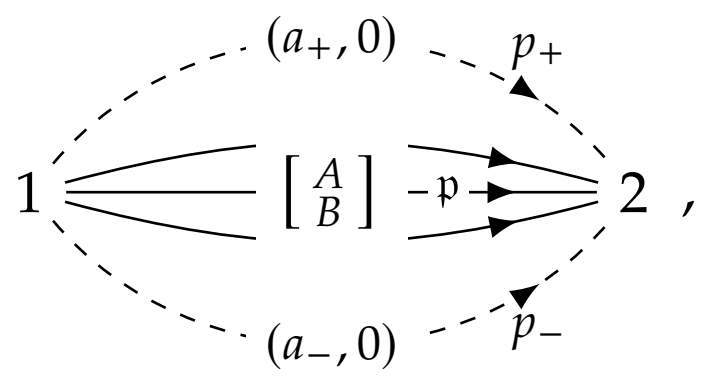

where the top and bottom dashed edges are purely holomorphic, carrying momenta $p_+$ and $p_-$, respectively, and the edge bundle $\left[\begin{smallmatrix} A \\ B \end{smallmatrix}\right]$ has total momentum $\mathfrak{p}$. For later convenience, we define $p_0 = p_+ + p_-$ and $a_0 = a_+ + a_-$. The corresponding lattice sum is then given by

$$\mathcal{C}\left[\begin{smallmatrix} a_+ & a_- & A \\ 0 & 0 & B \end{smallmatrix}\right] = \sideset{}{'}\sum_{p_1,\dots,p_R,p_+,p_-} \frac{1}{(p_+)^{a_+}(p_-)^{a_-}} \prod_{i=1}^{R} \frac{1}{(p_i)^{a_i}\,(\bar{p}_i)^{b_i}}\,\delta(p_0 + \mathfrak{p})\,, \tag{5.68}$$

where we assigned momenta $p_i$ to the edges in the bundle. In order for this sum to be absolutely convergent, we restrict to $a_0 \geq 3$. The basic strategy of holomorphic subgraph reduction is to isolate the two holomorphic edges, utilize the momentum-conserving delta-function to rewrite

$$\sideset{}{'}\sum_{p_+,p_-} \frac{1}{(p_+)^{a_+}(p_-)^{a_-}}\delta(p_0 + \mathfrak{p}) = \sideset{}{'}\sum_{p_+\neq -\mathfrak{p}} \frac{1}{(p_+)^{a_+}(-\mathfrak{p} - p_+)^{a_-}} \tag{5.69}$$

and then to perform a partial-fraction decomposition in $p_+$ using (5.65). Once this has been done, the summation over $p_+$ can be performed explicitly. The resulting expression then has one less momentum, and thus one less edge, than the original MGF.

A subtlety in this procedure is that by naively distributing the sum over the partial-fraction decomposition, conditionally convergent sums can be produced. In particular, sums of the form

$$Q_k(p_0) = \sideset{}{'}\sum_{p\neq p_0} \frac{1}{p^k}\,, \qquad k \geq 1 \tag{5.70}$$

arise, which are not absolutely convergent for $k = 1, 2$. To rectify this issue, we must find appropriate definitions for these sums. The definitions which were chosen in [16] are

$$Q_1(p_0) = -\frac{1}{p_0} - \frac{\pi}{2\tau_2}(p_0 - \bar{p}_0) \tag{5.71a}$$

$$Q_2(p_0) = -\frac{1}{{p_0}^2} + \widehat{\mathrm{G}}_2 + \frac{\pi}{\tau_2} \tag{5.71b}$$

$$Q_k(p_0) = -\frac{1}{{p_0}^k} + \mathrm{G}_k \qquad k \geq 3\,. \tag{5.71c}$$

The choice (5.71) is not unique since it depends on the summation prescription chosen to evaluate (5.70). In order to obtain (5.71), one has to use Eisenstein summation

$$\begin{aligned} \sum_{p\neq r+s\tau}{}_{\mathrm{E}} f(p) = &\lim_{N\to\infty} \sum_{\substack{n=-N\\ n\neq s}}^{N} \left( \lim_{M\to\infty} \sum_{m=-M}^{M} f(m+n\tau) \right) \\ &+ \lim_{M\to\infty} \sum_{\substack{m=-M\\ m\neq r}}^{M} f(m+s\tau)\,, \end{aligned} \tag{5.72}$$

as will be explained in detail in Section 5.4.2 and Appendix B.1. An important point is that the term $-\frac{\pi}{2\tau_2}p_0$ in $Q_1(p_0)$ and the term $\frac{\pi}{\tau_2}$ in $Q_2(p_0)$ have different modular weights than the sums on the respective LHSs. But when plugged into the full expression resulting from partial-fraction decomposition of (5.69), these terms of abnormal modular weight cancel out, leading to a total result with the expected modular properties.

The final result for the holomorphic subgraph reduction of dihedral MGFs can be written as a closed formula [16],

$$\begin{aligned} \mathcal{C}\left[\begin{smallmatrix} a_+ & a_- & A \\ 0 & 0 & B \end{smallmatrix}\right] = &(-1)^{a_+}\mathrm{G}_{a_0}\,\mathcal{C}\left[\begin{smallmatrix} A \\ B \end{smallmatrix}\right] - \binom{a_0}{a_-}\,\mathcal{C}\left[\begin{smallmatrix} a_0 & A \\ 0 & B \end{smallmatrix}\right] \\ &+ \sum_{k=4}^{a_+} \binom{a_0-1-k}{a_+-k} \mathrm{G}_k\,\mathcal{C}\left[\begin{smallmatrix} a_0-k & A \\ 0 & B \end{smallmatrix}\right] \\ &+ \sum_{k=4}^{a_-} \binom{a_0-1-k}{a_--k} \mathrm{G}_k\,\mathcal{C}\left[\begin{smallmatrix} a_0-k & A \\ 0 & B \end{smallmatrix}\right] \\ &+ \binom{a_0-2}{a_+-1} \left\{ \widehat{\mathrm{G}}_2\,\mathcal{C}\left[\begin{smallmatrix} a_0-2 & A \\ 0 & B \end{smallmatrix}\right] + \frac{\pi}{\tau_2}\,\mathcal{C}\left[\begin{smallmatrix} a_0-1 & A \\ -1 & B \end{smallmatrix}\right] \right\}. \end{aligned} \tag{5.73}$$

For instance, the two-loop graph $\mathcal{C}\left[\begin{smallmatrix} 1 & 2 & 2 \\ 0 & 0 & 1 \end{smallmatrix}\right]$ is decomposed into one-loop graphs by (5.73),

$$\mathcal{C}\left[\begin{smallmatrix} 1 & 2 & 2 \\ 0 & 0 & 1 \end{smallmatrix}\right] = 3\,\mathcal{C}\left[\begin{smallmatrix} 5 & 0 \\ 1 & 0 \end{smallmatrix}\right] - \widehat{\mathrm{G}}_2\,\mathcal{C}\left[\begin{smallmatrix} 3 & 0 \\ 1 & 0 \end{smallmatrix}\right] - \frac{\pi}{\tau_2}\mathrm{G}_4\,. \tag{5.74}$$

In the `Modular Graph Forms` `Mathematica` package, the dihedral HSR (5.73) is performed by the function **`DiCSimplify`**. With the default options, **`DiCSimplify`** also applies all known dihedral basis decompositions to the result and uses momentum conservation to remove negative entries where possible as will be detailed in Section 5.5.1. Both features can be disabled by setting the Boolean options **`momSimplify`** and **`useIds`** to **`False`** (they are **`True`** by default). Hence, in order to get just the result of the formula (5.73), we can run e.g.

In[37]:= $\texttt{DiCSimplify}\left[\texttt{c}\left[\begin{smallmatrix} 2 & 2 & 3 & 6 \\ 1 & 2 & 0 & 0 \end{smallmatrix}\right], \texttt{momSimplify} \to \texttt{False}, \texttt{useIds} \to \texttt{False}\right]$

Out[37]= $-84\,\mathrm{C}\left[\begin{smallmatrix} 2 & 2 & 9 \\ 1 & 2 & 0 \end{smallmatrix}\right] + 6\,\mathrm{C}\left[\begin{smallmatrix} 2 & 2 & 5 \\ 1 & 2 & 0 \end{smallmatrix}\right] \mathrm{G}_4 + \mathrm{C}\left[\begin{smallmatrix} 2 & 2 & 3 \\ 1 & 2 & 0 \end{smallmatrix}\right] \mathrm{G}_6 + 21\,\mathrm{C}\left[\begin{smallmatrix} 2 & 2 & 7 \\ 1 & 2 & 0 \end{smallmatrix}\right] \hat{\mathrm{G}}_2 + \dfrac{21\,\pi\,\mathrm{C}\left[\begin{smallmatrix} 2 & 2 & 8 \\ 1 & 2 & -1 \end{smallmatrix}\right]}{\tau_2}$ .

The function **`DiCSimplify`** applies the formula (5.73) always to the two leftmost holomorphic columns. It also performs antiholomorphic subgraph reduction by applying the complex conjugate of (5.73) to graphs with a closed antiholomorphic subgraph. In order to deactivate dihedral HSR in **`DiCSimplify`** or **`CSimplify`**, one can set the Boolean option **`diHSR`** to **`False`** (the default is **`True`**).

### 5.4.2 *Higher-point holomorphic subgraph reduction*

HSR for higher-point graphs was worked out in [I] and this section is a slightly rewritten version of Section 5 of the reference.

For higher-point graphs, the holomorphic subgraph has more edges and the sum (5.70) takes the form[7]

$$\sum_{p\neq p_1,\dots,p_n}' \frac{1}{p^{a_0}(p-p_1)^{a_1}\dots(p-p_n)^{a_n}} \tag{5.75}$$

for some external momenta $p_i$ and corresponding exponents $a_i$, $i = 1,\dots,n$. We will assume that all of the $p_i$ are distinct; if this is not the case, we can just increase the corresponding exponents. We will also exclude the case of $n = 1$, $a_0 = a_1 = 1$, since in that case the sum (5.75) is not absolutely convergent.

It suffices to specialize to the case

$$S_{a_0}(p_1,\dots,p_n) = \sum_{p\neq p_1,\dots,p_n}' \frac{1}{p^{a_0}(p-p_1)\dots(p-p_n)}\,, \tag{5.76}$$

since we may get a sum (5.75) with arbitrary $a_i$, $i = 1,\dots,n$ from (5.76) by differentiating with respect to the external momenta $p_i$. The

7 Note that unlike in the previous section, $a_0$ is now being used to refer to a single exponent, as opposed to a sum over them.

validity of this interchange of derivatives and sums follows by uniform convergence. For any $a_0, n \geq 1$ we may rewrite

$$S_{a_0}(p_1, \ldots, p_n) = \sum_{i=1}^{n} \frac{1}{p^{a_0}} \frac{1}{p - p_i} \prod_{\substack{j=1 \\ j \neq i}}^{n} \frac{1}{p_i - p_j}, \tag{5.77}$$

which can be verified by induction. Now we can use the partial-fraction identity (5.65) to decompose the term in front of the product and obtain

$$\begin{aligned} S_{a_0}(p_1, \ldots, p_n) = \sideset{}{'}\sum_{p \neq p_1, \ldots, p_n} \Bigg[ & \sum_{i=1}^{n} \left( \frac{1}{p_i^{a_0}(p - p_i)} \prod_{\substack{j=1 \\ j \neq i}}^{n} \frac{1}{p_i - p_j} \right) \\ & + (-1)^n \sum_{\ell=1}^{a_0} \frac{h_{\ell-1}(p_1, \ldots, p_n)}{p^{a_0 - \ell + 1} \prod_{i=1}^{n} p_i^{\ell}} \Bigg], \end{aligned} \tag{5.78}$$

where the $h_k(p_1, \ldots, p_n)$ are symmetric polynomials in $p_1, \ldots, p_n$ of homogeneous order $(n-1)k$, defined by

$$h_k(p_1, \ldots, p_n) = \sum_{\substack{a_1, \ldots, a_n = 0 \\ a = (n-1)k}}^{k} \prod_{i=1}^{n} p_i^{a_i} \tag{5.79}$$

with $a = a_1 + \cdots + a_n$. In (5.78) we used the identity

$$\sum_{i=1}^{n} \frac{1}{p_i^{\ell}} \prod_{\substack{j=1 \\ j \neq i}}^{n} \frac{1}{p_i - p_j} = (-1)^{n+1} h_{\ell-1}(p_1, \ldots, p_n) \prod_{i=1}^{n} \frac{1}{p_i^{\ell}}, \tag{5.80}$$

which can again be proven by induction.

We can carry out the sum over $p$ in (5.78) by choosing a summation prescription for which the sum over each individual term in the summand converges, and then distributing the sum over the individual terms. In particular, we may work with the Eisenstein summation prescription $\sum_{\mathrm{E}}$ defined in (5.72), and then distribute the sums in (5.78), yielding

$$\begin{aligned} S_{a_0}(p_1, \ldots, p_n) = & \sum_{i=1}^{n} \left( \sideset{}{'}\sum_{\mathrm{E}\; p \neq p_1, \ldots, p_n} \frac{1}{p - p_i} \frac{1}{p_i^{a_0}} \prod_{\substack{j=1 \\ j \neq i}}^{n} \frac{1}{p_i - p_j} \right) \\ & + (-1)^n \sum_{\ell=1}^{a_0} \sideset{}{'}\sum_{\mathrm{E}\; p \neq p_1, \ldots, p_n} \frac{1}{p^{a_0 - \ell + 1}} \frac{h_{\ell-1}(p_1, \ldots, p_n)}{\prod_{i=1}^{n} p_i^{\ell}}. \end{aligned} \tag{5.81}$$

In Appendix B.1, we derive for the sums over $p$

$$\sideset{}{'_{\mathrm{E}}}\sum_{p\neq p_1,\dots,p_n} \frac{1}{p-p_i} = \frac{1}{p_i} + \sum_{\substack{j=1\\ j\neq i}}^{n} \frac{1}{p_i-p_j} + \frac{\pi}{\tau_2}(p_i-\bar{p}_i) \tag{5.82a}$$

$$\sideset{}{'_{\mathrm{E}}}\sum_{p\neq p_1,\dots,p_n} \frac{1}{p} = -\sum_{i=1}^{n} \frac{1}{p_i} \tag{5.82b}$$

$$\sideset{}{'_{\mathrm{E}}}\sum_{p\neq p_1,\dots,p_n} \frac{1}{p^2} = \widehat{\mathrm{G}}_2 + \frac{\pi}{\tau_2} - \sum_{i=1}^{n} \frac{1}{p_i^2} \tag{5.82c}$$

$$\sideset{}{'_{\mathrm{E}}}\sum_{p\neq p_1,\dots,p_n} \frac{1}{p^k} = \mathrm{G}_k - \sum_{i=1}^{n} \frac{1}{p_i^k} \qquad k\geq 3\,. \tag{5.82d}$$

Restricting to terms due to (5.82a) and (5.82b) in (5.81) leads to the expression

$$\sum_{i=1}^{n}\left[\left(\frac{2}{p_i} + \sum_{\substack{j=1\\ i\neq j}}^{n}\left(\frac{1}{p_i-p_j} + \frac{1}{p_j}\right) + \frac{\pi}{\tau_2}(p_i-\bar{p}_i)\right)\frac{1}{p_i^{a_0}}\prod_{\substack{j=1\\ i\neq j}}^{n}\frac{1}{p_i-p_j}\right]. \tag{5.83}$$

Since the term $\frac{\pi}{\tau_2}p$ in (5.82a) and $\frac{\pi}{\tau_2}$ in (5.82c) do not have the modular weight of the LHSs, they have to cancel out when plugging (5.82) into (5.81). In order to check this explicitly, consider only the first term and the contribution $\ell = a_0 - 1$ to the second term in (5.81) and set

$$\sideset{}{'_{\mathrm{E}}}\sum_{p\neq p_1,\dots,p_n} \frac{1}{p-p_i} \longrightarrow \frac{\pi}{\tau_2}p_i \tag{5.84}$$

$$\sideset{}{'_{\mathrm{E}}}\sum_{p\neq p_1,\dots,p_n} \frac{1}{p^2} \longrightarrow \frac{\pi}{\tau_2}\,. \tag{5.85}$$

Then, (5.81) becomes

$$\frac{\pi}{\tau_2}\left(\sum_{i=1}^{n}\frac{1}{p_i^{a_0-1}}\prod_{\substack{j=1\\ j\neq i}}^{n}\frac{1}{p_i-p_j} + (-1)^n\frac{h_{a_0-2}(p_1,\dots,p_n)}{\prod_{i=1}^{n}p_i^{a_0-2}}\right), \tag{5.86}$$

which vanishes according to (5.80). This calculation is an important consistency check for (5.82).

Evidently, when performing HSR on an MGF, one encounters sums of the form

$$Q_k(p_1,\dots,p_n) = \sideset{}{'}\sum_{p\neq p_1,\dots,p_n}\frac{1}{p^k}\,, \qquad k\geq 1 \tag{5.87}$$

and shifted versions of these. As we saw above, one way to evaluate these is to use the Eisenstein summation prescription and for $k \geq 2$, we will therefore use

$$Q_2(p_1, \dots, p_n) = \widehat{G}_2 + \frac{\pi}{\tau_2} - \sum_{i=1}^{n} \frac{1}{p_i^2} \tag{5.88a}$$

$$Q_k(p_1, \dots, p_n) = G_k - \sum_{i=1}^{n} \frac{1}{p_i^k}, \quad k \geq 3. \tag{5.88b}$$

In the case $k = 1$ however, the Eisenstein prescription is shift-dependent (the extra term $\frac{\pi}{\tau_2}(p_i - \bar{p}_i)$ in (5.82a) does not arise from a shift of the RHS of (5.82b)). This is cumbersome and unintuitive and we would like to find an expression for $Q_1$ which we can use in the sum (5.78) as if $Q_1$ were shift-independent. I.e. we want to find a $Q_1$ such that if we replace

$$\begin{aligned} \sideset{}{'}\sum_{p \neq p_1, \dots, p_n} \frac{1}{p} &\to Q_1(p_1, \dots, p_n) \\ \sideset{}{'}\sum_{p \neq p_1, \dots, p_n} \frac{1}{p_i - p} &\to Q_1(p_i, \underbrace{p_i - p_1, \dots, p_i - p_n}_{\text{omit } p_i - p_i}) \end{aligned} \tag{5.89}$$

together with (5.88) in (5.78), we obtain the same result as if we had used (5.82). The appropriate definition to this end is

$$Q_1(p_1, \dots, p_n) = -\sum_{i=1}^{n} \frac{1}{p_i} - \frac{\pi}{(n+1)\tau_2} \sum_{i=1}^{n} (p_i - \bar{p}_i). \tag{5.90}$$

Using (5.90) together with the replacements (5.89) in (5.78) and restricting to terms due to $Q_1$ leads to

$$\begin{aligned} \sum_{i=1}^{n} &\left( Q_1(p_i, \underbrace{p_i - p_1, \dots, p_i - p_n}_{\text{omit } p_i - p_i}) \frac{1}{p_i^{a_0}} \prod_{\substack{j=1 \\ j \neq i}}^{n} \frac{1}{p_i - p_j} \right) \\ &+ (-1)^n Q_1(p_1, \dots, p_n) \frac{h_{a_0 - 1}(p_1, \dots, p_n)}{\prod_{i=1}^{n} p_i^{a_0}} \end{aligned} \tag{5.91}$$

$$\begin{aligned} = \sum_{i=1}^{n} &\left[ \left( \frac{1}{p_i} + \sum_{\substack{j=1 \\ i \neq j}}^{n} \frac{1}{p_i - p_j} + \frac{\pi}{(n+1)\tau_2} (p_i - \bar{p}_i + \sum_{\substack{j=1 \\ i \neq j}}^{n} (p_i - p_j - \bar{p}_i + \bar{p}_j)) \right.\right. \\ &\left.\left. + \sum_{j=1}^{n} \frac{1}{p_j} + \frac{\pi}{(n+1)\tau_2} \sum_{j=1}^{n} (p_j - \bar{p}_j) \right) \frac{1}{p_i^{a_0}} \prod_{\substack{j=1 \\ j \neq i}}^{n} \frac{1}{p_i - p_j} \right] \end{aligned} \tag{5.92}$$

$$= \sum_{i=1}^{n}\left[\left(\frac{2}{p_i} + \sum_{\substack{j=1\\ i\neq j}}^{n}\left(\frac{1}{p_i - p_j} + \frac{1}{p_j}\right) + \frac{\pi}{\tau_2}(p_i - \bar{p}_i)\right)\frac{1}{p_i^{a_0}}\prod_{\substack{j=1\\ i\neq j}}^{n}\frac{1}{p_i - p_j}\right], \tag{5.93}$$

which is exactly (5.83) and hence the prescription using the $Q_k$ is equivalent to the rigorous calculation with sums evaluated using the Eisenstein summation convention. In particular, the terms of incorrect modular weight cancel out. Note that the expressions (5.71) found in [16] for dihedral graphs are the special case $n = 2$ of (5.88) and (5.90).

With the expressions (5.88) and (5.90), any modular graph form with an $n$-point holomorphic subgraph can be decomposed. In the next section, we will work out the special case of HSR for trihedral graphs. As we will see, performing the HSR using the expressions for the $Q_i$ derived in this section is laborious and it may be challenging to write the final expression back into MGFs in the general case. For this reason, we provide a different procedure to compute $n$-point HSR in Section 5.4.4.

### 5.4.3 *Trihedral holomorphic subgraph reduction*

Since trihedral graphs have three vertices, closed holomorphic subgraphs can have two or three vertices and we will treat these cases separately.

For two-point holomorphic subgraphs, the trihedral graph takes the form

$$\mathcal{C}\left[\begin{smallmatrix} A_1 \\ B_1 \end{smallmatrix}\middle|\begin{smallmatrix} a_+ & a_- & A_2 \\ 0 & 0 & B_2 \end{smallmatrix}\middle|\begin{smallmatrix} A_3 \\ B_3 \end{smallmatrix}\right] =$$

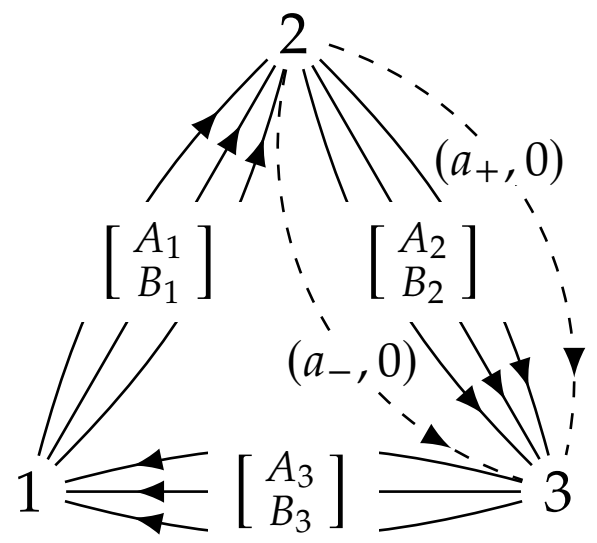

, (5.94)

where the dashed edges are purely holomorphic and we define again $a_0 = a_+ + a_-$. For absolute convergence, we restrict to $a_0 \geq 3$. The trihedral two-point HSR is a straightforward generalization of the dihedral HSR (5.73) and is explicitly given by [I]

$$
\begin{aligned}
&\mathcal{C}\left[\begin{smallmatrix} A_1 \\ B_1 \end{smallmatrix}\middle|\begin{smallmatrix} a_+ & a_- & A_2 \\ 0 & 0 & B_2 \end{smallmatrix}\middle|\begin{smallmatrix} A_3 \\ B_3 \end{smallmatrix}\right] \\
&= (-1)^{a_+} \mathrm{G}_{a_0}\, \mathcal{C}\left[\begin{smallmatrix} A_1 \\ B_1 \end{smallmatrix}\middle|\begin{smallmatrix} A_2 \\ B_2 \end{smallmatrix}\middle|\begin{smallmatrix} A_3 \\ B_3 \end{smallmatrix}\right] - \binom{a_0}{a_+} \mathcal{C}\left[\begin{smallmatrix} A_1 \\ B_1 \end{smallmatrix}\middle|\begin{smallmatrix} a_0 & A_2 \\ 0 & B_2 \end{smallmatrix}\middle|\begin{smallmatrix} A_3 \\ B_3 \end{smallmatrix}\right] \\
&\quad + \sum_{k=4}^{a_+} \binom{a_0 - k - 1}{a_+ - k} \mathrm{G}_k\, \mathcal{C}\left[\begin{smallmatrix} A_1 \\ B_1 \end{smallmatrix}\middle|\begin{smallmatrix} a_0-k & A_2 \\ 0 & B_2 \end{smallmatrix}\middle|\begin{smallmatrix} A_3 \\ B_3 \end{smallmatrix}\right] \\
&\quad + \sum_{k=4}^{a_-} \binom{a_0 - k - 1}{a_- - k} \mathrm{G}_k\, \mathcal{C}\left[\begin{smallmatrix} A_1 \\ B_1 \end{smallmatrix}\middle|\begin{smallmatrix} a_0-k & A_2 \\ 0 & B_2 \end{smallmatrix}\middle|\begin{smallmatrix} A_3 \\ B_3 \end{smallmatrix}\right] \\
&\quad + \binom{a_0 - 2}{a_+ - 1} \left( \widehat{\mathrm{G}}_2\, \mathcal{C}\left[\begin{smallmatrix} A_1 \\ B_1 \end{smallmatrix}\middle|\begin{smallmatrix} a_0-2 & A_2 \\ 0 & B_2 \end{smallmatrix}\middle|\begin{smallmatrix} A_3 \\ B_3 \end{smallmatrix}\right] + \frac{\pi}{\tau_2} \mathcal{C}\left[\begin{smallmatrix} A_1 \\ B_1 \end{smallmatrix}\middle|\begin{smallmatrix} a_0-1 & A_2 \\ -1 & B_2 \end{smallmatrix}\middle|\begin{smallmatrix} A_3 \\ B_3 \end{smallmatrix}\right] \right) .
\end{aligned}
\tag{5.95}
$$

We now proceed to holomorphic subgraph reduction of three-point holomorphic subgraphs in trihedral modular graph forms. This will not only yield a powerful formula for decomposing trihedral MGFs, but also serve as an example of the general higher-point HSR discussed in Section 5.4.2 above, in particular, we will make use of the expressions (5.88) and (5.90). The presentation is taken from Section 4.2 of [I]. The graphs in question are

$$
\mathcal{C}\left[\begin{smallmatrix} A_1 & a_2 \\ B_1 & 0 \end{smallmatrix}\middle|\begin{smallmatrix} A_3 & a_4 \\ B_3 & 0 \end{smallmatrix}\middle|\begin{smallmatrix} A_5 & a_6 \\ B_5 & 0 \end{smallmatrix}\right] =
\tag{5.96}
$$

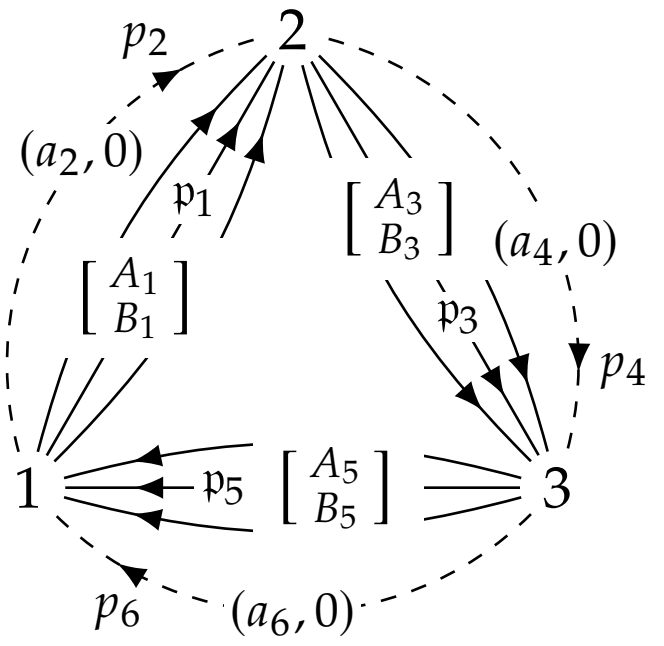

where $\mathfrak{p}_i$ is the total momentum of edge bundle $i$. The dashed holomorphic edges form a three-point subgraph, and the general lattice sum for such graphs is

$$
\mathcal{C}\left[\begin{smallmatrix} A_1 & a_2 \\ B_1 & 0 \end{smallmatrix}\middle|\begin{smallmatrix} A_3 & a_4 \\ B_3 & 0 \end{smallmatrix}\middle|\begin{smallmatrix} A_5 & a_6 \\ B_5 & 0 \end{smallmatrix}\right] = \sum_{\{p\}}' \left( \prod \frac{1}{\mathfrak{p}^A \bar{\mathfrak{p}}^B} \right) \frac{1}{p_2^{a_2} p_4^{a_4} p_6^{a_6}} \delta_{\mathfrak{p}_1 + p_2, \mathfrak{p}_3 + p_4} \delta_{\mathfrak{p}_3 + p_4, \mathfrak{p}_5 + p_6} \,,
\tag{5.97}
$$

with the summation being over all the momenta and

$$
\prod \frac{1}{\mathfrak{p}^A \bar{\mathfrak{p}}^B} = \prod_{i=1,3,5} \prod_{n_i=1}^{R_i} \frac{1}{(p_i^{(n_i)})^{a_i^{(n_i)}}} \frac{1}{(\bar{p}_i^{(n_i)})^{b_i^{(n_i)}}} \,.
\tag{5.98}
$$

In what follows, we will also use the notation $\mathfrak{p}_{ij} = \mathfrak{p}_i - \mathfrak{p}_j$ and $a_0 = a_2 + a_4 + a_6$. To evaluate (5.97), we may begin by using the delta functions to replace $p_2$ and $p_4$ by $p_6$ and the various external momenta $\mathfrak{p}_i$. In particular, we may rewrite

$$C\left[\begin{smallmatrix} A_1 & a_2 \\ B_1 & 0 \end{smallmatrix}\middle|\begin{smallmatrix} A_3 & a_4 \\ B_3 & 0 \end{smallmatrix}\middle|\begin{smallmatrix} A_5 & a_6 \\ B_5 & 0 \end{smallmatrix}\right] = \sideset{}{'}\sum_{\{p_i^{(n_i)}\}} \sideset{}{'}\sum_{p_6 \neq \mathfrak{p}_{15}, \mathfrak{p}_{35}} \left( \prod \frac{1}{\mathfrak{p}^A \bar{\mathfrak{p}}^B} \right) \frac{1}{p_6^{a_6} (p_6 - \mathfrak{p}_{15})^{a_2} (p_6 - \mathfrak{p}_{35})^{a_4}} \,. \tag{5.99}$$

Since $\prod \frac{1}{\mathfrak{p}^A \bar{\mathfrak{p}}^B}$ does not depend on $p_6$, we can focus on evaluating the sum

$$\mathcal{S} = \sideset{}{'}\sum_{p_6 \neq \mathfrak{p}_{15}, \mathfrak{p}_{35}} \frac{1}{p_6^{a_6} (p_6 - \mathfrak{p}_{15})^{a_2} (p_6 - \mathfrak{p}_{35})^{a_4}} \,. \tag{5.100}$$

In order to perform the sum (5.100), we first separate out all cases in which $\mathfrak{p}_{15}$ and $\mathfrak{p}_{35}$ are equal to each other or to zero. In particular, there are five cases to study,

$$\mathfrak{p}_{15} = \mathfrak{p}_{35} = 0 \qquad \mathcal{L}_1 = \sideset{}{'}\sum_{p_6} \frac{1}{p_6^{a_0}} \tag{5.101a}$$

$$\mathfrak{p}_{15} = \mathfrak{p}_{35} \neq 0 \qquad \mathcal{L}_2 = \sideset{}{'}\sum_{p_6 \neq \mathfrak{p}_{15}} \frac{1}{p_6^{a_6} (p_6 - \mathfrak{p}_{15})^{a_2 + a_4}} \tag{5.101b}$$

$$\mathfrak{p}_{15} \neq 0 \,, \quad \mathfrak{p}_{35} = 0 \qquad \mathcal{L}_3 = \sideset{}{'}\sum_{p_6 \neq \mathfrak{p}_{15}} \frac{1}{p_6^{a_6 + a_4} (p_6 - \mathfrak{p}_{15})^{a_2}} \tag{5.101c}$$

$$\mathfrak{p}_{15} = 0 \,, \quad \mathfrak{p}_{35} \neq 0 \qquad \mathcal{L}_4 = \sideset{}{'}\sum_{p_6 \neq \mathfrak{p}_{35}} \frac{1}{p_6^{a_6 + a_2} (p_6 - \mathfrak{p}_{35})^{a_4}} \tag{5.101d}$$

$$\mathfrak{p}_{15} \neq \mathfrak{p}_{35} \,, \quad \mathfrak{p}_{15}, \mathfrak{p}_{35} \neq 0 \quad \mathcal{L}_5 = \sideset{}{'}\sum_{p_6 \neq \mathfrak{p}_{15}, \mathfrak{p}_{35}} \frac{1}{p_6^{a_6} (p_6 - \mathfrak{p}_{15})^{a_2} (p_6 - \mathfrak{p}_{35})^{a_4}} \,, \tag{5.101e}$$

and the function $\mathcal{S}$ is the sum of the above five terms. We may now evaluate them one by one. The first sum is trivial,

$$\mathcal{L}_1 = \mathrm{G}_{a_0} \tag{5.102}$$

and to evaluate the second sum, we use the partial-fraction identity (5.65), which allows us to rewrite $\mathcal{L}_2$ as

$$(-1)^{a_2 + a_4} \mathcal{L}_2 = \sideset{}{'}\sum_{p_6 \neq \mathfrak{p}_{15}} \left[ \sum_{k=1}^{a_6} \binom{a_0 - k - 1}{a_6 - k} \frac{1}{p_6^k \, \mathfrak{p}_{15}^{a_0 - k}} \right.$$
$$\left. + \sum_{k=1}^{a_2 + a_4} \binom{a_0 - k - 1}{a_2 + a_4 - k} \frac{1}{(\mathfrak{p}_{15} - p_6)^k \mathfrak{p}_{15}^{a_0 - k}} \right]$$

$$= \sum_{k=1}^{a_6} \binom{a_0-k-1}{a_6-k} \frac{Q_k(\mathfrak{p}_{15})}{\mathfrak{p}_{15}^{a_0-k}} + \sum_{k=1}^{a_2+a_4} \binom{a_0-k-1}{a_2+a_4-k} \frac{Q_k(\mathfrak{p}_{15})}{\mathfrak{p}_{15}^{a_0-k}} . \tag{5.103}$$

We now use the expressions (5.71) for the $Q_k$. Upon applying the identities

$$\binom{a_1+a_2}{a_1} = \sum_{k=1}^{a_1} \binom{a_1+a_2-k-1}{a_1-k} + \sum_{k=1}^{a_2} \binom{a_1+a_2-k-1}{a_2-k} \tag{5.104a}$$

$$\binom{a_0-2}{a_6-1} = \binom{a_0-3}{a_2+a_4-2} + \binom{a_0-3}{a_6-2} \tag{5.104b}$$

between binomial coefficients, the sum $\mathcal{L}_2$ simplifies to

$$\begin{aligned}(-1)^{a_2+a_4}\mathcal{L}_2 = &\sum_{k=4}^{a_6} \binom{a_0-k-1}{a_6-k} \frac{\mathrm{G}_k}{\mathfrak{p}_{15}^{a_0-k}} + \sum_{k=4}^{a_2+a_4} \binom{a_0-k-1}{a_2+a_4-k} \frac{\mathrm{G}_k}{\mathfrak{p}_{15}^{a_0-k}} \\ &- \binom{a_0}{a_6} \frac{1}{\mathfrak{p}_{15}^{a_0}} + \binom{a_0-2}{a_6-1} \frac{1}{\mathfrak{p}_{15}^{a_0-1}} \left( \mathfrak{p}_{15}\widehat{\mathrm{G}}_2 + \frac{\pi}{\tau_2}\bar{\mathfrak{p}}_{15} \right) .\end{aligned} \tag{5.105}$$

Crucially, note that the $\frac{\pi}{\tau_2}$ terms in $Q_2(p_1)$ have canceled with the $\frac{\pi}{2\tau_2}p_0$ terms of $Q_1(p_1)$, just as in the dihedral case. Recall that this was necessary for obtaining a modular covariant final result, since such terms had different modular weight than the other terms.

The sum $\mathcal{L}_3$ can be obtained from (5.105) by replacing $a_6 \to a_4 + a_6$ and $a_2 + a_4 \to a_2$. $\mathcal{L}_4$ can be reached by similar relabelings, so we may now proceed directly to $\mathcal{L}_5$. To begin, we apply the decomposition formula (5.65) twice to obtain

$$\begin{aligned}(-)^{a_2+a_4}\mathcal{L}_5 = &\sum_{k=1}^{a_6}\sum_{\ell=1}^{k} \binom{a_2+a_6-k-1}{a_6-k}\binom{a_4+k-\ell-1}{k-\ell} \frac{Q_\ell(\mathfrak{p}_{15},\mathfrak{p}_{35})}{(\mathfrak{p}_{15})^{a_2+a_6-k}(\mathfrak{p}_{35})^{a_4+k-\ell}} \\ &+ \sum_{k=1}^{a_6}\sum_{\ell=1}^{a_4} \binom{a_2+a_6-k-1}{a_6-k}\binom{a_4+k-\ell-1}{a_4-\ell} \frac{Q_\ell(\mathfrak{p}_{31},\mathfrak{p}_{35})}{(\mathfrak{p}_{15})^{a_2+a_6-k}(\mathfrak{p}_{35})^{a_4+k-\ell}} \\ &+ \sum_{k=1}^{a_2}\sum_{\ell=1}^{a_4} \binom{a_2+a_6-k-1}{a_2-k}\binom{a_4+k-\ell-1}{a_4-\ell} (-)^k \frac{Q_\ell(\mathfrak{p}_{31},\mathfrak{p}_{35})}{(\mathfrak{p}_{15})^{a_2+a_6-k}(\mathfrak{p}_{31})^{a_4+k-\ell}} \\ &+ \sum_{k=1}^{a_2}\sum_{\ell=1}^{k} \binom{a_2+a_6-k-1}{a_2-k}\binom{a_4+k-\ell-1}{k-\ell} (-)^k \frac{Q_\ell(-\mathfrak{p}_{15},\mathfrak{p}_{31})}{(\mathfrak{p}_{15})^{a_2+a_6-k}(\mathfrak{p}_{31})^{a_4+k-\ell}} ,\end{aligned} \tag{5.106}$$

where we used the $Q_k(p_1, p_2)$ as in (5.87) and (5.89). Now, we use the expressions (5.88) and (5.90) which were derived for the general case in Section 5.4.2 to perform the sums $Q_k$, resulting in

$$\begin{aligned}(-)^{a_2+a_4}\mathcal{L}_5 = &\sum_{k=1}^{a_6}\binom{a_2+a_6-k-1}{a_6-k}\frac{1}{(\mathfrak{p}_{15})^{a_2+a_6-k}(\mathfrak{p}_{35})^{a_4+k}}\mathcal{X}_k(\mathfrak{p}_{15},\mathfrak{p}_{35}) \\ &+\sum_{k=1}^{a_2}\binom{a_2+a_6-k-1}{a_2-k}\frac{(-)^k}{(\mathfrak{p}_{15})^{a_2+a_6-k}(\mathfrak{p}_{31})^{a_4+k}}\mathcal{X}_k(-\mathfrak{p}_{15},\mathfrak{p}_{31}),\end{aligned} \tag{5.107}$$

where

$$\begin{aligned}\mathcal{X}_k(p,q) = &-\sum_{\ell=1}^{k}\binom{a_4+k-\ell-1}{k-\ell}\left(\frac{q}{p}\right)^{\ell} - \sum_{\ell=1}^{a_4}\binom{a_4+k-\ell-1}{a_4-\ell}\left(\frac{q}{q-p}\right)^{\ell} \\ &+\sum_{\ell=4}^{k}\binom{a_4+k-\ell-1}{k-\ell}q^{\ell}\mathrm{G}_\ell + \sum_{\ell=4}^{a_4}\binom{a_4+k-\ell-1}{a_4-\ell}q^{\ell}\mathrm{G}_\ell \\ &-\binom{a_4+k}{a_4} + \binom{a_4+k-2}{k-1}\left(q^2\widehat{\mathrm{G}}_2 + \frac{\pi}{\tau_2}q\bar{q}\right).\end{aligned} \tag{5.108}$$

With (5.107), we have completed the evaluation of the five sums $\mathcal{L}_i$ listed in (5.101) which make up the sum $\mathcal{S}$ in (5.100). In order to obtain our final formula for three-point HSR of the trihedral graph (5.99), we must now carry out the sums over the remaining momenta. This amounts to plugging the expressions obtained for the $\mathcal{L}_i$ above back into (5.99) and rewriting the result in terms of MGFs. We denote the completely summed versions of the $\mathcal{L}_i$ by $L_i$, such that our final answer is given by

$$\mathcal{C}\left[\begin{smallmatrix}A_1 & a_2\\ B_1 & 0\end{smallmatrix}\middle|\begin{smallmatrix}A_3 & a_4\\ B_3 & 0\end{smallmatrix}\middle|\begin{smallmatrix}A_5 & a_6\\ B_5 & 0\end{smallmatrix}\right] = \sum_{i=1}^{5} L_i\,. \tag{5.109}$$

Although obtaining the $L_i$ from the $\mathcal{L}_i$ is lengthy, no conceptual novelties arise. The details of the calculation and the final result for the $L_i$ are spelled out in Appendix B.2.

As an example, consider the graph[8]

$$\mathcal{C}\left[\begin{smallmatrix}1 & 1\\ 1 & 0\end{smallmatrix}\middle|\begin{smallmatrix}1 & 2\\ 1 & 0\end{smallmatrix}\middle|\begin{smallmatrix}1\\ 0\end{smallmatrix}\right] = \sum_{p_i}{}' \frac{1}{p_1\bar{p}_1 p_3\bar{p}_3}\frac{1}{p_2 p_4^2 p_6}\delta_{p_1+p_2,p_3+p_4}\delta_{p_1+p_2,p_6}\,, \tag{5.110}$$

which contains a three-point holomorphic subgraph. The expression (B.13) for $L_1$ in Appendix B.2 yields in this case

$$L_1 = \mathrm{G}_4\,\mathcal{C}\left[\varnothing\middle|\begin{smallmatrix}1\\ 1\end{smallmatrix}\middle|\begin{smallmatrix}1\\ 1\end{smallmatrix}\right] = \mathrm{G}_4\,\mathcal{C}\left[\begin{smallmatrix}1\\ 1\end{smallmatrix}\right]^2 = 0\,, \tag{5.111}$$

8 This graph is not in its canonical representation to indicate the assignments of the $A_i$, $B_i$ and $a_i$ according to (5.97).

where we used the topological simplifications (5.25) and (5.21). Similarly, $L_2$ is, according to (B.14), given by

$$\begin{aligned} L_2 &= 4\,\mathcal{C}\big[\begin{smallmatrix}1\\1\end{smallmatrix}\big|\begin{smallmatrix}1\\1\end{smallmatrix}\big|\begin{smallmatrix}4\\0\end{smallmatrix}\big] - \widehat{\mathrm{G}}_2\,\mathcal{C}\big[\begin{smallmatrix}1\\1\end{smallmatrix}\big|\begin{smallmatrix}1\\1\end{smallmatrix}\big|\begin{smallmatrix}2\\0\end{smallmatrix}\big] - \frac{\pi}{\tau_2}\,\mathcal{C}\big[\begin{smallmatrix}1\\1\end{smallmatrix}\big|\begin{smallmatrix}1\\1\end{smallmatrix}\big|\begin{smallmatrix}3\\-1\end{smallmatrix}\big] \\ &= 4\,\mathcal{C}\big[\begin{smallmatrix}6&0\\2&0\end{smallmatrix}\big] - \widehat{\mathrm{G}}_2\,\mathcal{C}\big[\begin{smallmatrix}4&0\\2&0\end{smallmatrix}\big] - \frac{\pi}{\tau_2}\,\mathcal{C}\big[\begin{smallmatrix}5&0\\1&0\end{smallmatrix}\big]\,, \end{aligned} \tag{5.112}$$

where we used the topological simplifications (5.24) and (5.23). Along the same lines, we find from the expressions (B.15), (B.16) and (B.18),

$$L_3 = L_4 = 0 \tag{5.113}$$

$$L_5 = -X_1 + \tilde{X}_1\,, \tag{5.114}$$

where

$$X_1 = 3\,\mathcal{C}\big[\begin{smallmatrix}6&0\\2&0\end{smallmatrix}\big] - \frac{\pi}{\tau_2}\,\mathcal{C}\big[\begin{smallmatrix}5&0\\1&0\end{smallmatrix}\big] - \widehat{\mathrm{G}}_2\,\mathcal{C}\big[\begin{smallmatrix}4&0\\2&0\end{smallmatrix}\big] - \mathcal{C}\big[\begin{smallmatrix}1&2&3\\1&0&1\end{smallmatrix}\big] \tag{5.115a}$$

$$\begin{aligned} \tilde{X}_1 &= -\,\mathcal{C}\big[\begin{smallmatrix}6&0\\2&0\end{smallmatrix}\big] + \mathcal{C}\big[\begin{smallmatrix}3&0\\1&0\end{smallmatrix}\big]^2 + \widehat{\mathrm{G}}_2\,\mathcal{C}\big[\begin{smallmatrix}1&1&2\\0&1&1\end{smallmatrix}\big] \\ &\quad + \frac{\pi}{\tau_2}\,\mathcal{C}\big[\begin{smallmatrix}1&2&2\\1&-1&1\end{smallmatrix}\big] - 3\,\mathcal{C}\big[\begin{smallmatrix}1&2&3\\1&1&0\end{smallmatrix}\big]\,. \end{aligned} \tag{5.115b}$$

In total, we have

$$\begin{aligned} \mathcal{C}\big[\begin{smallmatrix}1\\0\end{smallmatrix}\big|\begin{smallmatrix}1&1\\0&1\end{smallmatrix}\big|\begin{smallmatrix}1&2\\1&0\end{smallmatrix}\big] &= \mathcal{C}\big[\begin{smallmatrix}3&0\\1&0\end{smallmatrix}\big]^2 + \mathcal{C}\big[\begin{smallmatrix}1&2&3\\1&0&1\end{smallmatrix}\big] - 3\,\mathcal{C}\big[\begin{smallmatrix}1&2&3\\1&1&0\end{smallmatrix}\big] \\ &\quad + \widehat{\mathrm{G}}_2\,\mathcal{C}\big[\begin{smallmatrix}1&1&2\\0&1&1\end{smallmatrix}\big] + \frac{\pi}{\tau_2}\,\mathcal{C}\big[\begin{smallmatrix}1&2&2\\1&-1&1\end{smallmatrix}\big]\,. \end{aligned} \tag{5.116}$$

The two-loop graphs can be simplified further by using the momentum conservation identities (5.36) and the factorization identity (5.46) repeatedly. E.g. we have

$$\begin{aligned} 0 &= \mathcal{C}\big[\begin{smallmatrix}0&2&2\\0&1&1\end{smallmatrix}\big] + 2\,\mathcal{C}\big[\begin{smallmatrix}1&1&2\\0&1&1\end{smallmatrix}\big] \\ &= \mathcal{C}\big[\begin{smallmatrix}4&0\\2&0\end{smallmatrix}\big] + 2\,\mathcal{C}\big[\begin{smallmatrix}1&1&2\\0&1&1\end{smallmatrix}\big]\,. \end{aligned} \tag{5.117}$$

In this way, we obtain the decomposition

$$\begin{aligned} \mathcal{C}\big[\begin{smallmatrix}1\\0\end{smallmatrix}\big|\begin{smallmatrix}1&1\\0&1\end{smallmatrix}\big|\begin{smallmatrix}1&2\\1&0\end{smallmatrix}\big] &= -\frac{1}{2}\,\mathcal{C}\big[\begin{smallmatrix}6&0\\2&0\end{smallmatrix}\big] + \frac{3}{2}\,\mathcal{C}\big[\begin{smallmatrix}3&0\\1&0\end{smallmatrix}\big]^2 - \frac{1}{2}\widehat{\mathrm{G}}_2\,\mathcal{C}\big[\begin{smallmatrix}4&0\\2&0\end{smallmatrix}\big] \\ &\quad + 3\frac{\pi}{\tau_2}\,\mathcal{C}\big[\begin{smallmatrix}5&0\\1&0\end{smallmatrix}\big] - \frac{\pi}{\tau_2}\widehat{\mathrm{G}}_2\,\mathcal{C}\big[\begin{smallmatrix}3&0\\1&0\end{smallmatrix}\big] - \Big(\frac{\pi}{\tau_2}\Big)^2 \mathrm{G}_4 \end{aligned} \tag{5.118}$$

of a trihedral three-loop graph into one-loop graphs.

Although trihedral modular graph forms do not depend on the order of the blocks, the form (5.116) of the decomposition depends on the assignment of the blocks to the $\big[\begin{smallmatrix}A_i\\B_i\end{smallmatrix}\big]$, since we broke the permutation symmetry by solving the momentum conservation constraints and preforming the partial-fraction decomposition. E.g. for the representa-tion $\mathcal{C}\big[\begin{smallmatrix}1&1\\1&0\end{smallmatrix}\big|\begin{smallmatrix}1\\0\end{smallmatrix}\big|\begin{smallmatrix}1&2\\1&0\end{smallmatrix}\big]$ of the graph (5.110), the three-point HSR formula in Appendix B.2 yields

$$\begin{aligned}\mathcal{C}\big[\begin{smallmatrix}1\\0\end{smallmatrix}\big|\begin{smallmatrix}1&1\\0&1\end{smallmatrix}\big|\begin{smallmatrix}1&2\\1&0\end{smallmatrix}\big] = 4\,\mathcal{C}\big[\begin{smallmatrix}6&0\\2&0\end{smallmatrix}\big] - \widehat{\mathrm{G}}_2\,\mathcal{C}\big[\begin{smallmatrix}4&0\\2&0\end{smallmatrix}\big] + 3\,\mathcal{C}\big[\begin{smallmatrix}1&1&4\\0&1&1\end{smallmatrix}\big] - \mathcal{C}\big[\begin{smallmatrix}1&2&3\\1&1&0\end{smallmatrix}\big]\\ - \widehat{\mathrm{G}}_2\,\mathcal{C}\big[\begin{smallmatrix}1&1&2\\0&1&1\end{smallmatrix}\big] - \frac{\pi}{\tau_2}\mathcal{C}\big[\begin{smallmatrix}5&0\\1&0\end{smallmatrix}\big] - \frac{\pi}{\tau_2}\mathcal{C}\big[\begin{smallmatrix}1&1&3\\0&1&0\end{smallmatrix}\big].\end{aligned} \tag{5.119}$$

This can be simplified to (5.118) by using the identities from Section 5.3 and the dihedral holomorphic subgraph reduction (5.73). In more complicated cases, however, equating these different decompositions leads to valuable new identities between MGFs. As explained in more detail in Appendix B.2, even divergent graphs can arise in this decomposition. These divergences however cancel out upon further simplification of the result.

In the `Mathematica` package `Modular Graph Forms`, the trihedral two-point HSR formula (5.95) is implemented in the function **`TriCSimplify`**. Again, with the default options, negative entries are removed via momentum conservation and identities from the database are applied, so in order to just apply (5.95), we run

In[38]:= **TriCSimplify**$\Big[$**c**$\big[\begin{smallmatrix}2&2&1\\0&0&1\end{smallmatrix},\begin{smallmatrix}1&1\\1&1\end{smallmatrix},\begin{smallmatrix}1&1\\1&1\end{smallmatrix}\big]$**, momSimplify → False, useIds → False**$\Big]$

$$\text{Out[38]=}\ \mathrm{C}\big[\begin{smallmatrix}2&2\\0&0\end{smallmatrix}\big]\mathrm{C}\big[\begin{smallmatrix}1\\1\end{smallmatrix}\big|\begin{smallmatrix}1&1\\1&1\end{smallmatrix}\big|\begin{smallmatrix}1&1\\1&1\end{smallmatrix}\big] - 6\,\mathrm{C}\big[\begin{smallmatrix}1&1\\1&1\end{smallmatrix}\big|\begin{smallmatrix}1&1\\1&1\end{smallmatrix}\big|\begin{smallmatrix}1&4\\1&0\end{smallmatrix}\big] + 2\,\mathrm{C}\big[\begin{smallmatrix}1&1\\1&1\end{smallmatrix}\big|\begin{smallmatrix}1&1\\1&1\end{smallmatrix}\big|\begin{smallmatrix}1&2\\1&0\end{smallmatrix}\big]\hat{\mathrm{G}}_2 + \frac{2\pi\,\mathrm{C}\big[\begin{smallmatrix}1&1\\1&1\end{smallmatrix}\big|\begin{smallmatrix}1&1\\1&1\end{smallmatrix}\big|\begin{smallmatrix}1&3\\1&-1\end{smallmatrix}\big]}{\tau_2}\ .$$

The three-point HSR detailed in Appendix B.2 is also performed by the function **`TriCSimplify`**, although it is not implemented in exactly the same form as it is written in Appendix B.2. I.e. the momentum assignment and partial fraction were performed slightly differently, leading to different, but equivalent, expression for $L_5$ in (B.18). E.g. in the representation $\mathcal{C}\big[\begin{smallmatrix}1&1\\1&0\end{smallmatrix}\big|\begin{smallmatrix}1&2\\1&0\end{smallmatrix}\big|\begin{smallmatrix}1\\0\end{smallmatrix}\big]$, the graph $\mathcal{C}\big[\begin{smallmatrix}1\\0\end{smallmatrix}\big|\begin{smallmatrix}1&1\\0&1\end{smallmatrix}\big|\begin{smallmatrix}1&2\\1&0\end{smallmatrix}\big]$ was decomposed in (5.116). This computation can be performed by running

In[39]:= **DiCSimplify**$\Big[$**TriCSimplify**$\Big[$**c**$\big[\begin{smallmatrix}1&1\\1&0\end{smallmatrix},\begin{smallmatrix}1&2\\1&0\end{smallmatrix},\begin{smallmatrix}1\\0\end{smallmatrix}\big]\Big]$**, momSimplify → False, useIds → False**$\Big]$

$$\text{Out[39]=}\ \mathrm{C}\big[\begin{smallmatrix}3&0\\1&0\end{smallmatrix}\big]^2 + \mathrm{C}\big[\begin{smallmatrix}1&2&3\\1&0&1\end{smallmatrix}\big] - 3\,\mathrm{C}\big[\begin{smallmatrix}1&2&3\\1&1&0\end{smallmatrix}\big] + \mathrm{C}\big[\begin{smallmatrix}1&1&2\\0&1&1\end{smallmatrix}\big]\hat{\mathrm{G}}_2 + \frac{\pi\,\mathrm{C}\big[\begin{smallmatrix}1&2&2\\1&-1&1\end{smallmatrix}\big]}{\tau_2}.$$

**`TriCSimplify`** performs HSR on the first suitable holomorphic subgraph. It first performs the two-point version, then the three-point version, also antiholomorphic subgraphs are simplified. With the Boolean option **`triHSR`**, trihedral HSR can be deactivated (its default value is **`True`**) and with the Boolean options **`tri2ptHSR`** and **`tri3ptHSR`**, the two- and three-point versions can be deactivated individually.

As discussed in Appendix B.2, the result of the three-point HSR formula contains divergent graphs if the second block in the trihedral MGF to be reduced contains a $\left[\begin{smallmatrix}1&1\\0&1\end{smallmatrix}\right]$ subblock. In **`TriCSimplify`**, the blocks are rearranged automatically so that the result is never divergent. If this is not possible because every block has a $\left[\begin{smallmatrix}1&1\\0&1\end{smallmatrix}\right]$ subblock, the warning **`TriCSimplify::NoConvHSROrder`** is issued. If the Boolean option **`divHSR`** of **`TriCSimplify`** is set to **`True`** (the default), the expression containing divergent graphs is returned, otherwise, **`TriCSimplify`** just returns the input. E.g.

In[40]:= $\mathtt{DiCSimplify}\left[\mathtt{TriCSimplify}\left[\mathtt{c}\left[\begin{array}{cc}1&1\\0&1\end{array},\begin{array}{cc}1&1\\0&1\end{array},\begin{array}{ccc}1&1&2\\0&1&1\end{array}\right],\ \mathtt{momSimplify}\to\mathtt{False},\ \mathtt{useIds}\to\mathtt{False}\right],\ \mathtt{useIds}\to\mathtt{False}\right]$

TriCSimplify : No ordering of the blocks of $\mathrm{C}\left[\begin{array}{cc|cc|ccc}1&1&1&1&1&1&2\\0&1&0&1&0&1&1\end{array}\right]$ is suitable for convergent three–point HSR.

Out[40]= $$\mathrm{C}\left[\begin{array}{c|cc|ccc}1&1&1&1&2&2\\1&0&1&1&0&1\end{array}\right]+2\,\mathrm{C}\left[\begin{array}{c|cc|ccc}1&1&2&1&1&2\\1&1&0&0&1&1\end{array}\right]-3\,\mathrm{C}\left[\begin{array}{c|cc|cc}1&1&2&1&3\\1&1&1&1&0\end{array}\right]+$$
$$3\,\mathrm{C}\left[\begin{array}{cc|cc|cc}1&1&1&2&1&2\\0&1&1&0&1&1\end{array}\right]+\mathrm{C}\left[\begin{array}{cccc}1&1&2&2\\0&1&1&2\end{array}\right]\hat{\mathrm{G}}_2+2\,\mathrm{C}\left[\begin{array}{c|cc|cc}1&1&1&1&2\\1&0&1&1&1\end{array}\right]\hat{\mathrm{G}}_2-$$
$$\frac{\pi\,\mathrm{C}\left[\begin{array}{c|cc|ccc}1&1&1&1&1&2\\1&-1&1&0&1&1\end{array}\right]}{\tau_2}+\frac{\pi\,\mathrm{C}\left[\begin{array}{c|cc|cc}1&1&2&1&2\\1&1&-1&1&1\end{array}\right]}{\tau_2}-\frac{\pi\,\mathrm{C}\left[\begin{array}{cc|cc|cc}1&1&1&1&1&2\\-1&1&0&1&1&1\end{array}\right]}{\tau_2}.$$

The function **`TriCSimplify`** is called by the function **`CSimplify`** and **`CSimplify`** also inherits the options of **`TriCSimplify`**.

### 5.4.4 *Holomorphic subgraph reduction and Fay identities*

The discussion of holomorphic subgraph reduction has so far been exclusively in terms of the sum representation of the MGFs. In the integral representation, HSR corresponds to certain identities for products of the $f^{(n)}(z,\tau)$ (3.91). These descend from the *Fay identity* of the Kronecker–Eisenstein series [204, 211]

$$\begin{aligned}F(z_1,\eta_1,\tau)F(z_2,\eta_2,\tau) &= F(z_1{-}z_2,\eta_1,\tau)F(z_2,\eta_1{+}\eta_2,\tau)\\ &\quad + F(z_2{-}z_1,\eta_2,\tau)F(z_1,\eta_1{+}\eta_2,\tau)\end{aligned} \tag{5.120a}$$

$$\begin{aligned}\Omega(z_1,\eta_1,\tau)\Omega(z_2,\eta_2,\tau) &= \Omega(z_1{-}z_2,\eta_1,\tau)\Omega(z_2,\eta_1{+}\eta_2,\tau)\\ &\quad + \Omega(z_2{-}z_1,\eta_2,\tau)\Omega(z_1,\eta_1{+}\eta_2,\tau)\end{aligned} \tag{5.120b}$$

by means of the expansion (3.87) and are given by [28]

$$\begin{aligned}f_{12}^{(a_1)}f_{13}^{(a_2)} &= (-1)^{a_1-1}f_{23}^{(a_1+a_2)}+\sum_{j=0}^{a_1}\binom{a_2+j-1}{j}f_{32}^{(a_1-j)}f_{13}^{(a_2+j)}\\ &\quad+\sum_{j=0}^{a_2}\binom{a_1+j-1}{j}f_{12}^{(a_1+j)}f_{23}^{(a_2-j)}\,,\end{aligned} \tag{5.121}$$

where $a_1, a_2 \geq 0$. According to (3.119), a factor $f_{ij}^{(a)}$ in a Koba–Nielsen integral corresponds to a (holomorphic) $(a, 0)$-edge. Hence, when (5.121) is applied in a Koba–Nielsen integrand, it generates an identity between modular graph forms with holomorphic edges.

### HOLOMORPHIC SUBGRAPHS WITH MORE THAN TWO VERTICES

Consider an MGF with an $n$-point holomorphic subgraph ($n > 2$) given by a Koba–Nielsen integral over $C_{12}^{(a_1,0)} C_{13}^{(a_2,0)} = \pm f_{12}^{(a_1)} f_{13}^{(a_2)}$ and $n-2$ further factors $C_{ij}^{(a_k,0)}$, as well as additional non-holomorphic edges. In this case, by focusing on the holomorphic edges, the MGF-identity implied by (5.121) can be written graphically as

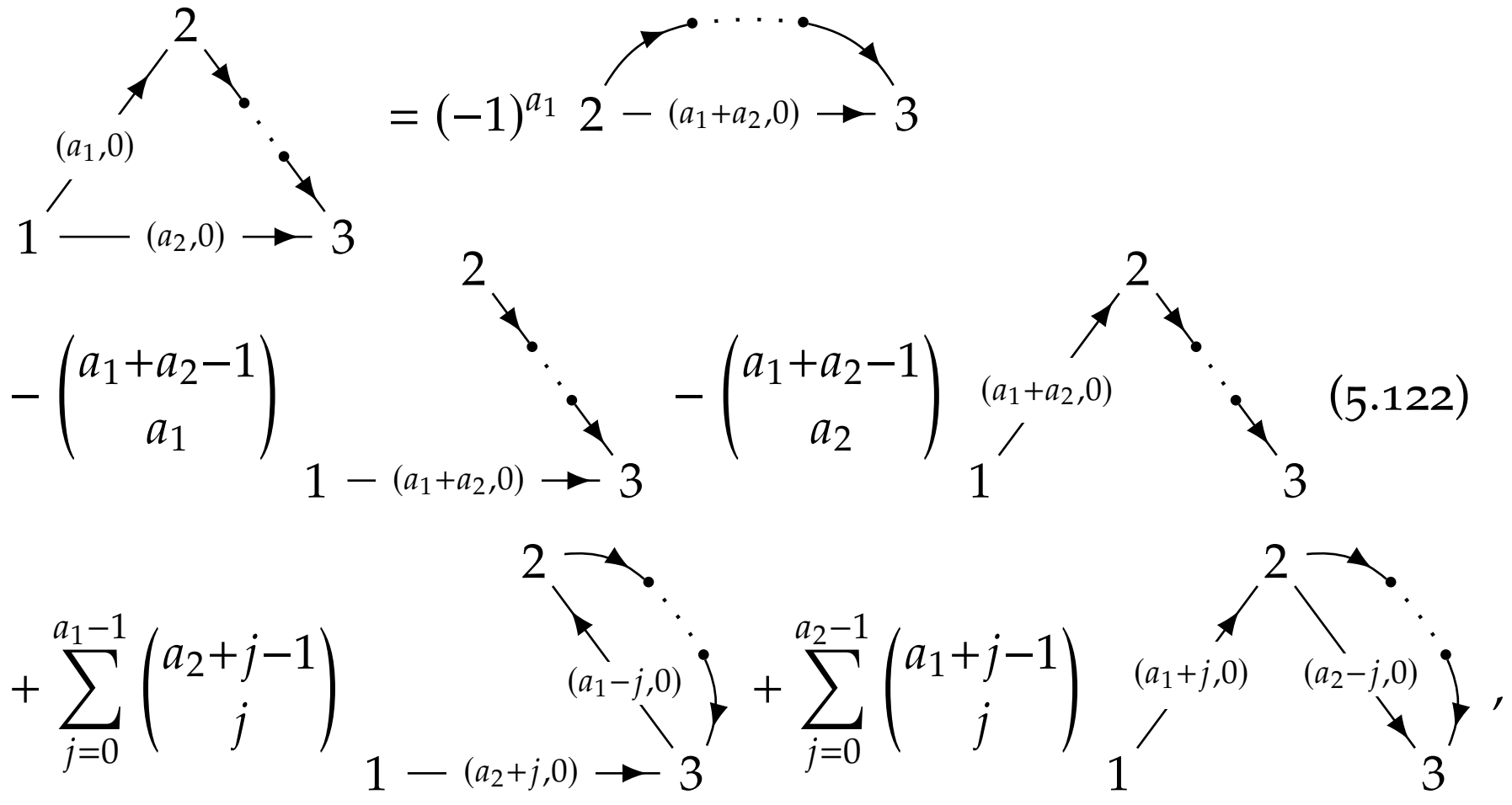

(5.122)

where the ellipsis denotes a sequence of holomorphic edges such that the LHS forms a closed $n$-point holomorphic graph. The full MGF has additional non-holomorphic edges which can in general connect any vertices and are not drawn in (5.122). In going from (5.121) to (5.122), we have separated the contributions form $f^{(0)} = 1$ since $f^{(0)} \neq -C^{(0,0)}$, according to (5.45). In the representation (5.122) it is clear that on the RHS each term has either one edge less than the LHS (terms two and three) or the closed holomorphic subgraph has one edge less (terms four and five) or both (the first term). If a term on the RHS has an edge less than the LHS, the associated MGF has one loop order less when accounting for the non-holomorphic edges suppressed in (5.122) as well. Hence, the Fay identities (5.121) allow to reduce $n$-point HSR to $(n-1)$-point HSR plus graphs of lower loop order.

As an example, consider the tetrahedral graph

$$\mathcal{C}\left[\begin{array}{c}1\\0\\\hline 1\\0\end{array}\middle\|\begin{array}{c}1\\0\\\hline 1\\0\end{array}\middle\|\begin{array}{c}1\\1\\\hline 1\\1\end{array}\right] = \tag{5.123}$$

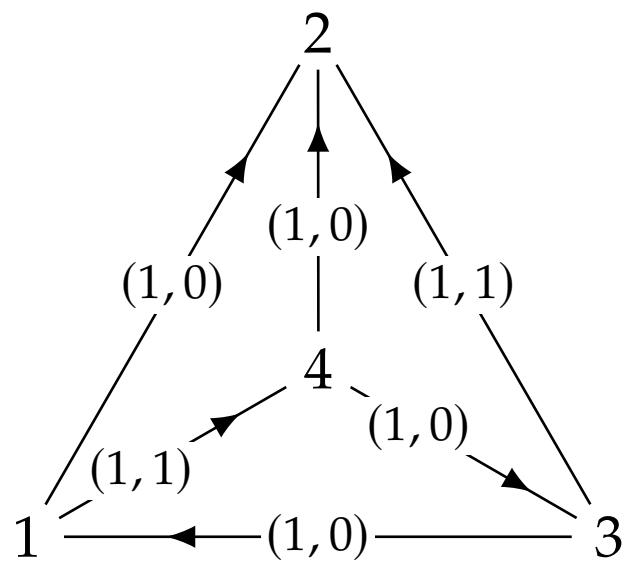

which has a four-point holomorphic subgraph and appears in the four-gluon amplitude in the heterotic string discussed in Chapter 6 at the order $\alpha'^2$, cf. (6.46) below. By applying (5.122) to the two holomorphic edges connected to vertex 4, we obtain the decomposition (with the graphs not yet in their canonical representation)

$$\begin{aligned}\mathcal{C}\left[\begin{array}{c}1\\0\\\hline 1\\0\end{array}\middle\|\begin{array}{c}1\\0\\\hline 1\\0\end{array}\middle\|\begin{array}{c}1\\1\\\hline 1\\1\end{array}\right] =& -\mathcal{C}\left[\begin{array}{c}1\\0\\\hline 1\\0\end{array}\middle\|\begin{array}{c}\varnothing\\\hline 1\\0\end{array}\middle\|\begin{array}{cc}1&2\\1&0\\\hline \multicolumn{2}{c}{1}\\\multicolumn{2}{c}{1}\end{array}\right] - \mathcal{C}\left[\begin{array}{c}1\\0\\\hline 2\\0\end{array}\middle\|\begin{array}{c}\varnothing\\\hline 1\\0\end{array}\middle\|\begin{array}{c}1\\1\\\hline 1\\1\end{array}\right] - \mathcal{C}\left[\begin{array}{c}1\\0\\\hline \varnothing\end{array}\middle\|\begin{array}{c}2\\0\\\hline 1\\0\end{array}\middle\|\begin{array}{c}1\\1\\\hline 1\\1\end{array}\right]\\ &+\mathcal{C}\left[\begin{array}{c}1\\0\\\hline 1\\0\end{array}\middle\|\begin{array}{c}\varnothing\\\hline 1\\0\end{array}\middle\|\begin{array}{cc}1&1\\0&1\\\hline \multicolumn{2}{c}{1}\\\multicolumn{2}{c}{1}\end{array}\right] - \mathcal{C}\left[\begin{array}{c}1\\0\\\hline \varnothing\end{array}\middle\|\begin{array}{c}1\\0\\\hline 1\\0\end{array}\middle\|\begin{array}{cc}1&1\\0&1\\\hline \multicolumn{2}{c}{1}\\\multicolumn{2}{c}{1}\end{array}\right].\end{aligned} \tag{5.124}$$

In this expression, every graph has one empty block and can be simplified using the topological simplifications of Section 5.3.2 to

$$\mathcal{C}\left[\begin{array}{c}1\\0\\\hline 1\\0\end{array}\middle\|\begin{array}{c}1\\0\\\hline 1\\0\end{array}\middle\|\begin{array}{c}1\\1\\\hline 1\\1\end{array}\right] = 2\,\mathcal{C}\left[\begin{smallmatrix}1&2&3\\0&1&1\end{smallmatrix}\right] + 2\,\mathcal{C}\left[\begin{smallmatrix}1\\0\end{smallmatrix}\middle|\begin{smallmatrix}1&1\\0&1\end{smallmatrix}\middle|\begin{smallmatrix}1&2\\0&1\end{smallmatrix}\right]. \tag{5.125}$$

In this way, the four-point HSR in the original graph was reduced to three-point HSR. This can be performed either via another Fay identity or via (5.109) and together with the basis decompositions to be discussed in Section 5.7, we obtain the final result

$$\begin{aligned}\mathcal{C}\left[\begin{array}{c}1\\0\\\hline 1\\0\end{array}\middle\|\begin{array}{c}1\\0\\\hline 1\\0\end{array}\middle\|\begin{array}{c}1\\1\\\hline 1\\1\end{array}\right] =& \, 2\,\mathcal{C}\left[\begin{smallmatrix}6&0\\2&0\end{smallmatrix}\right] - 4\,\mathcal{C}\left[\begin{smallmatrix}3&0\\1&0\end{smallmatrix}\right]^2 + 2\widehat{\mathrm{G}}_2\,\mathcal{C}\left[\begin{smallmatrix}4&0\\2&0\end{smallmatrix}\right]\\ &- 12\frac{\pi}{\tau_2}\,\mathcal{C}\left[\begin{smallmatrix}5&0\\1&0\end{smallmatrix}\right] + 4\frac{\pi}{\tau_2}\widehat{\mathrm{G}}_2\,\mathcal{C}\left[\begin{smallmatrix}3&0\\1&0\end{smallmatrix}\right] + 4\left(\frac{\pi}{\tau_2}\right)^2\mathrm{G}_4\,.\end{aligned} \tag{5.126}$$

In general, the closed holomorphic subgraph is of course not necessary for the identity (5.121) to hold. Hence, if we remove the dashed edges from (5.122), this generates identities between modular graph forms which have at least two non-parallel holomorphic edges both connected to the same vertex. For trihedral graphs, we have e.g.

$$
\begin{aligned}
\mathcal{C}\big[\begin{smallmatrix} A_1 & a_1 \\ B_1 & 0 \end{smallmatrix}\big|\begin{smallmatrix} A_2 & a_2 \\ B_2 & 0 \end{smallmatrix}\big|\begin{smallmatrix} A_3 \\ B_3 \end{smallmatrix}\big] = (-)^{a_1+a_2}\,&\mathcal{C}\big[\begin{smallmatrix} A_1 \\ B_1 \end{smallmatrix}\big|\begin{smallmatrix} A_2 \\ B_2 \end{smallmatrix}\big|\begin{smallmatrix} A_3 & a_1+a_2 \\ B_3 & 0 \end{smallmatrix}\big] \\
&+(-)^{a_1+1}\binom{a_1+a_2-1}{a_1}\,\mathcal{C}\big[\begin{smallmatrix} A_1 \\ B_1 \end{smallmatrix}\big|\begin{smallmatrix} A_2 & a_1+a_2 \\ B_2 & 0 \end{smallmatrix}\big|\begin{smallmatrix} A_3 \\ B_3 \end{smallmatrix}\big] \\
&+(-)^{a_2+1}\binom{a_1+a_2-1}{a_2}\,\mathcal{C}\big[\begin{smallmatrix} A_1 & a_1+a_2 \\ B_1 & 0 \end{smallmatrix}\big|\begin{smallmatrix} A_2 \\ B_2 \end{smallmatrix}\big|\begin{smallmatrix} A_3 \\ B_3 \end{smallmatrix}\big] \\
&+(-)^{a_1}\sum_{j=0}^{a_1-1}\binom{a_2+j-1}{j}\,\mathcal{C}\big[\begin{smallmatrix} A_1 \\ B_1 \end{smallmatrix}\big|\begin{smallmatrix} A_2 & a_2+j \\ B_2 & 0 \end{smallmatrix}\big|\begin{smallmatrix} A_3 & a_1-j \\ B_3 & 0 \end{smallmatrix}\big] \\
&+(-)^{a_2}\sum_{j=0}^{a_2-1}\binom{a_1+j-1}{j}\,\mathcal{C}\big[\begin{smallmatrix} A_1 & a_1+j \\ B_1 & 0 \end{smallmatrix}\big|\begin{smallmatrix} A_2 \\ B_2 \end{smallmatrix}\big|\begin{smallmatrix} A_3 & a_2-j \\ B_3 & 0 \end{smallmatrix}\big]\,.
\end{aligned}
\tag{5.127}
$$

This identity will be a key ingredient in deriving the basis decompositions for all dihedral and trihedral modular graph forms of total modular weight at most 12 is Section 5.7. If the $\big[\begin{smallmatrix} A_3 \\ B_3 \end{smallmatrix}\big]$-block contains a holomorphic edge, (5.127) is a reduction of three-point HSR to two-point HSR and graphs of lower loop order. In this case, the Fay identity could be used on any pair of non-parallel holomorphic edges and this choice corresponds to the different ways to perform the partial-fraction decomposition in Section 5.4.2, leading to interesting identities between MGFs in general. As an example, consider the graph $\mathcal{C}\big[\begin{smallmatrix} 1 \\ 0 \end{smallmatrix}\big|\begin{smallmatrix} 1 & 1 \\ 0 & 1 \end{smallmatrix}\big|\begin{smallmatrix} 1 & 2 \\ 1 & 0 \end{smallmatrix}\big]$ which was decomposed using the traditional HSR method in Section 5.4.3. Applying (5.127) to the first two holomorphic columns of this graph leads to

$$
\begin{aligned}
\mathcal{C}\big[\begin{smallmatrix} 1 \\ 0 \end{smallmatrix}\big|\begin{smallmatrix} 1 & 1 \\ 0 & 1 \end{smallmatrix}\big|\begin{smallmatrix} 1 & 2 \\ 1 & 0 \end{smallmatrix}\big] = \mathcal{C}\big[\varnothing\big|\begin{smallmatrix} 1 \\ 1 \end{smallmatrix}\big|\begin{smallmatrix} 1 & 2 & 2 \\ 1 & 0 & 0 \end{smallmatrix}\big] - \mathcal{C}\big[\varnothing\big|\begin{smallmatrix} 1 & 1 \\ 0 & 1 \end{smallmatrix}\big|\begin{smallmatrix} 1 & 1 & 2 \\ 0 & 1 & 0 \end{smallmatrix}\big] + \mathcal{C}\big[\varnothing\big|\begin{smallmatrix} 1 & 2 \\ 1 & 0 \end{smallmatrix}\big|\begin{smallmatrix} 1 & 2 \\ 1 & 0 \end{smallmatrix}\big] \\
- \mathcal{C}\big[\begin{smallmatrix} 1 \\ 0 \end{smallmatrix}\big|\begin{smallmatrix} 1 \\ 1 \end{smallmatrix}\big|\begin{smallmatrix} 1 & 1 & 2 \\ 0 & 1 & 0 \end{smallmatrix}\big] + \mathcal{C}\big[\begin{smallmatrix} 1 \\ 1 \end{smallmatrix}\big|\begin{smallmatrix} 2 \\ 0 \end{smallmatrix}\big|\begin{smallmatrix} 1 & 2 \\ 1 & 0 \end{smallmatrix}\big]\,,
\end{aligned}
\tag{5.128}
$$

which can be shown to be equal to the decomposition (5.116) upon using the topological simplifications from Section 5.3.2 and the dihedral HSR formula (5.73). On the other hand, we can also apply (5.127) to the second and third holomorphic edges, yielding

$$
\mathcal{C}\big[\begin{smallmatrix} 1 \\ 0 \end{smallmatrix}\big|\begin{smallmatrix} 1 & 1 \\ 0 & 1 \end{smallmatrix}\big|\begin{smallmatrix} 1 & 2 \\ 1 & 0 \end{smallmatrix}\big] = \mathcal{C}\big[\begin{smallmatrix} 1 \\ 0 \end{smallmatrix}\big|\begin{smallmatrix} 1 \\ 1 \end{smallmatrix}\big|\begin{smallmatrix} 1 & 3 \\ 1 & 0 \end{smallmatrix}\big] - \mathcal{C}\big[\begin{smallmatrix} 1 \\ 1 \end{smallmatrix}\big|\begin{smallmatrix} 1 \\ 1 \end{smallmatrix}\big|\begin{smallmatrix} 1 & 3 \\ 0 & 0 \end{smallmatrix}\big] + \mathcal{C}\big[\begin{smallmatrix} 1 \\ 1 \end{smallmatrix}\big|\begin{smallmatrix} 1 & 1 \\ 0 & 1 \end{smallmatrix}\big|\begin{smallmatrix} 1 & 2 \\ 0 & 0 \end{smallmatrix}\big]\,. \tag{5.129}
$$

This can be simplified to (5.119) by topological identities.

### HOLOMORPHIC SUBGRAPHS WITH TWO VERTICES

The restriction of (5.122) to holomorphic edges which are not parallel arises because the Fay identity for Kronecker–Eisenstein series (5.120) involves the three different arguments $z_1$, $z_2$ and $z_1-z_2$. As discussed in Appendix A of [38], by taking the limit $z_1 \to z_2$ of (5.120a), we obtain[9]

$$F(z,\eta_1,\tau)F(z,\eta_2,\tau) = F(z,\eta_1+\eta_2,\tau)\Big(g^{(1)}(\eta_1,\tau)+g^{(1)}(\eta_2,\tau)\Big) - \partial_z F(z,\eta_1+\eta_2,\tau)\,, \tag{5.130}$$

with the expansion coefficient $g^{(1)}$ defined in (3.80). This translates for the doubly-periodic version $\Omega$ (3.82) of the Kronecker–Eisenstein series into

$$\Omega(z,\eta_1,\tau)\Omega(z,\eta_2,\tau) = \Omega(z,\eta_1+\eta_2,\tau)\left(g^{(1)}(\eta_1,\tau)+g^{(1)}(\eta_2,\tau)+\frac{\pi}{\tau_2}(\eta_1+\eta_2)\right) - \partial_z\Omega(z,\eta_1+\eta_2,\tau)\,. \tag{5.131}$$

Expanding (5.131) in $\eta_1$ and $\eta_2$ and using

$$g^{(1)}(\eta,\tau) = \frac{1}{\eta} - \sum_{k=2}^{\infty}\eta^{k-1}\mathrm{G}_k(\tau) \tag{5.132}$$

yields [IV]

$$\begin{aligned} f^{(a_1)}(z)f^{(a_2)}(z) = (-1)^{a_2}\Theta(a_1+a_2-4)\mathrm{G}_{a_1+a_2} + \binom{a_1+a_2}{a_2} f^{(a_1+a_2)}(z) \\ - \sum_{k=4}^{a_1}\binom{a_1+a_2-1-k}{a_2-1}\mathrm{G}_k f^{(a_1+a_2-k)}(z) \\ - \sum_{k=4}^{a_2}\binom{a_1+a_2-1-k}{a_1-1}\mathrm{G}_k f^{(a_1+a_2-k)}(z) \\ - \binom{a_1+a_2-2}{a_2-1}\Big(\widehat{\mathrm{G}}_2 f^{(a_1+a_2-2)}(z) + \partial_z f^{(a_1+a_2-1)}(z)\Big)\,, \end{aligned} \tag{5.133}$$

where $a_1, a_2 > 0$ and $\Theta$ is the Heaviside step-function

$$\Theta(x) = \begin{cases} 1 & \text{if } x \geq 0 \\ 0 & \text{if } x < 0 \end{cases}. \tag{5.134}$$

Integrating (5.133) against a suitable product of $C^{(a,b)}$ functions yields two-point HSR upon using that

$$\partial_z f^{(a_1+a_2-1)}(z) = (-1)^{a_1+a_2+1}\frac{\pi}{\tau_2}C^{(a_1+a_2-1,-1)}(z) \tag{5.135}$$

according to (5.34). E.g. when (5.133) for $a_1+a_2 \geq 3$ is integrated against $\prod_{i=1}^{R} C^{(a_i,b_i)}$, we obtain the dihedral HSR identity (5.73).

Together, (5.121) and (5.133) allow to perform holomorphic subgraph reduction of holomorphic subgraphs with arbitrarily many vertices in

9 The following discussion follows Appendix A of [IV] closely.

a compact way. Note that when using Fay identities, we circumvent the need to evaluate conditionally convergent sums with the Eisenstein summation prescription as shown in Section 5.4.2. For trihedral three-point HSR, it was checked explicitly in many cases that a combination of (5.127) and two-point HSR yields an equivalent expression to the one obtained from the formula in Appendix B.2.

In the `Modular Graph Forms` package, the trihedral Fay identities (5.127) are implemented in the function **`TriFay`** which returns an equation. The first argument of this function is the trihedral MGF to be decomposed, the second (optional) argument has the form **`{{b1,c1},{b2,c2}}`**, where **`b1`** and **`b2`** are the blocks of the (anti)holomorphic edges to be used and **`c1`** and **`c2`** are the columns of those edges. If the second argument is omitted, the first suitable pair of (anti)holomorphic edges is selected automatically. As an example, we will consider the decomposition of the graph $C\left[\begin{smallmatrix}1\\0\end{smallmatrix}\middle|\begin{smallmatrix}1&1\\0&1\end{smallmatrix}\middle|\begin{smallmatrix}1&2\\1&0\end{smallmatrix}\right]$ as discussed around (5.128) and (5.129). In order to apply (5.127) to the first two holomorphic columns and then simplify the result to obtain (5.116), we run

In[41]:= $\texttt{TriFay}\left[\texttt{c}\left[\begin{smallmatrix}1\\0\end{smallmatrix},\begin{smallmatrix}1&1\\0&1\end{smallmatrix},\begin{smallmatrix}1&2\\1&0\end{smallmatrix}\right],\{\{1,1\},\{2,1\}\}\right]$

```
DiCSimplify[TriCSimplify[%[[2]]], useIds → False,
  momSimplify → False]
```

Out[41]= $C\left[\begin{smallmatrix}1\\0\end{smallmatrix}\middle|\begin{smallmatrix}1&1\\0&1\end{smallmatrix}\middle|\begin{smallmatrix}1&2\\1&0\end{smallmatrix}\right] == C\left[\{\}\middle|\begin{smallmatrix}1\\1\end{smallmatrix}\middle|\begin{smallmatrix}1&2&2\\1&0&0\end{smallmatrix}\right] - C\left[\{\}\middle|\begin{smallmatrix}1&1\\0&1\end{smallmatrix}\middle|\begin{smallmatrix}1&1&2\\0&1&0\end{smallmatrix}\right] + C\left[\{\}\middle|\begin{smallmatrix}1&2\\1&0\end{smallmatrix}\middle|\begin{smallmatrix}1&2\\1&0\end{smallmatrix}\right] - C\left[\begin{smallmatrix}1\\0\end{smallmatrix}\middle|\begin{smallmatrix}1\\1\end{smallmatrix}\middle|\begin{smallmatrix}1&1&2\\0&1&0\end{smallmatrix}\right]$

Out[42]= $C\left[\begin{smallmatrix}3&0\\1&0\end{smallmatrix}\right]^2 + C\left[\begin{smallmatrix}1&2&3\\1&0&1\end{smallmatrix}\right] - 3C\left[\begin{smallmatrix}1&2&3\\1&1&0\end{smallmatrix}\right] + C\left[\begin{smallmatrix}1&1&2\\0&1&1\end{smallmatrix}\right]\hat{G}_2 + \frac{\pi C\left[\begin{smallmatrix}1&2&2\\1&-1&1\end{smallmatrix}\right]}{\tau_2}$

reproducing (5.116). Similarly, (5.119) can be obtained by changing the second argument of **`TriFay`** in In[41] to **`{{2,1},{3,2}}`** and replacing the option **`momSimplify → False`** of **`DiCSimplify`** by **`diHSR → False`**.

As mentioned above, trihedral three-point HSR is performed by the function **`TriCSimplify`**, which implements the formula from Appendix B.2. If the Boolean option **`tri3ptFayHSR`** (which is inherited by **`CSimplify`**), is set to **`True`** (the default is **`False`**), the three-point HSR is instead performed using the Fay identity (5.127) and subsequent two-point HSR. The results of applying the two techniques may look different, if the basis decompositions from Section 5.7 are not applied, but they are equivalent, as can be seen when the basis decompositions are plugged in.

### 5.4.5 *Iterated holomorphic subgraph reduction*

If a graph contains several closed holomorphic subgraphs, one can iterate the holomorphic subgraph reduction. Of course, the end result should not depend on the order in which the HSRs were performed, but this is not always manifest and this fact can lead to interesting new identities for graphs involving negative entries.

In the case of purely holomorphic graphs, it is clear that the MGF can, as a holomorphic modular form, be written as a polynomial in $G_4$ and $G_6$. In order to see this explicitly, however, we have to apply the (generalized) Ramanujan identities from Section 5.3.5. E.g. the graph $\mathcal{C}\left[\begin{smallmatrix} a_1 & a_2 & a_3 \\ 0 & 0 & 0 \end{smallmatrix}\right]$ can be decomposed according to (5.73) into

$$
\begin{aligned}
(-)^{a_3}\,\mathcal{C}\left[\begin{smallmatrix} a_+ & a_- & a_3 \\ 0 & 0 & 0 \end{smallmatrix}\right] = &-\binom{a_1+a_2}{a_2} G_{a_0} \\
&+ \sum_{k=4}^{a_1} \binom{a_1+a_2-1-k}{a_1-k} G_k G_{a_0-k} \\
&+ \sum_{k=4}^{a_2} \binom{a_1+a_2-1-k}{a_2-k} G_k G_{a_0-k} \\
&+ \binom{a_1+a_2-2}{a_1-1} \left\{ \widehat{G}_2 G_{a_0-2} + \frac{1}{a_0-2}\frac{\pi}{\tau_2}\nabla^{(a_0-2)} G_{a_0-2} \right\},
\end{aligned}
\tag{5.136}
$$

where $a_0 = a_1 + a_2 + a_3$. The Cauchy–Riemann derivative of $G_{a_0-2}$ has to be decomposed into holomorphic Eisenstein series by means of Ramanujan identities. For purely holomorphic graphs of higher loop order, performing the HSR in different orders leads to different expressions and again, the identities from Section 5.3.5 have to be used to show explicitly that they agree. E.g. consider the graph $\mathcal{C}\left[\begin{smallmatrix} 1 & 2 & 2 & 3 \\ 0 & 0 & 0 & 0 \end{smallmatrix}\right]$. If we start from its canonical representation $\mathcal{C}\left[\begin{smallmatrix} 1 & 2 & 2 & 3 \\ 0 & 0 & 0 & 0 \end{smallmatrix}\right]$ and always perform the HSR on the two leftmost holomorphic columns, we obtain the decomposition

$$
\begin{aligned}
\mathcal{C}\left[\begin{smallmatrix} 1 & 2 & 2 & 3 \\ 0 & 0 & 0 & 0 \end{smallmatrix}\right] = {}& 12 G_6 \widehat{G}_2 - 30 G_8 - G_4 \widehat{G}_2^2 + \frac{\pi}{\tau_2}\left(19\,\mathcal{C}\left[\begin{smallmatrix} 7 & 0 \\ -1 & 0 \end{smallmatrix}\right] - 4\widehat{G}_2\,\mathcal{C}\left[\begin{smallmatrix} 5 & 0 \\ -1 & 0 \end{smallmatrix}\right]\right) \\
&- 3\left(\frac{\pi}{\tau_2}\right)^2 \mathcal{C}\left[\begin{smallmatrix} 6 & 0 \\ -2 & 0 \end{smallmatrix}\right].
\end{aligned}
\tag{5.137}
$$

Starting from the representation $\mathcal{C}\left[\begin{smallmatrix} 2 & 2 & 1 & 3 \\ 0 & 0 & 0 & 0 \end{smallmatrix}\right]$ of the graph and always performing the leftmost HSR yields instead

$$
\mathcal{C}\left[\begin{smallmatrix} 1 & 2 & 2 & 3 \\ 0 & 0 & 0 & 0 \end{smallmatrix}\right] = 24 G_8 - G_4^2 - 2 G_4 \widehat{G}_2^2 - 14\frac{\pi}{\tau_2}\,\mathcal{C}\left[\begin{smallmatrix} 7 & 0 \\ -1 & 0 \end{smallmatrix}\right] + 2\left(\frac{\pi}{\tau_2}\right)^2 \mathcal{C}\left[\begin{smallmatrix} 6 & 0 \\ -2 & 0 \end{smallmatrix}\right].
\tag{5.138}
$$

Using the identities (5.61) and (5.64), one can show that both (5.137) and (5.138) are equal to $-\frac{5}{7}G_4^2$. Similar calculations can of course also be done for purely holomorphic higher-point graphs.

If the graph under consideration has several holomorphic subgraphs but is not purely holomorphic, setting the different HSR orders equal leads to interesting identities between graphs with negative entries next to zero entries, which could be seen as even more general forms of the Ramanujan identities. E.g. the graph $\mathcal{C}\big[\begin{smallmatrix} 2 & 2 & 3 & 3 \\ 0 & 2 & 0 & 0 \end{smallmatrix}\big]$ has two closed holomorphic subgraphs. Performing first the HSR on the $\big[\begin{smallmatrix} 3 & 3 \\ 0 & 0 \end{smallmatrix}\big]$ columns leads to an expression containing $\mathcal{C}\big[\begin{smallmatrix} 2 & 2 & 5 \\ 0 & 2 & -1 \end{smallmatrix}\big]$. Performing first the HSR on the $\big[\begin{smallmatrix} 2 & 3 \\ 0 & 0 \end{smallmatrix}\big]$ columns leads to an expression containing $\mathcal{C}\big[\begin{smallmatrix} 2 & 3 & 4 \\ 2 & 0 & -1 \end{smallmatrix}\big]$. Setting both of these expressions equal leads to the relation

$$\begin{aligned}\frac{1}{2}\mathcal{C}\big[\begin{smallmatrix} 2 & 3 & 4 \\ 2 & 0 & -1 \end{smallmatrix}\big] - \mathcal{C}\big[\begin{smallmatrix} 2 & 2 & 5 \\ 0 & 2 & -1 \end{smallmatrix}\big] &= 5\,\mathcal{C}\big[\begin{smallmatrix} 9 & 0 \\ 1 & 0 \end{smallmatrix}\big] + \widehat{\mathrm{G}}_2\,\mathcal{C}\big[\begin{smallmatrix} 7 & 0 \\ 1 & 0 \end{smallmatrix}\big] - \frac{\tau_2}{\pi}\Big(\mathrm{G}_4\widehat{\mathrm{G}}_2\,\mathcal{C}\big[\begin{smallmatrix} 4 & 0 \\ 2 & 0 \end{smallmatrix}\big] \\ &\quad + \widehat{\mathrm{G}}_2^2\,\mathcal{C}\big[\begin{smallmatrix} 6 & 0 \\ 2 & 0 \end{smallmatrix}\big] - 5\mathrm{G}_4\,\mathcal{C}\big[\begin{smallmatrix} 6 & 0 \\ 2 & 0 \end{smallmatrix}\big] - 7\mathrm{G}_6\,\mathcal{C}\big[\begin{smallmatrix} 4 & 0 \\ 2 & 0 \end{smallmatrix}\big]\Big)\,.\end{aligned} \tag{5.139}$$

Note that although the two graphs on the LHS are related by momentum conservation, this does not allow one to reduce the number of independent graphs.

Since the function **`DiCSimplify`** in the `Modular Graph Forms` package can decompose any derivative of any holomorphic Eisenstein series when the option **`basisExpandG`** is set to **`True`**, the computations presented in this section can be easily performed by applying the **`CSimplify`** function to different representations of the same graph.

## 5.5 THE SIEVE ALGORITHM

With the techniques described in the last two sections, many valuable identities between modular graph forms can be derived. However, if one is interested in simplifying a particular MGF, e.g. one which has appeared as an expansion coefficient of a Koba–Nielsen integral, it is not always clear which techniques to combine to obtain the desired decomposition. In this situation, the sieve algorithm, first introduced in [16], can be used: It allows for a systematic decomposition (up to an overall constant) of arbitrary MGFs, as long as the basis for the decomposition is known.

### 5.5.1 *Constructing identities*

As a starting point, assume that we have a combination $F$ of MGFs of homogeneous modular weight $(|A|, |B|)$ and we want to check whether or not it vanishes. The idea behind the sieve algorithm is to repeatedly take derivatives of $F$ using the Maaß operator $\nabla^{(|A|)}$ defined in (3.51). Due to an intricate interplay between momentum conservation identities and HSR, every derivative can be expressed as a linear combination of products of holomorphic Eisenstein series, MGFs with non-negative antiholomorphic labels for each edge, $\tau_2$ with non-positive exponent, MGFs of the form $\mathcal{C}\big[\begin{smallmatrix} k & 0 \\ -n & 0 \end{smallmatrix}\big]$ with $k > n$ and modular invariant factors. After taking $|B|$ derivatives, the antiholomorphic modular weight vanishes according to (3.52) and hence each term in the derivative has to factorize, since any unfactorized MGFs would have to have vanishing antiholomorphic labels and therefore be amenable to HSR, leading to a factorized expression. Using the generalized Ramanujan identities from Section 5.3.5, the factors of the form $\mathcal{C}\left[\begin{smallmatrix} k & 0 \\ -n & 0 \end{smallmatrix}\right]$ can be decomposed as well. Since each term is factorized, the total modular weight $a+b$ of every leftover MGF is strictly less than $|A|+|B|$ and if we know all identities between MGFs of lower total modular weight, it is manifest if the $|B|^{\text{th}}$ derivative of $F$ vanishes or not. If $F$ has $|A| = |B|$, then Lemma 1 in [16] guarantees that if the derivative vanishes, $F = 0$ up to an overall constant. If $|A| \neq |B|$ and $F$ can be written as the derivative of an expression with $|A| = |B|$, this primitive vanishes up to a constant, so $F = 0$ as well. We conjecture that the same is true if $F$ cannot be written as the derivative of an expression with $|A| = |B|$, in line with all cases we tested. In this way, we can generate identities at progressively higher total modular weight.

We will now discuss in more detail how one can avoid negative antiholomorphic edge labels in the derivative of an MGF. First, note that a negative edge label in the derivative is due to a holomorphic edge in the original graph (assuming that the original graph did not already contain negative antiholomorphic labels): $\nabla^{(a)}$ maps the labels $(a, 0)$ to the labels $(a+1, -1)$. This $-1$ can be removed by the (antiholomorphic) momentum conservation identity which arises from the same MGF with the $(a+1, -1)$-edge replaced by a $(a+1, 0)$-edge. Since the $(a+1, -1)$-edge connects two vertices, both of these can be used to construct a momentum conservation identity to remove the $-1$. If however the vertex we use for the momentum conservation has other holomorphic edges attached to it, there will be contributions in the momentum conservation identity in which these other edges carry negative antiholomorphic labels. These negative labels can in turn be removed by momentum conservation and so on. There is only one case, in which this procedure does not work: If the seed (it is always the same) has a closed holomorphic subgraph, we can only move the $-1$ around this subgraph but never eliminate it entirely. Fortunately, this case only arises if the MGF we applied the derivative to in the first place had a holomorphic subgraph. Thus by performing HSR before taking the derivative, we can avoid this problem and can be sure to be able to remove all negative entries. To summarize, since HSR translates graphs with closed holomorphic subgraphs into combinations of graphs without closed holomorphic subgraphs and holomorphic Eisenstein series, we can use momentum conservation and HSR to trade negative antiholomorphic labels for holomorphic Eisenstein series.[10]

The Cauchy–Riemann derivative of a holomorphic Eisenstein series has the form $\mathcal{C}\left[\begin{smallmatrix} 2k+1 & 0 \\ -1 & 0 \end{smallmatrix}\right]$, i.e. it is a graph with one edge with negative

10 If we assume that all holomorphic labels of the original MGF are at least one (as in [16]) then the HSR is the only source of holomorphic Eisenstein series. If we also allow for vanishing holomorphic labels, as we want to do here, holomorphic Eisenstein series can also arise from factorizations, e.g. $\nabla^{(5)}\,\mathcal{C}\left[\begin{smallmatrix} 0 & 2 & 3 \\ 1 & 2 & 0 \end{smallmatrix}\right]$ contains a term $-3(\frac{\pi}{\tau_2})^2 \mathrm{E}_2 \mathrm{G}_4$ although no HSR was performed.

antiholomorphic weight. In this case, momentum conservation (and HSR) cannot be used to remove the negative entry and in the original version published in [16], this fact was used to *sieve* the space of MGFs for identities: After taking a derivative and trading negative antiholomorphic entries for holomorphic Eisenstein series, one subtracts the same derivative of an MGF in such a way that all holomorphic Eisenstein series cancel. Then, one can take the next derivative of the combined expression without generating irremovable negative antiholomorphic labels. After having taken $|B|$ derivatives, the result is purely holomorphic (and still modular), so we can expand it in the ring of holomorphic Eisenstein series. By subtracting one final MGF such that this derivative vanishes, one has constructed an identity up to an overall constant. In fact, if a combination of modular graph forms vanishes, then the holomorphic Eisenstein series have to cancel out in every derivative, as will be explained in Section 5.5.2. This can however only be verified, if the prefactors of the holomorphic Eisenstein series are linearly independent. Since they carry lower total modular weight than the complete expression, this means that we need to know all identities between graphs of lower total modular weight.

As an example, consider the dihedral MGF $C\left[\begin{smallmatrix}1&1&2\\1&2&1\end{smallmatrix}\right]$. In order to find a simplification for this graph, start by taking its Cauchy–Riemann derivative,

$$\nabla^{(4)}\, C\left[\begin{smallmatrix}1&1&2\\1&2&1\end{smallmatrix}\right] = 2\,C\left[\begin{smallmatrix}1&1&3\\1&2&0\end{smallmatrix}\right] + C\left[\begin{smallmatrix}1&2&2\\1&1&1\end{smallmatrix}\right] + C\left[\begin{smallmatrix}1&2&2\\2&0&1\end{smallmatrix}\right]. \tag{5.140}$$

Since no negative entries arise, we can directly take the next derivative,

$$\begin{aligned}\nabla^{(4)^2}\, C\left[\begin{smallmatrix}1&1&2\\1&2&1\end{smallmatrix}\right] = 6\,C\left[\begin{smallmatrix}1&1&4\\1&2&-1\end{smallmatrix}\right] + 2\,C\left[\begin{smallmatrix}1&2&3\\2&1&-1\end{smallmatrix}\right] + 4\,C\left[\begin{smallmatrix}1&2&3\\2&0&0\end{smallmatrix}\right]\\ + 6\,C\left[\begin{smallmatrix}1&2&3\\1&1&0\end{smallmatrix}\right] + 2\,C\left[\begin{smallmatrix}2&2&2\\0&1&1\end{smallmatrix}\right],\end{aligned} \tag{5.141}$$

where we used the notation (3.54) for the second Cauchy–Riemann derivative. The graph $C\left[\begin{smallmatrix}1&2&3\\2&0&0\end{smallmatrix}\right]$ has to be decomposed by HSR and the negative entries in the other two graphs in the first line can be removed by antiholomorphic momentum conservation of the seeds $C\left[\begin{smallmatrix}1&1&4\\1&2&0\end{smallmatrix}\right]$ and $C\left[\begin{smallmatrix}1&2&3\\2&1&0\end{smallmatrix}\right]$, respectively, and subsequent HSR. The result is

$$\begin{aligned}\nabla^{(4)^2}\, C\left[\begin{smallmatrix}1&1&2\\1&2&1\end{smallmatrix}\right] = 4\,C\left[\begin{smallmatrix}1&2&3\\1&1&0\end{smallmatrix}\right] - 6\,C\left[\begin{smallmatrix}1&1&4\\1&1&0\end{smallmatrix}\right] + 2\,C\left[\begin{smallmatrix}2&2&2\\0&1&1\end{smallmatrix}\right]\\ - 10\,C\left[\begin{smallmatrix}6&0\\2&0\end{smallmatrix}\right] + 6\left(\frac{\pi}{\tau_2}\right)^2 \mathrm{E}_2\mathrm{G}_4\,.\end{aligned} \tag{5.142}$$

In order to take one further derivative, we have to cancel the expression $\mathrm{E}_2\mathrm{G}_4$. Since

$$\nabla^{(4)^2}\left(\left(\frac{\pi}{\tau_2}\right)^4 \mathrm{E}_2^2\right) = 8\,C\left[\begin{smallmatrix}3&0\\1&0\end{smallmatrix}\right]^2 + 12\left(\frac{\pi}{\tau_2}\right)^2 \mathrm{E}_2\mathrm{G}_4\,, \tag{5.143}$$

we can take one further derivative of $\mathcal{C}\left[\begin{smallmatrix}1&1&2\\1&2&1\end{smallmatrix}\right]-\frac{1}{2}\left(\frac{\pi}{\tau_2}\right)^2\mathrm{E}_2\mathrm{G}_4$ without generating irremovable negative antiholomorphic labels. After momentum conservation and HSR, we obtain

$$\nabla^{(4)^3}\left(\mathcal{C}\left[\begin{smallmatrix}1&1&2\\1&2&1\end{smallmatrix}\right]-\frac{1}{2}\left(\frac{\pi}{\tau_2}\right)^2\mathrm{E}_2^2\right)=-168\,\mathcal{C}\left[\begin{smallmatrix}7&0\\1&0\end{smallmatrix}\right]+12\mathrm{G}_4\,\mathcal{C}\left[\begin{smallmatrix}3&0\\1&0\end{smallmatrix}\right]. \qquad (5.144)$$

Using

$$\nabla^{(4)^3}\,\mathcal{C}\left[\begin{smallmatrix}1&1&2\\1&1&2\end{smallmatrix}\right]=108\,\mathcal{C}\left[\begin{smallmatrix}7&0\\1&0\end{smallmatrix}\right]-12\mathrm{G}_4\,\mathcal{C}\left[\begin{smallmatrix}3&0\\1&0\end{smallmatrix}\right] \qquad (5.145)$$

we can cancel the term $\mathrm{G}_4\,\mathcal{C}\left[\begin{smallmatrix}3&0\\1&0\end{smallmatrix}\right]$ in (5.144) and take one final derivative,

$$\nabla^{(4)^4}\left(\mathcal{C}\left[\begin{smallmatrix}1&1&2\\1&2&1\end{smallmatrix}\right]+\mathcal{C}\left[\begin{smallmatrix}1&1&2\\1&1&2\end{smallmatrix}\right]-\frac{1}{2}\left(\frac{\pi}{\tau_2}\right)^2\mathrm{E}_2^2\right)=-420\mathrm{G}_8\,. \qquad (5.146)$$

The fourth derivative of $\left(\frac{\pi}{\tau_2}\right)^4\mathrm{E}_4$ is also proportional to $\mathrm{G}_8$ and we have

$$\nabla^{(4)^4}\left(\mathcal{C}\left[\begin{smallmatrix}1&1&2\\1&2&1\end{smallmatrix}\right]+\mathcal{C}\left[\begin{smallmatrix}1&1&2\\1&1&2\end{smallmatrix}\right]-\frac{1}{2}\left(\frac{\pi}{\tau_2}\right)^2\mathrm{E}_2^2+\frac{1}{2}\left(\frac{\pi}{\tau_2}\right)^4\mathrm{E}_4\right)=0\,. \qquad (5.147)$$

Lemma 1 in [16] now states that this implies

$$\mathcal{C}\left[\begin{smallmatrix}1&1&2\\1&2&1\end{smallmatrix}\right]+\mathcal{C}\left[\begin{smallmatrix}1&1&2\\1&1&2\end{smallmatrix}\right]-\frac{1}{2}\left(\frac{\pi}{\tau_2}\right)^2\mathrm{E}_2^2+\frac{1}{2}\left(\frac{\pi}{\tau_2}\right)^4\mathrm{E}_4=\left(\frac{\pi}{\tau_2}\right)^4\mathrm{const.} \qquad (5.148)$$

with some $\tau$-independent constant.[11] Using the techniques discussed in the previous sections, one can also decompose $\mathcal{C}\left[\begin{smallmatrix}1&1&2\\1&2&1\end{smallmatrix}\right]$ directly and finds that the constant vanishes in this case (as expected since there is no single-valued MZV at the expected transcendental weight 4, cf. (2.39)).

In general, finding MGFs with the correct Cauchy–Riemann derivatives to cancel the holomorphic Eisenstein series can be challenging but if we want to find a decomposition of an MGF into a set of basis MGFs, we can just take the derivatives of a linear combination of the basis elements and adjust the coefficients so that the holomorphic Eisenstein series cancel. This is what is done in the implementation of the sieve algorithm in the `ModularGraphForms` package.

Instead of canceling holomorphic Eisenstein series in every derivative as described above and in [16], one can also use the generalized Ramanujan identities discussed in Section 5.3.5 to perform the derivatives of the holomorphic Eisenstein series. In this way, the highest derivative of any MGF can be written in terms of holomorphic Eisenstein series and MGFs of lower total modular weight for which we assume that the relations are known, hence identities can be found explicitly.

11 Due to our normalization conventions, the graphs with equal total holomorphic and antiholomorphic edge labels are not modular invariant, hence the integration constant is multiplied by a suitable power of $\frac{\pi}{\tau_2}$.

In the example above, the resulting fourth derivatives are

$$\nabla^{(4)^4} \mathcal{C}\left[\begin{smallmatrix}1&1&2\\1&2&1\end{smallmatrix}\right] = 120\mathrm{E}_2\mathrm{G}_4\widehat{\mathrm{G}}_2^2 - 840\mathrm{E}_2\mathrm{G}_6\widehat{\mathrm{G}}_2 + 600\mathrm{E}_2\mathrm{G}_4^2 - 360\mathrm{G}_4^2$$
$$+ 840\mathrm{G}_6\frac{\tau_2}{\pi}\mathcal{C}\left[\begin{smallmatrix}3&0\\1&0\end{smallmatrix}\right] - 240\mathrm{G}_4\widehat{\mathrm{G}}_2\frac{\tau_2}{\pi}\mathcal{C}\left[\begin{smallmatrix}3&0\\1&0\end{smallmatrix}\right] \tag{5.149a}$$

$$\nabla^{(4)^4} \mathcal{C}\left[\begin{smallmatrix}1&1&2\\1&1&2\end{smallmatrix}\right] = 288\mathrm{G}_4^2 + 48\mathrm{G}_4\widehat{\mathrm{G}}_2\frac{\tau_2}{\pi}\mathcal{C}\left[\begin{smallmatrix}3&0\\1&0\end{smallmatrix}\right]$$
$$- 168\mathrm{G}_6\frac{\tau_2}{\pi}\mathcal{C}\left[\begin{smallmatrix}3&0\\1&0\end{smallmatrix}\right] \tag{5.149b}$$

$$\nabla^{(4)^4}\left(\left(\frac{\pi}{\tau_2}\right)^2\mathrm{E}_2^2\right) = 240\mathrm{E}_2\mathrm{G}_4\widehat{\mathrm{G}}_2^2 - 1680\mathrm{E}_2\mathrm{G}_6\widehat{\mathrm{G}}_2 + 1200\mathrm{E}_2\mathrm{G}_4^2 + 216\mathrm{G}_4^2$$
$$- 384\mathrm{G}_4\widehat{\mathrm{G}}_2\frac{\tau_2}{\pi}\mathcal{C}\left[\begin{smallmatrix}3&0\\1&0\end{smallmatrix}\right] + 1344\mathrm{G}_6\frac{\tau_2}{\pi}\mathcal{C}\left[\begin{smallmatrix}3&0\\1&0\end{smallmatrix}\right] \tag{5.149c}$$

$$\nabla^{(4)^4}\left(\left(\frac{\pi}{\tau_2}\right)^4\mathrm{E}_4\right) = 180\mathrm{G}_4^2\,. \tag{5.149d}$$

Setting a linear combination of these four expressions to zero and requiring the coefficients of the various terms on the RHS to vanish leaves (5.147) as the only solution. If no solution had existed, the four MGFs in (5.149) would have been proven to be linearly independent.

In the `Modular Graph Forms` package, the removal of edge labels −1 for dihedral and trihedral graphs is done by the functions **`DiCSimplify`** and **`TriCSimplify`**, if the option **`momSimplify`** is set to **`True`** (the default). The sieve algorithm itself is implemented in the function **`CSieveDecomp`**, which uses the traditional method of canceling holomorphic Eisenstein series in every step. If no further options are given, this function tries to decompose the graph given in its argument into the basis discussed in Section 5.7, e.g. for the graph $\mathcal{C}\left[\begin{smallmatrix}1&1&2\\1&2&1\end{smallmatrix}\right]$ we considered above, we can run

In[43]:= $\texttt{CSieveDecomp}\left[\texttt{c}\left[\begin{smallmatrix}1&1&2\\1&2&1\end{smallmatrix}\right]\right]$

Out[43]= $-\mathrm{C}\left[\begin{smallmatrix}1&1&2\\1&1&2\end{smallmatrix}\right] + \frac{\pi^4\,\mathrm{E}_2^2}{2\,\tau_2^4} - \frac{\pi^4\,\mathrm{E}_4}{2\,\tau_2^4} + \frac{\pi^4\,\texttt{intConst}\left[\begin{smallmatrix}1&1&2\\1&2&1\end{smallmatrix}\right]}{\tau_2^4}$ ,

reproducing (5.148). The last term in the output is an undetermined integration constant, labeled by the exponent matrix of the original graph. Such a constant is added for all graphs with equal holomorphic and antiholomorphic weight. Setting the Boolean option **`verbose`** of **`CSieveDecomp`** to **`True`** prints a detailed progress report into the notebook with the expressions appearing in each derivative and the prefactors of the holomorphic Eisenstein series which are set to zero. E.g. the output for the third derivative in the computation above is

`3rd derivative:`

$$-168\,\mathrm{C}\left[\begin{smallmatrix}7&0\\1&0\end{smallmatrix}\right] - 108\,\texttt{bCoeff[1]}\,\mathrm{C}\left[\begin{smallmatrix}7&0\\1&0\end{smallmatrix}\right] - 120\,\texttt{bCoeff[2]}\,\mathrm{C}\left[\begin{smallmatrix}7&0\\1&0\end{smallmatrix}\right] +$$
$$12\,\mathrm{C}\left[\begin{smallmatrix}3&0\\1&0\end{smallmatrix}\right]\mathrm{G}_4 + 12\,\texttt{bCoeff[1]}\,\mathrm{C}\left[\begin{smallmatrix}3&0\\1&0\end{smallmatrix}\right]\mathrm{G}_4$$

```
(Anti-)holomorphic Eisenstein series:

{G4}

Coefficients that should be zero:

{12 C[3 0; 1 0] + 12 bCoeff[1] C[3 0; 1 0]}

Find solution for all

{C[3 0; 1 0]}

Solutions:

{{bCoeff[1] → - 1}} .
```

This is the step described in (5.144) and (5.145) above. As one can see, **CSieveDecomp** forms a linear combination of the basis elements with coefficients **bCoeff** and subtracts it from the MGF which is decomposed. Then, derivatives are taken and in each step the coefficients of the holomorphic Eisenstein series are set to zero by fixing some of the **bCoeff**.

If for the modular weight of the MGF no basis is implemented, the error **CSieveDecomp::noBasis** is issued. In general, the basis used for the decomposition is determined by the option **basis** of **CSieveDecomp**. If **basis** is an empty list (the default), the basis is determined by the function **CBasis**, to be discussed in more detail in Section 5.7. Otherwise, one can also supply a list of MGFs of the same weight as the MGF to be decomposed. E.g. we can reproduce the momentum conservation identity of the seed $C\left[\begin{smallmatrix}1&2&2\\1&2&1\end{smallmatrix}\right]$ (up to an overall constant) by running

In[44]:= **CSieveDecomp**$\left[\mathtt{c}\left[\begin{smallmatrix}1&1&2\\1&2&1\end{smallmatrix}\right], \mathtt{basis}\to\left\{\mathtt{c}\left[\begin{smallmatrix}0&2&2\\1&1&2\end{smallmatrix}\right], \mathtt{c}\left[\begin{smallmatrix}1&1&2\\1&1&2\end{smallmatrix}\right]\right\}\right]$

Out[44]= $-\mathtt{C}\left[\begin{smallmatrix}0&2&2\\1&1&2\end{smallmatrix}\right]-\mathtt{C}\left[\begin{smallmatrix}1&1&2\\1&1&2\end{smallmatrix}\right]+\dfrac{\pi^4\,\mathtt{intConst}\left[\begin{smallmatrix}1&1&2\\1&2&1\end{smallmatrix}\right]}{\tau_2^4}$ .

If not all coefficients can be fixed (e.g. because the **basis** provided is not linearly independent), **bCoeff** will appear in the output. If no decomposition could be found, the error **CSieveDecomp::noSol** is issued and the derivative specified in which a holomorphic Eisenstein series could not be canceled. The calculation can then be investigated further with the **verbose** option. This can happen e.g. if the basis is not complete or if identities of lower weight for MGFs multiplying holomorphic Eisenstein series are missing. For further options and the meaning of other error messages, cf. Appendix A.

### 5.5.2 *Relation to iterated Eisenstein integrals*

The technique of constructing identities using the sieve algorithm is closely related to how modular graph forms were written in terms of the iterated Eisenstein integrals (4.15) in [34] as discussed in Section 4.3.1. In both cases, we take Cauchy–Riemann derivatives of the MGF in question and simplify them to remove negative antiholomorphic labels until we encounter holomorphic Eisenstein series.

In the sieve algorithm, we then try to cancel these holomorphic Eisenstein series by adding suitable MGFs and take further derivatives. If we want to write the MGF we started with in terms of iterated Eisenstein integrals, we take further derivatives only of the terms not involving the holomorphic Eisenstein series and (if it is not already known in terms of iterated Eisenstein integrals) of the coefficient of the holomorphic Eisenstein series. We repeat this until no more derivatives can be taken and integrate back using the definition (4.15) of iterated Eisenstein integrals. The integration constants (up to the one which is also left undetermined by the sieve algorithm) are fixed by requiring the correct modular transformation properties as mentioned in Section 4.3.1.

For the example of $\mathcal{C}\left[\begin{smallmatrix} 1 & 1 & 2 \\ 1 & 2 & 1 \end{smallmatrix}\right]$, we encounter a holomorphic Eisenstein series in the second derivative (5.142) and hence proceed by taking derivatives of the remaining terms $\mathcal{C}\left[\begin{smallmatrix} 1 & 2 & 3 \\ 1 & 1 & 0 \end{smallmatrix}\right]$, $\mathcal{C}\left[\begin{smallmatrix} 1 & 1 & 4 \\ 1 & 1 & 0 \end{smallmatrix}\right]$, $\mathcal{C}\left[\begin{smallmatrix} 2 & 2 & 2 \\ 0 & 1 & 1 \end{smallmatrix}\right]$ and $\mathcal{C}\left[\begin{smallmatrix} 6 & 0 \\ 2 & 0 \end{smallmatrix}\right]$ with the result

$$\nabla^{(6)}\,\mathcal{C}\left[\begin{smallmatrix} 1 & 2 & 3 \\ 1 & 1 & 0 \end{smallmatrix}\right] = 0 \tag{5.150a}$$

$$\nabla^{(6)}\,\mathcal{C}\left[\begin{smallmatrix} 1 & 1 & 4 \\ 1 & 1 & 0 \end{smallmatrix}\right] = 18\,\mathcal{C}\left[\begin{smallmatrix} 7 & 0 \\ 1 & 0 \end{smallmatrix}\right] - 6\mathrm{G}_4\,\mathcal{C}\left[\begin{smallmatrix} 3 & 0 \\ 1 & 0 \end{smallmatrix}\right] \tag{5.150b}$$

$$\nabla^{(6)}\,\mathcal{C}\left[\begin{smallmatrix} 2 & 2 & 2 \\ 0 & 1 & 1 \end{smallmatrix}\right] = 0 \tag{5.150c}$$

$$\nabla^{(6)^2}\,\mathcal{C}\left[\begin{smallmatrix} 6 & 0 \\ 2 & 0 \end{smallmatrix}\right] = 42\mathrm{G}_8\,. \tag{5.150d}$$

In (5.150b), we have to take two more derivatives,[12]

$$\nabla^{(7)}\,\mathcal{C}\left[\begin{smallmatrix} 7 & 0 \\ 1 & 0 \end{smallmatrix}\right] = 7\mathrm{G}_8 \tag{5.151a}$$

$$\nabla^{(3)}\,\mathcal{C}\left[\begin{smallmatrix} 3 & 0 \\ 1 & 0 \end{smallmatrix}\right] = 3\mathrm{G}_4\,. \tag{5.151b}$$

By integrating first (5.151), then (5.150) and finally (5.147), we obtain an expression for $\mathcal{C}\left[\begin{smallmatrix} 1 & 1 & 2 \\ 1 & 2 & 1 \end{smallmatrix}\right]$ in terms of iterated Eisenstein integrals. However, since determining the modular $S$-transformation of the iterated Eisenstein integrals is in general tedious (by the necessity of so-called *multiple modular values* [31]), modular properties have only been used to fix the integration constants in simple cases.

12 Since we know the expansion of $\mathrm{E}_k$ in terms of iterated Eisenstein integrals from [34], it would be much easier to obtain $\mathcal{C}\left[\begin{smallmatrix} 6 & 0 \\ 2 & 0 \end{smallmatrix}\right]$, $\mathcal{C}\left[\begin{smallmatrix} 7 & 0 \\ 1 & 0 \end{smallmatrix}\right]$ and $\mathcal{C}\left[\begin{smallmatrix} 3 & 0 \\ 1 & 0 \end{smallmatrix}\right]$ from (5.56) and the differential equation (4.17) of the iterated Eisenstein integrals. But here, we want to illustrate the general procedure for the case of MGFs of which the representation in terms of iterated Eisenstein integral is unknown.

The language of iterated Eisenstein integrals also shows why the holomorphic Eisenstein series in fact have to cancel out in every derivative of a vanishing combination of MGFs: If there is a left-over holomorphic Eisenstein series $G_k$ in the $n^{\text{th}}$ derivative (note that in order to be sure about this, one has to simplify first the prefactors of all holomorphic Eisenstein series and hence know the relations between MGFs of lower total modular weight), integrating this derivative produces $n$ zeros in the labels of the iterated Eisenstein integrals of the terms not containing the holomorphic Eisenstein series and a $k$ and $n-1$ zeros for the term containing the holomorphic Eisenstein series. Since iterated Eisenstein integrals with different labels are linearly independent [35], the original MGF cannot vanish.

## 5.6 DIVERGENT MODULAR GRAPH FORMS

So far, we have not discussed the convergence properties of the lattice sum (3.123) of MGFs, but, of course, if the edge labels become too low, the sum (3.123) is not absolutely convergent any more. Interestingly, conditionally convergent or even divergent sums can arise even when one applies the techniques above only to convergent sums. We saw an example of this in the context of three-point HSR in Section 5.4.3. However, unlike in this case, sometimes the divergence cannot be avoided, e.g. when using the sieve algorithm to find decompositions of certain convergent graphs. When deriving identities, one way to deal with this phenomenon is to just disregard all identities in which divergent graphs appear. This is the approach taken in Section 5.7 for (convergent) dihedral and trihedral modular graph forms of weight $a+b \leq 10$. However, in this way, one misses many valuable identities and hence it is desirable to have at least a partial understanding of how to interpret divergent MGFs. In this section, we will describe concrete results which go in this direction. Below, we will use these divergent techniques to obtain all dihedral and trihedral (convergent) basis decompositions for $a+b=12$.

### 5.6.1 *Divergence conditions*

In this section, we will give simple power-counting arguments to determine if a particular MGF is absolutely convergent or not, building on the behavior of holomorphic Eisenstein series, for which we know that

$$G_a = \sum_p{}' \frac{1}{p^a} \tag{5.152}$$

is absolutely convergent for $a \geq 3$, conditionally convergent for $a = 2$ and divergent for $a \leq 1$. Accordingly, we will call an MGF convergent if all momenta in the sum (3.123) have at least three powers in the denominator (adding powers of $p$ and $\bar{p}$) and divergent if any momentum appears with two or less powers in the denominator.[13]

In order to determine the total powers with which a momentum can appear, one has to perform some of the sums first by using the momentum-conserving delta functions (cf. e.g. (5.4)). Of course, there is considerable freedom in which sums we choose for this, hence different final expressions can result, with different total powers of the momenta. These expressions correspond to different rotations of the coordinate axes in the lattice spanned by the momenta. Since by counting the total exponents, we only test the convergence properties along the coordinate axes, we pick the representation with the lowest total power. To illustrate this, consider the dihedral graph

$$\mathcal{C}\left[\begin{smallmatrix} 1 & 1 & 2 \\ 0 & 0 & 2 \end{smallmatrix}\right] = \sum_{p_1,p_2,p_3}' \frac{\delta(p_1+p_2+p_3)}{p_1 p_2 |p_3|^4} \,. \tag{5.153}$$

We can use the delta function to perform either the $p_3$ sum or the $p_2$ sum, yielding the expressions

$$\mathcal{C}\left[\begin{smallmatrix} 1 & 1 & 2 \\ 0 & 0 & 2 \end{smallmatrix}\right] = \sum_{p_1,p_2}' \frac{1}{p_1 p_2 |p_1+p_2|^4} = -\sum_{p_1,p_3} \frac{1}{p_1(p_1+p_3)|p_3|^4} \,. \tag{5.154}$$

In the first of these expressions, $p_1$ and $p_2$ both come with a power of 5 in the denominator, hence according to our criterion above, $\mathcal{C}\left[\begin{smallmatrix} 1 & 1 & 2 \\ 0 & 0 & 2 \end{smallmatrix}\right]$ should be convergent. In the second expression in (5.154) however, $p_1$ comes with a power of 2, hence, $\mathcal{C}\left[\begin{smallmatrix} 1 & 1 & 2 \\ 0 & 0 & 2 \end{smallmatrix}\right]$ should be divergent. The reason that the first expression seems to be convergent is that the divergence lies in the direction of $p_1+p_2 = \text{const.}$, whereas by counting the powers of $p_1$ and $p_2$, we only probed the directions along those two momenta. Therefore, $\mathcal{C}\left[\begin{smallmatrix} 1 & 1 & 2 \\ 0 & 0 & 2 \end{smallmatrix}\right]$ is divergent.

To summarize, an MGF is only convergent if the powers of all momenta are at least three, in all possible ways to solve the delta functions. We will translate this in the following into conditions on the labels of the two-, three- and four-point graphs introduced in Section 5.2.

In dihedral graphs, if we perform the sum over momentum $p$ with the delta function, we will increase the total powers of all other momenta by the total power of $p$. Hence, our divergence criterion for dihedral

13 Note that this simple power-counting criterion does not constitute a proof of the convergence or divergence of the lattice sum of the MGF. As we will discuss below, this power-counting argument tends to underestimate the convergence of the sum since possible cancellations are not accounted for.

graphs, taking into account that we can use any of the momenta to solve the delta function, is

$$\mathcal{C}\left[\begin{smallmatrix} A \\ B \end{smallmatrix}\right] \text{ convergent} \quad \Leftrightarrow \quad \min_{\substack{i,j \\ i\neq j}}(c_i + c_j) > 2\,, \tag{5.155}$$

where $c_i = a_i + b_i$ and $i, j$ run over all edges. The basic criterion (5.155) will have to be satisfied for all edge bundles in higher-point graphs as well, but the global structure of these graphs adds further criteria.

In general, solving delta functions is equivalent to assigning loop momenta consistently to the edges of the graph. Hence, by going through the topologically distinct assignments, we can see to which edges a certain momentum can propagate and hence what the convergence conditions for this graph should be. When considering graphs with edge bundles between the vertices (like the graphs introduced in Section 5.2), we first assign the total momenta of the bundles consistently. Then, in a bundle of total momentum $\mathfrak{p}$, with edges carrying momenta $p_1 \dots p_R$, we can choose any edge to solve the momentum conservation constraint, e.g. we can drop momentum $p_1$ and assign momentum $\mathfrak{p} - \sum_{i=2}^{R} p_i$ to this edge. For the convergence conditions, the implications of this are twofold: First, each momentum can appear in any other edge of the same bundle, implying the condition (5.155) for each bundle. Second, the total momenta of the edge bundles can appear in any edge, hence we should count the lowest total power for each edge bundle when determining the convergence condition due to the total momenta. We will go through this procedure for the three-point and all four-point graphs in Section 5.2 in the following.

For trihedral graphs, there is just one way to assign the bundle momenta, namely

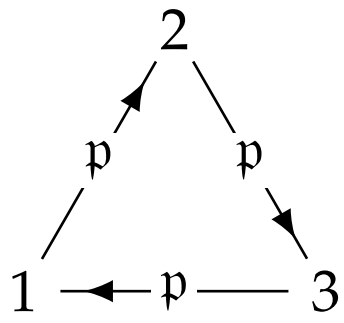

, (5.156)

i.e. the graph $\mathcal{C}\left[\begin{smallmatrix} A_1 \\ B_1 \end{smallmatrix}\middle|\begin{smallmatrix} A_2 \\ B_2 \end{smallmatrix}\middle|\begin{smallmatrix} A_3 \\ B_3 \end{smallmatrix}\right]$ is convergent iff

$$\begin{aligned} &\min_{\substack{i,j \\ i\neq j}} \left(c_i^{(k)} + c_j^{(k)}\right) > 2 \quad \forall\, k \in \{1,2,3\} \\ \text{and} \qquad &\check{c}_1 + \check{c}_2 + \check{c}_3 > 2\,, \end{aligned} \tag{5.157}$$

where $c_i^{(k)} = a_i^{(k)} + b_i^{(k)}$ for $i = 1, \dots, R_k$ and $\check{c}_k = \min_i(a_i^{(k)} + b_i^{(k)})$, where $i$ runs over all edges in block $k$. As described above, the first condition is due to the individual momenta in the edge bundles, whereas the second condition is due to the total bundle momentum $\mathfrak{p}$. If $\mathcal{C}\left[\begin{smallmatrix} A_1 \\ B_1 \end{smallmatrix}\middle|\begin{smallmatrix} A_2 \\ B_2 \end{smallmatrix}\middle|\begin{smallmatrix} A_3 \\ B_3 \end{smallmatrix}\right]$ carries only non-negative edge labels and does not contain a $(0,0)$-edge (i.e. is

not factorizable), then $c_i \geq 1$ for all edges and the second condition in (5.157) is always satisfied. The same will be true for all other conditions on top of (5.155) for every block in the following.

As a straightforward extension of the trihedral result, the box graph $\mathcal{C}\left[\begin{smallmatrix} A_1 \\ B_1 \end{smallmatrix}\middle|\begin{smallmatrix} A_2 \\ B_2 \end{smallmatrix}\middle|\begin{smallmatrix} A_3 \\ B_3 \end{smallmatrix}\middle|\begin{smallmatrix} A_4 \\ B_4 \end{smallmatrix}\right]$ is convergent iff

$$\begin{aligned} &\min_{\substack{i,j \\ i\neq j}} \left(c_i^{(k)} + c_j^{(k)}\right) > 2 \quad \forall\, k \in \{1,2,3,4\} \\ \text{and} \quad &\check{c}_1 + \check{c}_2 + \check{c}_3 + \check{c}_4 > 2\,, \end{aligned} \tag{5.158}$$

with the same notation as in (5.157).

In kite graphs, there are two topologically distinct ways of assigning the total momenta of the edge bundles. They are

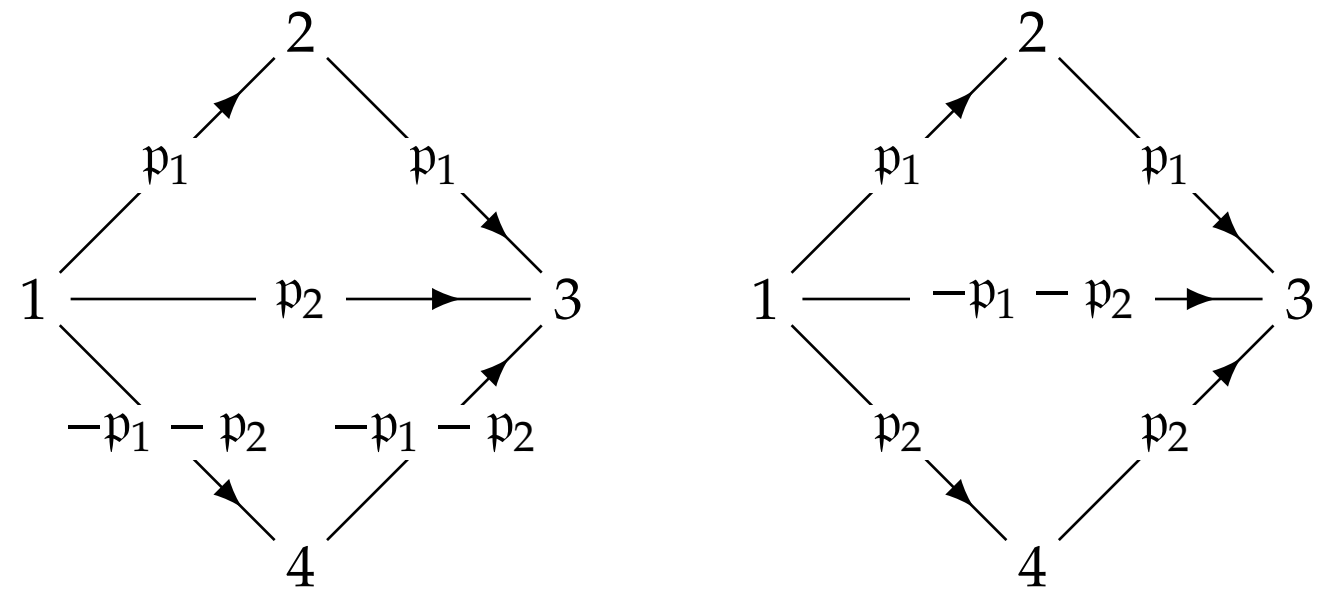

, (5.159)

implying that $\mathcal{C}\left[\begin{smallmatrix} A_1 \\ B_1 \end{smallmatrix}\middle|\begin{smallmatrix} A_2 \\ B_2 \end{smallmatrix}\middle\|\begin{smallmatrix} A_3 \\ B_3 \end{smallmatrix}\middle|\begin{smallmatrix} A_4 \\ B_4 \end{smallmatrix}\middle\|\begin{smallmatrix} A_5 \\ B_5 \end{smallmatrix}\right]$ is convergent iff

$$\begin{aligned} &\min_{\substack{i,j \\ i\neq j}} \left(c_i^{(k)} + c_j^{(k)}\right) > 2 \quad \forall\, k \in \{1,2,3,4,5\} \\ \text{and} \quad &\check{c}_i + \check{c}_j + \check{c}_5 > 2 \quad \forall\, (i,j) \in \{(1,2),(3,4)\} \\ \text{and} \quad &\check{c}_1 + \check{c}_2 + \check{c}_3 + \check{c}_4 > 2\,. \end{aligned} \tag{5.160}$$

For tetrahedral graphs, there are again two topologically distinct ways to assign the three independent total edge-bundle momenta,

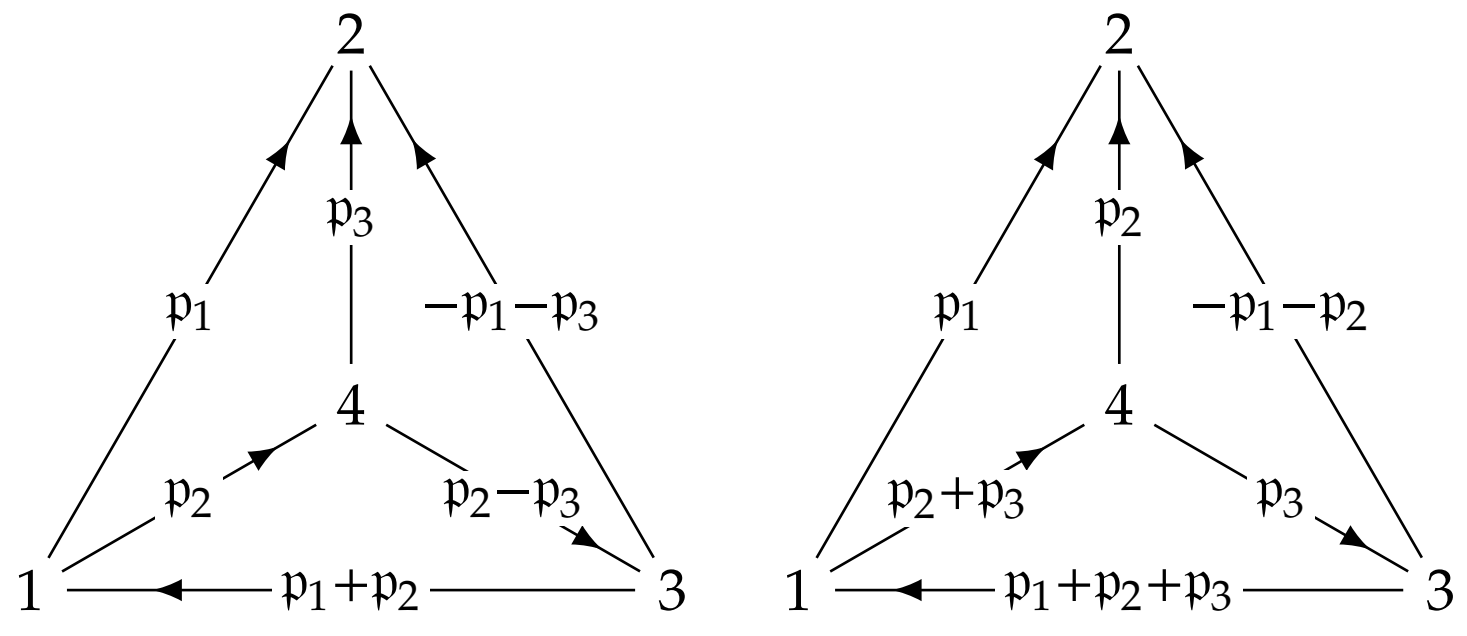

. (5.161)

This implies that the tetrahedral graph

$$C\left[\begin{array}{c|c|c} \frac{A_1}{B_1} & \frac{A_2}{B_2} & \frac{A_3}{B_3} \\ \hline \frac{A_4}{B_4} & \frac{A_5}{B_5} & \frac{A_6}{B_6} \end{array}\right] \tag{5.162}$$

is convergent iff

$$\begin{aligned} &\min_{\substack{i,j \\ i\neq j}} \left(c_i^{(k)} + c_j^{(k)}\right) > 2 \;\forall\; k \in \{1,2,3,4,5,6\} \qquad (5.163)\\ \text{and} \quad &\check{c}_i + \check{c}_j + \check{c}_k > 2 \;\forall\; (i,j,k) \in \{(1,2,6),(1,3,5),(2,3,4),(4,5,6)\} \\ \text{and} \quad &\check{c}_i + \check{c}_j + \check{c}_k + \check{c}_\ell > 2 \;\forall\; (i,j,k,\ell) \in \{(1,2,4,5),(1,3,4,6),(2,3,5,6)\}\,. \end{aligned}$$

Here, the penultimate line corresponds to all closed three-point subgraphs, the last line corresponds to all closed four-point subgraphs.

The convergence conditions discussed so far only depend on the sum of the holomorphic and antiholomorphic label. That this view tends to underestimate the convergence of the sum can be seen by considering the two one-loop graphs $C\left[\begin{smallmatrix}1 & 0\\ 0 & 1\end{smallmatrix}\right]$ and $C\left[\begin{smallmatrix}1 & 1\\ 0 & 0\end{smallmatrix}\right]$. According to our condition (5.155), both graphs should be equally divergent. But of course, while the sum $C\left[\begin{smallmatrix}1 & 0\\ 0 & 1\end{smallmatrix}\right]$ is divergent, the sum $C\left[\begin{smallmatrix}1 & 1\\ 0 & 0\end{smallmatrix}\right]$ is only conditionally convergent and we regularize it by introducing additional powers of the momentum as in (3.31), yielding $\widehat{\mathrm{G}}_2$. In general, graphs containing a $\left[\begin{smallmatrix}1 & 1\\ 0 & 0\end{smallmatrix}\right]$ subblock can be simplified using the divergent HSR discussed in Section 5.6.4.

In the integral representation, this can be seen as follows: $f^{(1)}(z,\tau) \sim \frac{1}{z}$ is the only one out of the $f^{(a)}$ which has a pole. The fact that $C\left[\begin{smallmatrix}1 & 1\\ 0 & 0\end{smallmatrix}\right]$ is conditionally convergent is reflected in the fact that the integral of $\frac{1}{z^2}$ over a ball around the origin vanishes, whereas the divergence of $C\left[\begin{smallmatrix}1 & 0\\ 0 & 1\end{smallmatrix}\right]$ is reflected in the divergence of the integral of $|f^{(1)}(z)|^2 \sim \frac{1}{|z|^2}$.

In the package `Modular Graph Forms`, the function **`CCheckConv`** checks for convergence of the argument using the criteria (5.155) and (5.157) on dihedral and trihedral graphs. The return value is either **`True`** for convergent MGFs or **`False`** for divergent MGFs, e.g.

```
In[45]:= CCheckConv[c[0 1 2 / 1 0 2]]

         CCheckConv[c[-1 2 / 0 2, 1 1 / 0 1, 1 1 / 0 1]]

Out[45]= False

Out[46]= False .
```

On top of dihedral and trihedral graphs, **`CCheckConv`** also checks for $\mathrm{E}_k$, $\mathrm{G}_k$ and $\overline{\mathrm{G}}_k$ with $k < 2$, all other expressions are treated as convergent. As soon as any divergent object is detected in the argument, **`CCheckConv`** returns **`False`**.

### 5.6.2 *Divergent modular graph forms from Koba–Nielsen integrals*

We study MGFs in order to expand Koba–Nielsen integrals comprising the Koba–Nielsen factor (3.72) and a polynomial in the functions $f^{(a)}(z,\tau)$ and $\overline{f^{(b)}(z,\tau)}$ given in (3.91) and (3.92). If this polynomial contains a factor $|f^{(1)}_{ij}|^2$ (where $f^{(1)}_{ij} = f^{(1)}(z_i - z_j)$), the MGFs in the expansion of the Koba–Nielsen integral are all divergent since $|f^{(1)}_{ij}|^2$ leads to a $\left[\begin{smallmatrix}1&0\\0&1\end{smallmatrix}\right]$ subblock, which violates the criterion (5.155).

However, the Koba–Nielsen factor regulates this divergence: Since the Jacobi theta function satisfies $\theta_1(z,\tau) \sim z$ for small $z$, $\exp(s_{ij}G_{ij}) \sim |z_{ij}|^{-2s_{ij}}$ for small $z_{ij}$. Using integration-by-parts identities for the Koba–Nielsen integral, one can in fact show that a Koba–Nielsen integral with a $|f^{(1)}_{ij}|^2$ prefactor has a pole in the Mandelstams. Hence, the appearance of divergent MGFs is merely a signal that one has tried to Taylor-expand around a pole.

As an example, consider the two-point Koba–Nielsen integral

$$\int \mathrm{d}\mu_1 |f^{(1)}_{12}|^2 \,\mathrm{KN}_2\,, \tag{5.164}$$

whose naive $\alpha'$ expansion

$$\frac{\pi}{\tau_2}\mathrm{E}_1 - s_{12}\frac{\tau_2}{\pi}\,\mathcal{C}\left[\begin{smallmatrix}0&1&1\\1&0&1\end{smallmatrix}\right] - \frac{1}{2}s_{12}^2\left(\frac{\tau_2}{\pi}\right)^2 \mathcal{C}\left[\begin{smallmatrix}0&1&1&1\\1&0&1&1\end{smallmatrix}\right] + \mathcal{O}(s_{12}^3) \tag{5.165}$$

exhibits divergent MGFs at every order in $s_{12}$. In order to make the pole in $s_{12}$ manifest, consider the derivative [IV]

$$\partial_{\bar z_2}\big(f^{(1)}_{12}\,\mathrm{KN}_2\big)\,. \tag{5.166}$$

We now use (3.90) and

$$\partial_{z_j}\mathrm{KN}_n = \sum_{i\neq j} s_{ij} f^{(1)}(z_{ij},\tau)\,\mathrm{KN}_n\,, \tag{5.167}$$

which follows from (3.89), to evaluate (5.166). With this, we obtain

$$\partial_{\bar z_2}\big(f^{(1)}_{12}\,\mathrm{KN}_2\big) = \left(\frac{\pi}{\tau_2} - \pi\delta^{(2)}(z_{12},\bar z_{12})\right)\mathrm{KN}_2 + s_{12}|f^{(1)}_{12}|^2\,\mathrm{KN}_2\,. \tag{5.168}$$

Integrating over $z_2$ and solving for (5.164) yields (since $\mathrm{KN}_2 \longrightarrow 0$ for $z_{12} \longrightarrow 0$ the term with the delta function does not contribute)

$$\int \mathrm{d}\mu_1 |f^{(1)}_{12}|^2\,\mathrm{KN}_2 = -\frac{1}{s_{12}}\frac{\pi}{\tau_2}\int \mathrm{d}\mu_1\,\mathrm{KN}_2\,, \tag{5.169}$$

making the pole in $s_{12}$ explicit. The remaining Koba–Nielsen integral in (5.169) has an expansion in convergent MGFs. Variations of this technique to expose the kinematic poles in Koba–Nielsen integrals can be found in countless examples in the literature.

At two points, the integral (5.169) is the only one with a pole in the Mandelstams and it is associated to the collision of the two punctures. At three point, several different poles can appear, including nested poles due to the collision of all three punctures. The rewriting of all relevant three-point integrals making the pole structure manifest and reducing divergent expansions to convergent ones as above, is summarized in Appendix B.3.

In general, we can use the Fay identity (5.121) to rewrite the $f_{ij}^{(1)}$ contributions to the integrand in terms of $f_{ij}^{(a)}$ with $a > 1$ and derivatives of the Koba–Nielsen factor as in (5.167). When integrating these expressions by parts, we make one pole explicit and obtain an expression with poles of lower multiplicity.

Aside from the integration-by-parts techniques discussed so far, there is an alternative way of dealing with Koba–Nielsen integrals with kinematic poles. It amounts to rewriting the integral into a sum so that the divergences cancel between the summands. Details about this *subtraction scheme* for two-particle poles are provided in Appendix B.4.

### 5.6.3 *Divergent modular graph forms from momentum conservation*

Apart from the expansion of Koba–Nielsen integrals, divergent modular graph forms can also appear in momentum-conservation identities of convergent graphs. In the sum representation (5.32) of momentum conservation, this means that the exchange of the sum over edges $e'$ and the sum over momenta $p_e$ is not allowed in this case. Performing it anyway leads to the decomposition of a convergent series into a sum of divergent series. As an example, consider the convergent seed $\mathcal{C}\left[\begin{smallmatrix} 0 & 1 & 2 \\ 1 & 1 & 2 \end{smallmatrix}\right]$, whose antiholomorphic momentum-conservation identity is

$$\mathcal{C}\left[\begin{smallmatrix} 0 & 1 & 2 \\ 1 & 0 & 2 \end{smallmatrix}\right] + \mathcal{C}\left[\begin{smallmatrix} 0 & 1 & 2 \\ 1 & 1 & 1 \end{smallmatrix}\right] + \left(\frac{\pi}{\tau_2}\right)^3 (\mathrm{E}_1\mathrm{E}_2 - \mathrm{E}_3) = 0\,, \tag{5.170}$$

after factorization. The graph $\mathcal{C}\left[\begin{smallmatrix} 0 & 1 & 2 \\ 1 & 0 & 2 \end{smallmatrix}\right]$ and the Eisenstein series $\mathrm{E}_1$ are both divergent.

When dealing only with convergent MGFs, momentum-conservation identities involving divergent graphs should be discarded. However, as we will discuss shortly, it is sometimes desirable to have identities between divergent MGFs and momentum-conservation identities involving divergent MGFs can be used to define those divergent MGFs. In this framework, we treat the divergent non-holomorphic Eisenstein series $\mathrm{E}_1$ as a basis element for divergent MGFs and find decompositions in the same way as we did for convergent MGFs. E.g. (5.170), together with the (convergent) identity

$$\mathcal{C}\left[\begin{smallmatrix} 0 & 1 & 2 \\ 1 & 1 & 1 \end{smallmatrix}\right] = -\frac{1}{2}\left(\frac{\pi}{\tau_2}\right)^3 (\mathrm{E}_3 - \zeta_3)\,, \tag{5.171}$$

can be used to decompose the divergent graph $\mathcal{C}\big[\begin{smallmatrix} 0 & 1 & 2 \\ 1 & 0 & 2 \end{smallmatrix}\big]$,

$$\mathcal{C}\big[\begin{smallmatrix} 0 & 1 & 2 \\ 1 & 0 & 2 \end{smallmatrix}\big] = \left(\frac{\pi}{\tau_2}\right)^3 \left(\frac{3}{2}\mathrm{E}_3 - \mathrm{E}_1\mathrm{E}_2 + \frac{1}{2}\zeta_3\right). \tag{5.172}$$

Note that this does not extend to momentum-conservation identities of divergent seeds which have to be treated separately, cf. Section 5.6.6 below.

In particular, momentum-conservation identities involving divergent graphs can appear in the sieve algorithm, when removing entries of $-1$ as described in Section 5.5.1. As an example for this phenomenon, consider the graph $\mathcal{C}\big[\begin{smallmatrix} 0 & 1 & 2 & 3 \\ 1 & 1 & 2 & 0 \end{smallmatrix}\big]$, whose Cauchy–Riemann derivative is given by

$$\nabla^{(6)}\mathcal{C}\big[\begin{smallmatrix} 0 & 1 & 2 & 3 \\ 1 & 1 & 2 & 0 \end{smallmatrix}\big] = 3\,\mathcal{C}\big[\begin{smallmatrix} 0 & 1 & 2 & 4 \\ 1 & 1 & 2 & -1 \end{smallmatrix}\big] + 2\,\mathcal{C}\big[\begin{smallmatrix} 0 & 1 & 3 & 3 \\ 1 & 1 & 1 & 0 \end{smallmatrix}\big] + \mathcal{C}\big[\begin{smallmatrix} 0 & 2 & 2 & 3 \\ 1 & 0 & 2 & 0 \end{smallmatrix}\big]. \tag{5.173}$$

The $-1$-entry in the first term can be removed by a momentum-conservation identity which yields, after factorization and divergent HSR (to be discussed below in Section 5.6.4),

$$\begin{aligned} \mathcal{C}\big[\begin{smallmatrix} 0 & 1 & 2 & 4 \\ 1 & 1 & 2 & -1 \end{smallmatrix}\big] = 5\,\mathcal{C}\big[\begin{smallmatrix} 0 & 2 & 5 \\ 1 & 2 & 0 \end{smallmatrix}\big] + \mathcal{C}\big[\begin{smallmatrix} 1 & 2 & 4 \\ 1 & 2 & 0 \end{smallmatrix}\big] - \mathcal{C}\big[\begin{smallmatrix} 0 & 1 & 2 & 4 \\ 1 & 1 & 1 & 0 \end{smallmatrix}\big] \\ - \mathrm{G}_4\,\mathcal{C}\big[\begin{smallmatrix} 0 & 1 & 2 \\ 1 & 0 & 2 \end{smallmatrix}\big] - \widehat{\mathrm{G}}_2\,\mathcal{C}\big[\begin{smallmatrix} 0 & 2 & 3 \\ 1 & 2 & 0 \end{smallmatrix}\big] \\ + \frac{\pi}{\tau_2}\left(\mathcal{C}\big[\begin{smallmatrix} 0 & 2 & 4 \\ 1 & 1 & 0 \end{smallmatrix}\big] - \mathcal{C}\big[\begin{smallmatrix} 6 & 0 \\ 2 & 0 \end{smallmatrix}\big]\right) + \left(\frac{\pi}{\tau_2}\right)^3 (\mathrm{E}_2 - \mathrm{E}_1\mathrm{E}_2)\,\mathrm{G}_4\,. \end{aligned} \tag{5.174}$$

As explained in detail in Section 5.5.1, when constructing identities with the sieve algorithm, we seek to cancel holomorphic Eisenstein series by adding suitable MGFs. In order to do this consistently, we need to know all relations for the MGFs in the prefactor of the holomorphic Eisenstein series. In the example (5.174), however, the prefactor of $\mathrm{G}_4$ is

$$-\mathcal{C}\big[\begin{smallmatrix} 0 & 1 & 2 \\ 1 & 0 & 2 \end{smallmatrix}\big] + \left(\frac{\pi}{\tau_2}\right)^3 (\mathrm{E}_2 - \mathrm{E}_1\mathrm{E}_2) \tag{5.175}$$

and hence in particular involves divergent MGFs. I.e. in this case, we need to know the decomposition (5.172) to see explicitly that the divergence cancels out and to continue with the sieve algorithm.

In general, since (according to (5.53)) the action of the Cauchy–Riemann operator on modular graph forms leaves the sum of holomorphic and antiholomorphic labels for each edge invariant and the divergence conditions in Section 5.6.1 are all functions of this sum only, each term in the derivative of an MGF $\mathcal{C}_\Gamma$ will have the same convergence properties as $\mathcal{C}_\Gamma$. Momentum conservation however increases the sum of the labels in one edge and decreases it in another edge in each term, therefore changing the convergence properties. But since the MGF decomposed in this way is convergent, the divergences have to cancel out upon plugging in identities for the divergent graphs.

For the remainder of this discussion, we will restrict to dihedral graphs, where the edge labels are written as columns in one block, but the arguments generalize straightforwardly to higher-point graphs. In [16], where the sieve algorithm was introduced, the authors restricted to the case of strictly positive holomorphic labels and non-negative antiholomorphic labels. In this case, the column sum for all edges is at least 2, with at most one $(1,0)$ edge since we assume that HSR is already performed. After taking the Cauchy–Riemann derivative, momentum conservation is only necessary in the term in which the $(1,0)$ edge is replaced by a $(2,-1)$ edge. In the momentum conservation identity, this edge will become a $(2,0)$ edge in each term, hence the column sum for each edge is again 2 with at most one edge of sum 1, i.e. each term is convergent. In this way, the problem of divergent MGFs in the sieve algorithm is avoided in [16] and the present discussion can therefore be regarded as an extension of the previously known techniques.

### 5.6.4 *Divergent holomorphic subgraph reduction*

On top of momentum conservation and factorization, holomorphic subgraph reduction is a central technique to derive identities for modular graph forms. It is therefore desirable to extend HSR to divergent graphs. To this end, we will distinguish the case in which the divergence appears within the holomorphic subgraph, i.e. the sum of the labels of the edges forming the holomorphic subgraph is at most 2, from the case in which the divergence appears outside the holomorphic subgraph, i.e. the sum of labels within the holomorphic subgraph is at least 3, but the entire MGF is still divergent.

In the case of a divergence outside the holomorphic subgraph, the sum over the loop momentum which is performed when doing HSR is convergent. I.e. the divergence acts merely as a spectator and the formulas for two- and three-point HSR discussed in Section 5.4 are still valid. E.g. dihedral graphs in which the divergence lies outside the holomorphic subgraph are given by $\mathcal{C}\left[\begin{smallmatrix} 0 & 1 & a & A \\ 1 & 0 & 0 & B \end{smallmatrix}\right]$ with $a \geq 2$ and all column sums in $\left[\begin{smallmatrix} A \\ B \end{smallmatrix}\right]$ at least two. In this case, we can apply the two-point HSR formula (5.73) and obtain results consistent with momentum conservation. For the graph $\mathcal{C}\left[\begin{smallmatrix} 0 & 1 & a & A \\ 1 & 0 & 0 & B \end{smallmatrix}\right]$ with $a \geq 3$ we can see this explicitly by using the holomorphic momentum-conservation identity of the convergent seed $\mathcal{C}\left[\begin{smallmatrix} 1 & 1 & a & A \\ 1 & 0 & 0 & B \end{smallmatrix}\right]$,

$$\begin{aligned} \mathcal{C}\left[\begin{smallmatrix} 0 & 1 & a & A \\ 1 & 0 & 0 & B \end{smallmatrix}\right] = &-\mathcal{C}\left[\begin{smallmatrix} 1 & 1 & a-1 & A \\ 1 & 0 & 0 & B \end{smallmatrix}\right] - \sum_{i=1}^{R} \mathcal{C}\left[\begin{smallmatrix} 1 & 1 & a & A-S_i \\ 1 & 0 & 0 & B \end{smallmatrix}\right] \\ &+ \mathcal{C}\left[\begin{smallmatrix} 1 & a & A \\ 1 & 0 & B \end{smallmatrix}\right] - \frac{\pi}{\tau_2} \mathrm{E}_1 \mathrm{G}_a \prod_{i=1}^{R} \mathcal{C}\left[\begin{smallmatrix} a_i & 0 \\ b_i & 0 \end{smallmatrix}\right], \end{aligned} \tag{5.176}$$

and applying the HSR formula (5.73) to the two convergent graphs on the RHS. Similar calculations can be done at three point and the extension of the HSR formulas to divergent graphs in this way was checked empirically for many cases.

If the holomorphic subgraph itself is divergent, the sum over the loop momentum which we perform when doing HSR is not convergent any more and hence we cannot use the usual HSR formulas in this case. If we restrict to only non-negative edge labels and assume that the graph under consideration has already been factorized (i.e. it does not contain any $(0,0)$ edges), then holomorphic subgraphs with more than two edges cannot be divergent. For this reason, we will restrict to the case of divergent two-point holomorphic subgraphs. In the sum representation, in which the two-point HSR formula (5.73) was derived first, it is unclear how to proceed in the case of divergent sums. In the integral representation, however, in which the two-point HSR formula was derived from the coincident limit (5.133) of the Fay identity, it is straightforward to generalize (5.73) to divergent holomorphic subgraphs: We can just take the $a_1 = a_2 = 1$ case of (5.133),

$$\left(f^{(1)}(z)\right)^2 = 2f^{(2)}(z) - \widehat{G}_2 - \partial_z f^{(1)}(z) \tag{5.177}$$

and integrate it against a product of $C^{(a,b)}(z)$ functions, as defined in (3.118), yielding

$$\mathcal{C}\left[\begin{smallmatrix} 1 & 1 & A \\ 0 & 0 & B \end{smallmatrix}\right] = -2\,\mathcal{C}\left[\begin{smallmatrix} 2 & A \\ 0 & B \end{smallmatrix}\right] - \widehat{G}_2\,\mathcal{C}\left[\begin{smallmatrix} A \\ B \end{smallmatrix}\right] + \frac{\pi}{\tau_2}\,\mathcal{C}\left[\begin{smallmatrix} 1 & A \\ -1 & B \end{smallmatrix}\right]. \tag{5.178}$$

Note that (5.73) has an additional term $\widehat{G}_2\,\mathcal{C}\left[\begin{smallmatrix} 0 & A \\ 0 & B \end{smallmatrix}\right]$ when naively extended to $a_+ = a_- = 1$. Empirically, we found that (5.178) is compatible with momentum conservation in a large number of cases. Furthermore, (5.178) agrees with the special cases

$$\mathcal{C}\left[\begin{smallmatrix} 1 & 1 & a \\ 0 & 0 & b \end{smallmatrix}\right] = -2\,\mathcal{C}\left[\begin{smallmatrix} a+2 & 0 \\ b & 0 \end{smallmatrix}\right] + \frac{\pi}{\tau_2}\,\mathcal{C}\left[\begin{smallmatrix} a+1 & 0 \\ b-1 & 0 \end{smallmatrix}\right] \tag{5.179}$$

$$\mathcal{C}\left[\begin{smallmatrix} 1 & 1 & 1 & 1 \\ 0 & 0 & 1 & 1 \end{smallmatrix}\right] = -2\,\mathcal{C}\left[\begin{smallmatrix} 4 & 0 \\ 2 & 0 \end{smallmatrix}\right] - \left(\frac{\pi}{\tau_2}\right)^2 \widehat{G}_2(E_2 + 2) + 4\frac{\pi}{\tau_2}\,\mathcal{C}\left[\begin{smallmatrix} 3 & 0 \\ 1 & 0 \end{smallmatrix}\right] \tag{5.180}$$

which were obtained in [III], where (5.177) was derived in a different way than from the coincident limit of Fay identities.

The divergent two-point HSR identity (5.178) has a straightforward generalization to trihedral (and higher-point graphs),

$$\begin{aligned} \mathcal{C}\left[\begin{smallmatrix} 1 & 1 & A_1 \\ 0 & 0 & B_1 \end{smallmatrix}\middle|\begin{smallmatrix} A_2 \\ B_2 \end{smallmatrix}\middle|\begin{smallmatrix} A_3 \\ B_3 \end{smallmatrix}\right] = -2\,\mathcal{C}\left[\begin{smallmatrix} 2 & A_1 \\ 0 & B_1 \end{smallmatrix}\middle|\begin{smallmatrix} A_2 \\ B_2 \end{smallmatrix}\middle|\begin{smallmatrix} A_3 \\ B_3 \end{smallmatrix}\right] - \widehat{G}_2\,\mathcal{C}\left[\begin{smallmatrix} A_1 \\ B_1 \end{smallmatrix}\middle|\begin{smallmatrix} A_2 \\ B_2 \end{smallmatrix}\middle|\begin{smallmatrix} A_3 \\ B_3 \end{smallmatrix}\right] \\ + \frac{\pi}{\tau_2}\,\mathcal{C}\left[\begin{smallmatrix} 1 & A_1 \\ -1 & B_1 \end{smallmatrix}\middle|\begin{smallmatrix} A_2 \\ B_2 \end{smallmatrix}\middle|\begin{smallmatrix} A_3 \\ B_3 \end{smallmatrix}\right]. \end{aligned} \tag{5.181}$$

The only kind of divergent HSR which cannot be treated in this way occurs if the holomorphic subgraph has a higher-point divergence,

since this necessarily means that the holomorphic subgraph involves negative labels.

One might be tempted to also extend the trihedral Fay identity (5.127) to divergent graphs. However, this was found to lead to contradictions, as illustrated in the following: Consider the divergent trihedral graph $\mathcal{C}\left[\begin{smallmatrix}1\\0\end{smallmatrix}\middle|\begin{smallmatrix}0&1\\1&0\end{smallmatrix}\middle|\begin{smallmatrix}0&1\\2&0\end{smallmatrix}\right]$ and simplify it once by performing three-point HSR and once by applying (5.127) to the first column and to the second column of the third block, yielding the decompositions

$$\begin{aligned}\mathcal{C}\left[\begin{smallmatrix}1\\0\end{smallmatrix}\middle|\begin{smallmatrix}0&1\\1&0\end{smallmatrix}\middle|\begin{smallmatrix}0&1\\2&0\end{smallmatrix}\right] &\overset{?}{=} \widehat{\mathrm{G}}_2\,\mathcal{C}\left[\begin{smallmatrix}1&0\\3&0\end{smallmatrix}\right]\\ &\quad+\frac{1}{2}\left(\frac{\pi}{\tau_2}\right)^3(\mathrm{E}_1(\mathrm{E}_1-4\mathrm{E}_2+2)-3\mathrm{E}_2+5\mathrm{E}_3-\zeta_3)\end{aligned} \tag{5.182a}$$

$$\begin{aligned}\mathcal{C}\left[\begin{smallmatrix}1\\0\end{smallmatrix}\middle|\begin{smallmatrix}0&1\\1&0\end{smallmatrix}\middle|\begin{smallmatrix}0&1\\2&0\end{smallmatrix}\right] &\overset{?}{=} \widehat{\mathrm{G}}_2\,\mathcal{C}\left[\begin{smallmatrix}1&0\\3&0\end{smallmatrix}\right]+\frac{\pi}{\tau_2}\mathcal{C}\left[\begin{smallmatrix}0&1&1\\1&-1&2\end{smallmatrix}\right]\\ &\quad+\frac{1}{2}\left(\frac{\pi}{\tau_2}\right)^3(4\mathrm{E}_1\mathrm{E}_2-5\mathrm{E}_3+\zeta_3)\,.\end{aligned} \tag{5.182b}$$

Applying the Fay identity (5.127) to any other pair of holomorphic or antiholomorphic columns also leads to (5.182a). Together, (5.182a) and (5.182b) imply

$$\mathcal{C}\left[\begin{smallmatrix}0&1&1\\1&-1&2\end{smallmatrix}\right] \overset{?}{=} \frac{1}{2}\left(\frac{\pi}{\tau_2}\right)^2(\mathrm{E}_1^2+2\mathrm{E}_1-3\mathrm{E}_2)\,. \tag{5.183}$$

Next, consider the divergent trihedral graph $\mathcal{C}\left[\begin{smallmatrix}0\\1\end{smallmatrix}\middle|\begin{smallmatrix}0&1\\2&0\end{smallmatrix}\middle|\begin{smallmatrix}1&1\\0&0\end{smallmatrix}\right]$ which can be decomposed via two-point HSR and Fay into

$$\mathcal{C}\left[\begin{smallmatrix}0\\1\end{smallmatrix}\middle|\begin{smallmatrix}0&1\\2&0\end{smallmatrix}\middle|\begin{smallmatrix}1&1\\0&0\end{smallmatrix}\right] \overset{?}{=} -\left(\frac{\pi}{\tau_2}\right)^3(\mathrm{E}_1-2\mathrm{E}_2+\mathrm{E}_3-\zeta_3) \tag{5.184a}$$

$$\begin{aligned}\mathcal{C}\left[\begin{smallmatrix}0\\1\end{smallmatrix}\middle|\begin{smallmatrix}0&1\\2&0\end{smallmatrix}\middle|\begin{smallmatrix}1&1\\0&0\end{smallmatrix}\right] &\overset{?}{=} -\frac{\pi}{\tau_2}\mathcal{C}\left[\begin{smallmatrix}0&1&1\\1&-1&2\end{smallmatrix}\right]\\ &\quad+\frac{1}{2}\left(\frac{\pi}{\tau_2}\right)^3(\mathrm{E}_1^2-2\mathrm{E}_1+\mathrm{E}_2-2\mathrm{E}_3+2\zeta_3)\,,\end{aligned} \tag{5.184b}$$

yielding the identity

$$\mathcal{C}\left[\begin{smallmatrix}0&1&1\\1&-1&2\end{smallmatrix}\right] \overset{?}{=} \frac{1}{2}\left(\frac{\pi}{\tau_2}\right)^2(\mathrm{E}_1^2-3\mathrm{E}_2)\,, \tag{5.185}$$

differing form (5.183) by a term $\frac{\pi}{\tau_2}\mathrm{E}_1$. For this reason, we will not apply the Fay identity (5.127) to divergent graphs.

In the `Mathematica` package `Modular Graph Forms`, divergent HSR is implemented in the functions **`DiCSimplify`** and **`TriCSimplify`**, along with the convergent HSR. If divergent HSR is performed or not, is controlled by the Boolean option **`divHSR`**. Dihedral and trihedral HSR can be activated and deactivated individually with the Boolean options **`diDivHSR`** and **`triDivHSR`**. The default values of all these options are **`True`**. If one of these options is set to **`False`** and for this reason a divergent HSR could not be performed, the warning **DiCSimplify::divHSRNotPossible** or **TriCSimplify::divHSRNotPossible** is issued.

### 5.6.5 *Taking derivatives of divergent graphs*

It would be desirable to apply the sieve algorithm discussed in Section 5.5 also to divergent MGFs to derive decompositions of divergent MGFs which are e.g. useful to perform the sieve algorithm on convergent MGFs. In order to do this, we have to take derivatives of divergent MGFs. Unfortunately, this is not straightforward and, if done naively, contradictions to momentum conservation identities can arise. As above, we will restrict in this section to two-point divergences occurring within one edge bundle since higher-point divergences are only relevant for graphs with negative entries.

Empirically, we found that taking derivatives of divergent MGFs using the formula (5.53) is consistent with momentum conservation if the divergence has the form $\left[\begin{smallmatrix}1&0\\0&1\end{smallmatrix}\right]$, however, a complete understanding of the structure of derivatives of these divergences is still lacking. If the divergence has the form $\left[\begin{smallmatrix}1&1\\0&0\end{smallmatrix}\right]$, we can first apply the divergent HSR formula (5.178), leading to a modification of the usual derivative expression (5.53). E.g. consider the graph $\mathcal{C}\left[\begin{smallmatrix}0&0&A\\1&1&B\end{smallmatrix}\right]$ with all column sums in $\left[\begin{smallmatrix}A\\B\end{smallmatrix}\right]$ at least 2. Using divergent HSR (5.178), it can be rewritten as

$$\mathcal{C}\left[\begin{smallmatrix}0&0&A\\1&1&B\end{smallmatrix}\right] = -2\,\mathcal{C}\left[\begin{smallmatrix}0&A\\2&B\end{smallmatrix}\right] - \widehat{\mathrm{G}}_2\,\mathcal{C}\left[\begin{smallmatrix}A\\B\end{smallmatrix}\right] + \frac{\pi}{\tau_2}\,\mathcal{C}\left[\begin{smallmatrix}-1&A\\1&B\end{smallmatrix}\right]. \tag{5.186}$$

Taking the derivative using (5.55) and using (5.178) to write the result back into a graph with a holomorphic subgraph yields

$$\nabla^{(|A|)}\,\mathcal{C}\left[\begin{smallmatrix}0&0&A\\1&1&B\end{smallmatrix}\right] = \sum_{i=1}^{R}\mathcal{C}\left[\begin{smallmatrix}0&0&A+S_i\\1&1&B-S_i\end{smallmatrix}\right] - \frac{\pi}{\tau_2}\prod_{i=1}^{R}\mathcal{C}\left[\begin{smallmatrix}a_i&0\\b_i&0\end{smallmatrix}\right], \tag{5.187}$$

with an additional term as compared to a naive application of (5.55) on $\mathcal{C}\left[\begin{smallmatrix}0&0&A\\1&1&B\end{smallmatrix}\right]$.

Aside from HSR, this additional term can also be understood as arising from the derivative of the regularization term implicitly contained in $\mathcal{C}\left[\begin{smallmatrix}0&0&A\\1&1&B\end{smallmatrix}\right]$. To see this, we first write the regularization term explicitly,

$$\mathcal{C}\left[\begin{smallmatrix}0&0&A\\1&1&B\end{smallmatrix}\right] = \lim_{s\to0}\mathcal{C}\left[\begin{smallmatrix}s&0&A\\s+1&1&B\end{smallmatrix}\right] \tag{5.188}$$

and exchange the limit and the differential, resulting in

$$\nabla^{(|A|)}\,\mathcal{C}\left[\begin{smallmatrix}0&0&A\\1&1&B\end{smallmatrix}\right] = \lim_{s\to0} s\,\mathcal{C}\left[\begin{smallmatrix}s+1&0&A\\s&1&B\end{smallmatrix}\right] + \sum_{i=1}^{R}\mathcal{C}\left[\begin{smallmatrix}0&0&A+S_i\\1&1&B-S_i\end{smallmatrix}\right]. \tag{5.189}$$

Next, we rewrite the first term using the momentum-conservation identity of the seed $\mathcal{C}\left[\begin{smallmatrix} s+1 & 0 & A \\ s+1 & 1 & B \end{smallmatrix}\right]$, which is convergent for all $s \geq 0$, and factorization, yielding

$$\lim_{s\to 0} s\,\mathcal{C}\left[\begin{smallmatrix} s+1 & 0 & A \\ s & 1 & B \end{smallmatrix}\right] = -\lim_{s\to 0} s\left(\mathcal{C}\left[\begin{smallmatrix} s+1 & 0 & A \\ s+1 & 0 & B \end{smallmatrix}\right] - \sum_{i=1}^{R} \mathcal{C}\left[\begin{smallmatrix} s+1 & 0 & A \\ s+1 & 0 & B-S_i \end{smallmatrix}\right]\right) \tag{5.190}$$

$$= -\lim_{s\to 0} s\left(\left(\frac{\pi}{\tau_2}\right)^{s+1} \mathrm{E}_{s+1} \prod_{i=1}^{R} \mathcal{C}\left[\begin{smallmatrix} a_i & 0 \\ b_i & 0 \end{smallmatrix}\right] - \mathcal{C}\left[\begin{smallmatrix} s+1 & A \\ s+1 & B \end{smallmatrix}\right]\right). \tag{5.191}$$

The last terms in (5.190) and (5.191) are convergent for all $s \geq 0$ and hence drop out after taking the limit. $\mathrm{E}_1$ however is divergent and with the first Kronecker limit formula

$$\mathrm{E}_{s+1} = \frac{1}{s} + \mathcal{O}(s^0)\,, \tag{5.192}$$

we obtain

$$\lim_{s\to 0} s\,\mathcal{C}\left[\begin{smallmatrix} s+1 & 0 & A \\ s & 1 & B \end{smallmatrix}\right] = -\frac{\pi}{\tau_2} \prod_{i=1}^{R} \mathcal{C}\left[\begin{smallmatrix} a_i & 0 \\ b_i & 0 \end{smallmatrix}\right]. \tag{5.193}$$

Plugging this into (5.189) yields (5.187), the result previously obtained from divergent HSR. Note that, to find this agreement, it is crucial that we do not simplify the last term in (5.186) using momentum conservation before applying the derivative. Consider e.g. the graph $\mathcal{C}\left[\begin{smallmatrix} 0 & 0 & 1 \\ 1 & 1 & 1 \end{smallmatrix}\right]$. In this case, (5.187) predicts

$$\nabla^{(1)}\,\mathcal{C}\left[\begin{smallmatrix} 0 & 0 & 1 \\ 1 & 1 & 1 \end{smallmatrix}\right] = \mathcal{C}\left[\begin{smallmatrix} 0 & 0 & 2 \\ 1 & 1 & 0 \end{smallmatrix}\right] - \frac{\pi}{\tau_2}\,\mathcal{C}\left[\begin{smallmatrix} 1 & 0 \\ 1 & 0 \end{smallmatrix}\right] = -2\left(\frac{\pi}{\tau_2}\right)^2 \mathrm{E}_2 \tag{5.194}$$

after applying divergent HSR (5.186). On the other hand, we have

$$\mathcal{C}\left[\begin{smallmatrix} 0 & 0 & 1 \\ 1 & 1 & 1 \end{smallmatrix}\right] = -2\,\mathcal{C}\left[\begin{smallmatrix} 1 & 0 \\ 3 & 0 \end{smallmatrix}\right] + \frac{\pi}{\tau_2}\,\mathcal{C}\left[\begin{smallmatrix} -1 & 1 \\ 1 & 1 \end{smallmatrix}\right], \tag{5.195}$$

according to (5.186). Acting with $\nabla^{(1)}$ directly on (5.195) leads to (5.194). If however we simplify $\mathcal{C}\left[\begin{smallmatrix} -1 & 1 \\ 1 & 1 \end{smallmatrix}\right] = \widehat{\overline{\mathrm{G}}}_2$, the derivative of the last term does not vanish any more since $\widehat{\overline{\mathrm{G}}}_2$ is not antiholomorphic, leading to a disagreement with (5.194).

Similarly to (5.187), we take the derivative of terms of the form $\mathcal{C}\left[\begin{smallmatrix} 1 & 1 & A \\ 0 & 0 & B \end{smallmatrix}\right]$ by first applying the formula (5.178) and then the usual expression (5.55) for the derivative. The generalization to higher-point graphs with two-point divergences is straightforward.

Since the techniques outlined in this section to take derivatives of divergent MGFs are conjectural and subtle, in the implementation in the `ModularGraphForms` package, a warning is issued whenever the functions **`CHolCR`** and **`CAHolCR`** encounter a divergent graph in their argument. If the Boolean option **`divDer`** of these functions is set to **`False`** (the default is **True**), **Nothing** is returned if it is divergent. If **divDer** is set to **True**, divergent derivatives are treated exactly like convergent ones, only (divergent) HSR is performed on the input (without momentum simplification) before the derivative is taken.

### 5.6.6 *Divergent momentum conservation and factorization*

Naively performing momentum conservation of divergent seeds and factorization leads to inconsistencies, e.g. consider the holomorphic momentum-conservation identity of the seed $C\left[\begin{smallmatrix}1&1&2\\0&0&3\end{smallmatrix}\right]$ which is naively[14]

$$C\left[\begin{smallmatrix}1&1&1\\0&0&3\end{smallmatrix}\right] \stackrel{?}{=} -2C\left[\begin{smallmatrix}0&1&2\\0&0&3\end{smallmatrix}\right] \stackrel{?}{=} -2C\left[\begin{smallmatrix}1&0\\0&0\end{smallmatrix}\right]C\left[\begin{smallmatrix}2&0\\0&3\end{smallmatrix}\right] - 2\left(\frac{\pi}{\tau_2}\right)^3 \mathrm{E}_3 \stackrel{?}{=} -2\left(\frac{\pi}{\tau_2}\right)^3 \mathrm{E}_3\,, \tag{5.196}$$

where the first term vanishes due to odd label sums in both MGFs. The divergent HSR formula (5.178) however (and also momentum conservation of the convergent seed $C\left[\begin{smallmatrix}1&1&2\\0&3&0\end{smallmatrix}\right]$) leads to

$$C\left[\begin{smallmatrix}1&1&1\\0&0&3\end{smallmatrix}\right] = \left(\frac{\pi}{\tau_2}\right)^3 (\mathrm{E}_2 - 2\mathrm{E}_3)\,, \tag{5.197}$$

contradicting (5.196). In this section, we will discuss some of the phenomena that arise in divergent momentum conservation and factorization but leave a complete understanding to the future.

The additional term in (5.197) can be understood in the integral representation of the MGF as follows: Consider the graph

$$C\left[\begin{smallmatrix}0&1&A\\0&0&B\end{smallmatrix}\right] = \int_\Sigma \frac{\mathrm{d}^2 z}{\tau_2} C^{(0,0)}(z) f^{(1)}(z) \prod_{i=1}^{R} C^{(a_i,b_i)}(z)\,, \tag{5.198}$$

where $\left[\begin{smallmatrix}A\\B\end{smallmatrix}\right]$ contains no $\left[\begin{smallmatrix}1\\0\end{smallmatrix}\right]$, $\left[\begin{smallmatrix}0\\1\end{smallmatrix}\right]$ or $\left[\begin{smallmatrix}1\\1\end{smallmatrix}\right]$ columns. We saw in (5.45) that $C^{(0,0)}(z) = \tau_2\delta(z,\bar z) - 1$, leading to the usual factorization rule. In (5.198), the delta function instructs to take the $z \to 0$ limit of $f^{(1)}(z)\prod_{i=1}^{R} C^{(a_i,b_i)}(z)$. But since $f^{(1)}(z)$ has Laurent expansion

$$f^{(1)}(z) = \frac{1}{z} - z\widehat{\mathrm{G}}_2 - \bar z\frac{\pi}{\tau_2} + O(z,\bar z)^3 \tag{5.199}$$

and in particular a pole at 0, we have to expand the product to first order to obtain

$$\lim_{z\to 0} f^{(1)}(z)\prod_{i=1}^{R} C^{(a_i,b_i)}(z) = \left(\partial_z \prod_{i=1}^{R} C^{(a_i,b_i)}(z)\right)_{z=0} = -\frac{\pi}{\tau_2}\sum_{i=1}^{R}\prod_{j=1}^{R} C\left[\begin{smallmatrix}a_j & 0\\ b_j-\delta_{ij} & 0\end{smallmatrix}\right], \tag{5.200}$$

14 We will see below that the first two equalities are not correct for divergent graphs. The last equality is too naive because $C\left[\begin{smallmatrix}1&0\\0&0\end{smallmatrix}\right]$ is conditionally convergent and can yield infinity, depending on the summation prescription used.

using (5.34) and the fact that the product vanishes at zero since $|A|+|B|$ is odd if $\mathcal{C}\left[\begin{smallmatrix}0 & 1 & A\\ 0 & 0 & B\end{smallmatrix}\right]$ is non-trivial. This yields the modified factorization rule

$$\mathcal{C}\left[\begin{smallmatrix}0 & 1 & A\\ 0 & 0 & B\end{smallmatrix}\right] = -\frac{\pi}{\tau_2}\sum_{i=1}^{R}\prod_{j=1}^{R}\mathcal{C}\left[\begin{smallmatrix}a_j & 0\\ b_j-\delta_{ij} & 0\end{smallmatrix}\right] - \mathcal{C}\left[\begin{smallmatrix}1 & A\\ 0 & B\end{smallmatrix}\right]. \tag{5.201}$$

If more $\left[\begin{smallmatrix}1\\0\end{smallmatrix}\right]$ columns are present, higher derivatives of the remaining graphs have to be taken. If $\left[\begin{smallmatrix}A\\B\end{smallmatrix}\right]$ contains a $\left[\begin{smallmatrix}1\\1\end{smallmatrix}\right]$ column, corresponding to a Green function in the integral, we have to iterate this procedure, since the derivative of the Green function is $f^{(1)}$ (cf. (3.89)) and hence contains again a pole. In this way we obtain for the MGF $\mathcal{C}\left[\begin{smallmatrix}0 & 1 & 1_n & A\\ 0 & 0 & 1_n & B\end{smallmatrix}\right]$, where $1_n$ is the row vector with $n$ entries of 1, the factorization rule

$$\begin{aligned}\mathcal{C}\left[\begin{smallmatrix}0 & 1 & 1_n & A\\ 0 & 0 & 1_n & B\end{smallmatrix}\right] = \left(\frac{\pi}{\tau_2}\right)^{n+1}\sum_{i=1}^{R}\prod_{j=1}^{R}\mathcal{C}\left[\begin{smallmatrix}a_j & 0\\ b_j-\delta_{ij} & 0\end{smallmatrix}\right]\sum_{k=0}^{n}(-1)^{k+1}\frac{n!}{(n-k)!}\mathrm{E}_1^{n-k}\\ - \mathcal{C}\left[\begin{smallmatrix}1 & 1_n & A\\ 0 & 1_n & B\end{smallmatrix}\right].\end{aligned} \tag{5.202}$$

For trihedral graphs, we have similarly

$$\begin{aligned}&\mathcal{C}\left[\begin{smallmatrix}0 & 1 & 1_n & A_1\\ 0 & 0 & 1_n & B_1\end{smallmatrix}\middle|\begin{smallmatrix}A_2\\B_2\end{smallmatrix}\middle|\begin{smallmatrix}A_3\\B_3\end{smallmatrix}\right]\\ &= (-1)^{|2|}\left(\frac{\pi}{\tau_2}\right)^{n+1}\mathcal{C}\left[\begin{smallmatrix}A_2 & A_3\\ B_2 & B_3\end{smallmatrix}\right]\sum_{i=1}^{R_1}\prod_{j=1}^{R_1}\mathcal{C}\left[\begin{smallmatrix}a_1^{(j)} & 0\\ b_1^{(j)}-\delta_{ij} & 0\end{smallmatrix}\right]\sum_{k=0}^{n}(-1)^{k+1}\frac{n!}{(n-k)!}\mathrm{E}_1^{n-k}\\ &\quad - \mathcal{C}\left[\begin{smallmatrix}1 & 1_n & A_1\\ 0 & 1_n & B_1\end{smallmatrix}\middle|\begin{smallmatrix}A_2\\B_2\end{smallmatrix}\middle|\begin{smallmatrix}A_3\\B_3\end{smallmatrix}\right].\end{aligned} \tag{5.203}$$

In general, the Laurent expansion of $f^{(n)}$ contains a term $\sim \frac{\bar{z}^{n-1}}{z}$, which vanishes at the origin for $n \geq 3$. The $z \to 0$ limit of $f^{(2)}$ depends on the direction in which the origin is approached, but $\frac{\bar{z}}{z}$ vanishes when integrated against a delta function due to the angular part of the integration. Therefore, only the case of $f^{(1)}$ yields additional terms as discussed above.

When (5.201) is used in (5.196), we obtain the correct additional term, up to a factor of 2, which arose in the momentum-conservation identity from the product rule of $\partial_{\bar{z}}$ acting on $f^{(1)}$ (cf. (5.35)). This spurious factor of 2 is again due to the pole in $f^{(1)}$, as can be understood by considering the integral

$$\int_{B_r(0)} \mathrm{d}^2 z\, \partial_{\bar{z}}\left(\frac{1}{z^2}\right) z\,, \tag{5.204}$$

where $B_r(0)$ is the ball of radius $r$ around 0. Evaluating (5.196) using $\partial_{\bar{z}}\left(\frac{1}{z}\right) = \pi\delta^{(2)}(z)$ and the product rule leads to

$$\int_{B_r(0)} \mathrm{d}^2 z\, \partial_{\bar{z}}\left(\frac{1}{z^2}\right) z \stackrel{?}{=} 2\int_{B_r(0)} \mathrm{d}^2 z\, \frac{1}{z}\partial_{\bar{z}}\left(\frac{1}{z}\right) z = 2\pi\,, \tag{5.205}$$

whereas the factor of 2 is absent if we apply Stokes' theorem,

$$\int_{B_r(0)} \mathrm{d}^2 z\, \partial_{\bar{z}} \left( \frac{1}{z^2} \right) z = \frac{1}{2i} \oint_{\partial B_r(0)} \mathrm{d}z\, \frac{1}{z} = \pi \operatorname*{Res}_{z=0} \left( \frac{1}{z} \right) = \pi\,. \tag{5.206}$$

Empirically, momentum-conservation identities of seeds with a divergence of the form $\left[\begin{smallmatrix} 1 & 0 \\ 0 & 1 \end{smallmatrix}\right]$ seem to be consistent, but we have not investigated them any further. For trihedral graphs, if the two blocks adjacent to the vertex used for momentum conservation are convergent and no three-point divergence appears in the graph, the resulting momentum-conservation identity is valid. If these conditions are not met, the same care has to be taken as with the dihedral graphs.

In the `Modular Graph Forms` package, the modified factorization rules (5.202) and (5.203) are implemented in the functions **`DiCSimplify`** and **`TriCSimplify`**, but since they are not tested as thoroughly as the convergent manipulations, a warning is issued if these special cases are encountered. If more than one $\left[\begin{smallmatrix} 1 \\ 0 \end{smallmatrix}\right]$ or $\left[\begin{smallmatrix} 0 \\ 1 \end{smallmatrix}\right]$ column appears next to a $\left[\begin{smallmatrix} 0 \\ 0 \end{smallmatrix}\right]$ column, the input is returned. The momentum conservation functions **`DiHolMomConsId`** and **`TriHolMomConsId`** and their complex conjugates issue a warning when the seed is divergent.

As we will see in the next section, the basis decompositions of MGFs obtained in this thesis rely on manipulations of divergent MGFs only for the modular weights $(6, 6)$ and $(7, 5)$ (and their complex conjugates). The expansion of the generating functions of two- and three-point Koba–Nielsen integrals (introduced in Chapter 7 below), involving these sectors was checked to satisfy the Cauchy–Riemann equations derived in Chapter 7. Furthermore, the Laurent polynomials of this expansion were checked against the closed formula for two-point Laurent polynomials given in Chapter 8.

## 5.7 BASIS DECOMPOSITIONS

By combining the techniques discussed in the sections above, we can systematically generate identities for modular graph forms, starting from a small number of known relations. In the end, we obtain decompositions of a large class of complicated MGFs into a small number of simple graphs. That these actually form a basis for all MGFs can be proven using techniques from iterated Eisenstein integrals discussed in Chapter 8.

In the `Modular Graph Forms` `Mathematica` package, decompositions for all dihedral and trihedral convergent MGFs with non-negative edge labels of modular weight $(a, b)$ with $a + b \leq 12$ are given, starting just from the dihedral identities

$$D_3 = \mathrm{E}_3 + \zeta_3 \tag{5.207}$$

$$D_5 = 60C_{1,1,3} + 10D_3\mathrm{E}_2 - 48\mathrm{E}_5 + 16\zeta_5\,, \tag{5.208}$$

| weight | di. non-HSR | di. HSR | tri. non-HSR | tri. HSR |
|---|---|---|---|---|
| $(1,1)$ | 0 | 0 | 0 | 0 |
| $(2,2)$ | 1 | 0 | 0 | 0 |
| $(3,1)$ | 1 | 0 | 0 | 0 |
| $(3,3)$ | 7 | 2 | 0 | 0 |
| $(4,2)$ | 5 | 3 | 0 | 0 |
| $(5,1)$ | 1 | 4 | 0 | 0 |
| $(4,4)$ | 27 | 10 | 28 | 20 |
| $(5,3)$ | 22 | 12 | 17 | 25 |
| $(6,2)$ | 11 | 16 | 0 | 29 |
| $(7,1)$ | 1 | 14 | 0 | 12 |
| $(5,5)$ | 83 | 40 | 326 | 248 |
| $(6,4)$ | 73 | 44 | 247 | 291 |
| $(7,3)$ | 47 | 50 | 91 | 322 |
| $(8,2)$ | 19 | 50 | 0 | 243 |
| $(9,1)$ | 1 | 35 | 0 | 94 |
| $(6,6)$ | 228 | 138 | 2236 | 2044 |
| $(7,5)$ | 206 | 142 | 1844 | 2191 |
| $(8,4)$ | 150 | 154 | 990 | 2359 |
| $(9,3)$ | 83 | 149 | 276 | 2008 |
| $(10,2)$ | 29 | 124 | 0 | 1207 |
| $(11,1)$ | 1 | 74 | 0 | 439 |
| total | 996 | 1061 | 6055 | 11532 |

Table 5.2: Number of convergent dihedral and trihedral MGFs with non-negative edge labels, excluding products. For graphs containing closed holomorphic subgraphs, no basis decompositions need to be found independently, they are implied by HSR and the basis decompositions of the non-HSR graphs.

where $D_\ell$ is defined in (3.115) and $C_{a,b,c}$ in (3.126). These two identities are also the only source of zeta-values in the basis decompositions.

### 5.7.1 *Systematic derivation of identities*

In order to apply the techniques discussed above systematically, we consider subspaces with total modular weight $a+b$ = const. of the space of all MGFs and derive all identities in one subspace before continuing to the next higher total weight.

Within each subspace, we start by considering weight $a = b$ which corresponds to MGFs which are modular invariant after multiplication by $\tau_2^a$. We generate identities in this space by combining momentum conservation with Fay identities:

- We write down all convergent dihedral and trihedral MGFs of weight $(a+1, a)$ and $(a, a+1)$ without closed holomorphic subgraphs and use them as seeds to generate holomorphic and antiholomorphic momentum-conservation identities, respectively. Closed Holomorphic subgraphs in the seeds would necessarily lead to negative labels in the identity which could not be removed by momentum conservation.

- We write down all convergent trihedral MGFs of weight $(a, a)$, including those which contain closed holomorphic subgraphs and apply the Fay identity (5.127) in all possible ways.

Afterwards, we remove all relations which contain divergent MGFs after topological simplifications and factorizations. Then, we simplify the remaining identities using HSR, the (generalized) Ramanujan identities discussed in Section 5.3.5 and identities known from lower total modular weight and expand holomorphic Eisenstein series in the ring spanned by $G_4$ and $G_6$. The resulting large system of linear equations, together with the identities (5.207) and (5.208) can then be solved for all convergent dihedral and trihedral MGFs which do not appear in the basis.

After the $a = b$ sector, we continue with the $a > b$ sectors with increasing $a$ (and the same total modular weight) as follows: In addition to the momentum conservation and Fay identities for these sectors, we also take the Cauchy–Riemann derivative of all basis decompositions in the $(a-1, b+1)$ sector (excluding MGFs containing closed holomorphic subgraphs), which were found before. Again, we remove all relations containing divergent MGFs. Finally, we take the complex conjugate of all identities obtained, to also cover the $a < b$ sectors.

In this way, basis decompositions for all convergent dihedral and trihedral MGFs can be found with total modular weight $a + b \leq 10$. The number of these MGFs is listed in Table 5.2. Note that we did not need to use the sieve algorithm in this process, hence we do not have undetermined integration constants in the basis decompositions.

Although the strategy outlined above is successful in the $a + b \leq 10$ sectors, at weight $(6, 6)$, it is not sufficient to decompose all trihedral MGFs. To obtain the decompositions of these graphs as well, we keep the momentum conservation identities containing divergent graphs and simplify them using the divergent HSR outlined in Section 5.6.4 if possible (both divergent holomorphic subgraphs and divergences outside of the holomorphic subgraph appear). In this way, we can decompose all graphs in the $(6, 6)$ and $(7, 5)$ sectors. For the remaining sectors in Table 5.2, the convergent identities are sufficient again.

In this way, basis decompositions for 1646 dihedral and 9520 trihedral convergent MGFs with non-negative edge labels and without closed holomorphic subgraphs were found and implemented in the functions `DiCSimplify` and `TriCSimplify` of the `Modular Graph Forms` package. Since **CSimplify** calls **DiCSimplify** and **TriCSimplify**, we have e.g.

```
In[47]:= CSimplify[c[1 1 1 1
                     1 1 1 1]]
         CSimplify[c[1, 1 1, 1 1
                     1, 1 1, 1 1]]
```

$$\text{Out[47]=}\ 24\, C\Big[\begin{smallmatrix} 1 & 1 & 2 \\ 1 & 1 & 2 \end{smallmatrix}\Big] + \frac{3\pi^4 E_2^2}{\tau_2^4} - \frac{18\pi^4 E_4}{\tau_2^4}$$

$$\text{Out[48]=}\ 2\, C\Big[\begin{smallmatrix} 1 & 1 & 3 \\ 1 & 1 & 3 \end{smallmatrix}\Big] - \frac{2\pi^5 E_5}{5\tau_2^5} + \frac{3\pi^5 \zeta_5}{10\tau_2^5}\,.$$

All the basis decompositions contained in the `Modular Graph Forms` package were checked to be compatible with the Cauchy–Riemann equation of the generating series of Koba–Nielsen integrals discussed in Chapter 7 at two- and three points. The decompositions of MGFs with $a+b \leq 10$ were used in [V] to find representations of MGFs in terms of iterated Eisenstein integrals via this generating series, as discussed in Chapter 8.

### 5.7.2 *Bases for modular graph forms*

Using the procedure outlined in Section 5.7.1, we obtain decompositions for many modular graph forms, which leave as independent MGFs only the ones listed in Table 5.3, which in fact form a basis, as can be proven using iterated Eisenstein integrals, as discussed in Chapter 8. The basis elements in the sector $(a,b)$ with $a<b$ are given by complex conjugation. Furthermore, basis elements containing a holomorphic Eisenstein series are not listed in Table 5.3, since they can be constructed from the bases at lower weights, e.g. the $(6,4)$ sector contains the additional basis elements $\mathrm{G}_4\, \mathcal{C}\big[\begin{smallmatrix} 2 & 0 \\ 4 & 0 \end{smallmatrix}\big]$ and $\mathrm{G}_6\overline{\mathrm{G}}_4$. In the following, we will refer to basis elements given as products as *reducible* and the remaining ones as *irreducible*. On top of various modular graph forms, we have included in Table 5.3 also the constants $\zeta_3$, $\zeta_5$ and $\zeta_3^2$ in the relevant sectors.

Note that starting from total modular weight 10, the sector with equal holomorphic and antiholomorphic weight contains cusp forms. Specifically, in the basis of the $(5,5)$ sector, the three cusp forms

$$\mathcal{C}\big[\begin{smallmatrix} 1 & 0 \\ 3 & 0 \end{smallmatrix}\big]\mathcal{C}\big[\begin{smallmatrix} 4 & 0 \\ 2 & 0 \end{smallmatrix}\big] - \mathcal{C}\big[\begin{smallmatrix} 3 & 0 \\ 1 & 0 \end{smallmatrix}\big]\mathcal{C}\big[\begin{smallmatrix} 2 & 0 \\ 4 & 0 \end{smallmatrix}\big] \tag{5.209a}$$

$$\mathcal{A}\big[\begin{smallmatrix} 0 & 2 & 3 \\ 3 & 0 & 2 \end{smallmatrix}\big] \tag{5.209b}$$

$$\mathcal{A}\big[\begin{smallmatrix} 0 & 1 & 2 & 2 \\ 1 & 1 & 0 & 3 \end{smallmatrix}\big] \tag{5.209c}$$

| weight | no. | basis elements |
|---|---|---|
| $(2,2)$ | 1 | $\left(\frac{\pi}{\tau_2}\right)^2\mathrm{E}_2$ |
| $(3,1)$ | 1 | $\mathcal{C}\left[\begin{smallmatrix}3&0\\1&0\end{smallmatrix}\right]$ |
| $(3,3)$ | 2 | $\left(\frac{\pi}{\tau_2}\right)^3\mathrm{E}_3$, $\left(\frac{\pi}{\tau_2}\right)^3\zeta_3$ |
| $(4,2)$ | 1 | $\mathcal{C}\left[\begin{smallmatrix}4&0\\2&0\end{smallmatrix}\right]$ |
| $(5,1)$ | 1 | $\mathcal{C}\left[\begin{smallmatrix}5&0\\1&0\end{smallmatrix}\right]$ |
| $(4,4)$ | 4 | $\left(\frac{\pi}{\tau_2}\right)^4\mathrm{E}_4$, $\mathcal{C}\left[\begin{smallmatrix}1&1&2\\1&1&2\end{smallmatrix}\right]$, $\left(\frac{\pi}{\tau_2}\right)^4\mathrm{E}_2^2$, $\mathcal{C}\left[\begin{smallmatrix}1&0\\3&0\end{smallmatrix}\right]\mathcal{C}\left[\begin{smallmatrix}3&0\\1&0\end{smallmatrix}\right]$ |
| $(5,3)$ | 3 | $\mathcal{C}\left[\begin{smallmatrix}5&0\\3&0\end{smallmatrix}\right]$, $\mathcal{C}\left[\begin{smallmatrix}1&1&3\\1&1&1\end{smallmatrix}\right]$, $\left(\frac{\pi}{\tau_2}\right)^2\mathrm{E}_2\,\mathcal{C}\left[\begin{smallmatrix}3&0\\1&0\end{smallmatrix}\right]$ |
| $(6,2)$ | 2 | $\mathcal{C}\left[\begin{smallmatrix}6&0\\2&0\end{smallmatrix}\right]$, $\mathcal{C}\left[\begin{smallmatrix}3&0\\1&0\end{smallmatrix}\right]^2$ |
| $(7,1)$ | 1 | $\mathcal{C}\left[\begin{smallmatrix}7&0\\1&0\end{smallmatrix}\right]$ |
| $(5,5)$ | 9 | $\left(\frac{\pi}{\tau_2}\right)^5\mathrm{E}_5$, $\mathcal{C}\left[\begin{smallmatrix}1&1&3\\1&1&3\end{smallmatrix}\right]$, $\mathcal{A}\left[\begin{smallmatrix}0&2&3\\3&0&2\end{smallmatrix}\right]$, $\mathcal{A}\left[\begin{smallmatrix}0&1&2&2\\1&1&0&3\end{smallmatrix}\right]$, $\left(\frac{\pi}{\tau_2}\right)^5\zeta_5$, $\left(\frac{\pi}{\tau_2}\right)^5\mathrm{E}_2\mathrm{E}_3$, $\left(\frac{\pi}{\tau_2}\right)^5\mathrm{E}_2\zeta_3$, $\mathcal{C}\left[\begin{smallmatrix}1&0\\3&0\end{smallmatrix}\right]\mathcal{C}\left[\begin{smallmatrix}4&0\\2&0\end{smallmatrix}\right]$, $\mathcal{C}\left[\begin{smallmatrix}3&0\\1&0\end{smallmatrix}\right]\mathcal{C}\left[\begin{smallmatrix}2&0\\4&0\end{smallmatrix}\right]$ |
| $(6,4)$ | 8 | $\mathcal{C}\left[\begin{smallmatrix}6&0\\4&0\end{smallmatrix}\right]$, $\mathcal{C}\left[\begin{smallmatrix}1&1&4\\1&1&2\end{smallmatrix}\right]$, $\mathcal{C}\left[\begin{smallmatrix}1&2&3\\1&0&3\end{smallmatrix}\right]$, $\mathcal{C}\left[\begin{smallmatrix}1&1&1&3\\0&1&1&2\end{smallmatrix}\right]$, $\left(\frac{\pi}{\tau_2}\right)^3\mathrm{E}_3\,\mathcal{C}\left[\begin{smallmatrix}3&0\\1&0\end{smallmatrix}\right]$, $\left(\frac{\pi}{\tau_2}\right)^3\mathrm{E}_2\,\mathcal{C}\left[\begin{smallmatrix}4&0\\2&0\end{smallmatrix}\right]$, $\mathcal{C}\left[\begin{smallmatrix}3&0\\1&0\end{smallmatrix}\right]\zeta_3$, $\mathcal{C}\left[\begin{smallmatrix}1&0\\3&0\end{smallmatrix}\right]\mathcal{C}\left[\begin{smallmatrix}5&0\\1&0\end{smallmatrix}\right]$ |
| $(7,3)$ | 5 | $\mathcal{C}\left[\begin{smallmatrix}7&0\\3&0\end{smallmatrix}\right]$, $\mathcal{C}\left[\begin{smallmatrix}1&1&5\\1&1&1\end{smallmatrix}\right]$, $\mathcal{C}\left[\begin{smallmatrix}0&2&5\\1&0&2\end{smallmatrix}\right]$, $\mathcal{C}\left[\begin{smallmatrix}3&0\\1&0\end{smallmatrix}\right]\mathcal{C}\left[\begin{smallmatrix}4&0\\2&0\end{smallmatrix}\right]$, $\left(\frac{\pi}{\tau_2}\right)^2\mathrm{E}_2\,\mathcal{C}\left[\begin{smallmatrix}5&0\\1&0\end{smallmatrix}\right]$ |
| $(8,2)$ | 3 | $\mathcal{C}\left[\begin{smallmatrix}8&0\\2&0\end{smallmatrix}\right]$, $\mathcal{C}\left[\begin{smallmatrix}0&3&5\\1&0&1\end{smallmatrix}\right]$, $\mathcal{C}\left[\begin{smallmatrix}3&0\\1&0\end{smallmatrix}\right]\mathcal{C}\left[\begin{smallmatrix}5&0\\1&0\end{smallmatrix}\right]$ |
| $(9,1)$ | 1 | $\mathcal{C}\left[\begin{smallmatrix}9&0\\1&0\end{smallmatrix}\right]$ |
| $(6,6)$ | 21 | $\left(\frac{\pi}{\tau_2}\right)^6\mathrm{E}_6$, $\mathcal{C}\left[\begin{smallmatrix}1&1&4\\1&1&4\end{smallmatrix}\right]$, $\mathcal{C}\left[\begin{smallmatrix}1&2&3\\1&2&3\end{smallmatrix}\right]$, $\mathcal{C}\left[\begin{smallmatrix}2&2&2\\2&2&2\end{smallmatrix}\right]$, $\mathcal{C}\left[\begin{smallmatrix}1&1&2&2\\1&1&2&2\end{smallmatrix}\right]$, $\mathcal{A}\left[\begin{smallmatrix}0&2&4\\5&0&1\end{smallmatrix}\right]$, $\mathcal{A}\left[\begin{smallmatrix}0&2&2&2\\3&0&1&2\end{smallmatrix}\right]$, $\mathcal{A}\left[\begin{smallmatrix}0&1&2&3\\2&1&3&0\end{smallmatrix}\right]$, $\left(\frac{\pi}{\tau_2}\right)^6\zeta_3^2$, $\left(\frac{\pi}{\tau_2}\right)^6\mathrm{E}_3^2$, $\left(\frac{\pi}{\tau_2}\right)^6\mathrm{E}_3\zeta_3$, $\left(\frac{\pi}{\tau_2}\right)^6\mathrm{E}_2\mathrm{E}_4$, $\left(\frac{\pi}{\tau_2}\right)^2\mathrm{E}_2\,\mathcal{C}\left[\begin{smallmatrix}1&1&2\\1&1&2\end{smallmatrix}\right]$, $\left(\frac{\pi}{\tau_2}\right)^6\mathrm{E}_2^3$, $\mathcal{C}\left[\begin{smallmatrix}4&0\\2&0\end{smallmatrix}\right]\mathcal{C}\left[\begin{smallmatrix}2&0\\4&0\end{smallmatrix}\right]$, $\mathcal{C}\left[\begin{smallmatrix}5&0\\1&0\end{smallmatrix}\right]\mathcal{C}\left[\begin{smallmatrix}1&0\\5&0\end{smallmatrix}\right]$, $\mathcal{C}\left[\begin{smallmatrix}3&0\\1&0\end{smallmatrix}\right]\mathcal{C}\left[\begin{smallmatrix}3&0\\5&0\end{smallmatrix}\right]$, $\mathcal{C}\left[\begin{smallmatrix}1&0\\3&0\end{smallmatrix}\right]\mathcal{C}\left[\begin{smallmatrix}5&0\\3&0\end{smallmatrix}\right]$, $\mathcal{C}\left[\begin{smallmatrix}3&0\\1&0\end{smallmatrix}\right]\mathcal{C}\left[\begin{smallmatrix}1&1&1\\1&1&3\end{smallmatrix}\right]$, $\mathcal{C}\left[\begin{smallmatrix}1&0\\3&0\end{smallmatrix}\right]\mathcal{C}\left[\begin{smallmatrix}1&1&3\\1&1&1\end{smallmatrix}\right]$, $\left(\frac{\pi}{\tau_2}\right)^2\mathrm{E}_2\,\mathcal{C}\left[\begin{smallmatrix}3&0\\1&0\end{smallmatrix}\right]\mathcal{C}\left[\begin{smallmatrix}1&0\\3&0\end{smallmatrix}\right]$ |
| $(7,5)$ | 18 | $\mathcal{C}\left[\begin{smallmatrix}7&0\\5&0\end{smallmatrix}\right]$, $\mathcal{C}\left[\begin{smallmatrix}0&1&6\\1&4&0\end{smallmatrix}\right]$, $\mathcal{C}\left[\begin{smallmatrix}0&1&6\\2&3&0\end{smallmatrix}\right]$, $\mathcal{C}\left[\begin{smallmatrix}0&2&5\\2&3&0\end{smallmatrix}\right]$, $\mathcal{C}\left[\begin{smallmatrix}0&3&4\\4&0&1\end{smallmatrix}\right]$, $\mathcal{C}\left[\begin{smallmatrix}1&1&2&3\\1&1&2&1\end{smallmatrix}\right]$, $\mathcal{C}\left[\begin{smallmatrix}1&2&2&2\\1&0&2&2\end{smallmatrix}\right]$, $\mathcal{C}\left[\begin{smallmatrix}0&1&2&4\\2&1&2&0\end{smallmatrix}\right]$, $\left(\frac{\pi}{\tau_2}\right)^3\mathrm{E}_3\,\mathcal{C}\left[\begin{smallmatrix}4&0\\2&0\end{smallmatrix}\right]$, $\left(\frac{\pi}{\tau_2}\right)^3\mathcal{C}\left[\begin{smallmatrix}4&0\\2&0\end{smallmatrix}\right]\zeta_3$, $\left(\frac{\pi}{\tau_2}\right)^4\mathrm{E}_4\,\mathcal{C}\left[\begin{smallmatrix}3&0\\1&0\end{smallmatrix}\right]$, $\left(\frac{\pi}{\tau_2}\right)^2\mathrm{E}_2\,\mathcal{C}\left[\begin{smallmatrix}5&0\\3&0\end{smallmatrix}\right]$, $\mathcal{C}\left[\begin{smallmatrix}3&0\\1&0\end{smallmatrix}\right]\mathcal{C}\left[\begin{smallmatrix}1&1&2\\1&1&2\end{smallmatrix}\right]$, $\left(\frac{\pi}{\tau_2}\right)^2\mathrm{E}_2\,\mathcal{C}\left[\begin{smallmatrix}1&1&3\\1&1&1\end{smallmatrix}\right]$, $\mathcal{C}\left[\begin{smallmatrix}3&0\\1&0\end{smallmatrix}\right]^2\mathcal{C}\left[\begin{smallmatrix}1&0\\3&0\end{smallmatrix}\right]$, $\mathcal{C}\left[\begin{smallmatrix}5&0\\1&0\end{smallmatrix}\right]\mathcal{C}\left[\begin{smallmatrix}2&0\\4&0\end{smallmatrix}\right]$, $\mathcal{C}\left[\begin{smallmatrix}1&0\\3&0\end{smallmatrix}\right]\mathcal{C}\left[\begin{smallmatrix}6&0\\2&0\end{smallmatrix}\right]$, $\left(\frac{\pi}{\tau_2}\right)^4\mathrm{E}_2^2\,\mathcal{C}\left[\begin{smallmatrix}3&0\\1&0\end{smallmatrix}\right]$ |
| $(8,4)$ | 14 | $\mathcal{C}\left[\begin{smallmatrix}8&0\\4&0\end{smallmatrix}\right]$, $\mathcal{C}\left[\begin{smallmatrix}0&2&6\\2&2&0\end{smallmatrix}\right]$, $\mathcal{C}\left[\begin{smallmatrix}0&3&5\\2&2&0\end{smallmatrix}\right]$, $\mathcal{C}\left[\begin{smallmatrix}0&4&4\\3&0&1\end{smallmatrix}\right]$, $\mathcal{C}\left[\begin{smallmatrix}1&2&2&3\\1&1&2&0\end{smallmatrix}\right]$, $\mathcal{C}\left[\begin{smallmatrix}1&2&2&3\\1&1&2&0\end{smallmatrix}\right]$ $\mathcal{C}\left[\begin{smallmatrix}4&0\\2&0\end{smallmatrix}\right]^2$, $\left(\frac{\pi}{\tau_2}\right)^3\mathrm{E}_3\,\mathcal{C}\left[\begin{smallmatrix}5&0\\1&0\end{smallmatrix}\right]$, $\left(\frac{\pi}{\tau_2}\right)^3\mathcal{C}\left[\begin{smallmatrix}5&0\\1&0\end{smallmatrix}\right]\zeta_3$, $\mathcal{C}\left[\begin{smallmatrix}3&0\\1&0\end{smallmatrix}\right]\mathcal{C}\left[\begin{smallmatrix}5&0\\3&0\end{smallmatrix}\right]$, $\left(\frac{\pi}{\tau_2}\right)^2\mathrm{E}_2\,\mathcal{C}\left[\begin{smallmatrix}6&0\\2&0\end{smallmatrix}\right]$, $\mathcal{C}\left[\begin{smallmatrix}3&0\\1&0\end{smallmatrix}\right]\mathcal{C}\left[\begin{smallmatrix}1&1&3\\1&1&1\end{smallmatrix}\right]$, $\left(\frac{\pi}{\tau_2}\right)^2\mathrm{E}_2\,\mathcal{C}\left[\begin{smallmatrix}3&0\\1&1\end{smallmatrix}\right]^2$, $\mathcal{C}\left[\begin{smallmatrix}3&0\\1&0\end{smallmatrix}\right]\mathcal{C}\left[\begin{smallmatrix}7&0\\1&0\end{smallmatrix}\right]$ |
| $(9,3)$ | 8 | $\mathcal{C}\left[\begin{smallmatrix}9&0\\3&0\end{smallmatrix}\right]$, $\mathcal{C}\left[\begin{smallmatrix}0&3&6\\1&2&0\end{smallmatrix}\right]$, $\mathcal{C}\left[\begin{smallmatrix}0&3&6\\2&1&0\end{smallmatrix}\right]$, $\mathcal{C}\left[\begin{smallmatrix}0&4&5\\2&1&0\end{smallmatrix}\right]$, $\mathcal{C}\left[\begin{smallmatrix}4&0\\2&0\end{smallmatrix}\right]\mathcal{C}\left[\begin{smallmatrix}5&0\\1&0\end{smallmatrix}\right]$, $\mathcal{C}\left[\begin{smallmatrix}3&0\\1&0\end{smallmatrix}\right]\mathcal{C}\left[\begin{smallmatrix}6&0\\2&0\end{smallmatrix}\right]$, $\left(\frac{\pi}{\tau_2}\right)^2\mathrm{E}_2\,\mathcal{C}\left[\begin{smallmatrix}7&0\\1&0\end{smallmatrix}\right]$, $\mathcal{C}\left[\begin{smallmatrix}3&0\\1&0\end{smallmatrix}\right]^2$ |
| $(10,2)$ | 4 | $\mathcal{C}\left[\begin{smallmatrix}10&0\\2&0\end{smallmatrix}\right]$, $\mathcal{C}\left[\begin{smallmatrix}0&4&6\\1&1&0\end{smallmatrix}\right]$, $\mathcal{C}\left[\begin{smallmatrix}5&0\\1&0\end{smallmatrix}\right]^2$, $\mathcal{C}\left[\begin{smallmatrix}3&0\\1&0\end{smallmatrix}\right]\mathcal{C}\left[\begin{smallmatrix}7&0\\1&0\end{smallmatrix}\right]$ |
| $(11,1)$ | 1 | $\mathcal{C}\left[\begin{smallmatrix}11&0\\1&0\end{smallmatrix}\right]$ |

Table 5.3: Basis elements used in the `Modular Graph Forms` package for (convergent) modular graph forms of weight $a + b \leq 12$, excluding holomorphic Eisenstein series. The second column gives the number of basis elements (including zeta values).

| weight | no. | basis elements |
|---|---|---|
| $(2,2)$ | 1 | $E_2$ |
| $(3,1)$ | 1 | $\nabla_0 E_2$ |
| $(3,3)$ | 2 | $E_3,\ \zeta_3$ |
| $(4,2)$ | 1 | $\nabla_0 E_3$ |
| $(5,1)$ | 1 | $\nabla_0^2 E_3$ |
| $(4,4)$ | 4 | $E_4,\ E_{2,2},\ E_2^2,\ \tau_2^{-2}\nabla_0 E_2 \overline{\nabla}_0 E_2$ |
| $(5,3)$ | 3 | $\nabla_0 E_4,\ \nabla_0 E_{2,2},\ E_2 \nabla_0 E_2$ |
| $(6,2)$ | 2 | $\nabla_0^2 E_4,\ (\nabla_0 E_2)^2$ |
| $(7,1)$ | 1 | $\nabla_0^3 E_4$ |
| $(5,5)$ | 9 | $E_5,\ E_{2,3},\ B_{2,3},\ B'_{2,3},\ \zeta_5,$<br>$E_2E_3,\ E_2\zeta_3,\ \tau_2^{-2}\overline{\nabla}_0 E_2 \nabla_0 E_3,\ \tau_2^{-2}\nabla_0 E_2 \overline{\nabla}_0 E_3$ |
| $(6,4)$ | 8 | $\nabla_0 E_5,\ \nabla_0 E_{2,3},\ \nabla_0 B_{2,3},\ \nabla_0 B'_{2,3},$<br>$\nabla_0 E_2 E_3,\ E_2 \nabla_0 E_3,\ \nabla_0 E_2 \zeta_3,\ \tau_2^{-2}\overline{\nabla}_0 E_2 \nabla_0^2 E_3$ |
| $(7,3)$ | 5 | $\nabla_0^2 E_5,\ \nabla_0^2 E_{2,3},\ \nabla_0^2 B'_{2,3},\ \nabla_0 E_2 \nabla_0 E_3,\ E_2 \nabla_0^2 E_3$ |
| $(8,2)$ | 3 | $\nabla_0^3 E_5,\ \nabla_0^3 B'_{2,3},\ \nabla_0 E_2 \nabla_0^2 E_3$ |
| $(9,1)$ | 1 | $\nabla_0^4 E_5$ |
| $(6,6)$ | 21 | $E_6,\ E_{2,4},\ E_{3,3},\ E'_{3,3},\ E_{2,2,2},\ B_{2,4},\ B'_{2,4},\ B_{2,2,2},\ \zeta_3^2,$<br>$E_3^2,\ E_3\zeta_3,\ E_2E_4,\ E_2E_{2,2},\ E_2^3,\ \tau_2^{-2}\nabla_0 E_3 \overline{\nabla}_0 E_3,\ \tau_2^{-4}\nabla_0^2 E_3 \overline{\nabla}_0^2 E_3$<br>$\tau_2^{-2}\nabla_0 E_2 \overline{\nabla}_0 E_4,\ \tau_2^{-2}\overline{\nabla}_0 E_2 \nabla_0 E_4,\ \tau_2^{-2}\nabla_0 E_2 \overline{\nabla}_0 E_{2,2},\ \tau_2^{-2}\overline{\nabla}_0 E_2 \nabla_0 E_{2,2},$<br>$\tau_2^{-2} E_2 \nabla_0 E_2 \overline{\nabla}_0 E_2$ |
| $(7,5)$ | 18 | $\nabla_0 E_6,\ \nabla_0 E_{2,4},\ \nabla_0 E_{3,3},\ \nabla_0 E'_{3,3},\ \nabla_0 E_{2,2,2},\ \nabla_0 B_{2,4},\ \nabla_0 B'_{2,4},\ \nabla_0 B_{2,2,2},$<br>$E_3 \nabla_0 E_3,\ \nabla_0 E_3 \zeta_3,\ \nabla_0 E_2 E_4,\ E_2 \nabla_0 E_4,\ \nabla_0 E_2 E_{2,2},\ E_2 \nabla_0 E_{2,2},\ E_2^2 \nabla_0 E_2,$<br>$\tau_2^{-2}\nabla_0^2 E_3 \overline{\nabla}_0 E_3,\ \tau_2^{-2}\overline{\nabla}_0 E_2 \nabla_0^2 E_4,\ \tau_2^{-2}(\nabla_0 E_2)^2 \overline{\nabla}_0 E_2$ |
| $(8,4)$ | 14 | $\nabla_0^2 E_6,\ \nabla_0^2 E_{2,4},\ \nabla_0^2 E'_{3,3},\ \nabla_0^2 B_{2,4},\ \nabla_0^2 B'_{2,4},\ \nabla_0^2 B_{2,2,2}$<br>$(\nabla_0 E_3)^2,\ E_3 \nabla_0^2 E_3,\ \nabla_0^2 E_3 \zeta_3,\ \nabla_0 E_2 \nabla_0 E_4,\ E_2 \nabla_0^2 E_4,\ \nabla_0 E_2 \nabla_0 E_{2,2},$<br>$E_2(\nabla_0 E_2)^2,\ \tau_2^{-2}\overline{\nabla}_0 E_2 \nabla_0^3 E_4$ |
| $(9,3)$ | 8 | $\nabla_0^3 E_6,\ \nabla_0^3 E'_{3,3},\ \nabla_0^3 B_{2,4},\ \nabla_0^3 B'_{2,4},\ \nabla_0 E_3 \nabla_0^2 E_3,\ \nabla_0 E_2 \nabla_0^2 E_4,$<br>$E_2 \nabla_0^3 E_4,\ (\nabla_0 E_2)^3$ |
| $(10,2)$ | 4 | $\nabla_0^4 E_6,\ \nabla_0^4 B'_{2,4},\ (\nabla_0^2 E_3)^2,\ \nabla_0 E_2 \nabla_0^3 E_4$ |
| $(11,1)$ | 1 | $\nabla_0^5 E_6$ |

Table 5.4: Basis of (convergent) modular graph forms of weight $a + b \leq 12$, excluding holomorphic Eisenstein series. The prefactors of $\tau_2$ were chosen such that the modular weight in the sector $(a+k, a-k)$ is $(0, -2k)$ for $0 \leq k < a$. The second column gives the number of basis elements (including zeta values).

appear. Similarly, the $(6,6)$ basis contains the cusp forms

$$\mathcal{C}\left[\begin{smallmatrix}3&0\\1&0\end{smallmatrix}\right]\mathcal{C}\left[\begin{smallmatrix}3&0\\5&0\end{smallmatrix}\right]-\mathcal{C}\left[\begin{smallmatrix}1&0\\3&0\end{smallmatrix}\right]\mathcal{C}\left[\begin{smallmatrix}5&0\\3&0\end{smallmatrix}\right] \tag{5.210a}$$

$$\mathcal{C}\left[\begin{smallmatrix}3&0\\1&0\end{smallmatrix}\right]\mathcal{C}\left[\begin{smallmatrix}1&1&1\\1&1&3\end{smallmatrix}\right]-\mathcal{C}\left[\begin{smallmatrix}1&0\\3&0\end{smallmatrix}\right]\mathcal{C}\left[\begin{smallmatrix}1&1&3\\1&1&1\end{smallmatrix}\right] \tag{5.210b}$$

$$\mathcal{A}\left[\begin{smallmatrix}0&2&4\\5&0&1\end{smallmatrix}\right] \tag{5.210c}$$

$$\mathcal{A}\left[\begin{smallmatrix}0&2&2&2\\3&0&1&2\end{smallmatrix}\right] \tag{5.210d}$$

$$\mathcal{A}\left[\begin{smallmatrix}0&1&2&3\\2&1&3&0\end{smallmatrix}\right]. \tag{5.210e}$$

The remaining basis elements in these sectors are real. Note that if we form antisymmetric combinations $\mathcal{A}\left[\begin{smallmatrix}A\\B\end{smallmatrix}\right]$ in the $(a,a)$ sectors with $a \leq 4$, these vanish since all basis elements are real. The cusp forms (5.209a) and (5.209b) were discussed in [185], whereas (5.209c) has higher loop order than the graphs studied in the reference. In the weight $(6,6)$ sector, the dimension of the space of two-loop imaginary cusp forms was found to be 2 in [185], in agreement with (5.210).

The basis of MGFs has an intricate structure which is closely related to the counting of iterated Eisenstein integrals, but this structure is not manifest in the basis given in Table 5.3. To make the relation to iterated Eisenstein integrals more transparent, we will use a second basis, summarized in Table 5.4. The basis has been multiplied by $\tau_2^{a+k}/\pi^a$ in the $(a+k, a-k)$ sector in Table 5.4 as compared to Table 5.3 for ease of notation. This means in particular that the basis elements given for the $a=b$ sectors are rendered modular invariant.

The structure of the basis in Table 5.4 is the following: In the modular invariant sectors, we split the irreducible basis elements into real and complex MGFs. The real ones are denoted by E, the complex ones by B, where the subscript refers to the holomorphic Eisenstein series appearing in the Cauchy–Riemann equations of the respective basis element. If several basis elements belong to the same sector w.r.t. these holomorphic Eisenstein series, we use a prime to distinguish them.

The non-holomorphic Eisenstein series $\mathrm{E}_k$ defined in (3.33) belong to the real basis elements. The remaining real basis elements were defined in [34] to streamline their Cauchy–Riemann equations and are written in terms of MGFs in (4.28). A subscript $k$ means in this notation that the holomorphic Eisenstein series $\mathrm{G}_{2k}$ appears in the Cauchy–Riemann equations, i.e. in the lowest Cauchy–Riemann derivative in which a holomorphic Eisenstein series appears. This determines the sector of iterated Eisenstein integrals that appear in the expansion of the basis element, as will be detailed in Section 8.4. E.g. the basis element $\mathrm{E}_{2,4}$ belongs to the $\mathrm{G}_4\mathrm{G}_8$ sector. The Cauchy–Riemann equations which make this manifest for the real irreducible basis elements are

$$\nabla_0^k \mathrm{E}_k = \frac{\tau_2^{2k}}{\pi^k}\frac{(2k-1)!}{(k-1)!}\mathrm{G}_{2k} \tag{5.211a}$$

$$\nabla_0^3 \mathrm{E}_{2,2} = -6\frac{\tau_2^4}{\pi^2}\mathrm{G}_4\nabla_0\mathrm{E}_2 \tag{5.211b}$$

$$\nabla_0^3 \mathrm{E}_{2,3} = -2\nabla_0 \mathrm{E}_2 \nabla_0^2 \mathrm{E}_3 - 4\frac{\tau_2^4}{\pi^2}\mathrm{G}_4 \nabla_0 \mathrm{E}_3 \tag{5.211c}$$

$$\nabla_0^5 \mathrm{E}_{3,3} = 180\frac{\tau_2^6}{\pi^3}\mathrm{G}_6 \nabla_0^2 \mathrm{E}_3 \tag{5.211d}$$

$$\nabla_0^4 \mathrm{E}'_{3,3} = -12\frac{\tau_2^6}{\pi^3}\mathrm{G}_6 \nabla_0 \mathrm{E}_3 \tag{5.211e}$$

$$\begin{aligned}\nabla_0^3 \mathrm{E}_{2,4} = &-\frac{27}{2}\nabla_0(\mathrm{E}_2\nabla_0^2\mathrm{E}_4) - \frac{27}{4}\nabla_0^3 \mathrm{B}_{2,4} - \frac{21}{40}\nabla_0^3 \mathrm{B}'_{2,4} \\ &- 27\frac{\tau_2^4}{\pi^2}\mathrm{G}_4\nabla_0\mathrm{E}_4\end{aligned} \tag{5.211f}$$

$$\nabla_0^3 \mathrm{E}_{2,2,2} = (\nabla_0 \mathrm{E}_2)^3 - 12\frac{\tau_2^4}{\pi^2}\mathrm{G}_4 \nabla_0 \mathrm{E}_{2,2}\,, \tag{5.211g}$$

where we use the Cauchy–Riemann operator defined in (3.55) and the complex basis elements $\mathrm{B}_{2,4}$ and $\mathrm{B}'_{2,4}$ are defined in (5.214). The right-hand sides in (5.211) all lie manifestly in the same sector of holomorphic Eisenstein series as indicated by the subscripts on the left-hand side. In [34], the real irreducible basis elements E were written in terms of the iterated Eisenstein integrals (4.15). From this, we can read off their Laurent polynomials [39, 40], namely

$$\mathrm{E}_k\big|_{q^0\bar{q}^0} = (-1)^{k-1}\frac{B_{2k}}{(2k)!}(4y)^k + 4\binom{2k-3}{k-1}\zeta_{2k-1}(4y)^{1-k} \tag{5.212a}$$

$$\mathrm{E}_{2,2}\big|_{q^0\bar{q}^0} = -\frac{y^4}{20250} + \frac{y\zeta_3}{45} + \frac{5\zeta_5}{12y} - \frac{\zeta_3^2}{4y^2} \tag{5.212b}$$

$$\mathrm{E}_{2,3}\big|_{q^0\bar{q}^0} = -\frac{4y^5}{297675} + \frac{2y^2\zeta_3}{945} - \frac{\zeta_5}{180} + \frac{7\zeta_7}{16y^2} - \frac{\zeta_3\zeta_5}{2y^3} \tag{5.212c}$$

$$\mathrm{E}_{3,3}\big|_{q^0\bar{q}^0} = \frac{2y^6}{6251175} + \frac{y\zeta_5}{210} + \frac{\zeta_7}{16y} - \frac{7\zeta_9}{64y^3} + \frac{9\zeta_5^2}{64y^4} \tag{5.212d}$$

$$\mathrm{E}'_{3,3}\big|_{q^0\bar{q}^0} = -\frac{y^6}{18753525} + \frac{y\zeta_5}{630} + \frac{3\zeta_7}{160y} - \frac{7\zeta_9}{480y^3} \tag{5.212e}$$

$$\mathrm{E}_{2,4}\big|_{q^0\bar{q}^0} = -\frac{y^6}{70875} + \frac{y^3\zeta_3}{525} + \frac{3\zeta_7}{40y} + \frac{25\zeta_9}{8y^3} - \frac{135\zeta_3\zeta_7}{32y^4} \tag{5.212f}$$

$$\mathrm{E}_{2,2,2}\big|_{q^0\bar{q}^0} = \frac{4y^6}{9568125} - \frac{2y^3\zeta_3}{10125} + \frac{y\zeta_5}{54} + \frac{\zeta_3^2}{90} + \frac{661\zeta_7}{1800y} - \frac{5\zeta_3\zeta_5}{12y^2} + \frac{\zeta_3^3}{6y^3}\,, \tag{5.212g}$$

where $y = \pi\tau_2$, and the Laurent polynomial of $\mathrm{E}_k$ can be read off from (3.34).

The cusp forms listed in (5.209) and (5.210) were all of the form $\mathcal{C}_\Gamma - \overline{\mathcal{C}_\Gamma}$ and hence purely imaginary. Using the Laurent polynomials (5.212), and the basis elements given in Table 5.4, it is easy to show that there are no real cusp forms in the space of MGFs at weight $(a, a)$ with $a \leq 5$ and that there are five real cusp forms at weight $(6,6)$ and transcendentality 6. A basis in this space of real cusp forms is given by

$$\begin{aligned} \mathrm{S}_1 = {} & \frac{8}{15}\mathrm{E}_{3,3} - 4\mathrm{E}'_{3,3} - \frac{1}{3}\mathrm{E}_2^3 + \mathrm{E}_4\mathrm{E}_2 + \frac{349}{875}\mathrm{E}_3^2 + \frac{2}{45}\zeta_3^2 + \frac{1}{3}\tau_2^{-2}\mathrm{E}_2\overline{\nabla}_0\mathrm{E}_2\nabla_0\mathrm{E}_2 \\ & - \frac{233}{1750}\tau_2^{-2}\overline{\nabla}_0\mathrm{E}_3\nabla_0\mathrm{E}_3 + \frac{1}{10500}\tau_2^{-4}\overline{\nabla}_0^2\mathrm{E}_3\nabla_0^2\mathrm{E}_3 \\ & - \frac{1}{6}\tau_2^{-2}\Big(\nabla_0\mathrm{E}_2\overline{\nabla}_0\mathrm{E}_4 + \overline{\nabla}_0\mathrm{E}_2\nabla_0\mathrm{E}_4\Big) \end{aligned} \tag{5.213a}$$

$$\begin{aligned} \mathrm{S}_2 = {} & \mathrm{E}_{2,4} + \frac{8748}{175}\mathrm{E}_{3,3} - \frac{5622}{35}\mathrm{E}'_{3,3} - \frac{269}{50}\mathrm{E}_3^2 - \frac{3739}{2100}\tau_2^{-2}\overline{\nabla}_0\mathrm{E}_3\nabla_0\mathrm{E}_3 \\ & + \frac{1}{840}\tau_2^{-4}\overline{\nabla}_0^2\mathrm{E}_3\nabla_0^2\mathrm{E}_3 + \frac{9}{8}\tau_2^{-2}\Big(\nabla_0\mathrm{E}_2\overline{\nabla}_0\mathrm{E}_4 + \overline{\nabla}_0\mathrm{E}_2\nabla_0\mathrm{E}_4\Big) \end{aligned} \tag{5.213b}$$

$$\begin{aligned} \mathrm{S}_3 = {} & \mathrm{E}_{2,2,2} + \frac{5288}{1125}\mathrm{E}_{3,3} - \frac{2644}{75}\mathrm{E}'_{3,3} + \mathrm{E}_2\mathrm{E}_{2,2} - \frac{1}{6}\mathrm{E}_2^3 + \frac{401}{17500}\mathrm{E}_3^2 \\ & + \frac{1}{4}\tau_2^{-2}\mathrm{E}_2\overline{\nabla}_0\mathrm{E}_2\nabla_0\mathrm{E}_2 - \frac{11801}{39375}\tau_2^{-2}\overline{\nabla}_0\mathrm{E}_3\nabla_0\mathrm{E}_3 \\ & + \frac{127}{630000}\tau_2^{-4}\overline{\nabla}_0^2\mathrm{E}_3\nabla_0^2\mathrm{E}_3 \end{aligned} \tag{5.213c}$$

$$\begin{aligned} \mathrm{S}_4 = {} & -2\mathrm{E}_2\mathrm{E}_{2,2} - \mathrm{E}_2^3 + \frac{8757}{1250}\mathrm{E}_3^2 + \frac{1}{5}\zeta_3^2 + \frac{3}{2}\tau_2^{-2}\mathrm{E}_2\overline{\nabla}_0\mathrm{E}_2\nabla_0\mathrm{E}_2 \\ & + \tau_2^{-2}\Big(\nabla_0\mathrm{E}_2\overline{\nabla}_0\mathrm{E}_{2,2} + \overline{\nabla}_0\mathrm{E}_2\nabla_0\mathrm{E}_{2,2}\Big) - \frac{3283}{1875}\tau_2^{-2}\overline{\nabla}_0\mathrm{E}_3\nabla_0\mathrm{E}_3 \\ & - \frac{7}{15000}\tau_2^{-4}\overline{\nabla}_0^2\mathrm{E}_3\nabla_0^2\mathrm{E}_3 \end{aligned} \tag{5.213d}$$

$$\begin{aligned} \mathrm{S}_5 = {} & -\frac{9}{5}\mathrm{E}_2\mathrm{E}_{2,2} - \frac{311}{350}\mathrm{E}_2^3 + \frac{26187}{12500}\mathrm{E}_3^2 + \mathrm{E}_3\zeta_3 + \frac{311}{2625}\zeta_3^2 \\ & + \frac{307}{700}\tau_2^{-2}\mathrm{E}_2\overline{\nabla}_0\mathrm{E}_2\nabla_0\mathrm{E}_2 - \frac{1638}{3125}\tau_2^{-2}\overline{\nabla}_0\mathrm{E}_3\nabla_0\mathrm{E}_3 \\ & + \frac{21}{50000}\tau_2^{-4}\overline{\nabla}_0^2\mathrm{E}_3\nabla_0^2\mathrm{E}_3\,. \end{aligned} \tag{5.213e}$$

The complex irreducible basis elements follow the same notation regarding the sectors of holomorphic Eisenstein series. They are defined in terms of lattice sums by

$$\mathrm{B}_{2,3} = \left(\frac{\tau_2}{\pi}\right)^5 \left(\mathcal{A}\left[\begin{smallmatrix} 0 & 1 & 2 & 2 \\ 1 & 1 & 0 & 3 \end{smallmatrix}\right] + \mathcal{C}\left[\begin{smallmatrix} 3 & 0 \\ 1 & 0 \end{smallmatrix}\right]\mathcal{C}\left[\begin{smallmatrix} 2 & 0 \\ 4 & 0 \end{smallmatrix}\right] - \mathcal{C}\left[\begin{smallmatrix} 1 & 0 \\ 3 & 0 \end{smallmatrix}\right]\mathcal{C}\left[\begin{smallmatrix} 4 & 0 \\ 2 & 0 \end{smallmatrix}\right]\right) \tag{5.214a}$$

$$\begin{aligned} \mathrm{B}'_{2,3} = {} & \left(\frac{\tau_2}{\pi}\right)^5 \left(\frac{1}{2}\mathcal{A}\left[\begin{smallmatrix} 0 & 2 & 3 \\ 3 & 0 & 2 \end{smallmatrix}\right] + \mathcal{A}\left[\begin{smallmatrix} 0 & 1 & 2 & 2 \\ 1 & 1 & 0 & 3 \end{smallmatrix}\right] + \mathcal{C}\left[\begin{smallmatrix} 3 & 0 \\ 1 & 0 \end{smallmatrix}\right]\mathcal{C}\left[\begin{smallmatrix} 2 & 0 \\ 4 & 0 \end{smallmatrix}\right] - \mathcal{C}\left[\begin{smallmatrix} 1 & 0 \\ 3 & 0 \end{smallmatrix}\right]\mathcal{C}\left[\begin{smallmatrix} 4 & 0 \\ 2 & 0 \end{smallmatrix}\right]\right) \\ & + \frac{129}{20}\mathrm{E}_5 - \frac{1}{2}\mathrm{E}_2\zeta_3 - \frac{21}{4}\mathrm{C}_{1,1,3} \end{aligned} \tag{5.214b}$$

$$\begin{aligned} \mathrm{B}_{2,4} = {} & \left(\frac{\tau_2}{\pi}\right)^6 \left(\mathcal{A}\left[\begin{smallmatrix} 0 & 2 & 4 \\ 5 & 0 & 1 \end{smallmatrix}\right] + 2\left(\mathcal{C}\left[\begin{smallmatrix} 3 & 0 \\ 1 & 0 \end{smallmatrix}\right]\mathcal{C}\left[\begin{smallmatrix} 3 & 0 \\ 5 & 0 \end{smallmatrix}\right] - \mathcal{C}\left[\begin{smallmatrix} 1 & 0 \\ 3 & 0 \end{smallmatrix}\right]\mathcal{C}\left[\begin{smallmatrix} 5 & 0 \\ 3 & 0 \end{smallmatrix}\right]\right)\right) \\ & + \mathrm{C}_{1,1,4} + \frac{1}{3}\mathrm{C}_{1,2,3} + \frac{1}{9}\mathrm{C}_{2,2,2} - \mathrm{E}_2\mathrm{E}_4 - \frac{13}{9}\mathrm{E}_6 \end{aligned} \tag{5.214c}$$

$$
\begin{aligned}
\mathrm{B}'_{2,4} &= \left(\frac{\tau_2}{\pi}\right)^6 \mathcal{A}\left[\begin{smallmatrix} 0 & 2 & 2 & 2 \\ 3 & 0 & 1 & 2 \end{smallmatrix}\right] - 30C_{1,1,4} - 10C_{1,2,3} - \frac{10}{3}C_{2,2,2} \\
&\quad - 3\mathrm{E}_3\zeta_3 + \frac{130}{3}\mathrm{E}_6
\end{aligned} \tag{5.214d}
$$

$$
\begin{aligned}
\mathrm{B}_{2,2,2} &= \left(\frac{\tau_2}{\pi}\right)^6 \Big( 4\mathcal{A}\left[\begin{smallmatrix} 0 & 1 & 2 & 3 \\ 2 & 1 & 3 & 0 \end{smallmatrix}\right] + \frac{121}{50}\mathcal{A}\left[\begin{smallmatrix} 0 & 2 & 2 & 2 \\ 3 & 0 & 1 & 2 \end{smallmatrix}\right] - \frac{113}{5}\mathcal{A}\left[\begin{smallmatrix} 0 & 2 & 4 \\ 5 & 0 & 1 \end{smallmatrix}\right] \\
&\qquad + \frac{266}{5}\left( \mathcal{C}\left[\begin{smallmatrix} 1 & 0 \\ 3 & 0 \end{smallmatrix}\right]\mathcal{C}\left[\begin{smallmatrix} 5 & 0 \\ 3 & 0 \end{smallmatrix}\right] - \mathcal{C}\left[\begin{smallmatrix} 3 & 0 \\ 1 & 0 \end{smallmatrix}\right]\mathcal{C}\left[\begin{smallmatrix} 3 & 0 \\ 5 & 0 \end{smallmatrix}\right] \right) \\
&\qquad + 4\left( \mathcal{C}\left[\begin{smallmatrix} 3 & 0 \\ 1 & 0 \end{smallmatrix}\right]\mathcal{C}\left[\begin{smallmatrix} 1 & 1 & 1 \\ 1 & 1 & 3 \end{smallmatrix}\right] - \mathcal{C}\left[\begin{smallmatrix} 1 & 0 \\ 3 & 0 \end{smallmatrix}\right]\mathcal{C}\left[\begin{smallmatrix} 1 & 1 & 3 \\ 1 & 1 & 1 \end{smallmatrix}\right] \right) \Big) \\
&\quad + 6C_{1,1,2}\mathrm{E}_2 - \frac{27}{5}\mathrm{E}_2\mathrm{E}_4 - \frac{63}{50}\mathrm{E}_3\zeta_3 \,,
\end{aligned} \tag{5.214e}
$$

where the real modular graph functions $C_{a,b,c}$ are defined in (3.126). Only the first of these basis elements is purely imaginary, the others contain imaginary and real contributions. The complex conjugates of the basis MGFs in (5.214) are

$$\overline{\mathrm{B}_{2,3}} = -\mathrm{B}_{2,3} \tag{5.215a}$$

$$\overline{\mathrm{B}'_{2,3}} = -\mathrm{B}'_{2,3} - \mathrm{E}_2\zeta_3 - \frac{21}{2}\mathrm{E}_{2,3} \tag{5.215b}$$

$$\overline{\mathrm{B}_{2,4}} = -\mathrm{B}_{2,4} - 2\mathrm{E}_2\mathrm{E}_4 + \frac{2}{9}\mathrm{E}_{2,4} \tag{5.215c}$$

$$\overline{\mathrm{B}'_{2,4}} = -\mathrm{B}'_{2,4} - 6\mathrm{E}_3\zeta_3 - \frac{20}{3}\mathrm{E}_{2,4} \tag{5.215d}$$

$$\overline{\mathrm{B}_{2,2,2}} = -\mathrm{B}_{2,2,2} - \frac{63}{25}\mathrm{E}_3\zeta_3 + 12\mathrm{E}_2\mathrm{E}_{2,2}\,. \tag{5.215e}$$

The definition of the basis elements E and B was guided by the maxim to delay the appearance of holomorphic Eisenstein series in the Cauchy–Riemann equations to higher derivatives and to separate the different sectors of holomorphic Eisenstein series at the same time. Although this does not fix the basis elements uniquely, the remaining freedom allows only for $\mathrm{B}_{2,3}$ for a complete splitting into real and imaginary basis elements. Similarly to (5.214), the first Cauchy–Riemann derivatives of the complex basis elements in which holomorphic Eisenstein series appear, are

$$
\begin{aligned}
\nabla_0^2\mathrm{B}_{2,3} &= \frac{2}{7}\nabla_0^2\mathrm{B}'_{2,3} + \frac{3}{2}\left(\nabla_0\mathrm{E}_2\nabla_0\mathrm{E}_3 - \mathrm{E}_2\nabla_0^2\mathrm{E}_3 + \nabla_0^2\mathrm{E}_{2,3}\right) \\
&\quad + \frac{\tau_2^4}{\pi^2}\mathrm{G}_4\left(9\mathrm{E}_3 + 3\zeta_3\right)
\end{aligned} \tag{5.216a}
$$

$$\nabla_0^4\mathrm{B}'_{2,3} = 1260\frac{\tau_2^6}{\pi^3}\mathrm{G}_6\nabla_0\mathrm{E}_2 \tag{5.216b}$$

$$\nabla_0^4\mathrm{B}_{2,4} = -\frac{7}{90}\nabla_0^4\mathrm{B}'_{2,4} - 1680\frac{\tau_2^8}{\pi^4}\mathrm{G}_8\mathrm{E}_2 \tag{5.216c}$$

$$\nabla_0^5\mathrm{B}'_{2,4} = 151200\frac{\tau_2^8}{\pi^4}\mathrm{G}_8\nabla_0\mathrm{E}_2 \tag{5.216d}$$

$$\nabla_0^3 \mathrm{B}_{2,2,2} = -9(\nabla_0 \mathrm{E}_2)^3 - \frac{\tau_2^4}{\pi^2} \mathrm{G}_4 \big(72 \mathrm{E}_2 \nabla_0 \mathrm{E}_2 + 36 \nabla_0 \mathrm{E}_{2,2}\big) \,. \tag{5.216e}$$

Since the complex basis elements are given in (5.214) in terms of real basis elements, for which the Laurent polynomials are listed in (5.212), and cusp forms with vanishing Laurent polynomials, we can assemble the Laurent polynomials of the B as well. They are given by

$$\mathrm{B}_{2,3}\big|_{q^0\bar{q}^0} = 0 \tag{5.217a}$$

$$\mathrm{B}'_{2,3}\big|_{q^0\bar{q}^0} = \frac{y^5}{14175} - \frac{y^2\zeta_3}{45} + \frac{7\zeta_5}{240} - \frac{\zeta_3^2}{2y} - \frac{147\zeta_7}{64y^2} + \frac{21\zeta_3\zeta_5}{8y^3} \tag{5.217b}$$

$$\mathrm{B}_{2,4}\big|_{q^0\bar{q}^0} = -\frac{4y^6}{637875} - \frac{\zeta_7}{180y} + \frac{25\zeta_9}{72y^3} - \frac{35\zeta_3\zeta_7}{32y^4} \tag{5.217c}$$

$$\mathrm{B}'_{2,4}\big|_{q^0\bar{q}^0} = \frac{2y^6}{42525} - \frac{4y^3\zeta_3}{315} - \frac{\zeta_7}{4y} - \frac{9\zeta_3\zeta_5}{4y^2} - \frac{125\zeta_9}{12y^3} + \frac{225\zeta_3\zeta_7}{16y^4} \tag{5.217d}$$

$$\mathrm{B}_{2,2,2}\big|_{q^0\bar{q}^0} = -\frac{y^6}{151875} + \frac{y\zeta_5}{18} + \frac{\zeta_3^2}{10} + \frac{311\zeta_3\zeta_5}{200y^2} - \frac{3\zeta_3^3}{2y^3} \,. \tag{5.217e}$$

The basis elements E and B span the irreducible sectors of the modular invariant subspaces of MGFs. For the subspaces with modular weight $(a, b)$ with $a > b$, we take the Cauchy–Riemann derivatives of the E and B as irreducible basis elements. Since the space of MGFs of weight $(a + k, a - k)$ shrinks with growing $k$, there are relations between the Cauchy–Riemann derivatives of the E and B, leading to dropouts in this pattern. In general, these dropouts are manifest in the Cauchy–Riemann equations (5.211) and (5.216), however some of the real basis elements satisfy relations at derivatives lower than the one in which the first holomorphic Eisenstein series appear as stated in (5.211). These additional relations are

$$\nabla_0^2 \mathrm{E}_{2,2} = -\frac{1}{2}(\nabla_0 \mathrm{E}_2)^2 \tag{5.218a}$$

$$\nabla_0^2 \mathrm{E}_{3,3} = \frac{3}{4}(\nabla_0 \mathrm{E}_3)^2 + \frac{15}{2}\nabla_0^2 \mathrm{E}'_{3,3} \tag{5.218b}$$

$$\nabla_0^2 \mathrm{E}_{2,2,2} = -2\nabla_0 \mathrm{E}_2 \nabla_0 \mathrm{E}_{2,2} \,. \tag{5.218c}$$

For the complex basis elements, there are no relations at lower derivatives than in (5.216).

On top of the irreducible basis elements E and B, there are reducible basis elements which are products of irreducible basis elements of lower weights. We also take derivatives of these reducible basis elements to generate the bases of weight $(a, b)$ with $a > b$. Again, this is constrained by the relations (5.211), (5.216) and (5.218). As for the irreducible basis elements, the Cauchy–Riemann derivatives of the reducible basis elements also contain terms with holomorphic Eisenstein series, which are not written in the basis. Furthermore, the derivative of terms of the form $\overline{\nabla}_0^n \mathrm{E}_k$ is (up to prefactors) $\overline{\nabla}_0^{n-1}\mathrm{E}_k$. The derivative of the only depth-two instance $\overline{\nabla}_0\mathrm{E}_{2,2}$ gives rise to $2\mathrm{E}_{2,2} - \mathrm{E}_2^2$, as follows from the Laplace equation $(\Delta - 2)\mathrm{E}_{2,2} = -\mathrm{E}_2^2$ [39].

Since the action of the derivative operators $\nabla_0$ and $\overline{\nabla}_0$ on $y$ is straightforwardly given by

$$\nabla_0 y = \overline{\nabla}_0 y = \frac{y^2}{\pi}\,, \tag{5.219}$$

using the decompositions into the basis of Table 5.4 and the known Laurent polynomials (5.212) and (5.217), we can easily assemble the Laurent polynomials of all dihedral and trihedral MGFs of total weight $a + b \leq 12$. These computations are made straightforward in the `Modular Graph Forms` package as outlined in the following.

Computations in the `Modular Graph Forms` package are performed in the basis listed in Table 5.3. Using the function **`CConvertToNablaE`**, an expression can be converted into the basis given in Table 5.4. The real basis elements are represented by e.g. **`e[2,2]`**, and **`ep[3,3]`** for the primed version. The complex basis elements are given by e.g. **`b[2,3]`** and **`bp[2,3]`**. The Cauchy–Riemann derivatives are denoted by the functions **`nablaE`**, **`nablaEp`**, **`nablaB`** and **`nablaBp`**. Their complex conjugates are **`nablaBarE`**, **`nablaBarEp`**, **`nablaBarBBar`** and **`nablaBarBpBar`**. The first arguments of these functions is always the order of the derivative, the second is a list with the subscripts of the basis element, e.g. $\overline{\nabla}_0^2\overline{\mathrm{B}}_{2,4}$ is denoted by **`nablaBarBBar[2,{2,4}]`**. These basis elements are translated back into the basis given in Table 5.3 by the function **`CConvertFromNablaE`**. Note that only the derivatives appearing in Table 5.4 can be converted in this way. As an example, the decomposition of the graph $C\left[\begin{smallmatrix}1 & 2 & 4\\ 2 & 2 & 1\end{smallmatrix}\right]$ can be performed by

In[49]:= $\texttt{CConvertToNablaE}\left[\texttt{CSimplify}\left[\texttt{c}\left[\begin{smallmatrix}1 & 2 & 4\\ 2 & 2 & 1\end{smallmatrix}\right]\right]\right]$

Out[49]= $\dfrac{3\,\pi^6\,\nabla\mathrm{E}_6}{28\,\tau_2^7} - \dfrac{5\,\pi^6\,\nabla\mathrm{E}_{3,3}}{9\,\tau_2^7} + \dfrac{5\,\pi^6\,\nabla\mathrm{E}'_{3,3}}{3\,\tau_2^7}\,.$

The derivative operator $\nabla_0$ is not implemented directly, but since it is given by $\nabla_0 = \tau_2\nabla^{(0)}$ (cf. (3.55)), it can be obtained by acting with **`tau[2]CHolCR`** on an MGF with vanishing modular weight. E.g. the Cauchy–Riemann equation (5.216c) is reproduced by

In[50]:= $\texttt{CConvertToNablaE}\left[\texttt{Nest}\left[\texttt{CSimplify}\left[\texttt{tau[2]CHolCR[\#]}\right]\texttt{\&,b[2,4],4}\right]\right]$

Out[50]= $-\dfrac{7}{90}\nabla^4\mathrm{B}'_{2,4} - \dfrac{1680\,\mathrm{E}_2\,\mathrm{G}_8\,\tau_2^8}{\pi^4}\,.$

The Laurent polynomials (5.212) and (5.217) are implemented in the function **`CLaurentPoly`**, which replaces each of the basis elements by its Laurent polynomial and performs the necessary Cauchy–Riemann derivatives. E.g. the Laurent polynomial of the graph $C\left[\begin{smallmatrix}1 & 2 & 4\\ 2 & 2 & 1\end{smallmatrix}\right]$ decomposed in Out[49] can be obtained via

```
In[51]:= CLaurentPoly[Out[49]]
```

Out[51]= $-\frac{19\,\pi^{12}}{91216125}+\frac{5\,\pi^{12}\,\zeta_5^2}{16\,y^{10}}+\frac{\pi^{12}\,\zeta_7}{288\,y^7}-\frac{7\,\pi^{12}\,\zeta_9}{64\,y^9}-\frac{135\,\pi^{12}\,\zeta_{11}}{512\,y^{11}}$ .

The basis elements at a certain weight are accessible via the function **CBasis**. If the option **basis** is set to the string "C" (the default value), the basis from Table 5.3 is returned, if it is set to the string "nablaE", the basis from Table 5.4 is returned, e.g.

```
In[52]:= CBasis[3, 5]

         CBasis[3, 5, basis → "nablaE"]
```

Out[52]= $\left\{\mathrm{C}\left[\begin{smallmatrix}1&1&1\\1&1&3\end{smallmatrix}\right],\mathrm{C}\left[\begin{smallmatrix}3&0\\5&0\end{smallmatrix}\right],\frac{\pi^2\,\mathrm{C}\left[\begin{smallmatrix}1&0\\3&0\end{smallmatrix}\right]\mathrm{E}_2}{\tau_2^2}\right\}$

Out[53]= $\left\{\bar{\nabla}\mathrm{E}_{2,2},\bar{\nabla}\mathrm{E}_4,\mathrm{E}_2\,\bar{\nabla}\mathrm{E}_2\right\}$ .

Together with the function **zIntegrate** described in Section 5.1.2, the basis decompositions available in the `Modular Graph Forms` package are sufficient to expand all two- and three-point Koba–Nielsen integrals to the orders which give rise to MGFs of total modular weight at most 12. This will be crucial for checking and solving the differential equation of the generating series of Koba–Nielsen integrals, discussed in Chapters 7 and 8 and in evaluating the one-loop heterotic amplitudes investigated in the next chapter. The arXiv submission of [II] includes the expansion of the two- and three-point version of (8.17) up to order 12 in the basis of Table 5.4 and for the three-point version also the Laurent polynomial of the generating series. At two-point, it was checked that the Laurent polynomial obtained using the basis decompositions agrees with the closed formula (8.66) derived from genus-zero integrals.

# 6

# HETEROTIC AMPLITUDES

In this chapter, we will apply the techniques for manipulations of MGFs derived in the previous chapter to the calculation of the low-energy expansion of four-gluon scattering in SO(32) heterotic string theory at genus one. Compared to the type-II string, the heterotic string carries less supersymmetry and is therefore less constraining. This is reflected in worldsheet integrands with non-trivial modular transformations,[1] giving rise to MGFs, as opposed to the modular graph functions which appeared in Section 3.3. We will also argue that any $n$-point heterotic amplitude of gauge-bosons and gravitons can be expanded in MGFs. The same is true for $n$-graviton amplitudes in type-II, as we saw in Section 3.2.3.

The simplifications for MGFs from the last chapter will allow us to perform the integral over the punctures and to express the amplitude in terms of iterated Eisenstein integrals, making a comparison of the SO(32) heterotic amplitude to the prediction from the single-valued prescription as discussed in Section 4.3.2 possible. To do this comparison, we will extend the mapping between integration cycles of the open string and Jacobi forms of the punctures in the closed string to non-abelian external states. In this way, we can identify many terms in the weight $(2,0)$ contribution, however, some terms are not reproduced by the esv map from Section 4.3.2, motivating a different approach in the next chapter.

As another consequence of the half-maximal supersymmetry of the heterotic string, the coefficients of a given MGF in massless one-loop[2] amplitudes usually mix different orders in $\alpha'$. In superstring amplitudes in turn, the order in the $\alpha'$ expansion correlates with the transcendental weights of the accompanying iterated integrals – (elliptic) MZVs or MGFs. This property known as *uniform transcendentality* can also be found in the context of some dimensionally regularized Feynman integrals with the regularization parameter $\varepsilon$ taking the rôle of $\alpha'$ [213–217]. We decompose the four-point gauge amplitude of the heterotic

1 The overall modular invariance of the integrand is restored by a $z$-independent modular form which cancels the modular weight of the $z$-dependent contributions.

2 See [100, 212] for the analogous phenomenon in tree-level amplitudes of the heterotic string, where the coefficients of a given multiple zeta value are usually geometric series in $\alpha'$.

string into integrals that we conjecture to be individually uniformly transcendental – both in the single-trace and the double-trace sector. In [IV], we show that these integrals are indeed uniformly transcendental if their asymptotic value at the cusp satisfies this property, cf. Section 8.5.3. The non-uniform transcendentality of the overall amplitude is then reflected by the coefficients of these basis integrals. The classification of uniformly transcendental moduli-space integrals is expected to give important clues about the mathematical properties of the underlying twisted cohomologies [126].

The results presented in this chapter were published in [III] and the present text has extensive overlap with the reference.

This chapter is structured as follows: Section 6.1 discusses the structure of the four-gluon genus-one amplitude in the heterotic string and explains how to rewrite the integral over the CFT correlator into a linear combination of Koba–Nielsen integrals. In Section 6.2 these Koba–Nielsen integrals are expanded in MGFs using the techniques discussed in Chapter 5. Furthermore, in Section 6.2.3 we will write the Koba–Nielsen integrals in terms of building blocks of uniform transcendentality (with some lengthy integration-by-parts manipulations moved to Appendix C) and in Section 6.2.4 we will perform the final integral over $\tau$ to the order $\alpha'^2$, thereby obtaining the complete analytic part of the amplitude to that order. In Section 6.3 we will use the elliptic single-valued prescription from [34], which was briefly discussed in Section 4.3, to reproduce the modular invariant building block (3.109) of the amplitude (known from type-IIB four-graviton scattering) from open strings. Then, we will extend this prescription to include a contribution of non-vanishing modular weight and describe the limitations of the prescription from [34] in this context. This extension provides an important hint towards the more general proposal of an elliptic single-valued map in [17].

## 6.1 CFT CORRELATORS FOR ONE-LOOP GAUGE AMPLITUDES

In this section, we will calculate the CFT correlator for four gluon scatting at genus one in the heterotic string and express in a way suitable for an expansion in terms of MGFs.

Up to an overall normalization factor, the prescription for the four-point function reads [218–221]

$$\mathcal{M}_4 = \int_{\mathcal{F}} \frac{d^2\tau}{\tau_2^5} \frac{1}{\eta^{24}(\tau)} \int_{\Sigma^3} d\mu_3 \langle \prod_{j=1}^{4} \mathcal{V}^{a_j}(z_j, \epsilon_j, k_j) \rangle^{\tau} , \qquad (6.1)$$

cf. (3.1) and (3.48). The inverse factors of the Dedekind eta function $\eta(\tau)$, defined in (3.20), arise as the partition function of the 26 worldsheet bosons in the non-supersymmetric sector, and $\mathcal{V}^a(z,\epsilon,k)$ denotes the vertex operator for an external gauge boson [218]

$$\mathcal{V}^a(z,\epsilon,k) = J^a(z) V_{\rm SUSY}(\bar z,\epsilon,k) e^{ik\cdot X(z,\bar z)}\,, \tag{6.2}$$

with polarization vector $\epsilon$, lightlike momentum $k$ and adjoint index $a$, as mentioned in (3.73). The correlation function $\langle \ldots \rangle^\tau$ in (6.1) is evaluated on a torus of modular parameter $\tau$ and allows factoring out the contribution from the Kac–Moody currents $J^a(z)$. The leftover correlator involving $V_{\rm SUSY}(\bar z,\epsilon,k)$ and $e^{ik\cdot X(z,\bar z)}$ matches a chiral half of type-II superstrings, and its four-point instance is completely determined by maximal supersymmetry [1] and was given in (3.74) in terms of a color-ordered tree-level amplitude $A^{\rm tree}_{\rm SYM}(1,2,3,4)$ of ten-dimensional super-Yang–Mills and the Koba–Nielsen factor. The manifestly supersymmetric calculation in the pure-spinor formalism [63] leads to the same conclusion for any combination of gauge bosons and gauginos.

### 6.1.1 *Structure of the Kac–Moody correlators*

We shall now focus on the correlation function of the Kac–Moody currents $J^a$ in (6.2), that carries all the dependence on the adjoint indices $a_1,\ldots,a_4$ of the external gauge bosons. In a fermionic representation $J^a(z) = t^a_{ij}\psi^i\psi^j(z)$ of the currents, the correlators receive contributions from different spin structures – the boundary conditions for the worldsheet spinors under $\psi^j(z+1) = \pm\psi^j(z)$ and $\psi^j(z+\tau) = \pm\psi^j(z)$. We will be mostly interested in the gauge group $\mathrm{Spin}(32)/\mathbb{Z}_2$ with Lie-algebra generators $t^a_{ij}$ and fundamental indices $i,j = 1,2,\ldots,32$.

For four-point functions, the only contributions come from the even spin structures that we label with an integer $\nu = 2,3,4$, and the corresponding fermionic two-point function or *Szegő kernel* [222] can be brought into the universal form

$$S_\nu(z,\tau) = \frac{\theta_1'(0,\tau)\theta_\nu(z,\tau)}{\theta_\nu(0,\tau)\theta_1(z,\tau)}\,, \tag{6.3}$$

where the Jacobi theta functions were defined in (3.63) and (3.64). In the fermionic realization of Kac–Moody currents, the contribution of a given spin structure to the correlation function reduces to (sums of products of) Szegő kernels (6.3), e.g. [174]

$$\langle J^{a_1}(z_1)J^{a_2}(z_2)\rangle^\tau_\nu = \mathrm{Tr}(t^{a_1}t^{a_2})S_\nu(z_{12})S_\nu(z_{21}) \tag{6.4}$$

$$\langle J^{a_1}(z_1)J^{a_2}(z_2)J^{a_3}(z_3)\rangle^\tau_\nu = \overleftrightarrow{\mathrm{Tr}}(t^{a_1}t^{a_2}t^{a_3})S_\nu(z_{12})S_\nu(z_{23})S_\nu(z_{31}) \tag{6.5}$$

and

$$\begin{aligned}
&\langle J^{a_1}(z_1)J^{a_2}(z_2)J^{a_3}(z_3)J^{a_4}(z_4)\rangle_\nu^\tau \\
&= \overset{\leftrightarrow}{\mathrm{Tr}}(t^{a_1}t^{a_2}t^{a_3}t^{a_4})S_\nu(z_{12})S_\nu(z_{23})S_\nu(z_{34})S_\nu(z_{41}) \\
&\quad + \mathrm{Tr}(t^{a_1}t^{a_2})\,\mathrm{Tr}(t^{a_3}t^{a_4})S_\nu(z_{12})S_\nu(z_{21})S_\nu(z_{34})S_\nu(z_{43}) + \mathrm{cyc}(2,3,4)\,,
\end{aligned} \tag{6.6}$$

where we use the following shorthand for parity-weighted traces relevant for $n \geq 3$ currents,

$$\overset{\leftrightarrow}{\mathrm{Tr}}(t^{a_1}t^{a_2}\cdots t^{a_n}) = \mathrm{Tr}(t^{a_1}t^{a_2}\cdots t^{a_n}) + (-1)^n \mathrm{Tr}(t^{a_n}\cdots t^{a_2}t^{a_1})\,. \tag{6.7}$$

The sum over cyclic permutations refers to both lines of (6.6), and it acts on both the adjoint indices $a_2, a_3, a_4$ and the punctures $z_2, z_3, z_4$. Furthermore, each of the spin-structure dependent current correlators is weighted by the fermionic partition function of the $\mathrm{Spin}(32)/\mathbb{Z}_2$ model,

$$Z_\nu^{\mathrm{het}}(\tau) := 2\zeta_4^2\theta_\nu^{16}(0,\tau)\,. \tag{6.8}$$

Since we do not track the overall normalization of the amplitude in (6.1), the prefactor $2\zeta_4^2$ is introduced along the way for later convenience. The end results for the current correlators in (6.1) are proportional to the spin-summed expressions,

$$\langle J^{a_1}(z_1)J^{a_2}(z_2)\cdots J^{a_n}(z_n)\rangle^\tau = \sum_{\nu=2}^{4} Z_\nu^{\mathrm{het}}(\tau)\langle J^{a_1}(z_1)J^{a_2}(z_2)\cdots J^{a_n}(z_n)\rangle_\nu^\tau\,, \tag{6.9}$$

and we will next construct convenient representations of (6.9) from the elliptic functions of Section 3.2.3.

### 6.1.2 *Spin-summed current correlators*

By the form of the spin-structure dependent correlators (6.4) and (6.6), we will decompose the spin sums (6.9) according to the traces of Lie-algebra generators. For $n$ gauge currents we get in general both single- and multi-trace contributions

$$\begin{aligned}
\mathcal{H}_{12\ldots n}(\tau) &= \langle J^{a_1}(z_1)J^{a_2}(z_2)\cdots J^{a_n}(z_n)\rangle^\tau \big|_{\mathrm{Tr}(t^{a_1}t^{a_2}\cdots t^{a_{n-1}}t^{a_n})} \\
&= \sum_{\nu=2}^{4} Z_\nu^{\mathrm{het}}(\tau)S_\nu(z_{12},\tau)S_\nu(z_{23},\tau)\cdots S_\nu(z_{n1},\tau)
\end{aligned} \tag{6.10}$$

$$\begin{aligned}
\mathcal{H}_{12\ldots p|p+1\ldots n}(\tau) &= \langle J^{a_1}(z_1)J^{a_2}(z_2)\cdots J^{a_n}(z_n)\rangle^\tau \big|_{\mathrm{Tr}(t^{a_1}t^{a_2}\cdots t^{a_p})\mathrm{Tr}(t^{a_{p+1}}\cdots t^{a_n})} \\
&= \sum_{\nu=2}^{4} Z_\nu^{\mathrm{het}}(\tau)S_\nu(z_{12},\tau)S_\nu(z_{23},\tau)\cdots S_\nu(z_{p1},\tau) \\
&\quad \times S_\nu(z_{p+1,p+2},\tau)S_\nu(z_{p+2,p+3},\tau)\cdots S_\nu(z_{n,p+1},\tau)\,.
\end{aligned} \tag{6.11}$$

The dependence of the Szegő kernels (6.3) on the spin structure $\nu$ can be simplified by relating them to Kronecker–Eisenstein series (3.79) with one of the half-periods

$$\omega_2 = \frac{1}{2}\,, \qquad \omega_3 = -\frac{1+\tau}{2}\,, \qquad \omega_4 = \frac{\tau}{2} \tag{6.12}$$

in the second argument $\eta$. Given that the $\theta_{\nu=1,2,3,4}$ functions can be mapped into each other by a half-period shift in the first argument,

$$\theta_2(z+\tfrac{1}{2},\tau) = -\theta_1(z,\tau) \tag{6.13a}$$

$$\theta_4(z+\tfrac{\tau}{2},\tau) = ie^{-i\pi z}q^{-1/8}\theta_1(z,\tau) \tag{6.13b}$$

$$\theta_4(z+\tfrac{1}{2},\tau) = \theta_3(z,\tau) \tag{6.13c}$$

$$\theta_3(z+\tfrac{\tau}{2},\tau) = e^{-i\pi z}q^{-1/8}\theta_2(z,\tau)\,, \tag{6.13d}$$

we have $S_\nu(z_{ij},\tau) \sim F(z_{ij},\omega_\nu,\tau)$ up to phase factors that drop out from the products of Szegő kernels in the spin sums (6.10) and (6.11) [223]

$$\begin{aligned} &S_\nu(z_{12},\tau)S_\nu(z_{23},\tau)\ldots S_\nu(z_{n1},\tau) \\ &= F(z_{12},\omega_\nu,\tau)F(z_{23},\omega_\nu,\tau)\ldots F(z_{n1},\omega_\nu,\tau)\,. \end{aligned} \tag{6.14}$$

This naturally introduces the elliptic functions $V_a$ generated by the cycles of Kronecker–Eisenstein series in (3.96). Given that the right-hand side of (6.14) defines an elliptic function of $\omega_\nu$, all the $\nu$-dependence can be absorbed into Weierstraß functions (3.5) of half-periods,

$$e_\nu(\tau) = \wp(\omega_\nu,\tau)\,. \tag{6.15}$$

The vanishing of $\partial_z\wp(z,\tau)$ at $z=\omega_\nu$ and the differential equation $\partial_z^2\wp(z,\tau) = 6(\wp(z,\tau))^2 - 30\mathrm{G}_4$ then lead to a polynomial appearance of the Weierstraß functions which carries the $\nu$-dependence [223, 224]

$$\begin{aligned} S_\nu(z_{12})S_\nu(z_{21}) &= V_2(1,2) + e_\nu \\ S_\nu(z_{12})S_\nu(z_{23})S_\nu(z_{31}) &= V_3(1,2,3) + e_\nu V_1(1,2,3) \\ S_\nu(z_{12})S_\nu(z_{23})S_\nu(z_{34})S_\nu(z_{41}) &= V_4(1,2,3,4) + e_\nu V_2(1,2,3,4) + e_\nu^2 - 6\mathrm{G}_4\,. \end{aligned} \tag{6.16}$$

Since the Weierstraß functions are furthermore related by $e_\nu^3 - 15\mathrm{G}_4 e_\nu - 35\mathrm{G}_6 = 0$, the spin sums of the $n$-point current correlators in (6.10) and (6.11) can be reduced to the three inequivalent cases[3] [220]

$$\sum_{\nu=2}^{4} Z_\nu^{\text{het}} = \mathrm{G}_4^2\,, \quad \sum_{\nu=2}^{4} Z_\nu^{\text{het}} e_\nu = -\frac{7}{2}\mathrm{G}_4\mathrm{G}_6\,, \quad \sum_{\nu=2}^{4} Z_\nu^{\text{het}} e_\nu^2 = \frac{49}{6}\mathrm{G}_6^2 + \frac{5}{3}\mathrm{G}_4^3\,. \tag{6.17}$$

3 These identities are a consequence of the relations $e_2+e_3+e_4 = 0$, $e_2e_3+e_3e_4+e_4e_2 = -15\mathrm{G}_4$ and $e_2e_3e_4 = 35\mathrm{G}_6$ among the Weierstraß invariants as well as the connection with theta functions via $\pi^2(\theta_4(0,\tau))^4 = e_2 - e_3$, $\pi^2(\theta_2(0,\tau))^4 = e_3 - e_4$ and $\pi^2(\theta_3(0,\tau))^4 = e_2 - e_4$.

As a bottom line, (6.16) and (6.17) lead us to the following representations

$$\mathcal{H}_{12} = \mathrm{G}_4^2 V_2(1,2) - \frac{7}{2}\mathrm{G}_4\mathrm{G}_6 \tag{6.18}$$

$$\mathcal{H}_{123} = \mathrm{G}_4^2 V_3(1,2,3) - \frac{7}{2}\mathrm{G}_4\mathrm{G}_6 V_1(1,2,3) \tag{6.19}$$

$$\mathcal{H}_{1234} = \mathrm{G}_4^2 V_4(1,2,3,4) - \frac{7}{2}\mathrm{G}_4\mathrm{G}_6 V_2(1,2,3,4) - \frac{13}{3}\mathrm{G}_4^3 + \frac{49}{6}\mathrm{G}_6^2 \tag{6.20}$$

$$\begin{aligned}\mathcal{H}_{12|34} &= \mathrm{G}_4^2 V_2(1,2)V_2(3,4) - \frac{7}{2}\mathrm{G}_4\mathrm{G}_6\big[V_2(1,2)+V_2(3,4)\big]\\ &\quad + \frac{5}{3}\mathrm{G}_4^3 + \frac{49}{6}\mathrm{G}_6^2\end{aligned} \tag{6.21}$$

for the spin sums in (6.10) and (6.11) which enter the trace decomposition of the four-point correlator,

$$\begin{aligned}&\langle J^{a_1}(z_1)J^{a_2}(z_2)J^{a_3}(z_3)J^{a_4}(z_4)\rangle^\tau\\ &\overset{\leftrightarrow}{=} \mathrm{Tr}(t^{a_1}t^{a_2}t^{a_3}t^{a_4})\mathcal{H}_{1234} + \mathrm{Tr}(t^{a_1}t^{a_2})\mathrm{Tr}(t^{a_3}t^{a_4})\mathcal{H}_{12|34} + \mathrm{cyc}(2,3,4)\,.\end{aligned} \tag{6.22}$$

As before, the sum over cyclic permutations refers to both terms in (6.22).

### 6.1.3 *The key integrals over torus punctures*

We shall now pinpoint the integrals over torus punctures that need to be performed in the four-point gauge amplitude (6.1) and whose low-energy expansion will be the main topic of the later sections. After factoring out the polarization dependent parts (3.74),[4]

$$\mathcal{M}_4 = (k_1\cdot k_2)(k_2\cdot k_3)A^{\mathrm{tree}}_{\mathrm{SYM}}(1,2,3,4)\int_{\mathcal{F}}\frac{\mathrm{d}^2\tau}{\tau_2^2\,\eta^{24}(\tau)}\,M_4(\tau)\,, \tag{6.23}$$

we will be interested in the following integral over the punctures,

$$M_4(\tau) = \int \mathrm{d}\mu_3\,\langle J^{a_1}(0)J^{a_2}(z_2)J^{a_3}(z_3)J^{a_4}(z_4)\rangle^\tau\ \mathrm{KN}_4\,. \tag{6.24}$$

For four massless particles, constraints (2.26) on the Mandelstam invariants $s_{ij}$ become

$$s_{34} = s_{12}\,,\qquad s_{14} = s_{23}\,,\qquad s_{13} = s_{24} = -s_{12} - s_{23}\,. \tag{6.25}$$

---

4 In order to reproduce the normalization conventions for the four-point gauge amplitude in [220], the right-hand side of (6.23) needs to be multiplied by $\frac{1}{(2\zeta_4)^2}\big(\frac{\alpha' g}{64\pi^5}\big)^2$, where $g$ denotes the gauge coupling. The inverse factor of $2\zeta_4^2$ compensates for our choice of normalization of the partition function in (6.8).

With the result (6.22) for the current correlators in terms of the spin sums (6.20) and (6.21), the right-hand side of (6.24) boils down to five inequivalent Koba–Nielsen integrals $\mathcal{I}^{(w,0)}_{\dots}$ over elliptic functions,

$$\begin{aligned} M_4(\tau) = \overset{\leftrightarrow}{\mathrm{Tr}}(t^{a_1}t^{a_2}t^{a_3}t^{a_4})\Big[&\mathrm{G}_4^2\mathcal{I}^{(4,0)}_{1234} - \frac{7}{2}\mathrm{G}_4\mathrm{G}_6\mathcal{I}^{(2,0)}_{1234} \\ &- \frac{13}{3}\mathrm{G}_4^3\mathcal{I}^{(0,0)} + \frac{49}{6}\mathrm{G}_6^2\mathcal{I}^{(0,0)}\Big] \\ + \mathrm{Tr}(t^{a_1}t^{a_2})\,\mathrm{Tr}(t^{a_3}t^{a_4})\Big[&\mathrm{G}_4^2\mathcal{I}^{(4,0)}_{12|34} - \frac{7}{2}\mathrm{G}_4\mathrm{G}_6\mathcal{I}^{(2,0)}_{12|34} \\ &+ \frac{5}{3}\mathrm{G}_4^3\mathcal{I}^{(0,0)} + \frac{49}{6}\mathrm{G}_6^2\mathcal{I}^{(0,0)}\Big] \\ + \mathrm{cyc}(2,3,4)& \end{aligned} \tag{6.26}$$

The notation '+ cyc(2, 3, 4)' refers to cyclic permutations of both terms w.r.t. the adjoint indices $a_2, a_3, a_4$ and the Mandelstam invariants $s_{ij}$. Two of the integrals in the single-trace or *planar* sector of (6.26) are defined by a cyclic ordering in the subscript

$$\mathcal{I}^{(4,0)}_{1234}(s_{ij},\tau) = \int \mathrm{d}\mu_3\, V_4(1,2,3,4)\;\mathrm{KN}_4 \tag{6.27}$$

$$\mathcal{I}^{(2,0)}_{1234}(s_{ij},\tau) = \int \mathrm{d}\mu_3\, V_2(1,2,3,4)\;\mathrm{KN}_4\;. \tag{6.28}$$

Furthermore, the permutation-invariant integral

$$\mathcal{I}^{(0,0)}(s_{ij},\tau) = \int \mathrm{d}\mu_3\;\mathrm{KN}_4 \tag{6.29}$$

is universal to the single- and double-trace sectors of (6.26), and it furthermore occurs in the four-point one-loop amplitude of type-II superstrings (cf. (3.102)) and was expanded in (3.138). In the double-trace or *non-planar* sector of (6.26), we have further instances of $\mathcal{I}^{(0,0)}$ and

$$\mathcal{I}^{(4,0)}_{12|34}(s_{ij},\tau) = \int \mathrm{d}\mu_3\, V_2(1,2)V_2(3,4)\;\mathrm{KN}_4 \tag{6.30}$$

$$\mathcal{I}^{(2,0)}_{12|34}(s_{ij},\tau) = \int \mathrm{d}\mu_3\, \big[V_2(1,2) + V_2(3,4)\big]\;\mathrm{KN}_4\;. \tag{6.31}$$

In all of (6.27) to (6.31), translation invariance has been used to fix $z_1 = 0$.

The superscripts in the notation for the integrals keep track of their modular weights: By the modular properties (3.100) of the elliptic $V_a$-functions, the integrals are easily checked to transform as modular forms of holomorphic weights $(a, 0)$,

$$\mathcal{I}^{(a,0)}_{\dots}\Big(s_{ij}, \frac{\alpha\tau+\beta}{\gamma\tau+\delta}\Big) = (\gamma\tau+\delta)^a\,\mathcal{I}^{(a,0)}_{\dots}(s_{ij},\tau)\,, \tag{6.32}$$

where the ellipsis may represent any permutation of 1234 and 12|34 (or be empty to incorporate modular invariance of $\mathcal{I}^{(0,0)}$). With the modular weight $(k,0)$ of $\mathrm{G}_k$, each term in the four-point integral (6.26) is a form of weight $(12,0)$. In the integrated amplitude (6.23), this compensates the weight $(-12,0)$ of the bosonic partition function $\eta^{-24}$ in agreement with modular invariance.

In the rest of this chapter, we will compute the low-energy expansion of the integrals (6.27), (6.28) and (6.30), (6.31) over the punctures by expanding the Koba–Nielsen factor in the Mandelstams and applying the methods discussed in Chapter 3. The resulting $\alpha'$ expansion will be simplified using the techniques from Chapter 5. The modular weights (6.32) apply to each order in the $\alpha'$ expansion of the integrals $\mathcal{I}^{(a,0)}_{\dots}$.

Note that all the integrals (6.27), (6.28) and (6.30), (6.31) defined above can be written in terms of the component integrals (7.5) of the generating function (7.1) of Koba–Nielsen integrals introduced in Chapter 7 and written in terms of iterated Eisenstein integrals in Chapter 8.

## 6.2 LOW-ENERGY EXPANSION AND MODULAR GRAPH FORMS

In Section 3.2.3, we saw that the $V_a$ functions can be written as polynomials in $f^{(a_i)}_{i,i+1} = f^{(a_i)}(z_i - z_{i+1})$ with $\sum_i a_i = a$, cf. the examples in (3.99). When the Koba–Nielsen factor in the $\mathcal{I}^{(a,0)}_{\dots}$ is expanded in $\alpha'$, we obtain additional factors $G_{ij}$ in the integrand and together with (3.119), this naturally leads to the integral representation of MGFs (3.120) for the expansion coefficients.

In order to track the different contributions from the expanded Koba–Nielsen factor to $\mathcal{I}^{(a,0)}_{1234}$, we introduce (with $w = 0,2,4$ and $V_0(1,2,3,4) = 1$),[5]

$$H^{(a,0)}_{1234}\Big[\prod_{i<j} G_{ij}^{n_{ij}}\Big] = \Big(\frac{\pi}{\tau_2}\Big)^{\sum_{i<j} n_{ij}} \int \mathrm{d}\mu_3\, V_a(1,2,3,4) \prod_{i<j} G_{ij}^{n_{ij}}\,. \qquad (6.33)$$

With this definition, the leading orders of the $\alpha'$ expansions of $\mathcal{I}^{(w,0)}_{1234}$ read

$$\begin{aligned}\mathcal{I}^{(a,0)}_{1234}(s_{ij},\tau) = H^{(a,0)}_{1234}[\varnothing] + \frac{\tau_2}{\pi}\sum_{1\le i<j}^{4} s_{ij} H^{(a,0)}_{1234}[G_{ij}] \\ + \frac{1}{2}\Big(\frac{\tau_2}{\pi}\Big)^2 \sum_{\substack{1\le i<j\\ 1\le k<l}}^{4} s_{ij}s_{kl} H^{(a,0)}_{1234}[G_{ij}G_{kl}] + O(\alpha'^3)\,.\end{aligned} \qquad (6.34)$$

5 As compared to [III], we have introduced here a prefactor $(\frac{\pi}{\tau_2})^{\sum_{i<j} n_{ij}}$ to keep resulting expressions compact. In [III], this prefactor was included into the definitions of the MGFs.

Analogous definitions can be made in the non-planar sector,

$$H^{(4,0)}_{12|34}\Big[\prod_{i<j} G_{ij}^{n_{ij}}\Big] = \Big(\frac{\pi}{\tau_2}\Big)^{\sum_{i<j} n_{ij}} \int \mathrm{d}\mu_3\, V_2(1,2)V_2(3,4) \prod_{i<j} G_{ij}^{n_{ij}} \tag{6.35}$$

$$H^{(2,0)}_{12|34}\Big[\prod_{i<j} G_{ij}^{n_{ij}}\Big] = \Big(\frac{\pi}{\tau_2}\Big)^{\sum_{i<j} n_{ij}} \int \mathrm{d}\mu_3 \big[V_2(1,2)+V_2(3,4)\big] \prod_{i<j} G_{ij}^{n_{ij}}\,, \tag{6.36}$$

and (6.34) with $1234 \to 12|34$ applies to the $\alpha'$ expansions of the non-planar integrals $\mathcal{I}^{(a,0)}_{12|34}$.

By analogy with the definition of modular graph functions through the $\alpha'$ expansion of $\mathcal{I}^{(0,0)}$, we will refer to the above $H^{(a,0)}_{1234}[\prod_{i<j} G_{ij}^{n_{ij}}]$ and $H^{(a,0)}_{12|34}[\prod_{i<j} G_{ij}^{n_{ij}}]$ as *heterotic graph forms*. They are MGFs of weight $(a+\sum_{i<j} n_{ij}, \sum_{i<j} n_{ij})$, as one can see from modular invariance of $G_{ij}$ and $\mathrm{d}\mu_3$ and the weight $(a,0)$ of $V_a$. Modular graph functions are recovered from the weight-zero instances $H^{(0,0)}_{1234}[\ldots]$.

Note that similar techniques have been applied in [10][6] to evaluate certain integrals over the punctures in five- and six-point gauge correlators at $\exp\Big(\sum_{i<j} s_{ij} G_{ij}\Big) \to 1$ that demonstrate the absence of $\mathrm{Tr}(F^5)$ and $\mathrm{Tr}(F^6)$ operators in the one-loop effective action of heterotic strings.

### 6.2.1 *Planar contributions*

In this section, we compute and simplify the planar heterotic graph forms that arise from the Koba–Nielsen integrals over $V_a(1,2,3,4)$ at the leading orders in $\alpha'$ and $a = 2,4$. These results follow from the representation of the $V_a$ functions in terms of the doubly-periodic $f^{(a)}$ as alluded to above. The relevant integrands are

$$V_2(1,2,3,4) = f^{(1)}_{12} f^{(1)}_{34} + f^{(1)}_{23} f^{(1)}_{41} + \big[f^{(2)}_{12} + f^{(1)}_{12} f^{(1)}_{23} + \mathrm{cyc}(1,2,3,4)\big] \tag{6.37}$$

$$\begin{aligned} V_4(1,2,3,4) = {}& f^{(1)}_{12} f^{(1)}_{23} f^{(1)}_{34} f^{(1)}_{41} + f^{(1)}_{12} f^{(3)}_{34} + f^{(2)}_{12} f^{(2)}_{34} + f^{(3)}_{12} f^{(1)}_{34} \\ &+ f^{(1)}_{23} f^{(3)}_{41} + f^{(2)}_{23} f^{(2)}_{41} + f^{(3)}_{23} f^{(1)}_{41} \\ &+ \big[f^{(1)}_{12} f^{(1)}_{23} f^{(2)}_{34} + f^{(1)}_{12} f^{(2)}_{23} f^{(1)}_{34} + f^{(2)}_{12} f^{(1)}_{23} f^{(1)}_{34} \\ &\quad + f^{(1)}_{12} f^{(3)}_{23} + f^{(2)}_{12} f^{(2)}_{23} + f^{(3)}_{12} f^{(1)}_{23} + f^{(4)}_{12} + \mathrm{cyc}(1,2,3,4)\big]\,. \end{aligned} \tag{6.38}$$

By permutation symmetry of the Koba–Nielsen factor, the dihedral symmetry (3.98) of the $V_a$ functions propagates to the integrals ($a = 2,4$),

$$\mathcal{I}^{(a,0)}_{1234} = \mathcal{I}^{(a,0)}_{4321}\,, \qquad \mathcal{I}^{(a,0)}_{1234} = \mathcal{I}^{(a,0)}_{2341}\,. \tag{6.39}$$

On these grounds, only a small fraction of heterotic graph forms at a given order in $\alpha'$ are inequivalent under dihedral symmetry, see Table

6 The integrations in [10] are performed before summing over the spin structure (6.9), based on a double Fourier expansion of the Szegő kernel similar to (3.84).

| order | inequivalent planar heterotic graph forms | | |
|---|---|---|---|
| $\alpha'^0$ | $H^{(a,0)}_{1234}[\varnothing]$ | | |
| $\alpha'^1$ | $H^{(a,0)}_{1234}[G_{12}]$, $H^{(a,0)}_{1234}[G_{13}]$ | | |
| $\alpha'^2$ | $H^{(a,0)}_{1234}[G^2_{12}]$, $H^{(a,0)}_{1234}[G_{13}G_{24}]$, $H^{(a,0)}_{1234}[G_{12}G_{13}]$ | | |
| | $H^{(a,0)}_{1234}[G^2_{13}]$, $H^{(a,0)}_{1234}[G_{12}G_{34}]$, $H^{(a,0)}_{1234}[G_{12}G_{23}]$ | | |
| $\alpha'^3$ | $H^{(a,0)}_{1234}[G^3_{12}]$, | $H^{(a,0)}_{1234}[G^2_{13}G_{24}]$, | $H^{(a,0)}_{1234}[G_{12}G_{23}G_{34}]$ |
| | $H^{(a,0)}_{1234}[G^3_{13}]$, | $H^{(a,0)}_{1234}[G^2_{12}G_{34}]$, | $H^{(a,0)}_{1234}[G_{12}G_{13}G_{14}]$ |
| | $H^{(a,0)}_{1234}[G^2_{12}G_{13}]$, | $H^{(a,0)}_{1234}[G_{12}G^2_{13}]$, | $H^{(a,0)}_{1234}[G_{12}G_{23}G_{31}]$ |
| | $H^{(a,0)}_{1234}[G^2_{12}G_{23}]$, | $H^{(a,0)}_{1234}[G_{12}G_{13}G_{34}]$, | $H^{(a,0)}_{1234}[G_{12}G_{13}G_{24}]$ |

Table 6.1: Inequivalent planar heterotic graph forms with respect to the dihedral symmetry.

6.1. In the remainder of this subsection, we will restrict our attention to the heterotic graph forms in the table.

### LEADING ORDERS $\alpha'^0$ AND $\alpha'^1$

In the absence of any $G_{ij}$ in the integrand, we obtain the simplest heterotic graph forms

$$H^{(2,0)}_{1234}[\varnothing] = 0\,, \qquad H^{(4,0)}_{1234}[\varnothing] = \mathrm{G}_4\,, \tag{6.40}$$

where the Eisenstein series $\mathrm{G}_4$ on the right-hand side can be traced back to the contribution of $f^{(1)}_{12}f^{(1)}_{23}f^{(1)}_{34}f^{(1)}_{41}$ in (6.38), see [178]. At first order in $\alpha'$, the two inequivalent heterotic graph forms in Table 6.1 are found to be

$$H^{(2,0)}_{1234}[G_{12}] = -\mathcal{C}\big[\begin{smallmatrix}3&0\\1&0\end{smallmatrix}\big]\,, \qquad H^{(2,0)}_{1234}[G_{13}] = 2\,\mathcal{C}\big[\begin{smallmatrix}3&0\\1&0\end{smallmatrix}\big]\,, \tag{6.41}$$

as well as

$$\begin{aligned} H^{(4,0)}_{1234}[G_{12}] &= -\mathcal{C}\big[\begin{smallmatrix}3&1&1\\0&0&1\end{smallmatrix}\big] - 4\,\mathcal{C}\big[\begin{smallmatrix}5&0\\1&0\end{smallmatrix}\big] \\ H^{(4,0)}_{1234}[G_{13}] &= \mathcal{C}\big[\begin{smallmatrix}2&2&1\\0&0&1\end{smallmatrix}\big] + 6\,\mathcal{C}\big[\begin{smallmatrix}5&0\\1&0\end{smallmatrix}\big]\,. \end{aligned} \tag{6.42}$$

The MGFs associated with three-edge graphs can be simplified via dihedral HSR (5.73), and the $H^{(4,0)}_{1234}[G_{ij}]$ can be expressed solely in terms of single lattice sums (just like the $H^{(2,0)}_{1234}[G_{ij}]$),

$$H^{(4,0)}_{1234}[G_{12}] = -\widehat{\mathrm{G}}_2\,\mathcal{C}\big[\begin{smallmatrix}3&0\\1&0\end{smallmatrix}\big] - \mathrm{G}_4 \tag{6.43}$$

$$H^{(4,0)}_{1234}[G_{13}] = 2\widehat{\mathrm{G}}_2\,\mathcal{C}\big[\begin{smallmatrix}3&0\\1&0\end{smallmatrix}\big] + 2\mathrm{G}_4\,. \tag{6.44}$$

### SUBLEADING ORDERS $\alpha'^2$ AND $\alpha'^3$

At the order $\alpha'^2$, the six inequivalent heterotic graph forms in Table 6.1 evaluate to

$$
\begin{aligned}
H^{(2,0)}_{1234}[G^2_{12}] &= -\,C\left[\begin{smallmatrix}1&1&2\\1&1&0\end{smallmatrix}\right]\\
H^{(2,0)}_{1234}[G^2_{13}] &= 2\,C\left[\begin{smallmatrix}1&1&2\\1&1&0\end{smallmatrix}\right]\\
H^{(2,0)}_{1234}[G_{13}G_{24}] &= -2\,C\left[\begin{smallmatrix}4&0\\2&0\end{smallmatrix}\right]\\
H^{(2,0)}_{1234}[G_{13}G_{12}] &= -\,C\left[\begin{smallmatrix}4&0\\2&0\end{smallmatrix}\right]-C\left[\begin{smallmatrix}1&1&2\\0&1&1\end{smallmatrix}\right]\\
H^{(2,0)}_{1234}[G_{12}G_{34}] &= C\left[\begin{smallmatrix}4&0\\2&0\end{smallmatrix}\right]\\
H^{(2,0)}_{1234}[G_{12}G_{23}] &= C\left[\begin{smallmatrix}4&0\\2&0\end{smallmatrix}\right],
\end{aligned}
\tag{6.45}
$$

as well as

$$
\begin{aligned}
H^{(4,0)}_{1234}[G^2_{12}] &= -4\,C\left[\begin{smallmatrix}1&1&4\\1&1&0\end{smallmatrix}\right]-C\left[\begin{smallmatrix}1&1&1&3\\0&1&1&0\end{smallmatrix}\right]\\
H^{(4,0)}_{1234}[G^2_{13}] &= 6\,C\left[\begin{smallmatrix}1&1&4\\1&1&0\end{smallmatrix}\right]+C\left[\begin{smallmatrix}1&1&2&2\\1&1&0&0\end{smallmatrix}\right]\\
H^{(4,0)}_{1234}[G_{13}G_{24}] &= -2\,C\left[\begin{smallmatrix}6&0\\2&0\end{smallmatrix}\right]-4\,C\left[\begin{smallmatrix}2&2&2\\0&1&1\end{smallmatrix}\right]+8\,C\left[\begin{smallmatrix}1&2&3\\0&1&1\end{smallmatrix}\right]-C\left[\begin{array}{c}1\\0\\\hline 1\\0\end{array}\middle\|\begin{array}{c}1\\0\\\hline 1\\0\end{array}\middle\|\begin{array}{c}1\\1\\\hline 1\\1\end{array}\right]\\
H^{(4,0)}_{1234}[G_{13}G_{12}] &= -\,C\left[\begin{smallmatrix}6&0\\2&0\end{smallmatrix}\right]-C\left[\begin{smallmatrix}1&1&4\\0&1&1\end{smallmatrix}\right]+C\left[\begin{smallmatrix}1&2&3\\1&1&0\end{smallmatrix}\right]-C\left[\begin{smallmatrix}3&0\\1&0\end{smallmatrix}\right]^2+C\left[\begin{smallmatrix}1\\0\end{smallmatrix}\middle|\begin{smallmatrix}1&1\\0&1\end{smallmatrix}\middle|\begin{smallmatrix}1&2\\1&0\end{smallmatrix}\right]\\
H^{(4,0)}_{1234}[G_{12}G_{34}] &= 3\,C\left[\begin{smallmatrix}6&0\\2&0\end{smallmatrix}\right]+4\,C\left[\begin{smallmatrix}1&1&4\\0&1&1\end{smallmatrix}\right]-2\,C\left[\begin{smallmatrix}1&2&3\\1&0&1\end{smallmatrix}\right]+C\left[\begin{smallmatrix}3&0\\1&0\end{smallmatrix}\right]^2+C\left[\begin{smallmatrix}2\\0\end{smallmatrix}\middle|\begin{smallmatrix}1&1\\0&1\end{smallmatrix}\middle|\begin{smallmatrix}1&1\\0&1\end{smallmatrix}\right]\\
H^{(4,0)}_{1234}[G_{12}G_{23}] &= 3\,C\left[\begin{smallmatrix}6&0\\2&0\end{smallmatrix}\right]+4\,C\left[\begin{smallmatrix}1&1&4\\0&1&1\end{smallmatrix}\right]-2\,C\left[\begin{smallmatrix}1&2&3\\1&0&1\end{smallmatrix}\right]+C\left[\begin{smallmatrix}3&0\\1&0\end{smallmatrix}\right]^2+C\left[\begin{smallmatrix}2\\0\end{smallmatrix}\middle|\begin{smallmatrix}1&1\\0&1\end{smallmatrix}\middle|\begin{smallmatrix}1&1\\0&1\end{smallmatrix}\right].
\end{aligned}
\tag{6.46}
$$

At this order, the contribution of $f^{(1)}_{12}f^{(1)}_{23}f^{(1)}_{34}f^{(1)}_{41}$ to $V_4(1,2,3,4)$ introduces trihedral MGFs as well as a tetrahedral representative in $H^{(4,0)}_{1234}[G_{13}G_{24}]$.

The complexity of the lattice sums in $H^{(a,0)}_{1234}[G_{ij}G_{kl}]$ can be considerably reduced: The three-edge sums at $w=2$ boil down to $C\left[\begin{smallmatrix}4&0\\2&0\end{smallmatrix}\right]$, e.g.

$$
H^{(2,0)}_{1234}[G^2_{12}] = -\,C\left[\begin{smallmatrix}4&0\\2&0\end{smallmatrix}\right], \qquad H^{(2,0)}_{1234}[G_{13}G_{12}] = -\frac{1}{2}\,C\left[\begin{smallmatrix}4&0\\2&0\end{smallmatrix}\right]. \tag{6.47}
$$

The trihedral graphs in (6.46) can be simplified using the three-point HSR formula (5.109), yielding

$$
\begin{aligned}
C\left[\begin{smallmatrix}2&1\\0&1\end{smallmatrix}\middle|\begin{smallmatrix}1&1\\1&0\end{smallmatrix}\middle|\begin{smallmatrix}1\\0\end{smallmatrix}\right] &= -\frac{1}{2}\,C\left[\begin{smallmatrix}6&0\\2&0\end{smallmatrix}\right]+\frac{3}{2}\,C\left[\begin{smallmatrix}3&0\\1&0\end{smallmatrix}\right]^2-\frac{1}{2}\widehat{\mathrm{G}}_2\,C\left[\begin{smallmatrix}4&0\\2&0\end{smallmatrix}\right]\\
&\quad +3\frac{\pi}{\tau_2}\,C\left[\begin{smallmatrix}5&0\\1&0\end{smallmatrix}\right]-\frac{\pi}{\tau_2}\widehat{\mathrm{G}}_2\,C\left[\begin{smallmatrix}3&0\\1&0\end{smallmatrix}\right]-\left(\frac{\pi}{\tau_2}\right)^2\mathrm{G}_4\\
C\left[\begin{smallmatrix}1&1\\1&0\end{smallmatrix}\middle|\begin{smallmatrix}1&1\\1&0\end{smallmatrix}\middle|\begin{smallmatrix}2\\0\end{smallmatrix}\right] &= 2\,C\left[\begin{smallmatrix}6&0\\2&0\end{smallmatrix}\right]-2\,C\left[\begin{smallmatrix}3&0\\1&0\end{smallmatrix}\right]^2+\widehat{\mathrm{G}}_2\,C\left[\begin{smallmatrix}4&0\\2&0\end{smallmatrix}\right]\\
&\quad -6\frac{\pi}{\tau_2}\,C\left[\begin{smallmatrix}5&0\\1&0\end{smallmatrix}\right]+2\frac{\pi}{\tau_2}\widehat{\mathrm{G}}_2\,C\left[\begin{smallmatrix}3&0\\1&0\end{smallmatrix}\right]+2\left(\frac{\pi}{\tau_2}\right)^2\mathrm{G}_4\,.
\end{aligned}
\tag{6.48}
$$

The tetrahedral graph in (6.46) has a closed four-point holomorphic subgraph and was decomposed in (5.126). Furthermore, dihedral HSR as in (5.73) and the basis decompositions from Section 5.7 reduce all the dihedral MGFs in $H^{(4,0)}_{1234}[G_{ij}G_{kl}]$ to one-loop graphs. For instance,

$$\begin{aligned} H^{(4,0)}_{1234}[G^2_{12}] &= -\widehat{\mathrm{G}}_2\, C\big[\begin{smallmatrix}4&0\\2&0\end{smallmatrix}\big] - 8\frac{\pi}{\tau_2}\, C\big[\begin{smallmatrix}5&0\\1&0\end{smallmatrix}\big] + 2\frac{\pi}{\tau_2}\widehat{\mathrm{G}}_2\, C\big[\begin{smallmatrix}3&0\\1&0\end{smallmatrix}\big] \\ &\quad + \Big(\frac{\pi}{\tau_2}\Big)^2 \mathrm{G}_4(2+\mathrm{E}_2) \end{aligned} \tag{6.49}$$

$$H^{(4,0)}_{1234}[G_{12}G_{34}] = \widehat{\mathrm{G}}_2\, C\big[\begin{smallmatrix}4&0\\2&0\end{smallmatrix}\big] - 6\frac{\pi}{\tau_2}\, C\big[\begin{smallmatrix}5&0\\1&0\end{smallmatrix}\big] + 2\frac{\pi}{\tau_2}\widehat{\mathrm{G}}_2\, C\big[\begin{smallmatrix}3&0\\1&0\end{smallmatrix}\big] + 2\Big(\frac{\pi}{\tau_2}\Big)^2 \mathrm{G}_4\,. \tag{6.50}$$

Once this simplification is performed for all heterotic graph forms in (6.46), the remaining $H^{(4,0)}_{1234}[G_{ij}G_{kl}]$ are found to be related to (6.49) and (6.50) via

$$\begin{aligned} H^{(4,0)}_{1234}[G^2_{13}] &= 3\mathrm{G}_4\mathrm{E}_2 - 2H^{(4,0)}_{1234}[G^2_{12}] \\ H^{(4,0)}_{1234}[G_{13}G_{24}] &= -2H^{(4,0)}_{1234}[G_{12}G_{34}] \\ H^{(4,0)}_{1234}[G_{13}G_{12}] &= -\frac{1}{2}H^{(4,0)}_{1234}[G_{12}G_{34}] \\ H^{(4,0)}_{1234}[G_{12}G_{23}] &= H^{(4,0)}_{1234}[G_{12}G_{34}]\,. \end{aligned} \tag{6.51}$$

Furthermore, the contributions of $\widehat{\mathrm{G}}_2\frac{\pi}{\tau_2}C\big[\begin{smallmatrix}3&0\\1&0\end{smallmatrix}\big]$ and $(\frac{\pi}{\tau_2})^2\mathrm{G}_4$ turn out to cancel when assembling the complete second order of $\mathcal{I}^{(4,0)}_{1234}$, see (6.56) below.

We have also evaluated the $\alpha'^3$-order of the integral $\mathcal{I}^{(2,0)}_{1234}$. The occurring MGFs are listed in Appendix B.1 of [III] and simplified in Appendix B.2 of the reference. Their specific combination selected by applying momentum conservation to $\mathcal{I}^{(2,0)}_{1234}$ will be given below.

## CHECKING AND ASSEMBLING THE RESULTS

The symmetry properties of the $V_a$ functions [174]

$$\begin{aligned} V_2(1,2,3,4) + V_2(1,3,4,2) + V_2(1,4,2,3) &= 0 \\ V_4(1,2,3,4) + V_4(1,3,4,2) + V_4(1,4,2,3) &= 3\mathrm{G}_4 \end{aligned} \tag{6.52}$$

impose the following constraints on heterotic graph forms

$$\begin{aligned} H^{(2,0)}_{1234}[\dots] + H^{(2,0)}_{1342}[\dots] + H^{(2,0)}_{1423}[\dots] &= 0 \\ H^{(4,0)}_{1234}[\dots] + H^{(4,0)}_{1342}[\dots] + H^{(4,0)}_{1423}[\dots] &= 3\mathrm{G}_4 H^{(0,0)}_{1234}[\dots]\,, \end{aligned} \tag{6.53}$$

where the ellipses represent arbitrary monomials in $G_{ij}$ (the same ones in each term of the respective equation), and all of our results for $H^{(a,0)}_{1234}[\dots]$ have been checked to satisfy these consistency conditions.

Examples for the modular graph functions on the right-hand side of (6.53) include

$$H^{(0,0)}_{1234}[\varnothing] = 1\,, \qquad H^{(0,0)}_{1234}[G_{ij}] = 0\,, \qquad H^{(0,0)}_{1234}[G^2_{ij}] = \Big(\frac{\pi}{\tau_2}\Big)^2 \mathrm{E}_2\,. \tag{6.54}$$

With the results for the planar heterotic graph forms $H^{(2,0)}_{1234}[\dots]$ at orders $\alpha'^{\leq 3}$ and $H^{(4,0)}_{1234}[\dots]$ at orders $\alpha'^{\leq 2}$, we arrive at the low-energy expansions (cf. (6.34))

$$\begin{aligned}
\mathcal{I}^{(2,0)}_{1234}(s_{ij},\tau) &= 6s_{13}\frac{\tau_2}{\pi}\,C\big[\begin{smallmatrix}3&0\\1&0\end{smallmatrix}\big]+2(s_{13}^2+2s_{12}s_{23})\Big(\frac{\tau_2}{\pi}\Big)^2 C\big[\begin{smallmatrix}4&0\\2&0\end{smallmatrix}\big] \qquad (6.55)\\
&\quad+4s_{13}(s_{13}^2-s_{12}s_{23})\Big(\frac{\tau_2}{\pi}\Big)^3\Big(3\,C\big[\begin{smallmatrix}3&1&1\\1&1&1\end{smallmatrix}\big]-4\,C\big[\begin{smallmatrix}5&0\\3&0\end{smallmatrix}\big]+3\Big(\frac{\pi}{\tau_2}\Big)^2\mathrm{E}_2\,C\big[\begin{smallmatrix}3&0\\1&0\end{smallmatrix}\big]\Big)+O(\alpha'^4)\\
\mathcal{I}^{(4,0)}_{1234}(s_{ij},\tau) &= \mathrm{G}_4+6s_{13}\Big(\mathrm{G}_4+\frac{\tau_2}{\pi}\widehat{\mathrm{G}}_2\,C\big[\begin{smallmatrix}3&0\\1&0\end{smallmatrix}\big]\Big)+2(s_{13}^2-s_{12}s_{23})\mathrm{G}_4\mathrm{E}_2 \qquad (6.56)\\
&\quad+2(s_{13}^2+2s_{12}s_{23})\Big(\frac{\tau_2}{\pi}\,C\big[\begin{smallmatrix}5&0\\1&0\end{smallmatrix}\big]+\Big(\frac{\tau_2}{\pi}\Big)^2\widehat{\mathrm{G}}_2\,C\big[\begin{smallmatrix}4&0\\2&0\end{smallmatrix}\big]\Big)+O(\alpha'^3)
\end{aligned}$$

after applying the Mandelstam identities (6.25). One can use

$$\begin{aligned}
&C\big[\begin{smallmatrix}3&0\\1&0\end{smallmatrix}\big] = \frac{\pi\nabla_0\mathrm{E}_2}{2\tau_2^2}\,,\quad C\big[\begin{smallmatrix}4&0\\2&0\end{smallmatrix}\big] = \frac{\pi\nabla_0\mathrm{E}_3}{3\tau_2^2}\,,\quad C\big[\begin{smallmatrix}5&0\\1&0\end{smallmatrix}\big] = \frac{(\pi\nabla_0)^2\mathrm{E}_3}{12\tau_2^4}\\
&C\big[\begin{smallmatrix}5&0\\3&0\end{smallmatrix}\big] = \frac{\pi\nabla_0\mathrm{E}_4}{4\tau_2^2}\,,\quad C\big[\begin{smallmatrix}1&1&3\\1&1&1\end{smallmatrix}\big]-\frac{8}{5}C\big[\begin{smallmatrix}5&0\\3&0\end{smallmatrix}\big] = \frac{\pi\nabla_0\mathrm{E}_{2,2}}{\tau_2^2}
\end{aligned} \tag{6.57}$$

to express these expansions via Cauchy–Riemann derivatives of modular invariant real basis elements E, yielding

$$\begin{aligned}
\mathcal{I}^{(2,0)}_{1234}(s_{ij},\tau) &= 3s_{13}\frac{\pi\nabla_0\mathrm{E}_2}{\tau_2^2}+\frac{2}{3}(s_{13}^2+2s_{12}s_{23})\frac{\pi\nabla_0\mathrm{E}_3}{\tau_2^2} \qquad (6.58)\\
&\quad+s_{13}(s_{13}^2-s_{12}s_{23})\Big(\frac{4}{5}\frac{\pi\nabla_0\mathrm{E}_4}{\tau_2^2}+12\frac{\pi\nabla_0\mathrm{E}_{2,2}}{\tau_2^2}+6\mathrm{E}_2\frac{\pi\nabla_0\mathrm{E}_2}{\tau_2^2}\Big)+O(\alpha'^4)\\
\mathcal{I}^{(4,0)}_{1234}(s_{ij},\tau) &= \mathrm{G}_4+3s_{13}\Big(2\mathrm{G}_4+\frac{\widehat{\mathrm{G}}_2\,\pi\nabla_0\mathrm{E}_2}{\tau_2^2}\Big)+2(s_{13}^2-s_{12}s_{23})\mathrm{G}_4\mathrm{E}_2 \qquad (6.59)\\
&\quad+(s_{13}^2+2s_{12}s_{23})\Big(\frac{(\pi\nabla_0)^2\mathrm{E}_3}{6\tau_2^4}+\frac{2\widehat{\mathrm{G}}_2\,\pi\nabla_0\mathrm{E}_3}{3\tau_2^2}\Big)+O(\alpha'^3)\,,
\end{aligned}$$

cf. (5.56) for the derivatives of the non-holomorphic Eisenstein series and see (4.28) for the definition of $\mathrm{E}_{2,2}$ and (3.138) for the analogous expansion of $\mathcal{I}^{(0,0)}$.

### 6.2.2 *Non-planar contributions*

We will now adapt the strategy of the previous section to the low-energy expansion of the Koba–Nielsen integrals $\mathcal{I}^{(a,0)}_{12|34}$ in the double-trace sector. The non-planar heterotic graph forms (6.35) and (6.36) in the $\alpha'$

| order | inequivalent non-planar heterotic graph forms |
|---|---|
| $\alpha'^0$ | $H^{(a,0)}_{12\|34}[\varnothing]$ |
| $\alpha'^1$ | $H^{(a,0)}_{12\|34}[G_{12}]$, $H^{(a,0)}_{12\|34}[G_{13}]$ |
| $\alpha'^2$ | $H^{(a,0)}_{12\|34}[G_{12}^2]$, $H^{(a,0)}_{12\|34}[G_{12}G_{34}]$, $H^{(a,0)}_{12\|34}[G_{13}G_{24}]$<br>$H^{(a,0)}_{12\|34}[G_{13}^2]$, $H^{(a,0)}_{12\|34}[G_{12}G_{13}]$, $H^{(a,0)}_{12\|34}[G_{13}G_{23}]$ |
| $\alpha'^3$ | $H^{(a,0)}_{12\|34}[G_{12}^3]$, $H^{(a,0)}_{12\|34}[G_{12}G_{13}^2]$, $H^{(a,0)}_{12\|34}[G_{12}G_{13}G_{34}]$<br>$H^{(a,0)}_{12\|34}[G_{13}^3]$, $H^{(a,0)}_{12\|34}[G_{13}^2G_{24}]$, $H^{(a,0)}_{12\|34}[G_{12}G_{13}G_{14}]$<br>$H^{(a,0)}_{12\|34}[G_{12}^2G_{34}]$, $H^{(a,0)}_{12\|34}[G_{13}^2G_{14}]$, $H^{(a,0)}_{12\|34}[G_{13}G_{14}G_{23}]$<br>$H^{(a,0)}_{12\|34}[G_{12}^2G_{13}]$, $H^{(a,0)}_{12\|34}[G_{12}G_{13}G_{24}]$, $H^{(a,0)}_{12\|34}[G_{13}G_{14}G_{34}]$ |

Table 6.2: Inequivalent non-planar heterotic graph forms with respect to the symmetries $H^{(a,0)}_{12|34} = H^{(a,0)}_{21|34} = H^{(a,0)}_{34|12}$.

expansion contain the function $V_2(i,j)$, which can be written in terms of the function $f^{(a)}$,

$$V_2(i,j) = 2f^{(2)}_{ij} - (f^{(1)}_{ij})^2\,, \tag{6.60}$$

similarly to (6.37) and (6.38) in the planar sector. As we will see momentarily, the $(f^{(1)})^2$-contribution yields divergent MGFs which can be simplified using the HSR identity (5.178).[7] Again, we exploit the symmetries

$$\mathcal{I}^{(a,0)}_{12|34} = \mathcal{I}^{(a,0)}_{21|34} = \mathcal{I}^{(a,0)}_{34|12} \tag{6.61}$$

of the integrals at each order in $\alpha'$ to reduce the number of heterotic graph forms that need to be calculated independently, see Table 6.2.

### LEADING ORDERS $\alpha'^0$ AND $\alpha'^1$

In the absence of Green functions in the integrand, only the second term in (6.60) contributes, leading to

$$H^{(2,0)}_{12|34}[\varnothing] = 2\widehat{G}_2\,, \qquad H^{(4,0)}_{12|34}[\varnothing] = \widehat{G}_2^2\,. \tag{6.62}$$

At first order in $\alpha'$, the $(f^{(1)})^2$ term in (6.60) yields conditionally convergent lattice sums

$$H^{(2,0)}_{12|34}[G_{12}] = -2\,C\Big[\begin{smallmatrix}3&0\\1&0\end{smallmatrix}\Big] - C\Big[\begin{smallmatrix}1&1&1\\0&0&1\end{smallmatrix}\Big] \tag{6.63}$$

$$H^{(4,0)}_{12|34}[G_{12}] = -2\widehat{G}_2\,C\Big[\begin{smallmatrix}3&0\\1&0\end{smallmatrix}\Big] - \widehat{G}_2\,C\Big[\begin{smallmatrix}1&1&1\\0&0&1\end{smallmatrix}\Big]\,. \tag{6.64}$$

7 In [III], regularized MGFs are defined by equating different representations of $V_2(i,j)$. As discussed below (5.178), these different representations are exactly related by the $a_1 = a_2 = 1$ incidence of the Fay identity (5.133), which is also the basis for (5.178) and therefore both approaches lead to the same results.

The divergent HSR identity (5.178) implies

$$C\left[\begin{smallmatrix}1&1&1\\0&0&1\end{smallmatrix}\right] = -2\,C\left[\begin{smallmatrix}3&0\\1&0\end{smallmatrix}\right] + \frac{\pi}{\tau_2}\widehat{\mathrm{G}}_2\,, \tag{6.65}$$

and hence we have for the inequivalent heterotic graph forms at the first order in $\alpha'$,

$$H^{(2,0)}_{12|34}[G_{12}] = -\frac{\pi}{\tau_2}\widehat{\mathrm{G}}_2, \qquad H^{(2,0)}_{12|34}[G_{13}] = 0 \tag{6.66}$$

$$H^{(4,0)}_{12|34}[G_{12}] = -\frac{\pi}{\tau_2}\widehat{\mathrm{G}}_2^2, \qquad H^{(4,0)}_{12|34}[G_{13}] = 0\,. \tag{6.67}$$

### SUBLEADING ORDERS $\alpha'^2$, $\alpha'^3$ AND BEYOND

At the second order in $\alpha'$, obtain the inequivalent heterotic graph forms

$$\begin{aligned}
H^{(2,0)}_{12|34}[G_{12}^2] &= 2\Big(\frac{\pi}{\tau_2}\Big)^2\widehat{\mathrm{G}}_2(\mathrm{E}_2+2) - 4\frac{\pi}{\tau_2}\,C\left[\begin{smallmatrix}3&0\\1&0\end{smallmatrix}\right]\\
H^{(2,0)}_{12|34}[G_{13}^2] &= 2\Big(\frac{\pi}{\tau_2}\Big)^2\widehat{\mathrm{G}}_2\mathrm{E}_2\\
H^{(2,0)}_{12|34}[G_{13}G_{23}] &= -\frac{\pi}{\tau_2}\,C\left[\begin{smallmatrix}3&0\\1&0\end{smallmatrix}\right]\\
H^{(2,0)}_{12|34}[G_{12}G_{13}] &= H^{(2,0)}_{12|34}[G_{12}G_{34}] = H^{(2,0)}_{12|34}[G_{13}G_{24}] = 0
\end{aligned}$$

and

$$\begin{aligned}
H^{(4,0)}_{12|34}[G_{12}^2] &= \Big(\frac{\pi}{\tau_2}\Big)^2\widehat{\mathrm{G}}_2^2(\mathrm{E}_2{+}2){-}4\frac{\pi}{\tau_2}\widehat{\mathrm{G}}_2\,C\left[\begin{smallmatrix}3&0\\1&0\end{smallmatrix}\right]\\
H^{(4,0)}_{12|34}[G_{13}^2] &= \Big(\frac{\pi}{\tau_2}\Big)^2\widehat{\mathrm{G}}_2^2\mathrm{E}_2\\
H^{(4,0)}_{12|34}[G_{13}G_{24}] &= \Big(\frac{\pi}{\tau_2}\Big)^2\mathrm{G}_4\\
H^{(4,0)}_{12|34}[G_{12}G_{13}] &= 0\\
H^{(4,0)}_{12|34}[G_{12}G_{34}] &= \Big(\frac{\pi}{\tau_2}\Big)^2\widehat{\mathrm{G}}_2^2\\
H^{(4,0)}_{12|34}[G_{13}G_{23}] &= -\frac{\pi}{\tau_2}\widehat{\mathrm{G}}_2\,C\left[\begin{smallmatrix}3&0\\1&0\end{smallmatrix}\right].
\end{aligned} \tag{6.68}$$

On top of (6.65), the third order in $\alpha'$ involves the divergent HSR identity

$$C\left[\begin{smallmatrix}1&1&1&1\\0&0&1&1\end{smallmatrix}\right] = -2\,C\left[\begin{smallmatrix}4&0\\2&0\end{smallmatrix}\right] + 4\frac{\pi}{\tau}\,C\left[\begin{smallmatrix}3&0\\1&0\end{smallmatrix}\right] - \Big(\frac{\pi}{\tau_2}\Big)^2\widehat{\mathrm{G}}_2(2+\mathrm{E}_2)\,. \tag{6.69}$$

The inequivalent heterotic graph forms at order $\alpha'^3$ from Table 6.2 are given in Appendix C of [III]. The four-point gauge amplitude only requires a specific combination of them that simplifies and is given below.

#### ASSEMBLING THE RESULTS

With the results for the non-planar heterotic graph forms at order $\alpha'^{\leq 3}$, the low-energy expansions of the integrals $\mathcal{I}^{(a,0)}_{12|34}$ are found to be

$$\begin{aligned}\mathcal{I}^{(2,0)}_{12|34} &= 2\widehat{G}_2 - 2s_{12}\widehat{G}_2 + 2s_{12}^2\Big(\widehat{G}_2(1{+}2E_2) - \frac{\pi\nabla_0 E_2}{\tau_2^2}\Big)\\ &\quad - 2s_{13}s_{23}\Big(2\widehat{G}_2E_2 + \frac{\pi\nabla_0 E_2}{\tau_2^2}\Big)\\ &\quad + 2s_{12}^3\Big(\frac{\pi\nabla_0 E_2}{\tau_2^2} - \frac{2}{3}\frac{\pi\nabla_0 E_3}{\tau_2^2} - (1{+}2E_2)\widehat{G}_2\Big)\\ &\quad + 2s_{12}s_{13}s_{23}\Big((2E_2{+}5E_3{+}\zeta_3)\widehat{G}_2 + \frac{\pi\nabla_0 E_2}{\tau_2^2} - \frac{2}{3}\frac{\pi\nabla_0 E_3}{\tau_2^2}\Big) + \mathcal{O}(\alpha'^4)\end{aligned} \tag{6.70}$$

$$\begin{aligned}\mathcal{I}^{(4,0)}_{12|34} &= \widehat{G}_2^2 - 2s_{12}\widehat{G}_2^2 + s_{12}^2\Big(G_4 + \widehat{G}_2^2(3{+}2E_2) - \frac{2\widehat{G}_2\,\pi\nabla_0 E_2}{\tau_2^2}\Big)\\ &\quad - 2s_{13}s_{23}\Big(G_4 + \widehat{G}_2^2E_2 + \frac{\widehat{G}_2\,\pi\nabla_0 E_2}{\tau_2^2}\Big)\\ &\quad + s_{12}s_{13}s_{23}\Big((4E_2{+}5E_3{+}\zeta_3)\widehat{G}_2^2 - 2G_4\\ &\qquad\qquad + \frac{(\pi\nabla_0)^2E_3}{\tau_2^4} + \frac{\widehat{G}_2\pi\nabla_0 E_2}{\tau_2^2} - \frac{4}{3}\frac{\widehat{G}_2\pi\nabla_0 E_3}{\tau_2^2}\Big)\\ &\quad + s_{12}^3\Big( - 2G_4 - 4\widehat{G}_2^2(1{+}E_2)\\ &\qquad\qquad + \frac{(\pi\nabla_0)^2E_3}{3\tau_2^4} + \frac{4\widehat{G}_2\pi\nabla_0 E_2}{\tau_2^2} - \frac{4}{3}\frac{\widehat{G}_2\pi\nabla_0 E_3}{\tau_2^2}\Big) + \mathcal{O}(\alpha'^4)\end{aligned} \tag{6.71}$$

after applying the basis decompositions from Section 5.7. Similar to the representations (6.58) and (6.59) of the planar integrals, we have used the substitutions (6.57). Together with the planar results and the expression (3.138) for $\mathcal{I}^{(0,0)}$, (6.70) and (6.71) complete the ingredients for the $\tau$ integrand (6.24) of the heterotic-string amplitude.

### 6.2.3 *Uniform transcendentality decomposition*

As mentioned in the introduction to this chapter, one of the remarkable features of type-II amplitudes is that they exhibit so-called uniform transcendentality at each order in $\alpha'$. In this section, we will study the transcendentality properties of the heterotic string by restricting to the salient points that require a rewriting of the basis integrals $\mathcal{I}^{(a,0)}_{\dots}$; additional details can be found in Appendix C.

In analogy with the superstring, we associate transcendental weights to the various objects appearing in the low-energy expansion of the heterotic integrals over the punctures as follows. The Eisenstein series $G_k$ and $E_k$ as well as $\zeta_k$ are assigned transcendental weight $k$, i.e. $\pi$ has transcendental weight one, whereas $\tau$ and $\nabla_0$ have transcendental weight zero. Accordingly, one finds transcendental weight one for both $\pi\nabla_0$ and $y = \pi\tau_2$, i.e. weight $k+p$ for $(\pi\nabla_0)^p \mathrm{E}_k$, and weight $4+p$ for $(\pi\nabla_0)^p \mathrm{E}_{2,2}$. A more general definition of transcendental weight in terms of iterated integrals is given in Appendix C, but the assignment above suffices for the discussion of this section.

Inspecting the $\alpha'$ expansions of the planar single-trace integrals $\mathcal{I}^{(0,0)}$ and $\mathcal{I}^{(2,0)}_{1234}$ of (3.138) and (6.58), one sees that their $k^{\text{th}}$ order consistently involves MGFs of weight $k$ and $k+2$, respectively. Thus, these two integrals are referred to as uniformly transcendental.

By contrast, $\mathcal{I}^{(4,0)}_{1234}$ in (6.59) violates uniform transcendentality since the same type of transcendental object appears at different orders in the $\alpha'$ expansion. For instance, $\mathrm{G}_4$ of transcendentality four appears with $1 + 6s_{13} + \ldots$ and thus at different orders in $\alpha'$. Similarly, the integrals $\mathcal{I}^{(a,0)}_{12|34}$ in (6.70) and (6.71) from the double-trace sector violate uniform transcendentality. This can for instance be seen from the terms $\sim (1+2\mathrm{E}_2)$ along with $s_{12}^2$ in $\mathcal{I}^{(2,0)}_{12|34}$ and $\sim (3+2\mathrm{E}_2)$ along with $s_{12}^2$ in $\mathcal{I}^{(4,0)}_{12|34}$, respectively.

This violation of uniform transcendentality can be traced back to the following phenomenon. For the planar integral $\mathcal{I}^{(4,0)}_{1234}$ we see from (6.38) that there is a leading contribution with a closed cycle of the form $f^{(1)}_{12} f^{(1)}_{23} f^{(1)}_{34} f^{(1)}_{41}$. This cycle exhibits purely holomorphic modular weight $(4,0)$ and is thus amenable to HSR as discussed in Section 5.4. However, the formula (5.73) for dihedral HSR generically produces explicit factors of $\widehat{\mathrm{G}}_2 = \mathrm{G}_2 - \frac{\pi}{\tau_2}$ which are clearly not of uniform transcendental weight. We therefore expect that *all* closed cycles $f^{(1)}_{12} f^{(1)}_{23} \ldots f^{(1)}_{k1}$ in the $n$-point integrand break uniform transcendentality, including those with $k=n$. This is in marked contrast with the genus-zero situation where only closed *sub*cycles $(z_{12}z_{23}\ldots z_{k1})^{-1}$ in the integrand with $k \leq n-2$ violate uniform transcendentality [99, 212, 225, 226].[8] The subcycles $f^{(1)}_{12} f^{(1)}_{21}$ in the integrands of the non-uniformly transcendental integrals $\mathcal{I}^{(2,0)}_{12|34}$ and $\mathcal{I}^{(4,0)}_{12|34}$ in the non-planar sector confirm the general expectation.

At genus zero, any non-uniformly transcendental disk or sphere integral over $n$ punctures can be expanded in a basis of uniformly transcendental integrals, see [126, 227, 228] for a general argument and [98, 212, 226, 229] for examples and methods. This basis, known as Parke–Taylor basis, consists of $(n-3)!$ elements [98, 227, 230] and spans the twisted cohomology defined by the Koba–Nielsen factor made out of $|z_{ij}|^{-s_{ij}}$ [126].

8 This point can be illustrated by considering a four-point integral at genus zero over closed subcycles with an integrand of the form $(z_{12}z_{21})^{-1}(z_{34}z_{43})^{-1}$. Integrating by parts in $z$ leads to the cyclic factor $(z_{12}z_{23}z_{34}z_{41})^{-1}$ subtending all four punctures that is called a Parke–Taylor factor. The integration by parts also generates the rational factor $\frac{s_{23}}{1+s_{12}}$ in Mandelstam invariants that mixes different orders in $\alpha'$. Genus-zero integrals over Parke–Taylor factors subtending all the punctures are known to be uniformly transcendental (which is for instance evident from their representation in terms of the Drinfeld associator [24]). Hence, the original subcycle expression with $(z_{12}z_{21})^{-1}(z_{34}z_{43})^{-1}$ must violate uniform transcendentality.

At genus one, a classification of integration-by-parts inequivalent half-integrands — i.e. chiral halves for torus integrands is conjectured in [37, 38]. While genus-one correlators of the open superstring exclude a variety of worldsheet functions by maximal supersymmetry [1, 63], Kac–Moody correlators such as (6.22) give a more accurate picture of the problem. In (7.1), we define a generating series of Koba–Nielsen integrals whose coefficients span all the above $\mathcal{I}^{(a,0)}_{\dots}$ (and all other Koba–Nielsen integrals appearing in string amplitudes) via integration-by-parts and Fay identities. These equivalence classes are again referred to as twisted cohomologies, where the twist is defined by the Koba–Nielsen factor $\mathrm{KN}_n$.

Therefore we shall now re-express the planar and non-planar integrands in a basis of uniformly transcendental integrals, hoping that this will also shed light on the question of a basis for twisted cohomologies at genus one. We present below candidate basis elements $\widehat{\mathcal{I}}^{(a,0)}_{\dots}$ of conjectured uniform transcendentality that appear suitable for the four-current correlator (6.22). Our explicit expressions at leading orders in $\alpha'$ and their different modular weights can be used to exclude relations among the $\widehat{\mathcal{I}}^{(a,0)}_{\dots}$. However, it is beyond the scope of this work to arrive at a reliable prediction for the basis dimension of uniform-transcendentality integrals at four points. At the level of the generating series (7.1), we will conjecture the basis dimension to be $(n-1)!$ at $n$ points for each chiral half, but this needs to be adjusted for the counting of the component integrals (7.5).

In the relation between the new quantities $\widehat{\mathcal{I}}^{(a,0)}_{\dots}$ and the genus-one integrals $\mathcal{I}^{(4,0)}_{1234}$, $\mathcal{I}^{(2,0)}_{12|34}$ and $\mathcal{I}^{(4,0)}_{12|34}$ all terms that break uniform transcendentality are contained in simple explicit coefficients like $\widehat{\mathrm{G}}_2$ or $(1+s_{12})^{-1}$. The manipulations necessary to arrive at the $\widehat{\mathcal{I}}^{(a,0)}_{\dots}$ are given in detail in Appendices C.1 and C.2 and driven by integration by parts, resulting again in series in MGFs which bypass the need for HSR and avoid the conditionally convergent or divergent lattice sums caused by integration over $V_2(i,j)$. Aspects of the computational complexity when using the $\mathcal{I}^{(a,0)}_{\dots}$ versus the $\widehat{\mathcal{I}}^{(a,0)}_{\dots}$ can be found in Appendix C.3.

## PLANAR UNIFORMLY TRANSCENDENTAL INTEGRALS

As derived in Appendix C.1, a decomposition of the single-trace part of the four-point gauge amplitude that exhibits uniform transcendentality is

$$M_4(\tau)\big|_{\mathrm{Tr}(t^{a_1}t^{a_2}t^{a_3}t^{a_4})} = \mathrm{G}_4^2\widehat{\mathcal{I}}^{(4,0)}_{1234} + \mathrm{G}_4\Big(\mathrm{G}_4\widehat{\mathrm{G}}_2 - \frac{7}{2}\mathrm{G}_6\Big)\mathcal{I}^{(2,0)}_{1234} + \Big(\frac{49}{6}\mathrm{G}_6^2 - \frac{10}{3}\mathrm{G}_4^3\Big)\mathcal{I}^{(0,0)}\,. \tag{6.72}$$

In this expression we introduced the following combination of modular weight $(4,0)$

$$\begin{aligned}\widehat{\mathcal{I}}^{(4,0)}_{1234}(s_{ij},\tau) &= \mathcal{I}^{(4,0)}_{1234}(s_{ij},\tau) - \mathrm{G}_4\mathcal{I}^{(0,0)}_{1234}(s_{ij},\tau) - \widehat{\mathrm{G}}_2\mathcal{I}^{(2,0)}_{1234}(s_{ij},\tau)\\ &= 6s_{13}\mathrm{G}_4 + (s_{13}^2+2s_{12}s_{23})\frac{(\pi\nabla_0)^2\mathrm{E}_3}{6\tau_2^4} + O(\alpha'^3)\,,\end{aligned} \tag{6.73}$$

that manifestly respects uniform transcendentality to the order given. In (C.12), we provide a closed integral form of $\widehat{\mathcal{I}}^{(4,0)}_{1234}(s_{ij},\tau)$ that we conjecture to be uniformly transcendental at every order in $\alpha'$, with weight $k{+}3$ at the order of $\alpha'^k$. As argued above, $\mathcal{I}^{(2,0)}_{1234}$ and $\mathcal{I}^{(0,0)}$ are uniformly transcendental and all non-uniformly transcendental terms in the above way of writing the planar amplitude are in the coefficients of the basis integrals. The coefficient of $s_{13}$ in (6.73) may also be written as $6\mathrm{G}_4 = (\pi\nabla_0)^2\mathrm{E}_2/\tau_2^4$ to highlight the parallel with the MGF $(\pi\nabla_0)^2\mathrm{E}_3/(6\tau_2^4)$ at the subleading order $\alpha'^2$.

Note that the coefficient of $\mathcal{I}^{(0,0)}$ in (6.72) can be recognized as

$$\frac{49}{6}\mathrm{G}_6^2 - \frac{10}{3}\mathrm{G}_4^3 = -\frac{128\pi^{12}}{2025}\eta^{24}\,. \tag{6.74}$$

At the level of the integrated amplitude (6.23), this cancels the factor of $\eta^{-24}$ due to the partition function. Hence, one can import the techniques of the type-II amplitude [39, 127] to perform the modular integrals $\int_{\mathcal{F}}\frac{\mathrm{d}^2\tau}{\tau_2^2}\mathcal{I}^{(0,0)}$ in (6.72) as we shall see in Section 6.2.4.

Similar to (6.74) the coefficient of $\mathcal{I}^{(2,0)}_{1234}$ in (6.72) exhibits a special relative factor in the combination $\mathrm{G}_4\mathrm{G}_2 - \frac{7}{2}\mathrm{G}_6$ that can therefore be written as a $\tau$-derivative

$$\mathrm{G}_4\widehat{\mathrm{G}}_2 - \frac{7}{2}\mathrm{G}_6 = -\frac{\pi}{\tau_2}\mathrm{G}_4 + \pi^2 q\frac{\mathrm{d}\mathrm{G}_4}{\mathrm{d}q}\,, \tag{6.75}$$

using the Ramanujan identities (5.63). Hence, the only contribution $\sim q^0$ to (6.75) stems from the non-holomorphic term $-\frac{\pi}{\tau_2}\mathrm{G}_4$.

### NON-PLANAR UNIFORMLY TRANSCENDENTAL INTEGRALS

Similarly, we can also rewrite the non-planar part of the amplitude in terms of combinations that exhibit uniform transcendentality as follows

$$\begin{aligned}&M_4(\tau)\big|_{\mathrm{Tr}(t^{a_1}t^{a_2})\,\mathrm{Tr}(t^{a_3}t^{a_4})}\\ &= \frac{\mathrm{G}_4^2\left[\widehat{\mathcal{I}}^{(4,0)}_{12|34} + s_{13}^2(\widehat{\mathcal{I}}^{(4,0)}_{1243}{+}\widehat{\mathrm{G}}_2\mathcal{I}^{(2,0)}_{1243}) + s_{23}^2(\widehat{\mathcal{I}}^{(4,0)}_{1234}{+}\widehat{\mathrm{G}}_2\mathcal{I}^{(2,0)}_{1234})\right]}{(1+s_{12})^2}\\ &\quad+\left(\frac{\mathrm{G}_4^2\widehat{\mathrm{G}}_2}{(1{+}s_{12})^2} - \frac{7\mathrm{G}_4\mathrm{G}_6}{2(1{+}s_{12})}\right)\widehat{\mathcal{I}}^{(2,0)}_{12|34}\\ &\quad+\left(\frac{\mathrm{G}_4^2\widehat{\mathrm{G}}_2^2 + \mathrm{G}_4^3(s_{13}^2{+}s_{23}^2)}{(1+s_{12})^2} - 7\frac{\mathrm{G}_4\mathrm{G}_6\widehat{\mathrm{G}}_2}{1+s_{12}} + \frac{5}{3}\mathrm{G}_4^3 + \frac{49}{6}\mathrm{G}_6^2\right)\mathcal{I}^{(0,0)}\,.\end{aligned} \tag{6.76}$$

The details of the derivation of this result are given in Appendix C.2. On top of a contribution from the planar integral given in (6.73), (6.76) contains the non-planar integrals of conjectured uniform transcendentality $\widehat{\mathcal{I}}^{(a,0)}_{12|34}$ with modular weights $(a,0)$ and leading orders

$$\begin{aligned}
\widehat{\mathcal{I}}^{(2,0)}_{12|34} &= -2(s_{12}^2+s_{13}s_{23})\frac{\pi\nabla_0 \mathrm{E}_2}{\tau_2^2} - 4(s_{12}^3+s_{12}s_{13}s_{23})\frac{\pi\nabla_0\mathrm{E}_3}{3\tau_2^2} + O(\alpha'^4)\\
\widehat{\mathcal{I}}^{(4,0)}_{12|34} &= (s_{12}^3+3s_{12}s_{13}s_{23})\frac{(\pi\nabla_0)^2\mathrm{E}_3}{3\tau_2^4} + O(\alpha'^4)\,.
\end{aligned} \tag{6.77}$$

For definitions of the $\widehat{\mathcal{I}}^{(a,0)}_{12|34}$ to all orders in $\alpha'$ via $z$-integrals, see (C.20) and (C.26).

### 6.2.4 *The integrated amplitude and the low-energy effective action*

Our main focus of this chapter lies on the structure of the $\tau$-integrand $M_4(\tau)$ appearing in the four-point gauge amplitude (6.23). However, using the results of [39, 127, 175, 178, 198, 199, 231] on the integrals of certain combinations of MGFs and Eisenstein series, it is possible to perform the integral over $\tau$ analytically up to second order in $\alpha'$ both in the single-trace and the double-trace sector. We present the resulting values, keeping in mind that these have to be considered in our normalization (6.23), see also footnote 4 on page 159.

#### PLANAR AMPLITUDE UP TO SECOND ORDER IN $\alpha'$

Upon collecting all terms of the same structure in Mandelstam invariants up to quadratic order from (6.72) we have to evaluate the integrals appearing in

$$\begin{aligned}
\mathcal{M}_4\big|_{\mathrm{Tr}(t^{a_1}t^{a_2}t^{a_3}t^{a_4})} &\sim \int_{\mathcal{F}} \frac{d^2\tau}{\tau_2^2\,\eta^{24}}\Big(\frac{49}{6}\mathrm{G}_6^2 - \frac{10}{3}\mathrm{G}_4^3\Big)\\
&+s_{13}\int_{\mathcal{F}} \frac{d^2\tau}{\tau_2^2\,\eta^{24}}\Big(6\mathrm{G}_4^3 + 3\mathrm{G}_4^2\widehat{\mathrm{G}}_2\frac{\pi\nabla_0\mathrm{E}_2}{\tau_2^2} - \frac{21}{2}\mathrm{G}_4\mathrm{G}_6\frac{\pi\nabla_0\mathrm{E}_2}{\tau_2^2}\Big)\\
&+(s_{13}^2 - s_{12}s_{23})\int_{\mathcal{F}} \frac{d^2\tau}{\tau_2^2\,\eta^{24}}\Big(\frac{49}{3}\mathrm{G}_4\mathrm{G}_6^2\mathrm{E}_2 - \frac{20}{3}\mathrm{G}_4^3\mathrm{E}_2\Big)\\
&+(s_{13}^2+2s_{12}s_{23})\int_{\mathcal{F}} \frac{d^2\tau}{\tau_2^2\,\eta^{24}}\Big(\frac{2}{3}\mathrm{G}_4^2\widehat{\mathrm{G}}_2\frac{\pi\nabla_0\mathrm{E}_3}{\tau_2^2} - \frac{7}{3}\mathrm{G}_4\mathrm{G}_6\frac{\pi\nabla_0\mathrm{E}_3}{\tau_2^2} + \frac{1}{6}\mathrm{G}_4^2\frac{(\pi\nabla_0)^2\mathrm{E}_3}{\tau_2^4}\Big)\\
&+O(\alpha'^3)\,.
\end{aligned} \tag{6.78}$$

These integrals can be performed using the following observations. The combination of $\mathrm{G}_4^3$ and $\mathrm{G}_6^2$ appearing in the first and third line is that of (6.74) leading to $\eta^{24}$, making the first line proportional to the volume of $\mathcal{F}$ that equals $\frac{\pi}{3}$ while the third line is proportional to the integral of $\mathrm{E}_2$ over $\mathcal{F}$ that vanishes [175]. Using furthermore the results of [178, 231] for the remaining lines, we end up with the integrated planar amplitude

$$\mathcal{M}_4\Big|_{\mathrm{Tr}(t^{a_1}t^{a_2}t^{a_3}t^{a_4})} \sim -\frac{256\pi^{13}}{6075} + s_{13}\frac{32\pi^{13}}{6075}\left(\frac{25}{6} + \gamma_{\mathrm{E}} + \log\pi - 2\frac{\zeta_4'}{\zeta_4}\right) - (s_{13}^2 + 2s_{12}s_{23})\frac{32\pi^{13}}{30375} + O(\alpha'^3)\,. \tag{6.79}$$

The appearance of terms $\log\pi$ and $\mathrm{d}\log\zeta$ is due to the method of cutting off the fundamental domain $\mathcal{F}$ as discussed in Sections 3.3.1 and 3.4. Note that these terms as well as the Euler–Mascheroni constant $\gamma_{\mathrm{E}}$ cancel at the second order in $\alpha'$, a feature that we shall discuss in more detail below. The interplay of $\mathrm{d}\log\zeta$ with uniform transcendentality was discussed in [128].

From the point of view of the low-energy effective action, the terms above correspond to single-trace higher-derivative corrections of the schematic form $\mathrm{Tr}(F^4)$, $\mathrm{Tr}(D^2F^4)$ and $\mathrm{Tr}(D^4F^4)$, respectively. The lowest-order term in the one-loop scattering amplitude was already analyzed in [178, 219, 220, 232, 233]. The structure of higher-derivative invariants in super Yang–Mills theory was studied for example in [234–236] and the three operators above are of 1/2-, 1/4- and non-BPS type, respectively. General references on the effective action of heterotic string theories include [10, 231, 237–244]

### NON-PLANAR AMPLITUDE UP TO SECOND ORDER IN $\alpha'$

The integrated contribution from the double-trace sector can be determined by similar methods by starting from (6.76). The $\tau$-integral to be performed to quadratic order in Mandelstam invariants is

$$\begin{aligned}
\mathcal{M}_4\Big|_{\mathrm{Tr}(t^{a_1}t^{a_2})\,\mathrm{Tr}(t^{a_3}t^{a_4})} &\sim \int_{\mathcal{F}} \frac{\mathrm{d}^2\tau}{\tau_2^2\,\eta^{24}}\left(\mathrm{G}_4^2\widehat{\mathrm{G}}_2^2 - 7\mathrm{G}_4\mathrm{G}_6\widehat{\mathrm{G}}_2 + \frac{5}{3}\mathrm{G}_4^3 + \frac{49}{6}\mathrm{G}_6^2\right)\\
&+s_{12}\int_{\mathcal{F}} \frac{\mathrm{d}^2\tau}{\tau_2^2\,\eta^{24}}\left(7\mathrm{G}_4\mathrm{G}_6\widehat{\mathrm{G}}_2 - 2\mathrm{G}_4^2\widehat{\mathrm{G}}_2^2\right)\\
&+s_{12}^2\int_{\mathcal{F}} \frac{\mathrm{d}^2\tau}{\tau_2^2\,\eta^{24}}\Bigg(\mathrm{G}_4^3 + \widehat{\mathrm{G}}_2^2\mathrm{G}_4^2(3+2\mathrm{E}_2) - \frac{2\widehat{\mathrm{G}}_2\mathrm{G}_4^2\pi\nabla_0\mathrm{E}_2}{\tau_2^2} - 7\mathrm{G}_4\mathrm{G}_6\widehat{\mathrm{G}}_2(1+2\mathrm{E}_2)\\
&\qquad + 7\mathrm{G}_4\mathrm{G}_6\frac{\pi\nabla_0\mathrm{E}_2}{\tau_2^2} + \frac{10}{3}\mathrm{G}_4^3\mathrm{E}_2 + \frac{49}{3}\mathrm{G}_6^2\mathrm{E}_2\Bigg) \qquad\qquad (6.80)\\
&+s_{13}s_{23}\int_{\mathcal{F}} \frac{\mathrm{d}^2\tau}{\tau_2^2\,\eta^{24}}\Bigg(-2\mathrm{G}_4^3 - 2\mathrm{G}_4^2\widehat{\mathrm{G}}_2^2\mathrm{E}_2 - 2\frac{\mathrm{G}_4^2\widehat{\mathrm{G}}_2\pi\nabla_0\mathrm{E}_2}{\tau_2^2} + 14\mathrm{G}_4\mathrm{G}_6\widehat{\mathrm{G}}_2\mathrm{E}_2\\
&\qquad + 7\frac{\mathrm{G}_4\mathrm{G}_6\pi\nabla_0\mathrm{E}_2}{\tau_2^2} - \frac{10}{3}\mathrm{G}_4^3\mathrm{E}_2 - \frac{49}{3}\mathrm{G}_6^2\mathrm{E}_2\Bigg) + O(\alpha'^3)\,.
\end{aligned}$$

These integrals can again be performed using the results of [39, 178, 231], and we obtain the integrated double-trace amplitude to second order in $\alpha'$ as

$$\begin{aligned}\mathcal{M}_4\big|_{\mathrm{Tr}(t^{a_1}t^{a_2})\,\mathrm{Tr}(t^{a_3}t^{a_4})} \sim -\frac{128\pi^{13}}{6075}s_{12} + s_{12}^2\frac{64\pi^{13}}{3645}\left(-\frac{91}{30}+\gamma_\mathrm{E}+\log\pi-2\frac{\zeta_4'}{\zeta_4}\right)\\ - s_{13}s_{23}\frac{256\pi^{13}}{18225}\left(-\frac{11}{6}+\gamma_\mathrm{E}+\log\pi-2\frac{\zeta_4'}{\zeta_4}\right) + \mathcal{O}(\alpha'^3)\,.\end{aligned} \tag{6.81}$$

We note that there is no lowest-order term in the non-planar sector. As pointed out in [10], this is in agreement with the duality between the heterotic string and the type-I string [9], where $(\mathrm{Tr}(F^2))^2$ is absent at tree level. The first non-trivial correction term for double-trace operators is then $(\mathrm{Tr}(DF^2))^2$, and the eight-derivative order admits two independent kinematic structures.

### CONSISTENCY WITH TREE-LEVEL AMPLITUDES

Given the expressions (6.79) and (6.81) for the integrals over $\tau$, the appearance of $\gamma_\mathrm{E} + \log\pi - 2\frac{\zeta_4'}{\zeta_4}$ signals an interplay with the non-analytic momentum dependence of the respective amplitude at the same $\alpha'$-order, cf. [39, 127]. The non-analytic part of the four-point gauge amplitude can be inferred to comprise factors of $\log s_{ij}$

- in the planar sector at the order of $\alpha'$ but not at the orders of $\alpha'^0$ or $\alpha'^2$, see (6.79)
- in the non-planar sector at the order of $\alpha'^2$ but not at the orders of $\alpha'^0$ or $\alpha'^1$, see (6.81)

These patterns in the discontinuity of the one-loop amplitude are consistent with the $\alpha'$ expansion of the respective tree amplitudes of the heterotic string [237]

$$\begin{aligned}\mathcal{M}_4^\mathrm{tree}\big|_{\mathrm{Tr}(t^{a_1}t^{a_2}t^{a_3}t^{a_4})} &\sim 1 + 2\zeta_3 s_{12}s_{13}s_{23} + \mathcal{O}(\alpha'^5)\\ \mathcal{M}_4^\mathrm{tree}\big|_{\mathrm{Tr}(t^{a_1}t^{a_2})\,\mathrm{Tr}(t^{a_3}t^{a_4})} &\sim s_{23} - s_{12}s_{23} + s_{12}^2 s_{23} + \mathcal{O}(\alpha'^4)\,,\end{aligned} \tag{6.82}$$

where a polarization-dependent factor $A^\mathrm{tree}_\mathrm{SYM}(1,2,3,4)$ has been subsumed in the $\sim$. Unitarity relates the $(s_{ij})^w$-order of (6.82) to the non-analytic terms in the one-loop amplitudes that are signaled by the $(s_{ij})^{w+1}$-order in (6.79) and (6.81). In particular, the absence of a subleading order $\alpha'^1$ in the planar sector of $\mathcal{M}_4^\mathrm{tree}$ ties in with the absence of $\gamma_\mathrm{E} + \log\pi - 2\frac{\zeta_4'}{\zeta_4}$ at the $\alpha'^2$-order of (6.79). This is analogous to the discontinuity structure of the massless type-II amplitude [39, 127], where unitarity relates the $\alpha'^{w+1}$-order beyond the one-loop low-energy limit to the $\alpha'^w$-order of the tree amplitude beyond its supergravity limit.

### 6.2.5 *Modular graph forms in the massless $n$-point function*

Although the main focus of this Chapter is on the four-point amplitude involving gauge bosons, we shall now explain how the above techniques can be extended to higher multiplicity and to external gravitons. As we will see, the integration over the punctures in $n$-point one-loop amplitudes of the heterotic string involving any combination of gauge bosons and gravitons boils down to MGFs – at any order in the $\alpha'$ expansion.

#### $n$ EXTERNAL GAUGE BOSONS

For the $n$-point generalization of the amplitude (6.1) among four gauge bosons, the structure of the correlation functions in the integrand is well known. The supersymmetric chiral halves of the vertex operators (6.2) exclusively contribute (complex conjugates of) $f_{ij}^{(a)}$ and holomorphic Eisenstein series $\mathrm{G}_k$ to the $n$-point correlators [28]. This has been manifested in this reference by expressing RNS spin sums[9] of the worldsheet fermions in terms of the $V_a$ functions (3.96) and $\mathrm{G}_k$, also see [105] for analogous results with two external gauginos and [10, 101] for earlier work on the spin sums. The contributions from the worldsheet bosons $\partial_{\bar z} X(z, \bar z)$ are even simpler, they can be straightforwardly integrated out using Wick contractions that yield $\overline{f_{ij}^{(1)}}$ or $\partial_{\bar z_i} \overline{f_{ij}^{(1)}}$. Likewise, the Kac–Moody currents of (6.2) exclusively contribute $V_a$ functions and $\mathrm{G}_k$ to the complementary chiral half of the $n$-point correlators [174].

Given these results on the $n$-point integrands, it is important to note that $\mathrm{G}_k$ are MGFs and that $f_{ij}^{(a)}$ functions admit the same type of lattice-sum representation (3.91) as the Green function (3.65). Then, the Fourier integrals over the $n$ punctures yield momentum-conserving delta functions as explained in Section 3.3.2 , and one is left with the kinds of nested lattice sums that define MGFs [16].

Starting from the five-point function, the singularities $\overline{f_{ij}^{(1)}} \sim \frac{1}{\bar z_{ij}} + O(z, \bar z)$ in the supersymmetric correlators introduce kinematic poles into the integrals over the punctures. Still, the residues of these kinematic poles reduce to lower-multiplicity results and therefore give MGFs by an inductive argument.

On these grounds, one-loop scattering of $n$ gauge bosons in the heterotic string boil down to MGFs at each order in the $\alpha'$ expansion after integrating over the punctures at fixed $\tau$.

#### ADJOINING EXTERNAL GRAVITONS

The vertex operators of gravitons and gauge bosons in heterotic string theories have the same supersymmetric chiral half. That is why for mixed $n$-point amplitudes involving external gauge bosons and gravitons, the

9 Also see [172, 245] for recent examples of $f^{(a)}$ functions in manifestly supersymmetric higher-point amplitudes in the pure-spinor formalism.

supersymmetric half of the correlator is identical to that of $n$ gauge bosons. Only the non-supersymmetric chiral half of the correlators is sensitive to the species of massless states in the external legs since the graviton vertex operator involves the worldsheet boson $\partial_z X(z,\bar{z})$ in the place of the Kac–Moody current [218].

These additional worldsheet bosons of the gravitons contribute (sums of products of) $f^{(1)}_{ij}$ and $\partial_{z_i} f^{(1)}_{ij}$ due to Wick contractions and decouple from the current correlators of the gauge bosons. Moreover, they admit zero-mode contractions $\partial_{z_i} X(z_i,\bar{z}_i)\partial_{\bar{z}_j} X(z_j,\bar{z}_j) \to \frac{\pi}{\tau_2}$ between left and right movers, known from type-II amplitudes [169, 170, 246, 247]. These kinds of cross-contractions are specific to amplitudes involving gravitons, and the resulting factors of $\frac{\pi}{\tau_2}$ have the same modular weight $(1,1)$ as the contributions $f^{(1)}_{ij}\overline{f^{(1)}_{kl}}$ due to separate Wick contractions of the left and right movers.

Hence, the additional contributions due to $\partial_z X(z,\bar{z})$ in the graviton vertex operator boil down to $f^{(1)}_{ij}$, $\partial_{z_i} f^{(1)}_{ij}$ or $\frac{\pi}{\tau_2}$. All of these factors line up with the above statements on the correlators of the supersymmetric chiral half and the currents: The $n$-point correlators for mixed graviton- and gauge-boson amplitudes in heterotic string theories exclusively depend on the punctures via the functions $C^{(a_{ij},b_{ij})}(z_{ij},\tau)$ defined in (3.40) with $a_{ij}, b_{ij} \geq -1$ that may be accompanied by powers of $\frac{\pi}{\tau_2}$ and yield simple Fourier integrals w.r.t. $z_2, \ldots, z_n$. Each term in the $\alpha'$ expansion of the Koba–Nielsen factor is bound to yield MGFs upon integrating over the $z_j$. By the arguments given for $n$ gauge bosons, the kinematic poles do not alter this result.

Note that cases with $a_{ij} = b_{ij} = -1$ are due to the spurious factors of $\partial_{z_i} f^{(1)}_{ij}$ and $\partial_{\bar{z}_i} \overline{f^{(1)}_{ij}}$ in the left- and right-moving contributions to the correlators. One can always remove any appearance of $\partial_{z_i} f^{(1)}_{ij}$ and $\partial_{\bar{z}_i} \overline{f^{(1)}_{ij}}$ via integration by parts and thereby improve the bound on $a_{ij}, b_{ij}$ towards $a_{ij}, b_{ij} \geq 0$.

The $\alpha'$ expansion of four-point amplitudes involving gravitons has been studied beyond the leading order in [231, 242]. Moreover, selected terms in the five- and six-point gauge amplitudes that are relevant to $\text{Tr}(F^5)$ and $\text{Tr}(F^6)$ interactions have been studied in [10].

## 6.3 HETEROTIC STRINGS VERSUS OPEN SUPERSTRINGS

In this section, we point out new relations between open-superstring amplitudes and the integral $\mathcal{I}^{(2,0)}_{1234}$ over the torus punctures in the planar sector of the heterotic-string amplitude, cf. (6.26). The observations of this section can be viewed as generalizing the construction of an elliptic single-valued map from maximal supersymmetry as introduced in [34] and reviewed in Section 4.3, to half-maximal supersymmetry.

### 6.3.1 *Open-superstring integrals at genus one*

The construction of open-string one-loop integrals was reviewed in Chapter 4 with the first four orders in the $\alpha'$ expansion of the four-gluon amplitude $\mathcal{I}^{\rm open}_{1234}$ given in (4.8). In this section, we will extract a suitable symmetry component from this integral for the comparison to $\mathcal{I}^{(2,0)}_{1234}$ and perform the modular S-transformation necessary for the application of the esv map as in (4.34).

#### DECOMPOSITION INTO SYMMETRY COMPONENTS

By the properties of the integration cycle and the open-string Green function, the open-string integral exhibits the same dihedral symmetries w.r.t. its labels $1,2,3,4$ as the $V_a$ functions at even values of $a$,

$$\mathcal{I}^{\rm open}_{1234}(s_{ij},\tau) = \mathcal{I}^{\rm open}_{2341}(s_{ij},\tau)\,, \qquad \mathcal{I}^{\rm open}_{4321}(s_{ij},\tau) = \mathcal{I}^{\rm open}_{1234}(s_{ij},\tau)\,, \tag{6.83}$$

cf. (3.98). In order to explore further connections with the $V_a$ functions, we decompose the integral $\mathcal{I}^{\rm open}_{1234} = \frac{1}{6}Z^{(0)} + Z^{(2)}_{1234}$ into components with different symmetry properties in $1,2,3,4$,

$$Z^{(0)}(s_{ij},\tau) = \sum_{\sigma\in S_3} \mathcal{I}^{\rm open}_{1\sigma(234)}(s_{ij},\tau) \tag{6.84}$$

$$Z^{(2)}_{1234}(s_{ij},\tau) = \frac{1}{3}\big[2\mathcal{I}^{\rm open}_{1234}(s_{ij},\tau) - \mathcal{I}^{\rm open}_{1342}(s_{ij},\tau) - \mathcal{I}^{\rm open}_{1423}(s_{ij},\tau)\big]\,. \tag{6.85}$$

While the permutation symmetric component $Z^{(0)}$ of the open-string integral has been studied in [34], we will here investigate the $\alpha'$ expansion of the second component $Z^{(2)}_{\dots}$ subject to

$$Z^{(2)}_{1234}(s_{ij},\tau) + Z^{(2)}_{1342}(s_{ij},\tau) + Z^{(2)}_{1423}(s_{ij},\tau) = 0\,. \tag{6.86}$$

The symmetry properties of the $Z^{(0)}$ and $Z^{(2)}_{\dots}$ tie in with those of $V_0(1,2,3,4) = 1$ and $V_2(1,2,3,4)$, respectively, see (6.52).

By inserting the $\alpha'$ expansion (4.8) along with momentum conservation $s_{12}+s_{13}+s_{23} = 0$ into (6.84) and (6.85), we arrive at the following representation in terms of eMZVs[10]

$$\begin{aligned} Z^{(0)}(s_{ij},\tau) = 1 + (s_{13}^2 - s_{12}s_{23})\Big(2\omega(0,0,2) + \frac{5\zeta_2}{3}\Big) \\ + 6 s_{12}s_{23}s_{13}\beta_{2,3} + \mathcal{O}(\alpha'^4) \end{aligned} \tag{6.87}$$

10 We have used the following relations among eMZVs in simplifying (6.87) [29]

$$\omega(0,1,1,0,0) = \frac{\zeta_2}{12} + \omega(0,0,0,0,2)\,, \quad \omega(0,1,0,1,0) = \frac{\zeta_2}{12} + \frac{1}{2}\omega(0,0,2) - 4\omega(0,0,0,0,2)\,.$$

$$
\begin{aligned}
Z^{(2)}_{1234}(s_{ij},\tau) = -2s_{13}\omega(0,1,0,0) \\
&- \frac{2}{3}(s_{13}^2 + 2s_{12}s_{23})\big[\omega(0,1,0,1,0) + \omega(0,1,1,0,0)\big] \\
&+ s_{13}(s_{13}^2 - s_{12}s_{23})\beta_5 + O(\alpha'^4)\,. \qquad (6.88)
\end{aligned}
$$

Note that the coefficient $\beta_{2,3}$ in (4.8) drops out from the definition of $Z^{(2)}_{1234}$ in (6.85), and we are only left with a specific linear combination of $\omega(0,1,0,1,0)$ and $\omega(0,1,1,0,0)$ at order $\alpha'^2$.

### MODULAR TRANSFORMATION

A connection between the symmetrized open-string integral $Z^{(0)}$ in (6.84) and closed-string integrals [34] is based on the modular S-transformation $\tau \to -\frac{1}{\tau}$ of the contributing eMZVs. Otherwise, the $q$-series representation of the A-cycle eMZVs in (6.87) and (6.88) would not exhibit any open-string analogue of the $q$ expansion of MGFs around the cusp, more specifically of their Laurent polynomials in $y = \pi\tau_2$.

In order to determine the modular S-transformation of the $Z^{(a)}_{\dots}$ integrals in (6.87) and (6.88), we express the A-cycle eMZVs in terms of iterated Eisenstein integrals (4.15) [29, 248]

$$
\begin{aligned}
Z^{(0)}(s_{ij},\tau) &= 1 + (s_{12}s_{23}-s_{13}^2)\big[12\,\mathcal{E}_0(4,0)-\zeta_2\big] \\
&- s_{12}s_{23}s_{13}\Big[12\,\mathcal{E}_0(4,0,0)+300\,\mathcal{E}_0(6,0,0)-\frac{5\zeta_3}{2}\Big] + O(\alpha'^4) \qquad (6.89)
\end{aligned}
$$

$$
\begin{aligned}
Z^{(2)}_{1234}(s_{ij},\tau) &= \frac{3s_{13}}{2\pi^2}\big[6\,\mathcal{E}_0(4,0,0)-\zeta_3\big] + \frac{s_{13}^2+2s_{12}s_{23}}{2\pi^2}\big[120\,\mathcal{E}_0(6,0,0,0)-\zeta_4\big] \\
&+ \frac{s_{13}(s_{13}^2-s_{12}s_{23})}{2\pi^2}\big[1296\,\mathcal{E}_0(4,4,0,0,0)+432\,\mathcal{E}_0(4,0,4,0,0) \qquad (6.90) \\
&\quad +\frac{6}{5}\,\mathcal{E}_0(4,0,0,0,0)+4032\,\mathcal{E}_0(8,0,0,0,0)-216\,\mathcal{E}_0(4,0)\,\mathcal{E}_0(4,0,0) \\
&\quad +36\zeta_3\,\mathcal{E}_0(4,0)-5\zeta_5\big] + O(\alpha'^4)\,.
\end{aligned}
$$

The modular properties of the holomorphic Eisenstein series give rise to S-transformations such as (4.20). The remaining modular transformations relevant to (6.90) are displayed in (5.20) of [III] and in Appendix E.1 of the reference. These expressions for $\mathcal{E}_0(k_1,\dots;-\frac{1}{\tau})$ yield the following modular $\tau \to -\frac{1}{\tau}$ image of (6.89) [34]

$$
\begin{aligned}
&Z^{(0)}(s_{ij},-\tfrac{1}{\tau}) \\
&= 1 + (s_{13}^2-s_{12}s_{23})\Big(-\frac{T^2}{90}+\frac{\pi^2}{9}+\frac{2i\zeta_3}{T}+\frac{\pi^4}{30T^2}-12\,\mathcal{E}_0(4,0)-\frac{12i}{T}\,\mathcal{E}_0(4,0,0)\Big) \\
&\quad + s_{12}s_{23}s_{13}\Big(\frac{iT^3}{756}-\frac{i\pi^2T}{45}+\frac{\zeta_3}{2}+\frac{7i\pi^4}{72T}+\frac{2\pi^2\zeta_3}{T^2}-\frac{15\zeta_5}{2T^2}-\frac{17i\pi^6}{1890T^3} \qquad (6.91) \\
&\quad -\frac{12\pi^2}{T^2}\,\mathcal{E}_0(4,0,0)-300\,\mathcal{E}_0(6,0,0)-\frac{900i}{T}\,\mathcal{E}_0(6,0,0,0) \\
&\quad +\frac{900}{T^2}\,\mathcal{E}_0(6,0,0,0,0)\Big) + O(\alpha'^4)
\end{aligned}
$$

and the following result for (6.90)

$$
\begin{aligned}
Z^{(2)}_{1234}(s_{ij},-\tfrac{1}{\tau}) &= s_{13}\Big(\frac{iT}{60}-\frac{3\zeta_3}{2T^2}-\frac{i\pi^2}{12T}+\frac{i\pi^4}{60T^3}+\frac{9}{T^2}\,\mathcal{E}_0(4,0,0)\Big)\\
&+(s_{13}^2+2s_{12}s_{23})\Big(\frac{T^2}{3780}-\frac{i\zeta_5}{T^3}-\frac{\pi^2}{216}+\frac{\pi^4}{360T^2}-\frac{\pi^6}{756T^4}\\
&\qquad+\frac{60}{T^2}\,\mathcal{E}_0(6,0,0,0)+\frac{120i}{T^3}\,\mathcal{E}_0(6,0,0,0,0)\Big)\\
&+s_{13}(s_{13}^2-s_{12}s_{23})\beta_5(-\tfrac{1}{\tau})+O(\alpha'^4)\,,
\end{aligned}
\tag{6.92}
$$

where the modular S transformation of $\beta_5(\tau)$ is given by

$$
\begin{aligned}
\beta_5(-\tfrac{1}{\tau}) &= -\frac{iT^3}{7560}+\frac{i\pi^2T}{540}-\frac{\zeta_3}{20}-\frac{i\pi^4}{120T}\\
&-\frac{5\zeta_5}{2T^2}+\frac{\pi^2\zeta_3}{12T^2}+\frac{29i\pi^6}{11340T^3}+\frac{\pi^4\zeta_3}{60T^4}+\frac{3\zeta_7}{T^4}-\frac{i\pi^8}{1800T^5}\\
&+\Big(-\frac{iT}{5}+\frac{i\pi^2}{T}+\frac{18\zeta_3}{T^2}-\frac{i\pi^4}{5T^3}\Big)\,\mathcal{E}_0(4,0)\\
&+\Big(\frac{3}{10}-\frac{\pi^2}{2T^2}-\frac{\pi^4}{10T^4}\Big)\,\mathcal{E}_0(4,0,0)\\
&-\frac{108}{T^2}\,\mathcal{E}_0(4,0)\,\mathcal{E}_0(4,0,0)\\
&+\frac{216}{T^2}\Big(\mathcal{E}_0(4,0,4,0,0)+3\,\mathcal{E}_0(4,4,0,0,0)+\frac{\mathcal{E}_0(4,0,0,0,0)}{360}\Big)\\
&+\frac{2016}{T^2}\,\mathcal{E}_0(8,0,0,0,0)+\frac{10080i}{T^3}\,\mathcal{E}_0(8,0,0,0,0,0)\\
&-\frac{15120}{T^4}\,\mathcal{E}_0(8,0,0,0,0,0,0)\,.
\end{aligned}
\tag{6.93}
$$

The $\mathcal{E}_0(\ldots)$ on the right-hand sides of (6.91) to (6.93) are understood to be evaluated at argument $\tau$ rather than $-\frac{1}{\tau}$. Note that from the results of [28], any order in the $\alpha'$ expansion of (6.91) and (6.92) is expressible in terms of the B-cycle eMZVs of Enriquez [30]. A general discussion of the asymptotic expansion of B-cycle eMZVs around the cusp can be found in [30, 34, 177, 249].

### 6.3.2 *A proposal for a single-valued map at genus one*

After modular transformation, the open-string expressions (6.91) to (6.93) resemble the expansion of MGFs around the cusp: The Laurent polynomials of (6.91) in $T=\pi\tau$ parallel the Laurent polynomials of MGFs in $y=\pi\tau_2$. For instance, with the representations of $\mathrm{E}_k$ in terms of iterated Eisenstein integrals as e.g. in (4.27a), the $\alpha'$ expansion (3.138) of the closed-string integral $\mathcal{I}^{(0,0)}$ takes the following form,

$$\begin{aligned}\mathcal{I}^{(0,0)}(s_{ij},\tau)=&1+2(s_{13}^2-s_{12}s_{23})\Big(\frac{y^2}{45}+\frac{\zeta_3}{y}-12\,\mathrm{Re}[\mathcal{E}_0(4,0)]-\frac{6}{y}\,\mathrm{Re}[\mathcal{E}_0(4,0,0)]\Big)\\&+s_{12}s_{23}s_{13}\Big(\frac{2y^3}{189}+\zeta_3+\frac{15\zeta_5}{4y^2}-600\,\mathrm{Re}[\mathcal{E}_0(6,0,0)]\\&-\frac{900}{y}\,\mathrm{Re}[\mathcal{E}_0(6,0,0,0)]-\frac{450}{y^2}\,\mathrm{Re}[\mathcal{E}_0(6,0,0,0,0)]\Big)+O(\alpha'^4)\,.\end{aligned} \tag{6.94}$$

The necessary expressions for the E in terms of iterated Eisenstein integrals on top of (4.27a) are listed in [III]. Note that the leading low-energy orders $\mathcal{I}^{(0,0)} = 1 + O(s_{ij}^2)$ line up with the symmetry component $Z^{(0)}$ of the open-string integral, cf. (6.87), whereas the expansion of $Z^{(2)}_{1234}$ starts at $O(s_{ij})$. That is why the expression (6.94) was compared with the modular S-transformation of the *symmetrized* open-string integral in (6.91) [34]. In fact, the coefficients of the Mandelstam polynomials $s_{13}^2-s_{12}s_{23}$ and $s_{12}s_{23}s_{13}$ on the open- and closed-string side were observed to be related via the esv map defined in (4.32) [34], where the first part $T \longrightarrow 2iy$ of the esv map does not apply to the exponents in the $q$-series representation (4.18) of iterated Eisenstein integrals. The factors of $2i$ and 2 in (4.32) ensure that the holomorphic derivatives of $\tau$ and $\mathcal{E}_0(\dots;\tau)$ are preserved under esv, and they were engineered in [34] to obtain

$$\begin{aligned}&\mathrm{esv}\ Z^{(0)}(s_{ij},-\tfrac{1}{\tau})\\&=1+2(s_{13}^2-s_{12}s_{23})\Big(\frac{y^2}{45}+\frac{\zeta_3}{y}-12\,\mathrm{Re}[\mathcal{E}_0(4,0)]-\frac{6}{y}\,\mathrm{Re}[\mathcal{E}_0(4,0,0)]\Big)\\&\quad+s_{12}s_{23}s_{13}\Big(\frac{2y^3}{189}+\zeta_3+\frac{15\zeta_5}{4y^2}-600\,\mathrm{Re}[\mathcal{E}_0(6,0,0)]\\&\quad-\frac{900}{y}\,\mathrm{Re}[\mathcal{E}_0(6,0,0,0)]-\frac{450}{y^2}\,\mathrm{Re}[\mathcal{E}_0(6,0,0,0,0)]\Big)+O(\alpha'^4)\end{aligned} \tag{6.95}$$

which exactly matches (6.94) to the orders shown,

$$\mathcal{I}^{(0,0)}(s_{ij},\tau) = \mathrm{esv}\ Z^{(0)}(s_{ij},-\tfrac{1}{\tau})\,, \tag{6.96}$$

cf. (4.34). In fact, this correspondence has been checked to persist up to and including the order of $\alpha'^6$ and is conjectural at higher orders [34].

The relation (6.96) between open- and closed-string $\alpha'$ expansions at genus one strongly resembles the tree-level relation (2.40) between disk and sphere integrals. Hence, the rules in (4.32) were proposed [34] to implement an elliptic analogue of the single-valued map (2.51) of MZVs.

However, the formulation of the esv rules in (6.96) is in general ill-defined as it is not compatible with the shuffle multiplication of iterated Eisenstein integrals (4.19) as mentioned in Section 4.3.2. For instance, applying the esv rules to the right-hand side of

$$\mathcal{E}_0(4,0,0)^2 = 2\,\mathcal{E}_0(4,0,0,4,0,0) + 6\,\mathcal{E}_0(4,0,4,0,0,0) + 12\,\mathcal{E}_0(4,4,0,0,0,0) \tag{6.97}$$

yields a different result than the square of esv $\mathcal{E}_0(4,0,0) = \mathcal{E}_0(4,0,0) + \overline{\mathcal{E}_0(4,0,0)}$. The ambiguity in applying esv to (6.97) is proportional to the cross-term $\mathcal{E}_0(4,0,0)\overline{\mathcal{E}_0(4,0,0)}$ which is of order $O(q\,\bar{q})$ by the $q$ expansion (4.18). More generally, terms of the form $q^n\bar{q}^0$ and $q^0\bar{q}^n$ with $n \in \mathbb{N}_0$ in the output of the esv rules (4.32) are well-defined, i.e. independent of the order of applying shuffle multiplication and esv. We will later on encounter a similar restriction on the powers of $q$ and $\bar{q}$ in the expansion of MGFs that can be reliably predicted from open-string input.

For the iterated Eisenstein integrals of depth one in (6.91), the convergent $\mathcal{E}_0(k,0,\ldots,0)$ with $k \geq 4$ cannot be rewritten via shuffle multiplication without introducing divergent examples $\mathcal{E}_0(0,\ldots)$. By restricting the esv rules (4.32) to convergent iterated Eisenstein integrals, the ambiguities due to shuffle multiplication are relegated to depth two. The relation (6.96) has been established to the order of $\alpha'^6$ (including $\mathcal{E}_0$ of depth three) by picking an ad hoc convention for the use of shuffle-multiplication in the open-string input, see Section 4.3.3 of [34] for details.

### 6.3.3 *The closed-string integral over* $V_2(1,2,3,4)$ *versus* esv $Z^{(2)}_{1234}$

Given the above significance of the symmetrized open-string integral $Z^{(0)}$, we will next apply the esv rules (4.32) to the symmetry component $Z^{(2)}_{1234}$ in (6.85). Starting from the low-energy expansion (6.92) and (6.93) of its modular S transformation, one arrives at

$$\begin{aligned}
\text{esv}\ Z^{(2)}_{1234}(s_{ij},-\tfrac{1}{\tau}) &= s_{13}\Big(-\frac{y}{30}+\frac{3\zeta_3}{4y^2}-\frac{9}{2y^2}\,\text{Re}[\mathcal{E}_0(4,0,0)]\Big)\\
&+ (s_{13}^2+2s_{12}s_{23})\Big(-\frac{y^2}{945}+\frac{\zeta_5}{4y^3}-\frac{30}{y^2}\,\text{Re}[\mathcal{E}_0(6,0,0,0)]-\frac{30}{y^3}\,\text{Re}[\mathcal{E}_0(6,0,0,0,0)]\Big)\\
&+ s_{13}(s_{13}^2-s_{12}s_{23})\Big(-\frac{y^3}{945}-\frac{\zeta_3}{10}+\frac{5\zeta_5}{4y^2}+\frac{3\zeta_7}{8y^4}\\
&\quad+\Big[\frac{4y}{5}-\frac{18\zeta_3}{y^2}\Big]\,\text{Re}[\mathcal{E}_0(4,0)]+\frac{3}{5}\,\text{Re}[\mathcal{E}_0(4,0,0)]\\
&\quad-\frac{108}{y^2}\,\text{Re}[\mathcal{E}_0(4,0,4,0,0)+3\,\mathcal{E}_0(4,4,0,0,0)+\frac{1}{360}\,\mathcal{E}_0(4,0,0,0,0)]\\
&\quad+\frac{108}{y^2}\,\text{Re}[\mathcal{E}_0(4,0)]\,\text{Re}[\mathcal{E}_0(4,0,0)]-\frac{1008}{y^2}\,\text{Re}[\mathcal{E}_0(8,0,0,0,0)]\\
&\quad-\frac{2520}{y^3}\,\text{Re}[\mathcal{E}_0(8,0,0,0,0,0)]-\frac{1890}{y^4}\,\text{Re}[\mathcal{E}_0(8,0,0,0,0,0,0)]\Big)+O(\alpha'^4)\,,
\end{aligned} \tag{6.98}$$

which will now be related to the $\alpha'$ expansion of closed-string integrals.

### THE CLOSED-STRING EXPANSION IN TERMS OF ITERATED EISENSTEIN INTEGRALS

Given that the symmetry properties (6.86) of $Z^{(2)}_{1234}$ have been tailored to match those of $V_2(1,2,3,4)$, it is natural to compare (6.98) with the integral $\mathcal{I}^{(2,0)}_{1234}$ from the heterotic string. The low-energy expansion (6.58) of $\mathcal{I}^{(2,0)}_{1234}$ is written in terms of Cauchy–Riemann derivatives of MGFs and can therefore be expressed in terms of iterated Eisenstein integrals. For instance, the representation (4.27a) of $\mathrm{E}_2$ yields [34]

$$\pi\nabla_0 \mathrm{E}_2 = \frac{2y^3}{45} - \zeta_3 + 24y^2\,\mathcal{E}_0(4) + 12y\,\mathcal{E}_0(4,0) + 6\,\mathrm{Re}[\mathcal{E}_0(4,0,0)]\,. \tag{6.99}$$

The remaining $\nabla_0\mathrm{E}$ appearing in (6.58) are listed in [III]. In this way, we can cast the leading orders of (6.58) into the following form

$$\begin{aligned}\mathcal{I}^{(2,0)}_{1234}(s_{ij},\tau) &= \pi^2 s_{13}\Big(\frac{2y}{15}-\frac{3\zeta_3}{y^2}+\frac{18}{y^2}\,\mathrm{Re}[\mathcal{E}_0(4,0,0)]+72\,\mathcal{E}_0(4)+\frac{36}{y}\,\mathcal{E}_0(4,0)\Big)\\ &+\pi^2(s_{13}^2+2s_{12}s_{23})\Big(\frac{4y^2}{945}-\frac{\zeta_5}{y^3}+\frac{120}{y^2}\,\mathrm{Re}[\mathcal{E}_0(6,0,0,0)]+\frac{120}{y^3}\,\mathrm{Re}[\mathcal{E}_0(6,0,0,0,0)]\\ &\quad+160\,\mathcal{E}_0(6,0)+\frac{240}{y}\,\mathcal{E}_0(6,0,0)+\frac{120}{y^2}\,\mathcal{E}_0(6,0,0,0)\Big)+O(\alpha'^3)\,.\end{aligned} \tag{6.100}$$

A similar expression for the $\alpha'^3$-order is displayed in Appendix E.2 of [III]. There is a notable difference between the terms involving real parts of iterated Eisenstein integrals $\mathrm{Re}[\mathcal{E}_0]$ and the terms without real parts. The real parts $\mathrm{Re}[\mathcal{E}_0]$ and the pure $y$-terms match the esv image of the open-string integral in (6.98) up to a global rescaling of esv $Z^{(2)}_{1234}(s_{ij},-\frac{1}{\tau})$. By contrast, the iterated Eisenstein integrals without real part – specifically, the above $72\,\mathcal{E}_0(4)$ and $\frac{36}{y}\,\mathcal{E}_0(4,0)$ as well as the last line of (6.100) – do not have any open-string counterpart in (6.98). The same mismatch also arises at the third order in $\alpha'$.

### THE $P_{\mathrm{Re}}$ PROJECTION

We shall now give a more precise description of the commonalities and differences of the expressions (6.98) and (6.100) for esv $Z^{(2)}_{1234}(s_{ij},-\frac{1}{\tau})$ and $\mathcal{I}^{(2,0)}_{1234}(s_{ij},\tau)$. The contributions to (6.100) which do not have any obvious open-string correspondent will be isolated by defining a formal projection $P_{\mathrm{Re}}$ via

$$\begin{aligned}P_{\mathrm{Re}}\big(\overline{\mathcal{E}_0(k_1,\dots,k_r)}\big) &= 2\,\mathrm{Re}[\mathcal{E}_0(k_1,\dots,k_r)]\\ P_{\mathrm{Re}}\big(\mathcal{E}_0(k_1,\dots,k_r)\big) &= 0\end{aligned} \tag{6.101}$$

with $k_1 \neq 0$ which acts factor-wise on a product. The projection $P_{\rm Re}$ is designed to only keep the real parts of iterated Eisenstein integrals, i.e. the cases where holomorphic and antiholomorphic terms pair up. Moreover, Laurent polynomials in $y$ and MZVs are taken to be inert

$$P_{\rm Re}\big(y^m \zeta_{n_1,n_2,\ldots,n_r}\big) = y^m \zeta_{n_1,n_2,\ldots,n_r} \,. \tag{6.102}$$

Similar to the esv rule (4.32), the action of $P_{\rm Re}$ on $\overline{\mathcal{E}_0(k_1,\ldots,k_r)}$ is incompatible with shuffle multiplication and necessitates ad-hoc conventions for the presentation of its input when two or more of the entries $k_j$ are non-zero. By the expansions $\mathcal{E}_0(k_1,\ldots)=O(q)$ and $\overline{\mathcal{E}_0(k_1,\ldots)}=O(\bar q)$ for $k_1 \neq 0$, the ambiguity in evaluating $P_{\rm Re}$ has again at least one factor of both $q$ and $\bar q$. For instance, the representation for $\nabla_0 \mathrm{E}_2$ given in (6.99) is mapped to

$$P_{\rm Re}(\pi \nabla_0 \mathrm{E}_2) = \frac{2y^3}{45} - \zeta_3 + 6\,\mathrm{Re}[\mathcal{E}_0(4,0,0)]\,. \tag{6.103}$$

The expression (4.27a) for $\mathrm{E}_2$ is invariant under the projection (6.101), so it naturally extends to the product

$$\begin{aligned} P_{\rm Re}(\mathrm{E}_2 \pi \nabla_0 \mathrm{E}_2) &= \frac{2y^5}{2025} + \frac{y^2\zeta_3}{45} - \frac{\zeta_3^2}{y} + \Big(12\zeta_3 - \frac{8y^3}{15}\Big)\,\mathrm{Re}[\mathcal{E}_0(4,0)] \\ &\quad + \Big(\frac{12\zeta_3}{y} - \frac{2y^2}{15}\Big)\,\mathrm{Re}[\mathcal{E}_0(4,0,0)] \\ &\quad - 72\,\mathrm{Re}[\mathcal{E}_0(4,0)]\,\mathrm{Re}[\mathcal{E}_0(4,0,0)] - \frac{36}{y}\,\mathrm{Re}[\mathcal{E}_0(4,0,0)]^2\,. \end{aligned} \tag{6.104}$$

The projections of the remaining $\nabla_0 \mathrm{E}$ appearing in (6.58) are listed in (5.35) of [III].

### the relation between $Z^{(2)}_{1234}$ and $\mathcal{I}^{(2,0)}_{1234}$

When applied to the low-energy expansion (6.100) of $\mathcal{I}^{(2,0)}$ (and its third order in $\alpha'$ given in (E.3) of [III]), the projection $P_{\rm Re}$ removes all standalone instances of $\mathcal{E}_0$ but preserves the real parts $P_{\rm Re}\,\mathrm{Re}[\mathcal{E}_0] = \mathrm{Re}[\mathcal{E}_0]$:

$$\begin{aligned} P_{\rm Re}\big(\mathcal{I}^{(2,0)}_{1234}(s_{ij},\tau)\big) &= \pi^2 s_{13}\Big(\frac{2y}{15} - \frac{3\zeta_3}{y^2} + \frac{18}{y^2}\,\mathrm{Re}[\mathcal{E}_0(4,0,0)]\Big) \\ &+ \pi^2(s_{13}^2 + 2s_{12}s_{23})\Big(\frac{4y^2}{945} - \frac{\zeta_5}{y^3} + \frac{120}{y^2}\,\mathrm{Re}[\mathcal{E}_0(6,0,0,0)] \\ &\qquad + \frac{120}{y^3}\,\mathrm{Re}[\mathcal{E}_0(6,0,0,0,0)]\Big) \end{aligned}$$

$$
\begin{aligned}
&+\pi^2 s_{13}(s_{13}^2 - s_{12}s_{23})\Big(\frac{4y^3}{945}+\frac{2\zeta_3}{5}-\frac{5\zeta_5}{y^2}-\frac{3\zeta_7}{2y^4}+\Big(\frac{72\zeta_3}{y^2}-\frac{16y}{5}\Big)\,\mathrm{Re}[\mathcal{E}_0(4,0)] \\
&\quad -\frac{12}{5}\,\mathrm{Re}[\mathcal{E}_0(4,0,0)] - \frac{432}{y^2}\,\mathrm{Re}[\mathcal{E}_0(4,0)]\,\mathrm{Re}[\mathcal{E}_0(4,0,0)] \qquad (6.105)\\
&\quad +\frac{432}{y^2}\,\mathrm{Re}[\mathcal{E}_0(4,0,4,0,0)+3\,\mathcal{E}_0(4,4,0,0,0)+\tfrac{1}{360}\,\mathcal{E}_0(4,0,0,0,0,0)] \\
&\quad +\frac{4032}{y^2}\,\mathrm{Re}[\mathcal{E}_0(8,0,0,0,0,0)]+\frac{10080}{y^3}\,\mathrm{Re}[\mathcal{E}_0(8,0,0,0,0,0,0)] \\
&\quad +\frac{7560}{y^4}\,\mathrm{Re}[\mathcal{E}_0(8,0,0,0,0,0,0,0)]\Big) + O(\alpha'^4)\,.
\end{aligned}
$$

Up to a global prefactor $(2\pi i)^2$, this expression agrees with the esv image (6.98) of the open-string integral $Z^{(2)}_{1234}$. Hence, we have checked to the order of $\alpha'^3$ that

$$
P_{\mathrm{Re}}\big(\mathcal{I}^{(2,0)}_{1234}(s_{ij},\tau)\big) = (2\pi i)^2 \ \mathrm{esv}\, Z^{(2)}_{1234}(s_{ij},-\tfrac{1}{\tau})\,, \qquad (6.106)
$$

and conjecture this relation between open- and closed-string integrals to hold at higher orders as well. In the order-$\alpha'^3$ contribution (6.93) to $Z^{(2)}_{1234}(s_{ij},-\frac{1}{\tau})$, the product in the third line is understood to be mapped to esv $(\mathcal{E}_0(4,0)\,\mathcal{E}_0(4,0,0)) = (\mathrm{esv}\ \mathcal{E}_0(4,0))(\mathrm{esv}\ \mathcal{E}_0(4,0,0))$, see (6.98), i.e. without shuffle multiplication prior to the application of esv. Similar ad-hoc convention are expected to be possible at higher orders of $\mathcal{I}^{(2,0)}_{1234}$ and $Z^{(2)}_{1234}$ such as to satisfy (6.106).

Given that the $\alpha'$ expansion of $\mathcal{I}^{(2,0)}_{1234}$ is expressible in terms of MGFs, its expansion around the cusp is expected to be of the type (3.27),

$$
\mathcal{I}^{(2,0)}_{1234}(s_{ij},\tau) = \sum_{m,n=0}^{\infty} j_{m,n}(s_{ij},y)q^m\bar{q}^n\,. \qquad (6.107)
$$

The coefficients $j_{m,n}(s_{ij},y)$ are series in $s_{ij}$ such that each $\alpha'$-order comprises Laurent polynomials in $y$. Since the ambiguities in the evaluation of esv and $P_{\mathrm{Re}}$ were pointed out to be $O(q\,\bar{q})$, one can turn (6.106) into a well-defined conjecture by dropping terms $\sim q$,

$$
\mathcal{I}^{(2,0)}_{1234}(s_{ij},\tau) = (2\pi i)^2 \ \mathrm{esv}\, Z^{(2)}_{1234}(s_{ij},-\tfrac{1}{\tau}) + O(q)\,. \qquad (6.108)
$$

This form of our conjecture predicts all the coefficients $j_{0,n}(s_{ij},y)$ of $q^0\bar{q}^n$ in (6.107) with $n\in\mathbb{N}_0$ including the zero mode $j_{0,0}(s_{ij},y)$ from the open-string quantity $Z^{(2)}_{1234}$. The omission of $O(q)$-contributions in (6.108) bypasses both the need for the $P_{\mathrm{Re}}$ projection in (6.106) and the incompatibility of esv with the shuffle multiplication.

The modular weight $(2,0)$ of $\mathcal{I}^{(2,0)}_{1234}(s_{ij},\tau)$ is not at all evident from the relations (6.106) and (6.108) with open-string integrals. Hence, it should be possible to infer the coefficients $j_{m,n}(s_{ij},y)$ in (6.107) with $m\geq 1$ that do not have any known open-string counterpart from $j_{0,n}(s_{ij},y)$ via modular properties. This approach is particularly tractable as long as an ansatz of MGFs of suitable transcendental weight is available for a given order in $\alpha'$: For instance, suppose the $\alpha'^3$ order of $\mathcal{I}^{(2,0)}_{1234}$ is known to involve a $\mathbb{Q}$-linear combination of $\pi\nabla_0 \mathrm{E}_4$, $\mathrm{E}_2\pi\nabla_0\mathrm{E}_2$ and $\pi\nabla_0\mathrm{E}_{2,2}$, cf. (6.58). Then, the coefficients $c_1, c_2, c_3 \in \mathbb{Q}$ in an ansatz

$$\mathcal{I}^{(2,0)}_{1234}\big|_{\alpha'^3} = \frac{s_{13}(s_{13}^2 - s_{12}s_{23})}{\tau_2^2}(c_1\pi\nabla_0\mathrm{E}_4 + c_2\mathrm{E}_2\pi\nabla_0\mathrm{E}_2 + c_3\pi\nabla_0\mathrm{E}_{2,2}) \tag{6.109}$$

are uniquely determined to be $(c_1, c_2, c_3) = (\frac{4}{5}, 6, 12)$ by (6.106) and (6.108). At the $\alpha'^4$-order of $\mathcal{I}^{(2,0)}_{1234}$, one could envision a $(4+4)$-parameter ansatz comprising $\pi\nabla_0\mathrm{E}_5$, $\mathrm{E}_2\pi\nabla_0\mathrm{E}_3$, $\mathrm{E}_3\pi\nabla_0\mathrm{E}_2$ and $\pi\nabla_0\mathrm{E}_{2,3}$ along with both $s_{12}s_{23}s_{13}^2$ and $s_{12}^4 - 4s_{12}^2 s_{23}^2 + s_{23}^4$.

## INTEGRATION CYCLES VERSUS ELLIPTIC FUNCTIONS

It is amusing to compare the single-valued relation between genus-zero integrals with our present evidence for an elliptic single-valued correspondence between open and closed strings. At tree level, the single-valued map of MZVs was found to relate integration cycles on a disk boundary to Parke–Taylor factors $(z_{12}z_{23}\ldots z_{n1})^{-1}$. At genus one, the two links (6.96) and (6.106) between open- and closed-string $\alpha'$ expansions suggest that integration cycles on a cylinder boundary translate into combinations of the elliptic functions $V_a$ in (3.96).

It would be interesting to explain the correspondence between symmetrized open-string cycles and $V_0(1,2,\ldots,n) = 1$ as well as the four-point cycles of $\mathrm{Z}^{(2)}_{1234}$ and $V_2(1,2,3,4)$ from the viewpoint of Betti-deRham duality [250, 251]. The general dictionary between $V_a(1,2,\ldots,n)$ functions in a closed-string integrand and formal sums of integration cycles $\{(z_1,\ldots,z_n) \in \mathbb{R}^n,\ 0<z_1<z_2<\ldots<z_n<1\}$ on the open-string side is explored in [17].

One might wonder if the integral $\mathcal{I}^{(4,0)}_{1234}$ over the elliptic function $V_4(1,2,3,4)$ also admits an open-string correspondent along the lines of (6.96) and (6.106). However, the independent permutations of $V_0(1,2,3,4) = 1$ and $V_2(1,2,3,4)$ already exhaust the three combinations of four-point cycles that share the invariance under reflection $z_j \to 1 - z_j$ of the even-weight $V_{2k}(1,2,3,4)$. Moreover, since

$$\mathcal{I}^{(4,0)}_{1234} = \mathrm{G}_4(1 + 6s_{13}) + \frac{3s_{13}\widehat{\mathrm{G}}_2\pi\nabla_0\mathrm{E}_2}{\tau_2^2} + O(\alpha'^2) \tag{6.110}$$

violates uniform transcendentality and the open-string integral (4.8) satisfies uniform transcendentality, it might be hard to identify a suitable open-string integral with the same property. But it might be a more tractable problem to identify open-string counterparts for the conjecturally uniformly transcendental integrals $\widehat{\mathcal{I}}^{(4,0)}_{1234}$, $\widehat{\mathcal{I}}^{(2,0)}_{12|34}$ and $\widehat{\mathcal{I}}^{(4,0)}_{12|34}$ in (C.12), (C.20) and (C.26), respectively.

# 7

# DIFFERENTIAL EQUATIONS FOR A GENERATING SERIES OF MODULAR GRAPH FORMS

In the last two chapters we have first developed a host of simplification techniques for MGFs in Chapter 5, which we used to derive basis decompositions of MGFs with total modular weight at most 12, and then applied this in Chapter 6 to four-gluon scattering in the heterotic string. Although this could prove the power of the techniques developed in Chapter 5, it is hard to derive general statements for MGFs from these results and the basis decompositions become quickly more laborious for higher weights and topologies, as the numbers of MGFs for the different weights in Table 5.2 show. For this reason, we take a different approach in this Chapter: We will define a generating function of Koba–Nielsen integrals which, since Koba–Nielsen integrals can be expanded in MGFs, is also a generating function for MGFs. We will derive the Cauchy–Riemann and Laplace equations in $\tau$ of this generating series and obtain in this way infinite towers of Cauchy–Riemann and Laplace equations for MGFs of arbitrary weight. Furthermore, we open the door towards a more systematic analysis of the space of MGFs via iterated Eisenstein integrals which will be performed in Chapter 8 and obtain a new perspective on the elliptic single-valued map via a comparison to a similar differential equation in the open string [37, 38]. The results exhibited in this chapter were published in [IV] and the present text has extensive overlap with this reference.

This chapter is structured as follows: We start in Section 7.1 by defining the generating function of Koba–Nielsen integrals whose differential equation we want to study in the remainder of this chapter and illustrate its expansion in MGFs via some two- and three-point examples. In Section 7.2, we first derive some prerequisite differential equations satisfied by the Kronecker–Eisenstein series (3.82) and the Koba–Nielsen factor (3.72) and use these to obtain the two-point instances of the Cauchy–Riemann and Laplace equations satisfied by the generating function. In Section 7.3 we derive the $n$-point Cauchy–Riemann equation of the generating function and exhibit the three- and four-point cases. We finish with a derivation of the $n$-point Laplace equation in Section 7.4 and a discussion of its three-point instance.

## 7.1 BASICS OF GENERATING FUNCTIONS FOR ONE-LOOP STRING INTEGRALS

In this section, we will define the generating function of Koba–Nielsen integrals whose differential equations we want to derive in this Chapter and discuss their basic properties.

### 7.1.1 *Introducing generating functions for world-sheet integrals*

The main results of this chapter concern the following $(n-1)! \times (n-1)!$ matrix of integrals over the punctures

$$\begin{aligned} W^{\tau}_{\vec{\eta}}(\sigma|\rho) &= W^{\tau}_{\vec{\eta}}(1,\sigma(2,\ldots,n)|1,\rho(2,\ldots,n)) \qquad (7.1)\\ &= \int \mathrm{d}\mu_{n-1}\, \mathrm{KN}_n\, \rho\Big[\Omega(z_{12},\eta_{23\ldots n},\tau)\,\Omega(z_{23},\eta_{34\ldots n},\tau)\cdots\Omega(z_{n-1,n},\eta_n,\tau)\Big]\\ &\quad\times\sigma\Big[\overline{\Omega(z_{12},\eta_{23\ldots n},\tau)}\,\overline{\Omega(z_{23},\eta_{34\ldots n},\tau)}\cdots\overline{\Omega(z_{n-1,n},\eta_n,\tau)}\Big]\,, \end{aligned}$$

which is defined by the Koba–Nielsen factor (3.72) and the doubly-periodic Kronecker–Eisenstein series (3.82). The matrix elements of $W^{\tau}_{\vec{\eta}}(\sigma|\rho)$ are parametrized by permutations $\rho,\sigma \in \mathcal{S}_{n-1}$ that act separately on the $\Omega(\ldots)$ and $\overline{\Omega(\ldots)}$. The parameters of the Kronecker–Eisenstein series are

$$\eta_{i,i+1\ldots n} = \eta_i + \eta_{i+1} + \ldots + \eta_n\,. \qquad (7.2)$$

The permutations $\rho$ and $\sigma$ act on the points $z_i$ and parameters $\eta_i$ by permutation of the indices.

Due to the phase $\exp(2\pi i\eta\frac{\mathrm{Im}\,z}{\tau_2})$ in its definition (3.82), the doubly-periodic Kronecker–Eisenstein series is not meromorphic in $z$ or $\tau$. Accordingly we will refer to $\Omega(z,\eta,\tau)$ and $\overline{\Omega(z,\eta,\tau)}$ as *chiral* and *anti-chiral*, respectively. Still, $\Omega(z,\eta,\tau)$ is a meromorphic function of its second argument $\eta$.

The open-string analogues of the $W$-integrals are given by [37, 38]

$$\begin{aligned} Z^{\tau}_{\vec{\eta}}(\sigma|\rho) &= \int_{C(\sigma)} \mathrm{d}z_2\mathrm{d}z_3\ldots\mathrm{d}z_n\, \exp\Big(\sum_{1\le i<j}^{n} s_{ij}G_A(z_i-z_j,\tau)\Big) \qquad (7.3)\\ &\quad\times\rho\big[\Omega(z_{12},\eta_{23\ldots n},\tau)\,\Omega(z_{23},\eta_{34\ldots n},\tau)\cdots\Omega(z_{n-1,n},\eta_n,\tau)\big]\,, \end{aligned}$$

with $\sigma,\rho\in\mathcal{S}_{n-1}$, and the planar open-string Green function $G_A$ on the $A$-cycle reads (cf. (4.2))

$$G_A(z,\tau) = -\log\left(\frac{\theta(z,\tau)}{\eta(\tau)}\right) + \frac{i\pi\tau}{6} + \frac{i\pi}{2}\,. \qquad (7.4)$$

The integration domain $C(\sigma)$ prescribes the cyclic ordering $0 = z_1 < z_{\sigma(2)} < z_{\sigma(3)} < \ldots < z_{\sigma(n)} < 1$ of the punctures on the $A$-cycle of a torus (cf. Figure 4.1), that is why the $Z^{\tau}_{\vec{\eta}}(\sigma|\rho)$ will be referred to as $A$-cycle integrals henceforth. With this restriction to $z \in \mathbb{R}$, the open-string Green function (7.4) shares the holomorphic derivative $\partial_z G_A(z,\tau) = \partial_z G(z,\tau) = -f^{(1)}(z,\tau)$ of its closed-string counterpart (3.61), and the addition of $\frac{i\pi\tau}{6}+\frac{i\pi}{2}$ enforces that $\int_0^1 G_A(z,\tau)\mathrm{d}z = 0$ [34, 249]. The above choice of $C(\sigma)$ allows to generate cylinder- and Möbius-strip contributions to planar one-loop open-string amplitudes from (7.3) by restricting $\tau \in i\mathbb{R}_+$ and $\tau \in \frac{1}{2} + i\mathbb{R}_+$, respectively [51], as discussed in Section 4.1. Generalization to non-planar $A$-cycle integrals can be found in [37, 38].

## COMPONENT INTEGRALS AND STRING AMPLITUDES

The $W$-integrals in (7.1) are engineered to generate the integrals over torus punctures in closed-string one-loop amplitudes upon expansion in the $\eta_j$ and $\bar{\eta}_j$ variables. The expansion (3.87) of the doubly-periodic Kronecker–Eisenstein integrands introduces component integrals

$$\begin{aligned} W^{\tau}_{(A|B)}(\sigma|\rho) &= W^{\tau}_{(a_2,a_3,\ldots,a_n|b_2,b_3,\ldots,b_n)}(\sigma|\rho) \\ &= \int \mathrm{d}\mu_{n-1}\, \mathrm{KN}_n\, \rho\big[f^{(a_2)}_{12} f^{(a_3)}_{23} \ldots f^{(a_n)}_{n-1,n}\big]\, \sigma\big[\overline{f^{(b_2)}_{12}}\, \overline{f^{(b_3)}_{23}} \ldots \overline{f^{(b_n)}_{n-1,n}}\big] \end{aligned} \tag{7.5}$$

with $a_i, b_i \geq 0$. Note that the $z_{ij}$ arguments of the $f^{(a_k)}_{ij}$ with weights from the first index set $A = a_2, a_3, \ldots, a_n$ are permuted with the permutation $\rho$ in the second slot of the argument of $W^{\tau}_{(A|B)}$ and vice versa. This is to ensure consistency with the notation (7.3) of open-string integrals and also to have $W^{\tau}_{(A|B)}$ carry modular weight $(|A|, |B|)$, where

$$|A| = \sum_{i=2}^{n} a_i\,, \qquad |B| = \sum_{i=2}^{n} b_i\,. \tag{7.6}$$

The component integral $W^{\tau}_{(A|B)}(\sigma|\rho)$ can be extracted from its generating series $W^{\tau}_{\vec{\eta}}(\sigma|\rho)$ by isolating the coefficients of the parameters (7.2)

$$\begin{aligned} W^{\tau}_{\vec{\eta}}(\sigma|\rho) = \sum_{A,B} W^{\tau}_{(A|B)}(\sigma|\rho)\, \rho\big[\eta^{a_2-1}_{234\ldots n}\eta^{a_3-1}_{34\ldots n} \cdots \eta^{a_n-1}_{n}\big] \\ \times \sigma\big[\bar{\eta}^{b_2-1}_{234\ldots n}\bar{\eta}^{b_3-1}_{34\ldots n} \cdots \bar{\eta}^{b_n-1}_{n}\big]\,. \end{aligned} \tag{7.7}$$

Here and in the rest of this thesis, we use the abbreviating notation

$$\sum_{A,B} = \sum_{a_2,a_3,\ldots,a_n=0}^{\infty} \; \sum_{b_2,b_3,\ldots,b_n=0}^{\infty}\,. \tag{7.8}$$

Note that the permutations $\rho, \sigma$ in (7.5) and (7.7) only act on the subscripts of $z_{ij}$ and $\eta_j$ but not on the superscripts $a_i$ and $b_j$. The component integrals satisfy the reality condition

$$\overline{W^{\tau}_{(A|B)}(\sigma|\rho)} = W^{\tau}_{(B|A)}(\rho|\sigma)\,. \tag{7.9}$$

Component integrals of the type in (7.5) arise from the CFT correlators underlying one-loop amplitudes of closed bosonic strings, heterotic strings and type-II superstrings [III]. More specifically, the $f_{ij}^{(a)}$ were found to appear naturally from the spin sums of the RNS formalism [28] and the current algebra of heterotic strings [174].[1] For these theories, the $(n-1)! \times (n-1)!$ matrix in (7.5) is in fact claimed to contain a basis of the integrals that arise in string theory[2] for any massless one-loop amplitude. Moreover, massive-state amplitudes are likely to fall into the same basis.

The massless four-point one-loop integrand of type-II superstrings [1] for instance is proportional to the four-point component integrals $W^{\tau}_{(0,0,0|0,0,0)}$. Similarly, the five-point type-II amplitude involves $W^{\tau}_{(0,0,0,0|0,0,0,0)}$ and various permutations of the integrals $W^{\tau}_{(1,0,0,0|1,0,0,0)}$ and $W^{\tau}_{(0,1,0,0|1,0,0,0)}$ [169, 170, 253]. For a specific example at reduced supersymmetry, consider the integral $\mathcal{I}^{(2,0)}_{1234}$ defined in (6.28). The prefactor $V_2(1,2,3,4)$ of the Koba–Nielsen factor is given in terms of $f_{ij}^{(1)}$ and $f_{ij}^{(2)}$ in (6.37). The integral corresponding to each of these terms can then be written as a component integral (7.5). This only fails for the contribution $f_{41}^{(1)} f_{12}^{(1)}$ in $V_2(1,2,3,4)$, since the puncture $z_1$ is repeated. The integral over this factor can be translated into $W^{\tau}_{(A|B)}$ by means of the Fay identity (5.121) which implies

$$f_{12}^{(1)} f_{41}^{(1)} = f_{14}^{(1)} f_{24}^{(1)} + f_{12}^{(1)} f_{42}^{(1)} - f_{24}^{(2)} - f_{14}^{(2)} - f_{12}^{(2)} , \tag{7.10}$$

with no repeated $z_1$ in any term on the RHS. The $n$-point systematics and the role of $W^{\tau}_{(A|B)}$ at higher $A, B$ in the context of reduced supersymmetry are detailed in Appendix D.1.

The main motivation of this chapter is to study the $\alpha'$ expansion of component integrals (7.5) via generating-function methods. As detailed in Section 7.1.2 below, the coefficients in such $\alpha'$ expansions are torus integrals over Green functions as well as products of $f_{ij}^{(a)}$ and $f_{ij}^{(b)}$. At each order in $\alpha'$, these integrals fall into the framework of modular graph functions / forms. As will be demonstrated in later sections, the $W$-integrals (7.1) allow for streamlined derivations of differential equations for infinite families of MGFs.

1 Also see e.g. [10, 101, 120] for earlier work on RNS spin sums, [102–104, 172, 245] for $g_{ij}^{(a)}$ and $f_{ij}^{(a)}$ in one-loop amplitudes in the pure-spinor formalism and [121, 122] for applications to RNS one-loop amplitudes with reduced supersymmetry.

2 String-theory integrals can also involve integrals over $\partial_{z_i} f_{ij}^{(a)}$ or $f_{ij}^{(a_1)} f_{ij}^{(a_2)}$ that are not in the form of (7.5) but can be reduced to the conjectural basis by means of Fay identities and integration by parts w.r.t. the punctures. Similar reductions should be possible for products of $\partial_{z_i} f_{ij}^{(a)}$ or cycles $f_{i_1 i_2}^{(a_1)} f_{i_2 i_3}^{(a_2)} \ldots f_{i_k i_1}^{(a_k)}$ by adapting the recursive techniques of [226, 229, 252] to a genus-one setup.

### RELATIONS BETWEEN COMPONENT INTEGRALS

It is important to stress that the component integrals $W^{\tau}_{(A|B)}(\sigma|\rho)$ are not all linearly independent. There are two simple mechanisms that lead to relations between certain special cases of component integrals. Still, component integrals $W_{(A|B)}$ with generic weights $A, B$ are not affected by the subsequent relations, that is why they do not propagate to relations between the $(n-1)! \times (n-1)!$ generating series in (7.1).

Firstly, there can be relations between different $W^{\tau}_{(A|B)}(\sigma|\rho)$ stemming from the fact that the functions $f^{(a)}_{ij}$ entering in (7.5) satisfy $f^{(a)}_{ij} = (-1)^a f^{(a)}_{ji}$ and similarly for the $\overline{f^{(b)}_{ij}}$. Since these parity properties interchange points they intertwine with the permutations $\rho$ and $\sigma$. For instance, if the last two entries of $A$ and $B$ are $A = (a_2, \ldots, a_{n-2}, 0, a_n)$ and $B = (b_2, \ldots, b_{n-2}, 0, b_n)$, respectively, then the only places where the points $z_{n-1}$ and $z_n$ appear are $f^{(a_n)}_{n-1,n}$, $\overline{f^{(b_n)}_{n-1,n}}$ and in the permutation invariant Koba–Nielsen factor. Applying the parity transformation to these factors of $f^{(a_n)}$ and $\overline{f^{(b_n)}}$ therefore can be absorbed by composing the permutations $\rho$ and $\sigma$ with the transposition $n-1 \leftrightarrow n$ and an overall sign $(-1)^{a_n+b_n}$. This yields a simple instance of an algebraic relation between the component integrals and we shall see an explicit instance of this for three points in Section 7.3.2 below.

The second mechanism is integration by parts – integrals of total $z$-derivatives (or $\bar{z}$-derivatives) vanish due to the presence of the Koba–Nielsen factor. Such derivatives produce sums over $s_{ij} f^{(1)}_{ij}$ from the Koba–Nielsen factor (see (5.167)) and may also involve $\partial_{z_i} f^{(a)}_{ij}$. Integration by parts in combination with the Fay identities (5.121) can first of all be used to eliminate derivatives of $f^{(a)}_{ij}$ and conjecturally any integrand that does not line up with the form of $W^{\tau}_{(A|B)}(\sigma|\rho)$. Moreover, component integrals with $a_i, b_j = 1$ for some of the weights can be related by the $s_{ij} f^{(1)}_{ij}$ from the Koba–Nielsen derivatives. We note that these integration-by-parts relations can mix component integrals of different modular weight as they can also contain explicit instances of $\tau_2$. A two-point instance of such an integration-by-parts identity among component integrals can be found in (7.14) below.

### 7.1.2 *Low-energy expansion of component integrals*

The component integrals $W^{\tau}_{(A|B)}(\sigma|\rho)$ introduced in (7.5) depend on the Mandelstam variables $s_{ij}$ through the Koba–Nielsen factor (3.72). In this section, we will study the expansion of the $W^{\tau}_{(A|B)}(\sigma|\rho)$ in the Mandelstam variables, similarly to the expansion of the heterotic Koba–Nielsen integrals in Section 6.2. Since the Mandelstams as defined in (2.25) carry a factor of $\alpha'$, this expansion is also an expansion in $\alpha'$ and therefore corresponds to the low-energy expansion of the corresponding amplitude in the string-theory context.

Expanding the integrands of the component integrals $W^\tau_{(A|B)}(\sigma|\rho)$ in the $s_{ij}$ leads to

$$W^\tau_{(A|B)}(\sigma|\rho) = \int \mathrm{d}\mu_{n-1} \Bigg[ \prod_{1\leq i<j}^{n} \sum_{k_{ij}=0}^{\infty} \frac{1}{k_{ij}!} \big[ s_{ij} G(z_{ij},\tau) \big]^{k_{ij}} \Bigg] \\ \times \rho[f^{(a_2)}_{12} f^{(a_3)}_{23} \ldots f^{(a_n)}_{n-1,n}] \, \sigma[\overline{f^{(b_2)}_{12}}\, \overline{f^{(b_3)}_{23}} \ldots \overline{f^{(b_n)}_{n-1,n}}] \,, \tag{7.11}$$

where the contributions to the $(\alpha')^w$-order satisfy $\sum^n_{1\leq i<j} k_{ij} = w$. Once the $f^{(a)}_{ij}, \overline{f^{(b)}_{ij}}$ and $G(z_{ij},\tau)$ in the integrand are identified with the doubly-periodic functions $C^{(a,b)}_{ij}$ via (3.119), each term in the $\alpha'$ expansion of (7.11) is lined up with the integral representation (3.120) of MGFs. This shows that the coefficients at the $(\alpha')^w$-order of $W^\tau_{(A|B)}(\sigma|\rho)$ are MGFs of weight $(|A|+w,|B|+w)$ multiplied by powers of $\tau_2$ to harmonize the modular weights, see below for $(n\leq 3)$-point examples. Since MGFs vanish if the sum of their holomorphic and antiholomorphic modular weight is odd, also $W^\tau_{(A|B)}(\sigma|\rho) = 0$ if $|A| + |B|$ is odd.

Conversely, any convergent MGF can be realized through the $\alpha'$ expansion (7.11) of suitably chosen component integrals. The topology of the defining graph determines the minimal multiplicity $n$ that admits such a realization. For instance, the spanning set $C\big[\begin{smallmatrix} a & b & 0 \\ c & 0 & d \end{smallmatrix}\big]$ of two-loop MGFs [185][3] with $a,b,c,d \neq 0$ and $(b,d) \neq (1,1)$ arises at the $\alpha'^0$-order of the three-point component integral $W^\tau_{(a,b|c,d)}(3{,}2|2{,}3)$ over $f^{(a)}_{12} f^{(b)}_{23} \overline{f^{(c)}_{13}}\, \overline{f^{(d)}_{32}}$ since the intermediate point 1 is two-valent and can be contracted using (3.132). Further examples of dihedral and trihedral topology can be found in Table 7.1.

However, the $\alpha'$ expansion (7.11) at the level of the integrand is not applicable in presence of singularities $|z_{ij}|^{-2}$, i.e. in case of real combinations $|f^{(1)}_{ij}|^2 = f^{(1)}_{ij} \overline{f^{(1)}_{ij}}$. By the local behavior $\mathrm{KN}_n \sim |z_{ij}|^{-2s_{ij}}$ of the Koba–Nielsen factor as $z_i \to z_j$, the integration region over $|z_{ij}| \ll 1$ yields kinematic poles $\sim s^{-1}_{ij}$. Still, component integrals $W^\tau_{(A|B)}(\sigma|\rho)$ with integrands $\sim |f^{(1)}_{ij}|^2$ can be Laurent-expanded via suitable integration-by-parts manipulations as reviewed in Section 5.6.2. Since the residues of the kinematic poles[4] are expressible in terms of Koba–Nielsen integrals at lower multiplicity, any contribution to such manipulations can be integrated in the framework of MGFs.

3 In particular, the odd two-loop MGFs $\mathcal{A}_{u,v;w} = ((\frac{\tau_2}{\pi})^w (C\big[\begin{smallmatrix} w-u & u & 0 \\ w-v & 0 & v \end{smallmatrix}\big] - C\big[\begin{smallmatrix} w-v & v & 0 \\ w-u & 0 & u \end{smallmatrix}\big])$ studied in [185] arise at the $\alpha' \to 0$ limit of the component integrals $\mathrm{Im}\, W^\tau_{(w-u,u|w-v,v)}(3{,}2|2{,}3)$. The simplest odd MGF $\mathcal{A}_{1,2;5}$ which is not expressible in terms of $\nabla^n_0 \mathrm{E}_k$ in (5.56) and their complex conjugates is generated by the component integral $\mathrm{Im}\, W^\tau_{(4,1|3,2)}(3{,}2|2{,}3)$ over $\mathrm{Im}\,(f^{(4)}_{12} f^{(1)}_{23} \overline{f^{(3)}_{13}}\, \overline{f^{(2)}_{32}})$.

4 In the same way as $|f^{(1)}_{ij}|^2\, \mathrm{KN}_n$ integrates to kinematic poles $\sim s^{-1}_{ij}$, integrands with $(p-1)$ factors of $f^{(1)}_{i_a i_b}$ and $\overline{f^{(1)}_{i_a i_b}}$ each in the range $1 \leq a < b \leq p$ give rise to poles $\sim s^{-1}_{i_1 i_2 \ldots i_p}$ in multiparticle channels, cf. (2.27). Pole structures of this type can still be accounted for via analogous integrations-by-parts, and the residues are again expressible in terms of lower-multiplicity integrals.

Note that the differential-equation approach of the next sections does not require any tracking of kinematic poles, and our results do not rely on any rewritings of the integrals.

### TWO-POINT EXAMPLES

At two points, the generating function and the component integrals do not involve any on permutations and are given by

$$W^\tau_\eta = \sum_{a,b=0}^{\infty} \eta^{a-1}\bar{\eta}^{b-1} W^\tau_{(a|b)}\,, \quad W^\tau_{(a|b)} = \int \frac{\mathrm{d}^2 z_2}{\tau_2} f^{(a)}_{12}\overline{f^{(b)}_{12}}\,\mathrm{KN}_2\,, \tag{7.12}$$

where we have denoted $\eta = \eta_2$ for simplicity. By identifying the Green function as $\frac{\tau_2}{\pi}C^{(1,1)}(z,\tau)$ in (7.11), one can easily arrive at closed formulae for the $\alpha'$ expansion of $W^\tau_{(a|b)}$ in terms of dihedral MGFs (3.122)

$$\begin{aligned}
W^\tau_{(0|0)} &= 1 + \sum_{k=2}^{\infty} \frac{s_{12}^k}{k!}\Big(\frac{\tau_2}{\pi}\Big)^k C\big[\begin{smallmatrix} 1_k \\ 1_k \end{smallmatrix}\big] \\
W^\tau_{(a|0)} &= -\sum_{k=1}^{\infty} \frac{s_{12}^k}{k!}\Big(\frac{\tau_2}{\pi}\Big)^k C\big[\begin{smallmatrix} a & 1_k \\ 0 & 1_k \end{smallmatrix}\big]\,, \qquad\qquad a>0 \\
W^\tau_{(a|b)} &= (-1)^a \sum_{k=0}^{\infty} \frac{s_{12}^k}{k!}\Big(\frac{\tau_2}{\pi}\Big)^k C\big[\begin{smallmatrix} a & 0 & 1_k \\ 0 & b & 1_k \end{smallmatrix}\big]\,, \quad a,b>0\,, \quad (a,b)\neq(1,1)\,,
\end{aligned} \tag{7.13}$$

where $1_k$ denotes the row vector with $k$ entries of 1. The expansion of $W^\tau_{(0|b)}$ can be obtained by complex conjugating the expansion of $W^\tau_{(a|0)}$.

The $\alpha'$ expansion of $W^\tau_{(1|1)}$ requires extra care since the singularity $|f^{(1)}_{12}|^2 \sim \frac{1}{|z_2|^2}$ of the integrand leads to a kinematic pole in $s_{12}$ as mentioned above. We can make these poles explicit through the integration-by-parts-identities

$$s_{12}W^\tau_{(a|1)} + \frac{2\pi i}{\tau-\bar{\tau}}W^\tau_{(a-1|0)} = 0\,, \quad s_{12}W^\tau_{(1|b)} + \frac{2\pi i}{\tau-\bar{\tau}}W^\tau_{(0|b-1)} = 0\,, \tag{7.14}$$

which can be checked by evaluating the total derivative $\partial_{\bar{z}_2}(f^{(a)}_{12}\,\mathrm{KN}_2)$ via (3.94) and (5.167) and result in

$$W^\tau_{(1|1)} = -\frac{\pi}{s_{12}\tau_2}W^\tau_{(0|0)} = -\frac{\pi}{s_{12}\tau_2} - \sum_{k=2}^{\infty}\frac{s_{12}^{k-1}}{k!}\Big(\frac{\tau_2}{\pi}\Big)^{k-1} C\big[\begin{smallmatrix} 1_k \\ 1_k \end{smallmatrix}\big]\,, \tag{7.15}$$

cf. (5.169). Note that similar integrations by parts should suffice to rewrite higher-multiplicity component integrals $W^\tau_{(A|B)}(\sigma|\rho)$ with kinematic poles in terms of regular representatives with a Taylor expansion in $s_{ij}$. The kinematic poles will then appear as the expansion coefficients such as the factor of $s_{12}^{-1}$ in the first step of (7.15), see Appendix B.3 for a complete analysis of the poles in three-point integrals.

Although the $\alpha'$ expansion in terms of lattice sums could be quickly generated from (7.11), the representation in (7.13) is not optimal since many non-trivial identities between MGFs exist [15, 16, 39, 40, 127], as discussed in Chapter 5. As we saw in Section 5.7, these identities can be used to reduce the above expansions into a basis of lattice sums. At the lowest orders, these bases are given by the non-holomorphic Eisenstein series $\mathrm{E}_k$, their higher-depth analogues defined in (4.28) and the derivatives $\nabla_0 \mathrm{E}$ with the Cauchy–Riemann operator defined in (3.55). We have for example

$$\begin{aligned}
W^\tau_{(0|0)} &= 1 + \frac{1}{2}s_{12}^2\mathrm{E}_2 + \frac{1}{6}s_{12}^3(\mathrm{E}_3+\zeta_3) + s_{12}^4\left(\frac{\mathrm{E}_2^2}{8} + \frac{3\mathrm{E}_4}{20} + \mathrm{E}_{2,2}\right)\\
&\quad + s_{12}^5\left(\frac{\mathrm{E}_{2,3}}{2} + \frac{\mathrm{E}_2}{12}(\mathrm{E}_3+\zeta_3) + \frac{3\mathrm{E}_5}{14} + \frac{2\zeta_5}{15}\right) + O(s_{12}^6)\\
W^\tau_{(2|0)} &= \frac{\pi}{\tau_2^2}\Bigg[-\frac{1}{2}s_{12}\nabla_0\mathrm{E}_2 - \frac{1}{6}s_{12}^2\nabla_0\mathrm{E}_3\\
&\quad + s_{12}^3\left(-\frac{1}{4}\mathrm{E}_2\nabla_0\mathrm{E}_2 - \frac{3}{20}\nabla_0\mathrm{E}_4 - \nabla_0\mathrm{E}_{2,2}\right) + O(s_{12}^4)\Bigg]\\
W^\tau_{(4|0)} &= \frac{\pi^2}{\tau_2^4}\left[-\frac{1}{12}s_{12}\nabla_0^2\mathrm{E}_3 + s_{12}^2\left(\frac{1}{8}(\nabla_0\mathrm{E}_2)^2 - \frac{3}{40}\nabla_0^2\mathrm{E}_4\right) + O(s_{12}^3)\right]\\
W^\tau_{(3|1)} &= \frac{\pi^2}{\tau_2^3}\Bigg[\frac{1}{2}\nabla_0\mathrm{E}_2 + \frac{1}{6}s_{12}\nabla_0\mathrm{E}_3\\
&\quad + s_{12}^2\left(\frac{1}{4}\mathrm{E}_2\nabla_0\mathrm{E}_2 + \frac{3}{20}\nabla_0\mathrm{E}_4 + \nabla_0\mathrm{E}_{2,2}\right) + O(s_{12}^3)\Bigg]\\
W^\tau_{(2|2)} &= \left(\frac{\pi}{\tau_2}\right)^2\Bigg[\mathrm{E}_2 + s_{12}\mathrm{E}_3\\
&\quad + s_{12}^2\left(\frac{1}{4}\frac{(\overline{\nabla}_0\mathrm{E}_2)(\nabla_0\mathrm{E}_2)}{\tau_2^2} - \frac{1}{2}\mathrm{E}_2^2 + \frac{9}{5}\mathrm{E}_4 + 2\mathrm{E}_{2,2}\right) + O(s_{12}^3)\Bigg]\\
W^\tau_{(4|2)} &= \frac{\pi^3}{\tau_2^4}\Bigg[\frac{1}{3}\nabla_0\mathrm{E}_3 + s_{12}\left(\frac{3}{4}\nabla_0\mathrm{E}_4 - \frac{1}{2}\mathrm{E}_2\nabla_0\mathrm{E}_2\right)\\
&\quad + s_{12}^2\left(\frac{27}{14}\nabla_0\mathrm{E}_5 + \nabla_0\mathrm{E}_{2,3} + \frac{1}{24}\frac{(\overline{\nabla}_0\mathrm{E}_2)\nabla_0^2\mathrm{E}_3}{\tau_2^2} - \frac{1}{2}\mathrm{E}_3\nabla_0\mathrm{E}_2 - \frac{2}{3}\mathrm{E}_2\nabla_0\mathrm{E}_3\right)\\
&\quad + O(s_{12}^3)\Bigg]\,,
\end{aligned} \tag{7.16}$$

where the simplified $\alpha'$ expansion of $W^\tau_{(1|1)}$ follows from inserting the first line of (7.16) into (7.15). Note that none of the MGFs on the right-hand sides of (7.13) is amenable to HSR as discussed in Section 5.4. In fact, since the $z_{ij}$ arguments of the chiral or anti-chiral Kronecker–Eisenstein integrands of the $W$-integrals do not form any cycles, none of the MGFs in the $n$-point $\alpha'$ expansions (7.11) will allow for HSR.

### THREE-POINT EXAMPLES

From three points onward, the generating functions and component integrals start depending on a chiral and an anti-chiral permutation. Following (7.5), we introduce three-point component integrals by

$$W^{\tau}_{\eta_2,\eta_3}(\sigma|\rho) = \sum_{a_2,a_3=0}^{\infty}\sum_{b_2,b_3=0}^{\infty} \rho[\eta_{23}^{a_2-1}\eta_3^{a_3-1}]\sigma[\bar{\eta}_{23}^{b_2-1}\bar{\eta}_3^{b_3-1}] \times W^{\tau}_{(a_2,a_3|b_2,b_3)}(\sigma|\rho) \tag{7.17}$$

$$W^{\tau}_{(a_2,a_3|b_2,b_3)}(\sigma|\rho) = \int \frac{\mathrm{d}^2 z_2}{\tau_2}\frac{\mathrm{d}^2 z_3}{\tau_2}\, \rho[f_{12}^{(a_2)} f_{23}^{(a_3)}]\sigma[\overline{f_{12}^{(b_2)}}\,\overline{f_{23}^{(b_3)}}]\,\mathrm{KN}_3\,, \tag{7.18}$$

where $\rho,\sigma \in \mathcal{S}_2$ act on the subscripts $i,j \in \{2,3\}$ of the $\eta$ and $\bar{\eta}$ in (7.17) and of $f^{(n)}$ and $\overline{f^{(n)}}$ in (7.18) but not on those of $a_i$ and $b_j$.

As in the two-point case, kinematic poles arise if the integrand develops a $\frac{1}{|z|^2}$ singularity in (some combination of) the punctures. The details of how to treat these poles using integration-by-parts-identities are spelled out in Appendix B.3.

In contrast to the two-point case, the three-point $\alpha'$ expansions also contain trihedral MGFs as defined in (5.8). Nevertheless, using the identities from Chapter 5, the leading orders displayed below can also be brought into the basis spanned by the $\mathrm{E}_k$, their higher-depth generalizations and derivatives:

$$\begin{aligned}
W^{\tau}_{(0,0|0,0)}(2,3|2,3) &= 1 + \frac{1}{2}(s_{12}^2+s_{13}^2+s_{23}^2)\mathrm{E}_2 + \frac{1}{6}(s_{12}^3+s_{13}^3+s_{23}^3)(\mathrm{E}_3+\zeta_3)\\
&\quad + s_{12}s_{13}s_{23}\mathrm{E}_3 + O(s_{12}^4)\\
W^{\tau}_{(2,0|2,0)}(2,3|2,3) &= \left(\frac{\pi}{\tau_2}\right)^2\Big[\mathrm{E}_2 + s_{12}\mathrm{E}_3 + \frac{1}{2}(s_{13}^2+s_{23}^2-s_{12}^2)\mathrm{E}_2^2\\
&\quad + s_{13}s_{23}\Big(\frac{3}{2}\mathrm{E}_2^2 - \frac{33}{10}\mathrm{E}_4 - 2\mathrm{E}_{2,2}\Big)\\
&\quad + s_{12}^2\Big(\frac{1}{4}\frac{\nabla_0\mathrm{E}_2\overline{\nabla}_0\mathrm{E}_2}{\tau_2^2} + \frac{9}{5}\mathrm{E}_4 + 2\mathrm{E}_{2,2}\Big) + O(s_{12}^3)\Big]\\
W^{\tau}_{(2,0|2,0)}(2,3|3,2) &= \left(\frac{\pi}{\tau_2}\right)^2\Big[s_{23}\mathrm{E}_3 + (s_{12}+s_{13})s_{23}\Big(-\frac{1}{2}\mathrm{E}_2^2 + \frac{39}{20}\mathrm{E}_4 + \frac{1}{2}\mathrm{E}_{2,2}\Big)\\
&\quad + s_{23}^2\Big(\frac{9}{20}\mathrm{E}_4 + \frac{1}{2}\mathrm{E}_{2,2}\Big) + s_{12}s_{13}\frac{1}{4}\frac{\nabla_0\mathrm{E}_2\overline{\nabla}_0\mathrm{E}_2}{\tau_2^2} + O(s_{12}^3)\Big]\\
W^{\tau}_{(1,2|1,2)}(2,3|3,2) &= \left(\frac{\pi}{\tau_2}\right)^3\Big[\mathrm{E}_3 - (s_{12}+s_{13}+s_{23})\mathrm{E}_2^2\\
&\quad + (s_{12}+s_{13}+2s_{23})\Big(\frac{9}{5}\mathrm{E}_4 + 2\mathrm{E}_{2,2} + \frac{1}{4}\frac{(\nabla_0\mathrm{E}_2)(\overline{\nabla}_0\mathrm{E}_2)}{\tau_2^2}\Big) + O(s_{12}^2)\Big]\,.
\end{aligned} \tag{7.19}$$

| MGF | Koba–Nielsen-prefactor | component integral |
|---|---|---|
| $C\left[\begin{smallmatrix} a & 1_k \\ b & 1_k \end{smallmatrix}\right]$ | $f^{(a)}_{12}\,\overline{f^{(b)}_{13}}$ | $W^{\tau}_{(a,0\mid b,0)}(3,2\mid 2,3)\Big\vert_{s^k_{23}}$ |
| $C\left[\begin{smallmatrix} a & b & 0 & 1_k \\ c & 0 & d & 1_k \end{smallmatrix}\right]$ | $f^{(a)}_{12}\,f^{(b)}_{23}\,\overline{f^{(c)}_{13}}\,\overline{f^{(d)}_{32}}$ | $W^{\tau}_{(a,b\mid c,d)}(3,2\mid 2,3)\Big\vert_{s^k_{23}}$ |
| $C\left[\begin{smallmatrix} a & b & 1_k \\ c & d & 1_k \end{smallmatrix}\right]$ | $f^{(a)}_{12}\,f^{(b)}_{24}\,\overline{f^{(c)}_{13}}\,\overline{f^{(d)}_{34}}$ | $W^{\tau}_{(a,b,0\mid c,d,0)}(3,4,2\mid 2,4,3)\Big\vert_{s^k_{23}}$ |
| $C\left[\begin{smallmatrix} a & 1_k \\ d & 1_k \end{smallmatrix}\middle\vert\begin{smallmatrix} b & 0 \\ 0 & v \end{smallmatrix}\middle\vert\begin{smallmatrix} c & 0 \\ 0 & u \end{smallmatrix}\right]$ | $f^{(a)}_{12}\,f^{(b)}_{23}\,f^{(c)}_{34}\,\overline{f^{(d)}_{14}}\,\overline{f^{(u)}_{43}}\,\overline{f^{(v)}_{32}}$ | $W^{\tau}_{(a,b,c\mid d,u,v)}(4,3,2\mid 2,3,4)\Big\vert_{s^k_{24}}$ |

Table 7.1: Component integrals giving rise to different MGFs.

Note in particular that although $W^{\tau}_{(2,0|2,0)}(2,3|2,3)$ and $W^{\tau}_{(2,0|2,0)}(2,3|3,2)$ differ just in their chiral permutations, their $\alpha'$ expansions are very different.

#### FROM MODULAR GRAPH FORMS TO COMPONENT INTEGRALS

The closed formulæ (7.13) for two-point component integrals allow to identify infinite families of MGFs within their $\alpha'$ expansion. Similarly, list in Table 7.1 possible realizations of more general MGFs in ($n \geq 3$)-point component integrals. Like this, the differential equations of the MGFs in Table 7.1 can be extracted from the differential equations of $W$-integrals in later sections. It is straightforward to extend the list to arbitrary graph topologies, where the multiplicity of the associated component integrals will grow with the complexity of the graph.

## 7.2 PREREQUISITES AND TWO-POINT WARM-UP

In this section, we list some derivatives of the Kronecker–Eisenstein series (3.79) and (3.82) and the Koba–Nielsen factor (3.72) necessary for the derivation of the differential equation of the generating series (7.1). We furthermore define variants of the Cauchy–Riemann and Laplace operators introduced in Section 3.1.3 which act naturally on the generating series. To illustrate the general structure of the Cauchy–Riemann and Laplace-equations of (7.1), we derive them for the two-point case (7.12) and demonstrate their implications for MGFs.

### 7.2.1 *Derivatives of Kronecker–Eisenstein series and the Koba–Nielsen factor*

Since $F(z,\eta,\tau)$ as defined in (3.79) is meromorphic in $z$, the derivative $\partial_{\bar z}$ of $\Omega$ is easy to evaluate and given by

$$\partial_{\bar z}\Omega(z,\eta,\tau) = -\frac{2\pi i\eta}{\tau-\bar\tau}\Omega(z,\eta,\tau) + \pi\delta^{(2)}(z,\bar z)\,. \tag{7.20}$$

The first contribution stems from the additional phase in (3.82) and the $\delta^{(2)}$ contribution is due to the simple pole of $F(z,\eta,\tau)$ at $z=0$. Expanding (7.20) in $\eta$ leads to (3.94).

When taking a derivative with respect to $\tau$, the meromorphic Kronecker–Eisenstein series satisfies the mixed heat equation [204]

$$2\pi i\partial_\tau F(z,\eta,\tau)=\partial_z\partial_\eta F(z,\eta,\tau)\qquad(\partial_\tau \text{ at fixed } z)\ . \tag{7.21}$$

There are two different forms of the corresponding equation for the doubly-periodic $\Omega$: For $\partial_\tau$ at fixed $z$, we have

$$2\pi i\partial_\tau\Omega(z,\eta,\tau)=(\partial_z+\partial_{\bar z})\partial_\eta\Omega(z,\eta,\tau)-2\pi i\frac{\operatorname{Im} z}{\tau_2}\partial_z\Omega(z,\eta,\tau)\,, \tag{7.22a}$$

while for $\partial_\tau$ at fixed $u,v\in\mathbb{R}$ (with $z=u\tau+v$),

$$2\pi i\partial_\tau\Omega(u\tau+v,\eta,\tau)=\partial_v\partial_\eta\Omega(u\tau+v,\eta,\tau)\,, \tag{7.22b}$$

where $\partial_v=\partial_z+\partial_{\bar z}$. Noting the corollary

$$(\tau-\bar\tau)\partial_{\bar z}\partial_\eta\Omega(z,\eta,\tau)=-2\pi i(1+\eta\partial_\eta)\Omega(z,\eta,\tau) \tag{7.23}$$

of (7.20), we can derive a third variant of the mixed heat equation for $\partial_\tau$ at fixed $u,v$ from (7.22b):

$$2\pi i\big((\tau-\bar\tau)\partial_\tau+1+\eta\partial_\eta\big)\Omega(u\tau+v,\eta,\tau)=(\tau-\bar\tau)\partial_z\partial_\eta\Omega(u\tau+v,\eta,\tau)\,. \tag{7.24}$$

This form will be used to derive Cauchy–Riemann equations of Koba–Nielsen integrals below.

After having discussed the $\bar z$- and $\tau$-derivatives of the Kronecker–Eisenstein series, we will now derive an identity involving the $\eta$-derivative of $\Omega$ which will be important in the simplification of $\partial_\tau W^\tau_{\vec\eta}$ later on. We start by specializing (5.133) to $a_1=1,2$ yielding

$$\begin{aligned}f^{(1)}(z)f^{(a)}(z)&=-\partial_z f^{(a)}(z)+(a+1)f^{(a+1)}(z)-\widehat{\mathrm{G}}_2f^{(a-1)}(z)\\&\qquad-\sum_{k=4}^{a+1}\mathrm{G}_k f^{(a+1-k)}(z)\,,\end{aligned} \tag{7.25a}$$

$$\begin{aligned}f^{(2)}(z)f^{(a)}(z)&=-a\partial_z f^{(a+1)}(z)+\frac{1}{2}(a+1)(a+2)f^{(a+2)}(z)-a\widehat{\mathrm{G}}_2f^{(a)}(z)\\&\qquad-\sum_{k=4}^{a+2}(a+1-k)\mathrm{G}_k f^{(a+2-k)}(z)\,.\end{aligned} \tag{7.25b}$$

Here, $a\geq 0$ and again we set $f^{(a)}=0$ for $a<0$. Using (7.25a) and (7.25b) we conclude that

$$\begin{aligned}&\Big(f^{(1)}\partial_\eta-f^{(2)}\Big)\Omega\\&=-\eta^{-2}f^{(1)}+\sum_{a\geq 0}\eta^{a-1}\Big(af^{(1)}f^{(a+1)}-f^{(2)}f^{(a)}\Big)\end{aligned}$$

$$
\begin{aligned}
&= -\eta^{-2} f^{(1)} + \sum_{a\geq 0} \eta^{a-1}\bigg(\frac{1}{2}(a+2)(a-1) f^{(a+2)} - \sum_{k=4}^{a+2}(k-1)\mathrm{G}_k f^{(a+2-k)}\bigg) \\
&= \Big(\frac{1}{2}\partial_\eta^2 - \wp(\eta,\tau)\Big)\Omega(z,\eta,\tau)\,. \qquad (7.26)
\end{aligned}
$$

Note that the terms involving $z$-derivatives of the $f^{(a)}$ cancel in this particular combination and we are left with a purely algebraic expression in these functions. In passing to the last line of (7.26), we used the expansion (3.19) of the Weierstraß function.

On top of the above derivatives of Kronecker–Eisenstein series, we also need the derivatives of the Koba–Nielsen factor (3.72). The derivative of $\mathrm{KN}_n$ w.r.t. a puncture $z_i$ (5.167) implies for any $1 \leq k \leq n$ that

$$
\sum_{j=k}^{n} \partial_{z_j} \mathrm{KN}_n = \sum_{i=1}^{k-1}\sum_{j=k}^{n} s_{ij} f_{ij}^{(1)}\, \mathrm{KN}_n\ . \qquad (7.27)
$$

From the $\tau$-derivative at fixed $u, v$ of the Green function (3.65),

$$
2\pi i \partial_\tau G(u\tau + v, \tau) = \sum_p{}' \frac{e^{2\pi i\langle p,z\rangle}}{p^2} = -f^{(2)}(z,\tau)\,, \qquad (7.28)
$$

we deduce

$$
2\pi i \partial_\tau \mathrm{KN}_n = -\sum_{1\leq i<j}^{n} s_{ij} f^{(2)}(z_{ij},\tau)\,\mathrm{KN}_n \qquad (\partial_\tau \text{ at fixed } u_j, v_j)\,. \qquad (7.29)
$$

### 7.2.2 *Differential operators on generating series*

Equipped with the differential operators introduced in Section 3.1.3, we will derive and study differential equations satisfied by the generating integrals $W^\tau_{\vec\eta}$ defined in (7.1) in the remainder of this work.

These differential equations describe the dependence of $W^\tau_{\vec\eta}$ on $\tau$ at all orders in $\alpha'$ and in the series parameters $\vec\eta = (\eta_2, \eta_3, \ldots, \eta_n)$. As $W^\tau_{\vec\eta}$ is defined as an integral over the world-sheet torus with complex structure parameter $\tau$ we first have to clarify how the $\tau$-derivative acts on such integrals. Our convention will always be to treat such torus integrals as

$$
\int \frac{\mathrm{d}^2 z}{\tau_2} = \int_{[0,1]^2} \mathrm{d}u\mathrm{d}v\,, \qquad (7.30)
$$

such that they are taken to not depend on $\tau$ when the torus coordinate $z = u\tau + v$ is written in terms of two real variables along the unit square. The $\tau$-derivative is then always taken *at constant $u$ and $v$* and will only act on the integrand. Therefore, we can employ the identities for derivatives at fixed $u, v$ derived in Section 7.2.1.

While the definitions (3.51) and (3.58) of the Maaß operators and the Laplace operator are tailored to functions of definite modular weights $(a, b)$, the $W$-integrands in (7.1) mix different modular weights in their expansion w.r.t. $\eta_j$ and $\bar{\eta}_j$. More precisely, the component integrals $W^\tau_{(A|B)}(\sigma|\rho)$ in (7.5) have modular weights $(|A|, |B|)$, using the notation (7.6).

Hence, it remains to find a representation of the holomorphic Maaß operator in (3.51) such that its action on the expansion (7.7) of $n$-point $W$-integrals is compatible with the modular weights of the component integrals. The modular weights of the $W^\tau_{(A|B)}(\sigma|\rho)$ correlate with the homogeneity degrees in the $\eta_j$ and $\bar{\eta}_j$ that is measured by the differential operators $\sum_{j=2}^n \eta_j \partial_{\eta_j}$ and $\sum_{j=2}^n \bar{\eta}_j \partial_{\bar{\eta}_j}$, respectively. We therefore define the following operators on functions depending on $\tau$ and $\vec{\eta}$

$$\nabla^{(k)}_{\vec{\eta}} = (\tau - \bar{\tau})\partial_\tau + k + \sum_{j=2}^n \eta_j \partial_{\eta_j} \tag{7.31a}$$

$$\overline{\nabla}^{(k)}_{\vec{\eta}} = (\bar{\tau} - \tau)\partial_{\bar{\tau}} + k + \sum_{j=2}^n \bar{\eta}_j \partial_{\bar{\eta}_j} \,. \tag{7.31b}$$

Due to the shift in the expansion of the component integrals (7.7), there is an offset between the eigenvalues of $(\sum_{j=2}^n \eta_j \partial_{\eta_j}, \sum_{j=2}^n \bar{\eta}_j \partial_{\bar{\eta}_j})$ and the weights $(|A|, |B|)$ according to

$$\sum_{j=2}^n \eta_j \partial_{\eta_j} \rho\big[\eta_{23\ldots n}^{a_2-1} \cdots \eta_n^{a_n-1}\big] = (|A|-(n-1))\, \rho\big[\eta_{23\ldots n}^{a_2-1} \cdots \eta_n^{a_n-1}\big] \tag{7.32a}$$

$$\sum_{j=2}^n \bar{\eta}_j \partial_{\bar{\eta}_j} \sigma\big[\bar{\eta}_{23\ldots n}^{b_2-1} \cdots \bar{\eta}_n^{b_n-1}\big] = (|B|-(n-1))\, \sigma\big[\bar{\eta}_{23\ldots n}^{b_2-1} \cdots \bar{\eta}_n^{b_n-1}\big] \,. \tag{7.32b}$$

Thus we have to set $k = n-1$ in (7.31a) in order to obtain

$$\nabla^{(n-1)}_{\vec{\eta}} W^\tau_{\vec{\eta}}(\sigma|\rho) = \sum_{A,B} \nabla^{(|A|)} W^\tau_{(A|B)}(\sigma|\rho) \rho\big[\eta_{234\ldots n}^{a_2-1} \eta_{34\ldots n}^{a_3-1} \cdots \eta_n^{a_n-1}\big] \times \sigma\big[\bar{\eta}_{234\ldots n}^{b_2-1} \bar{\eta}_{34\ldots n}^{b_3-1} \cdots \bar{\eta}_n^{b_n-1}\big] \tag{7.33}$$

that relates the raising operator on the generating series $W^\tau_{\vec{\eta}}$ correctly to the raising operator on the component integrals $W^\tau_{(A|B)}$ of definite modular weight $(|A|, |B|)$. In (7.32), we have also used the $\mathcal{S}_{n-1}$ permutation invariance of the raising and lowering operators (7.31) that descends to the specific sums (7.2) of the $\eta_j$-variables in the expansion of the $W$-integrals via

$$\sum_{j=2}^{n} \eta_j \partial_{\eta_j} = \sum_{j=2}^{n} \eta_{j,j+1\ldots n} \partial_{\eta_{j,j+1\ldots n}} = \sum_{j=2}^{n} \rho\big[\eta_{j,j+1\ldots n} \partial_{\eta_{j,j+1\ldots n}}\big] \,. \tag{7.34}$$

The Laplace operator can be defined in a similar fashion to (7.31) as

$$\Delta_{\vec{\eta}} = \overline{\nabla}_{\vec{\eta}}^{(n-2)} \nabla_{\vec{\eta}}^{(n-1)} - \Big(n - 1 + \sum_{j=2}^{n} \eta_j \partial_{\eta_j}\Big)\Big(n - 2 + \sum_{j=2}^{n} \bar{\eta}_j \partial_{\bar{\eta}_j}\Big) \tag{7.35}$$

such that it acts on component integrals $W^{\tau}_{(A|B)}(\sigma|\rho)$ via $\Delta^{(|A|,|B|)}$ in (3.58) with appropriate weights $(|A|, |B|)$:

$$\begin{aligned} \Delta_{\vec{\eta}} W^{\tau}_{\vec{\eta}}(\sigma|\rho) = \sum_{A,B} \Delta^{(|A|,|B|)} W^{\tau}_{(A|B)}(\sigma|\rho) \qquad\qquad (7.36) \\ \times \rho\big[\eta_{234\ldots n}^{a_2-1} \eta_{34\ldots n}^{a_3-1} \cdots \eta_n^{a_n-1}\big] \, \sigma\big[\bar{\eta}_{234\ldots n}^{b_2-1} \bar{\eta}_{34\ldots n}^{b_3-1} \cdots \bar{\eta}_n^{b_n-1}\big] \,. \end{aligned}$$

These expressions are invariant under permutation of $\eta_2, \eta_3, \ldots, \eta_n$, i.e. $\rho\big[\nabla^{(k)}_{\vec{\eta}}\big] = \nabla^{(k)}_{\vec{\eta}}$ for any $\rho \in \mathcal{S}_{n-1}$, and valid at any order in the $(\eta_j, \bar{\eta}_j)$- and $\alpha'$-expansions of $W^{\tau}_{\vec{\eta}}$ at $n$ points.

In the remainder of this section we work out the first-order Cauchy–Riemann equation and the second-order Laplace equation satisfied by $W^{\tau}_{\vec{\eta}}$ for two points in order to illustrate the basic manipulations. We will dedicate Sections 7.3 and 7.4 to the Cauchy–Riemann equations and Laplace equations of $n$-point $W$-integrals.

### 7.2.3 *Two-point warm-up for differential equations*

For the simplest case of $n = 2$, there are no permutations to consider, and the $(n-1)! \times (n-1)!$ matrix in (7.1) reduces to the real scalar

$$W^{\tau}_{\eta} = \int \frac{\mathrm{d}^2 z_2}{\tau_2} \, \Omega(z_{12}, \eta, \tau) \, \overline{\Omega(z_{12}, \eta, \tau)} \; \mathrm{KN}_2 \,, \tag{7.37}$$

where we have denoted $\eta = \eta_2$ for simplicity.

#### CAUCHY–RIEMANN EQUATION

Under the two-point instance $\nabla^{(1)}_{\eta} = (\tau - \bar{\tau})\partial_\tau + 1 + \eta \partial_\eta$ of the operator (7.31a), the two-point $W$-integral (7.37) satisfies

$$\begin{aligned} 2\pi i \nabla^{(1)}_{\eta} W^{\tau}_{\eta} = \int \frac{\mathrm{d}^2 z_2}{\tau_2} \Big[ &- (\tau - \bar{\tau})\big(\partial_{z_2} \partial_\eta \Omega(z_{12}, \eta, \tau)\big) \, \overline{\Omega(z_{12}, \eta, \tau)} \; \mathrm{KN}_2 \\ &- s_{12}(\tau - \bar{\tau}) f^{(2)}_{12} \Omega(z_{12}, \eta, \tau) \, \overline{\Omega(z_{12}, \eta, \tau)} \; \mathrm{KN}_2 \Big] \end{aligned}$$

$$= \int \frac{d^2 z_2}{\tau_2} \left[ 2\pi i \bar{\eta} \partial_\eta + s_{12}(\tau - \bar{\tau}) \Big( f_{12}^{(1)} \partial_\eta - f_{12}^{(2)} \Big) \right] \times \Omega(z_{12}, \eta, \tau) \, \overline{\Omega(z_{12}, \eta, \tau)} \, \mathrm{KN}_2$$

$$= \left[ 2\pi i \bar{\eta} \partial_\eta + s_{12}(\tau - \bar{\tau}) \left( \frac{1}{2} \partial_\eta^2 - \wp(\eta, \tau) \right) \right] W_\eta^\tau \,, \tag{7.38}$$

where the first line on the right-hand side stems from (7.24) (for fixed coordinates $u$ and $v$ in the torus integral) and the second one from the Koba–Nielsen derivative (7.29). In passing to the third line, we have integrated $\partial_{z_2}$ by parts in the first term[5] and used (7.20), (5.167) and the fact that $\partial_\eta$ only acts on the $\Omega$ factor in the product. For the next equality, one simplifies $(f_{12}^{(1)} \partial_\eta - f_{12}^{(2)}) \Omega(z_{12}, \eta, \tau)$ via (7.26) that produces a Weierstraß function $\wp(\eta, \tau)$. Since the differential operator in $\eta$ does not depend on $z_2$, we have moved it out of the integral. It is instructive to compare the resulting expression to the corresponding Cauchy–Riemann equation in the open string, which we will now review briefly.

For $n$ points, the integrals $Z_{\vec{\eta}}^\tau$ in (7.3) close under $\tau$-derivatives [37, 38]

$$2\pi i \partial_\tau Z_{\vec{\eta}}^\tau(\sigma|\rho) = \sum_{\alpha \in \mathcal{S}_{n-1}} D_{\vec{\eta}}^\tau(\rho|\alpha) Z_{\vec{\eta}}^\tau(\sigma|\alpha) \,. \tag{7.39}$$

The $(n-1)! \times (n-1)!$ matrix-valued differential operators $D_{\vec{\eta}}^\tau$ relate different permutations of the integrands. Its entries are linear in Mandelstam invariants (2.25) and comprise derivatives w.r.t. the auxiliary variables $\eta_j$ as well as Weierstraß functions of the latter. In fact, the entire $\tau$-dependence in the $\eta_j$-expansion of $D_{\vec{\eta}}^\tau$ is carried by holomorphic Eisenstein series. That is why (7.39) manifests the appearance of iterated Eisenstein integrals in the $\alpha'$-expansion of open-string integrals, a canonical representation of eMZVs exposing all their relations over $\mathbb{Q}$, MZVs and $(2\pi i)^{-1}$, cf. Section 4.2.2.

In analogy to the two-point instance

$$2\pi i \partial_\tau Z_\eta^\tau = D_\eta^\tau Z_\eta^\tau \,, \qquad D_\eta^\tau = s_{12} \left( \frac{1}{2} \partial_\eta^2 - \wp(\eta, \tau) - 2\zeta_2 \right) \tag{7.40}$$

of (7.39), we define the closed-string differential operator to be

$$\mathrm{sv}\, D_\eta^\tau = s_{12} \left( \frac{1}{2} \partial_\eta^2 - \wp(\eta, \tau) \right) . \tag{7.41}$$

Note that this operator is meromorphic in $\eta$ and $\tau$. The open-string differential operator $D_\eta^\tau$ only differs from its closed-string counterpart through the additional term $-2\zeta_2 s_{12}$. As a formal prescription to drop

5 There are no boundary terms arising in this process since they are suppressed by the Koba–Nielsen factor.

the $\zeta_2$-contribution to $D^\tau_\eta$, we refer to the single-valued map for (motivic) MZVs [18, 19] in the notation for sv $D^\tau_\eta$ in (7.41), cf. (2.39). Using (7.41), (7.38) becomes

$$2\pi i \nabla^{(1)}_\eta W^\tau_\eta = \Big[2\pi i \bar\eta \partial_\eta + (\tau-\bar\tau)\,\text{sv}\, D^\tau_\eta\Big] W^\tau_\eta\,. \tag{7.42}$$

## LAPLACE EQUATION

According to (7.35), the representation of the Laplacian on the two-point $W$-integral (7.37) is given by $\Delta_\eta = \overline{\nabla}^{(0)}_\eta \nabla^{(1)}_\eta - (1+\eta\partial_\eta)\bar\eta\partial_{\bar\eta}$. In order to evaluate the action of the Maaß operators, we introduce the short-hand

$$Q^\tau_\eta = 2\pi i \bar\eta \partial_\eta + (\tau-\bar\tau)\,\text{sv}\, D^\tau_\eta \tag{7.43}$$

for the operator in the Cauchy–Riemann equation $2\pi i \nabla^{(1)}_\eta W^\tau_\eta = Q^\tau_\eta W^\tau_\eta$ derived in (7.42). The action of $\overline{\nabla}^{(0)}_\eta = \overline{\nabla}^{\eta}_{(1)} - 1$ on $Q^\tau_\eta W^\tau_\eta$ can be conveniently inferred by means of the commutation relation

$$[\overline{\nabla}^{(1)}_\eta, Q^\tau_\eta] = Q^\tau_\eta \tag{7.44}$$

along with the complex conjugate $-2\pi i \overline{\nabla}^{(1)}_\eta W^\tau_\eta = \overline{Q^\tau_\eta} W^\tau_\eta$ of (7.42),

$$\begin{aligned}(2\pi i)^2 \overline{\nabla}^{(0)}_\eta \nabla^{(1)}_\eta W^\tau_\eta &= 2\pi i (\overline{\nabla}^{(1)}_\eta - 1) Q^\tau_\eta W^\tau_\eta \\ &= 2\pi i \Big( Q^\tau_\eta \overline{\nabla}^{(1)}_\eta + [\overline{\nabla}^{(1)}_\eta, Q^\tau_\eta] - Q^\tau_\eta \Big) W^\tau_\eta \\ &= 2\pi i Q^\tau_\eta (\overline{\nabla}^{(1)}_\eta W^\tau_\eta) = -Q^\tau_\eta \overline{Q^\tau_\eta} W^\tau_\eta \\ &= \Big(2\pi i \bar\eta\partial_\eta + (\tau-\bar\tau)\,\text{sv}\, D^\tau_\eta\Big)\Big(2\pi i \eta \partial_{\bar\eta} + (\tau-\bar\tau)\overline{\text{sv}\, D^\tau_\eta}\Big) W^\tau_\eta\,.\end{aligned} \tag{7.45}$$

While the operator $\overline{\text{sv}\, D^\tau_\eta} = s_{12}\Big(\frac{1}{2}\partial^2_{\bar\eta} - \wp(\bar\eta,\bar\tau)\Big)$ commutes with sv $D^\tau_\eta$, two extra contributions arise when reordering sv $D^\tau_\eta \eta\partial_{\bar\eta} = \eta\partial_{\bar\eta}\,\text{sv}\, D^\tau_\eta + s_{12}\partial_\eta\partial_{\bar\eta}$ and $\bar\eta\partial_\eta\eta\partial_{\bar\eta} = \bar\eta\partial_{\bar\eta} + \bar\eta\eta\partial_\eta\partial_{\bar\eta}$, cf. (7.41). Hence, we are led to the following alternative form of (7.45),

$$\begin{aligned}(2\pi i)^2 \overline{\nabla}^{(0)}_\eta \nabla^{(1)}_\eta W^\tau_\eta = \Big[&(2\pi i)^2 \bar\eta \partial_{\bar\eta} + (2\pi i)^2 \eta\bar\eta\partial_\eta\partial_{\bar\eta} + 2\pi i s_{12}(\tau-\bar\tau)\partial_\eta\partial_{\bar\eta} \\ &+ 2\pi i (\tau-\bar\tau)(\bar\eta\partial_\eta \overline{\text{sv}\, D^\tau_\eta} + \eta\partial_{\bar\eta}\,\text{sv}\, D^\tau_\eta) + (\tau-\bar\tau)^2\,\text{sv}\, D^\tau_\eta \overline{\text{sv}\, D^\tau_\eta}\Big] W^\tau_\eta\,.\end{aligned} \tag{7.46}$$

Finally, the first two terms in the square bracket cancel when completing the Laplacian:

$$(2\pi i)^2 \Delta_\eta W^\tau_\eta = (2\pi i)^2 \Big[\overline{\nabla}^{(0)}_\eta \nabla^{(1)}_\eta - (1+\eta\partial_\eta)\bar\eta\partial_{\bar\eta}\Big] W^\tau_\eta$$

$$= \Big[ 2\pi i s_{12}(\tau-\bar\tau)\partial_\eta\partial_{\bar\eta} + 2\pi i(\tau-\bar\tau)(\bar\eta\partial_\eta\overline{\mathrm{sv}\, D^\tau_\eta} + \eta\partial_{\bar\eta}\,\mathrm{sv}\, D^\tau_\eta) \\ + (\tau-\bar\tau)^2\,\mathrm{sv}\, D^\tau_\eta\overline{\mathrm{sv}\, D^\tau_\eta}\Big] W^\tau_\eta\,. \tag{7.47}$$

### 7.2.4 *Two-point warm-up for component integrals*

We shall now translate the Cauchy–Riemann- and Laplace equations (7.42) and (7.47) of the generating integral $W^\tau_\eta$ to the equations satisfied by its component integrals $W^\tau_{(a|b)}$ defined in (7.12).

#### CAUCHY–RIEMANN EQUATION

At the level of component integrals (7.12), the Cauchy–Riemann equations (7.42) are equivalent to

$$\nabla^{(a)} W^\tau_{(a|b)} = s_{12}\frac{\tau_2}{\pi}\Big[\frac{1}{2}(a+2)(a-1)W^\tau_{(a+2|b)} - \sum_{k=4}^{a+2}(k-1)\mathrm{G}_k W^\tau_{(a+2-k|b)}\Big] \\ + a W^\tau_{(a+1|b-1)} \tag{7.48}$$

with the understanding that $W^\tau_{(a|-1)} = W^\tau_{(-1|b)} = 0$ for all $a, b \geq 0$. The simplest examples for low weights $a, b$ include

$$\begin{aligned}
\nabla^{(0)} W^\tau_{(0|0)} &= -s_{12}\frac{\tau_2}{\pi} W^\tau_{(2|0)} \\
\nabla^{(2)} W^\tau_{(2|0)} &= -s_{12}\frac{\tau_2}{\pi}(3\mathrm{G}_4 W^\tau_{(0|0)} - 2W^\tau_{(4|0)}) \\
\nabla^{(1)} W^\tau_{(1|1)} &= W^\tau_{(2|0)} \\
\nabla^{(0)} W^\tau_{(0|2)} &= -s_{12}\frac{\tau_2}{\pi} W^\tau_{(2|2)} \\
\nabla^{(4)} W^\tau_{(4|0)} &= -s_{12}\frac{\tau_2}{\pi}(5\mathrm{G}_6 W^\tau_{(0|0)} + 3\mathrm{G}_4 W^\tau_{(2|0)} - 9W^\tau_{(6|0)}) \\
\nabla^{(3)} W^\tau_{(3|1)} &= -s_{12}\frac{\tau_2}{\pi}(3\mathrm{G}_4 W^\tau_{(1|1)} - 5W^\tau_{(5|1)}) + 3W^\tau_{(4|0)} \\
\nabla^{(2)} W^\tau_{(2|2)} &= -s_{12}\frac{\tau_2}{\pi}(3\mathrm{G}_4 W^\tau_{(0|2)} - 2W^\tau_{(4|2)}) + 2W^\tau_{(3|1)} \\
\nabla^{(1)} W^\tau_{(1|3)} &= W^\tau_{(2|2)} \\
\nabla^{(0)} W^\tau_{(0|4)} &= -s_{12}\frac{\tau_2}{\pi} W^\tau_{(2|4)}\,.
\end{aligned} \tag{7.49}$$

These equations are obtained by direct evaluation of (7.42) and are valid at all orders in $\alpha'$. Below, we shall analyze their $\alpha'$-expansion and relate them to equations for MGFs.

### LAPLACE EQUATION

The Laplace equation (7.47) of the generating integrals $W^\tau_\eta$ implies the following component relations for the $W^\tau_{(a|b)}$ of modular weight $(a,b)$ in (7.12):

$$\begin{aligned}\Delta^{(a,b)} W^\tau_{(a|b)} = \Big(\frac{\tau_2}{\pi}\Big)^2 s_{12}^2 \Big\{ &\frac{1}{4}(a+2)(a-1)(b+2)(b-1) W^\tau_{(a+2|b+2)} \\ &- \frac{1}{2}(a+2)(a-1)\sum_{k=4}^{b+2}(k-1)\overline{\mathrm{G}}_k W^\tau_{(a+2|b+2-k)} \\ &- \frac{1}{2}(b+2)(b-1)\sum_{\ell=4}^{a+2}(\ell-1)\mathrm{G}_\ell W^\tau_{(a+2-\ell|b+2)} \\ &+ \sum_{k=4}^{b+2}(k-1)\sum_{\ell=4}^{a+2}(\ell-1)\mathrm{G}_\ell\overline{\mathrm{G}}_k W^\tau_{(a+2-\ell|b+2-k)}\Big\} \\ + \frac{\tau_2}{\pi} s_{12}\Big\{ &\frac{1}{2}(ab-2)(a+b) W^\tau_{(a+1|b+1)} \\ &- b\sum_{k=4}^{a+1}(k-1)\mathrm{G}_k W^\tau_{(a+1-k|b+1)} - a\sum_{\ell=4}^{b+1}(\ell-1)\overline{\mathrm{G}}_\ell W^\tau_{(a+1|b+1-\ell)}\Big\}\,. \end{aligned} \tag{7.50}$$

The simplest examples include

$$\begin{aligned} \Delta^{(0,0)} W^\tau_{(0|0)} &= s_{12}^2 \Big(\frac{\tau_2}{\pi}\Big)^2 W^\tau_{(2|2)} \\ \Delta^{(1,1)} W^\tau_{(1|1)} &= -s_{12}\frac{\tau_2}{\pi} W^\tau_{(2|2)} \\ \Delta^{(2,0)} W^\tau_{(2|0)} &= s_{12}^2 \Big(\frac{\tau_2}{\pi}\Big)^2 (3\mathrm{G}_4 W^\tau_{(0|2)} - 2W^\tau_{(4|2)}) - 2s_{12}\frac{\tau_2}{\pi} W^\tau_{(3|1)} \\ \Delta^{(2,2)} W^\tau_{(2|2)} &= s_{12}^2 \Big(\frac{\tau_2}{\pi}\Big)^2 \big(9\overline{\mathrm{G}}_4\mathrm{G}_4 W^\tau_{(0|0)} - 6\mathrm{G}_4 W^\tau_{(0|4)} - 6\overline{\mathrm{G}}_4 W^\tau_{(4|0)} + 4W^\tau_{(4|4)}\big) \\ &\quad + 4s_{12}\frac{\tau_2}{\pi} W^\tau_{(3|3)} \\ \Delta^{(3,1)} W^\tau_{(3|1)} &= s_{12}\frac{\tau_2}{\pi}(-3\mathrm{G}_4 W^\tau_{(0|2)} + 2W^\tau_{(4|2)}) \\ \Delta^{(4,0)} W^\tau_{(4|0)} &= s_{12}^2 \Big(\frac{\tau_2}{\pi}\Big)^2 (5\mathrm{G}_6 W^\tau_{(0|2)} + 3\mathrm{G}_4 W^\tau_{(2|2)} - 9W^\tau_{(6|2)}) - 4s_{12}\frac{\tau_2}{\pi} W^\tau_{(5|1)}\,. \end{aligned} \tag{7.51}$$

These are again valid to all orders in $\alpha'$ and we shall discuss their $\alpha'$-expansion below.

We note that the evaluation of the differential operators in (7.47) a priori leads to double poles in $\eta$ or $\bar\eta$ on the right-hand side. These do not occur on the left-hand side and therefore have to cancel. While this is not necessarily manifest, their appearance can be traced back to the integration by parts that was used in the derivation (7.42) of the differential equation. For this reason, the residues of the putative double poles vanish by the integration-by-parts identities (7.14), that is why the $\eta$-expansions on both sides of (7.47) are identical.

### LESSONS FOR MODULAR GRAPH FORMS

In Section 7.1.2, we calculated the leading orders in the $\alpha'$-expansions of two-point component integrals in terms of MGFs. Using these expansions, (7.48) and (7.50) imply Cauchy–Riemann and Laplace equations for MGFs.

As an example, using (7.13) to expand the right-hand side of the Cauchy–Riemann equation (7.49) for $W^{\tau}_{(2|0)}$ leads to

$$\begin{aligned} \nabla^{(2)} W^{\tau}_{(2|0)} &= -s_{12}\frac{\tau_2}{\pi}(3\mathrm{G}_4 W^{\tau}_{(0|0)} - 2W^{\tau}_{(4|0)}) \\ &= -3s_{12}\frac{\tau_2}{\pi}\mathrm{G}_4 - 2s_{12}^2\left(\frac{\tau_2}{\pi}\right)^2 \mathcal{C}\big[\begin{smallmatrix} 5 & 0 \\ 1 & 0 \end{smallmatrix}\big] + \mathcal{O}(s_{12}^3)\,. \end{aligned} \tag{7.52}$$

Similarly, the right-hand side of the Laplace equation (7.51) expands to[6]

$$\begin{aligned} \Delta^{(2,0)} W^{\tau}_{(2|0)} &= s_{12}^2\left(\frac{\tau_2}{\pi}\right)^2(3\mathrm{G}_4 W^{\tau}_{(0|2)} - 2W^{\tau}_{(4|2)}) - 2s_{12}\frac{\tau_2}{\pi}W^{\tau}_{(3|1)} \\ &= -2s_{12}\frac{\tau_2}{\pi}\mathcal{C}\big[\begin{smallmatrix} 3 & 0 \\ 1 & 0 \end{smallmatrix}\big] + 2s_{12}^2\left(\frac{\tau_2}{\pi}\right)^2(\mathcal{C}\big[\begin{smallmatrix} 0 & 1 & 3 \\ 1 & 1 & 0 \end{smallmatrix}\big] - \mathcal{C}\big[\begin{smallmatrix} 4 & 0 \\ 2 & 0 \end{smallmatrix}\big]) + \mathcal{O}(s_{12}^3)\,. \end{aligned} \tag{7.53}$$

The right-hand sides of (7.52) and (7.53) do not manifestly match the direct action (5.53) of $\nabla^{(2)}$ and $\Delta^{(2,0)}$ on the MGFs in the expansion (7.13) of $W^{\tau}_{(2|0)}$,

$$\begin{aligned} \nabla^{(2)} W^{\tau}_{(2|0)} &= -\, s_{12}\nabla^{(2)}\left(\frac{\tau_2}{\pi}\mathcal{C}\big[\begin{smallmatrix} 3 & 0 \\ 1 & 0 \end{smallmatrix}\big]\right) - \frac{1}{2}s_{12}^2\nabla^{(2)}\left(\left(\frac{\tau_2}{\pi}\right)^2\mathcal{C}\big[\begin{smallmatrix} 1 & 1 & 2 \\ 1 & 1 & 0 \end{smallmatrix}\big]\right) + \mathcal{O}(s_{12}^3) \\ &= -\, 3s_{12}\frac{\tau_2}{\pi}\mathrm{G}_4 - s_{12}^2\left(\frac{\tau_2}{\pi}\right)^2(\mathcal{C}\big[\begin{smallmatrix} 1 & 1 & 3 \\ 1 & 1 & -1 \end{smallmatrix}\big] + \mathcal{C}\big[\begin{smallmatrix} 1 & 2 & 2 \\ 1 & 0 & 0 \end{smallmatrix}\big]) + \mathcal{O}(s_{12}^3) \end{aligned} \tag{7.54}$$

$$\begin{aligned} \Delta^{(2,0)} W^{\tau}_{(2|0)} &= -\, s_{12}\,\Delta^{(2,0)}\left(\frac{\tau_2}{\pi}\mathcal{C}\big[\begin{smallmatrix} 3 & 0 \\ 1 & 0 \end{smallmatrix}\big]\right) - \frac{1}{2}s_{12}^2\,\Delta^{(2,0)}\left(\left(\frac{\tau_2}{\pi}\right)^2\mathcal{C}\big[\begin{smallmatrix} 1 & 1 & 2 \\ 1 & 1 & 0 \end{smallmatrix}\big]\right) + \mathcal{O}(s_{12}^3) \\ &= -\, 2s_{12}\frac{\tau_2}{\pi}\mathcal{C}\big[\begin{smallmatrix} 3 & 0 \\ 1 & 0 \end{smallmatrix}\big] - s_{12}^2\left(\frac{\tau_2}{\pi}\right)^2(2\,\mathcal{C}\big[\begin{smallmatrix} 0 & 1 & 3 \\ 2 & 1 & -1 \end{smallmatrix}\big] + \mathcal{C}\big[\begin{smallmatrix} 0 & 2 & 2 \\ 2 & 0 & 0 \end{smallmatrix}\big]) + \mathcal{O}(s_{12}^3)\,. \end{aligned} \tag{7.55}$$

Comparing (7.52) and (7.54) as well as (7.53) and (7.55) order by order in $\alpha'$ yields infinitely many identities for MGFs, e.g.

$$\begin{aligned} 2\,\mathcal{C}\big[\begin{smallmatrix} 5 & 0 \\ 1 & 0 \end{smallmatrix}\big] &= \mathcal{C}\big[\begin{smallmatrix} 1 & 1 & 3 \\ 1 & 1 & -1 \end{smallmatrix}\big] + \mathcal{C}\big[\begin{smallmatrix} 1 & 2 & 2 \\ 1 & 0 & 0 \end{smallmatrix}\big] \\ 2(\mathcal{C}\big[\begin{smallmatrix} 0 & 1 & 3 \\ 1 & 1 & 0 \end{smallmatrix}\big] - \mathcal{C}\big[\begin{smallmatrix} 4 & 0 \\ 2 & 0 \end{smallmatrix}\big]) &= -2\,\mathcal{C}\big[\begin{smallmatrix} 0 & 1 & 3 \\ 2 & 1 & -1 \end{smallmatrix}\big] - \mathcal{C}\big[\begin{smallmatrix} 0 & 2 & 2 \\ 2 & 0 & 0 \end{smallmatrix}\big]\,. \end{aligned} \tag{7.56}$$

In particular, the differential equations (7.48) and (7.50) of the component integrals bypass the need to perform HSR as discussed in Section 5.4 to all orders in $\alpha'$. This is exemplified by the identities for MGFs $\mathcal{C}\big[\begin{smallmatrix} A & c & d \\ B & 0 & 0 \end{smallmatrix}\big]$

6 We have not yet inserted the simplified form (7.16) of the $\alpha'$-expansions which are obtained after using identities between MGFs, since we want to illustrate that (7.48) and (7.50) can be used to generate these kinds of identities.

in (7.56) and becomes particularly convenient at $n \geq 3$ points, where HSR becomes increasingly more laborious, cf. Section 5.4.

Since for low weights many identities between MGFs are known [15, 16, II, 39, 40, 127], the expansions above also allow for an explicit test of the differential equations. In particular, the identities such as (7.56) generated by $W^\tau_{(2|0)}$ can be confirmed by applying simple identities at low weights. In general, by applying identities for dihedral MGFs, (7.48) can be verified to all orders in $\alpha'$, as detailed in Appendix D.2. The Laplace equations (7.51) have been verified to the order $(\alpha')^5$ for $W^\tau_{(0|0)}$, to the order $(\alpha')^4$ for $W^\tau_{(1|1)}$ and $W^\tau_{(2|0)}$ and to the orders $(\alpha')^3$ for $W^\tau_{(2|2)}$, $W^\tau_{(3|1)}$ and $W^\tau_{(4|0)}$.

## 7.3 CAUCHY–RIEMANN DIFFERENTIAL EQUATIONS

In this section, we derive the general first-order differential equation in $\tau$ for the generating series $W^\tau_{\vec{\eta}}(\rho|\sigma)$ at $n$ points. The steps will generalize the two-point derivation in Section 7.2.3 with some additional steps due to the permutations $\rho, \sigma \in \mathcal{S}_{n-1}$. After deriving the general $n$-point formula we exemplify it by studying in detail the cases $n = 3$ and $n = 4$.

### 7.3.1 *Cauchy–Riemann differential equation at n points*

In order to act with $\nabla^{(n-1)}_{\vec{\eta}}$ on the generating series $W^\tau_{\vec{\eta}}(\sigma|\rho)$ defined in (7.1), we observe that the Maaß raising and lowering operator distributes correctly according to (3.53) and acts only on the product of chiral $\Omega$-series and on the Koba–Nielsen factor. Moreover, the differential operator and the Koba–Nielsen factor are invariant under $\rho \in \mathcal{S}_{n-1}$ as can be seen from the definition (3.72) and the property (7.34). Using the mixed heat equation (7.22b) for the $\tau$-derivative of $\Omega$ as well as (7.29) for the $\tau$-derivative of the Koba–Nielsen factor this leads to

$$\begin{aligned}
&2\pi i \nabla^{(n-1)}_{\vec{\eta}} W^\tau_{\vec{\eta}}(\sigma|\rho) \\
&= \int \mathrm{d}\mu_{n-1} \rho \Big[ 2\pi i \nabla^{(n-1)}_{\vec{\eta}} \Big\{ \mathrm{KN}_n \prod_{p=2}^{n} \Omega(z_{p-1,p}, \xi_p, \tau) \Big\} \Big] \sigma \Big[ \prod_{q=2}^{n} \overline{\Omega(z_{q-1,q}, \xi_q, \tau)} \Big] \\
&= (\tau - \bar{\tau}) \int \mathrm{d}\mu_{n-1}\, \rho \Big[ - \sum_{i=2}^{n} \big( \partial_{z_i} \partial_{\xi_i} \Omega(z_{i-1,i}, \xi_i, \tau) \big)\, \mathrm{KN}_n \prod_{\substack{p=2 \\ p \neq i}}^{n} \Omega(z_{p-1,p}, \xi_p, \tau) \\
&\quad - \sum_{1 \leq i < j}^{n} s_{ij} f^{(2)}_{ij}\, \mathrm{KN}_n \prod_{p=2}^{n} \Omega(z_{p-1,p}, \xi_p, \tau) \Big] \sigma \Big[ \prod_{q=2}^{n} \overline{\Omega(z_{q-1,q}, \xi_q, \tau)} \Big], \quad (7.57)
\end{aligned}$$

where we have introduced the following short-hand:[7]

$$\xi_i = \eta_{i,i+1,\dots,n} = \sum_{k=i}^{n} \eta_k \,. \tag{7.58}$$

In each of the terms in the $i$-sum in (7.57) one can replace $\partial_{z_i} \to \partial_{z_i} + \partial_{z_{i+1}} + \cdots + \partial_{z_n} = \sum_{j=i}^{n} \partial_{z_j}$ as the function it acts on does not depend on the other $z$-variables. This has the advantage that one can integrate by parts all $z$-derivatives without producing any contribution from the other chiral Kronecker–Eisenstein series since they all depend on differences such that the corresponding terms cancel. This leads to two contributions: In the first the $z$-derivatives act on the Koba–Nielsen factor, and the second contribution comes from the action on the anti-chiral Kronecker–Eisenstein series. These two contributions are of different kinds and we first focus on the one when the $z$-derivative acts on the anti-chiral $\overline{\Omega}$.

A partial $z$-derivative acting on a single anti-chiral $\overline{\Omega}$ was given in (7.20) and generates the corresponding $\bar\eta$. It can be checked that the combination of all terms does not depend on the permutations $\rho$ and $\sigma$ that one started with and in total produces the operator $2\pi i \sum_{i=2}^{n} \bar\eta_i \partial_{\eta_i}$ acting on the whole expression. We emphasize that this is the only part that does not involve only holomorphic or antiholomorphic $\eta$-parameters but mixes them. Carrying out the full integration by parts, we can therefore rewrite (7.57) as

$$\begin{aligned} &2\pi i \nabla^{(n-1)}_{\vec\eta} W^{\tau}_{\vec\eta}(\sigma|\rho) \\ &= 2\pi i \sum_{i=2}^{n} \bar\eta_i \partial_{\eta_i} W^{\tau}_{\vec\eta}(\sigma|\rho) \\ &\quad + (\tau-\bar\tau)\int \mathrm{d}\mu_{n-1}\, \rho\Big[ \sum_{i=2}^{n} \Big( \sum_{j=i}^{n} \partial_{z_j} \mathrm{KN}_n \Big) \partial_{\xi_i} \prod_{p=2}^{n} \Omega(z_{p-1,p}, \xi_p, \tau) \\ &\qquad - \sum_{1\le i<j}^{n} s_{ij} f^{(2)}_{ij} \mathrm{KN}_n \prod_{p=2}^{n} \Omega(z_{p-1,p}, \xi_p, \tau) \Big] \sigma\Big[ \prod_{q=2}^{n} \overline{\Omega(z_{q-1,q}, \xi_q, \tau)} \Big] \,. \end{aligned} \tag{7.59}$$

We next focus on analyzing the terms inside the $\rho$-permutation by using (7.27) for evaluating the $z$-derivative acting on the Koba–Nielsen factor:

$$\begin{aligned} &\sum_{i=2}^{n}\Big(\sum_{j=i}^{n}\partial_{z_j}\mathrm{KN}_n\Big)\partial_{\xi_i}\prod_{p=2}^{n}\Omega(z_{p-1,p},\xi_p,\tau) - \sum_{1\le i<j}^{n} s_{ij} f^{(2)}_{ij}\,\mathrm{KN}_n \prod_{p=2}^{n}\Omega(z_{p-1,p},\xi_p,\tau) \\ &= \sum_{1\le i<j}^{n} s_{ij} \Big[ \sum_{k=i+1}^{j} f^{(1)}_{ij} \partial_{\xi_k} - f^{(2)}_{ij} \Big] \mathrm{KN}_n \prod_{p=2}^{n} \Omega(z_{p-1,p}, \xi_p, \tau) \,. \end{aligned} \tag{7.60}$$

---

7 Note that the permutation $\rho$ does not act on the immediate indices of $\xi_i$ but on the indices of the constituent $\eta_i$.

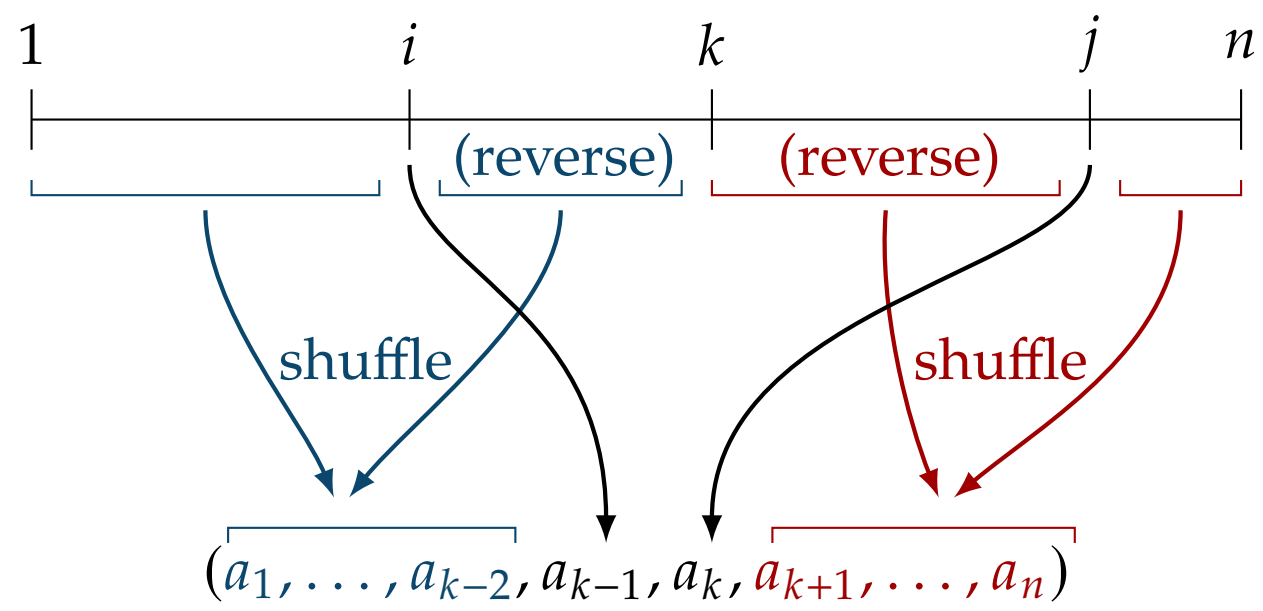

Figure 7.1: The shuffles appearing in the $s_{ij}$-form on the right-hand side of (7.62) for a fixed $i < k \leq j$. The values $i$ and $j$ are not included in the indicated ranges. The middle two intervals are reversed to descending order before the shuffles. The origin of this reversal is (D.18). All sequences obtained in this way constitute the set $S_n(i, j, k)$ defined in (7.61).

As shown in Appendix D.3, the cyclic product of Kronecker–Eisenstein series can be brought into a form such that the differential operator in square brackets has a simple action, see (D.14), generalizing (7.26). The result can be written in terms of certain shuffles (cf. (2.49)), explained in detail in Appendix D.3 and illustrated here in Figure 7.1. At $n$ points for $1 \leq i < j \leq n$ and $i < k \leq j$ one has to consider all sequences $(a_1, \ldots, a_n)$ in the set

$$S_n(i,j,k) = \Big\{ \big( \{1, \ldots, i-1\} \sqcup \{k-1, \ldots, i+1\}, i, j, \\ \{j-1, \ldots, k\} \sqcup \{j+1, \ldots, n\} \big) \Big\}, \tag{7.61}$$

where $i$ and $j$ are at fixed position determined by $k$ and the values to the left and right of them are given by shuffling very specific lists, half of which are reversed in order.

Using the result (D.14), equation (7.60) can then be evaluated to

$$\begin{aligned} &\sum_{1 \leq i < j}^{n} s_{ij} \left[ \sum_{k=i+1}^{j} f_{ij}^{(1)} \partial_{\xi_k} - f_{ij}^{(2)} \right] \mathrm{KN}_n \prod_{p=2}^{n} \Omega(z_{p-1,p}, \xi_p, \tau) \\ &= \sum_{1 \leq i < j}^{n} s_{ij} \Bigg[ \frac{1}{2} (\partial_{\eta_j} - \partial_{\eta_i})^2 \, \mathrm{KN}_n \prod_{p=2}^{n} \Omega(z_{p-1,p}, \xi_p, \tau) \\ &\quad - (-1)^{j-i+1} \sum_{k=i+1}^{j} \wp(\xi_k, \tau) \sum_{(a_1,\ldots,a_n) \in S_n(i,j,k)} \mathrm{KN}_n \prod_{p=2}^{n} \Omega\Big( z_{a_{p-1},a_p}, \sum_{\ell=p}^{n} \eta_{a_\ell}, \tau \Big) \Bigg] . \end{aligned} \tag{7.62}$$

Here, we have set $\partial_{\eta_1} = 0$ in the case $i = 1$ to avoid a separate bookkeeping of the terms $\sim \sum_{j=2}^{n} s_{1j} \partial_{\eta_j}^2$. We see that there is a 'diagonal term' containing the differential operators that goes back to the standard ordering of points. The terms including the Weierstraß functions mix the standard ordering with other orderings described by the set $S_n(i, j, k)$.

Since neither the differential operators nor the Weierstraß functions depend on $z$ they can be pulled out of the world-sheet integral.

We note that these operations are also related to the so-called $S$-map [254, 255] which enters the expressions of [38] for the $\tau$-derivatives of $A$-cycle integrals (7.3). In fact, the ($n \geq 6$)-point instances of the open-string differential operator $D^{\tau}_{\vec{\eta}}$ in (7.39) were conjectural in the reference, and (7.62) together with Appendix D.3 furnish the missing proof.

Equation (7.62) is expressed in an over-complete basis since a sequence $(a_1, \ldots, a_n) \in S_n(i, j, k)$ can have the index 1 at any place. Assume that the index 1 appears at position $m > 1$, i.e. $a_m = 1$ and write $(a_1, \ldots, a_n) = (A, 1, B)$ with $A = (a_1, \ldots, a_{m-1})$ and $B = (a_{m+1}, \ldots, a_n)$. The index 1 can be moved to the front using the fact that, as a consequence of the Fay identity (5.120b), products of Kronecker–Eisenstein series obey the shuffle identity [256]

$$\prod_{p=2}^{n} \Omega\Big(z_{a_{p-1},a_p}, \sum_{\ell=p}^{n} \eta_{a_\ell}, \tau\Big) = (-1)^{m-1} \sum_{(c_2,\ldots,c_n)\in A^t \sqcup\!\sqcup B} \prod_{p=2}^{n} \Omega\Big(z_{c_{p-1},c_p}, \sum_{\ell=p}^{n} \eta_{c_\ell}, \tau\Big)\,, \tag{7.63}$$

where $A^t = (a_{m-1}, a_{m-2}, \ldots, a_1)$ denotes the reversed sequence and we have set $c_1 = 1$ always. Applying this identity replaces one sequence $(a_1, \ldots, a_n) \in S_n(i, j, k)$ by a sum of sequences and the resulting integrals are then all of $W$-type but with different orderings of the $n-1$ unfixed points. This replaces the second term in (7.62) by a sum over all possible permutations $\alpha \in \mathcal{S}_{n-1}$ multiplying $W^{\tau}_{\vec{\eta}}(\sigma|\alpha)$ with coefficients $T^{\tau}_{\vec{\eta}}(\rho|\alpha)$ constructed out of Mandelstam invariants and Weierstraß functions. We write the total contribution of (7.62) to the Cauchy–Riemann derivative, up to an overall $(\tau - \bar{\tau})$, as the operator

$$\begin{aligned}\sum_{\alpha\in\mathcal{S}_{n-1}} \mathrm{sv}\, D^{\tau}_{\vec{\eta}}(\rho|\alpha) W^{\tau}_{\vec{\eta}}(\sigma|\alpha) = &\sum_{1\leq i<j}^{n} s_{ij} \Big[\frac{1}{2}(\partial_{\eta_j} - \partial_{\eta_i})^2\Big] W^{\tau}_{\vec{\eta}}(\sigma|\rho) \\ &+ \sum_{\alpha\in\mathcal{S}_{n-1}} T^{\tau}_{\vec{\eta}}(\rho|\alpha) W^{\tau}_{\vec{\eta}}(\sigma|\alpha)\,.\end{aligned} \tag{7.64}$$

Explicit expressions for $\mathrm{sv}\, D^{\tau}_{\vec{\eta}}(\rho|\alpha)$, detailing in particular the coefficients $T^{\tau}_{\vec{\eta}}(\rho|\alpha)$ will be given in Section 7.3.2 and 7.3.3 below for 3 and 4 points. At two points, one can identify $T^{\tau}_{\eta} = -s_{12}\wp(\eta, \tau)$ from the expression (7.41) for $\mathrm{sv}\, D^{\tau}_{\eta}$. The sv-notation here again instructs to drop the diagonal term $\sim -2\zeta_2 s_{12\ldots n}\delta_{\rho,\alpha}$ in the analogous open-string differential operator $D^{\tau}_{\vec{\eta}}(\rho|\alpha)$ in (7.39) [37, 38], i.e.

$$\mathrm{sv}\, D^{\tau}_{\vec{\eta}}(\alpha|\rho) = D^{\tau}_{\vec{\eta}}(\alpha|\rho)\Big|_{\zeta_2\to 0} = D^{\tau}_{\vec{\eta}}(\alpha|\rho) + 2\zeta_2 s_{12\ldots n}\delta_{\rho,\alpha}\,. \tag{7.65}$$

We have further separated $\text{sv}\, D^\tau_{\vec\eta}(\alpha|\rho)$ into a part that contains the holomorphic derivatives with respect to $\vec\eta$ and terms $T^\tau_{\vec\eta}(\alpha|\rho)$ that are completely meromorphic in $\vec\eta$ and $\tau$ and contain no derivatives.

Putting everything together we conclude that (7.1) obeys the Cauchy–Riemann equation

$$\begin{aligned} &2\pi i \nabla^{(n-1)}_{\vec\eta} W^\tau_{\vec\eta}(\sigma|\rho) \\ &= 2\pi i \sum_{i=2}^{n} \bar\eta_i \partial_{\eta_i} W^\tau_{\vec\eta}(\sigma|\rho) + (\tau-\bar\tau) \sum_{\alpha\in\mathcal{S}_{n-1}} \text{sv}\, D^\tau_{\vec\eta}(\rho|\alpha) W^\tau_{\vec\eta}(\sigma|\alpha) \\ &= \sum_{\alpha\in\mathcal{S}_{n-1}} Q^\tau_{\vec\eta}(\rho|\alpha) W^\tau_{\vec\eta}(\sigma|\alpha)\,, \end{aligned} \tag{7.66}$$

defining a short-hand for the action of the Maaß operator and with $\text{sv}\, D^\tau_{\vec\eta}$ given in (7.64). By expanding this equation in the $\eta$-parameters one can obtain systems of Cauchy–Riemann equations for the component integrals which in turn yield Cauchy–Riemann equations for MGFs.

### 7.3.2 *Three-point examples*

The three-point analogue of (7.37) is given by

$$\begin{aligned} W^\tau_{\eta_2,\eta_3}(\sigma|\rho) = \int \frac{d^2 z_2}{\tau_2}\frac{d^2 z_3}{\tau_2}\, \rho\Big[\Omega(z_{12},\eta_{23},\tau)\Omega(z_{23},\eta_3,\tau)\Big] \\ \times\, \sigma\Big[\overline{\Omega(z_{12},\eta_{23},\tau)}\,\overline{\Omega(z_{23},\eta_3,\tau)}\Big]\ \text{KN}_3\,. \end{aligned} \tag{7.67}$$

This is a (2×2) matrix of functions parametrized by the two permutations $\rho,\sigma\in\mathcal{S}_2$ that act on the indices 2 and 3 of the $z_j$ and $\eta_j$.

We first explain how to obtain the operators $\text{sv}\, D^\tau_{\vec\eta}(\rho|\alpha)$ in (7.64). Writing out (7.62) that is obtained from the combination of the $\tau$-derivative and the $\partial_z$-derivatives acting on the Koba–Nielsen factor yields for $n=3$

$$\begin{aligned} &\Big[\tfrac{1}{2}s_{12}\partial^2_{\eta_2}+\tfrac{1}{2}s_{13}\partial^2_{\eta_3}+\tfrac{1}{2}s_{23}\big(\partial^2_{\eta_2}-\partial^2_{\eta_3}\big)\Big]\Omega(z_{12},\eta_2+\eta_3,\tau)\Omega(z_{23},\eta_3,\tau)\,\text{KN}_3 \\ &-s_{12}\wp(\xi_2,\tau)\Omega(z_{12},\eta_2+\eta_3,\tau)\Omega(z_{23},\eta_3,\tau)\,\text{KN}_3 \\ &-s_{23}\wp(\xi_3,\tau)\Omega(z_{12},\eta_2+\eta_3,\tau)\Omega(z_{23},\eta_3,\tau)\,\text{KN}_3 \\ &+s_{13}\wp(\xi_2,\tau)\Omega(z_{13},\eta_2+\eta_3,\tau)\Omega(z_{32},\eta_2,\tau)\,\text{KN}_3 \\ &+s_{13}\wp(\xi_3,\tau)\Omega(z_{21},\eta_1+\eta_3,\tau)\Omega(z_{13},\eta_3,\tau)\,\text{KN}_3\,, \end{aligned} \tag{7.68}$$

where we have used $\partial_{\eta_1}=0$ and $\xi_2=\eta_2+\eta_3$ and $\xi_3=\eta_3$. The sequence sets $S_n(i,j,k)$ that were defined in (7.61) and appear in this case are $S_3(1,2,2)=\{(1,2,3)\}$ (for terms proportional to $s_{12}$), $S_3(2,3,3)=\{(1,2,3)\}$ (for $s_{23}$) and $S_3(1,3,2)=\{(1,3,2)\}$ as well as $S_3(1,3,3)=\{(2,1,3)\}$ (for $s_{13}$). The very last sequence $(2,1,3)$ does not start with the index 1 and needs to reordered using (7.63), yielding contributions to the sequences $(1,2,3)$ and $(1,3,2)$:

$$\begin{aligned}&\Omega(z_{21},\eta_1{+}\eta_3,\tau)\Omega(z_{13},\eta_3,\tau)\\ &= -\Omega(z_{12},\eta_2{+}\eta_3,\tau)\Omega(z_{23},\eta_3,\tau) - \Omega(z_{13},\eta_2{+}\eta_3,\tau)\Omega(z_{32},\eta_2,\tau)\,.\end{aligned} \tag{7.69}$$

Therefore we see that for $\rho(2,3) = (2,3)$ we obtain the following components of the first row of $\mathrm{sv}\,D^{\tau}_{\eta_2,\eta_3}(\rho|\alpha)$:

$$\begin{aligned}\mathrm{sv}\,D^{\tau}_{\eta_2,\eta_3}(2,3|2,3) &= s_{12}\big[\tfrac{1}{2}\partial^2_{\eta_2} - \wp(\eta_2{+}\eta_3,\tau)\big]\\ &\quad + s_{23}\big[\tfrac{1}{2}(\partial_{\eta_2}{-}\partial_{\eta_3})^2 - \wp(\eta_3,\tau)\big]\\ &\quad + s_{13}\big[\tfrac{1}{2}\partial^2_{\eta_3} - \wp(\eta_3,\tau)\big]\\ \mathrm{sv}\,D^{\tau}_{\eta_2,\eta_3}(2,3|3,2) &= s_{13}\big[\wp(\eta_2{+}\eta_3,\tau) - \wp(\eta_3,\tau)\big]\,.\end{aligned} \tag{7.70}$$

The second row associated with $\rho(2,3) = (3,2)$ follows from relabeling $s_{12} \leftrightarrow s_{13}$ and $\eta_2 \leftrightarrow \eta_3$,

$$\begin{aligned}\mathrm{sv}\,D^{\tau}_{\eta_2,\eta_3}(3,2|3,2) &= s_{13}\big[\tfrac{1}{2}\partial^2_{\eta_3} - \wp(\eta_2{+}\eta_3,\tau)\big]\\ &\quad + s_{23}\big[\tfrac{1}{2}(\partial_{\eta_2}{-}\partial_{\eta_3})^2 - \wp(\eta_2,\tau)\big]\\ &\quad + s_{12}\big[\tfrac{1}{2}\partial^2_{\eta_2} - \wp(\eta_2,\tau)\big]\\ \mathrm{sv}\,D^{\tau}_{\eta_2,\eta_3}(3,2|2,3) &= s_{12}\big[\wp(\eta_2{+}\eta_3,\tau) - \wp(\eta_2,\tau)\big]\,.\end{aligned} \tag{7.71}$$

By comparing these expressions with the open-string expression $D^{\tau}_{\vec{\eta}}(\sigma|\rho)$ given in [37, 38], we see that they agree up to a term $-2\zeta_2 s_{123}$ in the diagonal entries of the open-string operators. As the standard single-valued map for zeta values implies $\mathrm{sv}(\zeta_2) = 0$, our notation $\mathrm{sv}\,D^{\tau}_{\vec{\eta}}(\sigma|\rho)$ is consistent with the same operators in the open-string case.

One can then work out the three-point Cauchy–Riemann equations (7.66) that read for three points

$$\begin{aligned}2\pi i\nabla^{(2)}_{\eta_2,\eta_3} W^{\tau}_{\eta_2,\eta_3}(\sigma|\rho) = \sum_{\alpha\in\mathcal{S}_2}\Big[&2\pi i\delta_{\rho,\alpha}(\bar\eta_2\partial_{\eta_2}{+}\bar\eta_3\partial_{\eta_3})\\ &+ (\tau{-}\bar\tau)\,\mathrm{sv}\,D^{\tau}_{\eta_2,\eta_3}(\rho|\alpha)\Big]W^{\tau}_{\eta_2,\eta_3}(\sigma|\alpha)\,,\end{aligned} \tag{7.72}$$

by substituting in the matrix elements of $\mathrm{sv}\,D^{\tau}_{\eta_2,\eta_3}(\rho|\alpha)$ given in (7.70) and (7.71).

We carry out the derivation of the Cauchy–Riemann equations for component integrals (7.18) in detail in Appendix F of [IV] where we explain a subtlety in translating (7.72) to the component level: Both sides of (7.72) have to be expanded in the same $\eta$ variables (e.g. $\eta_{23} = \eta_2{+}\eta_3$ and $\eta_3$) but other permutations naturally come with different $\eta$ variables that have to be rearranged using the binomial theorem.

A general formula for the Cauchy-Riemann equation of the components (7.18) can be found in (F.6) of [IV]. It can be specialized to yield

$$\nabla^{(0)} W^\tau_{(0,0|b_2,b_3)}(\sigma|2{,}3) = -\frac{\tau_2}{\pi} s_{12} W^\tau_{(2,0|b_2,b_3)}(\sigma|2{,}3) - \frac{\tau_2}{\pi} s_{23} W^\tau_{(0,2|b_2,b_3)}(\sigma|2{,}3) - \frac{\tau_2}{\pi} s_{13}\Big(W^\tau_{(0,2|b_2,b_3)}(\sigma|2{,}3) + W^\tau_{(2,0|b_2,b_3)}(\sigma|3{,}2) - W^\tau_{(0,2|b_2,b_3)}(\sigma|3{,}2)\Big)$$

$$\begin{aligned}\nabla^{(1)} W^\tau_{(1,0|b_2,b_3)}(\sigma|2{,}3) &= W^\tau_{(2,0|b_2-1,b_3)}(\sigma|2{,}3) - \frac{\tau_2}{\pi} s_{23} W^\tau_{(1,2|b_2,b_3)}(\sigma|2{,}3) \\ &\quad + \frac{\tau_2}{\pi} s_{13}\Big(W^\tau_{(3,0|b_2,b_3)}(\sigma|2{,}3) - W^\tau_{(1,2|b_2,b_3)}(\sigma|2{,}3) \\ &\quad + 2W^\tau_{(0,3|b_2,b_3)}(\sigma|3{,}2) + W^\tau_{(1,2|b_2,b_3)}(\sigma|3{,}2) - W^\tau_{(3,0|b_2,b_3)}(\sigma|3{,}2)\Big)\end{aligned} \tag{7.73}$$

and further examples are listed in (F.7) of [IV].

The very simplest instance of this is for $(b_2, b_3) = (0, 0)$

$$\begin{aligned}&\nabla^{(0)} W^\tau_{(0,0|0,0)}(\sigma|2{,}3) \\ &= -\frac{\tau_2}{\pi} s_{12} W^\tau_{(2,0|0,0)}(\sigma|2{,}3) - \frac{\tau_2}{\pi} s_{23} W^\tau_{(0,2|0,0)}(\sigma|2{,}3) \\ &\quad - \frac{\tau_2}{\pi} s_{13}\Big(W^\tau_{(0,2|0,0)}(\sigma|2{,}3) + W^\tau_{(2,0|0,0)}(\sigma|3{,}2) - W^\tau_{(0,2|0,0)}(\sigma|3{,}2)\Big) \\ &= -\frac{\tau_2}{\pi}\left[s_{12} W^\tau_{(2,0|0,0)}(\sigma|2{,}3) + s_{13} W^\tau_{(2,0|0,0)}(\sigma|3{,}2) + s_{23} W^\tau_{(0,2|0,0)}(\sigma|2{,}3)\right] ,\end{aligned} \tag{7.74}$$

where we have used the corollary

$$W^\tau_{(0,2|0,0)}(\sigma|2{,}3) = W^\tau_{(0,2|0,0)}(\sigma|3{,}2) \tag{7.75}$$

of $f^{(2)}_{23} = f^{(2)}_{32}$. This is an example of the first type of linear dependence between component integrals mentioned in Section 7.1.1.

## LESSONS FOR MGFS

We now consider one instance of such a Cauchy–Riemann equation to probe its contents in the $\alpha'$-expansion. The example we shall look at involves the component integral

$$W^\tau_{(1,2|2,1)}(2,3|2,3) = -s_{13}\frac{\tau_2}{\pi} C\left[\begin{smallmatrix}1\\1\end{smallmatrix}\middle|\begin{smallmatrix}0&1\\2&0\end{smallmatrix}\middle|\begin{smallmatrix}0&2\\1&0\end{smallmatrix}\right] + O(s^2_{ij}) , \tag{7.76}$$

of modular weight $(3, 3)$. We have written out the $\alpha'$-expansion along the lines of Section 7.1.2 to the lowest non-trivial order which here contains a trihedral function. The Cauchy–Riemann equation in this case is

$$\begin{aligned}\nabla^{(3)} W^\tau_{(1,2|2,1)}(2,3|2,3) &= 2W^\tau_{(1,3|2,0)}(2,3|2,3) + W^\tau_{(2,2|1,1)}(2,3|2,3) \\ &\quad + \frac{\tau_2}{\pi}\Big\{2s_{13} W^\tau_{(1,4|2,1)}(2,3|2,3) + 2s_{23} W^\tau_{(1,4|2,1)}(2,3|2,3) \\ &\qquad - 2s_{13} W^\tau_{(1,4|2,1)}(2,3|3,2) + 2s_{13} W^\tau_{(2,3|2,1)}(2,3|2,3) \\ &\qquad - 2s_{13} W^\tau_{(2,3|2,1)}(2,3|3,2) + s_{13} W^\tau_{(3,2|2,1)}(2,3|2,3) \\ &\qquad - s_{13} W^\tau_{(3,2|2,1)}(2,3|3,2) - 3s_{13} \mathrm{G}_4 W^\tau_{(1,0|2,1)}(2,3|2,3) \\ &\qquad - 3s_{23} \mathrm{G}_4 W^\tau_{(1,0|2,1)}(2,3|2,3) + 3s_{13} \mathrm{G}_4 W^\tau_{(1,0|2,1)}(2,3|3,2)\Big\}\end{aligned}$$

$$
\begin{aligned}
&= \frac{\tau_2}{\pi} s_{13} \Big\{ 2\,\mathcal{C}\Big[\begin{smallmatrix}0&1&4\\2&0&1\end{smallmatrix}\Big] + \mathcal{C}\Big[\begin{array}{c|cc|cc}\scriptstyle 1&\scriptstyle 0&\scriptstyle 2&\scriptstyle 0&\scriptstyle 2\\ \scriptstyle 1&\scriptstyle 1&\scriptstyle 0&\scriptstyle 1&\scriptstyle 0\end{array}\Big] + 2\,\mathcal{C}\Big[\begin{smallmatrix}0&1&4\\1&2&0\end{smallmatrix}\Big] + \mathcal{C}\Big[\begin{smallmatrix}0&2&3\\1&0&2\end{smallmatrix}\Big] \\
&\qquad - 2\,\mathcal{C}\Big[\begin{smallmatrix}0&2&3\\1&2&0\end{smallmatrix}\Big] - 3\mathrm{G}_4\,\mathcal{C}\Big[\begin{smallmatrix}1&0\\3&0\end{smallmatrix}\Big] + 2\mathrm{E}_2\,\mathcal{C}\Big[\begin{smallmatrix}3&0\\1&0\end{smallmatrix}\Big] \Big\} + O(s_{ij}^2)\,,
\end{aligned}
\tag{7.77}
$$

where we have also written out the leading $\alpha'$-order of the right-hand side. Alternatively, we could have applied directly the differential operator to the expansion (7.76) using (5.53) which yields

$$
\begin{aligned}
\nabla^{(3)} W^{\tau}_{(1,2|2,1)}(2,3|2,3) &= -\frac{\tau_2}{\pi} s_{13} \Big\{ \mathcal{C}\Big[\begin{array}{c|cc|cc}\scriptstyle 1&\scriptstyle 0&\scriptstyle 2&\scriptstyle 0&\scriptstyle 2\\ \scriptstyle 1&\scriptstyle 1&\scriptstyle 0&\scriptstyle 2&\scriptstyle -1\end{array}\Big] + 2\,\mathcal{C}\Big[\begin{array}{c|cc|cc}\scriptstyle 1&\scriptstyle 0&\scriptstyle 1&\scriptstyle 0&\scriptstyle 3\\ \scriptstyle 1&\scriptstyle 2&\scriptstyle 0&\scriptstyle 1&\scriptstyle -1\end{array}\Big] \\
&\qquad + \mathcal{C}\Big[\begin{array}{c|cc|cc}\scriptstyle 2&\scriptstyle 0&\scriptstyle 1&\scriptstyle 0&\scriptstyle 2\\ \scriptstyle 0&\scriptstyle 2&\scriptstyle 0&\scriptstyle 1&\scriptstyle 0\end{array}\Big] \Big\} + O(s_{ij}^2)\,.
\end{aligned}
\tag{7.78}
$$

Equating this to (7.77) again leads to a non-trivial identity between MGFs, now mixing trihedral and dihedral type. This identity can be checked by using various identities from Chapter 5 to bring both (7.77) and (7.78) into the form

$$
s_{13}\left( \frac{5}{2}\frac{\pi^3 \mathrm{E}_2 \nabla_0 \mathrm{E}_2}{\tau_2^4} - 3\frac{\pi^3 \nabla_0 \mathrm{E}_4}{\tau_2^4} - 5\frac{\pi^3 \nabla_0 \mathrm{E}_{2,2}}{\tau_2^4} - \frac{3}{2}\mathrm{G}_4 \frac{\pi \overline{\nabla}_0 \mathrm{E}_2}{\tau_2^2} \right) + O(s_{ij}^2)\,. \tag{7.79}
$$

Similarly to the two-point results outlined in Section 7.2.4, also the three-point Cauchy–Riemann equations imply infinitely many relations between MGFs, now also including trihedral topologies. In particular, these identities allow to circumvent the lengthy three-point HSR (5.109). In the example above, when (7.78) is simplified by means of the factorization and momentum-conservation identities spelled out in the Sections 5.3.3 and 5.3.4, one obtains

$$
\begin{aligned}
\nabla^{(3)} W^{\tau}_{(1,2|2,1)}(2,3|2,3) &= \frac{\tau_2}{\pi} s_{13} \Big\{ -2\,\mathcal{C}\Big[\begin{array}{c|cc|cc}\scriptstyle 1&\scriptstyle 0&\scriptstyle 1&\scriptstyle 0&\scriptstyle 3\\ \scriptstyle 0&\scriptstyle 2&\scriptstyle 0&\scriptstyle 1&\scriptstyle 0\end{array}\Big] - \mathcal{C}\Big[\begin{array}{c|cc|cc}\scriptstyle 1&\scriptstyle 0&\scriptstyle 2&\scriptstyle 0&\scriptstyle 2\\ \scriptstyle 0&\scriptstyle 1&\scriptstyle 0&\scriptstyle 2&\scriptstyle 0\end{array}\Big] \\
&\qquad - \mathcal{C}\Big[\begin{array}{c|cc|cc}\scriptstyle 2&\scriptstyle 0&\scriptstyle 1&\scriptstyle 0&\scriptstyle 2\\ \scriptstyle 0&\scriptstyle 2&\scriptstyle 0&\scriptstyle 1&\scriptstyle 0\end{array}\Big] + \mathcal{C}\Big[\begin{array}{c|cc|cc}\scriptstyle 1&\scriptstyle 0&\scriptstyle 2&\scriptstyle 0&\scriptstyle 2\\ \scriptstyle 1&\scriptstyle 1&\scriptstyle 0&\scriptstyle 1&\scriptstyle 0\end{array}\Big] + 2\,\mathcal{C}\Big[\begin{smallmatrix}0&1&4\\2&0&1\end{smallmatrix}\Big] \Big\} + O(s_{ij}^2)\,.
\end{aligned}
\tag{7.80}
$$

In this expression, the first three trihedral MGFs have to be simplified using HSR. In (7.77), by contrast, no HSR is necessary which exemplifies a general feature of the Cauchy–Riemann equations generated by (7.72): They avoid a large number of iterated momentum conservations and all instances of HSR.

### 7.3.3 *Four-point examples*

At four points we restrict ourselves to providing the expression for the operators in (7.64). The following expressions for sv $D^{\tau}_{\vec{\eta}}(\sigma|\rho)$ can be obtained by applying the general method with the same steps as for three points:

$$\begin{aligned}
\mathrm{sv}\, D^{\tau}_{\vec\eta}(2,3,4|2,3,4) &= \frac{1}{2}\sum_{1\leq i<j}^{4} s_{ij}(\partial_{\eta_i}-\partial_{\eta_j})^2 - s_{12}\wp(\eta_2+\eta_3+\eta_4,\tau) \\
&\quad - (s_{13}+s_{23})\wp(\eta_3+\eta_4,\tau) - (s_{14}+s_{24}+s_{34})\wp(\eta_4,\tau) \\
\mathrm{sv}\, D^{\tau}_{\vec\eta}(2,3,4|2,4,3) &= (s_{14}+s_{24})\big[\wp(\eta_3+\eta_4,\tau) - \wp(\eta_4,\tau)\big] \\
\mathrm{sv}\, D^{\tau}_{\vec\eta}(2,3,4|3,2,4) &= s_{13}\big[\wp(\eta_2+\eta_3+\eta_4,\tau) - \wp(\eta_3+\eta_4,\tau)\big] \\
\mathrm{sv}\, D^{\tau}_{\vec\eta}(2,3,4|3,4,2) &= s_{13}\big[\wp(\eta_2+\eta_3+\eta_4,\tau) - \wp(\eta_3+\eta_4,\tau)\big] \\
\mathrm{sv}\, D^{\tau}_{\vec\eta}(2,3,4|4,2,3) &= s_{14}\big[\wp(\eta_3+\eta_4,\tau) - \wp(\eta_4,\tau)\big] \\
\mathrm{sv}\, D^{\tau}_{\vec\eta}(2,3,4|4,3,2) &= s_{14}\big[\wp(\eta_3+\eta_4,\tau) - \wp(\eta_2+\eta_3+\eta_4,\tau)\big]\,.
\end{aligned} \tag{7.81}$$

They agree with the corresponding open-string expressions $D^{\tau}_{\vec\eta}(\sigma|\rho)$ in [37, 38] after dropping the term $-2\zeta_2 s_{1234}$ in the diagonal entries which is annihilated by the single-valued map.

## 7.4 LAPLACE EQUATIONS

In this section, we extend the first-order Cauchy–Riemann equation (7.66) to a second-order Laplace equation. This is first done in general for $n$ points and then examples are worked out for a low number of points. The derivation follows the ideas in Section 7.2.3 about the two-point Laplace equations.

### 7.4.1 *Laplace equation at n points*

In order to extend the Cauchy–Riemann equation (7.66) to the Laplacian we need to act with $\overline{\nabla}^{(n-2)}_{\vec\eta}$ from (7.31b) on (7.66) and subtract an appropriate combination of weight terms according to (7.35). The action of $\overline{\nabla}^{(n-2)}_{\vec\eta} = \overline{\nabla}^{(n-1)}_{\vec\eta} - 1$ on (7.66) is simple since the differential operator passes through most terms in $Q^{\tau}_{\vec\eta}(\rho|\alpha)$ except for the explicit $\bar\eta_i$ in the diagonal term and the explicit $\bar\tau$ in front of $\mathrm{sv}\, D_{\vec\eta}(\rho|\alpha)$, leading to the simple commutation relation generalizing (7.44)

$$\Big[\overline{\nabla}^{(n-1)}_{\vec\eta}, Q^{\tau}_{\vec\eta}(\rho|\alpha)\Big] = Q^{\tau}_{\vec\eta}(\rho|\alpha)\,. \tag{7.82}$$

Taking the complex conjugate of (7.66) leads to[8]

$$2\pi i\overline{\nabla}^{(n-1)}_{\vec\eta} W^{\tau}_{\vec\eta}(\sigma|\alpha) = -\sum_{\beta\in\mathcal{S}_{n-1}} \overline{Q^{\tau}_{\vec\eta}}(\sigma|\beta) W^{\tau}_{\vec\eta}(\beta|\alpha)$$

8 Note that one has $\overline{W^{\tau}_{\vec\eta}(\sigma|\rho)} = W^{\tau}_{\vec\eta}(\rho|\sigma)$, leading to the summation over the first permutation labeling $W^{\tau}_{\vec\eta}$ in the complex conjugate equation.

$$= 2\pi i \sum_{i=2}^{n} \eta_i \partial_{\bar\eta_i} W^\tau_{\vec\eta}(\sigma|\alpha) \qquad (7.83)$$
$$+ (\tau - \bar\tau) \sum_{\beta\in\mathcal{S}_{n-1}} \overline{\mathrm{sv}\, D_{\vec\eta}}(\sigma|\beta) W^\tau_{\vec\eta}(\beta|\alpha)\,,$$

which implies (see (7.45) for the analogous two-point calculation)

$$\begin{aligned}
&(2\pi i)^2 \overline{\nabla}^{(n-2)}_{\vec\eta} \nabla^{(n-1)}_{\vec\eta} W^\tau_{\vec\eta}(\sigma|\rho) \\
&= 2\pi i \sum_{\alpha\in\mathcal{S}_{n-1}} \big(Q^\tau_{\vec\eta}(\rho|\alpha)\overline{\nabla}^{(n-1)}_{\vec\eta} + [\overline{\nabla}^{(n-1)}_{\vec\eta}, Q^\tau_{\vec\eta}(\rho|\alpha)] - Q^\tau_{\vec\eta}(\rho|\alpha)\big)\, W^\tau_{\vec\eta}(\sigma|\alpha) \\
&= \sum_{\alpha\in\mathcal{S}_{n-1}} Q^\tau_{\vec\eta}(\rho|\alpha)\big(2\pi i \overline{\nabla}^{(n-1)}_{\vec\eta} W^\tau_{\vec\eta}(\sigma|\alpha)\big) \\
&= -\sum_{\alpha,\beta\in\mathcal{S}_{n-1}} Q^\tau_{\vec\eta}(\rho|\alpha)\overline{Q^\tau_{\vec\eta}}(\sigma|\beta) W^\tau_{\vec\eta}(\beta|\alpha)\,. 
\end{aligned} \qquad (7.84)$$

This expression can be expanded further by moving all $\eta$-differential operators in $Q^\tau_{\vec\eta}$ to the right to act directly on $W^\tau_{\vec\eta}$ since most terms commute. The only extra contributions come from $\bar\eta_i \partial_{\eta_i}$ and $s_{ij}(\partial_{\eta_i} - \partial_{\eta_j})^2$ in $Q^\tau_{\vec\eta}$ acting on the $\eta_k \partial_{\bar\eta_k}$ in $\overline{Q^\tau_{\vec\eta}}$. The result is

$$\begin{aligned}
&\Big(2\pi i \sum_{i=2}^{n} \bar\eta_i \partial_{\eta_i} + \frac{1}{2}(\tau - \bar\tau) \sum_{1\le i<j}^{n} s_{ij}(\partial_{\eta_i} - \partial_{\eta_j})^2\Big) 2\pi i \sum_{k=2}^{n} \eta_k \partial_{\bar\eta_k} \\
&= (2\pi i)^2 \sum_{i=2}^{n} \bar\eta_i \partial_{\bar\eta_i} + (2\pi i)^2 \sum_{i,j=2}^{n} \eta_i \bar\eta_j \partial_{\eta_j} \partial_{\bar\eta_i} \\
&\quad + 2\pi i(\tau - \bar\tau) \sum_{1\le i<j}^{n} s_{ij}(\partial_{\eta_j} - \partial_{\eta_i})(\partial_{\bar\eta_j} - \partial_{\bar\eta_i}) \\
&\quad + (2\pi i) \sum_{k=2}^{n} \eta_k \partial_{\bar\eta_k} \frac{1}{2}(\tau - \bar\tau) \sum_{1\le i<j}^{n} s_{ij}(\partial_{\eta_j} - \partial_{\eta_i})^2\,,
\end{aligned} \qquad (7.85)$$

where the last line is part of $2\pi i(\tau - \bar\tau) \sum_{k=2}^{n} \eta_k \partial_{\bar\eta_k}\, \mathrm{sv}\, D^\tau_{\vec\eta}(\rho|\alpha)$, and therefore

$$\begin{aligned}
&(2\pi i)^2 \overline{\nabla}^{(n-2)}_{\vec\eta} \nabla^{(n-1)}_{\vec\eta} W^\tau_{\vec\eta}(\sigma|\rho) \\
&= \sum_{\alpha,\beta\in\mathcal{S}_{n-1}} \bigg\{ \delta_{\alpha,\rho}\delta_{\beta,\sigma} \bigg[ (2\pi i)^2 \Big( \sum_{i=2}^{n} \bar\eta_i \partial_{\bar\eta_i} + \sum_{i,j=2}^{n} \eta_i \bar\eta_j \partial_{\eta_j} \partial_{\bar\eta_i} \Big) \\
&\quad + 2\pi i(\tau - \bar\tau) \sum_{1\le i<j}^{n} s_{ij}(\partial_{\eta_j} - \partial_{\eta_i})(\partial_{\bar\eta_j} - \partial_{\bar\eta_i}) \bigg] \\
&\quad + 2\pi i(\tau - \bar\tau) \Big[ \delta_{\beta,\sigma} \sum_{i=2}^{n} \eta_i \partial_{\bar\eta_i}\, \mathrm{sv}\, D^\tau_{\vec\eta}(\rho|\alpha) + \delta_{\alpha,\rho} \sum_{i=2}^{n} \bar\eta_i \partial_{\eta_i} \overline{\mathrm{sv}\, D^\tau_{\vec\eta}}(\sigma|\beta) \Big] \\
&\quad + (\tau - \bar\tau)^2\, \mathrm{sv}\, D^\tau_{\vec\eta}(\rho|\alpha)\, \overline{\mathrm{sv}\, D^\tau_{\vec\eta}}(\sigma|\beta) \bigg\} W^\tau_{\vec\eta}(\beta|\alpha)\,.
\end{aligned} \qquad (7.86)$$

According to (7.35), the Laplacian differs from this by terms proportional to the weights that are also given by differential operators in $\eta$. The final result for the general Laplace equation is then

$$\begin{aligned}
&(2\pi i)^2 \Delta_{\vec\eta} W^{\tau}_{\vec\eta}(\sigma|\rho) \\
&= \sum_{\alpha,\beta\in\mathcal{S}_{n-1}} \Bigg\{ \delta_{\alpha,\rho}\delta_{\beta,\sigma}\Bigg[ (2\pi i)^2(2-n)\Big(n-1+\sum_{i=2}^{n}(\eta_i\partial_{\eta_i}+\bar\eta_i\partial_{\bar\eta_i})\Big) \\
&\quad + (2\pi i)^2 \sum_{2\le i<j}^{n} (\eta_i\bar\eta_j-\eta_j\bar\eta_i)(\partial_{\eta_j}\partial_{\bar\eta_i}-\partial_{\eta_i}\partial_{\bar\eta_j}) \\
&\quad + 2\pi i(\tau-\bar\tau)\sum_{1\le i<j}^{n} s_{ij}(\partial_{\eta_j}-\partial_{\eta_i})(\partial_{\bar\eta_j}-\partial_{\bar\eta_i})\Bigg] \\
&\quad + 2\pi i(\tau-\bar\tau)\Bigg[\delta_{\beta,\sigma}\sum_{i=2}^{n}\eta_i\partial_{\bar\eta_i}\,\mathrm{sv}\,D^{\tau}_{\vec\eta}(\rho|\alpha) + \delta_{\alpha,\rho}\sum_{i=2}^{n}\bar\eta_i\partial_{\eta_i}\overline{\mathrm{sv}\,D^{\tau}_{\vec\eta}(\sigma|\beta)}\Bigg] \\
&\quad + (\tau-\bar\tau)^2\,\mathrm{sv}\,D^{\tau}_{\vec\eta}(\rho|\alpha)\,\overline{\mathrm{sv}\,D^{\tau}_{\vec\eta}(\sigma|\beta)}\Bigg\} W^{\tau}_{\vec\eta}(\beta|\alpha)\,. 
\end{aligned} \tag{7.87}$$

The term in the second line is due to the fact that the second-derivative terms of (7.86) and $\Delta_{\vec\eta}-\overline{\nabla}^{(n-2)}_{\vec\eta}\nabla^{(n-1)}_{\vec\eta}$ comprise different contractions of the summation variables $i,j=2,3,\dots,n$: One is $(\eta\partial_{\bar\eta})(\bar\eta\partial_\eta)$ while the other is $(\eta\partial_\eta)(\bar\eta\partial_{\bar\eta})$, so that only the diagonal terms cancel and one is left with a rotation-type term that contributes for $n>2$, as does the first line. These terms were not visible in the two-point example (7.47). Above we still set $\partial_{\eta_1}=\partial_{\bar\eta_1}=0$. We note that the equation (7.87) has the correct reality property under complex conjugation associated with a real Laplacian at $n$ points.

Similar to the discussion in Section 7.2.3, the consistency of the $\eta$-expansions of the left-hand side and right-hand side of (7.87) follows from integration-by-parts-identities for the component integrals. The $\eta$-expansion around different variables is also analyzed in Appendix F of [IV] in a three-point example.

The general formula (7.87) can be evaluated for any number of points $n$, any permutations $\rho,\sigma\in\mathcal{S}_{n-1}$ and for any component integral $W^{\tau}_{(A|B)}(\sigma|\rho)$. The complexity of doing so grows very rapidly, therefore we restrict ourselves here to giving only a few low-weight examples.

### 7.4.2 *Three-point examples*

This section is dedicated to the three-point instance of the Laplace equation (7.87)

$$\begin{aligned}
&\Delta_{\eta_2,\eta_3} W^\tau_{\eta_2,\eta_3}(\sigma|\rho) \\
&= \sum_{\alpha,\beta\in\mathcal{S}_2} \Big\{ \delta_{\alpha,\rho}\delta_{\beta,\sigma}\Big[ (\eta_2\bar\eta_3 - \eta_3\bar\eta_2)(\partial_{\eta_3}\partial_{\bar\eta_2} - \partial_{\eta_2}\partial_{\bar\eta_3}) \\
&\qquad - \big(2 + \eta_2\partial_{\eta_2} + \bar\eta_2\partial_{\bar\eta_2} + \eta_3\partial_{\eta_3} + \bar\eta_3\partial_{\bar\eta_3}\big) \\
&\qquad + \frac{\tau_2}{\pi}\big(s_{12}\partial_{\eta_2}\partial_{\bar\eta_2} + s_{13}\partial_{\eta_3}\partial_{\bar\eta_3} + s_{23}(\partial_{\eta_2} - \partial_{\eta_3})(\partial_{\bar\eta_2} - \partial_{\bar\eta_3})\big)\Big] \\
&\quad + \frac{\tau_2}{\pi}\Big[\delta_{\beta,\sigma}(\eta_2\partial_{\bar\eta_2} + \eta_3\partial_{\bar\eta_3})\,\mathrm{sv}\,D^\tau_{\eta_2,\eta_3}(\rho|\alpha) \\
&\qquad + \delta_{\alpha,\rho}(\bar\eta_2\partial_{\eta_2} + \bar\eta_3\partial_{\eta_3})\overline{\mathrm{sv}\,D^\tau_{\eta_2,\eta_3}(\sigma|\beta)}\Big] \\
&\quad + \Big(\frac{\tau_2}{\pi}\Big)^2 \mathrm{sv}\,D^\tau_{\eta_2,\eta_3}(\rho|\alpha)\,\overline{\mathrm{sv}\,D^\tau_{\eta_2,\eta_3}(\sigma|\beta)}\Big\} W^\tau_{\eta_2,\eta_3}(\beta|\alpha)
\end{aligned} \tag{7.88}$$

and its implications for component integrals $W^\tau_{(a_2,a_3|b_2,b_3)}(\sigma|\rho)$ defined in (7.18). The matrix entries of $\mathrm{sv}\,D^\tau_{\eta_2,\eta_3}$ can be found in (7.70) and (7.71). In the simplest case with weights $(A|B) = (0,0|0,0)$ and $\rho(2,3) = \sigma(2,3) = (2,3)$, one obtains the following equation from (7.88):

$$\begin{aligned}
&\Delta W^\tau_{(0,0|0,0)}(2{,}3|2{,}3) \\
&= \Big(\frac{\tau_2}{\pi}\Big)^2 \Big[ s_{13}^2 W^\tau_{(0,2|0,2)}(2{,}3|2{,}3) + 2s_{13}s_{23} W^\tau_{(0,2|0,2)}(2{,}3|2{,}3) \\
&\quad + s_{23}^2 W^\tau_{(0,2|0,2)}(2{,}3|2{,}3) - s_{13}^2 W^\tau_{(0,2|0,2)}(2{,}3|3{,}2) - s_{13}s_{23} W^\tau_{(0,2|0,2)}(2{,}3|3{,}2) \\
&\quad - s_{13}^2 W^\tau_{(0,2|0,2)}(3{,}2|2{,}3) - s_{13}s_{23} W^\tau_{(0,2|0,2)}(3{,}2|2{,}3) + s_{13}^2 W^\tau_{(0,2|0,2)}(3{,}2|3{,}2) \\
&\quad + s_{13}^2 W^\tau_{(0,2|2,0)}(3{,}2|2{,}3) + s_{13}s_{23} W^\tau_{(0,2|2,0)}(3{,}2|2{,}3) - s_{13}^2 W^\tau_{(0,2|2,0)}(3{,}2|3{,}2) \\
&\quad + s_{13}^2 W^\tau_{(2,0|0,2)}(2{,}3|3{,}2) + s_{12}^2 W^\tau_{(2,0|2,0)}(2{,}3|2{,}3) + s_{12}s_{23} W^\tau_{(0,2|2,0)}(2{,}3|2{,}3) \\
&\quad - s_{13}^2 W^\tau_{(2,0|0,2)}(3{,}2|3{,}2) + s_{13}^2 W^\tau_{(2,0|2,0)}(3{,}2|3{,}2) + s_{12}s_{13} W^\tau_{(2,0|0,2)}(2{,}3|2{,}3) \\
&\quad + s_{12}s_{23} W^\tau_{(2,0|0,2)}(2{,}3|2{,}3) - s_{12}s_{13} W^\tau_{(2,0|0,2)}(3{,}2|2{,}3) \\
&\quad + s_{12}s_{13} W^\tau_{(2,0|2,0)}(2{,}3|3{,}2) + s_{12}s_{13} W^\tau_{(0,2|2,0)}(2{,}3|2{,}3) \\
&\quad + s_{12}s_{13} W^\tau_{(2,0|2,0)}(3{,}2|2{,}3) - s_{12}s_{13} W^\tau_{(0,2|2,0)}(2{,}3|3{,}2) \\
&\quad + s_{13}s_{23} W^\tau_{(2,0|0,2)}(2{,}3|3{,}2)\Big] \, .
\end{aligned} \tag{7.89}$$

This equation exhibits the non-trivial mixing of the permutations due to the sum over $\alpha$ and $\beta$ in (7.88). As was the case for the Cauchy–Riemann equation (7.74) at three points, there are relations between the various component integrals. For instance one has the relation

$$W^\tau_{(0,2|0,2)}(2{,}3|3{,}2) = W^\tau_{(0,2|0,2)}(3{,}2|2{,}3) \tag{7.90}$$

that can be deduced by looking at the world-sheet integral they represent and using that $f^{(2)}$ is an even function of its $z$-argument. Substituting in this and similar relations between the component integrals one arrives at

$$\begin{aligned}\Delta W^{\tau}_{(0,0|0,0)}(2,3|2,3) = \left(\frac{\tau_2}{\pi}\right)^2 \Big\{ & s_{12}s_{13}\big[W_{(2,0|2,0)}(3,2|2,3)+W_{(2,0|2,0)}(2,3|3,2)\big] \\ &+ s_{12}s_{23}\big[W_{(2,0|0,2)}(2,3|2,3) + W_{(0,2|2,0)}(2,3|2,3)\big] \\ &+ s_{13}s_{23}\big[W_{(2,0|0,2)}(2,3|3,2) + W_{(0,2|2,0)}(3,2|2,3)\big] \\ &+ s_{12}^2 W_{(2,0|2,0)}(2,3|2,3) + s_{13}^2 W_{(2,0|2,0)}(3,2|3,2) \\ &+ s_{23}^2 W_{(0,2|0,2)}(2,3|2,3)\Big\}\end{aligned} \tag{7.91}$$

that can be verified by explicitly acting with $\Delta = -(\tau-\bar\tau)^2\partial_\tau\partial_{\bar\tau}$ on the modular invariant pure Koba–Nielsen integral with

$$\Delta\,\mathrm{KN}_3 = \left(\frac{\tau_2}{\pi}\right)^2 (s_{12}f^{(2)}_{12}+s_{13}f^{(2)}_{13}+s_{23}f^{(2)}_{23})(s_{12}\overline{f^{(2)}_{12}}+s_{13}\overline{f^{(2)}_{13}}+s_{23}\overline{f^{(2)}_{23}})\,\mathrm{KN}_3\,. \tag{7.92}$$

The low-energy expansions of the above component integrals again translate into MGFs. The right-hand side of (7.91) to third order in $\alpha'$ expands as

$$\begin{aligned}\Delta W^{\tau}_{(0,0|0,0)}(2,3|2,3) &= (s_{12}^2+s_{23}^2+s_{13}^2)\left(\frac{\tau_2}{\pi}\right)^2 C\big[\begin{smallmatrix}2&0\\2&0\end{smallmatrix}\big] + 6s_{12}s_{23}s_{13}\left(\frac{\tau_2}{\pi}\right)^3 C\big[\begin{smallmatrix}3&0\\3&0\end{smallmatrix}\big] \\ &\quad + (s_{12}^3+s_{23}^3+s_{13}^3)\left(\frac{\tau_2}{\pi}\right)^3 C\big[\begin{smallmatrix}0&1&2\\2&1&0\end{smallmatrix}\big] + O(s_{ij}^4) \\ &= (s_{12}^2+s_{23}^2+s_{13}^2)\mathrm{E}_2 \\ &\quad + (s_{12}^3+s_{23}^3+s_{13}^3+6s_{12}s_{23}s_{13})\mathrm{E}_3 + O(s_{ij}^4)\,,\end{aligned} \tag{7.93}$$

where in the second step we have substituted in simplifications of MGFs of the type discussed in Chapter 5. The expansion of the component integral $W^{\tau}_{(0,0|0,0)}(2,3|2,3)$ itself is

$$\begin{aligned}W^{\tau}_{(0,0|0,0)}(2,3|2,3) = 1 &+ \frac{1}{2}(s_{12}^2+s_{23}^2+s_{13}^2)\left(\frac{\tau_2}{\pi}\right)^2 C\big[\begin{smallmatrix}2&0\\2&0\end{smallmatrix}\big] + s_{12}s_{23}s_{13}\left(\frac{\tau_2}{\pi}\right)^3 C\big[\begin{smallmatrix}3&0\\3&0\end{smallmatrix}\big] \\ &+ \frac{1}{6}(s_{12}^3+s_{23}^3+s_{13}^3)\left(\frac{\tau_2}{\pi}\right)^3 C\big[\begin{smallmatrix}1&1&1\\1&1&1\end{smallmatrix}\big] + O(s_{ij}^4)\,.\end{aligned} \tag{7.94}$$

Acting on this expression with the Laplacian using (5.53) leads again to non-trivial relations between MGFs including $\Delta(\frac{\tau_2}{\pi})^3 C\big[\begin{smallmatrix}1&1&1\\1&1&1\end{smallmatrix}\big] = 6\mathrm{E}_3$. Higher orders in $\alpha'$ reproduce Laplace equations such as [39]

$$(\Delta-2)\mathrm{E}_{2,2} = -\mathrm{E}_2^2\,, \qquad (\Delta-6)\mathrm{E}_{2,3} = \frac{\zeta_5}{10} - 4\mathrm{E}_2\mathrm{E}_3 \tag{7.95}$$

from generating-function methods.

Moreover, we have extracted the Laplace equations of various further three-point component integrals from (7.88) and verified consistency with the leading four or more orders in the $\alpha'$-expansion of

$$\begin{aligned} &W^\tau_{(1,0|1,0)}(2{,}3|2{,}3)\,, \quad W^\tau_{(1,0|0,1)}(2{,}3|2{,}3)\,, \quad W^\tau_{(2,0|0,0)}(2{,}3|2{,}3)\,, \\ &W^\tau_{(2,0|2,0)}(2{,}3|2{,}3)\,, \quad W^\tau_{(2,0|0,2)}(2{,}3|2{,}3)\,, \quad W^\tau_{(1,1|2,0)}(2{,}3|2{,}3)\,. \end{aligned} \tag{7.96}$$

Expressions for $\Delta^{(1,1)} W^\tau_{(1,0|0,1)}(2{,}3|2{,}3)$ and $\Delta^{(2,0)} W^\tau_{(2,0|0,0)}(2{,}3|2{,}3)$ in terms of component integrals similar to (7.91) can be found in Appendix F.3 of [IV]. We note that the general Laplace equation (7.87) does not produce any MGFs with negative entries on the edge labels and never requires using HSR.

### 7.4.3 *n-point examples*

We have seen for the simplest three-point integral $W^\tau_{(0,0|0,0)}(2{,}3|2{,}3)$ that the Laplace equation (7.91) derived from the generating function (7.87) can be alternatively obtained from the Koba–Nielsen derivative (7.92). Similarly, the Laplacian of the $n$-point Koba–Nielsen factor

$$\Delta\, \mathrm{KN}_n = \left(\frac{\tau_2}{\pi}\right)^2 \left( \sum_{1\le i<j}^{n} s_{ij} f^{(2)}_{ij} \right) \left( \sum_{1\le p<q}^{n} s_{pq} \overline{f^{(2)}_{pq}} \right) \mathrm{KN}_n \tag{7.97}$$

allows for a shortcut derivation of

$$\begin{aligned} \Delta W^\tau_{(0,0,\dots,0|0,0,\dots,0)}(2,3,\dots,n|2,3,\dots,n) &= \Delta \int \mathrm{d}\mu_{n-1}\, \mathrm{KN}_n \\ &= \left(\frac{\tau_2}{\pi}\right)^2 \int \mathrm{d}\mu_{n-1} \left( \sum_{1\le i<j}^{n} s_{ij} f^{(2)}_{ij} \right) \left( \sum_{1\le p<q}^{n} s_{pq} \overline{f^{(2)}_{pq}} \right) \mathrm{KN}_n\,. \end{aligned} \tag{7.98}$$

The component integral on the left-hand side generates the MGFs with only Green functions in the integrand. Hence, (7.98) reduces the Laplacian of arbitrary modular graph functions to (sums of) $\alpha'$-expansions of integrals over $f^{(2)}_{ij} \overline{f^{(2)}_{pq}}$. The latter can be straightforwardly lined up with $W^\tau_{(2,0,\dots,0|2,0,\dots,0)}(\sigma|\rho)$ and permutations of the subscripts 2 and 0.

In principle, this kind of direct computation involving the Koba–Nielsen derivative (7.97) can also be used beyond the simplest cases, e.g.

$$\begin{aligned} \Delta^{(2,0)} W^\tau_{(2,0,\dots,0|0,0,\dots,0)}(2,3,\dots,n|2,3,\dots,n) &= \Delta^{(2,0)} \int \mathrm{d}\mu_{n-1}\, f^{(2)}_{12}\, \mathrm{KN}_n \\ &= \int \mathrm{d}\mu_{n-1}\, \mathrm{KN}_n \left\{ \left(\frac{\tau_2}{\pi}\right)^2 f^{(2)}_{12} \left( \sum_{\substack{1\le i<j \\ (i,j)\neq(1,2)}}^{n} s_{ij} f^{(2)}_{ij} \right) \left( \sum_{1\le p<q}^{n} s_{pq} \overline{f^{(2)}_{pq}} \right) \right. \end{aligned}$$

$$
+ \left(\frac{\tau_2}{\pi}\right)^2 \Big[ s_{12}(3\mathrm{G}_4 - 2f^{(4)}_{12}) + 2f^{(3)}_{12} \sum_{j=3}^{n} s_{2j} f^{(1)}_{2j} \Big] \Big( \sum_{1\le p<q}^{n} s_{pq} \overline{f^{(2)}_{pq}} \Big)
$$

$$
+ 2\frac{\tau_2}{\pi} f^{(3)}_{12} \Big[ -s_{12}\overline{f^{(1)}_{12}} + \sum_{j=3}^{n} s_{2j} \overline{f^{(1)}_{2j}} \Big] \Big\} . \tag{7.99}
$$

Note that we needed to use component identities such as

$$
\begin{aligned}
((\tau-\bar\tau)\partial_\tau f^{(2)}_{12} + 2f^{(2)}_{12}) &= 2\frac{\tau_2}{\pi}\partial_{z_1} f^{(3)}_{12} , \\
f^{(2)}_{12} f^{(2)}_{12} - 2f^{(3)}_{12} f^{(1)}_{12} &= -2f^{(4)}_{12} + 3\mathrm{G}_4
\end{aligned} \tag{7.100}
$$

in intermediate steps to express the right-hand side in terms of the basis of $W^{\tau}_{(A|B)}(\sigma|\rho)$ (possibly after use of the Fay identity (5.121)). Manipulations of this type become increasingly complicated with additional factors of $f^{(a)}_{ij}$ and $f^{(b)}_{ij}$ in the integrand while the generating-function methods underlying (7.87) are insensitive to the choice of component integral under investigation. In summary, this section exemplifies the Laplacian action at the level of $n$-point component integrals and illustrates the kind of laborious manipulations that are bypassed in the generating-function approach.

# 8

# ALL MODULAR GRAPH FORMS FROM ITERATED EISENSTEIN INTEGRALS

In this chapter, we will solve the $n$-point Cauchy–Riemann equation (7.66) derived in the previous chapter for the generating function (7.1) of Koba–Nielsen integrals via Picard-iteration. A crucial step in this will be to find a suitable redefinition $\widehat{Y}^{\tau}_{\vec{\eta}}$ of $W^{\tau}_{\vec{\eta}}$ such that the $\tau$-derivative of the generating series can be expressed entirely in terms of holomorphic Eisenstein series and conjectural representations of the derivation algebra discussed in Section 4.2.2 acting on $\widehat{Y}^{\tau}_{\vec{\eta}}$. Solving this form of the Cauchy–Riemann equation naturally leads to an expansion of $\widehat{Y}^{\tau}_{\vec{\eta}}$ in terms of iterated Eisenstein integrals which satisfy constraints imposed by the derivation algebra. The initial value for the solution will be provided by the $\tau \to i\infty$ limit, which can be computed from tree-level integrals at two-points and from the basis decompositions and Laurent expansions of MGFs discussed in Chapter 5 at three points. A tree-level expression for the initial value for more than two-points is under investigation [257].

The differential equation w.r.t. $\tau$ leaves antiholomorphic integration constants undetermined and these are fixed by the reality properties of the generating series. We will perform this computation explicitly for modular weights $(a, b)$ with $a + b \leq 10$ for two- and three-point integrals, yielding iterated Eisenstein integrals of depth one and two. Furthermore, the modular properties of the generating series also fix the modular properties of the iterated Eisenstein integrals. Comparing the solution for $\widehat{Y}^{\tau}_{\vec{\eta}}$ in terms of iterated Eisenstein integrals to the one in terms of MGFs discussed in the last chapter leads to explicit relations between iterated Eisenstein integrals. Moreover, since the $\widehat{Y}^{\tau}_{\vec{\eta}}$ generate all MGFs and the iterated Eisenstein integrals are linearly independent for different labels [35], counting the iterated Eisenstein integrals subject to the constraints from the derivation algebra leads to a counting of basis dimensions of MGFs. In this way, we can confirm the basis dimensions found in Chapter 5 and predict the number of independent MGFs (and imaginary cusp forms) at higher weights.

Finally, comparing the iterated Eisenstein integrals obtained here to the ones obtained from a similar construction in the open string opens the door towards a new understanding of the elliptic single-valued map which is explored in [17].

The material discussed in this chapter was published in [V] and the present text has extensive overlap with the reference. The material in Section 8.1 is also partly taken from [IV].

This chapter is structured as follows: In Section 8.1, we review the solution of the open-string differential equation and define a new version $Y^\tau_{\vec{\eta}}$ of the generating series $W^\tau_{\vec{\eta}}$ introduced in Chapter 7, which satisfies a differential equation which is amenable to a solution via Picard iteration. In Section 8.2 we describe in detail, how this differential equation can be solved in terms of iterated Eisenstein integrals by performing a further redefinition of $Y^\tau_{\vec{\eta}}$ to $\widehat{Y}^\tau_{\vec{\eta}}$. We explicitly construct the solutions obtained in this way for two points in Section 8.3 and for three points in Section 8.4. We finish with Section 8.5, where we discuss general properties of the iterated Eisenstein integrals obtained, count the basis dimensions of MGFs for all modular weights $a + b \leq 14$ and discuss the uniform transcendentality of the $Y^\tau_{\vec{\eta}}$.

## 8.1 SETTING UP THE GENERATING FUNCTION

In this section, we give a brief review of the open-string differential equation and its solution in terms of iterated Eisenstein integrals via Picard iteration to illustrate the general idea. Furthermore, we define a version of the generating function (7.1) of Koba–Nielsen integrals which is amenable for a solution via Picard iteration.

### 8.1.1 *The open-string analogues*

We reiterate that the open-string integral is over the boundary of the cylinder with a certain ordering $\sigma$ of the punctures and we restrict to the planar case of all punctures on the same boundary for simplicity, cf. (7.3). As shown in [37, 38], these integrals satisfy the differential equation (7.39) with the differential operator $D^\tau_{\vec{\eta}}$ that is linear in the Mandelstam variables and whose single-valued version appears in (7.64). This homogeneous first-order differential equation can be solved formally by Picard iteration (with $q = e^{2\pi i\tau}$)

$$\begin{aligned} Z^\tau_{\vec{\eta}}(\sigma|\rho) &= Z^{i\infty}_{\vec{\eta}}(\sigma|\rho) + \frac{1}{2\pi i}\int_{i\infty}^{\tau} \mathrm{d}\tau_1 \sum_{\alpha\in\mathcal{S}_{n-1}} D^{\tau_1}_{\vec{\eta}}(\rho|\alpha) Z^{i\infty}_{\vec{\eta}}(\sigma|\alpha) \\ &\quad + \frac{1}{(2\pi i)^2}\int_{i\infty}^{\tau}\mathrm{d}\tau_1\int_{i\infty}^{\tau_1}\mathrm{d}\tau_2 \sum_{\alpha,\beta\in\mathcal{S}_{n-1}} D^{\tau_1}_{\vec{\eta}}(\rho|\alpha) D^{\tau_2}_{\vec{\eta}}(\alpha|\beta) Z^{i\infty}_{\vec{\eta}}(\sigma|\beta) + \ldots \\ &= \sum_{\ell=0}^{\infty} \frac{1}{(2\pi i)^{2\ell}} \int_{0<q_1<\ldots<q_\ell<q} \frac{\mathrm{d}q_1}{q_1}\cdots\frac{\mathrm{d}q_\ell}{q_\ell} \sum_{\alpha\in\mathcal{S}_{n-1}} (D^{\tau_\ell}_{\vec{\eta}}\cdots D^{\tau_1}_{\vec{\eta}})(\rho|\alpha) Z^{i\infty}_{\vec{\eta}}(\sigma|\alpha)\,. \end{aligned} \tag{8.1}$$

Note that here, $\tau_2$ is not $\operatorname{Im}\tau$, but just an integration variable. The summation variable $\ell$ in the last line tracks the orders of $\alpha'$ carried by the $D^{\tau_j}_{\vec\eta}$ matrices. The important point here is that the initial values $Z^{i\infty}_{\vec\eta}(\sigma|\alpha)$ are by themselves series in $\alpha'$ that have been identified with *disk* integrals of Parke–Taylor type at $n+2$ points [37, 38]. Their $\alpha'$-expansion is expressible in terms of MZVs [20, 22, 23, 153, 258], and the dependence on $s_{ij}$ can for instance be imported from the all-multiplicity methods of [24, 259]. Hence, any given $\alpha'$-order of the $A$-cycle integrals is accessible from finitely many terms in the sum over $\ell$ in (8.1), i.e. after finitely many steps of Picard iteration.

By expanding the Weierstraß functions in the matrix entries of $D^{\tau}_{\vec\eta}$ in terms of holomorphic Eisenstein series using (3.19), one can uniquely decompose

$$D^{\tau}_{\vec\eta} = \sum_{k=0}^{\infty}(1-k)\mathrm{G}_k(\tau) r_{\vec\eta}(\epsilon_k) \tag{8.2}$$

with $\mathrm{G}_0 = -1$. All the reference to $\eta_j$ resides in the differential operators $r_{\vec\eta}(\epsilon_k)$, where $\epsilon_k$ is a formal letter with $k = 0, 4, 6, 8, \ldots$ and we set $\epsilon_k = 0$ for all other $k$. Explicit expressions for the $r_{\vec\eta}(\epsilon_k)$ can be found in [37, 38] as well as (8.14) and (8.15) below ($R_{\vec\eta}(\epsilon_k) = r_{\vec\eta}(\epsilon_k)$ for $k \geq 4$) and they are believed to form matrix representations of Tsunogai's derivations dual to Eisenstein series [207], cf. Section 4.2.2. They inherit linearity in $s_{ij}$ from $D^{\tau}_{\vec\eta}$, and we have $r_{\vec\eta}(\epsilon_2) = 0$ at all multiplicities by the absence of $\mathrm{G}_2$ in the Laurent expansion (3.19) of $\wp(\eta, \tau)$.

Substituting (8.2) into (8.1), the entire $\tau$-dependence in the $\alpha'$-expansion of $A$-cycle integrals $Z^{\tau}_{\vec\eta}$ is carried by iterated Eisenstein integrals

$$\gamma(k_1, \ldots, k_\ell|\tau) = \frac{(-1)^\ell}{(2\pi i)^{2\ell}} \int\limits_{0<q_1<\ldots<q_\ell<q} \frac{\mathrm{d}q_1}{q_1} \ldots \frac{\mathrm{d}q_\ell}{q_\ell} \mathrm{G}_{k_1}(\tau_1)\ldots \mathrm{G}_{k_\ell}(\tau_\ell) \tag{8.3}$$

subject to tangential-base-point regularization [31] such that $\gamma(0|\tau) = \frac{\tau}{2\pi i}$.[1] We arrive at a sum over words $k_1, k_2, \ldots, k_\ell$ composed from the alphabet $k_j \in \{0, 4, 6, 8, \ldots\}$ [37, 38],

$$\begin{aligned} &Z^{\tau}_{\vec\eta}(\sigma|\rho) \\ &= \sum_{\ell=0}^{\infty} \sum_{\substack{k_1,\ldots,k_\ell\\=0,4,6,8,\ldots}} \Big[\prod_{j=1}^{\ell}(k_j-1)\Big] \gamma(k_1, \ldots, k_\ell|\tau) \sum_{\alpha\in\mathcal{S}_{n-1}} r_{\vec\eta}(\epsilon_{k_\ell}\ldots\epsilon_{k_1})_\rho{}^\alpha Z^{i\infty}_{\vec\eta}(\sigma|\alpha)\,, \end{aligned} \tag{8.4}$$

where we write $r_{\vec\eta}(\epsilon_{k_\ell}\ldots\epsilon_{k_2}\epsilon_{k_1}) = r_{\vec\eta}(\epsilon_{k_\ell})\ldots r_{\vec\eta}(\epsilon_{k_2}) r_{\vec\eta}(\epsilon_{k_1})$ for ease of notation.

The formula (8.4) provides an explicit evaluation of the one-loop $A$-cycle integrals in terms of iterated Eisenstein integrals and initial values that can be traced back to tree-level amplitudes. As all $\tau$-dependence

1 The iterated integrals $\gamma$ constitute a mere change of normalization compared to the $\mathcal{E}$ introduced in (1.5).

is carried by iterated integrals $\gamma(k_1,k_2,\ldots,k_\ell|\tau)$, the differential equation (7.39) is satisfied by the defining property

$$2\pi i\partial_\tau\gamma(k_1,k_2,\ldots,k_\ell|\tau) = -\mathrm{G}_{k_\ell}(\tau)\gamma(k_1,k_2,\ldots,k_{\ell-1}|\tau) \tag{8.5}$$

of iterated Eisenstein integrals, cf. (8.3). Moreover, (8.4) yields the correct initial condition for $Z^\tau_{\vec\eta}(\sigma|\rho)$ since $\lim_{\tau\to i\infty}\gamma(k_1,k_2,\ldots,k_\ell|\tau) = 0$ in the sense of a regularized value.

### 8.1.2 *An improved form the of the closed-string differential equation*

We shall now outline a starting point for the corresponding procedure to expand closed-string generating series $W^\tau_{\vec\eta}(\sigma|\rho)$. One can first rewrite the Cauchy–Riemann equation (7.66) as

$$\begin{aligned} 2\pi i(\tau-\bar\tau)\partial_\tau W^\tau_{\vec\eta}(\sigma|\rho) &= 2\pi i\Big[1-n+\sum_{i=2}^n(\bar\eta_i-\eta_i)\partial_{\eta_i}\Big]W^\tau_{\vec\eta}(\sigma|\rho)\\ &\quad+(\tau-\bar\tau)\sum_{\alpha\in\mathcal{S}_{n-1}}\mathrm{sv}\,D^\tau_{\vec\eta}(\rho|\alpha)W^\tau_{\vec\eta}(\sigma|\alpha)\,. \end{aligned} \tag{8.6}$$

The terms in the first line are independent of the Mandelstam variables and also mix the holomorphic and antiholomorphic orders in the variables of the generating series. This obstructs a direct link between Picard iteration and the $\alpha'$-expansion in analogy with the open-string construction. In the following, we will present a redefinition of the $W^\tau_{\vec\eta}$ integrals such that one can still obtain each order in the $\alpha'$-expansion of the component integrals through a finite number of elementary operations.

The contributions $1-n-\sum_{i=2}^n\eta_i\partial_{\eta_i}$ to the $\alpha'$-independent square-bracket in (8.6) can be traced back to the connection term in the Maaß operator (7.31a) which simply adjusts the modular weights. In general, one can suppress the connection term in (3.51) by enforcing vanishing holomorphic modular weight on the functions it acts on, which is always possible by multiplication with suitable powers of $(\tau-\bar\tau)$. Hence, we will consider a modified version of the $W$-integrals, where each component integral in (7.7) of modular weight $(|A|,|B|)$ is multiplied by $(\tau-\bar\tau)^{|A|}$ such as to attain the shifted modular weights $(0,|B|-|A|)$.

Since the component integrals $W^\tau_{(A|B)}$ defined in (7.5) have modular weights $(|A|,|B|)$, the desired modification of (7.7) is given by

$$\begin{aligned} &Y^\tau_{\vec\eta}(\sigma|\rho)\\ &=\sum_{A,B}(\tau-\bar\tau)^{|A|}W^\tau_{(A|B)}(\sigma|\rho)\,\sigma\big[\bar\eta_{234\ldots n}^{b_2-1}\cdots\bar\eta_n^{b_n-1}\big]\,\rho\big[\eta_{234\ldots n}^{a_2-1}\cdots\eta_n^{a_n-1}\big] \end{aligned}$$

$$
\begin{aligned}
&= \sum_{A,B} W^{\tau}_{(A|B)}(\sigma|\rho)\, \sigma\left[\bar\eta_{234\ldots n}^{b_2-1}\ldots \bar\eta_n^{b_n-1}\right] \\
&\qquad \times (\tau-\bar\tau)^{n-1}\, \rho\left[\big((\tau-\bar\tau)\eta_{234\ldots n}\big)^{a_2-1}\ldots\big((\tau-\bar\tau)\eta_n\big)^{a_n-1}\right] \\
&= (\tau-\bar\tau)^{n-1} W^{\tau}_{\vec\eta}(\sigma|\rho)\Big|_{\substack{\eta\to(\tau-\bar\tau)\eta \\ \bar\eta\to\bar\eta}} .
\end{aligned} \tag{8.7}
$$

Given that the entire $Y$-integral has holomorphic modular weight zero, the action of the Maaß operator (7.31a) reduces to $(\tau-\bar\tau)\partial_\tau$, and the Cauchy–Riemann equation (8.6) simplifies to

$$
\begin{aligned}
&2\pi i(\tau-\bar\tau)^2\partial_\tau Y^{\tau}_{\vec\eta}(\sigma|\rho) \\
&= 2\pi i\sum_{j=2}^{n}\bar\eta_j\partial_{\eta_j}Y^{\tau}_{\vec\eta}(\sigma|\rho) + (\tau-\bar\tau)^2\sum_{\alpha\in\mathcal{S}_{n-1}} \mathrm{sv}\, D^{\tau}_{(\tau-\bar\tau)\vec\eta}(\rho|\alpha) Y^{\tau}_{\vec\eta}(\sigma|\alpha) \\
&= \sum_{k=0}^{\infty}(1-k)(\tau-\bar\tau)^k \mathrm{G}_k(\tau)\sum_{\alpha\in\mathcal{S}_{n-1}} R_{\vec\eta}(\epsilon_k)_\rho{}^\alpha Y^{\tau}_{\vec\eta}(\sigma|\alpha)\,,
\end{aligned} \tag{8.8}
$$

where we have expanded the closed-string differential operator in terms of Eisenstein series in analogy with (8.2). For later convenience, we separate the $k=0$ contribution from the sum in (8.8) and obtain

$$
\begin{aligned}
2\pi i\partial_\tau Y^{\tau}_{\vec\eta}(\sigma|\rho) = \sum_{\alpha\in S_{n-1}}\Bigg\{ &-\frac{1}{(\tau-\bar\tau)^2}R_{\vec\eta}(\epsilon_0)_\rho{}^\alpha \\
&+\sum_{k=4}^{\infty}(1-k)(\tau-\bar\tau)^{k-2}\mathrm{G}_k(\tau)R_{\vec\eta}(\epsilon_k)_\rho{}^\alpha\Bigg\} Y^{\tau}_{\vec\eta}(\sigma|\alpha)\,.
\end{aligned} \tag{8.9}
$$

The operator $R_{\vec\eta}(\epsilon_0)$ contains also the $s_{ij}$-independent term $\sim \bar\eta_j\partial_{\eta_j}$ in its diagonal components

$$
R_{\vec\eta}(\epsilon_0)_\rho{}^\rho = \rho\Big[\sum_{1\le i<j}\frac{s_{ij}}{\eta^2_{j,j+1\ldots n}}\Big] - \frac{1}{2}\sum_{1\le i<j}s_{ij}(\partial_{\eta_i}-\partial_{\eta_j})^2 - 2\pi i\sum_{j=2}^{n}\bar\eta_j\partial_{\eta_j}\,, \tag{8.10}
$$

whereas $R_{\vec\eta}(\epsilon_0)_\rho{}^\alpha = r_{\vec\eta}(\epsilon_0)_\rho{}^\alpha$ for $\rho\neq\alpha$. The appearance of the term $\sim\bar\eta_j\partial_{\eta_j}$ in $R_{\vec\eta}(\epsilon_0)$ does not obstruct the $\alpha'$-expansion of component integrals from finitely many steps[2] in a formal solution with the structure of (8.4). For $k\geq 4$, we have agreement with the open-string expression $R_{\vec\eta}(\epsilon_k) = r_{\vec\eta}(\epsilon_k)$ and these matrices should again form a matrix representation of Tsunogai's derivations [207] as discussed in Section 4.2.2. We have checked that they preserve the commutation relations of the

2 This follows from the fact that none of the $R_{\vec\eta}(\epsilon_k)$ has a contribution that lowers the powers of $\bar\eta_j$. Hence, any component integral at a given order of $\bar\eta_j$ in (8.7) can only be affected by finitely many instances of the term $\bar\eta_j\partial_{\eta_j}$ in $R_{\vec\eta}(\epsilon_0)$ which is the only contribution to the $R_{\vec\eta}(\epsilon_k)$ without any factors of $s_{ij}$.

$\epsilon_k$, some examples of which are given in (4.24). For instance, we have checked that

$$R_{\vec{\eta}}\big(\,\mathrm{ad}_{\epsilon_0}^{k-1}(\epsilon_k)\big) = \mathrm{ad}_{R_{\vec{\eta}}(\epsilon_0)}^{k-1}\,\big(R_{\vec{\eta}}(\epsilon_k)\big) = 0\,, \qquad k \geq 2 \tag{8.11}$$

$$R_{\vec{\eta}}\big([\epsilon_{10},\epsilon_4]-3[\epsilon_8,\epsilon_6]\big) = \big[R_{\vec{\eta}}(\epsilon_{10}),R_{\vec{\eta}}(\epsilon_4)\big] - 3\big[R_{\vec{\eta}}(\epsilon_8),R_{\vec{\eta}}(\epsilon_6)\big] = 0\,. \tag{8.12}$$

in agreement with (4.24b) and (4.24c). Similarly, the $R_{\vec{\eta}}(\epsilon_k)$ at $n \leq 5$ points have been checked to preserve various generalizations of (4.24) that can be downloaded from [205]:

- relations among $[\epsilon_{k_1}, \epsilon_{k_2}]$ at $k_1+k_2 \leq 30$ and $n=2,3,4$ as well as $k_1+k_2 \leq 18$ and $n=5$,
- relations among $[\epsilon_{\ell_1}, [\epsilon_{\ell_2}, \epsilon_{\ell_3}]]$ at $\ell_1+\ell_2+\ell_3 \leq 30$ and $n = 2,3,4$,
- relations among $[\epsilon_{p_1}, [\epsilon_{p_2}, [\epsilon_{p_3}, \epsilon_{p_4}]]]$ at $p_1+p_2+p_3+p_4 \leq 26$, $n = 2,3$ as well as (partially relying on numerical methods) $p_1+p_2+p_3+p_4 \leq 18$, $n = 4$

We will see in Section 8.5.2 that relations like (4.24) will play a key role in the counting of independent MGFs at given modular weights, in the same way as they did for the counting of elliptic MZVs as discussed in Section 4.2.2.

Even though the operators $R_{\vec{\eta}}(\epsilon_k)$ satisfy the derivation-algebra relations (at least to the orders checked), their instances at given multiplicity $n$ are not a faithful representation of the derivation algebra. In other words, they can also satisfy more relations at fixed $n$. For instance, the two-point example (8.14) below implies that all $R_\eta(\epsilon_k)$ for $k \geq 4$ at $n = 2$ commute which is stronger than (4.24). As we shall use compositions of the operators $R_{\vec{\eta}}(\epsilon_k)$ in the rest of the thesis to solve (8.9), this means that their coefficients only occur in specific linear combinations in low-point results. This will lead to multiplicity-specific dropouts of MGFs in the $\alpha'$-expansion of $Y^\tau_{\vec{\eta}}$ at fixed $n$, in the same way as four-point closed-string tree-level amplitudes do not involve any MZVs of depth $\geq 2$.

Note that the asymmetric role of the $\eta_j$ and $\bar{\eta}_j$ variables in the definition (8.7) of the $Y$-integrals will modify the Laplace equation (7.87) and obscure its reality properties. For this reason we have worked with the generating series $W^\tau_{\vec{\eta}}(\sigma|\rho)$ in the previous Chapter.

At two-points, the permutations in (8.9) are trivial and we have

$$2\pi i\partial_\tau Y^\tau_\eta = \left\{-\frac{1}{(\tau-\bar{\tau})^2}R_\eta(\epsilon_0) + \sum_{k=4}^{\infty}(1-k)(\tau-\bar{\tau})^{k-2}\mathrm{G}_k(\tau)R_\eta(\epsilon_k)\right\}Y^\tau_\eta \tag{8.13}$$

with the following $\eta$- and $\bar\eta$-dependent operators

$$R_\eta(\epsilon_0) = s_{12}\Big(\frac{1}{\eta^2} - \frac{1}{2}\partial_\eta^2\Big) - 2\pi i\bar\eta\partial_\eta\,, \quad R_\eta(\epsilon_k) = s_{12}\eta^{k-2}\,, \quad k \geq 4\,. \tag{8.14}$$

The generalization to $(n{\geq}3)$ points requires $(n{-}1)! \times (n{-}1)!$ matrix-valued operators $R_{\vec\eta}(\epsilon_k)_\rho{}^\alpha$. At three points, for instance, the $R_{\vec\eta}(\epsilon_k)$ in (8.9) are $2 \times 2$ matrices

$$\begin{aligned}
R_{\eta_2,\eta_3}(\epsilon_0) &= \frac{1}{\eta_{23}^2}\begin{pmatrix} s_{12} & -s_{13} \\ -s_{12} & s_{13}\end{pmatrix} + \frac{1}{\eta_2^2}\begin{pmatrix} 0 & 0 \\ s_{12} & s_{12}{+}s_{23}\end{pmatrix} + \frac{1}{\eta_3^2}\begin{pmatrix} s_{13}{+}s_{23} & s_{13} \\ 0 & 0\end{pmatrix} \\
&\quad - \begin{pmatrix} 1 & 0 \\ 0 & 1\end{pmatrix}\Big(\frac{1}{2}s_{12}\partial_{\eta_2}^2 + \frac{1}{2}s_{13}\partial_{\eta_3}^2 + \frac{1}{2}s_{23}(\partial_{\eta_2}{-}\partial_{\eta_3})^2 + 2\pi i(\bar\eta_2\partial_{\eta_2}{+}\bar\eta_3\partial_{\eta_3})\Big)\,, \\
R_{\eta_2,\eta_3}(\epsilon_k) &= \eta_{23}^{k-2}\begin{pmatrix} s_{12} & -s_{13} \\ -s_{12} & s_{13}\end{pmatrix} + \eta_2^{k-2}\begin{pmatrix} 0 & 0 \\ s_{12} & s_{12}{+}s_{23}\end{pmatrix} \\
&\quad + \eta_3^{k-2}\begin{pmatrix} s_{13}{+}s_{23} & s_{13} \\ 0 & 0\end{pmatrix}\,, \qquad k \geq 4\,,
\end{aligned} \tag{8.15}$$

and their higher-multiplicity analogues following from [IV, 37, 38] are reviewed in Appendix B of [V].

We also note that $Y_{\vec\eta}^\tau$ satisfies the following equation when differentiated with respect to $\bar\tau$:

$$\begin{aligned}
-2\pi i\partial_{\bar\tau}Y_{\vec\eta}^\tau(\sigma|\rho) = \sum_{\alpha\in\mathcal{S}_{n-1}}\Big\{2\pi i\sum_{j=2}^n\Big[2\eta_j\partial_{\bar\eta_j} + \frac{\eta_j\partial_{\eta_j} - \bar\eta_j\partial_{\bar\eta_j}}{\tau-\bar\tau}\Big]\delta_\alpha^\sigma - \overline{R_{\vec\eta}(\epsilon_0)}_\alpha{}^\sigma \\
+ \sum_{k\geq4}(1{-}k)\overline{\mathrm{G}_k(\tau)}\,\overline{R_{\vec\eta}(\epsilon_k)}_\alpha{}^\sigma\Big\}Y_{\vec\eta}^\tau(\alpha|\rho)\,.
\end{aligned} \tag{8.16}$$

We shall not use this equation extensively but rather the holomorphic $\tau$-derivative (8.9) together with the reality properties of the component integrals, to be discussed in the next section. Similar to Brown's construction [32, 33, 74] of non-holomorphic modular forms, the series $Y_{\vec\eta}^\tau$ is engineered to simplify the holomorphic derivative (8.9) at the expense of the more lengthy expression (8.16) for the antiholomorphic one.

### 8.1.3 *Component integrals of* $Y^\tau_{\vec\eta}(\sigma|\rho)$

Following its definition in (8.7) in terms of component integrals of $W^\tau_{\vec\eta}$, the generating series $Y^\tau_{\vec\eta}$ can be written in closed form as (cf. (7.1))

$$\begin{aligned} Y^\tau_{\vec\eta}(\sigma|\rho) &= (\tau-\bar\tau)^{n-1}\int \mathrm{d}\mu_{n-1}\,\mathrm{KN}_n \\ &\times\sigma\Big[\overline{\Omega(z_{12},\eta_{23\dots n},\tau)}\,\overline{\Omega(z_{23},\eta_{34\dots n},\tau)}\cdots\overline{\Omega(z_{n-1,n},\eta_n,\tau)}\Big] \\ &\times\rho\Big[\Omega(z_{12},(\tau-\bar\tau)\eta_{23\dots n},\tau)\Omega(z_{23},(\tau-\bar\tau)\eta_{34\dots n},\tau)\cdots\Omega(z_{n-1,n},(\tau-\bar\tau)\eta_n,\tau)\Big]\,, \end{aligned} \tag{8.17}$$

where we have used the same shorthand $\eta_{i\dots j}=\eta_i+\dots+\eta_j$ as in Chapter 7. As in $W^\tau_{\vec\eta}$, the permutations $\sigma,\rho$ act on the subscripts of the generating parameters $\eta_i$ and insertion points $z_i$. We define the component integrals of (8.17) with an additional factor of $2\pi i$ (cf. (7.7)) to simplify some relations under complex conjugation below,

$$\begin{aligned} Y^\tau_{\vec\eta}(\sigma|\rho) = \sum_{A,B}(2\pi i)^{|B|}Y^\tau_{(A|B)}(\sigma|\rho)\,\rho\big[\eta^{a_2-1}_{234\dots n}\eta^{a_3-1}_{34\dots n}\cdots\eta^{a_n-1}_n\big] \\ \times\,\sigma\big[\bar\eta^{b_2-1}_{234\dots n}\bar\eta^{b_3-1}_{34\dots n}\cdots\bar\eta^{b_n-1}_n\big]\,, \end{aligned} \tag{8.18}$$

where we used the notations (7.6) and (7.8). This leads to the component integrals

$$\begin{aligned} Y^\tau_{(A|B)} &= Y^\tau_{(a_2,a_3,\dots,a_n|b_2,b_3,\dots,b_n)}(\sigma|\rho) \\ &= \frac{(\tau-\bar\tau)^{|A|}}{(2\pi i)^{|B|}}\int\mathrm{d}\mu_{n-1}\,\mathrm{KN}_n\,\rho\big[f^{(a_2)}_{12}f^{(a_3)}_{23}\cdots f^{(a_n)}_{n-1,n}\big]\,\sigma\big[\overline{f^{(b_2)}_{12}}\,\overline{f^{(b_3)}_{23}}\cdots\overline{f^{(b_n)}_{n-1,n}}\big]\,. \end{aligned} \tag{8.19}$$

Hence, $Y^\tau_{(A|B)}$ has modular weight

$$(0,|B|-|A|)\,, \tag{8.20}$$

with in particular vanishing holomorphic modular weight, which was the reason to construct the $Y^\tau_{\vec\eta}$ in the first place.

At two points, (8.17) becomes

$$Y^\tau_\eta = (\tau-\bar\tau)\int\frac{\mathrm{d}^2z_2}{\tau_2}\,\overline{\Omega(z_{12},\eta,\tau)}\Omega(z_{12},(\tau-\bar\tau)\eta,\tau)\,\mathrm{KN}_2\,, \tag{8.21}$$

with component integrals

$$Y^\tau_{(a|b)} = \frac{1}{(2\pi i)^b}Y^\tau_\eta\Big|_{\eta^{a-1}\bar\eta^{b-1}} = \frac{(\tau-\bar\tau)^a}{(2\pi i)^b}\int\frac{\mathrm{d}^2z_2}{\tau_2}\,\mathrm{KN}_2\,f^{(a)}_{12}\,\overline{f^{(b)}_{12}}\,. \tag{8.22}$$

We will later make essential use of the following reality properties: Complex conjugation of component integrals over $f^{(a)}_{12}\overline{f^{(b)}_{12}}$ exchanges $a \leftrightarrow b$, so we have

$$Y^\tau_{(a|b)} = (4y)^{a-b}\overline{Y^\tau_{(b|a)}}\,, \qquad \overline{Y^\tau_{(a|b)}} = (4y)^{a-b}Y^\tau_{(b|a)}\,, \tag{8.23}$$

where $y = \pi\tau_2$, and similarly,

$$Y^\tau_{(A|B)}(\sigma|\rho) = (4y)^{|A|-|B|}\overline{Y^\tau_{(B|A)}(\rho|\sigma)}\,. \tag{8.24}$$

As an example of how MGFs occur in the component integrals of the generating series $Y^\tau_{\vec\eta}$, we consider the two-point integrals (8.22). It can be checked by using identities for MGFs that the first few component integrals have the following $\alpha'$-expansions[3]

$$\begin{aligned} Y^\tau_{(0|0)} &= 1 + \frac{1}{2}s_{12}^2\mathrm{E}_2 + \frac{1}{6}s_{12}^3(\mathrm{E}_3+\zeta_3) + s_{12}^4\Big(\mathrm{E}_{2,2} + \frac{1}{8}\mathrm{E}_2^2 + \frac{3}{20}\mathrm{E}_4\Big) \\ &\quad + s_{12}^5\Big(\frac{1}{2}\mathrm{E}_{2,3} + \frac{1}{12}\mathrm{E}_2(\mathrm{E}_3+\zeta_3) + \frac{3}{14}\mathrm{E}_5 + \frac{2\zeta_5}{15}\Big) + O(s_{12}^6)\,, \end{aligned} \tag{8.25a}$$

$$\begin{aligned} Y^\tau_{(2|0)} &= 2s_{12}\pi\nabla_0\mathrm{E}_2 + \frac{2}{3}s_{12}^2\pi\nabla_0\mathrm{E}_3 \\ &\quad + s_{12}^3\Big(\frac{3}{5}\pi\nabla_0\mathrm{E}_4 + 4\pi\nabla_0\mathrm{E}_{2,2} + \mathrm{E}_2\pi\nabla_0\mathrm{E}_2\Big) \\ &\quad + s_{12}^4\Big(\frac{6}{7}\pi\nabla_0\mathrm{E}_5 + 2\pi\nabla_0\mathrm{E}_{2,3} + \frac{1}{3}\mathrm{E}_2\pi\nabla_0\mathrm{E}_3 + \frac{1}{3}\mathrm{E}_3\pi\nabla_0\mathrm{E}_2 \\ &\qquad + \frac{1}{3}\zeta_3\pi\nabla_0\mathrm{E}_2\Big) + O(s_{12}^5)\,, \end{aligned} \tag{8.25b}$$

$$\begin{aligned} Y^\tau_{(4|0)} &= -\frac{4}{3}s_{12}(\pi\nabla_0)^2\mathrm{E}_3 + s_{12}^2\Big(-\frac{6}{5}(\pi\nabla_0)^2\mathrm{E}_4 + 2(\pi\nabla_0\mathrm{E}_2)^2\Big) \\ &\quad + s_{12}^3\Big(-\frac{12}{7}(\pi\nabla_0)^2\mathrm{E}_5 - 4(\pi\nabla_0)^2\mathrm{E}_{2,3} - \frac{4}{3}(\pi\nabla_0\mathrm{E}_2)(\pi\nabla_0\mathrm{E}_3) \\ &\qquad - \frac{2}{3}\mathrm{E}_2(\pi\nabla_0)^2\mathrm{E}_3\Big) + O(s_{12}^4)\,. \end{aligned} \tag{8.25c}$$

## 8.2 SOLVING DIFFERENTIAL EQUATIONS FOR GENERATING SERIES

The goal of this section is to derive the form of the all-order $\alpha'$-expansion of the $Y^\tau_{\vec\eta}$ integrals (8.17) from their differential equation (8.9). As a first step we will rewrite the differential equation in a slightly different form using relations in the derivation algebra. This improved differential equation will allow for a formal solution whose properties we discuss in this section. In the next sections we make the formal solution fully explicit at certain orders by exploiting the reality properties of two- and three-point integrals.

3 When comparing with the $\alpha'$-expansions in (7.16), note that the component integrals (8.22) are related to the $W^\tau_{(a|b)}$ in Chapter 7 via $Y^\tau_{(a|b)} = \frac{(2i\tau_2)^a}{(2\pi i)^b}W^\tau_{(a|b)}$.

### 8.2.1 *Removing* $G_0$ *from the differential equation*

Given that the differential equation (8.9) is linear and of first order in $\tau$, it is tempting to solve it (up to antiholomorphic integration ambiguities) formally by line integrals over $\tau$ as was demonstrated in (8.1) for open-string integrals. In particular, the appearance of $(\tau-\bar\tau)^{k-2}G_k(\tau)$ on the right-hand side will introduce iterated integrals over holomorphic Eisenstein series in a formal solution. However, the differential equation features singular terms $\sim(\tau-\bar\tau)^{-2}$ that do not immediately line up with Brown's iterated Eisenstein integrals over $\tau^j G_k(\tau)$, $j=0,1,\ldots,k-2$ with well-studied modular transformations [31].

Therefore we first strive to remove the singular term $\sim(\tau-\bar\tau)^{-2}$ in (8.9) that does not have any accompanying Eisenstein series $G_{k\geq 4}$. This can be done by performing the invertible redefinition[4]

$$\widehat{Y}^\tau_{\vec\eta} = \exp\left(\frac{R_{\vec\eta}(\epsilon_0)}{4y}\right)Y^\tau_{\vec\eta} \quad\Leftrightarrow\quad Y^\tau_{\vec\eta} = \exp\left(-\frac{R_{\vec\eta}(\epsilon_0)}{4y}\right)\widehat{Y}^\tau_{\vec\eta}\,, \tag{8.26}$$

where the matrix multiplication w.r.t. the second index of $Y^\tau_{\vec\eta}(\sigma|\rho)$ is suppressed for ease of notation[5]. The redefined integrals obey a modified version of (8.9)

$$2\pi i\partial_\tau\widehat{Y}^\tau_{\vec\eta} = \sum_{k=4}^{\infty}(1-k)G_k(\tau)(\tau-\bar\tau)^{k-2}e^{-\frac{R_{\vec\eta}(\epsilon_0)}{2\pi i(\tau-\bar\tau)}}R_{\vec\eta}(\epsilon_k)e^{\frac{R_{\vec\eta}(\epsilon_0)}{2\pi i(\tau-\bar\tau)}}\widehat{Y}^\tau_{\vec\eta}\,, \tag{8.27}$$

where now the term without holomorphic Eisenstein series is absent and the $R_{\vec\eta}(\epsilon_k)$ are conjugated by exponentials of $R_{\vec\eta}(\epsilon_0)$. By the relations (8.11) in the derivation algebra, the exponentials along with a fixed $R_{\vec\eta}(\epsilon_k)$ truncate to a finite number of terms,

$$e^{-\frac{R_{\vec\eta}(\epsilon_0)}{2\pi i(\tau-\bar\tau)}}R_{\vec\eta}(\epsilon_k)e^{\frac{R_{\vec\eta}(\epsilon_0)}{2\pi i(\tau-\bar\tau)}} = \sum_{j=0}^{k-2}\frac{1}{j!}\left(\frac{-1}{2\pi i(\tau-\bar\tau)}\right)^j R_{\vec\eta}\big(\operatorname{ad}^j_{\epsilon_0}(\epsilon_k)\big)\,, \tag{8.28}$$

where we use the following shorthands here and below

$$R_{\vec\eta}\big(\operatorname{ad}^j_{\epsilon_0}(\epsilon_k)\big)=\operatorname{ad}^j_{R_{\vec\eta}(\epsilon_0)}R_{\vec\eta}(\epsilon_k)\,,\quad R_{\vec\eta}(\epsilon_{k_1}\epsilon_{k_2})=R_{\vec\eta}(\epsilon_{k_1})R_{\vec\eta}(\epsilon_{k_2})\,. \tag{8.29}$$

Hence, the differential equation (8.27) simplifies to

$$2\pi i\partial_\tau\widehat{Y}^\tau_{\vec\eta} = \sum_{k=4}^{\infty}(1-k)G_k(\tau)\sum_{j=0}^{k-2}\frac{1}{j!}\left(\frac{-1}{2\pi i}\right)^j(\tau-\bar\tau)^{k-2-j}R_{\vec\eta}\big(\operatorname{ad}^j_{\epsilon_0}(\epsilon_k)\big)\widehat{Y}^\tau_{\vec\eta}\,. \tag{8.30}$$

---

4 We are grateful to Nils Matthes and Erik Panzer for discussions that led to this redefinition.

5 More explicitly, $\widehat{Y}^\tau_{\vec\eta}(\sigma|\rho) = \sum_{\alpha\in\mathcal{S}_{n-1}}\exp\left(\frac{R_{\vec\eta}(\epsilon_0)}{4y}\right)_{\rho}{}^{\alpha}\,Y^\tau_{\vec\eta}(\sigma|\alpha)$.

Now, the operator on the right-hand side is manifestly free of singular terms in $(\tau-\bar\tau)$, and the sum over $k$ starts at $k=4$. All the integration kernels in this differential equation are of the form $(\tau-\bar\tau)^j \mathrm{G}_k(\tau)$ with $k \geq 4$ and $0 \leq j \leq k-2$. Hence, our kernels line up with those of Brown's holomorphic and single-valued iterated Eisenstein integrals [31–33].

### 8.2.2 *Formal expansion of the solution*

The form (8.30) bodes well for a representation in terms of $\widehat{Y}^\tau_{\vec\eta}$ as an (iterated) line integral from $\tau$ to some reference point that we take to be the cusp at $\tau \to i\infty$. In particular, the differential equation contains no negative powers of $(\tau-\bar\tau)$ or $y=\pi\tau_2$, and this property will propagate to the solution $\widehat{Y}^\tau_{\vec\eta}$, see Section 8.2.4 for further details. The original integrals $Y^\tau_{\vec\eta}$, in turn, involve combinations of MGFs with negative powers of $y$ from their Laurent polynomials. The absence of negative powers of $y$ in $\widehat{Y}^\tau_{\vec\eta}$ is a crucial difference as compared to $Y^\tau_{\vec\eta}$ and is due to the redefinition (8.26). We shall later make this more manifest when we discuss explicit examples obtained from low-point amplitudes.

A formal solution of (8.30), that also exposes the $\alpha'$-expansion of the integrals, is given by the series

$$\begin{aligned}\widehat{Y}^\tau_{\vec\eta} = \sum_{\ell=0}^{\infty} \sum_{\substack{k_1,k_2,\ldots,k_\ell\\=4,6,8,\ldots}} \sum_{j_1=0}^{k_1-2} \sum_{j_2=0}^{k_2-2} \cdots \sum_{j_\ell=0}^{k_\ell-2} \left( \prod_{i=1}^{\ell} \frac{(-1)^{j_i}(k_i-1)}{(k_i-j_i-2)!}\right) \mathcal{E}^{\rm sv}\Big[\begin{smallmatrix} j_1 & j_2 & \ldots & j_\ell \\ k_1 & k_2 & \ldots & k_\ell\end{smallmatrix};\tau\Big] \\ \times R_{\vec\eta}\big(\mathrm{ad}_{\epsilon_0}^{k_\ell-j_\ell-2}(\epsilon_{k_\ell})\ldots \mathrm{ad}_{\epsilon_0}^{k_2-j_2-2}(\epsilon_{k_2})\,\mathrm{ad}_{\epsilon_0}^{k_1-j_1-2}(\epsilon_{k_1})\big)\,\widehat{Y}^{i\infty}_{\vec\eta}\,,\end{aligned} \tag{8.31}$$

if the $\tau$-dependent constituents solve the initial-value problem

$$\begin{aligned}2\pi i\partial_\tau \mathcal{E}^{\rm sv}\Big[\begin{smallmatrix} j_1 & j_2 & \ldots & j_\ell \\ k_1 & k_2 & \ldots & k_\ell\end{smallmatrix};\tau\Big] &= -(2\pi i)^{2-k_\ell+j_\ell}(\tau-\bar\tau)^{j_\ell}\mathrm{G}_{k_\ell}(\tau) \\ &\quad\times \mathcal{E}^{\rm sv}\Big[\begin{smallmatrix} j_1 & j_2 & \ldots & j_{\ell-1} \\ k_1 & k_2 & \ldots & k_{\ell-1}\end{smallmatrix};\tau\Big]\,,\end{aligned} \tag{8.32a}$$

$$\lim_{\tau\to i\infty} \mathcal{E}^{\rm sv}\Big[\begin{smallmatrix} j_1 & j_2 & \ldots & j_\ell \\ k_1 & k_2 & \ldots & k_\ell\end{smallmatrix};\tau\Big] = 0\,. \tag{8.32b}$$

The vanishing at the cusp here is understood in terms of a regularized limit that will be discussed in more detail in Section 8.2.4 and is akin to the method of *tangential-base-point regularization* introduced in [31]. Its net effect can be summarized by assigning $\int_{i\infty}^{\tau} d\tau' = \tau$ which regularizes the $\tau \to i\infty$ limit of all (strictly) positive powers of $\tau$ and $\bar\tau$ to zero (in the absence of negative powers) and hence $\lim_{\tau\to i\infty}\tau_2^n = 0$ for all $n>0$.

The parameter $\ell$ in (8.31) will be referred to as *depth*, and we define at depth zero that $\mathcal{E}^{\rm sv}[\varnothing;\tau] = 1$. Since the sums over the $k_i$ start at $k_i=4$, and all the $R_{\vec\eta}(\epsilon_{k\geq 4})$ in (8.14), (8.15) and Appendix B of [V] are linear in $s_{ij}$, the depth-$\ell$ contributions to (8.31) involve at least $\ell$ powers of $\alpha'$. As we will see, any order in the $\alpha'$-expansion of the component integrals (8.19) can be obtained from a finite number of terms in (8.31) on the basis of elementary operations. Like this, the relation (8.31) reduces the $\alpha'$-expansion of the generating series $\widehat{Y}^{\tau}_{\vec{\eta}}$ to the way more tractable problem of determining the initial values at the cusp $\widehat{Y}^{i\infty}_{\vec{\eta}}$ and the objects $\mathcal{E}^{\rm sv}$:

- The initial values $\widehat{Y}^{i\infty}_{\vec{\eta}}$ are series in $\eta_i, \bar{\eta}_i, s_{ij}$ whose coefficients should be $\mathbb{Q}$-linear combinations of single-valued MZVs from genus-zero sphere integrals [257]. We shall give a closed formula at two points in Section 8.3.1. Given that $\widehat{Y}^{i\infty}_{\vec{\eta}}$ at higher points are still under investigation [257], we shall here use MGF techniques to determine the initial data at three points to certain orders, see Section 8.4. As will be detailed in Section 8.2.4, we exploit the absence of negative powers of $y$ in the expansion of $\widehat{Y}^{\tau}_{\vec{\eta}}$ around the cusp to extract a well-defined initial value $\widehat{Y}^{i\infty}_{\vec{\eta}}$.

- The objects $\mathcal{E}^{\rm sv}$ are partly determined by the differential equations (8.32) but, since the $\mathcal{E}^{\rm sv}$ are non-holomorphic, the $\partial_\tau$ derivative is not sufficient to determine them: As will be detailed in Section 8.2.5, one can add antiholomorphic functions of $\bar{\tau}$ that vanish at the cusp at every step in their iterative construction. The analysis in [IV] also provides a differential equation for the $\partial_{\bar{\tau}}$-derivative of $Y$, see (8.16). However, we shall be able to determine the $\mathcal{E}^{\rm sv}$ from the reality properties (8.24) of the component integrals, i.e. without making recourse to the differential equation with respect to $\partial_{\bar{\tau}}$.

As we shall see in the next section, it turns out to be useful for expressing the $Y^{\tau}_{\vec{\eta}}$ rather than the $\widehat{Y}^{\tau}_{\vec{\eta}}$ to redefine the $\mathcal{E}^{\rm sv}$ into specific linear combinations that satisfy differential equations that are advantageous for the analysis. The notation $\mathcal{E}^{\rm sv}$ is chosen due to the similarity to holomorphic and single-valued iterated Eisenstein integrals defined by Brown and obeying similar first-order differential equations [31–33]. Explicit expressions for the $\mathcal{E}^{\rm sv}$ in terms of holomorphic iterated Eisenstein integrals and their complex conjugates will be given in Section 8.2.5, where we also address the issue of the integration constants.

### 8.2.3 *Solution for the original integrals*

Our original goal was to expand the $Y^{\tau}_{\vec{\eta}}$-integrals (8.17) in $\alpha'$. In order to translate the formal solution (8.31) for the redefined integrals $\widehat{Y}^{\tau}_{\vec{\eta}}$ to the original ones $Y^{\tau}_{\vec{\eta}}$ we have to invert the exponentials in (8.26). In the first place, this introduces the exponential in the second line of the $\alpha'$-expansion (8.31)

$$Y^{\tau}_{\vec{\eta}} = \sum_{\ell=0}^{\infty} \sum_{\substack{k_1,k_2,\ldots,k_\ell\\=4,6,8,\ldots}} \sum_{j_1=0}^{k_1-2} \sum_{j_2=0}^{k_2-2} \cdots \sum_{j_\ell=0}^{k_\ell-2} \left( \prod_{i=1}^{\ell} \frac{(-1)^{j_i}(k_i-1)}{(k_i-j_i-2)!} \right) \mathcal{E}^{\rm sv}\big[\begin{smallmatrix} j_1 & j_2 & \ldots & j_\ell \\ k_1 & k_2 & \ldots & k_\ell \end{smallmatrix};\tau\big]$$
$$\times \exp\left(-\frac{R_{\vec{\eta}}(\epsilon_0)}{4y}\right) R_{\vec{\eta}}\big(\mathrm{ad}_{\epsilon_0}^{k_\ell-j_\ell-2}(\epsilon_{k_\ell})\cdots \mathrm{ad}_{\epsilon_0}^{k_1-j_1-2}(\epsilon_{k_1})\big)\, \widehat{Y}^{i\infty}_{\vec{\eta}} \tag{8.33}$$

that we then commute through adjoint derivation operators to act on the value $\widehat{Y}^{i\infty}_{\vec{\eta}}$ at the cusp. This amounts to conjugating the $\mathrm{ad}_{\epsilon_0}^{k_i-j_i-2}(\epsilon_{k_i})$ via

$$\begin{aligned}&\exp\Big(-\frac{R_{\vec{\eta}}(\epsilon_0)}{4y}\Big)R_{\vec{\eta}}\big(\mathrm{ad}_{\epsilon_0}^{k-j-2}(\epsilon_k)\big)\exp\Big(\frac{R_{\vec{\eta}}(\epsilon_0)}{4y}\Big)\\&=\sum_{p=0}^{j}\frac{1}{p!}\Big(-\frac{1}{4y}\Big)^p R_{\vec{\eta}}\big(\mathrm{ad}_{\epsilon_0}^{k-j+p-2}(\epsilon_k)\big)\,.\end{aligned}\tag{8.34}$$

The modified powers of $\mathrm{ad}_{\epsilon_0}$ regroup the $\mathcal{E}^{\mathrm{sv}}$ into the combination

$$\begin{aligned}\beta^{\mathrm{sv}}\big[\begin{smallmatrix}j_1 & j_2 & \dots & j_\ell\\ k_1 & k_2 & \dots & k_\ell\end{smallmatrix};\tau\big]=&\sum_{p_1=0}^{k_1-j_1-2}\sum_{p_2=0}^{k_2-j_2-2}\cdots\sum_{p_\ell=0}^{k_\ell-j_\ell-2}\prod_{q=1}^{\ell}\binom{k_q-j_q-2}{p_q}\\&\times\Big(\frac{1}{4y}\Big)^{p_1+p_2+\ldots+p_\ell}\mathcal{E}^{\mathrm{sv}}\big[\begin{smallmatrix}j_1+p_1 & j_2+p_2 & \dots & j_\ell+p_\ell\\ k_1 & k_2 & \dots & k_\ell\end{smallmatrix};\tau\big]\end{aligned}\tag{8.35}$$

with $0\le j_i\le k_i-2$ and $\beta^{\mathrm{sv}}[\varnothing;\tau]=1$, i.e. the $\alpha'$-expansion (8.33) can be compactly rewritten as

$$\begin{aligned}Y^\tau_{\vec{\eta}}=&\sum_{\ell=0}^{\infty}\sum_{\substack{k_1,k_2,\ldots,k_\ell\\=4,6,8,\ldots}}\sum_{j_1=0}^{k_1-2}\sum_{j_2=0}^{k_2-2}\cdots\sum_{j_\ell=0}^{k_\ell-2}\Big(\prod_{i=1}^{\ell}\frac{(-1)^{j_i}(k_i-1)}{(k_i-j_i-2)!}\Big)\beta^{\mathrm{sv}}\big[\begin{smallmatrix}j_1 & j_2 & \dots & j_\ell\\ k_1 & k_2 & \dots & k_\ell\end{smallmatrix};\tau\big]\\&\times R_{\vec{\eta}}\big(\mathrm{ad}_{\epsilon_0}^{k_\ell-j_\ell-2}(\epsilon_{k_\ell})\cdots\mathrm{ad}_{\epsilon_0}^{k_1-j_1-2}(\epsilon_{k_1})\big)\exp\Big(-\frac{R_{\vec{\eta}}(\epsilon_0)}{4y}\Big)\widehat{Y}^{i\infty}_{\vec{\eta}}\,.\end{aligned}\tag{8.36}$$

This is the formal solution of the $\alpha'$-expansion of the generating series $Y^\tau_{\vec{\eta}}$ of worldsheet integrals. As we reviewed in Section 8.1.3, the component integrals appearing in the Laurent expansion of $Y^\tau_{\vec{\eta}}$ with respect to the $\vec{\eta}$ variables can be represented in terms of MGFs. Hence, (8.36) results in a representation of arbitrary MGFs in terms of $\beta^{\mathrm{sv}}$ and the ingredients of $\exp(-R_{\vec{\eta}}(\epsilon_0)/(4y))\widehat{Y}^{i\infty}_{\vec{\eta}}$ — conjecturally $\mathbb{Q}[y^{-1}]$-linear combinations of single-valued MZVs. We stress that by this, all the relations among MGFs will be automatically exposed in view of the linear-independence result on holomorphic iterated Eisenstein integrals of [35].

The notations $\mathcal{E}^{\mathrm{sv}}$ and $\beta^{\mathrm{sv}}$ refer to their interpretations in [17] as single-valued images of certain holomorphic iterated Eisenstein integrals $\mathcal{E}$ and $\beta$, respectively. The $\mathcal{E}$ will be defined in (8.48) and their change of basis towards the $\beta$ is described in [17].

### PROPERTIES OF $\beta^{\rm sv}$

The simplest examples of the relation (8.35) at depths one and two read

$$\begin{aligned}\beta^{\rm sv}\big[\begin{smallmatrix} j_1 \\ k_1\end{smallmatrix};\tau\big] &= \sum_{p_1=0}^{k_1-j_1-2}\binom{k_1-j_1-2}{p_1}\Big(\frac{1}{4y}\Big)^{p_1}\mathcal{E}^{\rm sv}\big[\begin{smallmatrix} j_1+p_1 \\ k_1\end{smallmatrix};\tau\big]\,,\\ \beta^{\rm sv}\big[\begin{smallmatrix} j_1 & j_2 \\ k_1 & k_2\end{smallmatrix};\tau\big] &= \sum_{p_1=0}^{k_1-j_1-2}\sum_{p_2=0}^{k_2-j_2-2}\binom{k_1-j_1-2}{p_1}\binom{k_2-j_2-2}{p_2}\\ &\quad\times\Big(\frac{1}{4y}\Big)^{p_1+p_2}\mathcal{E}^{\rm sv}\big[\begin{smallmatrix} j_1+p_1 & j_2+p_2 \\ k_1 & k_2\end{smallmatrix};\tau\big]\,.\end{aligned} \tag{8.37}$$

One can easily invert the map between $\mathcal{E}^{\rm sv}$ and $\beta^{\rm sv}$ at any depth $\ell$

$$\begin{aligned}\mathcal{E}^{\rm sv}\big[\begin{smallmatrix} j_1 & j_2 & \dots & j_\ell \\ k_1 & k_2 & \dots & k_\ell\end{smallmatrix};\tau\big] &= \sum_{p_1=0}^{k_1-j_1-2}\sum_{p_2=0}^{k_2-j_2-2}\cdots\sum_{p_\ell=0}^{k_\ell-j_\ell-2}\prod_{q=1}^{\ell}\binom{k_q-j_q-2}{p_q}\\ &\quad\times\Big(-\frac{1}{4y}\Big)^{p_1+p_2+\ldots+p_\ell}\beta^{\rm sv}\big[\begin{smallmatrix} j_1+p_1 & j_2+p_2 & \dots & j_\ell+p_\ell \\ k_1 & k_2 & \dots & k_\ell\end{smallmatrix};\tau\big]\,.\end{aligned} \tag{8.38}$$

From (8.35) and (8.32) one can check that the differential equations obeyed by the $\beta^{\rm sv}$ is

$$\begin{aligned}-4\pi\nabla_0\beta^{\rm sv}\big[\begin{smallmatrix} j_1 & j_2 & \dots & j_\ell \\ k_1 & k_2 & \dots & k_\ell\end{smallmatrix};\tau\big] &= \sum_{i=1}^{\ell}(k_i-j_i-2)\beta^{\rm sv}\big[\begin{smallmatrix} j_1 & j_2 & \dots & j_{i-1} & j_i+1 & j_{i+1} & \dots & j_\ell \\ k_1 & k_2 & \dots & k_{i-1} & k_i & k_{i+1} & \dots & k_\ell\end{smallmatrix};\tau\big]\\ &\quad-\delta_{j_\ell,k_\ell-2}(\tau-\bar\tau)^{k_\ell}\mathrm{G}_{k_\ell}(\tau)\beta^{\rm sv}\big[\begin{smallmatrix} j_1 & j_2 & \dots & j_{\ell-1} \\ k_1 & k_2 & \dots & k_{\ell-1}\end{smallmatrix};\tau\big]\,,\end{aligned} \tag{8.39}$$

where we have used the differential operator $\nabla_0$ defined in (3.55) as it has a nice action on the MGFs appearing in the component integrals. Compared to (8.32), the differential equation produces holomorphic Eisenstein series only when the last pair $(j_\ell,k_\ell)$ of $\beta^{\rm sv}$ obeys $j_\ell = k_\ell-2$. As the $\beta^{\rm sv}$ are linear combinations of the $\mathcal{E}^{\rm sv}$, the boundary condition for the $\beta^{\rm sv}$ is still that

$$\lim_{\tau\to i\infty}\beta^{\rm sv}\big[\begin{smallmatrix} j_1 & j_2 & \dots & j_\ell \\ k_1 & k_2 & \dots & k_\ell\end{smallmatrix};\tau\big] = 0\,, \tag{8.40}$$

again in the sense of a regularized limit. As was the case for the $\mathcal{E}^{\rm sv}$, the $\partial_\tau$ derivative (8.39) and the boundary condition are not sufficient to determine the $\beta^{\rm sv}$ but the reality properties of the component $\Upsilon$-integrals will resolve the integration ambiguities.

At depths one and two the differential equation (8.39) specializes as follows

$$\begin{aligned}2-4\pi\nabla_0\beta^{\rm sv}\big[\begin{smallmatrix} j_1 \\ k_1\end{smallmatrix};\tau\big] &= (k_1-j_1-2)\beta^{\rm sv}\big[\begin{smallmatrix} j_1+1 \\ k_1\end{smallmatrix};\tau\big]\\ &\quad-\delta_{j_1,k_1-2}(\tau-\bar\tau)^{k_1}\mathrm{G}_{k_1}(\tau)\,,\end{aligned} \tag{8.41a}$$

$$\begin{aligned}-4\pi\nabla_0\beta^{\rm sv}\big[\begin{smallmatrix} j_1 & j_2 \\ k_1 & k_2\end{smallmatrix};\tau\big] &= (k_1-j_1-2)\beta^{\rm sv}\big[\begin{smallmatrix} j_1+1 & j_2 \\ k_1 & k_2\end{smallmatrix};\tau\big] + (k_2-j_2-2)\beta^{\rm sv}\big[\begin{smallmatrix} j_1 & j_2+1 \\ k_1 & k_2\end{smallmatrix};\tau\big]\\ &\quad-\delta_{j_2,k_2-2}(\tau-\bar\tau)^{k_2}\mathrm{G}_{k_2}(\tau)\beta^{\rm sv}\big[\begin{smallmatrix} j_1 \\ k_1\end{smallmatrix};\tau\big]\,.\end{aligned} \tag{8.41b}$$

#### CONSTRAINTS FROM THE DERIVATION ALGEBRA

The relations of the derivation algebra such as (8.12) imply that, starting from $\sum_i k_i \geq 14$, not all $\mathcal{E}^{\rm sv}$ and $\beta^{\rm sv}$ appear individually in the expansion of the generating series $Y^\tau_{\vec\eta}$ and $\widehat{Y}^\tau_{\vec\eta}$ but only certain linear combinations can arise. We currently do not have an independent definition of all $\mathcal{E}^{\rm sv}, \beta^{\rm sv}$ and such a definition is not needed for the component integrals in this thesis that conjecturally cover all closed-string one-loop amplitudes.

The simplest instance where the derivation-algebra relation (8.12) yields all-multiplicity dropouts of certain $\beta^{\rm sv}$ is in the weight-14 part of the expansion (8.36)

$$\begin{aligned}
Y^\tau_{\vec\eta} &= \Big[ \ldots + 27\beta^{\rm sv}\big[\begin{smallmatrix} 8 & 2 \\ 10 & 4 \end{smallmatrix}\big] R_{\vec\eta}(\epsilon_4\epsilon_{10}) + 27\beta^{\rm sv}\big[\begin{smallmatrix} 2 & 8 \\ 4 & 10 \end{smallmatrix}\big] R_{\vec\eta}(\epsilon_{10}\epsilon_4) \\
&\quad + 35\beta^{\rm sv}\big[\begin{smallmatrix} 6 & 4 \\ 8 & 6 \end{smallmatrix}\big] R_{\vec\eta}(\epsilon_6\epsilon_8) + 35\beta^{\rm sv}\big[\begin{smallmatrix} 4 & 6 \\ 6 & 8 \end{smallmatrix}\big] R_{\vec\eta}(\epsilon_8\epsilon_6) + \ldots \Big] \exp\Big(-\frac{R_{\vec\eta}(\epsilon_0)}{4y}\Big) \widehat{Y}^{i\infty}_{\vec\eta} \\
&= \Big[ \ldots + \Big\{ 27\beta^{\rm sv}\big[\begin{smallmatrix} 8 & 2 \\ 10 & 4 \end{smallmatrix}\big] + 27\beta^{\rm sv}\big[\begin{smallmatrix} 2 & 8 \\ 4 & 10 \end{smallmatrix}\big] \Big\} R_{\vec\eta}(\epsilon_4\epsilon_{10}) \\
&\quad + \Big\{ 35\beta^{\rm sv}\big[\begin{smallmatrix} 6 & 4 \\ 8 & 6 \end{smallmatrix}\big] - 81\beta^{\rm sv}\big[\begin{smallmatrix} 2 & 8 \\ 4 & 10 \end{smallmatrix}\big] \Big\} R_{\vec\eta}(\epsilon_6\epsilon_8) \\
&\quad + \Big\{ 35\beta^{\rm sv}\big[\begin{smallmatrix} 4 & 6 \\ 6 & 8 \end{smallmatrix}\big] + 81\beta^{\rm sv}\big[\begin{smallmatrix} 2 & 8 \\ 4 & 10 \end{smallmatrix}\big] \Big\} R_{\vec\eta}(\epsilon_8\epsilon_6) + \ldots \Big] \exp\Big(-\frac{R_{\vec\eta}(\epsilon_0)}{4y}\Big) \widehat{Y}^{i\infty}_{\vec\eta} \,,
\end{aligned} \tag{8.42}$$

where we have solved (8.12) for $R_{\vec\eta}(\epsilon_{10}\epsilon_4)$ in the second step. This shows that only a three-dimensional subspace of the four-dimensional span $\big\langle \beta^{\rm sv}\big[\begin{smallmatrix} 8 & 2 \\ 10 & 4 \end{smallmatrix}\big], \beta^{\rm sv}\big[\begin{smallmatrix} 2 & 8 \\ 4 & 10 \end{smallmatrix}\big], \beta^{\rm sv}\big[\begin{smallmatrix} 6 & 4 \\ 8 & 6 \end{smallmatrix}\big], \beta^{\rm sv}\big[\begin{smallmatrix} 4 & 6 \\ 6 & 8 \end{smallmatrix}\big] \big\rangle$ is realized by the generating series (8.36). The manipulations in (8.42) can be repeated for contributions to $Y^\tau_{\vec\eta}$ with $\mathrm{ad}^j_{\epsilon_0}$ acting on the $R_{\vec\eta}(\epsilon_4), R_{\vec\eta}(\epsilon_{10}), R_{\vec\eta}(\epsilon_6), R_{\vec\eta}(\epsilon_8)$. This implies similar dropouts among the $\beta^{\rm sv}\big[\begin{smallmatrix} j_1 & j_2 \\ 10 & 4 \end{smallmatrix}\big], \beta^{\rm sv}\big[\begin{smallmatrix} j_1 & j_2 \\ 4 & 10 \end{smallmatrix}\big], \beta^{\rm sv}\big[\begin{smallmatrix} j_1 & j_2 \\ 8 & 6 \end{smallmatrix}\big], \beta^{\rm sv}\big[\begin{smallmatrix} j_1 & j_2 \\ 6 & 8 \end{smallmatrix}\big]$ at all values of $j_1+j_2 \leq 10$ and modifies the counting of MGF at various modular weights, see Section 8.5.2 for details.

### 8.2.4 *Improved initial data and consistent truncations*

In this section, we illustrate the usefulness of the redefinition (8.26) from $Y^\tau_{\vec\eta}$ to $\widehat{Y}^\tau_{\vec\eta}$ further by discussing how it acts on and improves the initial data at the cusp $\tau \to i\infty$ that is contained in the Laurent polynomial. In this context, we also discuss practical aspects of extracting information on the component integrals by truncating the series $Y^\tau_{\vec\eta}$ and $\widehat{Y}^\tau_{\vec\eta}$ to specific orders in $s_{ij}$, $\eta_i$ and $\bar\eta_i$.

#### BEHAVIOR OF GENERATING SERIES NEAR THE CUSP

Another virtue of the redefinition (8.26) is that $\widehat{Y}^\tau_{\vec\eta}$ is better behaved at the cusp than $Y^\tau_{\vec\eta}$. While the Laurent polynomial of $Y^\tau_{\vec\eta}$ is known to feature both positive and negative powers of $y = \pi\tau_2$ (see e.g. (5.212) for Laurent polynomials of MGFs in the $\alpha'$-expansion), we shall see that the Laurent polynomial of $\widehat{Y}^{\tau}_{\vec{\eta}}$ only has non-negative powers. In order to define an initial value supplementing the differential equations, we will take a regularized limit of the Laurent polynomial with the convention to discard strictly positive powers of $y$ as $\tau \to i\infty$.[6]

However, such a regularized limit leads to inconsistencies with products such as $1 = y^n \cdot y^{-n}$, $n > 0$ when both positive and negative powers are present. This problem is relevant to $Y^{\tau}_{\vec{\eta}}$ but not to $\widehat{Y}^{\tau}_{\vec{\eta}}$, where negative powers of $y$ are absent. While their absence is not immediately obvious from the redefinition, we have already remarked above that it can be understood from the differential equations as we shall now explain in more detail.

The differential equation (8.30) relates $\partial_\tau \widehat{Y}^{\tau}_{\vec{\eta}}$ to products of the form $\mathrm{G}_k(\tau)(\tau-\bar{\tau})^{k-2-j} R_{\vec{\eta}}\big(\mathrm{ad}^j_{\epsilon_0}(\epsilon_k)\big)\widehat{Y}^{\tau}_{\vec{\eta}}$ with $k \geq 4$ and $j \leq k-2$. The lowest explicit power of $y = \pi\tau_2$ is therefore $y^0$ and in general only non-negative powers arise since $\mathrm{G}_k$ is holomorphic in $\tau$ and the derivations $R_{\vec{\eta}}\big(\mathrm{ad}^j_{\epsilon_0}(\epsilon_k)\big)$ do not depend on $\tau$ at all. The differential equation is therefore consistent with $\widehat{Y}^{\tau}_{\vec{\eta}}$ having only non-negative powers of $y$.

We note that the differential operator on the left-hand side of (8.30) lowers the $y$-power via $\partial_\tau y^{-m} = -m y^{-m-1}$ and therefore the presence of any negative power $y^{-m}$ in $\widehat{Y}^{\tau}_{\vec{\eta}}$ requires the presence of even more negative powers by the differential equation.[7] This is even true at any fixed order in the Mandelstam variables $s_{ij}$ and the parameters $\bar{\eta}_i$ since any operator on the right-hand side of (8.30) is either linear in $s_{ij}$ or in $\bar{\eta}_i$ by looking at the expressions in Section 8.1.2.

From the argument above we could still allow for an infinite series of negative powers in $y$ appearing in $\widehat{Y}^{\tau}_{\vec{\eta}}$. To rule this out we consider the component integrals arising in the original generating series $Y^{\tau}_{\vec{\eta}}$ defined in (8.17). The integrands of the $n$-point component integrals (8.19) have negative powers of $y$ bounded by $y^{\geq -(a+b)}$ at the order of $s^a_{ij}\bar{\eta}_i^{b-n+1}$. This bound follows from the fact that Green functions and $f^{(k)}$ or $\overline{f^{(k)}}$ contribute at most $y^{-1}$ and $y^{-k}$, respectively, as can for instance be seen from their lattice-sum representations (3.65) and (3.91).[8] Moreover, this bound is uniformly valid at all orders in $\eta_j$ since the latter are introduced in the combinations $(\tau-\bar{\tau})\eta_i$ by the Kronecker–Eisenstein integrands in (8.17). Finally, the bound of $y^{\geq -(a+b)}$ at the order of $s_{ij}^a \bar{\eta}_i^{b-n+1}$ can be transferred from $Y^\tau_{\vec{\eta}}$ to $\widehat{Y}^\tau_{\vec{\eta}}$ since they are related by the exponential of $R_{\vec{\eta}}(\epsilon_0)/y$ and the derivation in the numerator is linear in $(s_{ij}, \bar{\eta}_i)$. Therefore, we conclude that $\widehat{Y}^\tau_{\vec{\eta}}$ does not contain any negative powers of $y$ at any order in its $\alpha'$-expansion.

---

6 One can think of this regularized limit as realizing the $\tau \to i\infty$ limit of integrals $\int_{i\infty}^{\tau}$ that remove strictly positive powers of $\tau_2$ through their tangential-base-point regularization [31].

7 Since the derivative of a constant vanishes, this argument does not connect positive to negative powers.

8 In terms of the lattice sums, factors of $p = m\tau + n$ or $\bar{p} = m\bar{\tau} + n$ both count as a factor of $y$ when approaching the cusp as $\mathrm{Re}\,\tau$ does not matter there. Inspecting the powers of $p$ and $\bar{p}$ in the lattice-sum representations (3.65) of the Green function and (3.91) of $f^{(k)}$ leads to the claim. An alternative way of seeing this for $f^{(k)}$ is to note from (3.82) that the cuspidal behavior receives contributions from the exponential prefactor $\exp(2\pi i\eta\frac{\mathrm{Im}\,z}{\tau_2})$, and the order of $\eta^{k-1}$ is thus accompanied by up to $k$ inverse powers of $y$. For the Green function one may also inspect its explicit Laurent polynomial given for example in (2.15) of [127].

On these grounds, we define the initial value by the regularized limit,

$$\widehat{Y}^{i\infty}_{\vec{\eta}} = \widehat{Y}^\tau_{\vec{\eta}}\Big|_{q^0\bar{q}^0}\Big|_{y^0}\,, \tag{8.43}$$

which does not suffer from inconsistencies caused by products involving negative powers of $y$.

## EXPANSION AND TRUNCATION OF INITIAL DATA

The absence of negative powers of $y$ in the Laurent polynomials of $\widehat{Y}^\tau_{\vec{\eta}}$ can also be verified explicitly in examples at fixed order in the expansion variables. In practice, this is done by imposing cutoffs on the powers of $s_{ij}, \eta_j, \bar{\eta}_j$ in the expansion of $Y^\tau_{\vec{\eta}}$ or $\widehat{Y}^\tau_{\vec{\eta}}$. We will make our scheme of cutoffs more transparent by defining the *order* of a series in $\eta_i$ $\bar{\eta}_i$ and $s_{ij}$ through the assignment

$$\text{order}(\eta_i) = 1\,, \qquad \text{order}(\bar{\eta}_i) = 1\,, \qquad \text{order}(s_{ij}) = 2\,. \tag{8.44}$$

More precisely, the order of $Y^\tau_{\vec{\eta}}$ and $\widehat{Y}^\tau_{\vec{\eta}}$ is counted relative to the most singular term of homogeneity degree $\eta_j^{1-n}\bar{\eta}_j^{1-n}$ to make sure that the $\alpha' \to 0$ limit of the plain Koba–Nielsen integrals $Y^\tau_{(0,\ldots,0|0,\ldots,0)}(\sigma|\rho) = \int d\mu_{n-1}\,\text{KN}_n = 1 + O(\alpha'^2)$ has order zero. The assignment (8.44) is consistent with a counting of (inverse) lattice momenta: every factor of $\eta_i$ or $\bar{\eta}_i$ corresponds to an inverse momentum according to (3.91) while $s_{ij}$ always appears together with a Green function that contains two inverse momenta.

The notion of order in (8.44) ensures that order$(R_{\vec{\eta}}(\epsilon_0)) = 0$ by inspection of its explicit form in Section 8.1.2, i.e. that the operator $R_{\vec{\eta}}(\epsilon_0)/(4y)$ in the exponential preserves the order of an expression. As we have shown above, $\widehat{Y}^\tau_{\vec{\eta}}$ does not have any negative powers of $y$ at any order in the sense of (8.44) and, at the same time, it has bounded positive powers of $y$ at each order by inspection of the component integrals. Since $Y^\tau_{\vec{\eta}}$ has a bounded negative power of $y$ at any order, this implies that the exponential $\exp(R_{\vec{\eta}}(\epsilon_0)/(4y))$ entering the redefinition (8.26) terminates to a polynomial at any fixed order.

For instance, in the case of two points, (8.43) results in the following initial data

$$\begin{aligned}
\widehat{Y}^{i\infty}_{\eta} = \frac{1}{\bar\eta}\bigg\{&\frac{1}{\eta}\Big[1+\frac{1}{6}s_{12}^3\zeta_3+\frac{43}{360}s_{12}^5\zeta_5\Big]+\eta\Big[-2s_{12}\zeta_3-\frac{5}{3}s_{12}^3\zeta_5-\frac{1}{3}s_{12}^4\zeta_3^2\Big]\\
&+\eta^3\Big[-2s_{12}\zeta_5+2s_{12}^2\zeta_3^2-\frac{7}{2}s_{12}^3\zeta_7\Big]\\
&+\eta^5\big[-2s_{12}\zeta_7+4s_{12}^2\zeta_3\zeta_5\big]+\eta^7\big[-2s_{12}\zeta_9\big]\bigg\}\\
&+(2\pi i)\bigg\{-\Big[\frac{1}{s_{12}}+\frac{s_{12}^2}{6}\zeta_3+\frac{43s_{12}^4}{360}\zeta_5\Big]+\eta^2\Big[2\zeta_3+\frac{5}{3}s_{12}^2\zeta_5+\frac{s_{12}^3}{3}\zeta_3^2\Big]\\
&+\eta^4\Big[2\zeta_5-2s_{12}\zeta_3^2+\frac{7}{2}s_{12}^2\zeta_7\Big]+\eta^6\big[2\zeta_7-4s_{12}\zeta_3\zeta_5\big]+\eta^8\big[2\zeta_9\big]\bigg\}\\
&+(2\pi i)^2\bar\eta\bigg\{\frac{1}{\eta}\Big[\frac{s_{12}^3\zeta_3}{60}\Big]-\eta\Big[\frac{s_{12}^3\zeta_5}{30}\Big]\bigg\}-(2\pi i)^3\bar\eta^2\bigg\{\Big[\frac{s_{12}^2\zeta_3}{60}\Big]+\eta^2\Big[\frac{s_{12}^2\zeta_5}{30}\Big]\bigg\}\\
&-(2\pi i)^4\bar\eta^3\bigg\{\frac{1}{\eta}\Big[\frac{s_{12}^3\zeta_3}{1512}\Big]\bigg\}+(2\pi i)^5\bar\eta^4\bigg\{\Big[\frac{s_{12}^2\zeta_3}{1512}\Big]\bigg\}+(\text{order}\geq 12)\,, \qquad (8.45)
\end{aligned}$$

where we have given all terms of the form $\eta^{a-1}\bar\eta^{b-1}s_{12}^c$ up to the order $a+b+2c \leq 10$. The above expression has been obtained from a general formula for the two-point Laurent polynomial that we shall present in (8.66) below and the expansion (8.45) is also available in machine-readable form in an ancillary file within the arXiv submission of [V]. When disregarding the $(2\pi i)^k\bar\eta^{k-1}$, the all-order expansion of $\widehat{Y}^{i\infty}_{\eta}$ features no MZVs other than $\zeta_k^{\text{sv}}$, in agreement with the results of [188, 189] on the terms $\sim \eta^{-1}\bar\eta^{-1}$.

At low orders, (8.45) can be crosschecked by analyzing the MGFs in $Y^{\tau}_{\eta}$, inserting their Laurent polynomials in (5.212) and extracting the initial value according to (8.43). In both approaches, the redefinition (8.26) has been performed, the exponential $R_\eta(\epsilon_0)/(4y)$ truncates to a polynomial and one can verify order by order that all negative $y$-powers are eliminated from the Laurent polynomials.

With the assignments in (8.44), the operators $R_{\vec\eta}\big(\operatorname{ad}^j_{\epsilon_0}(\epsilon_k)\big)$ in (8.29) have order $k$ for any value of $j=0,1,\dots,k{-}2$. This is evident from the explicit examples given in Section 8.1.2 and this is important for truncations of the formal solution of the differential equation to a fixed order. The expansions of $Y^{\tau}_{\vec\eta}$ and $\widehat{Y}^{\tau}_{\vec\eta}$ to the $m$-th order can be related by the truncation of the exponential

$$\exp\Big(\pm\frac{R_{\vec\eta}(\epsilon_0)}{4y}\Big) \to \sum_{r=0}^{m}\frac{1}{r!}\Big(\pm\frac{R_{\vec\eta}(\epsilon_0)}{4y}\Big)^r \qquad (8.46)$$

when acting on the expansions.

Once the contributions of the operators $R_{\vec\eta}(\operatorname{ad}^{j_\ell}_{\epsilon_0}\epsilon_{k_\ell}\dots\operatorname{ad}^{j_1}_{\epsilon_0}\epsilon_{k_1})$ in (8.36) are computed to the order of $k_1+\ldots+k_\ell=m$, one can access the component integrals $Y^{\tau}_{A|B}(\sigma|\rho)$ up to and including homogeneity degree $\frac{1}{2}(m-|A|-|B|)$ in $s_{ij}$. Conversely, the $\beta^{\rm sv}[\begin{smallmatrix} j_1 & \dots & j_\ell \\ k_1 & \dots & k_\ell \end{smallmatrix}]$ appearing at homogeneity degree $s_{ij}^w$ in the $\alpha'$-expansion of $Y^\tau_{(A|B)}(\sigma|\rho)$ are bounded to feature $k_1+\ldots+k_\ell \leq 2w+|A|+|B|$.

The above bounds rely on the fact that, at $n$ points, the order of the series $\widehat{Y}^{i\infty}_{\vec\eta}$ is bounded by the most singular term $\eta_j^{1-n}\bar\eta_j^{1-n}$ exposed by the Kronecker–Eisenstein integrand in (8.17). At two points, for instance, the bound is saturated by the terms $\widehat{Y}^{i\infty}_\eta \to \frac{1}{\eta\bar\eta}-\frac{2\pi i}{s_{12}}$ without $\zeta_{2k+1}$ in (8.45). Their three-point analogues are given by

$$\begin{aligned}\widehat{Y}^{i\infty}_{\eta_2,\eta_3}(2,3|2,3) &= \frac{1}{\eta_{23}\eta_3\bar\eta_{23}\bar\eta_3}-\frac{2\pi i}{\eta_3\bar\eta_3 s_{12}}-\frac{2\pi i}{\eta_{23}\bar\eta_{23}s_{23}}\\ &\quad+\Big(\frac{1}{s_{12}}+\frac{1}{s_{23}}\Big)\frac{(2\pi i)^2}{s_{123}}+\ldots\,, \\ \widehat{Y}^{i\infty}_{\eta_2,\eta_3}(2,3|3,2) &= \frac{1}{\eta_{23}\eta_2\bar\eta_{23}\bar\eta_3}+\frac{2\pi i}{\eta_{23}\bar\eta_{23}s_{23}}-\frac{(2\pi i)^2}{s_{23}s_{123}}+\ldots\end{aligned} \tag{8.47}$$

and permutations in $2\leftrightarrow 3$, where $s_{123}=s_{12}+s_{13}+s_{23}$ as defined in (2.27), and all the terms in the ellipsis comprise MZVs and are higher order in the sense of (8.44).

### 8.2.5 *Real-analytic combinations of iterated Eisenstein integrals*

In this section, we relate the objects we called $\mathcal{E}^{\rm sv}$ in (8.31) to iterated integrals over holomorphic Eisenstein series. We will use Brown's holomorphic iterated Eisenstein integrals subject to tangential-base-point regularization [31],

$$\mathcal{E}[\begin{smallmatrix} j_1 & j_2 & \dots & j_\ell \\ k_1 & k_2 & \dots & k_\ell \end{smallmatrix};\tau] = -(2\pi i)^{1+j_\ell-k_\ell}\int_{i\infty}^{\tau}\mathrm{d}\tau'\,(\tau')^{j_\ell}\mathrm{G}_{k_\ell}(\tau')\mathcal{E}[\begin{smallmatrix} j_1 & j_2 & \dots & j_{\ell-1} \\ k_1 & k_2 & \dots & k_{\ell-1} \end{smallmatrix};\tau']\,, \tag{8.48}$$

which can be expressed straightforwardly in terms of the iterated Eisenstein integrals $\gamma_0(\ldots)$ or $\mathcal{E}_0(\ldots)$ seen in the $\alpha'$-expansion of open-string integrals [29, 34, 37, 38, 248], cf. Appendix G of [V]. The holomorphic iterated integrals (8.48) obey the following differential equations and initial conditions

$$\begin{aligned} 2\pi i\partial_\tau\mathcal{E}[\begin{smallmatrix} j_1 & j_2 & \dots & j_\ell \\ k_1 & k_2 & \dots & k_\ell \end{smallmatrix};\tau] &= -(2\pi i)^{2-k_\ell+j_\ell}\tau^{j_\ell}\mathrm{G}_{k_\ell}(\tau)\\ &\quad\times\mathcal{E}[\begin{smallmatrix} j_1 & j_2 & \dots & j_{\ell-1} \\ k_1 & k_2 & \dots & k_{\ell-1} \end{smallmatrix};\tau]\,,\end{aligned} \tag{8.49a}$$

$$\lim_{\tau\to i\infty}\mathcal{E}[\begin{smallmatrix} j_1 & j_2 & \dots & j_\ell \\ k_1 & k_2 & \dots & k_\ell \end{smallmatrix};\tau] = 0\,. \tag{8.49b}$$

These equations are similar to those of $\mathcal{E}^{\rm sv}$ in (8.32) but feature $\tau^{j_\ell}\mathrm{G}_{k_\ell}(\tau)$ in the place of $(\tau-\bar\tau)^{j_\ell}\mathrm{G}_{k_\ell}(\tau)$. The holomorphic iterated Eisenstein integrals (8.48) obey the standard shuffle identities

$$\begin{aligned}&\mathcal{E}[A_1,A_2,\ldots,A_\ell;\tau]\,\mathcal{E}[B_1,B_2,\ldots,B_m;\tau]\\ &=\mathcal{E}[(A_1,A_2,\ldots,A_\ell)\sqcup\!\sqcup(B_1,B_2,\ldots,B_m);\tau]\end{aligned} \tag{8.50}$$

with respect to the combined letters $A_i = {j_i \atop k_i}$, e.g. $\mathcal{E}[{j_1 \atop k_1};\tau]\mathcal{E}[{j_2 \atop k_2};\tau] = \mathcal{E}[{j_1 \atop k_1}{j_2 \atop k_2};\tau] + \mathcal{E}[{j_2 \atop k_2}{j_1 \atop k_1};\tau]$ and where $\sqcup\!\sqcup$ denotes the standard shuffle product defined in (2.49). There are no linear relations among the $\mathcal{E}$ with different entries [35].

It is tempting to define a solution to our differential equations (8.32) by starting from (8.48) and simply replacing the holomorphic integration kernels $\tau^{j_\ell}\mathrm{G}_{k_\ell}(\tau)$ by the non-holomorphic expressions $(\tau-\bar\tau)^{j_\ell}\mathrm{G}_{k_\ell}(\tau)$:

$$\begin{aligned}&\mathcal{E}^{\mathrm{sv}}_{\mathrm{min}}\left[{j_1 \atop k_1}{j_2 \atop k_2}{\dots \atop \dots}{j_\ell \atop k_\ell};\tau\right]\\ &= -(2\pi i)^{1+j_\ell-k_\ell}\int_{i\infty}^{\tau}\mathrm{d}\tau'\,(\tau'-\bar\tau)^{j_\ell}\mathrm{G}_{k_\ell}(\tau')\mathcal{E}^{\mathrm{sv}}_{\mathrm{min}}\left[{j_1 \atop k_1}{j_2 \atop k_2}{\dots \atop \dots}{j_{\ell-1} \atop k_{\ell-1}};\tau'\right]\end{aligned} \tag{8.51}$$

Since $\bar\tau$ is not the complex conjugate of the integration variables $\tau'$, these integrals are homotopy invariant. We call (8.51) the *minimal solution* of (8.32), and it also obeys the standard shuffle relations (8.50) with $\mathcal{E}^{\mathrm{sv}}_{\mathrm{min}}$ in the place of $\mathcal{E}$. Binomial expansion of the integration kernels straightforwardly relates this minimal solution to the holomorphic iterated Eisenstein integrals (8.48)

$$\begin{aligned}\mathcal{E}^{\mathrm{sv}}_{\mathrm{min}}\left[{j_1 \atop k_1}{j_2 \atop k_2}{\dots \atop \dots}{j_\ell \atop k_\ell};\tau\right] &= \sum_{r_1=0}^{j_1}\sum_{r_2=0}^{j_2}\dots\sum_{r_\ell=0}^{j_\ell}\binom{j_1}{r_1}\binom{j_2}{r_2}\cdots\binom{j_\ell}{r_\ell}\\ &\quad\times(-2\pi i\bar\tau)^{r_1+r_2+\dots+r_\ell}\mathcal{E}\left[{j_1-r_1 \atop k_1}{j_2-r_2 \atop k_2}{\dots \atop \dots}{j_\ell-r_\ell \atop k_\ell};\tau\right].\end{aligned} \tag{8.52}$$

However, (8.31) is supposed to generate real-analytic modular forms such as $\mathrm{E}_k$ and its Cauchy–Riemann derivatives, as e.g. in (8.25). Hence, the minimal solutions (8.51) need to be augmented by antiholomorphic functions $\overline{f\left[{j_1 \atop k_1}{\dots \atop \dots}{j_\ell \atop k_\ell};\tau\right]}$ that vanish at the cusp, and we shall solve (8.32) via

$$\mathcal{E}^{\mathrm{sv}}\left[{j_1 \atop k_1}{j_2 \atop k_2}{\dots \atop \dots}{j_\ell \atop k_\ell};\tau\right] = \sum_{i=0}^{\ell}\overline{f\left[{j_1 \atop k_1}{j_2 \atop k_2}{\dots \atop \dots}{j_i \atop k_i};\tau\right]}\mathcal{E}^{\mathrm{sv}}_{\mathrm{min}}\left[{j_{i+1} \atop k_{i+1}}{\dots \atop \dots}{j_\ell \atop k_\ell};\tau\right] \tag{8.53}$$

with $\overline{f[\emptyset;\tau]} = \mathcal{E}^{\mathrm{sv}}_{\mathrm{min}}[\emptyset;\tau] = 1$. The functions $\overline{f\left[{j_1 \atop k_1}{\dots \atop \dots}{j_\ell \atop k_\ell};\tau\right]}$ will be determined systematically by extracting the $\alpha'$-expansion of the component integrals (8.19) and imposing their reality properties (8.24). In particular, these reality properties imply that the $\overline{f}$ must be expressible in terms of antiholomorphic iterated Eisenstein integrals (with $\mathbb{Q}$-linear combinations of MZVs and powers of $\bar\tau$ in its coefficients): Referring back to (8.16), we see that the antiholomorphic derivative $\partial_{\bar\tau}Y^{\tau}_{\vec\eta}$ contains only $(\tau-\bar\tau)^{-1}$ and the kernels $\overline{\mathrm{G}_k}$ of antiholomorphic iterated Eisenstein integrals, thus excluding any other objects in $\overline{f}$.

### DEPTH ONE

As will be derived in detail in Section 8.3.5, the appropriate choice of integration constants at depth $\ell = 1$ is given by the purely antiholomorphic expression

$$\overline{f\big[\begin{smallmatrix} j_1 \\ k_1 \end{smallmatrix};\tau\big]} = \sum_{r_1=0}^{j_1} (-2\pi i\bar\tau)^{r_1} \binom{j_1}{r_1} (-1)^{j_1-r_1} \overline{\mathcal{E}\big[\begin{smallmatrix} j_1-r_1 \\ k_1 \end{smallmatrix};\tau\big]} \,. \tag{8.54}$$

Hence, for $\mathcal{E}^{\rm sv}$ at depth one, we obtain,

$$\mathcal{E}^{\rm sv}\big[\begin{smallmatrix} j_1 \\ k_1 \end{smallmatrix};\tau\big] = \sum_{r_1=0}^{j_1} (-2\pi i\bar\tau)^{r_1} \binom{j_1}{r_1} \left( \mathcal{E}\big[\begin{smallmatrix} j_1-r_1 \\ k_1 \end{smallmatrix};\tau\big] + (-1)^{j_1-r_1} \overline{\mathcal{E}\big[\begin{smallmatrix} j_1-r_1 \\ k_1 \end{smallmatrix};\tau\big]} \right) , \tag{8.55}$$

where the contributions $\sim (-2\pi i\bar\tau)^{r_1}\mathcal{E}\big[\begin{smallmatrix} j_1-r_1 \\ k_1 \end{smallmatrix};\tau\big]$ match the minimal solution (8.52) while the additional terms are due to (8.54). Such expressions should be contained in Brown's generating series of single-valued iterated Eisenstein integrals [31–33], and similar objects have been discussed in [260] as building blocks for a single-valued map at depth one. The reality properties of our component integrals yield an independent construction of (8.55) that will be detailed in Section 8.3.5.

The simplest instances of (8.55) are given by

$$\begin{aligned}
\mathcal{E}^{\rm sv}\big[\begin{smallmatrix} 0 \\ 4 \end{smallmatrix};\tau\big] &= \mathcal{E}\big[\begin{smallmatrix} 0 \\ 4 \end{smallmatrix};\tau\big] + \overline{\mathcal{E}\big[\begin{smallmatrix} 0 \\ 4 \end{smallmatrix};\tau\big]} \\
\mathcal{E}^{\rm sv}\big[\begin{smallmatrix} 1 \\ 4 \end{smallmatrix};\tau\big] &= \mathcal{E}\big[\begin{smallmatrix} 1 \\ 4 \end{smallmatrix};\tau\big] - \overline{\mathcal{E}\big[\begin{smallmatrix} 1 \\ 4 \end{smallmatrix};\tau\big]} + (-2\pi i\bar\tau)\Big(\mathcal{E}\big[\begin{smallmatrix} 0 \\ 4 \end{smallmatrix};\tau\big] + \overline{\mathcal{E}\big[\begin{smallmatrix} 0 \\ 4 \end{smallmatrix};\tau\big]}\Big) \\
\mathcal{E}^{\rm sv}\big[\begin{smallmatrix} 2 \\ 4 \end{smallmatrix};\tau\big] &= \mathcal{E}\big[\begin{smallmatrix} 2 \\ 4 \end{smallmatrix};\tau\big] + \overline{\mathcal{E}\big[\begin{smallmatrix} 2 \\ 4 \end{smallmatrix};\tau\big]} + 2(-2\pi i\bar\tau)\Big(\mathcal{E}\big[\begin{smallmatrix} 1 \\ 4 \end{smallmatrix};\tau\big] - \overline{\mathcal{E}\big[\begin{smallmatrix} 1 \\ 4 \end{smallmatrix};\tau\big]}\Big) \\
&\quad + (-2\pi i\bar\tau)^2\Big(\mathcal{E}\big[\begin{smallmatrix} 0 \\ 4 \end{smallmatrix};\tau\big] + \overline{\mathcal{E}\big[\begin{smallmatrix} 0 \\ 4 \end{smallmatrix};\tau\big]}\Big)
\end{aligned} \tag{8.56}$$

and it is easy to check from (8.49) that $\mathcal{E}^{\rm sv}\big[\begin{smallmatrix} j \\ 4 \end{smallmatrix}\big]$ at $j = 0, 1, 2$ satisfy (8.32). As the holomorphic iterated Eisenstein integrals (and their complex conjugates) are homotopy-invariant, these expressions represent well-defined real-analytic functions, and one can straightforwardly obtain their $(q, \bar q)$-expansion from (4.18) as detailed in Appendix G.1 of [V].

### DEPTH TWO

We next elaborate on the general form of the depth-two $\mathcal{E}^{\rm sv}$. Starting from the minimal solution (8.52), the reality properties of the component integrals dictate the following integration constant at depth $\ell = 2$

$$\begin{aligned}
\overline{f\big[\begin{smallmatrix} j_1 & j_2 \\ k_1 & k_2 \end{smallmatrix};\tau\big]} &= \sum_{r_1=0}^{j_1} \sum_{r_2=0}^{j_2} (2\pi i\bar\tau)^{r_1+r_2} (-1)^{j_1+j_2} \binom{j_1}{r_1}\binom{j_2}{r_2} \overline{\mathcal{E}\big[\begin{smallmatrix} j_2-r_2 & j_1-r_1 \\ k_2 & k_1 \end{smallmatrix};\tau\big]} \\
&\quad + \overline{\alpha\big[\begin{smallmatrix} j_1 & j_2 \\ k_1 & k_2 \end{smallmatrix};\tau\big]} \,,
\end{aligned} \tag{8.57}$$

where $\overline{\mathcal{E}\big[\begin{smallmatrix} j_2-r_2 & j_1-r_1 \\ k_2 & k_1 \end{smallmatrix};\tau\big]}$ and $\overline{\alpha\big[\begin{smallmatrix} j_1 & j_2 \\ k_1 & k_2 \end{smallmatrix};\tau\big]}$ are purely antiholomorphic and individually vanish at the cusp in the regularized limit $\tau \to i\infty$. Together with the depth-one expression (8.54), the decomposition (8.53) into $\mathcal{E}^{\rm sv}_{\rm min}$ then implies

$$\begin{aligned}\mathcal{E}^{\rm sv}\big[\begin{smallmatrix} j_1 & j_2 \\ k_1 & k_2 \end{smallmatrix};\tau\big] = \sum_{r_1=0}^{j_1}\sum_{r_2=0}^{j_2}(-2\pi i\bar\tau)^{r_1+r_2}\binom{j_1}{r_1}\binom{j_2}{r_2}\Big\{&\mathcal{E}\big[\begin{smallmatrix} j_1-r_1 & j_2-r_2 \\ k_1 & k_2 \end{smallmatrix};\tau\big] \\ &+(-1)^{j_1-r_1}\overline{\mathcal{E}\big[\begin{smallmatrix} j_1-r_1 \\ k_1 \end{smallmatrix};\tau\big]}\,\mathcal{E}\big[\begin{smallmatrix} j_2-r_2 \\ k_2 \end{smallmatrix};\tau\big] \\ &+(-1)^{j_1+j_2-r_1-r_2}\overline{\mathcal{E}\big[\begin{smallmatrix} j_2-r_2 & j_1-r_1 \\ k_2 & k_1 \end{smallmatrix};\tau\big]}\Big\} \\ &+\overline{\alpha\big[\begin{smallmatrix} j_1 & j_2 \\ k_1 & k_2 \end{smallmatrix};\tau\big]}\,,\end{aligned} \tag{8.58}$$

where the first term is the minimal solution (8.52). We expect similar expressions to follow from Brown's generating series of single-valued iterated Eisenstein integrals [31–33]. Moreover, the first three lines of (8.58) with lower-depth corrected versions of $\mathcal{E}$, $\overline{\mathcal{E}}$ and the need for further antiholomorphic corrections have featured in discussions about finding an explicit form of a single-valued map at depth two [260]. As we shall see in Sections 8.3.5 and 8.4.4, the reality properties of our component integrals yield an independent construction of (8.58).

We have separated the two terms in (8.57) for the following reasons:

- The $\overline{\mathcal{E}\big[\begin{smallmatrix} j_2-r_2 & j_1-r_1 \\ k_2 & k_1 \end{smallmatrix};\tau\big]}$ exhaust the antiholomorphic iterated Eisenstein integrals at depth two within $\mathcal{E}^{\rm sv}\big[\begin{smallmatrix} j_1 & j_2 \\ k_1 & k_2 \end{smallmatrix};\tau\big]$ which are necessary to satisfy the required reality properties. The $\overline{\alpha\big[\begin{smallmatrix} j_1 & j_2 \\ k_1 & k_2 \end{smallmatrix};\tau\big]}$ in turn conjecturally comprise $\zeta_{2k+1}$ and antiholomorphic iterated Eisenstein integrals of depth one. They are determined on a case-by-case basis for $(k_1,k_2) = (4,4),(6,4),(4,6)$ in this chapter, see (8.95) and (8.107), and we leave a general discussion for the future.

  We also note that, since the derivation-algebra relations such as (4.24) imply that at higher weight only certain linear combinations of the $\mathcal{E}^{\rm sv}$ arise in the solution of $Y^\tau_{\vec\eta}$, not all integration constants can be determined individually from the component integrals. For instance, (8.12) implies that certain linear combinations of $\overline{\alpha\big[\begin{smallmatrix} j_1 & j_2 \\ k_1 & k_2 \end{smallmatrix};\tau\big]}$ with $(k_1,k_2) \in \{(10,4),(4,10),(8,6),(6,8)\}$ and $j_i \leq k_i-2$ do not occur in the expansion of $Y^\tau_{\vec\eta}$ and are inaccessible with the methods of this work.

- Even in absence of $\overline{\alpha\big[\begin{smallmatrix} j_1 & j_2 \\ k_1 & k_2 \end{smallmatrix};\tau\big]}$, the right-hand side of (8.58) is invariant under the modular T-transformation $\tau \to \tau+1$. As will be argued in Section 8.5.1 the $\mathcal{E}^{\rm sv}$ must be T-invariant as well, so the unknown $\overline{\alpha\big[\begin{smallmatrix} j_1 & j_2 \\ k_1 & k_2 \end{smallmatrix};\tau\big]}$ need to be individually T-invariant (on top of being antiholomorphic and vanishing at the cusp).

An exemplary expression resulting from (8.58) is

$$\begin{aligned}\mathcal{E}^{\mathrm{sv}}\big[\begin{smallmatrix}2&0\\4&4\end{smallmatrix};\tau\big] &= \mathcal{E}\big[\begin{smallmatrix}2&0\\4&4\end{smallmatrix};\tau\big] + \overline{\mathcal{E}\big[\begin{smallmatrix}2\\4\end{smallmatrix};\tau\big]}\,\mathcal{E}\big[\begin{smallmatrix}0\\4\end{smallmatrix};\tau\big] + \overline{\mathcal{E}\big[\begin{smallmatrix}0&2\\4&4\end{smallmatrix};\tau\big]}\\ &\quad + 2(-2\pi i\bar\tau)\Big\{\mathcal{E}\big[\begin{smallmatrix}1&0\\4&4\end{smallmatrix};\tau\big] - \overline{\mathcal{E}\big[\begin{smallmatrix}1\\4\end{smallmatrix};\tau\big]}\,\mathcal{E}\big[\begin{smallmatrix}0\\4\end{smallmatrix};\tau\big] - \overline{\mathcal{E}\big[\begin{smallmatrix}0&1\\4&4\end{smallmatrix};\tau\big]}\Big\}\\ &\quad + (-2\pi i\bar\tau)^2\Big\{\mathcal{E}\big[\begin{smallmatrix}0&0\\4&4\end{smallmatrix};\tau\big] + \overline{\mathcal{E}\big[\begin{smallmatrix}0\\4\end{smallmatrix};\tau\big]}\,\mathcal{E}\big[\begin{smallmatrix}0\\4\end{smallmatrix};\tau\big] + \overline{\mathcal{E}\big[\begin{smallmatrix}0&0\\4&4\end{smallmatrix};\tau\big]}\Big\}\\ &\quad + \frac{2\zeta_3}{3}\bigg(\overline{\mathcal{E}\big[\begin{smallmatrix}0\\4\end{smallmatrix};\tau\big]} - \frac{i\pi\bar\tau}{360}\bigg)\,, \end{aligned}\tag{8.59}$$

where the last line corresponds to $\overline{\alpha\big[\begin{smallmatrix}2&0\\4&4\end{smallmatrix};\tau\big]}$ that will be determined in (8.95).

## HIGHER DEPTH AND SHUFFLE

The $\mathcal{E}^{\mathrm{sv}}$ with depth $\ell \geq 3$ will introduce additional antiholomorphic $\overline{f\big[\begin{smallmatrix}j_1&j_2&\dots&j_\ell\\k_1&k_2&\dots&k_\ell\end{smallmatrix};\tau\big]}$ that vanish at the cusp. These antiholomorphic integration constants will preserve the shuffle relations

$$\begin{aligned}&\mathcal{E}^{\mathrm{sv}}[A_1,A_2,\dots,A_\ell;\tau]\,\mathcal{E}^{\mathrm{sv}}[B_1,B_2,\dots,B_m;\tau]\\ &= \mathcal{E}^{\mathrm{sv}}[(A_1,A_2,\dots,A_\ell)\sqcup\!\sqcup(B_1,B_2,\dots,B_m);\tau]\end{aligned}\tag{8.60}$$

analogous to those of the holomorphic counterparts (8.50). For $\ell = 2$ two, the last terms of (8.58) are then constrained to obey $\overline{\alpha\big[\begin{smallmatrix}j_1&j_2\\k_1&k_2\end{smallmatrix};\tau\big]} + \overline{\alpha\big[\begin{smallmatrix}j_2&j_1\\k_2&k_1\end{smallmatrix};\tau\big]} = 0$. We expect that the decomposition (8.53) of $\mathcal{E}^{\mathrm{sv}}$ is related to Brown's construction of single-valued iterated Eisenstein integrals [31–33] by composing holomorphic and antiholomorphic generating series. A discussion of depth-($\ell \geq 3$) instances and more detailed connections with the work of Brown are left to the future.

Given the expressions (8.55) and (8.58) for the simplest $\mathcal{E}^{\mathrm{sv}}$, also the $\beta^{\mathrm{sv}}$ at depth $\ell \leq 2$ can be reduced to iterated Eisenstein integrals via (8.35). More specifically, this completely determines the $\beta^{\mathrm{sv}}$ at depth one and fixes their depth-two examples up to the antiholomorphic $\overline{\alpha\big[\begin{smallmatrix}j_1&j_2\\k_1&k_2\end{smallmatrix};\tau\big]}$ in (8.58). The latter will later be exemplified to comprise antiholomorphic iterated Eisenstein integrals at depth one and powers of $\bar\tau$. Note that the relation (8.35) between $\mathcal{E}^{\mathrm{sv}}$ and $\beta^{\mathrm{sv}}$ preserve the expected shuffle property and therefore

$$\begin{aligned}&\beta^{\mathrm{sv}}[A_1,A_2,\dots,A_\ell;\tau]\,\beta^{\mathrm{sv}}[B_1,B_2,\dots,B_m;\tau]\\ &= \beta^{\mathrm{sv}}[(A_1,A_2,\dots,A_\ell)\sqcup\!\sqcup(B_1,B_2,\dots,B_m);\tau]\,.\end{aligned}\tag{8.61}$$

### EXPANSION AROUND THE CUSP

The expansion of the above $\mathcal{E}^{\rm sv}$ around the cusp takes the form (3.27). Tangential-base-point regularization of the holomorphic iterated Eisenstein integrals leads to the behavior [31]

$$\begin{aligned} \mathcal{E}\big[\begin{smallmatrix} j_1 \\ k_1 \end{smallmatrix};\tau\big] &= \frac{B_{k_1}}{k_1!}\frac{(2\pi i\tau)^{j_1+1}}{j_1+1} + O(q)\,, \\ \mathcal{E}\big[\begin{smallmatrix} j_1 & j_2 \\ k_1 & k_2 \end{smallmatrix};\tau\big] &= \frac{B_{k_1}B_{k_2}}{k_1!k_2!}\frac{(2\pi i\tau)^{j_1+j_2+2}}{(j_1+1)(j_1+j_2+2)} + O(q) \end{aligned} \tag{8.62}$$

with Bernoulli numbers $B_{k_i}$ defined in (2.35). As a consequence of (8.55) and (8.58), the Laurent monomial at the order of $q^0\bar{q}^0$ in $\mathcal{E}^{\rm sv}$ at depth $\leq 2$ can be given in closed form,

$$\begin{aligned} \mathcal{E}^{\rm sv}\big[\begin{smallmatrix} j_1 \\ k_1 \end{smallmatrix};\tau\big] &= \frac{B_{k_1}}{k_1!}\frac{(-4y)^{j_1+1}}{j_1+1} + O(q,\bar{q})\,, \\ \mathcal{E}^{\rm sv}\big[\begin{smallmatrix} j_1 & j_2 \\ k_1 & k_2 \end{smallmatrix};\tau\big] &= \frac{B_{k_1}B_{k_2}}{k_1!k_2!}\frac{(-4y)^{j_1+j_2+2}}{(j_1+1)(j_1+j_2+2)} + O(q,\bar{q})\,. \end{aligned} \tag{8.63}$$

The $\overline{\alpha\big[\begin{smallmatrix} j_1 & j_2 \\ k_1 & k_2 \end{smallmatrix};\tau\big]}$ which are currently unknown at $k_1+k_2 \geq 14$ cannot contribute to the Laurent monomial since they need to be antiholomorphic, T-invariant and vanishing at the cusp. Note that the regime (8.63) of $\mathcal{E}^{\rm sv}$ can be formally obtained from (8.62) for $\mathcal{E}$ by replacing $\tau \to \tau - \bar{\tau}$, in line with the proposal for an elliptic single-valued map in (4.32).

The Laurent monomials of the $\beta^{\rm sv}$ at depth $\leq 2$ resulting from (8.37) and (8.63) read

$$\beta^{\rm sv}\big[\begin{smallmatrix} j_1 \\ k_1 \end{smallmatrix};\tau\big] = \frac{B_{k_1} j_1!(k_1-2-j_1)!(-4y)^{j_1+1}}{k_1!\,(k_1-1)!} + O(q,\bar{q})\,, \tag{8.64a}$$

$$\begin{aligned} \beta^{\rm sv}\big[\begin{smallmatrix} j_1 & j_2 \\ k_1 & k_2 \end{smallmatrix};\tau\big] &= \frac{B_{k_1}B_{k_2}(j_1+j_2+1)!(k_2-2-j_2)!(-4y)^{j_1+j_2+2}}{(j_1+1)k_1!k_2!(k_2+j_1)!} \\ &\quad \times {}_3F_2\Big[\begin{smallmatrix} 1+j_1,\, 2+j_1+j_2,\, 2+j_1-k_1 \\ 2+j_1,\, 1+j_1+k_2 \end{smallmatrix};1\Big] + O(q,\bar{q})\,. \end{aligned} \tag{8.64b}$$

## 8.3 EXPLICIT FORMS AT TWO POINTS

In this section, we evaluate explicitly the generating function $Y^{\tau}_{\eta}$ at two points as given in (8.21) up to order 10 and use this to determine several $\beta^{\rm sv}$ and $\mathcal{E}^{\rm sv}$ that were introduced in the previous section. The starting point is an explicit determination of the Laurent polynomial to obtain the initial data $\widehat{Y}^{i\infty}_{\eta}$ for equation (8.36) where we present an all-order result for two points. By exploiting the reality properties of the resulting two-point component integrals, we can find the integration constants in various $\beta^{\rm sv}$ and $\mathcal{E}^{\rm sv}$.

### 8.3.1 *Laurent polynomials and initial data*

The general idea is to obtain the initial data at $n$ points by reducing the one-loop calculation in the degeneration limit $\tau \to i\infty$ of the torus to an $(n+2)$-point tree-level calculation on the sphere.[9] At $n = 2$, mild generalizations of the techniques of [188, 189] lead to a closed formula involving the usual Virasoro–Shapiro four-point amplitude on the sphere,

$$\frac{\Gamma(1-a)\Gamma(1-b)\Gamma(1-c)}{\Gamma(1+a)\Gamma(1+b)\Gamma(1+c)} = \exp\left(2\sum_{k=1}^{\infty}\frac{\zeta_{2k+1}}{2k+1}\left[a^{2k+1}+b^{2k+1}+c^{2k+1}\right]\right), \quad (8.65)$$

where $a+b+c=0$. Its specific combinations that generate the two-point Laurent polynomial of (8.21) can be written in the following form [257], using the shorthand $\xi = i\pi\bar\eta/(2y)$,

$$\begin{aligned}
Y^{\tau}_{\eta}\big|_{q^0\bar q^0} &= i\pi\exp\left(\frac{s_{12}y}{3}\right)\Bigg\{\Big[\cot(2i\eta y)-i\Big]\Big[\cot(\pi\bar\eta)+i\Big] \\
&\quad\times\exp\left(\frac{s_{12}}{8y}\partial_\eta^2\right)\frac{1}{s_{12}+2\eta+2\xi}\left[\frac{\Gamma(1+\frac{s_{12}}{2}+\eta+\xi)\Gamma(1-s_{12})\Gamma(1+\frac{s_{12}}{2}-\eta-\xi)}{\Gamma(1-\frac{s_{12}}{2}+\eta+\xi)\Gamma(1+s_{12})\Gamma(1-\frac{s_{12}}{2}-\eta-\xi)}\right. \\
&\quad\left.-\,e^{-y(s_{12}+2\eta+2\xi)}\right] \\
&\quad+\Big[\cot(2i\eta y)+i\Big]\Big[\cot(\pi\bar\eta)-i\Big] \\
&\quad\times\exp\left(\frac{s_{12}}{8y}\partial_\eta^2\right)\frac{1}{s_{12}-2\eta-2\xi}\left[\frac{\Gamma(1+\frac{s_{12}}{2}+\eta+\xi)\Gamma(1-s_{12})\Gamma(1+\frac{s_{12}}{2}-\eta-\xi)}{\Gamma(1-\frac{s_{12}}{2}+\eta+\xi)\Gamma(1+s_{12})\Gamma(1-\frac{s_{12}}{2}-\eta-\xi)}\right. \\
&\quad\left.-\,e^{-y(s_{12}-2\eta-2\xi)}\right] \\
&\quad-\frac{2}{s_{12}}\exp\left(\frac{s_{12}}{8y}\partial_\eta^2\right)\frac{\Gamma(1+\frac{s_{12}}{2}+\eta+\xi)\Gamma(1-s_{12})\Gamma(1+\frac{s_{12}}{2}-\eta-\xi)}{\Gamma(1-\frac{s_{12}}{2}+\eta+\xi)\Gamma(1+s_{12})\Gamma(1-\frac{s_{12}}{2}-\eta-\xi)}\Bigg\}. \quad (8.66)
\end{aligned}$$

By tracking the coefficients of $\eta^{a-1}\bar\eta^{b-1}$, this results in the Laurent polynomials of the component integrals $Y^{\tau}_{(a|b)}$ defined in (8.22). Some exemplary instances are

$$\begin{aligned}
Y^{\tau}_{(0|0)}\big|_{q^0\bar q^0} &= 1 + s_{12}^2\left(\frac{y^2}{90}+\frac{\zeta_3}{2y}\right) + s_{12}^3\left(\frac{y^3}{2835}+\frac{\zeta_3}{6}+\frac{\zeta_5}{8y^2}\right) \\
&\quad+ s_{12}^4\left(\frac{y^4}{22680}+\frac{y\zeta_3}{36}+\frac{5\zeta_5}{12y}-\frac{\zeta_3^2}{8y^2}+\frac{3\zeta_7}{32y^3}\right) \\
&\quad+ s_{12}^5\left(\frac{y^5}{561330}+\frac{y^2\zeta_3}{324}+\frac{19\zeta_5}{144}+\frac{\zeta_3^2}{12y}+\frac{7\zeta_7}{32y^2}\right. \\
&\quad\left.-\frac{3\zeta_3\zeta_5}{16y^3}+\frac{15\zeta_9}{128y^4}\right) + O(s_{12}^6)
\end{aligned} \quad (8.67a)$$

9 For open-string integrals over the $A$-cycle, the $\tau \to i\infty$ limit at $n$ points has been reduced to explicitly known combinations of $(n+2)$-point disk integrals in [37, 38].

$$
\begin{aligned}
Y^\tau_{(2|0)}\big|_{q^0\bar q^0} &= s_{12}\Big(\frac{4y^3}{45}-2\zeta_3\Big)+s_{12}^2\Big(\frac{4y^4}{945}-\frac{\zeta_5}{y}\Big)\\
&\quad+s_{12}^3\Big(\frac{2y^5}{2835}+\frac{y^2\zeta_3}{9}-\frac{5\zeta_5}{3}+\frac{\zeta_3^2}{y}-\frac{9\zeta_7}{8y^2}\Big) \qquad (8.67\text{b})\\
&\quad+s_{12}^4\Big(\frac{2y^6}{56133}+\frac{2y^3\zeta_3}{81}-\frac{\zeta_3^2}{3}-\frac{7\zeta_7}{4y}+\frac{9\zeta_3\zeta_5}{4y^2}-\frac{15\zeta_9}{8y^3}\Big)+O(s_{12}^5)\\
Y^\tau_{(4|2)}\big|_{q^0\bar q^0} &= -\frac{8y^4}{945}+\frac{2\zeta_5}{y}+s_{12}\Big(-\frac{8y^5}{14175}+\frac{2y^2\zeta_3}{45}-\frac{2\zeta_3^2}{y}+\frac{45\zeta_7}{8y^2}\Big) \qquad (8.67\text{c})\\
&\quad+s_{12}^2\Big(-\frac{4y^6}{22275}-\frac{y\zeta_5}{30}+\frac{7\zeta_7}{2y}-\frac{45\zeta_3\zeta_5}{4y^2}+\frac{135\zeta_9}{8y^3}\Big)+O(s_{12}^3)\,,
\end{aligned}
$$

see Appendix C.1 of [V] for similar expressions for the Laurent polynomials of $Y^\tau_{(0|2)}$, $Y^\tau_{(4|0)}$ and $Y^\tau_{(3|5)}$. These expressions have been consistently expanded up to total order 10: According to the discussion around (8.44), two-point component integrals $Y^\tau_{(a|b)}$ are said to be expanded to the order $2k$ if the coefficients up to and including $s_{12}^{k-(a+b)/2}$ are worked out.

Clearly, the Laurent polynomials (8.67) of the $Y^\tau_{\vec\eta}$-integrals contain negative powers of $y=\pi\tau_2$. Passing to $\widehat{Y}^\tau_{\vec\eta}$ via the redefinition (8.26), the negative powers of $y$ disappear, and we extract the initial value already given in (8.45) from the zeroth power in $y$.

### 8.3.2 *Component integrals in terms of* $\beta^{\text{sv}}$

Having obtained the initial value (8.45), we now need to apply the series of operators in (8.36) and extract the coefficients of $\eta^{a-1}\bar\eta^{b-1}$ to identify the component integrals $Y^\tau_{(a|b)}$ defined in (8.22). The two-point representation (8.14) of the derivation algebra is not faithful and realizes fewer linear combinations of $\beta^{\text{sv}}[\,\vdots\,;\tau]$ as compared to the $R_{\vec\eta}(\epsilon_k)$ at $(n\geq 3)$ points: Since the operators $R_\eta(\epsilon_{k\geq 4})$ at two points are multiplicative ($\partial_\eta$ only occurs in $R_\eta(\epsilon_0)$), all the commutators $[R_\eta(\epsilon_{k_1}),R_\eta(\epsilon_{k_2})]$ with $k_1,k_2\geq 4$ vanish.

Given that $[R_\eta(\epsilon_4),R_\eta(\epsilon_6)]=0$, for instance, only a restricted set of $\beta^{\text{sv}}\big[\begin{smallmatrix}j_1 & j_2\\ 4 & 6\end{smallmatrix};\tau\big]$ and $\beta^{\text{sv}}\big[\begin{smallmatrix}j_1 & j_2\\ 6 & 4\end{smallmatrix};\tau\big]$ can be found in (8.36). In particular, $\beta^{\text{sv}}\big[\begin{smallmatrix}2 & 4\\ 4 & 6\end{smallmatrix};\tau\big]$ and $\beta^{\text{sv}}\big[\begin{smallmatrix}4 & 2\\ 6 & 4\end{smallmatrix};\tau\big]$ do not show up individually but always appear in the symmetric combination $\beta^{\text{sv}}\big[\begin{smallmatrix}2 & 4\\ 4 & 6\end{smallmatrix};\tau\big]+\beta^{\text{sv}}\big[\begin{smallmatrix}4 & 2\\ 6 & 4\end{smallmatrix};\tau\big]=\beta^{\text{sv}}\big[\begin{smallmatrix}4\\ 6\end{smallmatrix};\tau\big]\beta^{\text{sv}}\big[\begin{smallmatrix}2\\ 4\end{smallmatrix};\tau\big]$. In order to determine all the $\beta^{\text{sv}}\big[\begin{smallmatrix}j_1 & j_2\\ 4 & 6\end{smallmatrix};\tau\big]$ individually, we shall study three-point integrals and their reality properties in Section 8.4.

Applying the operators in (8.36), we extract for example the following expressions for the simplest component integrals in terms of the initial data following from (8.66) and the $\beta^{\text{sv}}$:

$$
\begin{aligned}
Y^{\tau}_{(0|0)} &= 1 + s_{12}^2\Big(-3\beta^{\rm sv}\big[\begin{smallmatrix}1\\4\end{smallmatrix};\tau\big] + \frac{\zeta_3}{2y}\Big) + s_{12}^3\Big(-5\beta^{\rm sv}\big[\begin{smallmatrix}2\\6\end{smallmatrix};\tau\big] + \frac{\zeta_3}{6} + \frac{\zeta_5}{8y^2}\Big)\\
&\quad + s_{12}^4\Big(-21\beta^{\rm sv}\big[\begin{smallmatrix}3\\8\end{smallmatrix};\tau\big] + 9\beta^{\rm sv}\big[\begin{smallmatrix}1&1\\4&4\end{smallmatrix};\tau\big] - 18\beta^{\rm sv}\big[\begin{smallmatrix}2&0\\4&4\end{smallmatrix};\tau\big]\\
&\qquad + 12\zeta_3\beta^{\rm sv}\big[\begin{smallmatrix}0\\4\end{smallmatrix};\tau\big] - \frac{3\zeta_3}{2y}\beta^{\rm sv}\big[\begin{smallmatrix}1\\4\end{smallmatrix};\tau\big] - \frac{\zeta_3^2}{8y^2} + \frac{5\zeta_5}{12y} + \frac{3\zeta_7}{32y^3}\Big)\\
&\quad + s_{12}^5\Big(-135\beta^{\rm sv}\big[\begin{smallmatrix}4\\10\end{smallmatrix};\tau\big] - 60\beta^{\rm sv}\big[\begin{smallmatrix}3&0\\6&4\end{smallmatrix};\tau\big] + 15\beta^{\rm sv}\big[\begin{smallmatrix}1&2\\4&6\end{smallmatrix};\tau\big] + 15\beta^{\rm sv}\big[\begin{smallmatrix}2&1\\6&4\end{smallmatrix};\tau\big]\\
&\qquad - 60\beta^{\rm sv}\big[\begin{smallmatrix}2&1\\4&6\end{smallmatrix};\tau\big] - \frac{1}{2}\zeta_3\beta^{\rm sv}\big[\begin{smallmatrix}1\\4\end{smallmatrix};\tau\big] + \frac{6\zeta_5}{y}\beta^{\rm sv}\big[\begin{smallmatrix}0\\4\end{smallmatrix};\tau\big] - \frac{3\zeta_5}{8y^2}\beta^{\rm sv}\big[\begin{smallmatrix}1\\4\end{smallmatrix};\tau\big]\\
&\qquad + 40\zeta_3\beta^{\rm sv}\big[\begin{smallmatrix}1\\6\end{smallmatrix};\tau\big] - \frac{5\zeta_3}{2y}\beta^{\rm sv}\big[\begin{smallmatrix}2\\6\end{smallmatrix};\tau\big] + \frac{43\zeta_5}{360} + \frac{\zeta_3^2}{12y} + \frac{7\zeta_7}{32y^2}\\
&\qquad - \frac{3\zeta_3\zeta_5}{16y^3} + \frac{15\zeta_9}{128y^4}\Big) + O(s_{12}^6)\,,
\end{aligned}
\tag{8.68a}
$$

$$
\begin{aligned}
Y^{\tau}_{(2|0)} &= s_{12}\big(3\beta^{\rm sv}\big[\begin{smallmatrix}2\\4\end{smallmatrix};\tau\big] - 2\zeta_3\big) + s_{12}^2\Big(10\beta^{\rm sv}\big[\begin{smallmatrix}3\\6\end{smallmatrix};\tau\big] - \frac{\zeta_5}{y}\Big)\\
&\quad + s_{12}^3\Big(63\beta^{\rm sv}\big[\begin{smallmatrix}4\\8\end{smallmatrix};\tau\big] - 9\beta^{\rm sv}\big[\begin{smallmatrix}1&2\\4&4\end{smallmatrix};\tau\big] + 27\beta^{\rm sv}\big[\begin{smallmatrix}2&1\\4&4\end{smallmatrix};\tau\big]\\
&\qquad 18\zeta_3\beta^{\rm sv}\big[\begin{smallmatrix}1\\4\end{smallmatrix};\tau\big] + \frac{3\zeta_3}{2y}\beta^{\rm sv}\big[\begin{smallmatrix}2\\4\end{smallmatrix};\tau\big] - \frac{5\zeta_5}{3} + \frac{\zeta_3^2}{y} - \frac{9\zeta_7}{8y^2}\Big)\\
&\quad + s_{12}^4\Big(540\beta^{\rm sv}\big[\begin{smallmatrix}5\\10\end{smallmatrix};\tau\big] - 30\beta^{\rm sv}\big[\begin{smallmatrix}1&3\\4&6\end{smallmatrix};\tau\big] + 165\beta^{\rm sv}\big[\begin{smallmatrix}2&2\\4&6\end{smallmatrix};\tau\big] - 15\beta^{\rm sv}\big[\begin{smallmatrix}2&2\\6&4\end{smallmatrix};\tau\big]\\
&\qquad + 90\beta^{\rm sv}\big[\begin{smallmatrix}3&1\\6&4\end{smallmatrix};\tau\big] + 60\beta^{\rm sv}\big[\begin{smallmatrix}4&0\\6&4\end{smallmatrix};\tau\big] + \frac{1}{2}\zeta_3\beta^{\rm sv}\big[\begin{smallmatrix}2\\4\end{smallmatrix};\tau\big] - 24\zeta_5\beta^{\rm sv}\big[\begin{smallmatrix}0\\4\end{smallmatrix};\tau\big]\\
&\qquad - \frac{9\zeta_5}{y}\beta^{\rm sv}\big[\begin{smallmatrix}1\\4\end{smallmatrix};\tau\big] + \frac{3\zeta_5}{8y^2}\beta^{\rm sv}\big[\begin{smallmatrix}2\\4\end{smallmatrix};\tau\big] - 110\zeta_3\beta^{\rm sv}\big[\begin{smallmatrix}2\\6\end{smallmatrix};\tau\big] + \frac{5\zeta_3}{y}\beta^{\rm sv}\big[\begin{smallmatrix}3\\6\end{smallmatrix};\tau\big]\\
&\qquad - \frac{\zeta_3^2}{3} - \frac{7\zeta_7}{4y} + \frac{9\zeta_3\zeta_5}{4y^2} - \frac{15\zeta_9}{8y^3}\Big) + O(s_{12}^5)\,,
\end{aligned}
\tag{8.68b}
$$

$$
\begin{aligned}
Y^{\tau}_{(0|2)} &= s_{12}\Big(3\beta^{\rm sv}\big[\begin{smallmatrix}0\\4\end{smallmatrix};\tau\big] - \frac{\zeta_3}{8y^2}\Big) + s_{12}^2\Big(10\beta^{\rm sv}\big[\begin{smallmatrix}1\\6\end{smallmatrix};\tau\big] - \frac{\zeta_5}{16y^3}\Big)\\
&\quad + s_{12}^3\Big(63\beta^{\rm sv}\big[\begin{smallmatrix}2\\8\end{smallmatrix};\tau\big] - 9\beta^{\rm sv}\big[\begin{smallmatrix}0&1\\4&4\end{smallmatrix};\tau\big] + 27\beta^{\rm sv}\big[\begin{smallmatrix}1&0\\4&4\end{smallmatrix};\tau\big] - \frac{9\zeta_3}{2y}\beta^{\rm sv}\big[\begin{smallmatrix}0\\4\end{smallmatrix};\tau\big]\\
&\qquad + \frac{3\zeta_3}{8y^2}\beta^{\rm sv}\big[\begin{smallmatrix}1\\4\end{smallmatrix};\tau\big] + \frac{\zeta_3}{60} - \frac{5\zeta_5}{48y^2} + \frac{\zeta_3^2}{16y^3} - \frac{9\zeta_7}{128y^4}\Big)\\
&\quad + s_{12}^4\Big(540\beta^{\rm sv}\big[\begin{smallmatrix}3\\10\end{smallmatrix};\tau\big] - 15\beta^{\rm sv}\big[\begin{smallmatrix}0&2\\4&6\end{smallmatrix};\tau\big] + 90\beta^{\rm sv}\big[\begin{smallmatrix}1&1\\4&6\end{smallmatrix};\tau\big] - 30\beta^{\rm sv}\big[\begin{smallmatrix}1&1\\6&4\end{smallmatrix};\tau\big]\\
&\qquad + 60\beta^{\rm sv}\big[\begin{smallmatrix}2&0\\4&6\end{smallmatrix};\tau\big] + 165\beta^{\rm sv}\big[\begin{smallmatrix}2&0\\6&4\end{smallmatrix};\tau\big] + \frac{\zeta_3}{2}\beta^{\rm sv}\big[\begin{smallmatrix}0\\4\end{smallmatrix};\tau\big] - 40\zeta_3\beta^{\rm sv}\big[\begin{smallmatrix}0\\6\end{smallmatrix};\tau\big]\\
&\qquad - \frac{15\zeta_3}{y}\beta^{\rm sv}\big[\begin{smallmatrix}1\\6\end{smallmatrix};\tau\big] + \frac{5\zeta_3}{8y^2}\beta^{\rm sv}\big[\begin{smallmatrix}2\\6\end{smallmatrix};\tau\big] - \frac{33\zeta_5}{8y^2}\beta^{\rm sv}\big[\begin{smallmatrix}0\\4\end{smallmatrix};\tau\big] + \frac{3\zeta_5}{16y^3}\beta^{\rm sv}\big[\begin{smallmatrix}1\\4\end{smallmatrix};\tau\big]\\
&\qquad + \frac{\zeta_5}{120y} - \frac{\zeta_3^2}{48y^2} - \frac{7\zeta_7}{64y^3} + \frac{9\zeta_3\zeta_5}{64y^4} - \frac{15\zeta_9}{128y^5}\Big) + O(s_{12}^5)\,.
\end{aligned}
\tag{8.68c}
$$

Further expansions of component integrals to order 10 can be found in Appendix C.1 of [V].

### 8.3.3 $\beta^{\rm sv}$ *versus modular graph forms*

As exemplified by (8.25), the $\alpha'$-expansion of component integrals $Y^\tau_{(a|b)}$ is expressible in terms of MGFs. By comparing the expansion of various component integrals in terms of MGFs with those in terms of the $\beta^{\rm sv}$ as derived above, we arrive at a dictionary between the two types of objects. More specifically, the two-point component integrals $Y^\tau_{(a|b)}$ are sufficient to express all $\beta^{\rm sv}$ at depth one and all depth-two $\beta^{\rm sv}$ with $(k_1,k_2)=(4,4)$ in terms of MGFs. Depth-two instances with $(k_1,k_2)=(6,4)$ or $(k_1,k_2)=(4,6)$ are not individually accessible at two points as explained at the beginning of Section 8.3.2 and will be fixed from three-point considerations in Section 8.4.

The resulting expressions one obtains in this way at depth one are[10]

$$\begin{aligned}
&& \beta^{\rm sv}\big[\begin{smallmatrix} 0\\ 6\end{smallmatrix}\big] &= -\frac{(\pi\overline{\nabla}_0)^2\mathrm{E}_3}{960y^4}+\frac{\zeta_5}{640y^4}\\
\beta^{\rm sv}\big[\begin{smallmatrix} 0\\ 4\end{smallmatrix}\big] &= \frac{\pi\overline{\nabla}_0\mathrm{E}_2}{24y^2}+\frac{\zeta_3}{24y^2} & \beta^{\rm sv}\big[\begin{smallmatrix} 1\\ 6\end{smallmatrix}\big] &= \frac{\pi\overline{\nabla}_0\mathrm{E}_3}{240y^2}+\frac{\zeta_5}{160y^3}\\
\beta^{\rm sv}\big[\begin{smallmatrix} 1\\ 4\end{smallmatrix}\big] &= -\frac{1}{6}\mathrm{E}_2+\frac{\zeta_3}{6y} & \beta^{\rm sv}\big[\begin{smallmatrix} 2\\ 6\end{smallmatrix}\big] &= -\frac{1}{30}\mathrm{E}_3+\frac{\zeta_5}{40y^2}\\
\beta^{\rm sv}\big[\begin{smallmatrix} 2\\ 4\end{smallmatrix}\big] &= \frac{2}{3}\pi\nabla_0\mathrm{E}_2+\frac{2\zeta_3}{3} & \beta^{\rm sv}\big[\begin{smallmatrix} 3\\ 6\end{smallmatrix}\big] &= \frac{1}{15}\pi\nabla_0\mathrm{E}_3+\frac{\zeta_5}{10y}\\
&& \beta^{\rm sv}\big[\begin{smallmatrix} 4\\ 6\end{smallmatrix}\big] &= -\frac{4}{15}(\pi\nabla_0)^2\mathrm{E}_3+\frac{2\zeta_5}{5}
\end{aligned} \tag{8.69a}$$

as well as

$$\beta^{\rm sv}\big[\begin{smallmatrix} 3\\ 8\end{smallmatrix}\big] = -\frac{1}{140}\mathrm{E}_4+\frac{\zeta_7}{224y^3} \qquad \beta^{\rm sv}\big[\begin{smallmatrix} 4\\ 10\end{smallmatrix}\big] = -\frac{1}{630}\mathrm{E}_5+\frac{\zeta_9}{1152y^4} \tag{8.69b}$$

and similar expressions for the remaining $\beta^{\rm sv}\big[\begin{smallmatrix} j\\ 8\end{smallmatrix}\big], \beta^{\rm sv}\big[\begin{smallmatrix} j\\ 10\end{smallmatrix}\big]$ in terms of Cauchy–Riemann derivatives of $\mathrm{E}_4, \mathrm{E}_5$ can be found in Appendix C.2 of [V].

At depth two, we find the modular graph function $\mathrm{E}_{2,2}$ defined in (4.28) and its derivatives:

$$\begin{aligned}
\beta^{\rm sv}\big[\begin{smallmatrix} 0&0\\ 4&4\end{smallmatrix}\big] &= \frac{(\pi\overline{\nabla}_0\mathrm{E}_2)^2}{1152y^4}+\frac{\zeta_3\pi\overline{\nabla}_0\mathrm{E}_2}{576y^4}+\frac{\zeta_3^2}{1152y^4} = \frac{1}{2}\Big(\beta^{\rm sv}\big[\begin{smallmatrix} 0\\ 4\end{smallmatrix}\big]\Big)^2\\
\beta^{\rm sv}\big[\begin{smallmatrix} 0&1\\ 4&4\end{smallmatrix}\big] &= -\frac{\pi\overline{\nabla}_0\mathrm{E}_{2,2}}{144y^2}-\frac{\mathrm{E}_2\pi\overline{\nabla}_0\mathrm{E}_2}{144y^2}-\frac{\zeta_3\mathrm{E}_2}{144y^2}+\frac{\zeta_3}{2160}-\frac{5\zeta_5}{1728y^2}+\frac{\zeta_3^2}{288y^3}\\
\beta^{\rm sv}\big[\begin{smallmatrix} 0&2\\ 4&4\end{smallmatrix}\big] &= \frac{\mathrm{E}_{2,2}}{18}+\frac{(\pi\overline{\nabla}_0\mathrm{E}_2)\pi\nabla_0\mathrm{E}_2}{36y^2}+\frac{\zeta_3\pi\nabla_0\mathrm{E}_2}{36y^2}-\frac{5\zeta_5}{216y}+\frac{\zeta_3^2}{72y^2}
\end{aligned}$$

---

10 We will no longer spell out the argument $\tau$ of $\beta^{\rm sv}[\dots]$ in (8.69) and later equations unless the argument is transformed. The same notation applies to $\mathcal{E}^{\rm sv}[\dots]$ (which is real-analytic like the $\beta^{\rm sv}[\dots]$) and the holomorphic quantities $\mathcal{E}[\dots]$, $\alpha[\dots]$, $f[\dots]$.

$$
\begin{aligned}
\beta^{\text{sv}}\left[\begin{smallmatrix} 1 & 0 \\ 4 & 4 \end{smallmatrix}\right] &= \frac{\pi\overline{\nabla}_0 \mathrm{E}_{2,2}}{144y^2} + \frac{\zeta_3 \pi\overline{\nabla}_0 \mathrm{E}_2}{144y^3} - \frac{\zeta_3}{2160} + \frac{5\zeta_5}{1728y^2} + \frac{\zeta_3^2}{288y^3} \\
\beta^{\text{sv}}\left[\begin{smallmatrix} 1 & 1 \\ 4 & 4 \end{smallmatrix}\right] &= \frac{\mathrm{E}_2^2}{72} - \frac{\zeta_3 \mathrm{E}_2}{36y} + \frac{\zeta_3^2}{72y^2} = \frac{1}{2}\Big(\beta^{\text{sv}}\left[\begin{smallmatrix} 1 \\ 4 \end{smallmatrix}\right]\Big)^2 \\
\beta^{\text{sv}}\left[\begin{smallmatrix} 1 & 2 \\ 4 & 4 \end{smallmatrix}\right] &= -\frac{\pi\nabla_0 \mathrm{E}_{2,2}}{9} - \frac{\mathrm{E}_2 \pi\nabla_0 \mathrm{E}_2}{9} + \frac{\zeta_3 \pi\nabla_0 \mathrm{E}_2}{9y} - \frac{5\zeta_5}{108} + \frac{\zeta_3^2}{18y} \\
\beta^{\text{sv}}\left[\begin{smallmatrix} 2 & 0 \\ 4 & 4 \end{smallmatrix}\right] &= -\frac{\mathrm{E}_{2,2}}{18} + \frac{\zeta_3 \pi\overline{\nabla}_0 \mathrm{E}_2}{36y^2} + \frac{5\zeta_5}{216y} + \frac{\zeta_3^2}{72y^2} \\
\beta^{\text{sv}}\left[\begin{smallmatrix} 2 & 1 \\ 4 & 4 \end{smallmatrix}\right] &= \frac{\pi\nabla_0 \mathrm{E}_{2,2}}{9} - \frac{\zeta_3 \mathrm{E}_2}{9} + \frac{5\zeta_5}{108} + \frac{\zeta_3^2}{18y} \\
\beta^{\text{sv}}\left[\begin{smallmatrix} 2 & 2 \\ 4 & 4 \end{smallmatrix}\right] &= \frac{2(\pi\nabla_0 \mathrm{E}_2)^2}{9} + \frac{4\zeta_3 \pi\nabla_0 \mathrm{E}_2}{9} + \frac{2\zeta_3^2}{9} = \frac{1}{2}\Big(\beta^{\text{sv}}\left[\begin{smallmatrix} 2 \\ 4 \end{smallmatrix}\right]\Big)^2 .
\end{aligned}
\tag{8.69c}
$$

Similar expressions arise when the associated $\mathcal{E}^{\text{sv}}$ are expressed in terms of MGFs via (8.38), see Appendix E.2 of [V]. From the expressions above one can verify the shuffle property (8.61) of the $\beta^{\text{sv}}$ in a straightforward manner, e.g.

$$
\beta^{\text{sv}}\left[\begin{smallmatrix} 0 & 2 \\ 4 & 4 \end{smallmatrix}\right] + \beta^{\text{sv}}\left[\begin{smallmatrix} 2 & 0 \\ 4 & 4 \end{smallmatrix}\right] = \beta^{\text{sv}}\left[\begin{smallmatrix} 0 \\ 4 \end{smallmatrix}\right] \beta^{\text{sv}}\left[\begin{smallmatrix} 2 \\ 4 \end{smallmatrix}\right] . \tag{8.70}
$$

### MODULAR GRAPH FORMS IN TERMS OF $\beta^{\text{sv}}$

These relations can also be inverted to obtain expressions for the MGFs in terms of the $\beta^{\text{sv}}$. At depth one they are

$$
\begin{aligned}
& & \frac{(\pi\overline{\nabla}_0)^2 \mathrm{E}_3}{y^4} &= -960\beta^{\text{sv}}\left[\begin{smallmatrix} 0 \\ 6 \end{smallmatrix}\right] + \frac{3\zeta_5}{2y^4} \\
\frac{\pi\overline{\nabla}_0 \mathrm{E}_2}{y^2} &= 24\beta^{\text{sv}}\left[\begin{smallmatrix} 0 \\ 4 \end{smallmatrix}\right] - \frac{\zeta_3}{y^2} & \frac{\pi\overline{\nabla}_0 \mathrm{E}_3}{y^2} &= 240\beta^{\text{sv}}\left[\begin{smallmatrix} 1 \\ 6 \end{smallmatrix}\right] - \frac{3\zeta_5}{2y^3} \\
\mathrm{E}_2 &= -6\beta^{\text{sv}}\left[\begin{smallmatrix} 1 \\ 4 \end{smallmatrix}\right] + \frac{\zeta_3}{y} & \mathrm{E}_3 &= -30\beta^{\text{sv}}\left[\begin{smallmatrix} 2 \\ 6 \end{smallmatrix}\right] + \frac{3\zeta_5}{4y^2} \\
\pi\nabla_0 \mathrm{E}_2 &= \frac{3}{2}\beta^{\text{sv}}\left[\begin{smallmatrix} 2 \\ 4 \end{smallmatrix}\right] - \zeta_3 & \pi\nabla_0 \mathrm{E}_3 &= 15\beta^{\text{sv}}\left[\begin{smallmatrix} 3 \\ 6 \end{smallmatrix}\right] - \frac{3\zeta_5}{2y} \\
& & (\pi\nabla_0)^2 \mathrm{E}_3 &= -\frac{15}{4}\beta^{\text{sv}}\left[\begin{smallmatrix} 4 \\ 6 \end{smallmatrix}\right] + \frac{3\zeta_5}{2}
\end{aligned}
\tag{8.71a}
$$

as well as

$$
\mathrm{E}_4 = -140\beta^{\text{sv}}\left[\begin{smallmatrix} 3 \\ 8 \end{smallmatrix}\right] + \frac{5\zeta_7}{8y^3} \qquad \mathrm{E}_5 = -630\beta^{\text{sv}}\left[\begin{smallmatrix} 4 \\ 10 \end{smallmatrix}\right] + \frac{35\zeta_9}{64y^4} \tag{8.71b}
$$

and similar expressions for the Cauchy–Riemann derivatives of $E_4$ and $E_5$ are given in Appendix C.2 of [V]. Inverting the depth-two relations (8.69c) leads to the shuffle-irreducible MGFs

$$\begin{aligned}
\frac{\pi\overline{\nabla}_0 E_{2,2}}{y^2} &= 144\beta^{\mathrm{sv}}\left[\begin{smallmatrix}1&0\\4&4\end{smallmatrix}\right] - \frac{24\zeta_3}{y}\beta^{\mathrm{sv}}\left[\begin{smallmatrix}0\\4\end{smallmatrix}\right] + \frac{\zeta_3}{15} - \frac{5\zeta_5}{12y^2} + \frac{\zeta_3^2}{2y^3}\\
E_{2,2} &= -18\beta^{\mathrm{sv}}\left[\begin{smallmatrix}2&0\\4&4\end{smallmatrix}\right] + 12\zeta_3\beta^{\mathrm{sv}}\left[\begin{smallmatrix}0\\4\end{smallmatrix}\right] + \frac{5\zeta_5}{12y} - \frac{\zeta_3^2}{4y^2}\\
\pi\nabla_0 E_{2,2} &= 9\beta^{\mathrm{sv}}\left[\begin{smallmatrix}2&1\\4&4\end{smallmatrix}\right] - 6\zeta_3\beta^{\mathrm{sv}}\left[\begin{smallmatrix}1\\4\end{smallmatrix}\right] - \frac{5\zeta_5}{12} + \frac{\zeta_3^2}{2y}\,.
\end{aligned} \tag{8.71c}$$

At two points, one can still derive expressions for the modular graph function $E_{2,3}$ in (4.28) and its Cauchy–Riemann derivatives:

$$\begin{aligned}
\frac{(\pi\overline{\nabla}_0)^2 E_{2,3}}{y^4} &= -3840\beta^{\mathrm{sv}}\left[\begin{smallmatrix}0&1\\4&6\end{smallmatrix}\right] - 7680\beta^{\mathrm{sv}}\left[\begin{smallmatrix}1&0\\4&6\end{smallmatrix}\right] - 11520\beta^{\mathrm{sv}}\left[\begin{smallmatrix}1&0\\6&4\end{smallmatrix}\right]\\
&\quad + \frac{1280\zeta_3}{y}\beta^{\mathrm{sv}}\left[\begin{smallmatrix}0\\6\end{smallmatrix}\right] + \frac{160\zeta_3}{y^2}\beta^{\mathrm{sv}}\left[\begin{smallmatrix}1\\6\end{smallmatrix}\right]\\
&\quad + \frac{72\zeta_5}{y^3}\beta^{\mathrm{sv}}\left[\begin{smallmatrix}0\\4\end{smallmatrix}\right] + \frac{8\zeta_3}{189} - \frac{2\zeta_5}{15y^2} + \frac{7\zeta_7}{8y^4} - \frac{3\zeta_3\zeta_5}{y^5}\\
\frac{\pi\overline{\nabla}_0 E_{2,3}}{y^2} &= 960\beta^{\mathrm{sv}}\left[\begin{smallmatrix}1&1\\4&6\end{smallmatrix}\right] + 480\beta^{\mathrm{sv}}\left[\begin{smallmatrix}2&0\\4&6\end{smallmatrix}\right] + 1440\beta^{\mathrm{sv}}\left[\begin{smallmatrix}2&0\\6&4\end{smallmatrix}\right]\\
&\quad - 320\zeta_3\beta^{\mathrm{sv}}\left[\begin{smallmatrix}0\\6\end{smallmatrix}\right] - \frac{160\zeta_3}{y}\beta^{\mathrm{sv}}\left[\begin{smallmatrix}1\\6\end{smallmatrix}\right] - \frac{36\zeta_5}{y^2}\beta^{\mathrm{sv}}\left[\begin{smallmatrix}0\\4\end{smallmatrix}\right]\\
&\quad + \frac{\zeta_5}{15y} - \frac{7\zeta_7}{8y^3} + \frac{3\zeta_3\zeta_5}{2y^4}\\
E_{2,3} &= -120\beta^{\mathrm{sv}}\left[\begin{smallmatrix}2&1\\4&6\end{smallmatrix}\right] - 120\beta^{\mathrm{sv}}\left[\begin{smallmatrix}3&0\\6&4\end{smallmatrix}\right] + \frac{12\zeta_5}{y}\beta^{\mathrm{sv}}\left[\begin{smallmatrix}0\\4\end{smallmatrix}\right] + 80\zeta_3\beta^{\mathrm{sv}}\left[\begin{smallmatrix}1\\6\end{smallmatrix}\right]\\
&\quad - \frac{\zeta_5}{36} + \frac{7\zeta_7}{16y^2} - \frac{\zeta_3\zeta_5}{2y^3}\\
\pi\nabla_0 E_{2,3} &= 90\beta^{\mathrm{sv}}\left[\begin{smallmatrix}2&2\\4&6\end{smallmatrix}\right] + 60\beta^{\mathrm{sv}}\left[\begin{smallmatrix}3&1\\6&4\end{smallmatrix}\right] + 30\beta^{\mathrm{sv}}\left[\begin{smallmatrix}4&0\\6&4\end{smallmatrix}\right]\\
&\quad - 60\zeta_3\beta^{\mathrm{sv}}\left[\begin{smallmatrix}2\\6\end{smallmatrix}\right] - 12\zeta_5\beta^{\mathrm{sv}}\left[\begin{smallmatrix}0\\4\end{smallmatrix}\right] - \frac{6\zeta_5}{y}\beta^{\mathrm{sv}}\left[\begin{smallmatrix}1\\4\end{smallmatrix}\right] - \frac{7\zeta_7}{8y} + \frac{3\zeta_3\zeta_5}{2y^2}\\
(\pi\nabla_0)^2 E_{2,3} &= -45\beta^{\mathrm{sv}}\left[\begin{smallmatrix}2&3\\4&6\end{smallmatrix}\right] - 15\beta^{\mathrm{sv}}\left[\begin{smallmatrix}3&2\\6&4\end{smallmatrix}\right] - 30\beta^{\mathrm{sv}}\left[\begin{smallmatrix}4&1\\6&4\end{smallmatrix}\right]\\
&\quad + 30\zeta_3\beta^{\mathrm{sv}}\left[\begin{smallmatrix}3\\6\end{smallmatrix}\right] + 12\zeta_5\beta^{\mathrm{sv}}\left[\begin{smallmatrix}1\\4\end{smallmatrix}\right] + \frac{3\zeta_5}{2y}\beta^{\mathrm{sv}}\left[\begin{smallmatrix}2\\4\end{smallmatrix}\right] + \frac{7\zeta_7}{8} - \frac{3\zeta_3\zeta_5}{y}\,.
\end{aligned} \tag{8.71d}$$

However, we will need three-point input to solve for the individual $\beta^{\mathrm{sv}}$ in terms of MGFs. We will furthermore fix the antiholomorphic integrations constants $\alpha\left[\begin{smallmatrix}j_1&j_2\\6&4\end{smallmatrix}\right]$ or $\alpha\left[\begin{smallmatrix}j_1&j_2\\4&6\end{smallmatrix}\right]$ in Section 8.4.

### CLOSED FORMULAE AT DEPTH ONE

As detailed in Appendix E, one can compute the ($s_{12} \to 0$)-limit of the component integrals $Y^{\tau}_{(a|b)}$ with $a+b \geq 4$ in closed form. By comparing the leading order of $Y^{\tau}_{(k|k)}$ resulting from (8.36) with the lattice-sum representations (3.33) of non-holomorphic Eisenstein series, one obtains

$$\mathrm{E}_k = \frac{(2k-1)!}{[(k-1)!]^2}\left\{-\beta^{\mathrm{sv}}\left[\begin{smallmatrix} k-1 \\ 2k \end{smallmatrix}\right] + \frac{2\zeta_{2k-1}}{(2k-1)(4y)^{k-1}}\right\}. \tag{8.72}$$

Similarly, the lattice-sum representations (5.56) of their Cauchy–Riemann derivatives arise at the $s_{12}^0$ order of $Y^{\tau}_{(a|b)}$ with $a \neq b$, and comparison with (8.36) implies $(0 \leq m \leq k-1)$

$$(\pi\nabla_0)^m \mathrm{E}_k = \left(-\frac{1}{4}\right)^m \frac{(2k-1)!}{(k-1)!(k-1-m)!}\left\{-\beta^{\mathrm{sv}}\left[\begin{smallmatrix} k-1+m \\ 2k \end{smallmatrix}\right] + \frac{2\zeta_{2k-1}}{(2k-1)(4y)^{k-1-m}}\right\}, \tag{8.73a}$$

$$\frac{(\pi\overline{\nabla}_0)^m \mathrm{E}_k}{y^{2m}} = \frac{(-4)^m(2k-1)!}{(k-1)!(k-1-m)!}\left\{-\beta^{\mathrm{sv}}\left[\begin{smallmatrix} k-1-m \\ 2k \end{smallmatrix}\right] + \frac{2\zeta_{2k-1}}{(2k-1)(4y)^{k-1+m}}\right\}. \tag{8.73b}$$

By solving these relations for the $\beta^{\mathrm{sv}}$, one arrives at

$$\beta^{\mathrm{sv}}\left[\begin{smallmatrix} k-1 \\ 2k \end{smallmatrix}\right] = -\frac{[(k-1)!]^2}{(2k-1)!}\mathrm{E}_k + \frac{2\zeta_{2k-1}}{(2k-1)(4y)^{k-1}} \tag{8.74}$$

as well as $(0 \leq m \leq k-1)$

$$\begin{aligned}\beta^{\mathrm{sv}}\left[\begin{smallmatrix} k-1+m \\ 2k \end{smallmatrix}\right] &= -\frac{(-4)^m(k-1)!\,(k-1-m)!\,(\pi\nabla_0)^m\mathrm{E}_k}{(2k-1)!} \\ &\quad + \frac{2\zeta_{2k-1}}{(2k-1)(4y)^{k-1-m}},\end{aligned} \tag{8.75a}$$

$$\begin{aligned}\beta^{\mathrm{sv}}\left[\begin{smallmatrix} k-1-m \\ 2k \end{smallmatrix}\right] &= -\frac{(k-1)!\,(k-1-m)!\,(\pi\overline{\nabla}_0)^m\mathrm{E}_k}{(-4)^m(2k-1)!\,y^{2m}} \\ &\quad + \frac{2\zeta_{2k-1}}{(2k-1)(4y)^{k-1+m}}.\end{aligned} \tag{8.75b}$$

### 8.3.4 *Simplifying modular graph forms*

By the linear-independence result on iterated Eisenstein integrals [35], the $\beta^{\mathrm{sv}}$ are suitable for obtaining relations between MGFs which are hard to see from their lattice-sum representation. In the following, we will illustrate this with the relation

$$D_3 = \mathrm{E}_3 + \zeta_3 \tag{8.76}$$

due to Zagier (cf. (5.1)), where the *banana graph functions* $D_\ell$ were defined in (3.112) and arise as the coefficients in the $\alpha'$-expansion of the component integral $Y^\tau_{(0|0)}$ via

$$Y^\tau_{(0|0)} = \sum_{n=0}^{\infty} \frac{1}{n!}(s_{12})^n D_n(\tau)\,. \tag{8.77}$$

The simplest non-trivial banana graph function is $D_2 = \mathrm{E}_2$. (One has $D_0 = 1$ and $D_1 = 0$, cf. (5.22) and (5.21).)

The identity (8.76) was first proven by explicitly performing one of the sums in $D_3$ and was one of the only two identities from which the basis decompositions in Section 5.7 were generated, cf. (5.207). To prove (8.76) independently using the $\beta^{\mathrm{sv}}$, we have to identify both MGFs in the relation as coefficients in the $\alpha'$-expansion of component integrals $Y^\tau_{(a|b)}$ which we can write in terms of $\beta^{\mathrm{sv}}$ using (8.36). Hence, (8.76) follows from comparing

$$D_3 = 6Y^\tau_{(0|0)}\Big|_{s_{12}^3} = -30\beta^{\mathrm{sv}}\big[\begin{smallmatrix} 2 \\ 6 \end{smallmatrix}\big] + \zeta_3 + \frac{3\zeta_5}{4y^2} \tag{8.78}$$

$$\mathrm{E}_3 = Y^\tau_{(3|3)}\Big|_{s_{12}^0} = -30\beta^{\mathrm{sv}}\big[\begin{smallmatrix} 2 \\ 6 \end{smallmatrix}\big] + \frac{3\zeta_5}{4y^2}\,. \tag{8.79}$$

Higher-loop generalizations of (8.76) are known from MGF techniques and are e.g. contained in the basis decompositions of Section 5.7,

$$D_4 = 24\mathrm{E}_{2,2} + 3\mathrm{E}_2^2 + \frac{18}{5}\mathrm{E}_4 \tag{8.80a}$$

$$D_5 = 60\mathrm{E}_{2,3} + 10\mathrm{E}_3\mathrm{E}_2 + \frac{180}{7}\mathrm{E}_5 + 10\zeta_3\mathrm{E}_2 + 16\zeta_5\,, \tag{8.80b}$$

see also [188, 189] for all-order results on the Laurent polynomials of banana graph functions. Relations among MGFs like (8.80) can be proven in the same way as (8.76): They become manifest once all MGFs in the relation are identified as $\alpha'$ coefficients of component integrals and expressed in terms of $\beta^{\mathrm{sv}}$ via (8.36). This will in fact expose all the relations among MGFs since the $\beta^{\mathrm{sv}}\big[\begin{smallmatrix} j_1 & j_2 & \dots & j_\ell \\ k_1 & k_2 & \dots & k_\ell \end{smallmatrix}\big]$ with different entries $j_i, k_i$ are linearly independent.

Of course, the reach of this procedure depends on the multiplicity of the $Y^\tau_{\vec\eta}$-integrals under consideration. For instance, $\mathrm{E}_{2,2}$ and $\mathrm{E}_{2,3}$ in (8.80) contain the two-loop graphs $C\big[\begin{smallmatrix} 1 & 1 & 2 \\ 1 & 1 & 2 \end{smallmatrix}\big]$ and $C\big[\begin{smallmatrix} 1 & 1 & 3 \\ 1 & 1 & 3 \end{smallmatrix}\big]$ which do not appear in any two-point component integral $Y^\tau_{(a|b)}$.[11] Instead, the $C\big[\begin{smallmatrix} 1 & 1 & k \\ 1 & 1 & k \end{smallmatrix}\big]$ first appear as the coefficient of $s_{23}^2$ in the three-point component integral $Y^\tau_{(k,0|k,0)}(2,3|3,2)$ discussed in Section 8.4.

11 The $\alpha'$-expansion of two-point component integral $Y^\tau_{(a|b)}$ only involves the lattice sums $C\big[\begin{smallmatrix} a & 0 & 1 & 1 & \dots & 1 \\ 0 & b & 1 & 1 & \dots & 1 \end{smallmatrix}\big]$, cf. (7.13).

As a reference, we express the lowest-loop banana graphs $D_n$ in terms of $\beta^{\rm sv}$, by comparing (8.77) with (8.68a), yielding

$$D_2 = -6\beta^{\rm sv}\big[\begin{smallmatrix}1\\4\end{smallmatrix}\big] + \frac{\zeta_3}{y}\,, \tag{8.81a}$$

$$D_3 = -30\beta^{\rm sv}\big[\begin{smallmatrix}2\\6\end{smallmatrix}\big] + \zeta_3 + \frac{3\zeta_5}{4y^2}\,, \tag{8.81b}$$

$$\begin{aligned} D_4 &= 216\beta^{\rm sv}\big[\begin{smallmatrix}1&1\\4&4\end{smallmatrix}\big] - 432\beta^{\rm sv}\big[\begin{smallmatrix}2&0\\4&4\end{smallmatrix}\big] - 504\beta^{\rm sv}\big[\begin{smallmatrix}3\\8\end{smallmatrix}\big] \\ &\quad + 288\zeta_3\beta^{\rm sv}\big[\begin{smallmatrix}0\\4\end{smallmatrix}\big] - \frac{36\zeta_3}{y}\beta^{\rm sv}\big[\begin{smallmatrix}1\\4\end{smallmatrix}\big] + \frac{10\zeta_5}{y} - \frac{3\zeta_3^2}{y^2} + \frac{9\zeta_7}{4y^3}\,, \end{aligned} \tag{8.81c}$$

$$\begin{aligned} D_5 &= 1800\beta^{\rm sv}\big[\begin{smallmatrix}1&2\\4&6\end{smallmatrix}\big] - 7200\beta^{\rm sv}\big[\begin{smallmatrix}2&1\\4&6\end{smallmatrix}\big] + 1800\beta^{\rm sv}\big[\begin{smallmatrix}2&1\\6&4\end{smallmatrix}\big] - 7200\beta^{\rm sv}\big[\begin{smallmatrix}3&0\\6&4\end{smallmatrix}\big] \\ &\quad - 16200\beta^{\rm sv}\big[\begin{smallmatrix}4\\10\end{smallmatrix}\big] - 60\zeta_3\beta^{\rm sv}\big[\begin{smallmatrix}1\\4\end{smallmatrix}\big] + 4800\zeta_3\beta^{\rm sv}\big[\begin{smallmatrix}1\\6\end{smallmatrix}\big] \\ &\quad - \frac{300\zeta_3}{y}\beta^{\rm sv}\big[\begin{smallmatrix}2\\6\end{smallmatrix}\big] + \frac{720\zeta_5}{y}\beta^{\rm sv}\big[\begin{smallmatrix}0\\4\end{smallmatrix}\big] - \frac{45\zeta_5}{y^2}\beta^{\rm sv}\big[\begin{smallmatrix}1\\4\end{smallmatrix}\big] \\ &\quad + \frac{43\zeta_5}{3} + \frac{10\zeta_3^2}{y} + \frac{105\zeta_7}{4y^2} - \frac{45\zeta_3\zeta_5}{2y^3} + \frac{225\zeta_9}{16y^4}\,, \end{aligned} \tag{8.81d}$$

see Appendix C.4 of [V] for similar $\beta^{\rm sv}$-representations of $D_6$ and $D_7$. As expected, these expressions satisfy the relations (8.80) if we plug in the $\beta^{\rm sv}$ representations (8.71) of the modular graph functions on the right-hand sides.

### 8.3.5 *Explicit $\beta^{\rm sv}$ from reality properties at two points*

In this section, we derive the antiholomorphic integration constants in certain instances of $\mathcal{E}^{\rm sv}$ in (8.55) and (8.58) from reality properties (8.23) of two-point component integrals. This will make the iterated-Eisenstein-integral representation of the associated $\beta^{\rm sv}$ and MGFs fully explicit.

#### DEPTH ONE

From reality of $y = \pi\tau_2$ and $Y^\tau_{(0|0)}$, the orders $s_{12}^2$ and $s_{12}^3$ of its $\alpha'$-expansion (8.68a) immediately imply that $\beta^{\rm sv}\big[\begin{smallmatrix}1\\4\end{smallmatrix}\big]$ and $\beta^{\rm sv}\big[\begin{smallmatrix}2\\6\end{smallmatrix}\big]$ are real. Similarly, from the instance $Y^\tau_{(2|0)} = 16y^2\overline{Y^\tau_{(0|2)}}$ of (8.23), the $s_{12}$ and $s_{12}^2$ orders of (8.68b) and (8.68c) imply that

$$\overline{\beta^{\rm sv}\big[\begin{smallmatrix}2\\4\end{smallmatrix}\big]} = (4y)^2\beta^{\rm sv}\big[\begin{smallmatrix}0\\4\end{smallmatrix}\big] \qquad \overline{\beta^{\rm sv}\big[\begin{smallmatrix}3\\6\end{smallmatrix}\big]} = (4y)^2\beta^{\rm sv}\big[\begin{smallmatrix}1\\6\end{smallmatrix}\big]\,. \tag{8.82}$$

By combining (8.23) with the $s_{ij}^0$-order of general $Y^\tau_{(a|b)}$ with $a+b \geq 4$ derived in appendix E, one arrives at the closed depth-one formula

$$\overline{\beta^{\rm sv}\big[\begin{smallmatrix}j\\k\end{smallmatrix}\big]} = (4y)^{2+2j-k}\beta^{\rm sv}\big[\begin{smallmatrix}k-2-j\\k\end{smallmatrix}\big]\,. \tag{8.83}$$

By (8.38), this also determines the complex conjugation properties of $\mathcal{E}^{\rm sv}$

$$\overline{\mathcal{E}^{\rm sv}\big[\begin{smallmatrix} j \\ k \end{smallmatrix}\big]} = (-1)^j \sum_{p=0}^{j} \binom{j}{p} (4y)^p \mathcal{E}^{\rm sv}\big[\begin{smallmatrix} j-p \\ k \end{smallmatrix}\big] \, . \tag{8.84}$$

This is crucial extra information beyond the initial-value problem (8.32): The latter only determines $\mathcal{E}^{\rm sv}\big[\begin{smallmatrix} j_1 & j_2 & \dots & j_\ell \\ k_1 & k_2 & \dots & k_\ell \end{smallmatrix}\big]$ up to antiholomorphic integration constants that vanish at the cusp, denoted by $\overline{f\big[\begin{smallmatrix} j_1 & j_2 & \dots & j_\ell \\ k_1 & k_2 & \dots & k_\ell \end{smallmatrix}\big]}$ in (8.53). The complex-conjugation property (8.84) in turn relates these integration constants to the holomorphic ingredients $\mathcal{E}^{\rm sv}_{\rm min}$ that are fixed by their differential equation and can be read off from its minimal solution (8.52). At $k = 4$, for instance, (8.84) reads

$$\begin{aligned} \overline{\mathcal{E}^{\rm sv}\big[\begin{smallmatrix} 0 \\ 4 \end{smallmatrix}\big]} &= \mathcal{E}^{\rm sv}\big[\begin{smallmatrix} 0 \\ 4 \end{smallmatrix}\big] \, , \\ \overline{\mathcal{E}^{\rm sv}\big[\begin{smallmatrix} 1 \\ 4 \end{smallmatrix}\big]} &= -4y\mathcal{E}^{\rm sv}\big[\begin{smallmatrix} 0 \\ 4 \end{smallmatrix}\big] - \mathcal{E}^{\rm sv}\big[\begin{smallmatrix} 1 \\ 4 \end{smallmatrix}\big] \, , \\ \overline{\mathcal{E}^{\rm sv}\big[\begin{smallmatrix} 2 \\ 4 \end{smallmatrix}\big]} &= 16y^2\mathcal{E}^{\rm sv}\big[\begin{smallmatrix} 0 \\ 4 \end{smallmatrix}\big] + 8y\mathcal{E}^{\rm sv}\big[\begin{smallmatrix} 1 \\ 4 \end{smallmatrix}\big] + \mathcal{E}^{\rm sv}\big[\begin{smallmatrix} 2 \\ 4 \end{smallmatrix}\big] \end{aligned} \tag{8.85}$$

and selects the antiholomorphic completion in $\mathcal{E}^{\rm sv}\big[\begin{smallmatrix} j \\ 4 \end{smallmatrix}\big] = \mathcal{E}^{\rm sv}_{\rm min}\big[\begin{smallmatrix} j \\ 4 \end{smallmatrix}\big] + \overline{f\big[\begin{smallmatrix} j \\ 4 \end{smallmatrix}\big]}$: By inserting the expansion (8.52) of the minimal $\mathcal{E}^{\rm sv}_{\rm min}$ in terms of holomorphic iterated integrals (8.48)

$$\begin{aligned} \mathcal{E}^{\rm sv}_{\rm min}\big[\begin{smallmatrix} 0 \\ 4 \end{smallmatrix}\big] &= \mathcal{E}\big[\begin{smallmatrix} 0 \\ 4 \end{smallmatrix}\big] \, , \\ \mathcal{E}^{\rm sv}_{\rm min}\big[\begin{smallmatrix} 1 \\ 4 \end{smallmatrix}\big] &= \mathcal{E}\big[\begin{smallmatrix} 1 \\ 4 \end{smallmatrix}\big] - 2\pi i\bar\tau\mathcal{E}\big[\begin{smallmatrix} 0 \\ 4 \end{smallmatrix}\big] \, , \\ \mathcal{E}^{\rm sv}_{\rm min}\big[\begin{smallmatrix} 2 \\ 4 \end{smallmatrix}\big] &= \mathcal{E}\big[\begin{smallmatrix} 2 \\ 4 \end{smallmatrix}\big] - 4\pi i\bar\tau\mathcal{E}\big[\begin{smallmatrix} 1 \\ 4 \end{smallmatrix}\big] + (2\pi i\bar\tau)^2\mathcal{E}\big[\begin{smallmatrix} 0 \\ 4 \end{smallmatrix}\big] \end{aligned} \tag{8.86}$$

into (8.85) and isolating the purely antiholomorphic terms, one is uniquely led to

$$\begin{aligned} \overline{f\big[\begin{smallmatrix} 0 \\ 4 \end{smallmatrix}\big]} &= \overline{\mathcal{E}\big[\begin{smallmatrix} 0 \\ 4 \end{smallmatrix}\big]} \, , \\ \overline{f\big[\begin{smallmatrix} 1 \\ 4 \end{smallmatrix}\big]} &= -\overline{\mathcal{E}\big[\begin{smallmatrix} 1 \\ 4 \end{smallmatrix}\big]} - 2\pi i\bar\tau\overline{\mathcal{E}\big[\begin{smallmatrix} 0 \\ 4 \end{smallmatrix}\big]} \, , \\ \overline{f\big[\begin{smallmatrix} 2 \\ 4 \end{smallmatrix}\big]} &= \overline{\mathcal{E}\big[\begin{smallmatrix} 2 \\ 4 \end{smallmatrix}\big]} + 4\pi i\bar\tau\overline{\mathcal{E}\big[\begin{smallmatrix} 1 \\ 4 \end{smallmatrix}\big]} + (2\pi i\bar\tau)^2\overline{\mathcal{E}\big[\begin{smallmatrix} 0 \\ 4 \end{smallmatrix}\big]} \, . \end{aligned} \tag{8.87}$$

This reasoning results in the expressions (8.56) for $\mathcal{E}^{\rm sv}\big[\begin{smallmatrix} j \\ 4 \end{smallmatrix}\big]$ and can be repeated straightforwardly at $k \geq 6$: The reality properties (8.84) completely fix the $\overline{\mathcal{E}\big[\begin{smallmatrix} j-r \\ k \end{smallmatrix}\big]}$ in (8.55) and uniquely determine $\mathcal{E}^{\rm sv}\big[\begin{smallmatrix} j \\ k \end{smallmatrix}\big]$ in terms of iterated Eisenstein integrals and their complex conjugates.

By combining the expression (8.55) for $\mathcal{E}^{\rm sv}\big[\begin{smallmatrix} j \\ k \end{smallmatrix}\big]$ with the dictionaries (8.37) and (8.71) to $\beta^{\rm sv}\big[\begin{smallmatrix} j \\ k \end{smallmatrix}\big]$ and MGFs, both $\mathrm{E}_k$ and their Cauchy–Riemann derivatives can then be reduced to holomorphic iterated Eisenstein integrals and their complex conjugates, e.g.

$$\begin{aligned}\pi\nabla_0 \mathrm{E}_2 &= \frac{3}{2}\mathcal{E}^{\mathrm{sv}}\big[\begin{smallmatrix}2\\4\end{smallmatrix}\big] - \zeta_3 \\ &= -12\pi^2\bar\tau^2\,\mathrm{Re}\,\mathcal{E}\big[\begin{smallmatrix}0\\4\end{smallmatrix}\big] + 12\pi\bar\tau\,\mathrm{Im}\,\mathcal{E}\big[\begin{smallmatrix}1\\4\end{smallmatrix}\big] + 3\,\mathrm{Re}\,\mathcal{E}\big[\begin{smallmatrix}2\\4\end{smallmatrix}\big] - \zeta_3\,,\end{aligned} \tag{8.88}$$

$$\begin{aligned}\mathrm{E}_2 &= -6\mathcal{E}^{\mathrm{sv}}\big[\begin{smallmatrix}1\\4\end{smallmatrix}\big] - \frac{3\mathcal{E}^{\mathrm{sv}}\big[\begin{smallmatrix}2\\4\end{smallmatrix}\big]}{2y} + \frac{\zeta_3}{y} \\ &= \frac{12\pi^2\tau\bar\tau\,\mathrm{Re}\,\mathcal{E}\big[\begin{smallmatrix}0\\4\end{smallmatrix}\big] - 6\pi(\tau+\bar\tau)\,\mathrm{Im}\,\mathcal{E}\big[\begin{smallmatrix}1\\4\end{smallmatrix}\big] - 3\,\mathrm{Re}\,\mathcal{E}\big[\begin{smallmatrix}2\\4\end{smallmatrix}\big] + \zeta_3}{y}\,,\end{aligned} \tag{8.89}$$

$$\begin{aligned}\pi\overline{\nabla}_0 \mathrm{E}_2 &= 24y^2\mathcal{E}^{\mathrm{sv}}\big[\begin{smallmatrix}0\\4\end{smallmatrix}\big] + 12y\mathcal{E}^{\mathrm{sv}}\big[\begin{smallmatrix}1\\4\end{smallmatrix}\big] + \frac{3}{2}\mathcal{E}^{\mathrm{sv}}\big[\begin{smallmatrix}2\\4\end{smallmatrix}\big] - \zeta_3 \\ &= -12\pi^2\tau^2\,\mathrm{Re}\,\mathcal{E}\big[\begin{smallmatrix}0\\4\end{smallmatrix}\big] + 12\pi\tau\,\mathrm{Im}\,\mathcal{E}\big[\begin{smallmatrix}1\\4\end{smallmatrix}\big] + 3\,\mathrm{Re}\,\mathcal{E}\big[\begin{smallmatrix}2\\4\end{smallmatrix}\big] - \zeta_3\,.\end{aligned} \tag{8.90}$$

At depth one, these iterated-Eisenstein-integral representations of $\mathrm{E}_k$ are well-known [15, 36] and serve as a cross-check for the expansion methods of this work. At higher depth, however, only a small number of MGFs has been expressed in terms of iterated Eisenstein integrals [32, 34], and we will later provide new representations for non-holomorphic imaginary cusp forms. Most importantly, the reality properties of component integrals determine the integration constants in higher-depth $\mathcal{E}^{\mathrm{sv}}$ and $\beta^{\mathrm{sv}}$ without referring to the MGFs in the $\alpha'$-expansion.

### DEPTH TWO

Based on the $\alpha'$-expansions (8.68) of two-point component integrals, the $s_{12}^4$-order of $Y^\tau_{(0|0)} = \overline{Y^\tau_{(0|0)}}$ and the $s_{12}^3$-order $Y^\tau_{(2|0)} = 16y^2\overline{Y^\tau_{(0|2)}}$ imply

$$\overline{\beta^{\mathrm{sv}}\big[\begin{smallmatrix}1&1\\4&4\end{smallmatrix}\big]} = \beta^{\mathrm{sv}}\big[\begin{smallmatrix}1&1\\4&4\end{smallmatrix}\big]\,, \qquad \overline{\beta^{\mathrm{sv}}\big[\begin{smallmatrix}0&0\\4&4\end{smallmatrix}\big]} = \frac{\beta^{\mathrm{sv}}\big[\begin{smallmatrix}2&2\\4&4\end{smallmatrix}\big]}{256y^4} \tag{8.91a}$$

$$\overline{\beta^{\mathrm{sv}}\big[\begin{smallmatrix}2&0\\4&4\end{smallmatrix}\big]} = \beta^{\mathrm{sv}}\big[\begin{smallmatrix}2&0\\4&4\end{smallmatrix}\big] - \frac{2\zeta_3}{3}\beta^{\mathrm{sv}}\big[\begin{smallmatrix}0\\4\end{smallmatrix}\big] + \frac{\zeta_3}{24y^2}\beta^{\mathrm{sv}}\big[\begin{smallmatrix}2\\4\end{smallmatrix}\big] \tag{8.91b}$$

$$\overline{\beta^{\mathrm{sv}}\big[\begin{smallmatrix}0&2\\4&4\end{smallmatrix}\big]} = \beta^{\mathrm{sv}}\big[\begin{smallmatrix}0&2\\4&4\end{smallmatrix}\big] + \frac{2\zeta_3}{3}\beta^{\mathrm{sv}}\big[\begin{smallmatrix}0\\4\end{smallmatrix}\big] - \frac{\zeta_3}{24y^2}\beta^{\mathrm{sv}}\big[\begin{smallmatrix}2\\4\end{smallmatrix}\big] \tag{8.91c}$$

$$\overline{\beta^{\mathrm{sv}}\big[\begin{smallmatrix}1&0\\4&4\end{smallmatrix}\big]} = \frac{\beta^{\mathrm{sv}}\big[\begin{smallmatrix}2&1\\4&4\end{smallmatrix}\big]}{16y^2} - \frac{\zeta_3}{24y^2}\beta^{\mathrm{sv}}\big[\begin{smallmatrix}1\\4\end{smallmatrix}\big] + \frac{\zeta_3}{96y^3}\beta^{\mathrm{sv}}\big[\begin{smallmatrix}2\\4\end{smallmatrix}\big] - \frac{\zeta_3}{2160} \tag{8.91d}$$

$$\overline{\beta^{\mathrm{sv}}\big[\begin{smallmatrix}0&1\\4&4\end{smallmatrix}\big]} = \frac{\beta^{\mathrm{sv}}\big[\begin{smallmatrix}1&2\\4&4\end{smallmatrix}\big]}{16y^2} + \frac{\zeta_3}{24y^2}\beta^{\mathrm{sv}}\big[\begin{smallmatrix}1\\4\end{smallmatrix}\big] - \frac{\zeta_3}{96y^3}\beta^{\mathrm{sv}}\big[\begin{smallmatrix}2\\4\end{smallmatrix}\big] + \frac{\zeta_3}{2160}\,. \tag{8.91e}$$

These simplest depth-two examples illustrate that $\overline{\beta^{\mathrm{sv}}\big[\begin{smallmatrix}j_1&j_2\\k_1&k_2\end{smallmatrix}\big]}$ introduce admixtures of single-valued MZVs and $\beta^{\mathrm{sv}}$ of lower-depth. There is no analogue of this feature at depth one in the expression (8.83) for $\overline{\beta^{\mathrm{sv}}\big[\begin{smallmatrix}j\\k\end{smallmatrix}\big]}$. In Section 8.4.4, three-point $\alpha'$-expansions will be used to extract similar complex-conjugation properties for all the individual $\beta^{\rm sv}\big[\begin{smallmatrix} j_1 & j_2 \\ 4 & 6\end{smallmatrix}\big]$ and $\beta^{\rm sv}\big[\begin{smallmatrix} j_1 & j_2 \\ 6 & 4\end{smallmatrix}\big]$. Our examples will line up with the conjectural closed depth-two formula

$$\overline{\beta^{\rm sv}\big[\begin{smallmatrix} j_1 & j_2 \\ k_1 & k_2\end{smallmatrix}\big]} = (4y)^{4+2j_1+2j_2-k_1-k_2}\beta^{\rm sv}\big[\begin{smallmatrix} k_2-2-j_2 & k_1-2-j_1 \\ k_2 & k_1\end{smallmatrix}\big] \text{ mod depth} < 2 \tag{8.92}$$

which translates as follows to the $\mathcal{E}^{\rm sv}$

$$\begin{aligned}\overline{\mathcal{E}^{\rm sv}\big[\begin{smallmatrix} j_1 & j_2 \\ k_1 & k_2\end{smallmatrix}\big]} &= (-1)^{j_1+j_2}\sum_{p_1=0}^{j_1}\sum_{p_2=0}^{j_2}\binom{j_1}{p_1}\binom{j_2}{p_2} \\ &\quad\times (4y)^{p_1+p_2}\mathcal{E}^{\rm sv}\big[\begin{smallmatrix} j_2-p_2 & j_1-p_1 \\ k_2 & k_1\end{smallmatrix}\big] \text{ mod depth} < 2\,.\end{aligned} \tag{8.93}$$

These complex-conjugation properties of $\mathcal{E}^{\rm sv}$ are consistent with the general depth-two expression (8.58), assuming the conjecture that the iterated Eisenstein integrals in $\overline{\alpha[\cdots]}$ have depth one and zero.

The $\zeta_3$-admixtures in (8.91) propagate to the following shuffle-inequivalent $\overline{\mathcal{E}^{\rm sv}\big[\begin{smallmatrix} j_1 & j_2 \\ 4 & 4\end{smallmatrix}\big]}$,

$$\begin{aligned}\overline{\mathcal{E}^{\rm sv}\big[\begin{smallmatrix} 1 & 0 \\ 4 & 4\end{smallmatrix}\big]} &= -4y\mathcal{E}^{\rm sv}\big[\begin{smallmatrix} 0 & 0 \\ 4 & 4\end{smallmatrix}\big] - \mathcal{E}^{\rm sv}\big[\begin{smallmatrix} 0 & 1 \\ 4 & 4\end{smallmatrix}\big] \\ \overline{\mathcal{E}^{\rm sv}\big[\begin{smallmatrix} 2 & 0 \\ 4 & 4\end{smallmatrix}\big]} &= 16y^2\mathcal{E}^{\rm sv}\big[\begin{smallmatrix} 0 & 0 \\ 4 & 4\end{smallmatrix}\big] + 8y\mathcal{E}^{\rm sv}\big[\begin{smallmatrix} 0 & 1 \\ 4 & 4\end{smallmatrix}\big] + \mathcal{E}^{\rm sv}\big[\begin{smallmatrix} 0 & 2 \\ 4 & 4\end{smallmatrix}\big] - \frac{y\zeta_3}{270} + \frac{2\zeta_3}{3}\mathcal{E}^{\rm sv}\big[\begin{smallmatrix} 0 \\ 4\end{smallmatrix}\big] \\ \overline{\mathcal{E}^{\rm sv}\big[\begin{smallmatrix} 2 & 1 \\ 4 & 4\end{smallmatrix}\big]} &= -64y^3\mathcal{E}^{\rm sv}\big[\begin{smallmatrix} 0 & 0 \\ 4 & 4\end{smallmatrix}\big] - 32y^2\mathcal{E}^{\rm sv}\big[\begin{smallmatrix} 0 & 1 \\ 4 & 4\end{smallmatrix}\big] - 4y\mathcal{E}^{\rm sv}\big[\begin{smallmatrix} 0 & 2 \\ 4 & 4\end{smallmatrix}\big] - 16y^2\mathcal{E}^{\rm sv}\big[\begin{smallmatrix} 1 & 0 \\ 4 & 4\end{smallmatrix}\big] \\ &\quad - 8y\mathcal{E}^{\rm sv}\big[\begin{smallmatrix} 1 & 1 \\ 4 & 4\end{smallmatrix}\big] - \mathcal{E}^{\rm sv}\big[\begin{smallmatrix} 1 & 2 \\ 4 & 4\end{smallmatrix}\big] + \frac{y^2\zeta_3}{135} - \frac{8y\zeta_3}{3}\mathcal{E}^{\rm sv}\big[\begin{smallmatrix} 0 \\ 4\end{smallmatrix}\big] - \frac{2\zeta_3}{3}\mathcal{E}^{\rm sv}\big[\begin{smallmatrix} 1 \\ 4\end{smallmatrix}\big]\,.\end{aligned} \tag{8.94}$$

These equations uniquely fix all the integration constants $\overline{\alpha\big[\begin{smallmatrix} j_1 & j_2 \\ 4 & 4\end{smallmatrix}\big]}$: One has to first express the $\mathcal{E}^{\rm sv}$ in terms of holomorphic iterated Eisenstein integrals $\mathcal{E}$ and their complex conjugates via (8.55) and (8.58). Then by comparing the purely holomorphic terms $\sim \tau, \mathcal{E}, \alpha\big[\begin{smallmatrix} j_1 & j_2 \\ 4 & 4\end{smallmatrix}\big]$ on the two sides of (8.94), one can read off

$$\begin{aligned}\alpha\big[\begin{smallmatrix} 1 & 0 \\ 4 & 4\end{smallmatrix}\big] &= \alpha\big[\begin{smallmatrix} 0 & 1 \\ 4 & 4\end{smallmatrix}\big] = 0 \\ \alpha\big[\begin{smallmatrix} 2 & 0 \\ 4 & 4\end{smallmatrix}\big] &= \frac{2\zeta_3}{3}\left(\mathcal{E}\big[\begin{smallmatrix} 0 \\ 4\end{smallmatrix}\big] + \frac{i\pi\tau}{360}\right) = -\alpha\big[\begin{smallmatrix} 0 & 2 \\ 4 & 4\end{smallmatrix}\big] \\ \alpha\big[\begin{smallmatrix} 2 & 1 \\ 4 & 4\end{smallmatrix}\big] &= \frac{2\zeta_3}{3}\left(2\pi i\tau\mathcal{E}\big[\begin{smallmatrix} 0 \\ 4\end{smallmatrix}\big] - \mathcal{E}\big[\begin{smallmatrix} 1 \\ 4\end{smallmatrix}\big] - \frac{\pi^2\tau^2}{360}\right) = -\alpha\big[\begin{smallmatrix} 1 & 2 \\ 4 & 4\end{smallmatrix}\big]\,.\end{aligned} \tag{8.95}$$

The expressions for $\overline{\alpha\big[\begin{smallmatrix} j_1 & j_2 \\ 4 & 4\end{smallmatrix}\big]}$ that enter the actual $\mathcal{E}^{\rm sv}\big[\begin{smallmatrix} j_1 & j_2 \\ 4 & 4\end{smallmatrix}\big]$ follow from complex conjugation, and we have used the shuffle relations (8.60) to infer the $\alpha\big[\begin{smallmatrix} j_1 & j_2 \\ 4 & 4\end{smallmatrix}\big]$ with $j_1 < j_2$. Moreover, the $\alpha\big[\begin{smallmatrix} j_1 & j_2 \\ 4 & 4\end{smallmatrix}\big]$ in (8.95) are invariant under the modular $T : \tau \longrightarrow \tau+1$ transformation, as exhibited in Appendix G of [V], in line with the discussion in Section 8.2.5. In an ancillary file to the arXiv submission of [17], all $\alpha\big[\begin{smallmatrix} j_1 & j_2 \\ k_1 & k_2\end{smallmatrix}\big]$ with $k_1 + k_2 \leq 12$ are given in a machine-readable form.

## 8.4 EXPLICIT FORMS AT THREE POINTS

An analysis similar to the one of Section 8.3 can be done at three points. Unlike formula (8.66) we do not have a closed expression for the all-order Laurent polynomial at three points to obtain the initial data directly. For this reason, we expand the component integrals into the MGFs-basis discussed in Section 5.7. From this expansion and the knowledge of the Laurent polynomials of the basis-MGFs as given in Section 5.7.2, we can construct the initial data $\widehat{Y}^{i\infty}_{\vec{\eta}}$ and solve for the remaining $\beta^{\mathrm{sv}}$ at depth two having $(k_1, k_2) = (4, 6)$ or $(k_1, k_2) = (6, 4)$. The expansion of the initial data to order 10 is available in machine-readable form in an ancillary file within the arXiv submission of [V]. As a consistency check of our procedure the instances of $\beta^{\mathrm{sv}}$ that were determined from the two-point analysis in Section 8.3 are consequences of the three-point considerations.

The detailed discussion of $\beta^{\mathrm{sv}}\big[\begin{smallmatrix} j_1 & j_2 \\ 4 & 6 \end{smallmatrix}\big]$, $\beta^{\mathrm{sv}}\big[\begin{smallmatrix} j_1 & j_2 \\ 6 & 4 \end{smallmatrix}\big]$ and the associated MGFs in this section is motivated as follows: The depth-two integrals $\beta^{\mathrm{sv}}\big[\begin{smallmatrix} j_1 & j_2 \\ 4 & 4 \end{smallmatrix}\big]$ have been described in terms of real MGFs $\mathrm{E}_2$ and $\mathrm{E}_{2,2}$, see (8.69c) and (8.71c), and their reality is a particularity of having the same Eisenstein series $\mathrm{G}_4$ in both integration kernels. Generic $\beta^{\mathrm{sv}}\big[\begin{smallmatrix} j_1 & j_2 \\ k_1 & k_2 \end{smallmatrix}\big]$ with $k_1 \neq k_2$, by contrast, introduce complex MGFs. As we saw in Section 5.7.2, the first complex basis elements appear at total modular weight $a + b = 10$ and hence, the $\beta^{\mathrm{sv}}\big[\begin{smallmatrix} j_1 & j_2 \\ 4 & 6 \end{smallmatrix}\big]$, $\beta^{\mathrm{sv}}\big[\begin{smallmatrix} j_1 & j_2 \\ 6 & 4 \end{smallmatrix}\big]$ in this section are the simplest non-trivial window into the generic properties of depth-two MGFs.

### 8.4.1 *Bases of modular graph forms up to order 10*

At two points, the expansion of any component integral $Y^{\tau}_{(a|b)}$ to order 10 is entirely expressible in terms of the modular graph functions $\mathrm{E}_{k\leq 5}, \mathrm{E}_{2,2}, \mathrm{E}_{2,3}$ as well as their Cauchy–Riemann derivatives, cf. (8.68) and Appendix C.2 of [V]. At three points, this is no longer the case: The $\alpha'$-expansion of various component integrals (8.19) introduces additional MGFs that are not expressible in terms of the real quantities $\mathrm{E}_{k\leq 5}, \mathrm{E}_{2,2}$ and $\mathrm{E}_{2,3}$. This resonates with the comments in early Section 8.3.2 that the operators $R_{\eta}(\epsilon_k)$ in the two-point differential equations obey relations that no longer hold for their three-point analogues $R_{\eta_2,\eta_3}(\epsilon_k)$ in (8.15).

The additional MGFs that start appearing at three points can be understood from the perspective of lattice sums. Expanding three-point component integrals $Y^{\tau}_{(a_2,a_3|b_2,b_3)}$ to order 10 introduces a large variety of dihedral and trihedral MGFs whose modular weight adds up to $\leq 10$.

As we saw in Section 5.7.2, already the two-loop graphs $C\big[\begin{smallmatrix} a_1 & a_2 & a_3 \\ b_1 & b_2 & b_3 \end{smallmatrix}\big]$ with $|A| + |B| = 10$ introduce irreducible cusp forms $\mathcal{A}\big[\begin{smallmatrix} a_1 & a_2 & a_3 \\ b_1 & b_2 & b_3 \end{smallmatrix}\big]$ as defined in (5.7) with vanishing Laurent polynomials. As summarized in Table 5.3, the known types of relations among dihedral and trihedral

MGFs leave three independent cusp forms built from $\mathcal{C}\left[\begin{smallmatrix}A\\B\end{smallmatrix}\right]$ with $|A| = |B| = 5$. One of them is expressible as the antisymmetrized product

$$\frac{(\nabla_0\mathrm{E}_2)\overline{\nabla}_0\mathrm{E}_3-(\overline{\nabla}_0\mathrm{E}_2)\nabla_0\mathrm{E}_3}{\tau_2^2} = 6\left(\frac{\tau_2}{\pi}\right)^5\left\{\mathcal{C}\left[\begin{smallmatrix}3&0\\1&0\end{smallmatrix}\right]\mathcal{C}\left[\begin{smallmatrix}2&0\\4&0\end{smallmatrix}\right]-\mathcal{C}\left[\begin{smallmatrix}1&0\\3&0\end{smallmatrix}\right]\mathcal{C}\left[\begin{smallmatrix}4&0\\2&0\end{smallmatrix}\right]\right\}, \quad (8.96)$$

and we additionally have two irreducible cusp forms that can be taken to be $\mathcal{A}\left[\begin{smallmatrix}0&2&3\\3&0&2\end{smallmatrix}\right]$ and $\mathcal{A}\left[\begin{smallmatrix}0&1&2&2\\1&1&0&3\end{smallmatrix}\right]$. While (8.96) and $\mathcal{A}\left[\begin{smallmatrix}0&2&3\\3&0&2\end{smallmatrix}\right]$ have been discussed in [185], the cusp form $\mathcal{A}\left[\begin{smallmatrix}0&1&2&2\\1&1&0&3\end{smallmatrix}\right]$ exceeds the loop orders studied in the reference.

For MGFs with different holomorphic and antiholomorphic modular weights $a \neq b$, one can construct basis elements from Cauchy–Riemann derivatives of modular invariants. As detailed in Table 5.4, the bases for $a + b \leq 8$ can be assembled from Cauchy–Riemann derivatives of $\mathrm{E}_{k\leq 4}$ and $\mathrm{E}_{2,2}$ (including products of $\mathrm{E}_2$, $\nabla_0\mathrm{E}_2$ and $\overline{\nabla}_0\mathrm{E}_2$). For $a + b = 10$ in turn, one needs to adjoin combinations of $\mathrm{E}_5$, $\mathrm{E}_{2,3}$, $\mathcal{A}\left[\begin{smallmatrix}0&2&3\\3&0&2\end{smallmatrix}\right]$, $\mathcal{A}\left[\begin{smallmatrix}0&1&2&2\\1&1&0&3\end{smallmatrix}\right]$ and their Cauchy–Riemann derivatives to obtain complete lattice-sum bases.

In order to obtain simple expressions for the full range of $\beta^{\mathrm{sv}}\left[\begin{smallmatrix}j_1&j_2\\4&6\end{smallmatrix}\right]$ and $\beta^{\mathrm{sv}}\left[\begin{smallmatrix}j_1&j_2\\6&4\end{smallmatrix}\right]$ in terms of lattice sums, it is convenient to delay the appearance of holomorphic Eisenstein series in the Cauchy–Riemann equations. This can be achieved by taking the modular invariant combinations $\mathrm{B}_{2,3}$ and $\mathrm{B}'_{2,3}$ defined in (5.214) as basis elements for the lattice sums with $a = b = 5$. The lowest-order Cauchy–Riemann derivatives that contain holomorphic Eisenstein series are listed in (5.216). While the MGF $\mathrm{B}_{2,3}$ is also an imaginary cusp form, the second form $\mathrm{B}'_{2,3}$ is neither real nor a cusp form, and its Laurent polynomial is determined by the known Laurent polynomials (5.212) of $\mathrm{E}_{2,3}$ and $\mathrm{E}_2$ and given in (5.217). The complex-conjugation properties are given in (5.215). The imaginary cusp forms with $a = b = 5$ studied in [185] were denoted by $\mathcal{A}_{1,2;5}$ and $\mathcal{A}_{1,4;5}$ there and can be rewritten in our basis as follows[12]

$$\mathcal{A}_{1,2;5} = \frac{1}{3}\left(\frac{\tau_2}{\pi}\right)^5\mathcal{A}\left[\begin{smallmatrix}0&2&3\\3&0&2\end{smallmatrix}\right] = \frac{2}{3}\left(\mathrm{B}'_{2,3} - \mathrm{B}_{2,3} + \frac{21}{4}\mathrm{E}_{2,3} + \frac{\zeta_3}{2}\mathrm{E}_2\right), \quad (8.97a)$$

$$\mathcal{A}_{1,4;5} = \frac{(\nabla_0\mathrm{E}_2)\overline{\nabla}_0\mathrm{E}_3 - (\overline{\nabla}_0\mathrm{E}_2)\nabla_0\mathrm{E}_3}{6\tau_2^2}\,. \quad (8.97b)$$

The extra cusp form $\mathcal{A}\left[\begin{smallmatrix}0&1&2&2\\1&1&0&3\end{smallmatrix}\right]$ entering the definition of $\mathrm{B}'_{2,3}$ did not arise in [185] as its lattice-sum representation requires three-loop graphs on the worldsheet.

Since the bases in Table 5.4 exclude factors of $\mathrm{G}_k$, the counting of basis elements without MZVs matches the number of $\beta^{\mathrm{sv}}$ that can enter the $\alpha'$-expansion of $Y^{\tau}_{\vec{\eta}}$ at the relevant order. As will be detailed in the

12 Note that the normalization conventions of [185] for $\mathcal{C}\left[\begin{smallmatrix}A\\B\end{smallmatrix}\right]$ and $\mathcal{A}\left[\begin{smallmatrix}A\\B\end{smallmatrix}\right]$ differ from ours in (5.4) and (5.7) by an additional factor of $\left(\frac{\tau_2}{\pi}\right)^{\frac{1}{2}(|A|+|B|)}$.

following sections, see in particular (8.101) and (8.103) to (8.105), the correspondence between MGFs- and iterated-integral bases is

$$
\begin{aligned}
\left.\begin{array}{r} \mathrm{E}_{2,3},\ \mathrm{E}_2\mathrm{E}_3,\ \tau_2^{-2}\overline{\nabla}_0\mathrm{E}_2\nabla_0\mathrm{E}_3 \\ \tau_2^{-2}\nabla_0\mathrm{E}_2\overline{\nabla}_0\mathrm{E}_3,\ \mathrm{B}_{2,3},\ \mathrm{B}'_{2,3} \end{array}\right\} &\leftrightarrow \left\{\begin{array}{l} \beta^{\rm sv}\big[\begin{smallmatrix}0&3\\4&6\end{smallmatrix}\big],\ \beta^{\rm sv}\big[\begin{smallmatrix}1&2\\4&6\end{smallmatrix}\big],\ \beta^{\rm sv}\big[\begin{smallmatrix}2&1\\4&6\end{smallmatrix}\big] \\ \beta^{\rm sv}\big[\begin{smallmatrix}3&0\\6&4\end{smallmatrix}\big],\ \beta^{\rm sv}\big[\begin{smallmatrix}2&1\\6&4\end{smallmatrix}\big],\ \beta^{\rm sv}\big[\begin{smallmatrix}1&2\\6&4\end{smallmatrix}\big] \end{array}\right. \\
\left.\begin{array}{r} \nabla_0\mathrm{E}_{2,3},\ \mathrm{E}_3\nabla_0\mathrm{E}_2,\ \mathrm{E}_2\nabla_0\mathrm{E}_3 \\ \tau_2^{-2}\overline{\nabla}_0\mathrm{E}_2\nabla_0^2\mathrm{E}_3,\ \nabla_0\mathrm{B}_{2,3},\ \nabla_0\mathrm{B}'_{2,3} \end{array}\right\} &\leftrightarrow \left\{\begin{array}{l} \beta^{\rm sv}\big[\begin{smallmatrix}0&4\\4&6\end{smallmatrix}\big],\ \beta^{\rm sv}\big[\begin{smallmatrix}1&3\\4&6\end{smallmatrix}\big],\ \beta^{\rm sv}\big[\begin{smallmatrix}2&2\\4&6\end{smallmatrix}\big] \\ \beta^{\rm sv}\big[\begin{smallmatrix}4&0\\6&4\end{smallmatrix}\big],\ \beta^{\rm sv}\big[\begin{smallmatrix}3&1\\6&4\end{smallmatrix}\big],\ \beta^{\rm sv}\big[\begin{smallmatrix}2&2\\6&4\end{smallmatrix}\big] \end{array}\right. \\
\left.\begin{array}{r} \nabla_0^2\mathrm{E}_{2,3},\ \nabla_0\mathrm{E}_2\nabla_0\mathrm{E}_3 \\ \mathrm{E}_2\nabla_0^2\mathrm{E}_3,\ \nabla_0^2\mathrm{B}'_{2,3} \end{array}\right\} &\leftrightarrow \left\{\begin{array}{l} \beta^{\rm sv}\big[\begin{smallmatrix}1&4\\4&6\end{smallmatrix}\big],\ \beta^{\rm sv}\big[\begin{smallmatrix}2&3\\4&6\end{smallmatrix}\big] \\ \beta^{\rm sv}\big[\begin{smallmatrix}4&1\\6&4\end{smallmatrix}\big],\ \beta^{\rm sv}\big[\begin{smallmatrix}3&2\\6&4\end{smallmatrix}\big] \end{array}\right. \\
\nabla_0\mathrm{E}_2\nabla_0^2\mathrm{E}_3,\ \nabla_0^3\mathrm{B}'_{2,3} &\leftrightarrow \beta^{\rm sv}\big[\begin{smallmatrix}2&4\\4&6\end{smallmatrix}\big],\ \beta^{\rm sv}\big[\begin{smallmatrix}4&2\\6&4\end{smallmatrix}\big]\,,
\end{aligned}
\tag{8.98}
$$

where the powers of $\tau_2$ were inserted to harmonize the modular weights. Similarly, we have $\nabla_0^m\mathrm{E}_5 \leftrightarrow \beta^{\rm sv}\big[\begin{smallmatrix}4+m\\10\end{smallmatrix}\big]$, $\tau_2^{-2m}\overline{\nabla}_0^m\mathrm{E}_5 \leftrightarrow \beta^{\rm sv}\big[\begin{smallmatrix}4-m\\10\end{smallmatrix}\big]$ with $m \leq 4$ according to Appendix C.2 of [V]. All the $\beta^{\rm sv}$ in (8.98) are understood to carry admixtures of lower depth analogous to the terms involving $\zeta_k$ in (8.71). Generalizations of (8.98) to higher weight will be discussed in Section 8.5.2.

### 8.4.2 *Three-point component integrals and cusp forms*

Based on the generating series (8.36), we have expanded all three-point component integrals $Y^\tau_{(a_2,a_3|b_2,b_3)}(\sigma|\rho)$ to order 10. Similar to the two-point case, the leading orders of the simplest cases $Y^\tau_{(0,0|0,0)}(\sigma|\rho)$ or $Y^\tau_{(1,0|1,0)}(\sigma|\rho)$, $Y^\tau_{(1,0|0,1)}(\sigma|\rho)$ are still expressible in terms of $\mathrm{E}_k$ and $\mathrm{E}_{p,q}$. Higher orders contain non-trivial trihedral MGFs. The simplest instances of cusp forms or the complex basis elements in (5.214) occur in the $\alpha'$-expansion of the following component integrals:

$$
\begin{aligned}
&Y^\tau_{(2,0|0,2)}(2,3|2,3) - Y^\tau_{(0,2|2,0)}(2,3|2,3) \\
&\quad = (s_{12}{-}s_{23})(2s_{12}s_{13}{-}s_{12}s_{23}{+}2s_{13}s_{23})\frac{(\nabla_0\mathrm{E}_2)\overline{\nabla}_0\mathrm{E}_3 - (\overline{\nabla}_0\mathrm{E}_2)\nabla_0\mathrm{E}_3}{12\tau_2^2} \\
&\qquad + s_{13}(s_{23}{-}s_{12})(s_{12}{+}s_{13}{+}s_{23})\mathrm{B}_{2,3} + O(s_{ij}^4)\,,
\end{aligned}
\tag{8.99a}
$$

$$
\begin{aligned}
&Y^\tau_{(2,1|3,0)}(2,3|2,3) - Y^\tau_{(3,0|2,1)}(2,3|2,3) \\
&\quad = s_{12}s_{13}\Big(\frac{4\mathrm{B}'_{2,3}}{3} + 7\mathrm{E}_{2,3} + \frac{2}{3}\zeta_3\mathrm{E}_2\Big) - \frac{1}{3}s_{13}(4s_{12}{+}3s_{13}{+}3s_{23})\mathrm{B}_{2,3} \\
&\qquad + s_{13}(3s_{12}{+}2s_{13}{+}2s_{23})\frac{(\nabla_0\mathrm{E}_2)\overline{\nabla}_0\mathrm{E}_3{-}(\overline{\nabla}_0\mathrm{E}_2)\nabla_0\mathrm{E}_3}{12\tau_2^2} + O(s_{ij}^3)\,.
\end{aligned}
\tag{8.99b}
$$

The expressions in (8.99) have been obtained by simplifying the lattice sums via the basis decompositions from Section 5.7. By matching these results with the $\alpha'$-expansions due to (8.36),

$$
Y^\tau_{(2,1|3,0)}(2,3|2,3) - Y^\tau_{(3,0|2,1)}(2,3|2,3)\Big|_{s_{13}^2} = -60\beta^{\rm sv}\big[\begin{smallmatrix}0&3\\4&6\end{smallmatrix}\big] + 270\beta^{\rm sv}\big[\begin{smallmatrix}1&2\\4&6\end{smallmatrix}\big]
$$

$$
\begin{aligned}
&+60\beta^{\text{sv}}\big[\begin{smallmatrix}1&2\\6&4\end{smallmatrix}\big]-390\beta^{\text{sv}}\big[\begin{smallmatrix}2&1\\4&6\end{smallmatrix}\big]-270\beta^{\text{sv}}\big[\begin{smallmatrix}2&1\\6&4\end{smallmatrix}\big]+390\beta^{\text{sv}}\big[\begin{smallmatrix}3&0\\6&4\end{smallmatrix}\big]+3\zeta_3\beta^{\text{sv}}\big[\begin{smallmatrix}1\\4\end{smallmatrix}\big]\\
&+260\zeta_3\beta^{\text{sv}}\big[\begin{smallmatrix}1\\6\end{smallmatrix}\big]-\frac{45\zeta_3}{y}\beta^{\text{sv}}\big[\begin{smallmatrix}2\\6\end{smallmatrix}\big]+\frac{5\zeta_3}{2y^2}\beta^{\text{sv}}\big[\begin{smallmatrix}3\\6\end{smallmatrix}\big]-\frac{39\zeta_5}{y}\beta^{\text{sv}}\big[\begin{smallmatrix}0\\4\end{smallmatrix}\big]\\
&+\frac{27\zeta_5}{4y^2}\beta^{\text{sv}}\big[\begin{smallmatrix}1\\4\end{smallmatrix}\big]-\frac{3\zeta_5}{8y^3}\beta^{\text{sv}}\big[\begin{smallmatrix}2\\4\end{smallmatrix}\big]+\frac{13\zeta_5}{120}
\end{aligned}\tag{8.100a}
$$

$$
\begin{aligned}
&Y^{\tau}_{(2,1|3,0)}(2,3|2,3)-Y^{\tau}_{(3,0|2,1)}(2,3|2,3)\Big|_{s_{12}s_{13}}=-90\beta^{\text{sv}}\big[\begin{smallmatrix}0&3\\4&6\end{smallmatrix}\big]+360\beta^{\text{sv}}\big[\begin{smallmatrix}1&2\\4&6\end{smallmatrix}\big]\\
&+90\beta^{\text{sv}}\big[\begin{smallmatrix}1&2\\6&4\end{smallmatrix}\big]+330\beta^{\text{sv}}\big[\begin{smallmatrix}2&1\\4&6\end{smallmatrix}\big]-360\beta^{\text{sv}}\big[\begin{smallmatrix}2&1\\6&4\end{smallmatrix}\big]-330\beta^{\text{sv}}\big[\begin{smallmatrix}3&0\\6&4\end{smallmatrix}\big]\\
&-220\zeta_3\beta^{\text{sv}}\big[\begin{smallmatrix}1\\6\end{smallmatrix}\big]-\frac{60\zeta_3}{y}\beta^{\text{sv}}\big[\begin{smallmatrix}2\\6\end{smallmatrix}\big]+\frac{15\zeta_3}{4y^2}\beta^{\text{sv}}\big[\begin{smallmatrix}3\\6\end{smallmatrix}\big]+\frac{33\zeta_5}{y}\beta^{\text{sv}}\big[\begin{smallmatrix}0\\4\end{smallmatrix}\big]\\
&+\frac{9\zeta_5}{y^2}\beta^{\text{sv}}\big[\begin{smallmatrix}1\\4\end{smallmatrix}\big]-\frac{9\zeta_5}{16y^3}\beta^{\text{sv}}\big[\begin{smallmatrix}2\\4\end{smallmatrix}\big]-\frac{\zeta_5}{90}
\end{aligned}\tag{8.100b}
$$

one can extract the following $\beta^{\text{sv}}$-representation of $\mathrm{B}_{2,3}$ and $\mathrm{B}'_{2,3}$

$$
\begin{aligned}
\mathrm{B}_{2,3}&=450\beta^{\text{sv}}\big[\begin{smallmatrix}2&1\\4&6\end{smallmatrix}\big]-450\beta^{\text{sv}}\big[\begin{smallmatrix}3&0\\6&4\end{smallmatrix}\big]+270\beta^{\text{sv}}\big[\begin{smallmatrix}2&1\\6&4\end{smallmatrix}\big]-270\beta^{\text{sv}}\big[\begin{smallmatrix}1&2\\4&6\end{smallmatrix}\big]\\
&\quad-3\zeta_3\beta^{\text{sv}}\big[\begin{smallmatrix}1\\4\end{smallmatrix}\big]-300\zeta_3\beta^{\text{sv}}\big[\begin{smallmatrix}1\\6\end{smallmatrix}\big]+\frac{45\zeta_3\beta^{\text{sv}}\big[\begin{smallmatrix}2\\6\end{smallmatrix}\big]}{y}\\
&\quad+\frac{45\zeta_5\beta^{\text{sv}}\big[\begin{smallmatrix}0\\4\end{smallmatrix}\big]}{y}-\frac{27\zeta_5\beta^{\text{sv}}\big[\begin{smallmatrix}1\\4\end{smallmatrix}\big]}{4y^2}-\frac{13\zeta_5}{120}\,,
\end{aligned}\tag{8.101a}
$$

$$
\begin{aligned}
\mathrm{B}'_{2,3}&=1260\beta^{\text{sv}}\big[\begin{smallmatrix}2&1\\4&6\end{smallmatrix}\big]-840\zeta_3\beta^{\text{sv}}\big[\begin{smallmatrix}1\\6\end{smallmatrix}\big]\\
&\quad+\frac{7\zeta_5}{240}-\frac{\zeta_3^2}{2y}-\frac{147\zeta_7}{64y^2}+\frac{21\zeta_3\zeta_5}{8y^3}\,.
\end{aligned}\tag{8.101b}
$$

Similarly, (5.214) implies $\beta^{\text{sv}}$-representations of the cusp forms

$$
\begin{aligned}
&\Big(\frac{\tau_2}{\pi}\Big)^5\mathcal{A}\big[\begin{smallmatrix}0&1&2&2\\1&1&0&3\end{smallmatrix}\big]=60\beta^{\text{sv}}\big[\begin{smallmatrix}0&3\\4&6\end{smallmatrix}\big]-60\beta^{\text{sv}}\big[\begin{smallmatrix}1&2\\6&4\end{smallmatrix}\big]+270\beta^{\text{sv}}\big[\begin{smallmatrix}2&1\\6&4\end{smallmatrix}\big]-270\beta^{\text{sv}}\big[\begin{smallmatrix}1&2\\4&6\end{smallmatrix}\big]\\
&+390\beta^{\text{sv}}\big[\begin{smallmatrix}2&1\\4&6\end{smallmatrix}\big]-390\beta^{\text{sv}}\big[\begin{smallmatrix}3&0\\6&4\end{smallmatrix}\big]-3\zeta_3\beta^{\text{sv}}\big[\begin{smallmatrix}1\\4\end{smallmatrix}\big]-260\zeta_3\beta^{\text{sv}}\big[\begin{smallmatrix}1\\6\end{smallmatrix}\big]+\frac{45\zeta_3}{y}\beta^{\text{sv}}\big[\begin{smallmatrix}2\\6\end{smallmatrix}\big]\\
&]-\frac{5\zeta_3}{2y^2}\beta^{\text{sv}}\big[\begin{smallmatrix}3\\6\end{smallmatrix}\big]+\frac{39\zeta_5}{y}\beta^{\text{sv}}\big[\begin{smallmatrix}0\\4\end{smallmatrix}\big]-\frac{27\zeta_5}{4y^2}\beta^{\text{sv}}\big[\begin{smallmatrix}1\\4\end{smallmatrix}\big]+\frac{3\zeta_5}{8y^3}\beta^{\text{sv}}\big[\begin{smallmatrix}2\\4\end{smallmatrix}\big]-\frac{13\zeta_5}{120}\,,
\end{aligned}\tag{8.102a}
$$

$$
\begin{aligned}
&\Big(\frac{\tau_2}{\pi}\Big)^5\mathcal{A}\big[\begin{smallmatrix}0&2&3\\3&0&2\end{smallmatrix}\big]=540\beta^{\text{sv}}\big[\begin{smallmatrix}1&2\\4&6\end{smallmatrix}\big]-540\beta^{\text{sv}}\big[\begin{smallmatrix}2&1\\6&4\end{smallmatrix}\big]+360\beta^{\text{sv}}\big[\begin{smallmatrix}2&1\\4&6\end{smallmatrix}\big]-360\beta^{\text{sv}}\big[\begin{smallmatrix}3&0\\6&4\end{smallmatrix}\big]\\
&-240\zeta_3\beta^{\text{sv}}\big[\begin{smallmatrix}1\\6\end{smallmatrix}\big]-\frac{90\zeta_3}{y}\beta^{\text{sv}}\big[\begin{smallmatrix}2\\6\end{smallmatrix}\big]+\frac{36\zeta_5}{y}\beta^{\text{sv}}\big[\begin{smallmatrix}0\\4\end{smallmatrix}\big]+\frac{27\zeta_5}{2y^2}\beta^{\text{sv}}\big[\begin{smallmatrix}1\\4\end{smallmatrix}\big]-\frac{\zeta_5}{60}\,,
\end{aligned}\tag{8.102b}
$$

where the vanishing of their Laurent polynomials can be crosschecked through the asymptotics (8.64) of the $\beta^{\text{sv}}$. Once we have fixed the antiholomorphic integration constants of the $\beta^{\text{sv}}\big[\begin{smallmatrix}j_1&j_2\\4&6\end{smallmatrix}\big]$ and $\beta^{\text{sv}}\big[\begin{smallmatrix}j_1&j_2\\6&4\end{smallmatrix}\big]$ in Section 8.4.4, one can extract the $q$-expansions of the MGFs from their new representations (8.101) and (8.102).

### 8.4.3 *Cauchy–Riemann derivatives of cusp forms and $\beta^{\rm sv}$*

The above procedure to relate the new basis elements $\mathrm{B}_{2,3}$ and $\mathrm{B}'_{2,3}$ to cusp forms can be repeated based on component integrals $Y^{\tau}_{(a_2,a_3|b_2,b_3)}(\sigma|\rho)$ of non-vanishing modular weight $(0, b_2{+}b_3{-}a_2{-}a_3)$. Their expansion in terms of $\beta^{\rm sv}$ to order 10 is available in an ancillary file within the arXiv submission of [V]. On top of (8.101), we find

$$\begin{aligned}\pi\nabla_0\mathrm{B}_{2,3} &= 135\beta^{\rm sv}\big[\begin{smallmatrix}1&3\\4&6\end{smallmatrix}\big]-270\beta^{\rm sv}\big[\begin{smallmatrix}2&2\\4&6\end{smallmatrix}\big]-\frac{135}{2}\beta^{\rm sv}\big[\begin{smallmatrix}2&2\\6&4\end{smallmatrix}\big]+90\beta^{\rm sv}\big[\begin{smallmatrix}3&1\\6&4\end{smallmatrix}\big]\\ &+\frac{225}{2}\beta^{\rm sv}\big[\begin{smallmatrix}4&0\\6&4\end{smallmatrix}\big]+\frac{3\zeta_3}{4}\beta^{\rm sv}\big[\begin{smallmatrix}2\\4\end{smallmatrix}\big]+180\zeta_3\beta^{\rm sv}\big[\begin{smallmatrix}2\\6\end{smallmatrix}\big]-\frac{45\zeta_3}{2y}\beta^{\rm sv}\big[\begin{smallmatrix}3\\6\end{smallmatrix}\big]-45\zeta_5\beta^{\rm sv}\big[\begin{smallmatrix}0\\4\end{smallmatrix}\big]\\ &-\frac{9\zeta_5}{y}\beta^{\rm sv}\big[\begin{smallmatrix}1\\4\end{smallmatrix}\big]+\frac{27\zeta_5}{16y^2}\beta^{\rm sv}\big[\begin{smallmatrix}2\\4\end{smallmatrix}\big]\,,\end{aligned} \tag{8.103a}$$

$$\begin{aligned}\frac{\pi\overline{\nabla}_0\,\overline{\mathrm{B}}_{2,3}}{y^2} &= 1440\beta^{\rm sv}\big[\begin{smallmatrix}1&1\\4&6\end{smallmatrix}\big]-1080\beta^{\rm sv}\big[\begin{smallmatrix}0&2\\4&6\end{smallmatrix}\big]+2160\beta^{\rm sv}\big[\begin{smallmatrix}1&1\\6&4\end{smallmatrix}\big]+1800\beta^{\rm sv}\big[\begin{smallmatrix}2&0\\4&6\end{smallmatrix}\big]\\ &-4320\beta^{\rm sv}\big[\begin{smallmatrix}2&0\\6&4\end{smallmatrix}\big]-12\zeta_3\beta^{\rm sv}\big[\begin{smallmatrix}0\\4\end{smallmatrix}\big]-1200\zeta_3\beta^{\rm sv}\big[\begin{smallmatrix}0\\6\end{smallmatrix}\big]-\frac{240\zeta_3}{y}\beta^{\rm sv}\big[\begin{smallmatrix}1\\6\end{smallmatrix}\big]+\frac{45\zeta_3}{y^2}\beta^{\rm sv}\big[\begin{smallmatrix}2\\6\end{smallmatrix}\big]\\ &+\frac{108\zeta_5}{y^2}\beta^{\rm sv}\big[\begin{smallmatrix}0\\4\end{smallmatrix}\big]-\frac{27\zeta_5}{2y^3}\beta^{\rm sv}\big[\begin{smallmatrix}1\\4\end{smallmatrix}\big]-\frac{\zeta_5}{4y}\,,\end{aligned} \tag{8.103b}$$

as well as

$$\begin{aligned}\frac{(\pi\overline{\nabla}_0)^3\overline{\mathrm{B}}'_{2,3}}{y^6} &= -483840\beta^{\rm sv}\big[\begin{smallmatrix}0&0\\6&4\end{smallmatrix}\big] + \frac{756\zeta_5}{y^4}\beta^{\rm sv}\big[\begin{smallmatrix}0\\4\end{smallmatrix}\big]\\ &\quad - \frac{8\zeta_3}{15y} - \frac{7\zeta_5}{5y^3} - \frac{63\zeta_3\zeta_5}{4y^6}\,,\end{aligned} \tag{8.104a}$$

$$\begin{aligned}\frac{(\pi\overline{\nabla}_0)^2\overline{\mathrm{B}}'_{2,3}}{y^4} &= 120960\beta^{\rm sv}\big[\begin{smallmatrix}1&0\\6&4\end{smallmatrix}\big] - \frac{756\zeta_5}{y^3}\beta^{\rm sv}\big[\begin{smallmatrix}0\\4\end{smallmatrix}\big]\\ &\quad - \frac{2\zeta_3}{15} + \frac{7\zeta_5}{5y^2} - \frac{147\zeta_7}{32y^4} + \frac{63\zeta_3\zeta_5}{4y^5}\,,\end{aligned} \tag{8.104b}$$

$$\begin{aligned}\frac{\pi\overline{\nabla}_0\,\overline{\mathrm{B}}'_{2,3}}{y^2} &= -15120\beta^{\rm sv}\big[\begin{smallmatrix}2&0\\6&4\end{smallmatrix}\big] - 24\zeta_3\beta^{\rm sv}\big[\begin{smallmatrix}0\\4\end{smallmatrix}\big] + \frac{378\zeta_5}{y^2}\beta^{\rm sv}\big[\begin{smallmatrix}0\\4\end{smallmatrix}\big]\\ &\quad - \frac{7\zeta_5}{10y} + \frac{\zeta_3^2}{2y^2} + \frac{147\zeta_7}{32y^3} - \frac{63\zeta_3\zeta_5}{8y^4}\end{aligned} \tag{8.104c}$$

and

$$\begin{aligned}\pi\nabla_0\mathrm{B}'_{2,3} &= -945\beta^{\rm sv}\big[\begin{smallmatrix}2&2\\4&6\end{smallmatrix}\big]+630\zeta_3\beta^{\rm sv}\big[\begin{smallmatrix}2\\6\end{smallmatrix}\big]\\ &\quad+\frac{\zeta_3^2}{2}+\frac{147\zeta_7}{32y}-\frac{63\zeta_3\zeta_5}{8y^2}\,,\end{aligned} \tag{8.105a}$$

$$(\pi\nabla_0)^2\mathrm{B}'_{2,3} = \frac{945}{2}\beta^{\rm sv}\big[\begin{smallmatrix}2&3\\4&6\end{smallmatrix}\big]-315\zeta_3\beta^{\rm sv}\big[\begin{smallmatrix}3\\6\end{smallmatrix}\big]-\frac{147\zeta_7}{32}+\frac{63\zeta_3\zeta_5}{4y}\,, \tag{8.105b}$$

$$(\pi\nabla_0)^3\mathrm{B}'_{2,3} = -\frac{945}{8}\beta^{\rm sv}\big[\begin{smallmatrix}2&4\\4&6\end{smallmatrix}\big]+\frac{315\zeta_3}{4}\beta^{\rm sv}\big[\begin{smallmatrix}4\\6\end{smallmatrix}\big]-\frac{63}{4}\zeta_3\zeta_5\,. \tag{8.105c}$$

Higher derivatives in turn involve holomorphic Eisenstein series, see (5.216). These relations can be inverted to express all $\beta^{\text{sv}}\left[\begin{smallmatrix} j_1 & j_2 \\ k_1 & k_2 \end{smallmatrix}\right]$ with $k_1 + k_2 = 10$ in terms of MGFs. The full expressions are given in Appendix E.1 of [V].

### 8.4.4 *Explicit $\beta^{\text{sv}}$ from reality properties at three points*

We shall now outline the computation of the antiholomorphic integration constants $\overline{\alpha\left[\begin{smallmatrix} j_1 & j_2 \\ 6 & 4 \end{smallmatrix}\right]}$ that enter the key quantities $\beta^{\text{sv}}\left[\begin{smallmatrix} j_1 & j_2 \\ 6 & 4 \end{smallmatrix}\right]$ of this section via (8.37) and (8.58). Similar to the steps in section 8.3.5, we first determine the complex conjugate $\overline{\beta^{\text{sv}}\left[\begin{smallmatrix} j_1 & j_2 \\ 6 & 4 \end{smallmatrix}\right]}$ from the reality properties $Y^{\tau}_{(a_2,a_3|b_2,b_3)}(\sigma|\rho) = (4y)^{a_2+a_3-b_2-b_3}\overline{Y^{\tau}_{(b_2,b_3|a_2,a_3)}(\rho|\sigma)}$ of the component integrals,

$$\overline{\beta^{\text{sv}}\left[\begin{smallmatrix} 0 & 0 \\ 6 & 4 \end{smallmatrix}\right]} = \frac{\beta^{\text{sv}}\left[\begin{smallmatrix} 2 & 4 \\ 4 & 6 \end{smallmatrix}\right]}{4096y^6} - \frac{\zeta_3}{6144y^6}\beta^{\text{sv}}\left[\begin{smallmatrix} 4 \\ 6 \end{smallmatrix}\right] + \frac{\zeta_5}{10240y^6}\beta^{\text{sv}}\left[\begin{smallmatrix} 2 \\ 4 \end{smallmatrix}\right] - \frac{\zeta_3}{907200y} - \frac{\zeta_5}{345600y^3} \tag{8.106a}$$

$$\overline{\beta^{\text{sv}}\left[\begin{smallmatrix} 1 & 0 \\ 6 & 4 \end{smallmatrix}\right]} = \frac{\beta^{\text{sv}}\left[\begin{smallmatrix} 2 & 3 \\ 4 & 6 \end{smallmatrix}\right]}{256y^4} - \frac{\zeta_3}{384y^4}\beta^{\text{sv}}\left[\begin{smallmatrix} 3 \\ 6 \end{smallmatrix}\right] + \frac{\zeta_5}{2560y^5}\beta^{\text{sv}}\left[\begin{smallmatrix} 2 \\ 4 \end{smallmatrix}\right] + \frac{\zeta_3}{907200} - \frac{\zeta_5}{86400y^2} \tag{8.106b}$$

$$\overline{\beta^{\text{sv}}\left[\begin{smallmatrix} 0 & 1 \\ 6 & 4 \end{smallmatrix}\right]} = \frac{\beta^{\text{sv}}\left[\begin{smallmatrix} 1 & 4 \\ 4 & 6 \end{smallmatrix}\right]}{256y^4} - \frac{\zeta_3}{1536y^5}\beta^{\text{sv}}\left[\begin{smallmatrix} 4 \\ 6 \end{smallmatrix}\right] + \frac{\zeta_5}{640y^4}\beta^{\text{sv}}\left[\begin{smallmatrix} 1 \\ 4 \end{smallmatrix}\right] - \frac{\zeta_3}{226800} + \frac{\zeta_5}{172800y^2} \tag{8.106c}$$

$$\overline{\beta^{\text{sv}}\left[\begin{smallmatrix} 2 & 0 \\ 6 & 4 \end{smallmatrix}\right]} = \frac{\beta^{\text{sv}}\left[\begin{smallmatrix} 2 & 2 \\ 4 & 6 \end{smallmatrix}\right]}{16y^2} - \frac{\zeta_3}{10080y^2}\beta^{\text{sv}}\left[\begin{smallmatrix} 2 \\ 4 \end{smallmatrix}\right] - \frac{\zeta_3}{24y^2}\beta^{\text{sv}}\left[\begin{smallmatrix} 2 \\ 6 \end{smallmatrix}\right] + \frac{\zeta_5}{640y^4}\beta^{\text{sv}}\left[\begin{smallmatrix} 2 \\ 4 \end{smallmatrix}\right] - \frac{\zeta_5}{21600y} \tag{8.106d}$$

$$\overline{\beta^{\text{sv}}\left[\begin{smallmatrix} 1 & 1 \\ 6 & 4 \end{smallmatrix}\right]} = \frac{\beta^{\text{sv}}\left[\begin{smallmatrix} 1 & 3 \\ 4 & 6 \end{smallmatrix}\right]}{16y^2} + \frac{\zeta_3}{6720y^2}\beta^{\text{sv}}\left[\begin{smallmatrix} 2 \\ 4 \end{smallmatrix}\right] - \frac{\zeta_3}{96y^3}\beta^{\text{sv}}\left[\begin{smallmatrix} 3 \\ 6 \end{smallmatrix}\right] + \frac{\zeta_5}{160y^3}\beta^{\text{sv}}\left[\begin{smallmatrix} 1 \\ 4 \end{smallmatrix}\right] + \frac{\zeta_5}{43200y} \tag{8.106e}$$

$$\overline{\beta^{\text{sv}}\left[\begin{smallmatrix} 0 & 2 \\ 6 & 4 \end{smallmatrix}\right]} = \frac{\beta^{\text{sv}}\left[\begin{smallmatrix} 0 & 4 \\ 4 & 6 \end{smallmatrix}\right]}{16y^2} - \frac{\zeta_3}{1680y^2}\beta^{\text{sv}}\left[\begin{smallmatrix} 2 \\ 4 \end{smallmatrix}\right] - \frac{\zeta_3}{384y^4}\beta^{\text{sv}}\left[\begin{smallmatrix} 4 \\ 6 \end{smallmatrix}\right] + \frac{\zeta_5}{40y^2}\beta^{\text{sv}}\left[\begin{smallmatrix} 0 \\ 4 \end{smallmatrix}\right] - \frac{\zeta_5}{21600y} \tag{8.106f}$$

$$\overline{\beta^{\text{sv}}\left[\begin{smallmatrix} 3 & 0 \\ 6 & 4 \end{smallmatrix}\right]} = \beta^{\text{sv}}\left[\begin{smallmatrix} 2 & 1 \\ 4 & 6 \end{smallmatrix}\right] - \frac{\zeta_3}{210}\beta^{\text{sv}}\left[\begin{smallmatrix} 1 \\ 4 \end{smallmatrix}\right] - \frac{2\zeta_3}{3}\beta^{\text{sv}}\left[\begin{smallmatrix} 1 \\ 6 \end{smallmatrix}\right] + \frac{\zeta_5}{160y^3}\beta^{\text{sv}}\left[\begin{smallmatrix} 2 \\ 4 \end{smallmatrix}\right] - \frac{\zeta_5}{5400} \tag{8.106g}$$

$$\overline{\beta^{\text{sv}}\left[\begin{smallmatrix} 2 & 1 \\ 6 & 4 \end{smallmatrix}\right]} = \beta^{\text{sv}}\left[\begin{smallmatrix} 1 & 2 \\ 4 & 6 \end{smallmatrix}\right] + \frac{\zeta_3}{315}\beta^{\text{sv}}\left[\begin{smallmatrix} 1 \\ 4 \end{smallmatrix}\right] - \frac{\zeta_3}{6y}\beta^{\text{sv}}\left[\begin{smallmatrix} 2 \\ 6 \end{smallmatrix}\right] + \frac{\zeta_5}{40y^2}\beta^{\text{sv}}\left[\begin{smallmatrix} 1 \\ 4 \end{smallmatrix}\right] + \frac{\zeta_5}{10800} \tag{8.106h}$$

$$\overline{\beta^{\text{sv}}\left[\begin{smallmatrix} 1 & 2 \\ 6 & 4 \end{smallmatrix}\right]} = \beta^{\text{sv}}\left[\begin{smallmatrix} 0 & 3 \\ 4 & 6 \end{smallmatrix}\right] - \frac{\zeta_3}{210}\beta^{\text{sv}}\left[\begin{smallmatrix} 1 \\ 4 \end{smallmatrix}\right] - \frac{\zeta_3}{24y^2}\beta^{\text{sv}}\left[\begin{smallmatrix} 3 \\ 6 \end{smallmatrix}\right] + \frac{\zeta_5}{10y}\beta^{\text{sv}}\left[\begin{smallmatrix} 0 \\ 4 \end{smallmatrix}\right] - \frac{\zeta_5}{5400} \tag{8.106i}$$

$$\begin{aligned}\overline{\beta^{\text{sv}}\left[\begin{smallmatrix} 4 & 0 \\ 6 & 4 \end{smallmatrix}\right]} &= 16y^2\beta^{\text{sv}}\left[\begin{smallmatrix} 2 & 0 \\ 4 & 6 \end{smallmatrix}\right] - \frac{16\zeta_3 y^2}{105}\beta^{\text{sv}}\left[\begin{smallmatrix} 0 \\ 4 \end{smallmatrix}\right] - \frac{32\zeta_3 y^2}{3}\beta^{\text{sv}}\left[\begin{smallmatrix} 0 \\ 6 \end{smallmatrix}\right] \\ &\quad + \frac{\zeta_5}{40y^2}\beta^{\text{sv}}\left[\begin{smallmatrix} 2 \\ 4 \end{smallmatrix}\right] - \frac{y\zeta_5}{1350}\end{aligned} \tag{8.106j}$$

$$\begin{aligned}\overline{\beta^{\text{sv}}\left[\begin{smallmatrix} 3 & 1 \\ 6 & 4 \end{smallmatrix}\right]} &= 16y^2\beta^{\text{sv}}\left[\begin{smallmatrix} 1 & 1 \\ 4 & 6 \end{smallmatrix}\right] + \frac{4\zeta_3 y^2}{105}\beta^{\text{sv}}\left[\begin{smallmatrix} 0 \\ 4 \end{smallmatrix}\right] - \frac{8\zeta_3 y}{3}\beta^{\text{sv}}\left[\begin{smallmatrix} 1 \\ 6 \end{smallmatrix}\right] \\ &\quad + \frac{\zeta_5}{10y}\beta^{\text{sv}}\left[\begin{smallmatrix} 1 \\ 4 \end{smallmatrix}\right] + \frac{y\zeta_5}{2700}\end{aligned} \tag{8.106k}$$

$$\begin{aligned}\overline{\beta^{\text{sv}}\left[\begin{smallmatrix} 2 & 2 \\ 6 & 4 \end{smallmatrix}\right]} &= 16y^2\beta^{\text{sv}}\left[\begin{smallmatrix} 0 & 2 \\ 4 & 6 \end{smallmatrix}\right] - \frac{8\zeta_3 y^2}{315}\beta^{\text{sv}}\left[\begin{smallmatrix} 0 \\ 4 \end{smallmatrix}\right] - \frac{2\zeta_3}{3}\beta^{\text{sv}}\left[\begin{smallmatrix} 2 \\ 6 \end{smallmatrix}\right] \\ &\quad + \frac{2\zeta_5}{5}\beta^{\text{sv}}\left[\begin{smallmatrix} 0 \\ 4 \end{smallmatrix}\right] - \frac{\zeta_5 y}{1350}\end{aligned} \tag{8.106l}$$

$$\begin{aligned}\overline{\beta^{\text{sv}}\left[\begin{smallmatrix} 4 & 1 \\ 6 & 4 \end{smallmatrix}\right]} &= 256y^4\beta^{\text{sv}}\left[\begin{smallmatrix} 1 & 0 \\ 4 & 6 \end{smallmatrix}\right] - \frac{128\zeta_3 y^3}{3}\beta^{\text{sv}}\left[\begin{smallmatrix} 0 \\ 6 \end{smallmatrix}\right] + \frac{2\zeta_5}{5}\beta^{\text{sv}}\left[\begin{smallmatrix} 1 \\ 4 \end{smallmatrix}\right] \\ &\quad - \frac{16y^4\zeta_3}{14175} + \frac{y^2\zeta_5}{675}\end{aligned} \tag{8.106m}$$

$$\begin{aligned}\overline{\beta^{\text{sv}}\left[\begin{smallmatrix} 3 & 2 \\ 6 & 4 \end{smallmatrix}\right]} &= 256y^4\beta^{\text{sv}}\left[\begin{smallmatrix} 0 & 1 \\ 4 & 6 \end{smallmatrix}\right] - \frac{32\zeta_3 y^2}{3}\beta^{\text{sv}}\left[\begin{smallmatrix} 1 \\ 6 \end{smallmatrix}\right] + \frac{8\zeta_5 y}{5}\beta^{\text{sv}}\left[\begin{smallmatrix} 0 \\ 4 \end{smallmatrix}\right] \\ &\quad + \frac{4y^4\zeta_3}{14175} - \frac{2\zeta_5 y^2}{675}\end{aligned} \tag{8.106n}$$

$$\begin{aligned}\overline{\beta^{\text{sv}}\left[\begin{smallmatrix} 4 & 2 \\ 6 & 4 \end{smallmatrix}\right]} &= 4096y^6\beta^{\text{sv}}\left[\begin{smallmatrix} 0 & 0 \\ 4 & 6 \end{smallmatrix}\right] - \frac{512\zeta_3 y^4}{3}\beta^{\text{sv}}\left[\begin{smallmatrix} 0 \\ 6 \end{smallmatrix}\right] + \frac{32\zeta_5 y^2}{5}\beta^{\text{sv}}\left[\begin{smallmatrix} 0 \\ 4 \end{smallmatrix}\right] \\ &\quad - \frac{64y^5\zeta_3}{14175} - \frac{8y^3\zeta_5}{675}\,.\end{aligned} \tag{8.106o}$$

We emphasize that this reasoning does not rely on any MGF representation and can be applied at higher orders $k_1 + k_2 \geq 14$, where a basis of lattice sums may not be explicitly available. These results line up with the closed depth-two formula (8.92) modulo admixtures of lower depth and determine $\overline{\beta^{\text{sv}}\left[\begin{smallmatrix} j_1 & j_2 \\ 4 & 6 \end{smallmatrix}\right]}$ via shuffle relations and (8.83).

In close analogy with (8.94), one can now solve (8.106) for the $\overline{\mathcal{E}^{\text{sv}}\left[\begin{smallmatrix} j_1 & j_2 \\ 6 & 4 \end{smallmatrix}\right]}$ and introduce the desired integration constants via (8.58). By comparing the purely holomorphic terms, we arrive at

$$\alpha\left[\begin{smallmatrix} 0 & 0 \\ 6 & 4 \end{smallmatrix}\right] = \alpha\left[\begin{smallmatrix} 1 & 0 \\ 6 & 4 \end{smallmatrix}\right] = \alpha\left[\begin{smallmatrix} 0 & 1 \\ 6 & 4 \end{smallmatrix}\right] = 0 \tag{8.107a}$$

$$\alpha\left[\begin{smallmatrix} 2 & 0 \\ 6 & 4 \end{smallmatrix}\right] = -\frac{i\pi\tau\zeta_3}{226800} - \frac{\zeta_3}{630}\mathcal{E}\left[\begin{smallmatrix} 0 \\ 4 \end{smallmatrix}\right] \tag{8.107b}$$

$$\alpha\left[\begin{smallmatrix} 1 & 1 \\ 6 & 4 \end{smallmatrix}\right] = \frac{i\pi\tau\zeta_3}{151200} + \frac{\zeta_3}{420}\mathcal{E}\left[\begin{smallmatrix} 0 \\ 4 \end{smallmatrix}\right] \tag{8.107c}$$

$$\alpha\left[\begin{smallmatrix}0&2\\6&4\end{smallmatrix}\right] = \frac{i\pi\tau\zeta_3}{56700} - \frac{\zeta_3}{105}\mathcal{E}\left[\begin{smallmatrix}0\\4\end{smallmatrix}\right] - \frac{2\zeta_3}{3}\mathcal{E}\left[\begin{smallmatrix}0\\6\end{smallmatrix}\right] \tag{8.107d}$$

$$\alpha\left[\begin{smallmatrix}3&0\\6&4\end{smallmatrix}\right] = \frac{\pi^2\tau^2\zeta_3}{75600} - \frac{i\pi\tau\zeta_3}{105}\mathcal{E}\left[\begin{smallmatrix}0\\4\end{smallmatrix}\right] + \frac{\zeta_3}{210}\mathcal{E}\left[\begin{smallmatrix}1\\4\end{smallmatrix}\right] \tag{8.107e}$$

$$\alpha\left[\begin{smallmatrix}2&1\\6&4\end{smallmatrix}\right] = -\frac{\pi^2\tau^2\zeta_3}{113400} + \frac{2i\pi\tau\zeta_3}{315}\mathcal{E}\left[\begin{smallmatrix}0\\4\end{smallmatrix}\right] - \frac{\zeta_3}{315}\mathcal{E}\left[\begin{smallmatrix}1\\4\end{smallmatrix}\right] \tag{8.107f}$$

$$\alpha\left[\begin{smallmatrix}1&2\\6&4\end{smallmatrix}\right] = -\frac{\pi^2\tau^2\zeta_3}{32400} - \frac{i\pi\tau\zeta_3}{105}\mathcal{E}\left[\begin{smallmatrix}0\\4\end{smallmatrix}\right] + \frac{\zeta_3}{210}\mathcal{E}\left[\begin{smallmatrix}1\\4\end{smallmatrix}\right] - \frac{4i\pi\tau\zeta_3}{3}\mathcal{E}\left[\begin{smallmatrix}0\\6\end{smallmatrix}\right] + \frac{2\zeta_3}{3}\mathcal{E}\left[\begin{smallmatrix}1\\6\end{smallmatrix}\right] \tag{8.107g}$$

$$\begin{aligned}\alpha\left[\begin{smallmatrix}4&0\\6&4\end{smallmatrix}\right] &= \frac{i\pi^3\tau^3\zeta_3}{28350} + \frac{4\pi^2\tau^2\zeta_3}{105}\mathcal{E}\left[\begin{smallmatrix}0\\4\end{smallmatrix}\right] + \frac{4i\pi\tau\zeta_3}{105}\mathcal{E}\left[\begin{smallmatrix}1\\4\end{smallmatrix}\right] - \frac{\zeta_3}{105}\mathcal{E}\left[\begin{smallmatrix}2\\4\end{smallmatrix}\right] \\ &\quad + \frac{i\pi\tau\zeta_5}{900} + \frac{2\zeta_5}{5}\mathcal{E}\left[\begin{smallmatrix}0\\4\end{smallmatrix}\right]\end{aligned} \tag{8.107h}$$

$$\alpha\left[\begin{smallmatrix}3&1\\6&4\end{smallmatrix}\right] = -\frac{i\pi^3\tau^3\zeta_3}{113400} - \frac{\pi^2\tau^2\zeta_3}{105}\mathcal{E}\left[\begin{smallmatrix}0\\4\end{smallmatrix}\right] - \frac{i\pi\tau\zeta_3}{105}\mathcal{E}\left[\begin{smallmatrix}1\\4\end{smallmatrix}\right] + \frac{\zeta_3}{420}\mathcal{E}\left[\begin{smallmatrix}2\\4\end{smallmatrix}\right] \tag{8.107i}$$

$$\begin{aligned}\alpha\left[\begin{smallmatrix}2&2\\6&4\end{smallmatrix}\right] &= -\frac{i\pi^3\tau^3\zeta_3}{18900} + \frac{2\pi^2\tau^2\zeta_3}{315}\mathcal{E}\left[\begin{smallmatrix}0\\4\end{smallmatrix}\right] + \frac{2i\pi\tau\zeta_3}{315}\mathcal{E}\left[\begin{smallmatrix}1\\4\end{smallmatrix}\right] - \frac{\zeta_3}{630}\mathcal{E}\left[\begin{smallmatrix}2\\4\end{smallmatrix}\right] \\ &\quad + \frac{8\pi^2\tau^2\zeta_3}{3}\mathcal{E}\left[\begin{smallmatrix}0\\6\end{smallmatrix}\right] + \frac{8i\pi\tau\zeta_3}{3}\mathcal{E}\left[\begin{smallmatrix}1\\6\end{smallmatrix}\right] - \frac{2\zeta_3}{3}\mathcal{E}\left[\begin{smallmatrix}2\\6\end{smallmatrix}\right]\end{aligned} \tag{8.107j}$$

$$\alpha\left[\begin{smallmatrix}4&1\\6&4\end{smallmatrix}\right] = -\frac{\pi^2\tau^2\zeta_5}{900} + \frac{4i\pi\tau\zeta_5}{5}\mathcal{E}\left[\begin{smallmatrix}0\\4\end{smallmatrix}\right] - \frac{2\zeta_5}{5}\mathcal{E}\left[\begin{smallmatrix}1\\4\end{smallmatrix}\right] \tag{8.107k}$$

$$\begin{aligned}\alpha\left[\begin{smallmatrix}3&2\\6&4\end{smallmatrix}\right] &= \frac{\pi^4\tau^4\zeta_3}{11340} + \frac{16i\pi^3\tau^3\zeta_3}{3}\mathcal{E}\left[\begin{smallmatrix}0\\6\end{smallmatrix}\right] - 8\pi^2\tau^2\zeta_3\mathcal{E}\left[\begin{smallmatrix}1\\6\end{smallmatrix}\right] \\ &\quad - 4i\pi\tau\zeta_3\mathcal{E}\left[\begin{smallmatrix}2\\6\end{smallmatrix}\right] + \frac{2\zeta_3}{3}\mathcal{E}\left[\begin{smallmatrix}3\\6\end{smallmatrix}\right]\end{aligned} \tag{8.107l}$$

$$\begin{aligned}\alpha\left[\begin{smallmatrix}4&2\\6&4\end{smallmatrix}\right] &= \frac{2i\pi^5\tau^5\zeta_3}{14175} - \frac{32\pi^4\tau^4\zeta_3}{3}\mathcal{E}\left[\begin{smallmatrix}0\\6\end{smallmatrix}\right] - \frac{64i\pi^3\tau^3\zeta_3}{3}\mathcal{E}\left[\begin{smallmatrix}1\\6\end{smallmatrix}\right] \\ &\quad + 16\pi^2\tau^2\zeta_3\mathcal{E}\left[\begin{smallmatrix}2\\6\end{smallmatrix}\right] + \frac{16i\pi\tau\zeta_3}{3}\mathcal{E}\left[\begin{smallmatrix}3\\6\end{smallmatrix}\right] - \frac{2\zeta_3}{3}\mathcal{E}\left[\begin{smallmatrix}4\\6\end{smallmatrix}\right] \\ &\quad - \frac{i\pi^3\tau^3\zeta_5}{675} - \frac{8\pi^2\tau^2\zeta_5}{5}\mathcal{E}\left[\begin{smallmatrix}0\\4\end{smallmatrix}\right] - \frac{8i\pi\tau\zeta_5}{5}\mathcal{E}\left[\begin{smallmatrix}1\\4\end{smallmatrix}\right] + \frac{2\zeta_5}{5}\mathcal{E}\left[\begin{smallmatrix}2\\4\end{smallmatrix}\right] .\end{aligned} \tag{8.107m}$$

One can furthermore check that the relation $\alpha\left[\begin{smallmatrix}j_1&j_2\\4&6\end{smallmatrix}\right] = -\alpha\left[\begin{smallmatrix}j_2&j_1\\6&4\end{smallmatrix}\right]$ holds, confirming the expected shuffle relations. Manifestly T-invariant representations can be found in Appendix G.2 of [V].

### 8.4.5 *Laplace equations of cusp forms*

In this section, we discuss the Laplace equations of the extra basis MGFs corresponding to $\beta^{\text{sv}}\left[\begin{smallmatrix}j_1&j_2\\6&4\end{smallmatrix}\right]$ and $\beta^{\text{sv}}\left[\begin{smallmatrix}j_1&j_2\\4&6\end{smallmatrix}\right]$. Their representatives $\mathrm{B}_{2,3}$ and $\mathrm{B}'_{2,3}$ in (5.214) satisfy

$$(\Delta+2)\mathrm{B}_{2,3} = 4\mathrm{B}'_{2,3}+21\mathrm{E}_{2,3}+\frac{3\Big((\nabla_0\mathrm{E}_2)\overline{\nabla}_0\mathrm{E}_3-(\overline{\nabla}_0\mathrm{E}_2)\nabla_0\mathrm{E}_3\Big)}{2\tau_2^2}+2\zeta_3\mathrm{E}_2 \tag{8.108a}$$

$$(\Delta-16)\mathrm{B}'_{2,3} = -14\mathrm{B}_{2,3}+\frac{105}{2}\mathrm{E}_{2,3}+21\mathrm{E}_2\mathrm{E}_3+7\zeta_3\mathrm{E}_2-\frac{21}{40}\zeta_5\,, \tag{8.108b}$$

as can be shown by combining their $\beta^{\mathrm{sv}}$ representations in (8.101) with the differential equations (8.39) obeyed by the $\beta^{\mathrm{sv}}$.[13] This system can be diagonalised to

$$(\Delta-12)(-\mathrm{B}_{2,3}+\mathrm{B}'_{2,3}) = \frac{63}{2}\mathrm{E}_{2,3}+21\mathrm{E}_2\mathrm{E}_3+5\zeta_3\mathrm{E}_2-\frac{21}{40}\zeta_5 - \frac{3\Big((\nabla_0\mathrm{E}_2)\overline{\nabla}_0\mathrm{E}_3-(\overline{\nabla}_0\mathrm{E}_2)\nabla_0\mathrm{E}_3\Big)}{2\tau_2^2} \tag{8.109a}$$

$$(\Delta-2)(-7\mathrm{B}_{2,3}+2\mathrm{B}'_{2,3}) = -42\mathrm{E}_{2,3}+42\mathrm{E}_2\mathrm{E}_3-\frac{21}{20}\zeta_5 - \frac{21\Big((\nabla_0\mathrm{E}_2)\overline{\nabla}_0\mathrm{E}_3-(\overline{\nabla}_0\mathrm{E}_2)\nabla_0\mathrm{E}_3\Big)}{2\tau_2^2}\,. \tag{8.109b}$$

It is rewarding to rewrite these Laplace equations in terms of cusp forms, i.e. eliminate $\mathrm{B}'_{2,3}$ in favor of the three-column cusp form $\mathcal{A}_{1,2;5}$ in the normalization conventions of (8.97):

$$(\Delta-12)\mathcal{A}_{1,2;5} = \frac{(\nabla_0\mathrm{E}_3)\overline{\nabla}_0\mathrm{E}_2-(\overline{\nabla}_0\mathrm{E}_3)\nabla_0\mathrm{E}_2}{\tau_2^2}\,, \tag{8.110a}$$

$$(\Delta-2)\mathrm{B}_{2,3} = 6\mathcal{A}_{1,2;5} - \frac{3\big((\nabla_0\mathrm{E}_3)\overline{\nabla}_0\mathrm{E}_2-(\overline{\nabla}_0\mathrm{E}_3)\nabla_0\mathrm{E}_2\big)}{2\tau_2^2}\,. \tag{8.110b}$$

Note that (8.110a) is a special case of the Laplace equation among two-loop MGFs studied in [185]. The system (8.110) can be diagonalised through the following linear combination of cusp forms

$$(\Delta-2)\Big(\mathrm{B}_{2,3}-\frac{3}{5}\mathcal{A}_{1,2;5}\Big) = -\frac{21\big((\nabla_0\mathrm{E}_3)\overline{\nabla}_0\mathrm{E}_2-(\overline{\nabla}_0\mathrm{E}_3)\nabla_0\mathrm{E}_2\big)}{10\tau_2^2}\,. \tag{8.111}$$

Even though they diagonalise the Laplacian, $\mathcal{A}_{1,2;5}$ and $\mathrm{B}_{2,3}-\frac{3}{5}\mathcal{A}_{1,2;5}$ have not been chosen as basis elements in table 5.4 since their Cauchy–Riemann derivatives yield holomorphic Eisenstein series in earlier steps than $\mathrm{B}_{2,3}$ and $\mathrm{B}'_{2,3}$, see (5.216).

## 8.5 PROPERTIES OF THE $\beta^{\mathrm{sv}}$ AND THEIR GENERATING SERIES $Y^{\tau}_{\vec{\eta}}$

In this section, we study the central objects $\beta^{\mathrm{sv}}$ and $Y^{\tau}_{\vec{\eta}}$ in more detail. Based on the modular properties of their generating series $Y^{\tau}_{\vec{\eta}}$, we

13 To obtain these Laplace equations, one first expresses $\overline{\nabla}_0\mathrm{B}_{2,3}$ and $\overline{\nabla}_0\mathrm{B}'_{2,3}$ through a combination of $\beta^{\mathrm{sv}}$ as in (8.103) and then acts with $\nabla_0$. The resulting expression is then converted back into MGFs by using the inverse relations shown in e.g. (8.69). The same result can also be obtained by acting with the derivatives on the lattice sum representations of $\mathrm{B}_{2,3}$ and $\mathrm{B}'_{2,3}$ and decomposing the result into the basis summarized in Table 5.4.

will determine the SL$(2,\mathbb{Z})$ transformations of the $\beta^{\rm sv}$ and assign a modular weight modulo corrections by $\beta^{\rm sv}$ of lower depth. This will be used to infer the counting of independent MGFs at various modular weights from the entries $j_i, k_i$ of the $\beta^{\rm sv}\big[\begin{smallmatrix} j_1 & j_2 & \dots & j_\ell \\ k_1 & k_2 & \dots & k_\ell \end{smallmatrix};\tau\big]$ that occur in the expansion of $Y^\tau_{\vec\eta}$. Finally, based on the transcendental weights of the $\beta^{\rm sv}$ and the accompanying combinations of $y$ and MZVs, we prove that the $\alpha'$-expansion of $Y^\tau_{\vec\eta}$ is uniformly transcendental if the initial values $\widehat{Y}^{i\infty}_{\vec\eta}$ are.

### 8.5.1 *Modular properties*

We first explore the modular properties of the $\beta^{\rm sv}$ that can be written in more compact form than those of the $\mathcal{E}^{\rm sv}$. The modular T- and S-transformation of the $\beta^{\rm sv}$ will be inferred from their appearance (8.36) in the generating function $Y^\tau_{\vec\eta}$. The torus-integral representation (8.17) of $Y^\tau_{\vec\eta}$ and the modular properties of its ingredients imply the SL$(2,\mathbb{Z})$ transformation

$$Y^{\frac{\alpha\tau+\beta}{\gamma\tau+\delta}}_{\vec\eta}(\sigma|\rho)\Big|^{\eta_j\to(\gamma\bar\tau+\delta)\eta_j}_{\bar\eta_j\to\frac{\bar\eta_j}{(\gamma\bar\tau+\delta)}} = Y^\tau_{\vec\eta}(\sigma|\rho)\,. \tag{8.112}$$

The asymmetric transformation law for the $\bar\eta_j$ and $\eta_j$ stems from the different choices of arguments for $\Omega$ and $\overline{\Omega}$ in the definition (8.17) of the generating series $Y^\tau_{\vec\eta}$. By the series expansion (8.36) of both sides of (8.112) in terms of $\beta^{\rm sv}\big[\begin{smallmatrix} j_1 & j_2 & \dots & j_\ell \\ k_1 & k_2 & \dots & k_\ell \end{smallmatrix};\tau\big]$ and $\beta^{\rm sv}\big[\begin{smallmatrix} j_1 & j_2 & \dots & j_\ell \\ k_1 & k_2 & \dots & k_\ell \end{smallmatrix};\frac{\alpha\tau+\beta}{\gamma\tau+\delta}\big]$, respectively, we can aim to infer the SL$(2,\mathbb{Z})$-properties of $\beta^{\rm sv}$.

#### T- AND S-TRANSFORMATIONS

The T-modular transformation $\tau\to\tau{+}1$ is an invariance of both $Y^\tau_{\vec\eta}$ and the operator $\exp(-R_{\vec\eta}(\epsilon_0)/(4y))$ acting on the initial values $\widehat{Y}^{i\infty}_{\vec\eta}$ in (8.36). Hence, the T-invariance of the closed-string integrals can be transferred to the $\beta^{\rm sv}$,

$$\beta^{\rm sv}\big[\begin{smallmatrix} j_1 & j_2 & \dots & j_\ell \\ k_1 & k_2 & \dots & k_\ell \end{smallmatrix};\tau{+}1\big] = \beta^{\rm sv}\big[\begin{smallmatrix} j_1 & j_2 & \dots & j_\ell \\ k_1 & k_2 & \dots & k_\ell \end{smallmatrix};\tau\big]\,, \tag{8.113}$$

as is also evident from the explicit low-depth examples worked out in the previous sections.

Under an S-modular transformation $\tau\to-1/\tau$, by contrast, we also have to take into account the (asymmetric) transformation of the $\eta_j,\bar\eta_j$ and that the imaginary part $\tau_2=y/\pi$ appears explicitly in the operator $\exp(-\frac{R_{\vec\eta}(\epsilon_0)}{4y})\widehat{Y}^{i\infty}_{\vec\eta}$ in (8.36). Hence, the S-modular transformations of the $\beta^{\rm sv}$ can be obtained by inserting

$$Y^{-1/\tau}_{\vec\eta}\Big|^{\eta_j\to\bar\tau\eta_j}_{\bar\eta_j\to\bar\eta_j/\bar\tau} = \sum_{\ell=0}^{\infty}\sum_{\substack{k_1,\ldots,k_\ell\\=4,6,8,\ldots}}\sum_{j_1=0}^{k_1-2}\sum_{j_2=0}^{k_2-2}\cdots\sum_{j_\ell=0}^{k_\ell-2}\left(\prod_{i=1}^{\ell}\frac{(-1)^{j_i}(k_i-1)}{(k_i-j_i-2)!}\right)$$
$$\times\beta^{\rm sv}\big[\begin{smallmatrix}j_1 & j_2 & \ldots & j_\ell\\ k_1 & k_2 & \ldots & k_\ell\end{smallmatrix};-\tfrac{1}{\tau}\big] \tag{8.114}$$
$$\times R_{\vec\eta}\big(\mathrm{ad}_{\epsilon_0}^{k_\ell-j_\ell-2}(\epsilon_{k_\ell})\ldots\mathrm{ad}_{\epsilon_0}^{k_2-j_2-2}(\epsilon_{k_2})\,\mathrm{ad}_{\epsilon_0}^{k_1-j_1-2}(\epsilon_{k_1})\big)$$
$$\times\exp\left(-|\tau|^2\frac{R_{\vec\eta}(\epsilon_0)}{4y}\right)\widehat{Y}^{i\infty}_{\vec\eta}\Big|^{\eta_j\to\bar\tau\eta_j}_{\bar\eta_j\to\bar\eta_j/\bar\tau}$$

into the left-hand side of (8.112), where the substitution on the $\eta$ variables applies to all occurrences on the right-hand side of (8.114).

Once a given instance of $\beta^{\rm sv}$ has been expressed in terms of MGFs, its S-modular properties can alternatively be inferred from the well-known transformation laws of the MGFs. Both approaches lead to the following exemplary transformations of the $\beta^{\rm sv}$:

$$\beta^{\rm sv}\big[\begin{smallmatrix}0\\4\end{smallmatrix};-\tfrac{1}{\tau}\big] = \bar\tau^2\Big\{\beta^{\rm sv}\big[\begin{smallmatrix}0\\4\end{smallmatrix};\tau\big] + \frac{\zeta_3}{24y^2}(\tau^2-1)\Big\}\,, \tag{8.115a}$$

$$\beta^{\rm sv}\big[\begin{smallmatrix}1\\4\end{smallmatrix};-\tfrac{1}{\tau}\big] = \beta^{\rm sv}\big[\begin{smallmatrix}1\\4\end{smallmatrix};\tau\big] + \frac{\zeta_3(|\tau|^2-1)}{6y}\,, \tag{8.115b}$$

$$\beta^{\rm sv}\big[\begin{smallmatrix}2\\4\end{smallmatrix};-\tfrac{1}{\tau}\big] = \frac{1}{\bar\tau^2}\Big\{\beta^{\rm sv}\big[\begin{smallmatrix}2\\4\end{smallmatrix};\tau\big] + \frac{2\zeta_3}{3}(\bar\tau^2-1)\Big\}\,, \tag{8.115c}$$

$$\begin{aligned}\beta^{\rm sv}\big[\begin{smallmatrix}2 & 0\\4 & 4\end{smallmatrix};-\tfrac{1}{\tau}\big] &= \beta^{\rm sv}\big[\begin{smallmatrix}2 & 0\\4 & 4\end{smallmatrix};\tau\big] + \frac{2\zeta_3}{3}(\bar\tau^2-1)\beta^{\rm sv}\big[\begin{smallmatrix}0\\4\end{smallmatrix};\tau\big]\\ &\quad+\frac{5\zeta_5(|\tau|^2-1)}{216y} + \frac{\zeta_3^2(1-2\bar\tau^2+|\tau|^4)}{72y^2}\,,\end{aligned} \tag{8.115d}$$

$$\begin{aligned}\beta^{\rm sv}\big[\begin{smallmatrix}1 & 2\\6 & 4\end{smallmatrix};-\tfrac{1}{\tau}\big] &= \beta^{\rm sv}\big[\begin{smallmatrix}1 & 2\\6 & 4\end{smallmatrix};\tau\big] + \frac{\zeta_5(\tau^3\bar\tau-1)}{160y^3}\beta^{\rm sv}\big[\begin{smallmatrix}2\\4\end{smallmatrix};\tau\big] - \frac{\zeta_3^2(|\tau|^2-1)}{2520y}\\ &\quad-\frac{7\zeta_7(|\tau|^4-1)}{3840y^2} + \frac{\zeta_3\zeta_5(|\tau|^6-2\tau^3\bar\tau+1)}{480y^3}\,.\end{aligned} \tag{8.115e}$$

Based on the general relations (8.74) and (8.75) to non-holomorphic Eisenstein series, modular S-transformations at depth one can be given in closed form

$$\beta^{\rm sv}\big[\begin{smallmatrix}j\\k\end{smallmatrix};-\tfrac{1}{\tau}\big] = \bar\tau^{k-2-2j}\beta^{\rm sv}\big[\begin{smallmatrix}j\\k\end{smallmatrix};\tau\big] - \frac{2\zeta_{k-1}\bar\tau^{k-2-2j}}{(k-1)(4y)^{k-2-j}} + \frac{2\zeta_{k-1}|\tau|^{2(k-2-j)}}{(k-1)(4y)^{k-2-j}}\,, \tag{8.116}$$

and their analogues at depth two and $k_1+k_2 \leq 10$ can be found in Appendix F of [V].

One important immediate consequence of (8.114) is that the maximal-depth term of any S-modular transformation is

$$\begin{aligned}&\beta^{\rm sv}\big[\begin{smallmatrix}j_1 & j_2 & \ldots & j_\ell\\ k_1 & k_2 & \ldots & k_\ell\end{smallmatrix};-\tfrac{1}{\tau}\big] \\ &= \bar\tau^{-2\ell-2(j_1+j_2+\ldots+j_\ell)+k_1+k_2+\ldots+k_\ell}\beta^{\rm sv}\big[\begin{smallmatrix}j_1 & j_2 & \ldots & j_\ell\\ k_1 & k_2 & \ldots & k_\ell\end{smallmatrix};\tau\big]\ \text{mod depth} \leq \ell-1\,,\end{aligned} \tag{8.117}$$

where the terms of subleading depth are illustrated by the examples in (8.115). This follows from taking the terms without MZVs in the initial values which determine the maximal-depth contributions and whose two- and three-point instances $\widehat{Y}^{i\infty}_{\eta} \rightarrow \frac{1}{\eta\bar\eta} - \frac{2\pi i}{s_{12}}$ and (8.47) are invariant under $\eta_j \rightarrow \bar\tau \eta_j$ and $\bar\eta_j \rightarrow \bar\eta_j/\bar\tau$. These terms in the initial values are annihilated by $R_{\vec\eta}(\epsilon_0)$ and therefore unaffected by its exponential. Hence, for the analysis of maximal-depth terms, it is sufficient to consider the rescaling of the $\eta_j, \bar\eta_j$ in the operators $R_{\vec\eta}(\mathrm{ad}^{k_i-j_i-2}_{\epsilon_0}(\epsilon_{k_i}))$ in (8.114) that have finite adjoint powers of $\epsilon_0$. Referring back to Section 8.1.2, we see that $R_{\vec\eta}(\epsilon_k) \sim s_{ij}\eta_j^{k-2}$ picks up a factor of $\bar\tau^{k-2}$ under the transformation $(\eta_j, \bar\eta_k) \rightarrow (\bar\tau\eta_j, \bar\eta_j/\bar\tau)$ of (8.114). In particular, since $R_{\vec\eta}(\epsilon_0)$ picks up a factor of $\bar\tau^{-2}$, the operators $R_{\vec\eta}(\mathrm{ad}^{k_i-j_i-2}_{\epsilon_0}(\epsilon_{k_i}))$ in (8.114) transform by $\bar\tau^{2+2j_i-k_i}$.

Demanding the maximal-depth terms in the S-transformation of the $\beta^{\rm sv}$ to cancel all of these factors or $\bar\tau^{2+2j_i-k_i}$ leads to (8.117). The argument is based on the modular invariance of the terms in $\widehat{Y}^{i\infty}_{\vec\eta}$ without MZVs which amounts to invariance under $\eta_j \rightarrow \bar\tau\eta_j$ and $\bar\eta_j \rightarrow \bar\eta_j/\bar\tau$. This is manifest in the two- and three-point examples and we present a conjecture for the MZV-free part of $\widehat{Y}^{i\infty}_{\vec\eta}$ for $n$=4 in Appendix D of [V]. The modular transformation (8.117) is thus firmly established for the combinations of $\beta^{\rm sv}$ that occur in the $Y^{\tau}_{\vec\eta}$-series at ($n$≤3) points. Since the counting of independent MGFs in the next subsection will rely on (8.117), we have checked that all the $\beta^{\rm sv}$ entering the weights under consideration there and admitted by the derivation algebra do occur in the three-point $Y^{\tau}_{\vec\eta}$. Our counting of MGFs in this work is therefore not tied to conjectural properties of ($n$≥ 4)-point initial values.

As a consequence of (8.117), even though the $\beta^{\rm sv}$ are not genuine modular forms, they can be assigned leading modular weights given by

$$\beta^{\rm sv}\big[\begin{smallmatrix} j_1 & j_2 & \dots & j_\ell \\ k_1 & k_2 & \dots & k_\ell \end{smallmatrix};\tau\big] \quad \leftrightarrow \quad \text{'modular weight'} \quad \Big(0, -2\ell + \sum_{i=1}^{\ell}(k_i - 2j_i)\Big)$$
$$\text{mod depth } \leq \ell - 1\,, \tag{8.118}$$

and these will be the modular weights of MGFs associated with the given $\beta^{\rm sv}$ as their leading-depth contributions. In order to compensate for the lower-depth corrections to the transformation (8.117) and attain a genuine modular form, expressions like (8.71d), (8.73) for MGFs comprise a tail of $\beta^{\rm sv}$ of lower depth. Note that there are only non-holomorphic weights just as for the component integrals in (8.20) as the generating function $Y^{\tau}_{\vec\eta}$ was rescaled by $\tau_2$ to absorb all holomorphic modular weights.

#### A CAVEAT FROM THE DERIVATION-ALGEBRA RELATIONS

An important qualification of the above arguments is that the derivation-algebra relations such as (8.12) imply that the generating series $Y^{\tau}_{\vec\eta}$ will not contain each possible $\beta^{\rm sv}$ with $j_i \leq k_i-2$ individually but certain combinations always appear together. The first instance of this implied by (8.12) occurs at $\ell = 2$, $k_1+k_2 = 14$ and was spelled out in (8.42). Therefore, even though $Y^{\tau}_{\vec\eta}$ has a perfectly well-defined modular transformation given by (8.112), this does not uniquely fix the modular behavior of all the individual $\beta^{\rm sv}$. Instead, from weight $\sum_i k_i \geq 14$ onward, only the specific combinations of $\beta^{\rm sv}$ realized in the $\alpha'$-expansion of $Y^{\tau}_{\vec\eta}$ (see for instance (8.42)) have to obey the modular properties (8.113) and (8.117). In principle, there is the freedom for the individual $\beta^{\rm sv}$ in these combinations to depart from the above T- and S-transformations, as long as these departures cancel from the $Y^{\tau}_{\vec\eta}$.

Fortunately, this ambiguity does not affect the closed-string integrals or the MGFs in its $\alpha'$-expansion. For the combinations of $\beta^{\rm sv}$ that drop out from $Y^{\tau}_{\vec\eta}$ (and therefore all component integrals) by derivation-algebra relations, we do not need or give an independent definition in this work.[14] Hence, (8.117) can be used as an effective modular transformation that holds for all combinations of $\beta^{\rm sv}$ relevant to this work. When studying the implications on MGFs in the next section, the dropouts of $\beta^{\rm sv}$ at given $\sum_i j_i$ and $\sum_i k_i$ will be taken into account, so the counting of MGFs can be safely based on (8.117) and the relations in the derivation algebra.

### 8.5.2 *Counting of modular graph forms*

The modular properties (8.117) of the $\beta^{\rm sv}$ can be used to count the number of independent MGFs of a given weight. This will lend further support to our basis of MGFs in Table 5.4. The modular weights $(a, b)$ of general lattice sums $C_\Gamma(\tau)$ (cf. (3.123)) are related to the entries of the highest-depth terms $\beta^{\rm sv}\big[\begin{smallmatrix} j_1 & j_2 & \dots & j_\ell \\ k_1 & k_2 & \dots & k_\ell \end{smallmatrix}\big]$ in their integral representation via

$$a+b = \sum_{i=1}^{\ell} k_i \, , \qquad a-b = 2\ell + \sum_{i=1}^{\ell} (2j_i - k_i) \, . \tag{8.119}$$

Note that our convention of modular weights is implied by the lattice-sum conventions (3.123) and differs by the factor of $(\frac{\tau_2}{\pi})^{\frac{1}{2}(a+b)}$ from [16, 40, 185].

While the second correspondence involving $a-b$ is simply a consequence of (8.118), the first one $a+b = \sum_{i=1}^{\ell} k_i$ requires further justification since $C\big[\begin{smallmatrix} a & \dots \\ b & \dots \end{smallmatrix}\big]$ and $\tau_2 C\big[\begin{smallmatrix} a+1 & \dots \\ b+1 & \dots \end{smallmatrix}\big]$ have the same total modular weight.

14 For combinations of $\beta^{\rm sv}$ that drop out from the $\alpha'$-expansion (8.36), we cannot determine the antiholomorphic integration constants from the reality properties (8.24) of component integrals either.

It can be understood by comparing the integral-representation (8.17) of $Y^{\tau}_{\vec\eta}$ with its $\alpha'$-expansion (8.36) in terms of $\beta^{\rm sv}$.

The integrals can be performed order by order in $\alpha'$ and $\eta_j, \bar\eta_j$, where the lattice-sum representations of $G(z,\tau)$, $f^{(k)}(z,\tau)$ and its complex conjugate yield MGFs according to (3.120). The respective contributions to the expansion variables and the modular weights of the lattice sums are

$$\begin{aligned} G(z,\tau) \quad &\leftrightarrow \quad s_{ij} \ \&\ \text{modular weights } (1,1) \\ f^{(k)}(z,\tau) \quad &\leftrightarrow \quad (\eta_j)^k \ \&\ \text{modular weights } (k,0) \\ \overline{f^{(k)}(z,\tau)} \quad &\leftrightarrow \quad (\bar\eta_j)^k \ \&\ \text{modular weights } (0,k)\,. \end{aligned} \tag{8.120}$$

We are disregarding powers of $\tau_2$ and overall prefactors $\sim (\eta_j\bar\eta_j)^{1-n}$ of the $Y^{\tau}_{\vec\eta}$, i.e. the modular weight $(1,1)$ of the Green function refers to its contributions to the lattice sums.

In the $\alpha'$-expansion (8.36), in turn, the correlation between powers of $s_{ij}, \eta_j, \bar\eta_j$ and the entries of $\beta^{\rm sv}$ is governed by the derivations. Their homogeneity degrees are $R_{\vec\eta}(\epsilon_0) \sim s_{ij}/\eta_j^2 + \bar\eta_j/\eta_j$ and $R_{\vec\eta}(\epsilon_k) \sim s_{ij}\eta_j^{k-2}$, which correspond to modular weights $R_{\vec\eta}(\epsilon_0) \leftrightarrow (-1,1)$ and $R_{\vec\eta}(\epsilon_k) \leftrightarrow (k{-}1,1)$ from the lattice-sum viewpoint (8.120). Hence, the $(s_{ij}, \eta_j, \bar\eta_j)$-counting of any operator $R_{\vec\eta}(\mathrm{ad}^j_{\epsilon_0}\, \epsilon_k)$ is the same as having an extra $a{+}b = k$ in lattice sums, regardless of the power $j$ of $\mathrm{ad}_{\epsilon_0}$. This explains why $a{+}b$ has to grow with $\sum_{i=1}^{\ell} k_i$.

Finally, the absence of a $k_i$-independent offset $a{+}b - \sum_{i=1}^{\ell} k_i$ can be checked by comparing the overall powers of $(s_{ij}, \eta_j, \bar\eta_j)$ in the initial value $\widehat{Y}^{i\infty}_{\vec\eta}$ and the integral representation of $Y^{\tau}_{\vec\eta}$. This is most conveniently done by noting the low-energy limit $Y^{\tau}_{(0,\dots,0|0,\dots,0)}(\sigma|\rho) = 1 + \mathcal{O}(\alpha'^2)$ of the simplest component integral at the leading order $\sim (\eta_j\bar\eta_j)^{1-n}$.

On these grounds, we will perform a counting of independent MGFs on the basis of (8.119) in the rest of this section. Our counting only refers to MGFs that do not evaluate to MZVs or products involving MZVs or holomorphic Eisenstein series. We explain our methods in most detail for modular invariant objects, where we also distinguish between real and imaginary invariants, but these methods also cover weights $(a,b)$ with $a \neq b$ that we list in Table 8.1. For all the values of $(a,b)$ where we perform the counting below we have verified explicitly that the relevant action of the operators $R_{\vec\eta}(\epsilon_k)$ on the MZV-free part of $\widehat{Y}^{i\infty}_{\vec\eta}$ at three points does not produce accidental linear dependences. Hence, (8.118) is firmly established in these cases, and the counting is accurate.

### REVIEWING WEIGHT $a+b \leq 8$

Up to total weight $\sum_i k_i < 8$ the only possible basis elements stem from $\beta^{\rm sv}\big[\begin{smallmatrix} j \\ k \end{smallmatrix}\big]$ of depth one. At fixed $k$, each choice of $0 \leq j \leq k{-}2$ leads to a different modular weight according to (8.118). This is in agreement with Table 5.4 featuring only a single basis element for all total weights $a{+}b < 8$. For instance, modular invariants are obtained for $\beta^{\mathrm{sv}}\big[\begin{smallmatrix} j \\ k \end{smallmatrix}\big]$ whenever $j = (k-2)/2$, and they are related to the $\mathrm{E}_{k/2}$ shown in the $(\frac{k}{2}, \frac{k}{2})$ rows of Table 5.4, see also the explicit formula (8.72).

Starting from lattice sums of total weight $a + b = 8$, there can also be invariant combinations of depth-two $\beta^{\mathrm{sv}}$. The condition for modular invariance implied by (8.117) becomes

$$\beta^{\mathrm{sv}}\big[\begin{smallmatrix} j_1 & j_2 \\ 4 & 4 \end{smallmatrix}\big] \quad \text{weight } (0,0) \quad \Leftrightarrow \quad 2 = j_1 + j_2\,, \quad 0 \leq j_1, j_2 \leq 2\,, \tag{8.121}$$

and there are three solutions to this condition given by $(j_1, j_2) \in \{(0,2), (1,1), (2,0)\}$, leading to three additional modular invariants of total weight 8 besides $\mathrm{E}_4$. Two linear combinations of such $\beta^{\mathrm{sv}}\big[\begin{smallmatrix} j_1 & j_2 \\ 4 & 4 \end{smallmatrix}\big]$ with $j_1{+}j_2{=}2$ can be realized by the shuffles $\beta^{\mathrm{sv}}\big[\begin{smallmatrix} 1 \\ 4 \end{smallmatrix}\big]^2$ and $\beta^{\mathrm{sv}}\big[\begin{smallmatrix} 0 \\ 4 \end{smallmatrix}\big]\beta^{\mathrm{sv}}\big[\begin{smallmatrix} 2 \\ 4 \end{smallmatrix}\big]$ which correspond to $\mathrm{E}_2^2$ and $(\nabla_0 \mathrm{E}_2)\overline{\nabla}_0 \mathrm{E}_2$ by (8.69a).[15] Hence, there is a single shuffle-irreducible modular invariant at depth two which can be chosen to be $\mathrm{E}_{2,2}$, expressed through $\beta^{\mathrm{sv}}$ in (8.71c). Together with $\mathrm{E}_4 \leftrightarrow \beta^{\mathrm{sv}}\big[\begin{smallmatrix} 3 \\ 8 \end{smallmatrix}\big]$ at depth one, this reasoning agrees with the total of four entries at weight $(4,4)$ in Table 5.4.

The same counting strategy can be applied at non-zero modular weight. Let us consider the example of $(a,b) = (5,3)$ in Table 5.4 which translates into modular weight $(0,-2)$ after multiplication by $\tau_2^5$. The relevant $\beta^{\mathrm{sv}}\big[\begin{smallmatrix} j_1 & j_2 \\ 4 & 4 \end{smallmatrix}\big]$ at depth two with antiholomorphic weight $-2$ have $j_1{+}j_2 = 3$ by (8.118) and this leaves the two options $(j_1, j_2) \in \{(1,2), (2,1)\}$. One of them is the shuffle $\mathrm{E}_2 \nabla_0 \mathrm{E}_2$, and the irreducible representative is $\nabla_0 \mathrm{E}_{2,2}$, see (8.71c). The connection with the irreducible modular invariant $\mathrm{E}_{2,2}$ can be anticipated by comparing the differential equation (8.41b) of the $\beta^{\mathrm{sv}}$ with the equations satisfied by the MGFs.

In general, the appearance of holomorphic Eisenstein series in Cauchy–Riemann derivatives or relations to shuffles as in $\nabla_0^2 \mathrm{E}_{2,2} = -\frac{1}{2}(\nabla_0 \mathrm{E}_2)^2$ implies that the number of basis MGFs with weights $(a{+}k, a{-}k)$ decreases with $|k|$. An overview of the MGFs and irreducible representatives at $a{+}b \leq 14$ can be found in Table 8.1 below.

### REVIEWING WEIGHT $a + b = 10$

Continuing to total weight 10, there are now additional possibilities at depth two coming from $(k_1, k_2) = (4,6)$ or $(6,4)$. The condition for modular invariant $\beta^{\mathrm{sv}}\big[\begin{smallmatrix} j_1 & j_2 \\ 4 & 6 \end{smallmatrix}\big]$ and $\beta^{\mathrm{sv}}\big[\begin{smallmatrix} j_2 & j_1 \\ 6 & 4 \end{smallmatrix}\big]$ becomes $j_1{+}j_2 = 3$ (with $0{\leq}j_1{\leq}2$ and $0{\leq}j_2{\leq}4$). Both cases lead to three solutions each, and thus there is a total of six modular invariants contributing to the lattice sums of weights $(a,b) = (5,5)$ that can be expressed through depth-two $\beta^{\mathrm{sv}}$. Together with the single contribution $\mathrm{E}_5 \leftrightarrow \beta^{\mathrm{sv}}\big[\begin{smallmatrix} 4 \\ 10 \end{smallmatrix}\big]$ from depth one,

15 For ease of notation, we will suppress here and in the following the overall factors of $\tau_2$. They are always implicit and understood to be such that the holomorphic modular weight vanishes, cf. (8.118). Therefore, $\tau_2^{-2}(\nabla_0 \mathrm{E}_2)\overline{\nabla}_0 \mathrm{E}_2$ will be just written as $(\nabla_0 \mathrm{E}_2)\overline{\nabla}_0 \mathrm{E}_2$.

we find seven modular invariant combinations of $\beta^{\mathrm{sv}}$ which matches the number of basis elements in the $(5,5)$ sector in Table 5.4.

Three combinations of the modular invariant $\beta^{\mathrm{sv}}\big[\begin{smallmatrix} j_1 & j_2 \\ 4 & 6 \end{smallmatrix}\big]$ and $\beta^{\mathrm{sv}}\big[\begin{smallmatrix} j_2 & j_1 \\ 6 & 4 \end{smallmatrix}\big]$ can be realized as a shuffle of $\beta^{\mathrm{sv}}\big[\begin{smallmatrix} j \\ 4 \end{smallmatrix}\big]\beta^{\mathrm{sv}}\big[\begin{smallmatrix} 3-j \\ 6 \end{smallmatrix}\big]$ with $j = 0,1,2$. This translates into modular invariant products $E_2 E_3$, $(\nabla_0 E_2)\overline{\nabla}_0 E_3$, $(\overline{\nabla}_0 E_2)\nabla_0 E_3$ and leaves three irreducible modular invariants at depth two that can be chosen to be $E_{2,3}$ and $B_{2,3}, B'_{2,3}$ in Table 5.4, see (8.71d) and (8.101) for their expressions in terms of $\beta^{\mathrm{sv}}$. Alternatively, one can trade $B_{2,3}, B'_{2,3}$ for the imaginary cusp forms $\mathcal{A}\big[\begin{smallmatrix} 0 & 2 & 3 \\ 3 & 0 & 2 \end{smallmatrix}\big], \mathcal{A}\big[\begin{smallmatrix} 0 & 1 & 2 & 2 \\ 1 & 1 & 0 & 3 \end{smallmatrix}\big]$ and organize the modular invariants according to their reality properties: three real basis elements $E_2 E_3$, $\mathrm{Re}[(\nabla_0 E_2)\overline{\nabla}_0 E_3]$, $E_{2,3}$ (one of them irreducible) and three imaginary basis elements $\mathrm{Im}[(\nabla_0 E_2)\overline{\nabla}_0 E_3]$, $\mathcal{A}\big[\begin{smallmatrix} 0 & 2 & 3 \\ 3 & 0 & 2 \end{smallmatrix}\big]$, $\mathcal{A}\big[\begin{smallmatrix} 0 & 1 & 2 & 2 \\ 1 & 1 & 0 & 3 \end{smallmatrix}\big]$ (two of them irreducible).

The counting of real and imaginary forms can also be obtained based on the reality properties (8.83) and (8.92) of the $\beta^{\mathrm{sv}}$: in the modular-invariant case, complex conjugation is only an operation on the labels of the $\beta^{\mathrm{sv}}$ at leading order in depth. Therefore one has to form combinations of the $\beta^{\mathrm{sv}}$ that are mapped to themselves or minus themselves under complex conjugation. For instance, since

$$\overline{\beta^{\mathrm{sv}}\big[\begin{smallmatrix} 2 & 1 \\ 4 & 6 \end{smallmatrix}\big]} = \beta^{\mathrm{sv}}\big[\begin{smallmatrix} 3 & 0 \\ 6 & 4 \end{smallmatrix}\big] \quad \text{modulo lower depth,} \tag{8.122}$$

the combinations $\beta^{\mathrm{sv}}\big[\begin{smallmatrix} 2 & 1 \\ 4 & 6 \end{smallmatrix}\big] \pm \beta^{\mathrm{sv}}\big[\begin{smallmatrix} 3 & 0 \\ 6 & 4 \end{smallmatrix}\big]$ give real and imaginary MGFs modulo lower depth, respectively. See e.g. the real $E_{2,3}$ in (8.71d) and the imaginary cusp form $B_{2,3}$ in (8.101a).

The analogous counting of MGFs with $a+b = 10$ and $a \neq b$ based on the $\beta^{\mathrm{sv}}$ can be found in Table 8.1 below.

### PREDICTIONS FOR WEIGHT $a + b = 12$

For lattice sums of weight $a + b = 12$, a basis of 19 modular invariants can be anticipated from $\beta^{\mathrm{sv}}$ at depth $\ell = 1,2,3$:

1. a single depth-one invariant $E_6 \leftrightarrow \beta^{\mathrm{sv}}\big[\begin{smallmatrix} 5 \\ 12 \end{smallmatrix}\big]$
2. 5 depth-two invariants $\beta^{\mathrm{sv}}\big[\begin{smallmatrix} j_1 & j_2 \\ 6 & 6 \end{smallmatrix}\big]$ with $j_1 + j_2 = 4$ and $0 \leq j_1, j_2 \leq 4$
3. 6 depth-two invariants $\beta^{\mathrm{sv}}\big[\begin{smallmatrix} j_1 & j_2 \\ 4 & 8 \end{smallmatrix}\big]$ & $\beta^{\mathrm{sv}}\big[\begin{smallmatrix} j_2 & j_1 \\ 8 & 4 \end{smallmatrix}\big]$ with $j_1 + j_2 = 4$ and $0 \leq j_1 \leq 2$ & $0 \leq j_2 \leq 6$
4. 7 depth-three invariants $\beta^{\mathrm{sv}}\big[\begin{smallmatrix} j_1 & j_2 & j_3 \\ 4 & 4 & 4 \end{smallmatrix}\big]$ with $j_1 + j_2 + j_3 = 3$ and $0 \leq j_i \leq 2$

We will analyze the shuffle- and reality properties separately in each sector 2, 3 and 4 and connect with known irreducible modular graph functions.

Sector 2 contains the shuffles $\beta^{\rm sv}\big[\begin{smallmatrix}2\\6\end{smallmatrix}\big]^2$, $\beta^{\rm sv}\big[\begin{smallmatrix}1\\6\end{smallmatrix}\big]\beta^{\rm sv}\big[\begin{smallmatrix}3\\6\end{smallmatrix}\big]$ and $\beta^{\rm sv}\big[\begin{smallmatrix}0\\6\end{smallmatrix}\big]\beta^{\rm sv}\big[\begin{smallmatrix}4\\6\end{smallmatrix}\big]$ that correspond to $\mathrm{E}_3^2$, $(\nabla_0\mathrm{E}_3)\overline{\nabla}_0\mathrm{E}_3$ and $(\overline{\nabla}_0^2\mathrm{E}_3)\nabla_0^2\mathrm{E}_3$ according to (8.69). This leaves two irreducibles which can be taken to be the quantities

$$\mathrm{E}_{3,3} = 450\beta^{\rm sv}\big[\begin{smallmatrix}4&0\\6&6\end{smallmatrix}\big] - 180\zeta_5\beta^{\rm sv}\big[\begin{smallmatrix}0\\6\end{smallmatrix}\big] + \frac{\zeta_7}{16y} - \frac{7\zeta_9}{64y^3} + \frac{9\zeta_5^2}{64y^4}\,, \tag{8.123}$$

$$\begin{aligned}\mathrm{E}'_{3,3} &= 120(\beta^{\rm sv}\big[\begin{smallmatrix}4&0\\6&6\end{smallmatrix}\big] - \beta^{\rm sv}\big[\begin{smallmatrix}3&1\\6&6\end{smallmatrix}\big]) - 48\zeta_5\beta^{\rm sv}\big[\begin{smallmatrix}0\\6\end{smallmatrix}\big] + \frac{12\zeta_5}{y}\beta^{\rm sv}\big[\begin{smallmatrix}1\\6\end{smallmatrix}\big]\\ &\quad + \frac{3\zeta_7}{160y} - \frac{7\zeta_9}{480y^3}\end{aligned} \tag{8.124}$$

corresponding to the lattice sums given in (4.28). The $\beta^{\rm sv}$ representations have been inferred from the differential equations and the Laurent polynomials of the real MGFs $\mathrm{E}_{3,3}$ and $\mathrm{E}'_{3,3}$.

Sector 3 also admits three shuffles $\beta^{\rm sv}\big[\begin{smallmatrix}0\\4\end{smallmatrix}\big]\beta^{\rm sv}\big[\begin{smallmatrix}4\\8\end{smallmatrix}\big]$, $\beta^{\rm sv}\big[\begin{smallmatrix}1\\4\end{smallmatrix}\big]\beta^{\rm sv}\big[\begin{smallmatrix}3\\8\end{smallmatrix}\big]$ and $\beta^{\rm sv}\big[\begin{smallmatrix}2\\4\end{smallmatrix}\big]\beta^{\rm sv}\big[\begin{smallmatrix}2\\8\end{smallmatrix}\big]$ corresponding to $(\overline{\nabla}_0\mathrm{E}_2)\nabla_0\mathrm{E}_4$, $\mathrm{E}_2\mathrm{E}_4$ and $(\nabla_0\mathrm{E}_2)\overline{\nabla}_0\mathrm{E}_4$, respectively. Two of them are real $\mathrm{E}_2\mathrm{E}_4$, $\mathrm{Re}[(\nabla_0\mathrm{E}_2)\overline{\nabla}_0\mathrm{E}_4]$ whereas a third one $\mathrm{Im}[(\nabla_0\mathrm{E}_2)\overline{\nabla}_0\mathrm{E}_4]$ is imaginary. The remaining three invariants are shuffle irreducible, and one real representative

$$\begin{aligned}\mathrm{E}_{2,4} &= -5670\beta^{\rm sv}\big[\begin{smallmatrix}4&0\\8&4\end{smallmatrix}\big] - 5670\beta^{\rm sv}\big[\begin{smallmatrix}2&2\\4&8\end{smallmatrix}\big] + 3780\zeta_3\beta^{\rm sv}\big[\begin{smallmatrix}2\\8\end{smallmatrix}\big]\\ &\quad + \frac{405\zeta_7}{4y^2}\beta^{\rm sv}\big[\begin{smallmatrix}0\\4\end{smallmatrix}\big] - \frac{9\zeta_7}{80y} + \frac{25\zeta_9}{8y^3} - \frac{135\zeta_3\zeta_7}{32y^4}\end{aligned} \tag{8.125}$$

corresponds to the lattice sum given in (4.28). As will be argued below, the remaining two shuffle irreducibles can be chosen to be imaginary.

Sector 4 admits 2+3 shuffles $\mathrm{E}_2^3$, $\mathrm{E}_2(\nabla_0\mathrm{E}_2)\overline{\nabla}_0\mathrm{E}_2$ and $\mathrm{E}_2\mathrm{E}_{2,2}$, $(\nabla_0\mathrm{E}_2)\overline{\nabla}_0\mathrm{E}_{2,2}$, $(\overline{\nabla}_0\mathrm{E}_2)\nabla_0\mathrm{E}_{2,2}$. Among the leftover two shuffle-irreducibles, one real representative

$$\begin{aligned}\mathrm{E}_{2,2,2} &= -216\beta^{\rm sv}\big[\begin{smallmatrix}2&1&0\\4&4&4\end{smallmatrix}\big] + 144\zeta_3\beta^{\rm sv}\big[\begin{smallmatrix}1&0\\4&4\end{smallmatrix}\big] + 10\zeta_5\beta^{\rm sv}\big[\begin{smallmatrix}0\\4\end{smallmatrix}\big]\\ &\quad - \frac{12\zeta_3^2}{y}\beta^{\rm sv}\big[\begin{smallmatrix}0\\4\end{smallmatrix}\big] + \frac{\zeta_3^2}{30} + \frac{661\zeta_7}{1800y} - \frac{5\zeta_3\zeta_5}{12y^2} + \frac{\zeta_3^3}{6y^3}\end{aligned} \tag{8.126}$$

corresponds to the lattice sum given in (4.28). As will be argued below, the second shuffle irreducible is imaginary.

In order to anticipate the number of real and imaginary irreducible modular invariants, the known types of relations among MGFs have been exhaustively applied to all dihedral and trihedral graph topologies at weights $(a,b)=(6,6)$ as described in Section 5.7.1. The solution to the large equation system identifies 14 real and 5 imaginary independent modular invariants, again excluding MZVs and $\mathrm{G}_k$ from our counting conventions, cf. Table 5.3. Given that modular invariant combinations of the known $\mathrm{E}_{\dots}$ already exhaust the 14 real invariants, the remaining shuffle irreducibles must admit imaginary representatives. This conclusion lends support to extending the reality properties of the $\beta^{\rm sv}$ given in (8.92) beyond $k_1{+}k_2 > 10$, and it is tempting to extrapolate it to arbitrary depth

$$\overline{\beta^{\rm sv}\big[\begin{smallmatrix} j_1 & j_2 & \dots & j_\ell \\ k_1 & k_2 & \dots & k_\ell \end{smallmatrix}\big]} = (4y)^{2\ell+\sum_{i=1}^{\ell}(2j_i-k_i)}\beta^{\rm sv}\big[\begin{smallmatrix} k_\ell-2-j_\ell & \dots & k_2-2-j_2 & k_1-2-j_1 \\ k_\ell & \dots & k_2 & k_1 \end{smallmatrix}\big] \quad \text{mod depth} \leq \ell{-}1\,. \tag{8.127}$$

This conjecture leads to the same counting of imaginary representatives, and the power of $4y$ therein vanishes exactly if the modular weight of $\beta^{\rm sv}$ in (8.118) does.

Hence, the 5 imaginary invariants at $(a,b) = (6,6)$ are $\mathrm{Im}[(\nabla_0 \mathrm{E}_2)\overline{\nabla}_0 \mathrm{E}_4]$, $\mathrm{Im}[(\nabla_0 \mathrm{E}_2)\overline{\nabla}_0 \mathrm{E}_{2,2}]$, two irreducible cusp forms from 3 and one irreducible cusp form from 4. The paper [185] identified two cusp forms at $(a,b) = (6,6)$ among the two-loop graphs on the worldsheet. Accordingly, three out of the five cusp forms in our counting require lattice sums associated with ($L \geq 3$)-loop graphs. Indeed, a detailed analysis of the relations between dihedral and trihedral MGFs shows that $\mathcal{A}\big[\begin{smallmatrix} 0 & 2 & 4 \\ 5 & 0 & 1 \end{smallmatrix}\big], \mathcal{A}\big[\begin{smallmatrix} 0 & 1 & 2 & 3 \\ 2 & 1 & 3 & 0 \end{smallmatrix}\big]$ and $\mathcal{A}\big[\begin{smallmatrix} 0 & 2 & 2 & 2 \\ 3 & 0 & 1 & 2 \end{smallmatrix}\big]$ qualify as a basis of shuffle-irreducible cusp forms at $(a,b) = (6,6)$, and $\mathrm{Im}[(\nabla_0 \mathrm{E}_2)\overline{\nabla}_0 \mathrm{E}_{2,2}]$ also exceeds the two-loop graphs when written in terms of lattice sums, cf. Table 5.3.

In summary, the 19 modular invariant lattice sums of weight $(a,b) = (6,6)$ comprise 11 shuffles (3 from 2, 3 from 3 and 5 from 4) and 8 shuffle irreducibles. The irreducibles admit 5 real representatives known in the literature ($\mathrm{E}_6$ from 1, $\mathrm{E}_{3,3}, \mathrm{E}'_{3,3}$ from 2, $\mathrm{E}_{2,4}$ from 3, $\mathrm{E}_{2,2,2}$ from 4) and 3 imaginary cusp forms (two from 3 and one from 4) generalizing $\mathcal{A}\big[\begin{smallmatrix} 0 & 2 & 3 \\ 3 & 0 & 2 \end{smallmatrix}\big]$, $\mathcal{A}\big[\begin{smallmatrix} 0 & 1 & 2 & 2 \\ 1 & 1 & 0 & 3 \end{smallmatrix}\big]$ described in Section 8.4. This counting agrees exactly with the findings of Section 5.7.

The analogous counting of MGFs with $a{+}b = 12$ and $a \neq b$ can be found in Table 8.1 below.

### WEIGHT $a+b = 14$ AND THE DERIVATION ALGEBRA

By extending the above counting method to weight $a{+}b = 14$, one is naïvely led to 44 modular invariants (26 of them shuffles). If all the $\beta^{\rm sv}$ were realized independently in the expansion (8.36) of $Y^{\tau}_{\vec\eta}$, the total of 44 would arise from the following sectors:

1. a single depth-one invariant $\mathrm{E}_7 \leftrightarrow \beta^{\rm sv}\big[\begin{smallmatrix} 6 \\ 14 \end{smallmatrix}\big]$
2. 6 depth-two invariants $\beta^{\rm sv}\big[\begin{smallmatrix} j_1 & j_2 \\ 4 & 10 \end{smallmatrix}\big]$ & $\beta^{\rm sv}\big[\begin{smallmatrix} j_2 & j_1 \\ 10 & 4 \end{smallmatrix}\big]$ with $j_1{+}j_2 = 5$ and $0{\leq}j_1{\leq}2$ and $0{\leq}j_2{\leq}8$
3. 10 depth-two invariants $\beta^{\rm sv}\big[\begin{smallmatrix} j_1 & j_2 \\ 6 & 8 \end{smallmatrix}\big]$ & $\beta^{\rm sv}\big[\begin{smallmatrix} j_2 & j_1 \\ 8 & 6 \end{smallmatrix}\big]$ with $j_1{+}j_2 = 5$ and $0{\leq}j_1{\leq}4$ and $0{\leq}j_2{\leq}6$
4. 27 depth-three invariants $\beta^{\rm sv}\big[\begin{smallmatrix} j_1 & j_2 & j_3 \\ 6 & 4 & 4 \end{smallmatrix}\big]$ with $j_1 + j_2 + j_3 = 4$ and $0 \leq j_1 \leq 4$ as well as $0 \leq j_2, j_3 \leq 2$ and permutations of $(k_1, k_2, k_3)$

However, weight $\sum_{i=1}^{\ell} k_i = 14$ is the first instance where the derivation algebra exhibits relations beyond the nilpotency properties in (8.11) that we have already used in the derivation of (8.36). The simplest instance was exhibited in (8.42).

More generally, the relation (8.12) implies additional relations under the adjoint $\epsilon_0$ action according to[16]

$$\begin{aligned} 0 &= R_{\vec\eta}\Big[\,\mathrm{ad}^{j}_{\epsilon_0}\Big(\big[\epsilon_{10},\epsilon_4\big]-3\big[\epsilon_8,\epsilon_6\big]\Big)\Big] \qquad (8.128)\\ &= \sum_{r=0}^{j}\binom{j}{r}R_{\vec\eta}\Big(\big[\,\mathrm{ad}^{r}_{\epsilon_0}(\epsilon_{10}),\mathrm{ad}^{j-r}_{\epsilon_0}(\epsilon_4)\big]-3\big[\,\mathrm{ad}^{r}_{\epsilon_0}(\epsilon_8),\mathrm{ad}^{j-r}_{\epsilon_0}(\epsilon_6)\big]\Big)\,, \end{aligned}$$

and similar relations arise at higher weight and depth, see (4.24) and [29, 209, 261]. In passing to the second line, we have rewritten the relation in terms of the quantities $R_{\vec\eta}\big(\,\mathrm{ad}^{j_1}_{\epsilon_0}(\epsilon_{k_1})\,\mathrm{ad}^{j_2}_{\epsilon_0}(\epsilon_{k_2})\big)$ that occur in the expansion (8.36) of $Y^{\tau}_{\vec\eta}$ (setting $j{\le}10$ in (8.128) and using $R_{\vec\eta}(\mathrm{ad}^{k-1}_{\epsilon_0}(\epsilon_k))=0$). As a consequence, the $\beta^{\mathrm{sv}}$ in the sectors 2 and 3 cannot all appear independently in the generating series $Y^{\tau}_{\vec\eta}$ of MGFs.

More specifically, (8.128) implies exactly one dropout among the $\beta^{\mathrm{sv}}\big[\begin{smallmatrix} j_1 & j_2\\ k_1 & k_2\end{smallmatrix}\big]$ with $k_1{+}k_2=14$ for each value $j=j_1{+}j_2$ with $0\le j\le 10$. At $j=5$, this reduces the total number of independent modular invariants with weight $(a,b)=(7,7)$ by one, leading to 43 rather than 44. The commutators in (8.128) imply that this reduction affects the shuffle-irreducible MGFs, and the dropout at $(a,b)=(7,7)$ concerns an imaginary modular invariant when the combinations of $\beta^{\mathrm{sv}}$ are organized into real and imaginary ones. Further details and the analogous counting of forms with $w\neq\bar w$ can be found in Table 8.1. We have checked that all $\beta^{\mathrm{sv}}$ noted in the MGF-column of the table occur in the $Y^{\tau}_{\vec\eta}$-series at three points (without accidental dropouts) and are therefore known to satisfy (8.117) without relying on the conjectural four-point data from Appendix D of [V]. The conjectural relation (8.127) implies that the basis at $(a,b)=(7,7)$ can be spanned by 24 real and 19 imaginary invariants.[17] It would be interesting to study at the level of the Laurent polynomials if our basis of real MGFs at this weight contains a cusp form.

16 We have checked that more general relations of the form

$$\big(R_{\vec\eta}(\epsilon_0)\big)^{j_1}\Big(\big[R_{\vec\eta}(\epsilon_{10}),R_{\vec\eta}(\epsilon_4)\big]-3\big[R_{\vec\eta}(\epsilon_8),R_{\vec\eta}(\epsilon_6)\big]\Big)\big(R_{\vec\eta}(\epsilon_0)\big)^{j_2}=0\,,$$

do not yield any further relations among the operators $R_{\vec\eta}\big(\,\mathrm{ad}^{j_1}_{\epsilon_0}(\epsilon_{k_1})\,\mathrm{ad}^{j_2}_{\epsilon_0}(\epsilon_{k_2})\big)$ in the expansion (8.36) of $Y^{\tau}_{\vec\eta}$.

17 As an immediate consequence of (8.127), we have a single real invariant $\mathrm{E}_7$ in sector 1 as well as 15 real and 12 imaginary invariants in sector 4. The sectors 2 and 3 are coupled through the relations (8.128) in the derivation algebra. It follows from (8.127) that the 15 independent instances of $R_{\vec\eta}\big(\,\mathrm{ad}^{j_1}_{\epsilon_0}(\epsilon_{k_1})\,\mathrm{ad}^{j_2}_{\epsilon_0}(\epsilon_{k_2})\big)$ in (8.36) are accompanied by 8 real and 7 imaginary linear combinations of $\beta^{\mathrm{sv}}\big[\begin{smallmatrix} j_1 & j_2\\ 4 & 10\end{smallmatrix}\big]$, $\beta^{\mathrm{sv}}\big[\begin{smallmatrix} j_2 & j_1\\ 10 & 4\end{smallmatrix}\big]$, $\beta^{\mathrm{sv}}\big[\begin{smallmatrix} j_1 & j_2\\ 6 & 8\end{smallmatrix}\big]$, $\beta^{\mathrm{sv}}\big[\begin{smallmatrix} j_2 & j_1\\ 8 & 6\end{smallmatrix}\big]$ at $j_1{+}j_2=5$.

| weight | # $\beta^{\text{sv}}$ | # MGFs | irred. MGFs | real MGFs | imag. MGFs |
|---|---|---|---|---|---|
| (2,2) | 1 | 1 | 1 | 1 | 0 |
| (3,1) | 1 | 1 | 1 | – | – |
| (3,3) | 1 | 1 | 1 | 1 | 0 |
| (4,2) | 1 | 1 | 1 | – | – |
| (5,1) | 1 | 1 | 1 | – | – |
| (4,4) | 4 | 4 | 2 | 4 | 0 |
| (5,3) | 3 | 3 | 2 | – | – |
| (6,2) | 2 | 2 | 1 | – | – |
| (7,1) | 1 | 1 | 1 | – | – |
| (5,5) | 7 | 7 | 4 | 4 | 3 |
| (6,4) | 7 | 7 | 4 | – | – |
| (7,3) | 5 | 5 | 3 | – | – |
| (8,2) | 3 | 3 | 2 | – | – |
| (9,1) | 1 | 1 | 1 | – | – |
| (6,6) | 19 | 19 | 8 | 14 | 5 |
| (7,5) | 17 | 17 | 8 | – | – |
| (8,4) | 13 | 13 | 6 | – | – |
| (9,3) | 8 | 8 | 4 | – | – |
| (10,2) | 4 | 4 | 2 | – | – |
| (11,1) | 1 | 1 | 1 | – | – |
| (7,7) | 44 | 43 | 17 | 24 | 19 |
| (8,6) | 41 | 40 | 16 | – | – |
| (9,5) | 33 | 32 | 13 | – | – |
| (10,4) | 22 | 21 | 9 | – | – |
| (11,3) | 12 | 11 | 5 | – | – |
| (12,2) | 5 | 4 | 2 | – | – |
| (13,1) | 1 | 1 | 1 | – | – |

Table 8.1: Counting of MGFs up to total weight $a+b=14$ based on the number of $\beta^{\text{sv}}$. The entries list the total number of MGFs (excluding holomorphic Eisenstein series and zeta values), the number of shuffle-irreducible MGFs as well as the number of real and imaginary MGFs in the modular invariant sectors. Up to total weight $a+b\leq 12$, the counting has been confirmed by the independent methods for dihedral and trihedral MGFs detailed in Chapter 5. For $a+b=14$, the derivation algebra imposes the additional constraint (8.12) on the combinations of the $\beta^{\text{sv}}$ that can appear in the generating function $Y^{\tau}_{\vec{\eta}}$, leading to a mismatch of the number of $\beta^{\text{sv}}$ and MGFs.

### WEIGHT $a + b \geq 16$ AND THE DERIVATION ALGEBRA

We have not performed a similarly detailed analysis at higher weight and only offer some general comments. At weight $a + b = 16$, similar dropouts in the naïve count of MGFs via $\beta^{\rm sv}$ arise from the depth-three relation (4.24e), obstructing for instance the independent appearance of all the $\beta^{\rm sv}\left[\begin{smallmatrix} j_1 & j_2 & j_3 \\ 8 & 4 & 4 \end{smallmatrix}\right]$ with $0 \leq j_1 \leq 6$ and $0 \leq j_2, j_3 \leq 2$. In case of modular invariants with $a = b = 8$, this leads to the dropout of a real MGF, leaving in total 108 MGFs, out of which 42 are imaginary cusp forms.

Weight $a + b = 18$ even allows for three sources of dropouts:

- the irreducible depth-two relation (4.24d) involving $(k_i, k_j) \in \{(4, 14), (6, 12), (8, 10)\}$
- left- and right-multiplication of the $(k_1{+}k_2 = 14)$-relation (4.24c) by a single $\epsilon_4$ and arbitrary powers of $\epsilon_0$
- an irreducible depth-four relation first seen in [209] and available for download at [205]

The systematics of relations in the derivation algebra is governed by the counting of holomorphic cusp forms [209]. The propagation of irreducible relations to higher depth and weight by multiplication with additional $\epsilon_k$ has been discussed in detail in [29]. The latter reference is dedicated to classifying relations among elliptic MZVs and counting their irreducible representatives at various lengths and depths. In this way [29] can be viewed as the open-string prototype of the present counting of MGFs.

### DEPTH VERSUS GRAPH DATA

We emphasize that the above counting of MGFs applies to closed-string integrals of arbitrary multiplicity and therefore to arbitrary graph topologies. The reason is that the $R_{\vec{\eta}}(\epsilon_k)$ were assumed to obey no further relations besides those in the derivation algebra, i.e. multiplicity-specific relations such as commutativity of the $R_{\eta}(\epsilon_{k\geq 4})$ at two points were disregarded.

Our bases of MGFs at $a{+}b \leq 12$ were built from dihedral representatives. Hence, the results of this section imply that any MGF at these weights associated with arbitrarily complicated graph topologies can be reduced to dihedral MGFs (possibly with $\mathbb{Q}$-linear combinations of MZVs in their coefficients), extending the explicit calculations for dihedral and trihedral graphs in Chapter 5. It would be interesting to determine the first combination of weights, where the appearance of a trihedral basis MGF is inevitable.

We have not found any general correlation between the loop order of an MGF and the maximum depth of the associated $\beta^{\rm sv}$. On the one hand, one-loop MGFs are still in one-to-one correspondence with $\beta^{\rm sv}$ at depth one by (8.73). On the other hand, a basis of MGFs with $a+b=10$ requires at least one three-loop graph (e.g. $\mathcal{A}\big[\begin{smallmatrix}0&1&2&2\\1&1&0&3\end{smallmatrix}\big]$) while the associated $\beta^{\rm sv}$ cannot exceed depth two. Up to $a+b=12$ all examples satisfy that the loop order of an irreducible $\beta^{\rm sv}$ of depth $\ell$ is at least $\ell$, i.e., the depth of an irreducible $\beta^{\rm sv}$ appears to be a lower bound for the loop order.

### 8.5.3 *Towards uniform transcendentality*

This section is dedicated to the transcendentality properties of the generating series $Y^{\tau}_{\vec{\eta}}$ that become manifest from our results. For a brief general discussion of uniform transcendentality, cf. Section 6.2.3. We will show that the component integrals (8.19) are uniformly transcendental provided that the same is true for the initial values $\widehat{Y}^{i\infty}_{\vec{\eta}}$.[18] In other words, the matrix- and operator-valued series

$$\Lambda^{\tau}_{\vec{\eta}} = \sum_{\ell=0}^{\infty} \sum_{\substack{k_1,k_2,\ldots,k_\ell\\=4,6,8,\ldots}} \sum_{j_1=0}^{k_1-2}\sum_{j_2=0}^{k_2-2}\cdots\sum_{j_\ell=0}^{k_\ell-2}\bigg(\prod_{i=1}^{\ell}\frac{(-1)^{j_i}(k_i-1)}{(k_i-j_i-2)!}\bigg)\mathcal{E}^{\rm sv}\big[\begin{smallmatrix}j_1&j_2&\ldots&j_\ell\\k_1&k_2&\ldots&k_\ell\end{smallmatrix};\tau\big]$$
$$\times\exp\Big(-\frac{R_{\vec{\eta}}(\epsilon_0)}{4y}\Big)R_{\vec{\eta}}\big(\mathrm{ad}_{\epsilon_0}^{k_\ell-j_\ell-2}(\epsilon_{k_\ell})\ldots\mathrm{ad}_{\epsilon_0}^{k_1-j_1-2}(\epsilon_{k_1})\big) \qquad (8.129)$$

relating $Y^{\tau}_{\vec{\eta}}=\Lambda^{\tau}_{\vec{\eta}}\widehat{Y}^{i\infty}_{\vec{\eta}}$ by (8.33) will be demonstrated to enjoy uniform transcendentality. Our reasoning closely follows the lines of Section 7.1 in [38], where the open-string analogues of the $Y^{\tau}_{\vec{\eta}}$ are shown to be uniformly transcendental.

In particular, this proves that the integrals which were conjectured to be uniformly transcendental in Section 6.2.3 are indeed of uniform transcendentality if their Laurent polynomials at the cusp are, since they can be written as a linear combination of the $Y^{\tau}_{(A|B|)}$ component integrals by means of integration-by-parts identities and the Fay identity (5.121). On the other hand, the integrals in Section 6.1.3 which were explicitly seen to be of non-uniform transcendentality have subcycles in their integrands and hence require the use of the coincident Fay identity (5.133) when they are written in terms of the $Y^{\tau}_{(A|B)}$. Since this identity violates uniform transcendentality, so do the corresponding integrals in Section 6.1.3.

#### WEIGHT ASSIGNMENTS AND UNIFORM TRANSCENDENTALITY OF THE GENERATING SERIES

We assign the following transcendental weights to the holomorphic building blocks in the $\alpha'$-expansion of open- and closed-string integrals,

18 It will be the main goal of [257] to express the initial values $\widehat{Y}^{i\infty}_{\vec{\eta}}$ in terms of uniformly transcendental sphere integrals as done in (8.65) and (8.66) for the two-point example.

| quantity | $\zeta_{n_1,n_2,\ldots,n_r}$ | $\pi$ | $\mathcal{E}[\begin{smallmatrix} j_1 & j_2 & \ldots & j_\ell \\ k_1 & k_2 & \ldots & k_\ell \end{smallmatrix}]$ | $\tau$ |
|---|---|---|---|---|
| transcendental weight | $\sum_{j=1}^{r} n_j$ | 1 | $\ell + \sum_{i=1}^{\ell} j_i$ | 0 |

leading to weight 0 for $\nabla_0$ and for instance weight $j_1{+}1$ for $\mathcal{E}[\begin{smallmatrix} j_1 \\ k_1 \end{smallmatrix}]$.

Moreover, complex conjugation is taken to preserve the weight, which leads to weight 0 for $\bar\tau$, weight 1 for $y$ and weight $\ell + \sum_{i=1}^{\ell} j_i$ for $\overline{\mathcal{E}[\begin{smallmatrix} j_1 & j_2 & \ldots & j_\ell \\ k_1 & k_2 & \ldots & k_\ell \end{smallmatrix}]}$. The weights of the holomorphic iterated Eisenstein integrals are inherited from those of eMZVs [29].

In order to infer the weight of the real-analytic $\mathcal{E}^{\rm sv}$ in the $\alpha'$-expansion (8.129), we first note that their building blocks $\mathcal{E}^{\rm sv}_{\rm min}[\begin{smallmatrix} j_1 & j_2 & \ldots & j_\ell \\ k_1 & k_2 & \ldots & k_\ell \end{smallmatrix}]$ involving only holomorphic $\mathcal{E}[\ldots]$ have transcendental weight $\ell + \sum_{i=1}^{\ell} j_i$ as is manifest in their representation (8.52). We will demonstrate in Section 8.5.3 that this propagates to the antiholomorphic integration constants $\overline{f[\begin{smallmatrix} j_1 & j_2 & \ldots & j_\ell \\ k_1 & k_2 & \ldots & k_\ell \end{smallmatrix}]}$ in the decomposition (8.53) of $\mathcal{E}^{\rm sv}[\begin{smallmatrix} j_1 & j_2 & \ldots & j_\ell \\ k_1 & k_2 & \ldots & k_\ell \end{smallmatrix}]$.

With these definitions we will show that the component integrals carry uniform transcendental weight

$$Y^\tau_{(A|B)}(\sigma|\rho) \text{ at order } \alpha'^w \quad \leftrightarrow \quad \text{trans. weight } w + |A|\,. \tag{8.130}$$

In order to give a uniform transcendental weight to the whole generating series $Y^\tau_{\vec\eta}$ we have to assign

$$s_{ij}, \eta_j, \bar\eta_j \quad \leftrightarrow \quad \text{transcendental weight } -1\,. \tag{8.131}$$

With this convention and the inverse factors of $(2\pi i)$ in the definition (8.19) of component integrals, (8.130) is equivalent to having

$$\text{claim:} \quad Y^\tau_{\vec\eta} \quad \leftrightarrow \quad \text{transcendental weight } 2(n{-}1) \tag{8.132}$$

for the generating series at $n$ points. This will be shown under the assumption that the initial data has uniform transcendental weight,

$$\text{assumption:} \quad \widehat{Y}^{i\infty}_{\vec\eta} \quad \leftrightarrow \quad \text{transcendental weight } 2(n{-}1)\,. \tag{8.133}$$

### TRANSCENDENTALITY OF THE SERIES IN $\beta^{\rm sv}$

We start by inspecting the constituents of the series $\Lambda^\tau_{\vec\eta}$ in (8.129). By the homogeneity degrees $R_{\vec\eta}(\epsilon_0) \sim s_{ij}/\eta_j^2 + 2\pi i \bar\eta_j/\eta_j$ and $R_{\vec\eta}(\epsilon_{k\geq 4}) \sim s_{ij}\eta_j^{k-2}$ of the derivations in Section 8.1.2, we get

$$R_{\vec\eta}(\epsilon_k) \quad \leftrightarrow \quad \text{transcendental weight } 1{-}k\,, \quad k \geq 0\,. \tag{8.134}$$

As an immediate consequence, the operators in the series (8.129) are assigned

$$\begin{aligned} \exp\Big(-\frac{R_{\vec\eta}(\epsilon_0)}{4y}\Big) \quad &\leftrightarrow \quad \text{transcendental weight } 0 \\ R_{\vec\eta}\big(\,{\rm ad}_{\epsilon_0}^{k-j-2}(\epsilon_k)\big) \quad &\leftrightarrow \quad \text{transcendental weight } -(j{+}1)\,. \end{aligned} \tag{8.135}$$

Hence, the transcendental weight we have found for the $\mathcal{E}^{\rm sv}_{\rm min}$ cancels that of the accompanying derivations,

$$\mathcal{E}^{\rm sv}_{\rm min}\big[\begin{smallmatrix} j_1 & \ldots & j_\ell \\ k_1 & \ldots & k_\ell \end{smallmatrix}\big]\, R_{\vec\eta}\big(\,{\rm ad}_{\epsilon_0}^{k_\ell-j_\ell-2}(\epsilon_{k_\ell})\ldots {\rm ad}_{\epsilon_0}^{k_1-j_1-2}(\epsilon_{k_1})\big) \ \leftrightarrow \ \text{trans. weight } 0\,. \tag{8.136}$$

We shall now argue that this has to extend to the full $\mathcal{E}^{\rm sv}_{\rm min} \to \mathcal{E}^{\rm sv}$: By the vanishing transcendental weight of $\exp(-R_{\vec\eta}(\epsilon_0)/(4y))$, it follows from (8.136) that the $\mathcal{E}^{\rm sv}_{\rm min}$ contributions to the series (8.129) have weight zero, i.e. $\Lambda^\tau_{\vec\eta}$ can only depart from vanishing transcendental weight via $\mathcal{E}^{\rm sv} - \mathcal{E}^{\rm sv}_{\rm min}$. The latter reduce to antiholomorphic $\overline{f\big[\begin{smallmatrix} j_1 & j_2 & \ldots & j_\ell \\ k_1 & k_2 & \ldots & k_\ell \end{smallmatrix}\big]}$, so by our assumption (8.133) on the initial values, the claims (8.132) on the the series $Y^\tau_{\vec\eta}$ and (8.130) on the component integrals can only be violated by antiholomorphic quantities.

However, a purely antiholomorphic violation of uniform transcendentality is incompatible with the reality properties (8.24) of the component integrals: The contributions from holomorphic iterated Eisenstein integrals are uniformly transcendental by (8.136), so the same must be true for those of the antiholomorphic ones. More precisely, by induction in the depth $\ell$ (which can be separated by isolating a fixed order in $\alpha'$), one can show that the antiholomorphic integration constants $\overline{f\big[\begin{smallmatrix} j_1 & j_2 & \ldots & j_\ell \\ k_1 & k_2 & \ldots & k_\ell \end{smallmatrix}\big]}$ must share the transcendental weights of the $\mathcal{E}^{\rm sv}_{\rm min}$, i.e.

$$\begin{aligned} \mathcal{E}^{\rm sv}\big[\begin{smallmatrix} j_1 & j_2 & \ldots & j_\ell \\ k_1 & k_2 & \ldots & k_\ell \end{smallmatrix}\big],\, \beta^{\rm sv}\big[\begin{smallmatrix} j_1 & j_2 & \ldots & j_\ell \\ k_1 & k_2 & \ldots & k_\ell \end{smallmatrix}\big] \ &\leftrightarrow \ \text{trans. weight } \ell + \sum_{i=1}^{\ell} j_i \\ \mathcal{E}^{\rm sv}\big[\begin{smallmatrix} j_1 & \ldots & j_\ell \\ k_1 & \ldots & k_\ell \end{smallmatrix}\big]\, R_{\vec\eta}\big(\,{\rm ad}_{\epsilon_0}^{k_\ell-j_\ell-2}(\epsilon_{k_\ell})\ldots {\rm ad}_{\epsilon_0}^{k_1-j_1-2}(\epsilon_{k_1})\big) \ &\leftrightarrow \ \text{trans. weight } 0\,. \end{aligned} \tag{8.137}$$

The matching transcendental weights of $\mathcal{E}^{\rm sv}$ and $\beta^{\rm sv}$ follow from their relation (8.35) and $y$ having weight 1. Based on (8.137) and (8.135), each term in the series (8.129) has transcendental weight zero, and the weight of $Y^\tau_{\vec\eta}$ agrees with that of the initial value $\widehat{Y}^{i\infty}_{\vec\eta}$. Hence, the claim (8.132) follows from the assumption (8.133).

At two points, the initial value following from the Laurent polynomials (8.66) has transcendental weight 2, where we again use the vanishing weight of $\exp(-R_{\vec\eta}(\epsilon_0)/(4y))$. This confirms the claims (8.130) and (8.132) at $n = 2$ since the series in $\mathcal{E}^{\rm sv}$ preserves the weight by (8.137).

At $n \geq 3$ points, the dictionary between $\widehat{Y}^{i\infty}_{\vec\eta}$ and $(n{+}2)$-point sphere integrals is under investigation [257]. From a variety of Laurent-polynomials in ($n{\geq}3$)-point MGFs [39, 40, 183] and preliminary studies of their generating series, there is substantial evidence that the transcendental weight of $\widehat{Y}^{i\infty}_{\vec\eta}$ is $2(n{-}1)$.

#### BASIS INTEGRALS VERSUS ONE-LOOP STRING AMPLITUDES

We emphasize that the discussion of this section is tailored to the conjectural basis $Y^\tau_{\vec\eta}$ of torus integrals. In order to extract the transcendentality properties of one-loop string amplitudes, it remains to

- express their torus integral in terms of component integrals (8.130), where the expansion coefficients[19] may involve $\mathbb{Q}$-linear combinations of $\mathrm{G}_k$ [28, 174], cf. also Chapter 6
- study the kinematic factors accompanying the component integrals
- integrate the modular parameter $\tau$ over the fundamental domain.

The subtle interplay of $\tau$-integration with the transcendental weights has been explored in [128] along with a powerful all-order result for the integrated four-point integral $Y^\tau_{(0,0,0|0,0,0)}$ that was shown to enjoy a natural extension of uniform transcendentality. In [262] it was argued that uniform transcendentality is violated starting from two loops.

The kinematic coefficients of $Y^\tau_{(A|B)}$ or $\mathrm{G}_k Y^\tau_{(A|B)}$ may feature different transcendentality properties, depending on the string theory under investigation. For the $\tau$-integrands of type-II superstrings, these kinematic factors should be independent of $\alpha'$ in a suitable normalization of the overall one-loop amplitude. This can for instance be seen from the explicit ($4 \leq n \leq 7$)-point results in [1, 104, 170, 172] and the worldsheet supersymmetry in the RNS formalism, even in case of reduced spacetime supersymmetry [121, 122]. Hence, the $\tau$-integrands of $n$-point type-II amplitudes at one loop are expected to be uniformly transcendental.

Heterotic and bosonic strings in turn are known to involve tachyon poles in their chiral halves due to factors like $\partial f^{(k)}_{ij}$ and $f^{(k)}_{ij} f^{(\ell)}_{ij}$ in their CFT correlators. They can still be rewritten in terms of $Y^\tau_{(A|B)}$ via integration by parts, as we saw in Section 6.2.3, but the expansion coefficients may involve factors like $(1 + s_{ij})^{-1}$ that break uniform transcendentality upon geometric-series expansion. Hence, even if one-loop amplitudes of heterotic and bosonic strings can be expanded in a uniformly transcendental integral basis, the overall $\tau$-integrand will generically lose this property through the kinematic factors. This effect is well-known from tree-level amplitudes in these theories [100, 212, 226].

---

19 The reduction of ($n{\geq}4$)-point gauge amplitudes of the heterotic string to a basis of $Y^\tau_{(A|B)}$ also involves the modular version $\widehat{\mathrm{G}}_2$ of $\mathrm{G}_2$ among the expansion coefficients, cf. Chapter 6.

# 9

# CONCLUSION AND OUTLOOK

In this thesis, we have studied genus-one closed-string integrals using modular graph forms. These are non-holomorphic modular forms associated to a graph and given either as a nested lattice sum or as an integral over several copies of the torus. In the following, we review the new results obtained in this work and name open questions and possible directions for future research.

## 9.1 SUMMARY OF RESULTS

The main results obtained in Chapters 5–8 are:

- We derived basis decompositions for all dihedral and trihedral MGFs of total modular weight at most 12. These, together with other simplification techniques for MGFs, were automatized in the `Mathematica` package `Modular Graph Forms`.
- Using these decompositions, we computed the first orders of the low-energy expansion of the genus-one four-gluon amplitude in the heterotic string and decomposed the amplitude into building blocks which we conjectured to be of uniform transcendentality.
- We defined a generating function for Koba–Nielsen integrals and derived its Cauchy–Riemann and Laplace equations at $n$ points. The Cauchy–Riemann equation could be identified to be the single-valued image of the corresponding differential equation in the open string.
- We solved this Cauchy–Riemann equation perturbatively in terms of iterated Eisenstein integrals, leading to a dictionary between MGFs and iterated Eisenstein integrals and a counting of independent MGFs of total modular weight at most 14.

We will now give further details about these points.

In Chapter 5, we discussed various techniques to derive relations between MGFs, including a systematic treatment of four-point graphs. In particular, we could make holomorphic subgraph reduction (HSR) mathematically rigorous, provide an algorithmic procedure for $n$-point holomorphic subgraphs and derive a closed formula for the three-point case. Furthermore, we could show that in the integral representation, Fay identities between Kronecker–Eisenstein series imply HSR and lead to a powerful iterative procedure for higher-point HSR. We found that divergent MGFs appear naturally from momentum-conservation identities of convergent graphs and we could track their occurrence in Koba–Nielsen integrals to poles in the kinematic variables. Using the integral representation, HSR could be extended to divergent graphs as well.

By combining the techniques discussed in this way, we were able to find basis decompositions for all dihedral and trihedral MGFs of total modular weight $a + b \leq 12$. The basis elements consist only of dihedral MGFs and could be split into real and complex MGFs with known Laurent polynomials and Cauchy–Riemann equations. Together with the basis decompositions, we can therefore easily assemble the Laurent polynomials of all dihedral and trihedral MGFs with $a + b \leq 12$. A convenient `Mathematica` implementation for these manipulations is provided. A reference to all functions and symbols used can be found in Appendix A.

In Chapter 6, we studied genus-one four-gluon scattering in heterotic string theory as a concrete example of a string amplitude in which modular graph forms of non-trivial modular weight arise. Using the techniques discussed in Chapter 5, the first orders of the planar and non-planar contributions to the amplitude could be brought into a compact form, making the evaluation of the integral over $\tau$ for the first three orders possible, yielding the complete analytic contribution to the amplitude for these orders. Furthermore, the amplitude could be decomposed into building blocks of conjectured uniform transcendentality.

We were able to match different symmetry components of open-string integration cycles to building blocks of the planar part of the heterotic amplitude. Using the single-valued prescription conjectured in [34] for four-graviton scattering in type-II, we could reproduce a number of terms in the heterotic amplitude (including the Laurent polynomial encoding the asymptotics at the cusp) from open-string expressions. However, we could also show that in the case of non-trivial modular transformation properties, the prescription from [III] could not reproduce all terms in the heterotic integrals. This shortcoming is remedied in the proposal for an elliptic single-valued map in [17].

In Chapter 7, we started a more systematic study of Koba–Nielsen integrals by considering the generating function of these integrals. This function does not contain all possible Koba–Nielsen integrals directly, but all torus integrals appearing in one-loop string amplitudes can be written in terms of the expansion coefficients of the generating function via integration-by-parts manipulations and Fay identities. We could determine the Cauchy–Riemann and Laplace equations for the generating function at $n$ points and relate the Cauchy–Riemann equation to the single-valued image of the corresponding open-string differential equation. Closed expressions for the component integrals of the generating series at two- and three-points were given. In this way, it is easy to derive Cauchy–Riemann and Laplace equations for MGFs, without the need for holomorphic subgraph reduction, removal of negative edge labels or identities for divergent MGFs, tremendously improving on the previously available methods.

The solution of the differential equation derived in Chapter 7 was discussed in Chapter 8. A reformulation made the differential equation amenable for a perturbative solution via Picard iteration, which naturally led to iterated Eisenstein integrals. Using the reality properties of the Koba–Nielsen integrals, these could be expressed through Brown's holomorphic iterated Eisenstein integrals [33] and their complex conjugates. Similarly, the modular $S$-transformation, which is notoriously hard to determine for iterated Eisenstein integrals, could be fixed from the modular transformation properties of the generating function.

Since the generating series of Koba–Nielsen integrals can also be expanded in MGFs, we could find a dictionary between basis-MGFs of total modular weight $a+b \leq 10$ and iterated Eisenstein integrals $\beta^{\text{sv}}\left[\begin{smallmatrix} j_1 & \cdots & j_\ell \\ k_1 & \cdots & k_\ell \end{smallmatrix}\right]$ up to depth two and $k_1 + k_2 \leq 10$. Furthermore, because iterated Eisenstein integrals with different labels are linearly independent, by counting iterated Eisenstein integrals, we can count basis dimensions of MGFs. The results agree with the explicit calculations from Chapter 5 and make predictions for higher weights. From the reality properties of the iterated Eisenstein integrals, we can furthermore deduce the number of imaginary cusp forms, which also agrees with the calculations from Chapter 5 for the weights available there. Finally, all relations between MGFs found in Chapter 5 could in principle be derived from the generating function, although it is laborious in practice to expand the generating function to sufficiently high order.

## 9.2 OPEN QUESTIONS AND FUTURE DIRECTIONS

The results outlined above open several door for interesting further investigations, the most important of which we will discuss in this section.

The basis dimensions found in Chapter 5 were confirmed to be sufficient for MGFs of arbitrary topology by the counting of iterated Eisenstein integrals in Chapter 8. This raises the question, at what weight the basis of MGFs cannot be reduced to only dihedral graphs any more and more complicated topologies have to be included. It would also be interesting to extend the implementation of the techniques presented here to four-point graphs. Furthermore, it is striking that Fay identities and momentum conservation are in fact enough to construct all the basis decompositions found in this investigation and it would be interesting to find a proof that this has to be the case.

In the construction for the single-valued map in the heterotic string in Chapter 6, we found a way to map an integration cycle from the open string to the Koba–Nielsen integrand $V_2(1,2,3,4)$ in the closed string by considering different symmetry components of the integration cycle. This mapping is extended to more general Koba–Nielsen integrands in [17], giving rise to a concrete proposal for one-loop integrals which are mapped into each other by the elliptic single-valued map.

The solution obtained for the differential equation of the generating function of Koba–Nielsen integrals discussed in Chapter 8 depends on an initial value at $\tau \to i\infty$ which for the two-point case can be obtained from genus-zero integrals. At three-points, it was supplied by the basis decompositions of Chapter 5, which allow to extract the Laurent polynomial of the Koba–Nielsen integrals at hand. The determination of the initial value from genus-zero integrals also for more than two points is an ongoing project [257].

The expressions for the generating function of Koba–Nielsen integrals in terms of iterated Eisenstein integrals obtained in Chapter 8 defines a notion of single-valued iterated Eisenstein integrals. We could write these in many cases in terms of holomorphic iterated Eisenstein integrals and their complex conjugates. This construction should be closely related to Brown's generating series of single-valued iterated Eisenstein integrals [33]. It would be rewarding to make this relation more explicit and see if our examples constitute a concrete realization of Brown's more abstract construction.

On top of these points directly relating to the results of this thesis, there are further interesting directions for future studies. In particular, many of the structures discussed in this thesis generalize to genus two: The modular parameter becomes a $2 \times 2$ period matrix in the Siegel upper half-plane and the Schottky–Klein prime form generalizes the Jacobi theta functions, giving rise to the Arakelov Green function. In this way, modular graph functions [116, 117, 263] and modular graph tensors [264] can be defined for genus two surfaces whose non-separating degenerations involve elliptic generalizations of MGFs [117, 263, 265, 266]. In genus-two amplitudes of type-II superstrings, it was shown in [192] that the contribution due to one Green function is proportional to the Zhang–Kawazumi invariant, which subsequently was integrated over the genus-two fundamental domain [193]. However, the evaluation of the CFT correlators becomes much more challenging at higher genus [13, 106–115, 267, 268].

It would be interesting to see, if the construction discussed in Chapter 8 can also be performed at higher genus. Since the natural generalization of Jacobi theta functions to genus two are prime forms, these could provide a starting point for a construction of a genus two generalization of the Kronecker–Eisenstein series. Maybe, this version of the Kronecker–Eisenstein series can be used to define a generating series of Koba–Nielsen integrals similar to the one we defined at genus one, which obeys differential equations amenable to a solution via Picard iteration. In this way, one might be able to arrive at a genus-two version of (single-valued) iterated Eisenstein integrals (possibly with Siegel modular forms as integration kernels). Since the single-valued integration defined in [269] is valid for periods of surfaces with arbitrary genus, it would be interesting to see if the above construction at genus two would connect to this general framework.

Another interesting generalization of MGFs would be to consider string amplitudes on curved backgrounds. A particularly tractable case is the plane-wave background, in which the worldsheet fields become massive [270]. In [123], a massive deformation of the Green function on the torus was defined, which solves the Helmholtz equation and reduces to the Green function (3.62) for vanishing mass. In the reference, the authors computed the Fourier coefficients of this Green function, which involve Bessel functions in their numerators but have otherwise the same structure as the massless Fourier coefficients. This Fourier expansion can be used for a straightforward definition of massive modular graph functions, along the lines of the construction in Section 3.3.2. It would be interesting to investigate these massive deformations in detail, extend this framework to modular graph forms with non-trivial modular weight and compute amplitudes explicitly in a plane-wave background.

We will finish this outlook with a few thoughts on the possible implications of the single-valued map for string theory as a whole: The single-valued map has been proven rigorously at tree-level [25–27] and, as we saw above, at one-loop an explicit proposal of an elliptic single-valued map is given in [17]. Furthermore, there are concrete starting points how a genus-two version could be approached. This suggests that the single-valued map is not just a mere coincidence due to the structure of tree-level amplitudes, but might actually be a sign of a deeper duality between string amplitudes. It is however unclear how far-reaching the consequences are. In particular, it is hard to guess whether this correspondence extends to high genera for which the moduli space of super Riemann surfaces is not split any more [119] and how it relates to the non-perturbative sectors of string theory. Answering these questions will be the subject of this exciting area of research in the years to come.

# A

# COMPLETE REFERENCE FOR THE MODULAR GRAPH FORMS PACKAGE

In this appendix, we give a complete reference of all symbols defined in the `Modular Graph Forms` package, all functions and their options and detailed instructions how to load the package. In Section A.4, we show how the integrals appearing in the four-gluon amplitude of the heterotic string discussed in Chapter 6 can be computed using the `Modular Graph Forms` package.

Within `Mathematica`, short descriptions of the various symbols, functions and options can be displayed using the **`Information`** function, e.g. by running **`?g`**. A list of all the symbols defined in the package is printed by running **`?ModularGraphForms`**`` `* ``. The options and default values for a function are accessible via the **`Options`** function, e.g.

```
In[54]:= Options[CBasis]

Out[54]={basis → C} .
```

## A.1 FILES AND LOADING THE PACKAGE

The `Mathematica` package `Modular Graph Forms` includes the three files `ModularGraphForms.m`, `DiIds.txt` and `TriIds.txt`. The first one provides the package itself, whereas the two text files contain the basis decompositions described in Section 5.7 for dihedral and trihedral graphs, respectively. The package loads the latter files automatically and expects them in the same directory, in which also the `ModularGraphForms.m` file is saved. However, the text files can also be imported into `Mathematica` using the **`Get`** function and can be used independently of the `Modular Graph Forms` package.

To load the package, call the **`Get`** function on the `ModularGraphForms.m` file. Either the full path can be provided,

```
In[55]:= Get["/home/user/ModularGraphForms.m"]
```

or, if the files are placed in one of the directories in `Mathematica`'s search path, it is sufficient to run

```
In[56]:= Get["ModularGraphForms.m"] .
```

A list of the directories in Mathematica's search path is available in the global variable **$Path** and includes the current directory, which by default is the directory in which the current Notebook is saved.

## A.2 SYMBOLS

The Modular Graph Forms package defines a number of symbols used for the various objects defined in this thesis. For most of these symbols, a 2d-notation is implemented which makes the output easier to read. E.g. $\tau_2$ is represented by **tau[2]**, but printed as

```
In[57]:= tau[2]

Out[57]= τ2 .
```

These 2d-outputs can be copied to input cells and used for further computations. The input form of the 2d-output can be accessed by the function **InputForm**, e.g.

```
In[58]:= InputForm[τ2]

Out[58]= tau[2] .
```

Using the **$Assumptions** variable, the Modular Graph Forms package sets the global assumption that $\tau_2 > 0$. This is helpful e.g. when simplifying equations.

### A.2.1 *General symbols*

Five general symbols used by the Modular Graph Forms package are

| Mathematica symbol | description |
|---|---|
| **tau** | modular parameter $\tau$ |
| **tauBar** | $\bar{\tau}$ |
| **tau[2]** | $\tau_2 = \operatorname{Im} \tau$ |
| **y** | $y = \pi\tau_2$ |
| **zeta[k]** | $\zeta_k$ as defined in (2.41) |
| **bCoeff** | coefficient in the sieve algorithm, cf. **CSieveDecomp** |

### A.2.2 *Modular graph forms*

The conventions for two-, three- and four-point modular graph forms were introduced in detail in Section 5.2. The symbols used to represent MGFs, (non-)holomorphic Eisenstein series and real and complex basis elements are

| Mathematica symbol | description |
|---|---|
| **c[…]** | MGF, cf. Section 5.2 |
| **a[…]** | $\mathcal{A}\left[\begin{smallmatrix} A \\ B \end{smallmatrix}\right]$ as defined in (5.7) |
| **intConst[…]** | integration constant, cf. **CSieveDecomp** |
| **intConstBar[…]** | complex conjugate of **intConst** |
| **g[k]** | $G_k$ as defined in (3.18) |
| **gBar[k]** | $\overline{G}_k$ |
| **gHat[2]** | $\widehat{G}_2$ as defined in (3.31) |
| **gBarHat[2]** | $\overline{\widehat{G}}_2$ |
| **e[k₁,…,kᵣ]** | $E_k$ as defined in (3.33) and $E_{k_1,\dots,k_r}$ as defined in (4.28) |
| **ep[k₁,…,kᵣ]** | $E'_{k_1,\dots,k_r}$ as defined in (4.28) |
| **b[k₁,…,kᵣ]** | $B_{k_1,\dots,k_r}$ as defined in (5.214) |
| **bp[k₁,…,kᵣ]** | $B'_{k_1,\dots,k_r}$ as defined in (5.214) |

Note that MGFs are represented by the symbol **c**, but are printed with a capital `C`. When copying this output into an input cell, the capital **C** should not be changed into a lowercase **c**. Furthermore, the basis elements listed here are meaningful only for the indices defined in (4.28) and (5.214).

The `Mathematica` symbols used to represent Cauchy–Riemann derivatives of real and complex basis elements of MGFs are

| Mathematica symbol | description |
|---|---|
| **nablaE[n,{k₁,…,kᵣ}]** | $\nabla_0^n E_{k_1,\dots,k_r}$ |
| **nablaBarE[n,{k₁,…,kᵣ}]** | $\overline{\nabla}_0^n E_{k_1,\dots,k_r}$ |
| **nablaEp[n,{k₁,…,kᵣ}]** | $\nabla_0^n E'_{k_1,\dots,k_r}$ |
| **nablaBarEp[n,{k₁,…,kᵣ}]** | $\overline{\nabla}_0^n E'_{k_1,\dots,k_r}$ |
| **nablaB[n,{k₁,…,kᵣ}]** | $\nabla_0^n B_{k_1,\dots,k_r}$ |
| **nablaBarBBar[n,{k₁,…,kᵣ}]** | $\overline{\nabla}_0^n \overline{B_{k_1,\dots,k_r}}$ |
| **nablaBp[n,{k₁,…,kᵣ}]** | $\nabla_0^n B'_{k_1,\dots,k_r}$ |
| **nablaBarBpBar[n,{k₁,…,kᵣ}]** | $\overline{\nabla}_0^n \overline{B'_{k_1,\dots,k_r}}$ |

The derivative operator $\nabla_0$ and its complex conjugate are defined in (3.55). The zeroth derivative returns the argument, e.g.

```
In[59]:= nablaE[0, {5}]

Out[59]= E5 .
```

### A.2.3 *Iterated Eisenstein integrals*

For compatibility with the data provided in the ancillary file of [V], the `Modular Graph Forms` package defines the following symbols for iterated Eisenstein integrals, although no manipulations of these objects can be performed within this package.

| `Mathematica` symbol | description |
|---|---|
| **esv**$\left[\begin{smallmatrix} j_1 & \cdots & j_\ell \\ k_1 & \cdots & k_\ell \end{smallmatrix}\right]$ | $\mathcal{E}^{\mathrm{sv}}\left[\begin{smallmatrix} j_1 & \cdots & j_\ell \\ k_1 & \cdots & k_\ell \end{smallmatrix};\tau\right]$ as in (8.53) |
| **esvS**$\left[\begin{smallmatrix} j_1 & \cdots & j_\ell \\ k_1 & \cdots & k_\ell \end{smallmatrix}\right]$ | $\mathcal{E}^{\mathrm{sv}}\left[\begin{smallmatrix} j_1 & \cdots & j_\ell \\ k_1 & \cdots & k_\ell \end{smallmatrix};-\frac{1}{\tau}\right]$ |
| **esvBar**$\left[\begin{smallmatrix} j_1 & \cdots & j_\ell \\ k_1 & \cdots & k_\ell \end{smallmatrix}\right]$ | $\overline{\mathcal{E}^{\mathrm{sv}}\left[\begin{smallmatrix} j_1 & \cdots & j_\ell \\ k_1 & \cdots & k_\ell \end{smallmatrix};\tau\right]}$ |
| **betasv**$\left[\begin{smallmatrix} j_1 & \cdots & j_\ell \\ k_1 & \cdots & k_\ell \end{smallmatrix}\right]$ | $\beta^{\mathrm{sv}}\left[\begin{smallmatrix} j_1 & \cdots & j_\ell \\ k_1 & \cdots & k_\ell \end{smallmatrix};\tau\right]$ as in (8.35) |
| **betasvS**$\left[\begin{smallmatrix} j_1 & \cdots & j_\ell \\ k_1 & \cdots & k_\ell \end{smallmatrix}\right]$ | $\beta^{\mathrm{sv}}\left[\begin{smallmatrix} j_1 & \cdots & j_\ell \\ k_1 & \cdots & k_\ell \end{smallmatrix};-\frac{1}{\tau}\right]$ |
| **betasvBar**$\left[\begin{smallmatrix} j_1 & \cdots & j_\ell \\ k_1 & \cdots & k_\ell \end{smallmatrix}\right]$ | $\overline{\beta^{\mathrm{sv}}\left[\begin{smallmatrix} j_1 & \cdots & j_\ell \\ k_1 & \cdots & k_\ell \end{smallmatrix};\tau\right]}$ |

### A.2.4 *Koba–Nielsen integrals*

For the evaluation and representation of Koba–Nielsen integrals and their generating series, the following symbols are defined.

| `Mathematica` symbol | description |
|---|---|
| **eta[k₁,…,kᵣ]** | $\eta_{k_1,\ldots,k_r}$ as in (7.1) |
| **etaBar[k₁,…,kᵣ]** | $\bar{\eta}_{k_1,\ldots,k_r}$ |
| **s[k₁,…,kᵣ]** | $s_{k_1,\ldots,k_r}$ as defined in (2.27) |
| **fz[a, i, j]** | $f^{(a)}_{ij}$ as defined in (3.91) |
| **fBarz[b, i, j]** | $\overline{f^{(b)}_{ij}}$ as defined in (3.92) |
| **gz[i, j]** | $G_{ij}$ as defined in (3.65) |
| **cz[a, b, i, j]** | $C^{(a,b)}_{ij}$ as defined in (3.118) |
| **vz[a,{k₁,…,kᵣ}]** | $V_a(k_1,\ldots,k_r)$ as defined in (3.96) |
| **vBarz[b,{k₁,…,kᵣ}]** | $\overline{V_b(k_1,\ldots,k_r)}$ |

Symbols which represent functions which can appear in the integrand of a Koba–Nielsen integral have the suffix **z**.

## A.3 FUNCTIONS

The functions in the `Modular Graph Forms` package are sorted into three main categories: Dihedral functions only manipulate dihedral MGFs and carry the prefix **`Di`**. Trihedral functions only manipulate trihedral MGFs and carry the prefix **`Tri`**. General functions act on MGFs of all supported graph topologies or perform other tasks which are not specific to any graph topology. They carry a prefix **`C`**. On top of these, there is limited support for four-point manipulations in the form of the function **`TetCSimplify`** and a function to expand Koba–Nielsen integrals in MGFs.

### A.3.1 *General functions*

#### `CBasis`

The function **`CBasis`** returns a list of basis elements for MGFs.

ARGUMENTS **`CBasis`** accepts two arguments, corresponding to the holomorphic and antiholomorphic modular weight of the basis.

RETURN VALUE **`CBasis`** returns the basis of MGFs at the modular weight passed as the arguments as listed in Tables 5.3 and 5.4. Note that at weight $(a+k, a-k)$, the basis elements in Table 5.3 have weight $(a+k, a-k)$, whereas in Table 5.4, they have weight $(0, -2k)$.

OPTIONS If the option **`basis`** is set to the string `"C"` (the default) the basis from Table 5.3 is returned, if the option **`basis`** is set to the string `"nablaE"`, the basis from Table 5.4 is returned. No other values for **`basis`** are admissible.

WARNINGS

- If the sum of the holomorphic- and antiholomorphic modular weights passed in the arguments is odd, the warning **`CBasis::`**`incorrModWeight` is issued and **`CBasis`** returns an empty list.
- If the sum of the holomorphic- and antiholomorphic modular weights passed in the arguments is less than four, the warning **`CBasis::`**`tooLowWeight` is issued and **`CBasis`** returns an empty list.
- If the basis for the modular weight passed to **`CBasis`** is not implemented, the warning **`CBasis::`**`noBasis` is issued and **`CBasis`** returns an empty list.

EXAMPLES

In[60]:= **CBasis[3, 7]**

Out[60]= $\left\{C\begin{bmatrix}1&1&1\\1&1&5\end{bmatrix}, C\begin{bmatrix}3&0\\7&0\end{bmatrix}, C\begin{bmatrix}1&0\\3&0\end{bmatrix}C\begin{bmatrix}2&0\\4&0\end{bmatrix}, \frac{\pi^2 C\begin{bmatrix}1&0\\5&0\end{bmatrix}E_2}{\tau_2^2}, C\begin{bmatrix}0&1&2\\2&0&5\end{bmatrix}\right\}$

In[61]:= **CBasis[3, 7, basis → "nablaE"]**

Out[61]= $\{\bar{\nabla}^2 E_{2,3}, \bar{\nabla}^2 E_5, \bar{\nabla}E_2\ \bar{\nabla}E_3, E_2\ \bar{\nabla}^2 E_3, \bar{\nabla}^2 \bar{B}'_{2,3}\}$

## CCheckConv

The function **CCheckConv** tests if MGFs are convergent or divergent.

ARGUMENT **CCheckConv** accepts one argument which is an arbitrary expression, possibly containing MGFs of any topology and Eisenstein series.

RETURN VALUE **CCheckConv** returns **True** or **False**. If the argument contains an MGF which is divergent according to the conditions discussed in Section 5.6.1 or a $\mathrm{E}_k$, $\mathrm{G}_k$ or $\overline{\mathrm{G}}_k$ with $k < 2$, the function returns **False**, otherwise it returns **True**.

EXAMPLES

In[62]:= **CCheckConv[e[1] c[ $\begin{smallmatrix}2&0\\3&0\end{smallmatrix}$ ]]**

Out[62]= False

Since $\mathrm{E}_1$ is divergent, the return value is **False**, even though $C\begin{bmatrix}2&0\\3&0\end{bmatrix} = 0$.

In[63]:= **CCheckConv[c[ $\begin{smallmatrix}1&2\\-2&2\end{smallmatrix}$, $\begin{smallmatrix}1&2\\1&2\end{smallmatrix}$, $\begin{smallmatrix}1&2\\-2&2\end{smallmatrix}$, $\begin{smallmatrix}1&2\\1&2\end{smallmatrix}$, $\begin{smallmatrix}2&2\\2&2\end{smallmatrix}$ ]]**

Out[63]= False

Since the last condition in (5.160) is violated, the return value is **False**.

In[64]:= **CCheckConv[c[ $\begin{smallmatrix}1&2\\-2&2\end{smallmatrix}$, $\begin{smallmatrix}1&2\\1&2\end{smallmatrix}$, $\begin{smallmatrix}1&2\\-2&2\end{smallmatrix}$, $\begin{smallmatrix}2&3\\2&1\end{smallmatrix}$, $\begin{smallmatrix}2&2\\2&2\end{smallmatrix}$ ]]**

Out[64]= True

Since here $\check{c}_4$ as defined below (5.157) is increased, the last condition in (5.160) is also satisfied and the return value is **True**.

## CComplexConj

The function **CComplexConj** computes the complex conjugate of an expression.

ARGUMENT **CComplexConj** accepts one arbitrary argument.

RETURN VALUE **CComplexConj** returns its argument with all MGFs complex conjugated and written in their canonical representation. This includes Eisenstein series, complex basis elements (according to (5.215)) Cauchy–Riemann derivatives of basis elements and integration constants, unless the MGF in the argument is real.

EXAMPLE

```
In[65]:= CComplexConj[{g[4], b[2, 4], intConst[1 2 1; 1 1 4],
           nablaB[1, {2, 4}]}]
```

Out[65]= $\left\{\overline{\mathrm{G}}_4,\ -\mathrm{B}_{2,4} - 2\,\mathrm{E}_2\,\mathrm{E}_4 + \frac{2\,\mathrm{E}_{2,4}}{9},\ \overline{\mathrm{intConst}\left[\begin{smallmatrix}1&2&1\\1&1&4\end{smallmatrix}\right]},\ \overline{\nabla}\mathrm{B}_{2,4}\right\}$

### **CConvertToNablaE** and **CConvertFromNablaE**

The functions **CConvertToNablaE** and **CConvertFromNablaE** convert an expression between the bases given in Tables 5.3 and 5.4.

ARGUMENT Both **CConvertToNablaE** and **CConvertFromNablaE** accept one arbitrary argument.

RETURN VALUE **CConvertToNablaE** replaces all of the basis elements in Table 5.3 in its argument with their expansions in the basis of Table 5.4. **CConvertFromNablaE** replaces all of the basis elements in Table 5.4 in its argument with their expansions in the basis of Table 5.3. On top of the elements listed explicitly in these tables, $\nabla_0^n \mathrm{E}_k$ and $\mathcal{C}\left[\begin{smallmatrix}k+n & 0\\ k-n & 0\end{smallmatrix}\right]$ are rewritten according to (5.56) for any $n$ and $k$. The results are not manipulated any further and MGFs in the argument which are not in the basis to be converted are left untouched.

EXAMPLES

```
In[66]:= CConvertToNablaE[c[1 1 4; 1 1 2]]
```

Out[66]= $\frac{\pi^5\,\nabla\mathrm{B}_{2,3}}{18\,\tau_2^6} - \frac{\pi^5\,\nabla\mathrm{B}'_{2,3}}{18\,\tau_2^6} - \frac{\pi^5\,\mathrm{E}_3\,\nabla\mathrm{E}_2}{12\,\tau_2^6} + \frac{\pi^5\,\mathrm{E}_2\,\nabla\mathrm{E}_3}{12\,\tau_2^6} + \frac{41\,\pi^5\,\nabla\mathrm{E}_5}{140\,\tau_2^6} + \frac{\pi^5\,\nabla\mathrm{E}_{2,3}}{24\,\tau_2^6} - \frac{\pi^5\,\nabla\mathrm{E}_2\,\zeta_3}{36\,\tau_2^6}$

```
In[67]:= CConvertToNablaE[c[1 2 3; 1 1 2]]
```

Out[67]= $\mathrm{C}\left[\begin{smallmatrix}1&2&3\\1&1&2\end{smallmatrix}\right]$

```
In[68]:= CConvertFromNablaE[Out[66]]
```

Out[68]= $\mathrm{C}\left[\begin{smallmatrix}1&1&4\\1&1&2\end{smallmatrix}\right]$

### **CHolCR** and **CAHolCR**

The functions **CHolCR** and **CAHolCR** compute the holomorphic- and antiholomorphic Cauchy–Riemann derivative, respectively.

ARGUMENT Both **CHolCR** and **CAHolCR** accept one argument which should be a functional expression (e.g. a polynomial) involving MGFs and Eisenstein series.

RETURN VALUE **CHolCR** returns the holomorphic Cauchy–Riemann derivative of its argument, using the derivative operator defined in (3.51), by applying (5.53). The result is always given in terms of lattice sums, even if the argument involves Cauchy–Riemann derivatives of basis elements. The generalized Ramanujan identities from Section 5.3.5 are not applied. If the argument contains a divergent graph with a closed holomorphic subgraph, HSR is applied before the derivative is taken, while $C\left[\begin{smallmatrix}2&0\\0&0\end{smallmatrix}\right]$ is not replaced by $\widehat{G}_2$. The output is not manipulated any further. **CAHolCR** returns the antiholomorphic Cauchy–Riemann derivative.

OPTIONS The Boolean option **divDer** specifies if derivatives of divergent graphs are taken or not. If it is set to **False** (the default is **True**) and a divergent MGF appears in the argument, **CHolCR** and **CAHolCR** return **Nothing**.

WARNINGS

- If the argument of **CHolCR** contains a divergent MGF, the warning **CHolCR::derOfDiv** is issued (and c.c.).
- The argument is passed to **CModWeight** (see below), to check if it has homogeneous modular weight. If it does not, the warning **CModWeight::WeightNotHom** is issued and **Nothing** is returned.

EXAMPLES

In[69]:= **CHolCR**$\left[\left\{\mathbf{nablaE[1,\{3\}],gBarHat[2],c}\left[\begin{smallmatrix}\mathbf{1}&\mathbf{1}&\mathbf{1}\\\mathbf{1}&\mathbf{1}&\mathbf{1}\end{smallmatrix}\right]\right\}\right]$

Out[69]= $\left\{\frac{12\,C\left[\begin{smallmatrix}5&0\\1&0\end{smallmatrix}\right]\tau_2^4}{\pi^3},\ \frac{\pi}{\tau_2^2},\ C\left[\begin{smallmatrix}1&1&2\\1&1&0\end{smallmatrix}\right]+C\left[\begin{smallmatrix}1&2&1\\1&0&1\end{smallmatrix}\right]+C\left[\begin{smallmatrix}2&1&1\\0&1&1\end{smallmatrix}\right]\right\}$

In[70]:= **CHolCR**$\left[\mathbf{c}\left[\begin{smallmatrix}\mathbf{0}&\mathbf{0}&\mathbf{3}\\\mathbf{1}&\mathbf{1}&\mathbf{3}\end{smallmatrix}\right]\right]$

CHolCR : Warning: You are generating the holomorphic Cauchy–Riemann derivative of the divergent expression $C\left[\begin{smallmatrix}0&0&3\\1&1&3\end{smallmatrix}\right]$. This may be problematic.

Out[70]= $-6\,C\left[\begin{smallmatrix}4&0\\4&0\end{smallmatrix}\right]+\frac{2\,\pi\,C\left[\begin{smallmatrix}3&0\\3&0\end{smallmatrix}\right]}{\tau_2}$

## CLaurentPoly

The function **CLaurentPoly** replaces basis elements by their Laurent polynomials.

ARGUMENT **CLaurentPoly** accepts one arbitrary argument.

RETURN VALUE **CLaurentPoly** returns its argument with the real basis elements (4.28), the complex basis elements (5.214), their Cauchy–Riemann derivatives and complex conjugates, as well as all non-holomorphic- and holomorphic Eisenstein series (including $\widehat{G}_2$) replaced by their Laurent polynomials. The Laurent polynomials of the real and complex basis elements are given in (5.212) and (5.217), respectively, those of the holomorphic and non-holomorphic Eisenstein series in (3.24) and (3.34). Their derivatives are obtained using (5.219).

OPTIONS The Boolean option **usey** specifies, if the output is given in terms of $\tau_2$ (**False**) or $y = \pi\tau_2$ (**True**, the default).

EXAMPLES

In[71]:= **CLaurentPoly[nablaBarBBar[2, {2, 4}]]**

$$\text{Out[71]=}\ -\frac{8\,y^8}{30375\,\pi^2} - \frac{105\,\zeta_3\,\zeta_7}{8\,\pi^2\,y^2} + \frac{25\,\zeta_9}{12\,\pi^2\,y}$$

In[72]:= **CLaurentPoly[{g[6], gHat[2], e[7]}, usey → False]**

$$\text{Out[72]=}\ \left\{\frac{2\,\pi^6}{945}, \frac{\pi^2}{3} - \frac{\pi}{\tau_2}, \frac{4\,\pi^7\,\tau_2^7}{18243225} + \frac{231\,\zeta_{13}}{512\,\pi^6\,\tau_2^6}\right\}$$

## CListHSRs

The function **CListHSRs** lists MGFs with closed holomorphic subgraphs in an expression.

ARGUMENTS **CListHSRs** accepts one arbitrary argument.

RETURN VALUE **CListHSRs** returns a list with all dihedral and trihedral graphs with closed holomorphic subgraphs appearing somewhere in its argument. If the argument does not contain any dihedral or trihedral graphs, **CListHSRs** returns the empty list.

EXAMPLES

$$\text{In[73]:=}\ \mathbf{CListHSRs}\left[\mathbf{c}\left[\begin{matrix}1\\0\end{matrix}, \begin{matrix}1&2\\0&1\end{matrix}, \begin{matrix}1&2\\2&0\end{matrix}\right] + \mathbf{c}\left[\begin{matrix}7&0\\3&0\end{matrix}\right]\right]$$

$$\text{Out[73]=}\ \left\{\mathrm{C}\left[\begin{matrix}1\\0\end{matrix}\middle|\begin{matrix}1&2\\0&1\end{matrix}\middle|\begin{matrix}1&2\\2&0\end{matrix}\right]\right\}$$

## CModWeight

The function **CModWeight** determines the modular weight of an expression.

ARGUMENT **CModWeight** accepts one argument which can be either a modular form (possibly of trivial modular weight), a product of modular forms or a sum of products of modular forms.

RETURN VALUE **CModWeight** returns a list with two elements, corresponding to the holomorphic and antiholomorphic modular weight, respectively.

WARNINGS

- If a sum is passed to **CModWeight** and the modular weights of the summands do not agree, **CModWeight** returns **Null** and the warning **CModWeight::WeightNotHom**, containing a list of the modular weights appearing in the sum, is issued.
- If symbols appear in the argument of **CModWeight**, for which no modular weight is implemented, **CModWeight** returns the modular weight which the expression would have if all symbols of unknown weight were modular invariant. The warning **CModWeight::UnknownExp**, containing a list of the terms whose weight could not be determined, is issued.

EXAMPLES

```
In[74]:= CModWeight[tau[2]^-2 nablaBarE[1, {2}] nablaE[2, {4}] +
           nablaBp[1, {2, 4}]]

Out[74]={0, − 2}

In[75]:= CModWeight[e[2] + C[2 0; 2 0]]

         CModWeight : The modular weight of the argument is not
         homogeneous, the weights {2,2}, {0,0} appear.

In[76]:= CModWeight[g[2] g[4]]

         CModWeight : Expression(s) {G2} found whose modular weight
         could not be determined. The returned weight assumes them to be
         modular invariant.

In[77]:={4, 0}
```

### CSieveDecomp

The function **CSieveDecomp** decomposes an MGF using the sieve algorithm.

ARGUMENTS **CSieveDecomp** accepts one argument which can be either a dihedral or a trihedral MGF without closed holomorphic subgraph.

RETURN VALUE **CSieveDecomp** performs the sieve algorithm on its argument as discussed in Section 5.5 and returns the decomposition obtained. If the holomorphic modular weight is larger than the antiholomorphic one, **CSieveDecomp** takes holomorphic derivatives, otherwise antiholomorphic ones. If both modular weights of the argument are equal, an integration constant **intConst** labeled by the exponent matrix of the argument and dressed with an appropriate factor of $\frac{\pi}{\tau_2}$ is added to the final decomposition. If the basis into which the argument is decomposed is not linearly independent, the output contains free parameters with head **bCoeff**.

OPTIONS

| option | possible values | default value | description |
|---|---|---|---|
| **verbose** | **True** **False** | **False** | activates verbose output |
| **divDer** | **True** **False** | **False** | activates decomposition of divergent graphs |
| **basis** | list of MGFs | **{}** | basis elements for decomposition |
| **addIds** | list of replacement rules for MGFs | **{}** | additional replacement rules applied to each derivative |
| **CSimplifyOpts** | option assignments of **CSimplify** | see below | options passed to **CSimplify** when simplifying the derivatives |

The default value of **CSimplifyOpts** is **{basisExpandG → True}**. If the option **basis** is set to the empty list, the appropriate basis is determined automatically using **CBasis**. Since this basis does not contain powers of $E_1$, it is not sufficient for the decomposition of divergent graphs. The basis elements have to be MGFs without closed holomorphic subgraphs of the same modular weight as the argument. Divergent basis elements are only admissible if **divDer** is set to **True**.

WARNINGS

- If the argument passed to **CSieveDecomp** is a divergent graph, the warning **CSieveDecomp::**divArg is issued. The decomposition proceeds only if **divDer** is set to **True**.
- If one of the basis elements is divergent, but the argument is not, the warning **CSieveDecomp::**divBasis is issued.
- If a holomorphic Eisenstein series could not be canceled in one of the derivatives, the warning **CSieveDecomp::**noSol is issued. This happens e.g. if the basis is not large enough.
- If in one of the derivatives, an undecomposed graph appears in the coefficient of a holomorphic Eisenstein series, the warning **CSieveDecomp::**holEisenCoeffNoBasis is issued and the algorithm interrupted. MGFs are considered decomposed if they appear in the basis given by **CBasis**. For modular weight $a + b \leq 12$, these undecomposed graphs will be divergent.

EXAMPLES

In[78]:= **CSieveDecomp**$\left[c\left[\begin{smallmatrix}1\\1\end{smallmatrix}, \begin{smallmatrix}1&1\\1&1\end{smallmatrix}, \begin{smallmatrix}1&1\\1&1\end{smallmatrix}\right]\right]$

Out[78]= $2\,C\left[\begin{smallmatrix}1&1&3\\1&1&3\end{smallmatrix}\right] - \frac{2\,\pi^5\,E_5}{5\,\tau_2^5} + \frac{\pi^5 \mathtt{intConst}\left[\begin{smallmatrix}1\\1\end{smallmatrix}\middle|\begin{smallmatrix}1&1\\1&1\end{smallmatrix}\middle|\begin{smallmatrix}1&1\\1&1\end{smallmatrix}\right]}{\tau_2^5}$

In[79]:= **CSieveDecomp**$\left[c\left[\begin{smallmatrix}1&1&1\\1&1&1\end{smallmatrix}\right], \mathbf{basis} \to \left\{c\left[\begin{smallmatrix}0&1&2\\1&1&1\end{smallmatrix}\right], \frac{\tau_2^3}{\mathbf{tau[2]}^3}\,\mathbf{e[3]}\right\}\right]$

Out[79]= $-C\left[\begin{smallmatrix}0&1&2\\1&1&1\end{smallmatrix}\right] + 2\,\mathtt{bCoeff[2]}\,C\left[\begin{smallmatrix}0&1&2\\1&1&1\end{smallmatrix}\right] + \frac{\pi^3\,\mathtt{bCoeff[2]}\,E_3}{\tau_2^3} + \frac{\tau^3\,\mathtt{intConst}\left[\begin{smallmatrix}1&1&1\\1&1&1\end{smallmatrix}\right]}{\tau_2^3}$

In[80]:= **CSieveDecomp**$\left[c\left[\begin{smallmatrix}0&1&1&1\\3&0&1&1\end{smallmatrix}\right]\right]$

CSieveDecomp : The 1st derivative contains the undecomposed graph(s) $\left\{C\left[\begin{smallmatrix}0&1&1\\1&0&1\end{smallmatrix}\right]\right\}$ as a coefficient of a holomorphic Eisenstein series.

Out[80]= $C\left[\begin{smallmatrix}0&1&1&1\\3&0&1&1\end{smallmatrix}\right]$

In[81]:= **CSieveDecomp**$\left[c\left[\begin{smallmatrix}0&1&1&1\\3&0&1&1\end{smallmatrix}\right], \mathbf{addIds} \to \left\{c\left[\begin{smallmatrix}0&1&1\\1&0&1\end{smallmatrix}\right] \to -\frac{\pi^2\,\mathbf{E}_1^2}{2\,\mathbf{tau[2]}^2} + \frac{\pi^2\,\mathbf{E}_2}{2\,\mathbf{tau[2]}^2}\right\}\right]$

Out[81]= $2\,C\left[\begin{smallmatrix}3&0\\5&0\end{smallmatrix}\right] - 2\,C\left[\begin{smallmatrix}1&1&1\\1&1&3\end{smallmatrix}\right] - \frac{\pi^2\,C\left[\begin{smallmatrix}1&0\\3&0\end{smallmatrix}\right]E_2}{\tau_2^2}$

### `CSimplify`

The function **`CSimplify`** performs all known simplifications for MGFs.

ARGUMENT **`CSimplify`** accepts one arbitrary argument.

RETURN VALUE **`CSimplify`** applies, in this order, the specialized functions **`TetCSimplify`**, **`TriCSimplify`** and **`DiCSimplify`** to its argument until it no longer changes and returns the result.

OPTIONS **`CSimplify`** accepts all the options of both **`TriCSimplify`** and **`DiCSimplify`** and passes them to these functions when they are called.

EXAMPLES

$$\texttt{In[82]:=}\ \mathtt{CSimplify}\left[\mathtt{c}\left[\begin{smallmatrix}1\\0\end{smallmatrix},\begin{smallmatrix}1&1\\0&1\end{smallmatrix},\begin{smallmatrix}1&2\\1&0\end{smallmatrix}\right]\right]$$

$$\texttt{Out[82]=}\ \frac{3}{2}C\begin{bmatrix}3&0\\1&0\end{bmatrix}^2-\frac{1}{2}C\begin{bmatrix}6&0\\2&0\end{bmatrix}-\frac{1}{2}C\begin{bmatrix}4&0\\2&0\end{bmatrix}\hat{G}_2-\frac{\pi^2 G_4}{\tau_2^2}+\frac{3\pi C\begin{bmatrix}5&0\\1&0\end{bmatrix}}{\tau_2}-\frac{\pi C\begin{bmatrix}3&0\\1&0\end{bmatrix}\hat{G}_2}{\tau_2}$$

$$\texttt{In[83]:=}\ \mathtt{CSimplify}\left[\mathtt{c}\left[\begin{smallmatrix}1\\0\end{smallmatrix},\begin{smallmatrix}1&1\\0&1\end{smallmatrix},\begin{smallmatrix}1&2\\1&0\end{smallmatrix}\right],\mathtt{tri3ptFayHSR}\to\mathtt{True}\right]$$

$$\texttt{Out[83]=}\ \frac{3}{2}C\begin{bmatrix}3&0\\1&0\end{bmatrix}^2-\frac{1}{2}C\begin{bmatrix}6&0\\2&0\end{bmatrix}-\frac{1}{2}C\begin{bmatrix}4&0\\2&0\end{bmatrix}\hat{G}_2-\frac{\pi^2 G_4}{\tau_2^2}+\frac{3\pi C\begin{bmatrix}5&0\\1&0\end{bmatrix}}{\tau_2}-\frac{\pi C\begin{bmatrix}3&0\\1&0\end{bmatrix}\hat{G}_2}{\tau_2}$$

$$\texttt{In[84]:=}\ \mathtt{CSimplify}\left[\mathtt{c}\left[\{\},\begin{smallmatrix}1\\1\end{smallmatrix},\begin{smallmatrix}1\\1\end{smallmatrix},\begin{smallmatrix}1\\1\end{smallmatrix},\{\},\begin{smallmatrix}1\\1\end{smallmatrix}\right]\right]$$

$$\texttt{Out[84]=}\ 0$$

## CSort

The function **CSort** sorts MGFs into their canonical representation.

ARGUMENT **CSort** accepts one arbitrary argument.

RETURN VALUE **CSort** returns its argument with all MGFs written in their canonical representation as discussed in Section 5.3.1.

EXAMPLE

$$\texttt{In[85]:=}\ \mathtt{CSort}\left[\left\{\mathtt{c}\left[\begin{smallmatrix}2&2\\1&1\end{smallmatrix},\begin{smallmatrix}1&1\\1&1\end{smallmatrix},\begin{smallmatrix}1&1\\1&1\end{smallmatrix},\begin{smallmatrix}1&2\\1&1\end{smallmatrix},\begin{smallmatrix}2&2\\1&1\end{smallmatrix},\begin{smallmatrix}1&1\\1&2\end{smallmatrix}\right]\right\},\ \mathtt{c}\left[\begin{smallmatrix}1&1\\1&0\end{smallmatrix},\begin{smallmatrix}1\\0\end{smallmatrix},\begin{smallmatrix}1&2\\1&0\end{smallmatrix}\right]\right]$$

$$\texttt{Out[85]=}\ \left\{C\left[\begin{array}{cc|cc|cc}1&1&1&1&2&2\\1&1&1&1&1&1\\\hline 1&2&2&2&1&1\\1&1&1&1&1&2\end{array}\right],C\left[\begin{array}{c|cc|cc}1&1&1&1&2\\0&0&1&1&0\end{array}\right]\right\}$$

### A.3.2 *Dihedral functions*

#### DiHolMomConsId and DiAHolMomConsId

The functions **DiHolMomConsId** and **DiAHolMomConsId** generate holomorphic and antiholomorphic dihedral momentum-conservation identities, respectively.

ARGUMENT Both **DiHolMomConsId** and **DiAHolMomConsId** accept a dihedral MGF as their only argument.

RETURN VALUE **DiHolMomConsId** returns the holomorphic momentum conservation identity (5.36) of the seeds given in the argument as an equation with RHS 0. **DiAHolMomConsId** returns the antiholomorphic momentum-conservation identity. No further manipulation as e.g. sorting into the canonical representation are performed on the output.

WARNINGS If the argument is divergent according to **CCheckConv**, the warning **DiHolMomConsId::divDiHolMomCons** (and c.c.) is issued.

EXAMPLES

In[86]:= **DiHolMomConsId**$\left[\mathbf{c}\left[\begin{smallmatrix}1&1&2\\1&1&1\end{smallmatrix}\right]\right]$

**DiAHolMomConsId**$\left[\mathbf{c}\left[\begin{smallmatrix}1&1&2\\1&1&1\end{smallmatrix}\right]\right]$

Out[86]= $C\left[\begin{smallmatrix}0&1&2\\1&1&1\end{smallmatrix}\right]+C\left[\begin{smallmatrix}1&0&2\\1&1&1\end{smallmatrix}\right]+C\left[\begin{smallmatrix}1&1&1\\1&1&1\end{smallmatrix}\right]==0$

Out[87]= $C\left[\begin{smallmatrix}1&1&2\\0&1&1\end{smallmatrix}\right]+C\left[\begin{smallmatrix}1&1&2\\1&0&1\end{smallmatrix}\right]+C\left[\begin{smallmatrix}1&1&2\\1&1&0\end{smallmatrix}\right]==0$

In[88]:= **DiHolMomConsId**$\left[\mathbf{c}\left[\begin{smallmatrix}0&1&2\\1&0&2\end{smallmatrix}\right]\right]$

DiHolMomConsId : You are generating the holomorphic momentum−conservation identity of the divergent seed $C\left[\begin{smallmatrix}0&1&2\\1&0&2\end{smallmatrix}\right]$. Divergent seeds can lead to inconsistent identities.

Out[88]= $C\left[\begin{smallmatrix}-1&1&2\\1&0&2\end{smallmatrix}\right]+C\left[\begin{smallmatrix}0&0&2\\1&0&2\end{smallmatrix}\right]+C\left[\begin{smallmatrix}0&1&1\\1&0&2\end{smallmatrix}\right]==0$

**DiCSimplify**

The function **DiCSimplify** performs all known dihedral simplifications.

ARGUMENT **DiCSimplify** accepts one arbitrary argument.

RETURN VALUE **DiCSimplify** returns the expression given as the argument with all dihedral MGFs (including one-loop graphs such as Eisenstein series) rewritten in a simplified form, if possible. This is done by performing the following manipulations on all dihedral graphs, until the result does not change any more.

1. Apply HSR (5.73) and (5.178)
2. Set $C[\varnothing]=1$, cf. (5.22)
3. Factorize on $\left[\begin{smallmatrix}0\\0\end{smallmatrix}\right]$ columns according to (5.46) and (5.202)
4. Set $C\left[\begin{smallmatrix}a\\b\end{smallmatrix}\right]=0$, cf. (5.21)
5. Remove entries of $-1$ by using momentum conservation as described in Section 5.5
6. Sort dihedral MGFs into their canonical representation as described in Section 5.3.1
7. Set graphs with odd $|A|+|B|$ to zero
8. Rewrite $C\left[\begin{smallmatrix}k&0\\0&0\end{smallmatrix}\right]=\mathrm{G}_k$ and c.c., cf. (3.129)
9. Rewrite $C\left[\begin{smallmatrix}2&0\\0&0\end{smallmatrix}\right]=\widehat{\mathrm{G}}_2$ and c.c., cf. (3.131)
10. Set $\mathrm{G}_k$ with $k$ odd to zero and c.c.
11. Rewrite $C\left[\begin{smallmatrix}k&0\\k&0\end{smallmatrix}\right]=\left(\frac{\pi}{\tau_2}\right)^k\mathrm{E}_k$, cf. (3.130)

12. Apply generalized Ramanujan identities discussed in Section 5.3.5 and expand holomorphic Eisenstein series in the ring of $G_4$ and $G_6$
13. Apply basis decompositions discussed in Section 5.7, in the basis listed in Table 5.3.

Within this process, the steps 2 to 12 are repeated until the result no longer changes, before step 13 is executed.

OPTIONS

| option | possible values | default value | description |
|---|---|---|---|
| **basisExpandG** | **True**<br>**False** | **False** | activates step 12 |
| **momSimplify** | **True**<br>**False** | **True** | deactivates step 5 |
| **repGHat2** | **True**<br>**False** | **True** | deactivates step 9 |
| **useIds** | **True**<br>**False** | **True** | deactivates step 13 |
| **diHSR** | **True**<br>**False** | **True** | deactivates step 1 |
| **divHSR** | **True**<br>**False** | **True** | deactivates step 1<br>for divergent graphs |
| **diDivHSR** | **True**<br>**False** | **True** | deactivates step 1<br>for divergent graphs |

Both options **divHSR** and **diDivHSR** have to be **True** for divergent graphs to be included in step 1.

WARNINGS

- If a graph in the argument contains a $\begin{bmatrix} 0 \\ 0 \end{bmatrix}$ column next to a $\begin{bmatrix} 1 \\ 0 \end{bmatrix}$ or $\begin{bmatrix} 1 \\ 0 \end{bmatrix}$ column, the warning **DiCSimplify::dangerousFact** is issued and the modified factorization rule (5.202) applied.
- If a divergent graph with a holomorphic subgraph is encountered but HSR cannot be performed because either one of the options **divHSR** or **diDivHSR** is set to **False**, the warning **DiCSimplify::divHSRNotPossible** is issued.

EXAMPLES

In[89]:= **DiCSimplify**$\left[\mathtt{c}\begin{bmatrix}1&2&2&2\\0&0&1&2\end{bmatrix}\right]$

Out[89]= $3\,C\begin{bmatrix}3&0\\1&0\end{bmatrix}C\begin{bmatrix}4&0\\2&0\end{bmatrix}-15\,C\begin{bmatrix}7&0\\3&0\end{bmatrix}-9C\begin{bmatrix}0&2&5\\1&0&2\end{bmatrix}+\frac{21}{2}C\begin{bmatrix}1&1&5\\1&1&1\end{bmatrix}-C\begin{bmatrix}5&0\\3&0\end{bmatrix}\hat{G}_2+\frac{1}{2}C\begin{bmatrix}1&1&3\\1&1&1\end{bmatrix}\hat{G}_2-\frac{2\,\pi^2\,C\begin{bmatrix}5&0\\1&0\end{bmatrix}}{\tau_2^2}+\frac{6\,\pi\,C\begin{bmatrix}6&0\\2&0\end{bmatrix}}{\tau}-\frac{2\,\pi\,C\begin{bmatrix}4&0\\2&0\end{bmatrix}\hat{G}_2}{\tau_2}$

In[90]:= **DiCSimplify**$\left[\mathtt{c}\begin{bmatrix}1&2&2&2\\0&0&1&2\end{bmatrix}\right.$, **momSimplify → False,**
**useIds → False**$\Big]$

Out[90]= $-3\,C\begin{bmatrix}2&2&3\\1&2&0\end{bmatrix}+C\begin{bmatrix}1&2&2\\0&1&2\end{bmatrix}\hat{G}_2+\frac{\pi\,C\begin{bmatrix}2&2&2\\-1&1&2\end{bmatrix}}{\tau_2}$

In[91]:= **DiCSimplify**$\left[\mathtt{c}\begin{bmatrix}0&0&1&1&1&2&4\\0&1&1&1&1&3&4\end{bmatrix}\right]$ **//Simplify**

DiCSimplify : The graph $C\begin{bmatrix}0&0&1&1&1&2&4\\0&1&1&1&1&3&4\end{bmatrix}$ is factorized and contains a (1,0) or (0,1) column. This may be problematic.

Out[91]= $-\frac{\pi^8\,C\begin{bmatrix}1&0\\3&0\end{bmatrix}(-6+6\,E_1-3\,E_1^2+E_1^3)\,E_4+C\begin{bmatrix}0&1&1&1&2&4\\1&1&1&1&3&4\end{bmatrix}\tau_2^8}{\tau_2^8}$

### A.3.3 *Trihedral functions*

**TriHolMomConsId** and **TriAHolMomConsId**

The functions **TriHolMomConsId** and **TriAHolMomConsId** generate trihedral holomorphic and antiholomorphic momentum-conservation identities, respectively.

ARGUMENTS Both **TriHolMomConsId** and **TriAHolMomConsId** accept two arguments: The first is a trihedral MGF, the second is one of the lists **{1,2}**, **{2,3}** or **{1,3}**, where the order of the elements in the list does not matter.

RETURN VALUE **TriHolMomConsId** returns the holomorphic trihedral momentum-conservation identity (5.37) as an equation with RHS zero. Due to the permutation symmetry of the three blocks in a trihedral MGF, any two blocks can be involved in the momentum-conservation identity (i.e. have their holomorphic weight reduced) and the second argument of **TriHolMomConsId** specifies which two blocks should be used to generate the identity. **TriAHolMomConsId** generates the antiholomorphic momentum conservation identity. No further manipulation is performed on the output.

WARNINGS If either one of the blocks in the second argument is divergent as a dihedral MGF or if the trihedral MGF in the first argument has a three-point divergence (cf. (5.157)), the warning **`TriHolMomConsId::`**`divTriHolMomCons` (and c.c.) is issued.

EXAMPLES

In[92]:= $\texttt{TriHolMomConsId}\left[c\left[\begin{array}{c}1\\1\end{array},\begin{array}{cc}1&1\\1&1\end{array},\begin{array}{cc}1&1\\1&1\end{array}\right],\{1,2\}\right]$

Out[92]= $\mathrm{C}\left[\begin{array}{c|cc|cc}0&1&1&1&1\\1&1&1&1&1\end{array}\right]-\mathrm{C}\left[\begin{array}{c|cc|cc}1&0&1&1&1\\1&1&1&1&1\end{array}\right]-\mathrm{C}\left[\begin{array}{c|cc|cc}1&1&0&1&1\\1&1&1&1&1\end{array}\right]==0$

In[93]:= $\texttt{TriAHolMomConsId}\left[c\left[\begin{array}{c}1\\1\end{array},\begin{array}{cc}1&1\\1&1\end{array},\begin{array}{cc}1&1\\1&1\end{array}\right],\{3,2\}\right]$

Out[93]= $-\mathrm{C}\left[\begin{array}{c|cc|cc}1&1&1&1&1\\1&0&1&1&1\end{array}\right]-\mathrm{C}\left[\begin{array}{c|cc|cc}1&1&1&1&1\\1&1&0&1&1\end{array}\right]+\mathrm{C}\left[\begin{array}{c|cc|cc}1&1&1&1&1\\1&1&1&0&1\end{array}\right]+$

$\mathrm{C}\left[\begin{array}{c|cc|cc}1&1&1&1&1\\1&1&1&1&0\end{array}\right]==0$

In[94]:= $\texttt{TriHolMomConsId}\left[c\left[\begin{array}{c}1\\1\end{array},\begin{array}{cc}0&1\\1&0\end{array},\begin{array}{cc}1&1\\1&1\end{array}\right],\{1,3\}\right]$

Out[94]= $c\left[\begin{array}{c}0\\1\end{array},\begin{array}{cc}0&1\\1&0\end{array},\begin{array}{cc}1&1\\1&1\end{array}\right]-c\left[\begin{array}{c}1\\1\end{array},\begin{array}{cc}0&1\\1&0\end{array},\begin{array}{cc}0&1\\1&1\end{array}\right]-c\left[\begin{array}{c}1\\1\end{array},\begin{array}{cc}0&1\\1&0\end{array},\begin{array}{cc}1&0\\1&1\end{array}\right]==0$

## TriFay

The function **`TriFay`** generates trihedral Fay identities.

ARGUMENTS **`TriFay`** accepts up to two arguments. The first (mandatory) argument is a trihedral MGF, the second (optional) argument is a list of the form **`{{`**$\mathbf{b_1}$**`,`**$\mathbf{c_1}$**`},{`**$\mathbf{b_2}$**`,`**$\mathbf{c_2}$**`}}`**, where $\mathbf{c_i}$ is a column number in the $\mathbf{b_i}^{\text{th}}$ block and the list selects two columns, both of the form $\left[\begin{smallmatrix}a\\0\end{smallmatrix}\right]$ with $a \geq 1$ or $\left[\begin{smallmatrix}0\\b\end{smallmatrix}\right]$ with $b \geq 1$ in the trihedral graph. If the second argument is omitted, **`TriFay`** selects the first suitable pair of columns automatically, starting from the left and trying holomorphic column pairs first.

RETURN VALUE **`TriFay`** returns an equation in which the LHS is the graph specified in the first argument and the RHS is given by (5.127), with the columns $\left[\begin{smallmatrix}a_1\\0\end{smallmatrix}\right]$ and $\left[\begin{smallmatrix}a_2\\0\end{smallmatrix}\right]$ selected by the second argument or determined automatically. No further manipulations are performed on the output.

WARNINGS If no second argument is passed to **`TriFay`** and no suitable pair of columns could be found, the warning **`TriFay::`**`noFayCols` is issued.

EXAMPLES

In[95]:= $\texttt{TriFay}\left[c\left[\begin{array}{c}1\\0\end{array},\begin{array}{cc}1&2\\0&2\end{array},\begin{array}{cc}1&2\\1&0\end{array}\right]\right]$

Out[95]= $\mathrm{C}\left[\begin{array}{c|cc|cc}1&1&2&1&2\\0&0&2&1&0\end{array}\right]==\mathrm{C}\left[\{\}\,\middle|\,\begin{array}{c|ccc}2&1&2&2\\2&1&0&0\end{array}\right]-\mathrm{C}\left[\{\}\,\middle|\,\begin{array}{cc|ccc}1&2&1&1&2\\0&2&0&1&0\end{array}\right]+$

$\mathrm{C}\left[\{\}\,\middle|\,\begin{array}{cc|cc}1&2&2&2\\1&0&0&2\end{array}\right]-\mathrm{C}\left[\begin{array}{c|c|ccc}1&2&1&1&2\\0&2&0&1&0\end{array}\right]+\mathrm{C}\left[\begin{array}{c|c|cc}2&2&1&2\\0&2&1&0\end{array}\right]$

In[96]:= $\texttt{TriFay}\left[c\left[\begin{array}{c}0\\1\end{array},\begin{array}{cc}0&1\\2&1\end{array},\begin{array}{cc}0&2\\1&2\end{array}\right],\{\{1,1\},\{3,1\}\}\right]$

Out[96]= $c\left[\begin{array}{c}0\\1\end{array},\begin{array}{cc}0&1\\2&1\end{array},\begin{array}{cc}0&2\\1&2\end{array}\right]==\mathrm{C}\left[\{\}\,\middle|\,\begin{array}{c|ccc}2&0&0&1\\2&2&2&1\end{array}\right]+\mathrm{C}\left[\{\}\,\middle|\,\begin{array}{cc|cc}0&1&0&2\\2&1&2&2\end{array}\right]-$

$\mathrm{C}\left[\{\}\,\middle|\,\begin{array}{cc|ccc}0&2&0&0&1\\1&2&1&2&1\end{array}\right]-\mathrm{C}\left[\begin{array}{c|c|ccc}0&2&0&0&1\\1&2&1&2&1\end{array}\right]+\mathrm{C}\left[\begin{array}{c|c|cc}0&2&0&1\\2&2&2&1\end{array}\right]$

### `TriCSimplify`

The function **`TriCSimplify`** applies all known trihedral simplifications.

ARGUMENT **`TriCSimplify`** accepts one arbitrary argument.

RETURN VALUE **`TriCSimplify`** returns the expression given as the argument with all trihedral MGFs rewritten in a simplified form, if possible. This is done by performing the following manipulations on all trihedral graphs, until the result no more changes.

1. Apply two-point HSR (5.95)
2. Apply three-point HSR (5.109)
3. Set graphs with odd total modular weight $a + b$ to zero
4. Apply the topological simplification (5.25)
5. Apply the topological simplification (5.24)
6. Factorize on $\begin{bmatrix} 0 \\ 0 \end{bmatrix}$ columns according to (5.47) and (5.203)
7. Remove entries of $-1$ by using momentum conservation as described in Section 5.5
8. Sort trihedral MGFs into their canonical representation as described in Section 5.3.1
9. Apply basis decompositions discussed in Section 5.7, in the basis listed in Table 5.3.

Within this process, the steps 3 to 8 are repeated until the result no longer changes, before step 9 is executed.

OPTIONS

| option | possible values | default value | description |
|---|---|---|---|
| **`momSimplify`** | **`True`** **`False`** | **`True`** | deactivates step 7 |
| **`useIds`** | **`True`** **`False`** | **`True`** | deactivates step 9 |
| **`triHSR`** | **`True`** **`False`** | **`True`** | deactivates steps 1 and 2 |
| **`tri2ptHSR`** | **`True`** **`False`** | **`True`** | deactivates step 1 |
| **`tri3ptHSR`** | **`True`** **`False`** | **`True`** | deactivates step 2 |
| **`tri3ptFayHSR`** | **`True`** **`False`** | **`False`** | activates three-point HSR via the Fay identity (5.127) instead of (5.109) |
| **`divHSR`** | **`True`** **`False`** | **`True`** | deactivates steps 1 and 2 for divergent graphs |
| **triDivHSR** | **True**<br>**False** | **True** | deactivates steps 1 and 2 for divergent graphs |

Both options **divHSR** and **triDivHSR** have to be **True** for divergent graphs to be included in steps 1 and 2. Furthermore, if **tri3ptFayHSR** is set to **True**, setting **tri2ptHSR** to **False** also deactivates three-point HSR since, according to (5.122), a two- and three-point closed holomorphic subgraph together cannot be simplified with the Fay identity (5.127).

WARNINGS

- If a graph in the argument contains a $\left[\begin{smallmatrix}0\\0\end{smallmatrix}\right]$ column next to a $\left[\begin{smallmatrix}1\\0\end{smallmatrix}\right]$ or $\left[\begin{smallmatrix}1\\0\end{smallmatrix}\right]$ column, the warning **TriCSimplify::**dangerousFact is issued and the modified factorization rule (5.203) applied.
- If a divergent graph with a holomorphic subgraph is encountered but HSR cannot be performed because either one of the options **divHSR** or **triDivHSR** is set to **False**, the warning **TriCSimplify::**divHSRNotPossible is issued.
- If three-point HSR is performed on a divergent graph using Fay identities by setting the option **tri3ptFayHSR** to **True**, the warning **TriCSimplify::**div3ptFay is issued.
- If three-point HSR is performed via (5.109) and there is no ordering of the blocks which prevents a divergent expression in the result (cf. discussion below (5.109)), the warning **TriCSimplify::**noConvHSROrder is issued. If one of the options **divHSR** or **triDivHSR** is set to **False**, the warning **TriCSimplify::**divHSRNotPossible is issued and the HSR is not performed.

EXAMPLES

In[97]:= **TriCSimplify**$\left[\mathbf{c}\left[\begin{smallmatrix}1\\0\end{smallmatrix},\begin{smallmatrix}1&1\\0&1\end{smallmatrix},\begin{smallmatrix}1&2\\1&0\end{smallmatrix}\right]\right]$

Out[97]= $-6\,\mathcal{C}\left[\begin{smallmatrix}2&4\\2&0\end{smallmatrix}\right]+2\,\mathcal{C}\left[\begin{smallmatrix}3&1&2\\1&1&0\end{smallmatrix}\right]-6\,\mathcal{C}\left[\begin{smallmatrix}4&1&1\\1&0&1\end{smallmatrix}\right]+\mathcal{C}\left[\begin{smallmatrix}1\\1\end{smallmatrix}\right]^2 G_4+$
$2\,\mathcal{C}\left[\begin{smallmatrix}2&1&1\\1&0&1\end{smallmatrix}\right]\hat{G}_2+\frac{2\pi\,\mathcal{C}\left[\begin{smallmatrix}2&3\\2&-1\end{smallmatrix}\right]}{\tau_2}+\frac{2\pi\,\mathcal{C}\left[\begin{smallmatrix}3&1&1\\0&0&1\end{smallmatrix}\right]}{\tau_2}$

In[98]:= **TriCSimplify**$\left[\mathbf{c}\left[\begin{smallmatrix}1\\0\end{smallmatrix},\begin{smallmatrix}1&1\\0&1\end{smallmatrix},\begin{smallmatrix}1&2\\1&0\end{smallmatrix}\right]\right.$, **tri3ptFayHSR** → **True**$\left.\right]$

Out[98]= $\mathcal{C}\left[\begin{smallmatrix}1\\1\end{smallmatrix}\right]\mathcal{C}\left[\begin{smallmatrix}1&1\\0&1\end{smallmatrix}\right]-\mathcal{C}\left[\begin{smallmatrix}2&1&3\\1&1&0\end{smallmatrix}\right]-\mathcal{C}\left[\begin{smallmatrix}2&1&3\\2&0&0\end{smallmatrix}\right]+$
$3\,\mathcal{C}\left[\begin{smallmatrix}4&1&1\\1&0&1\end{smallmatrix}\right]-\mathcal{C}\left[\begin{smallmatrix}2&1&1\\1&0&1\end{smallmatrix}\right]\hat{G}_2-\frac{\pi\,\mathcal{C}\left[\begin{smallmatrix}3&1&1\\0&0&1\end{smallmatrix}\right]}{\tau_2}$

In[99]:= **DiCSimplify[Out[97]-Out[98]]**

Out[99]= 0

In[100]:= **TriCSimplify**$\left[c\left[\begin{smallmatrix}1\\0\end{smallmatrix},\begin{smallmatrix}1&2\\0&0\end{smallmatrix},\begin{smallmatrix}1&2\\1&0\end{smallmatrix}\right],\right.$

**tri2ptHSR → False, tri3ptFayHSR → True**$\left.\right]$

Out[100]= $\mathrm{C}\left[\begin{smallmatrix}1\\0\end{smallmatrix}\middle|\begin{smallmatrix}1&2\\0&0\end{smallmatrix}\middle|\begin{smallmatrix}1&2\\1&0\end{smallmatrix}\right]$

### A.3.4 *Four-point simplification*

#### TetCSimplify

The function **TetCSimplify** applies topological simplifications on four-point graphs.

ARGUMENT **TetCSimplify** accepts one arbitrary argument.

RETURN VALUE **TetCSimplify** returns the expression given as the argument with all four-point MGFs rewritten in a simplified form, if possible. This is done by performing the following manipulations on all four-point graphs (not only tetrahedral ones), until the result does not change any more.

1. Set graphs with odd total modular weight $a + b$ to zero
2. Apply the topological simplification (5.26)
3. Apply the topological simplifications (5.27) and (5.28)
4. Apply the topological simplifications (5.29) and (5.30)
5. Set four-point MGFs to zero which vanish by symmetry, cf. (5.17)
6. Sort four-point MGFs into their canonical representation as described in Section 5.3.1

EXAMPLES

In[101]:= **TetCSimplify**$\left[c\left[\begin{smallmatrix}1&1\\1&1\end{smallmatrix},\begin{smallmatrix}1&1\\1&1\end{smallmatrix},\begin{smallmatrix}1&2\\1&1\end{smallmatrix},\begin{smallmatrix}2&2\\1&1\end{smallmatrix},\begin{smallmatrix}2&2\\1&1\end{smallmatrix},\begin{smallmatrix}1&1\\1&2\end{smallmatrix}\right]\right]$

Out[101]= 0

In[102]:= **TetCSimplify**$\left[c\left[\{\},\begin{smallmatrix}1\\1\end{smallmatrix},\begin{smallmatrix}1\\1\end{smallmatrix},\begin{smallmatrix}1\\1\end{smallmatrix},\{\},\begin{smallmatrix}1\\1\end{smallmatrix}\right]\right]$

Out[102]= $\mathrm{C}\left[\begin{smallmatrix}1\\1\end{smallmatrix}\right]\mathrm{C}\left[\begin{smallmatrix}1\\1\end{smallmatrix}\middle|\begin{smallmatrix}1\\1\end{smallmatrix}\middle|\begin{smallmatrix}1\\1\end{smallmatrix}\right]$

### A.3.5 *Koba–Nielsen integration*

#### zIntegrate

The function **zIntegrate** expands Koba–Nielsen integrals in terms of MGFs.

ARGUMENTS **zIntegrate** represents a Koba–Nielsen integral and accepts three arguments. The first argument should be a polynomial in the objects with suffix **z** introduced in Section A.2.4, specifying the prefactor of the Koba–Nielsen factor. The second argument should be a natural number specifying the number of punctures in the Koba–Nielsen factor or a list of pairs of natural numbers **{{i, j}, {k, l}, …}**, specifying the Green functions (and associated Mandelstam variables) appearing in the Koba–Nielsen factor. The third argument should be a natural number specifying the order to which the Koba–Nielsen integral is to be expanded.

RETURN VALUE **zIntegrate** returns the order specified by the last argument of the Koba–Nielsen integral specified by the first two arguments. The resulting MGFs are simplified using the general properties listed below (3.131) and all the techniques implemented in **CSimplify**, apart from HSR and the application of the basis decompositions from Section 5.7. If the resulting MGFs require graphs with more than four vertices, for which no notation was defined, a graphical representation of those graphs is printed. No constraints are placed on the Mandelstam variables.

EXAMPLES

```
In[103]:= zIntegrate[vz[2, {1, 2}] + vz[2, {3, 4}], 4, 1] // Simplify
          CSimplify[%]
```

Out[103]= $-\dfrac{\left(2\,\mathrm{C}\left[\begin{smallmatrix}3&0\\1&0\end{smallmatrix}\right]+\mathrm{C}\left[\begin{smallmatrix}1&1&1\\0&0&1\end{smallmatrix}\right]\right)(s_{1,2}+s_{3,4})\tau_2}{\pi}$

Out[104]= $-\hat{G}_2\, s_{1,2}-\hat{G}_2\, s_{3,4}$

```
In[105]:= zIntegrate[vz[2, {1, 2}] vz[2, {3, 4}], 4, 1] // Simplify
          CSimplify[%]
```

Out[105]= $-\dfrac{\left(2\,\mathrm{C}\left[\begin{smallmatrix}3&0\\1&0\end{smallmatrix}\right]+\mathrm{C}\left[\begin{smallmatrix}1&1&1\\0&0&1\end{smallmatrix}\right]\right)\hat{G}_2\,(s_{1,2}+s_{3,4})\tau_2}{\pi}$

Out[106]= $-\hat{G}_2^2\, s_{1,2}-\hat{G}_2^2\, s_{3,4}$

```
In[107]:= zIntegrate[fz[1, 1, 2] fBarz[1, 1, 3], 3, 2] // Simplify
          CSimplify[%] // Simplify
```

Out[107]= $\dfrac{s_{2,3}\left(-2\,\mathrm{C}\left[\begin{smallmatrix}1&1&1\\0&1&2\end{smallmatrix}\right]s_{1,2}-2\,\mathrm{C}\left[\begin{smallmatrix}0&1&2\\1&1&1\end{smallmatrix}\right]s_{1,3}+\mathrm{C}\left[\begin{smallmatrix}1&1&1\\1&1&1\end{smallmatrix}\right]s_{2,3}\right)\tau_2^2}{2\,\pi^2}$

Out[108]= $\dfrac{\pi\ s_{2,3}\ (s_{1,2}+s_{1,3}+s_{2,3})\ (E_3+\zeta_3)}{2\ \tau_2}$

## A.4 EXAMPLE: FOUR-GLUON SCATTERING IN THE HETEROTIC STRING

In this section, we use the functions introduced above to reproduce the expansions (6.58) and (6.59) for the integrals $\mathcal{I}^{(2,0)}_{1234}$ and $\mathcal{I}^{(4,0)}_{1234}$ defined in (6.27) and (6.28) which appear in the planar sector of four-gluon scattering in the heterotic string.

All of the steps in the calculation are automatized, with one exception: The four-point HSR-identity (5.124) has to be added by hand. To this end, we first define the replacement rule

In[109]:=
$$\texttt{tetrule} = c\left[\begin{smallmatrix}1\\0\end{smallmatrix},\begin{smallmatrix}1\\0\end{smallmatrix},\begin{smallmatrix}1\\1\end{smallmatrix},\begin{smallmatrix}1\\0\end{smallmatrix},\begin{smallmatrix}1\\0\end{smallmatrix},\begin{smallmatrix}1\\1\end{smallmatrix}\right] \to -c\left[\begin{smallmatrix}1\\0\end{smallmatrix},\{\},\begin{smallmatrix}1&2\\1&0\end{smallmatrix},\begin{smallmatrix}1\\0\end{smallmatrix},\begin{smallmatrix}1\\0\end{smallmatrix},\begin{smallmatrix}1\\1\end{smallmatrix}\right] - c\left[\begin{smallmatrix}1\\0\end{smallmatrix},\{\},\begin{smallmatrix}1\\1\end{smallmatrix},\begin{smallmatrix}2\\0\end{smallmatrix},\begin{smallmatrix}1\\0\end{smallmatrix},\begin{smallmatrix}1\\1\end{smallmatrix}\right] - c\left[\begin{smallmatrix}1\\0\end{smallmatrix},\begin{smallmatrix}2\\0\end{smallmatrix},\begin{smallmatrix}1\\1\end{smallmatrix},\{\},\begin{smallmatrix}1\\0\end{smallmatrix},\begin{smallmatrix}1\\1\end{smallmatrix}\right] + c\left[\begin{smallmatrix}1\\0\end{smallmatrix},\{\},\begin{smallmatrix}1&1\\1&0\end{smallmatrix},\begin{smallmatrix}1\\0\end{smallmatrix},\begin{smallmatrix}1\\0\end{smallmatrix},\begin{smallmatrix}1\\1\end{smallmatrix}\right] - c\left[\begin{smallmatrix}1\\0\end{smallmatrix},\begin{smallmatrix}1\\0\end{smallmatrix},\begin{smallmatrix}1&1\\1&0\end{smallmatrix},\{\},\begin{smallmatrix}1\\0\end{smallmatrix},\begin{smallmatrix}1\\1\end{smallmatrix}\right];$$

In order to bring the output into a nice form, we furthermore define the helper function

```
In[110]:= prettify[poly_] := Block[{ap, mandOrd, result},
          mandOrd = MonomialList[poly, s[1,2], s[2,3]]
             /.List[x__] :> Plus[x];
          result = DeleteCases[DeleteDuplicates[Flatten[
             CoefficientList[mandOrd, s[1,2], s[2,3]]]], 0];
          result = Collect[mandOrd, result];
          result = (SortBy[({Exponent[#/. s[i_, j_] :> ap s[i,j],
             ap], #}&)/@(List@@result), First][[All, 2]])
             /. List[x__] :> HoldForm[Plus[x]];
          Return[result]];
```

The integral $\mathcal{I}^{(4,0)}_{1234}$ can now be expanded to second order by running

```
In[111]:= Sum[zIntegrate[vz[4, {1, 2, 3, 4}], 4, i], {i, 0, 2}];
```

To this we apply the four-point HSR-rule from above, decompose all resulting MGFs into the basis from Table 5.3 and change the basis to Table 5.4,

```
In[112]:= %/. tetrule // CSimplify // CConvertToNablaE;
```

Since **zIntegrate** does not apply momentum conservation to the Mandelstam variables, we do this explicitly,

```
In[113]:= %//. {s[3, 4] → s[1, 2], s[1, 4] → s[2, 3],
          s[2, 4] → s[1, 3], s[2, 3] → - s[1, 2] - s[1, 3]};
```

Finally, we apply the function **prettify** defined in In[110] to rearrange the output

```
In[114]:= prettify[%]
```

Out[114]=
$$G_4 + (s_{1,2} + s_{2,3})\left(-6\,G_4 - \frac{3\,\pi\,\hat{G}_2\,\nabla E_2}{\tau_2^2}\right) + (s_{1,2}^2 + s_{2,3}^2)\left(2\,E_2\,G_4 + \frac{\pi^2\,\nabla^2 E_3}{6\,\tau_2^4} + \frac{2\,\pi\,\hat{G}_2\,\nabla E_3}{3\,\tau_2^2}\right) + s_{1,2}\,s_{2,3}\left(2\,E_2\,G_4 + \frac{2\,\pi^2\,\nabla^2 E_3}{3\,\tau_2^4} + \frac{8\,\pi\,\hat{G}_2\,\nabla E_3}{3\,\tau_2^2}\right),$$

which agrees with (6.59). The Laurent polynomial of the first orders of $\mathcal{I}^{(4,0)}_{1234}$ can now easily be obtained by

```
In[115]:= prettify[CLaurentPoly[ReleaseHold[%]]]
```

Out[115]=
$$\frac{\pi^4}{45}+(s_{1,2}+s_{2,3})\left(-\frac{2\pi^4 y}{45}-\frac{3\pi^4\zeta_3}{y^3}+\frac{\pi^4\zeta_3}{y^2}\right)+ s_{1,2}s_{2,3}\left(\frac{94\pi^4 y^2}{14175}+\frac{2\pi^4\zeta_3}{45\,y}+\frac{5\pi^4\zeta_5}{y^4}-\frac{\pi^4 4\zeta_5}{3y^3}\right)+ (s_{1,2}^2+s_{2,3}^2)\left(\frac{34\pi^4 y^2}{14175}+\frac{2\pi^4\zeta_3}{45\,y}+\frac{5\pi^4\zeta_5}{y^4}-\frac{\pi^4\zeta_5}{3y^3}\right).$$

Similarly, we can expand $\mathcal{I}^{(2,0)}_{1234}$ to third order by running

```
In[116]:= Sum[zIntegrate[vz[2, {1, 2, 3, 4}], 4, i], {i, 0, 3}];
          %//CSimplify//CConvertToNablaE;
          %//.{s[3, 4] → s[1, 2], s[1, 4] → s[2, 3],
            s[2, 4] → s[1, 3], s[2, 3] → - s[1, 2] - s[1, 3]};
          prettify[%]
```

Out[116]=
$$-\frac{3\pi\nabla E_2(s_{1,2}+s_{2,3})}{\tau_2^2}+\frac{8\pi\nabla E_3\, s_{1,2}s_{2,3}}{3\tau_2^2}+\frac{2\pi\nabla E_3(s_{1,2}^2+s_{2,3}^2)}{3\tau_2^2}+ (s_{1,2}^2 s_{2,3}+s_{1,2}s_{2,3}^2)\left(-\frac{12\pi E_2\nabla E_2}{\tau_2^2}-\frac{8\pi\nabla E_4}{5\tau_2^2}-\frac{24\pi\nabla E_{2,2}}{\tau_2^2}\right)+ (s_{1,2}^3+s_{2,3}^3)\left(-\frac{6\pi E_2\nabla E_2}{\tau_2^2}-\frac{4\pi\nabla E_4}{5\tau_2^2}-\frac{12\pi\nabla E_{2,2}}{\tau_2^2}\right),$$

in agreement with (6.58). The next higher order in $\alpha'$ of $\mathcal{I}^{(2,0)}_{1234}$ contains two tetrahedral graphs. One of them vanishes by symmetry, the other one can be reduced to trihedral graphs by means of the Fay identity (5.122),

$$\mathcal{C}\left[\begin{smallmatrix}1\\0\\\hline 1\\0\end{smallmatrix}\middle|\middle|\begin{smallmatrix}1\\1\\\hline 1\\1\end{smallmatrix}\middle|\middle|\begin{smallmatrix}1\\1\\\hline 1\\1\end{smallmatrix}\right]=0 \tag{A.1}$$

$$\mathcal{C}\left[\begin{smallmatrix}1\\0\\\hline 1\\1\end{smallmatrix}\middle|\middle|\begin{smallmatrix}1\\0\\\hline 1\\1\end{smallmatrix}\middle|\middle|\begin{smallmatrix}1\\1\\\hline 1\\1\end{smallmatrix}\right]=-\mathcal{C}\left[\begin{smallmatrix}\varnothing\\\hline 1\\1\end{smallmatrix}\middle|\middle|\begin{smallmatrix}\varnothing\\\hline 1\\1\end{smallmatrix}\middle|\middle|\begin{smallmatrix}1\\1\\\hline 1\,2\\1\,0\end{smallmatrix}\right]-\mathcal{C}\left[\begin{smallmatrix}\varnothing\\\hline 1\\1\end{smallmatrix}\middle|\middle|\begin{smallmatrix}2\\0\\\hline 1\\1\end{smallmatrix}\middle|\middle|\begin{smallmatrix}1\\1\\\hline 1\\1\end{smallmatrix}\right]-\mathcal{C}\left[\begin{smallmatrix}2\\0\\\hline 1\\1\end{smallmatrix}\middle|\middle|\begin{smallmatrix}\varnothing\\\hline 1\\1\end{smallmatrix}\middle|\middle|\begin{smallmatrix}1\\1\\\hline 1\\1\end{smallmatrix}\right] +\mathcal{C}\left[\begin{smallmatrix}\varnothing\\\hline 1\\1\end{smallmatrix}\middle|\middle|\begin{smallmatrix}1\\0\\\hline 1\\1\end{smallmatrix}\middle|\middle|\begin{smallmatrix}1\\1\\\hline 1\,1\\0\,1\end{smallmatrix}\right]-\mathcal{C}\left[\begin{smallmatrix}1\\0\\\hline 1\\1\end{smallmatrix}\middle|\middle|\begin{smallmatrix}\varnothing\\\hline 1\\1\end{smallmatrix}\middle|\middle|\begin{smallmatrix}1\\1\\\hline 1\,1\\0\,1\end{smallmatrix}\right]. \tag{A.2}$$

If the Fay identity (A.2) is added by hand, similarly to how (5.124) was added above, the expansion of $\mathcal{I}^{(2,0)}_{1234}$ can be extended to the order $\alpha'^4$.

# B

## FURTHER DETAILS ABOUT CHAPTER 5

This appendix contains various details about the properties of MGFs discussed in Chapter 5 and has extensive text overlap with [I].

### B.1 EISENSTEIN SUMMATION

In this Appendix, we apply the Eisenstein summation prescription to sums which are needed to evaluate the expression (5.81). It is a slightly extended version of Appendix B of [I]. As mentioned in (5.72), the Eisenstein summation prescription is defined as

$$\begin{aligned}\sum_{p\neq r+s\tau}^{\mathrm{E}} f(p) = \lim_{N\to\infty} \sum_{\substack{n=-N\\ n\neq s}}^{N} \left( \lim_{M\to\infty} \sum_{m=-M}^{M} f(m+n\tau) \right) \\ + \lim_{M\to\infty} \sum_{\substack{m=-M\\ m\neq r}}^{M} f(m+s\tau)\,,\end{aligned} \tag{B.1}$$

where $f$ is assumed to have a pole at $r+s\tau$ with $(r,s)\in\mathbb{Z}^2\setminus\{0\}$ and takes on finite values at all other lattice points. If a finite number of additional points are excluded from the sum, they have to be subtracted from the right-hand side.

We first consider the case $f(p) = 1/p$ with the points $0$ and $P = \{p_i = m_i + n_i\tau \,|\, i = 1,\dots,n\}$ being excluded from the sum. Then the second term in (B.1) is as sum over $1/m$ and vanishes by antisymmetry. For the first term, we use the trigonometric identity

$$\lim_{M\to\infty} \sum_{m=-M}^{M} \frac{1}{m+n\tau} = -i\pi \frac{1+q^n}{1-q^n}\,. \tag{B.2}$$

The sum of this over $n$ also vanishes by antisymmetry. Hence, the only remaining term is due to the excluded points in $P$ and we obtain

$$\sum_{p\notin P}^{\prime\,\mathrm{E}} \frac{1}{p} = -\sum_{p\in P} \frac{1}{p} \tag{B.3}$$

In a similar fashion, we may now consider the case in which $f(p) = 1/(p_i - p)$ and the points in $P \cup \{0\}$ are excluded. Now, the second term in (B.1) is

$$-\sum_{\substack{m=-M-m_i \\ m\neq 0}}^{M-m_i} \frac{1}{m} = -\sum_{\substack{m=-M \\ m\neq 0}}^{M} \frac{1}{m} - \sum_{m=-M-m_i}^{-M-1} \frac{1}{m} + \sum_{m=M-m_i+1}^{M} \frac{1}{m}, \tag{B.4}$$

where we set $p_i = m_i + n_i\tau$ and shifted the summation range. The first sum in (B.4) vanishes due to antisymmetry and in the limit $M \longrightarrow \infty$, and the last two vanish since they scale as $\mathcal{O}(1/M)$. For the first term in (B.1), we use again (B.2) and obtain

$$-i\pi \sum_{\substack{n=-N+n_i \\ n\neq 0}}^{N+n_i} \frac{1+q^n}{1-q^n} = -i\pi \left[ \sum_{\substack{n=-N \\ n\neq 0}}^{N} + \sum_{n=-N+1}^{N+n_i} - \sum_{n=-N}^{-N+n_i-1} \right] \frac{1+q^n}{1-q^n}. \tag{B.5}$$

Again, the first sum vanishes by antisymmetry as above, but since

$$\frac{1+q^n}{1-q^n} \longrightarrow \pm 1 \quad (n \longrightarrow \pm\infty), \tag{B.6}$$

the last two sums do not vanish in the limit $N \longrightarrow \infty$, but instead we get

$$-i\pi \lim_{N\to\infty} \sum_{\substack{n=-N+n_i \\ n\neq 0}}^{N+n_i} \frac{1+q^n}{1-q^n} = -\frac{\pi}{\tau_2}(p_i - \bar{p}_i). \tag{B.7}$$

Taking also the excluded terms into account, the final result is

$$\sum_{p\notin P}{}'_{\mathrm{E}} \frac{1}{p_i - p} = -\frac{1}{p_i} - \sum_{\substack{p\in P \\ p\neq p_i}} \frac{1}{p_i - p} - \frac{\pi}{\tau_2}(p_i - \bar{p}_i). \tag{B.8}$$

Note that this illustrates an important point: upon Eisenstein summation the sum $\sum_p'(1/p)$ is not invariant under shifts of the summation variable $p \longrightarrow p - p_i$. This effect does not occur for sums of the form $\sum_p'(1/p^k)$ with $k \geq 3$ since they are absolutely convergent. So in these cases, we have just

$$\sum_{p\notin P}{}'_{\mathrm{E}} \frac{1}{p^k} = \mathrm{G}_k - \sum_{p\in P} \frac{1}{p^k} \qquad k \geq 3. \tag{B.9}$$

For the case $k = 2$, the sum is conditionally convergent. But because

$$\lim_{M\to\infty} \sum_{m=-M}^{M} \frac{1}{(m+n\tau)^2} = -4\pi^2 \frac{q^n}{(1-q^n)^2} \longrightarrow 0 \quad (n \longrightarrow \pm\infty), \tag{B.10}$$

the Eisenstein summation in this case is found to not be shift dependent, leaving us with the result (cf. (3.32))

$$\sum_{p\notin P}{}^{\prime}{}_{\mathrm{E}}\,\frac{1}{p^2}=\widehat{\mathrm{G}}_2+\frac{\pi}{\tau_2}-\sum_{p\in P}\frac{1}{p^2}\,. \tag{B.11}$$

## B.2 TRIHEDRAL HOLOMORPHIC SUBGRAPH REDUCTION

In Section 5.4.3, we considered trihedral holomorphic subgraph reduction of graphs with three-point holomorphic subgraphs, i.e. decompositions of graphs of the form

$$\mathcal{C}\left[\begin{smallmatrix}A_1 & a_2\\ B_1 & 0\end{smallmatrix}\middle|\begin{smallmatrix}A_3 & a_4\\ B_3 & 0\end{smallmatrix}\middle|\begin{smallmatrix}A_5 & a_6\\ B_5 & 0\end{smallmatrix}\right]=\sum_{\{p_i^{(n_i)}\}}{}^{\prime}\;\sum_{p_6\neq\mathfrak{p}_{15},\mathfrak{p}_{35}}{}^{\prime}\left(\prod\frac{1}{\mathfrak{p}^A\bar{\mathfrak{p}}^B}\right)\frac{1}{p_6^{a_6}(p_6-\mathfrak{p}_{15})^{a_2}(p_6-\mathfrak{p}_{35})^{a_4}}\,, \tag{B.12}$$

with the notation explained in (5.96) and (5.98). We decomposed the sum over $p_6$ into the six contributions $\mathcal{L}_1,\dots,\mathcal{L}_6$, defined in (5.101), and performed the sum over $p_6$ in each of these, resulting in the expressions (5.102), (5.105) and (5.107) and relabelings of (5.105). These have to be plugged back into (B.12) and summed over the remaining momenta (which just means they have to be expressed in terms of MGFs) in order to find the final decomposition formula. These steps are discussed in this appendix, which is largely identical to the Sections 4.2.3 and 4.2.4 of [I].

To begin, we have (recall that $a_0 = a_2 + a_4 + a_6$)

$$L_1=\sum_{\{p_i^{(n_i)}\}}{}^{\prime}\left(\prod\frac{1}{\mathfrak{p}^A\bar{\mathfrak{p}}^B}\right)\mathcal{L}_1\,\delta_{\mathfrak{p}_1,\mathfrak{p}_3}\delta_{\mathfrak{p}_3,\mathfrak{p}_5}=\mathrm{G}_{a_0}\,\mathcal{C}\left[\begin{smallmatrix}A_1\\ B_1\end{smallmatrix}\middle|\begin{smallmatrix}A_3\\ B_3\end{smallmatrix}\middle|\begin{smallmatrix}A_5\\ B_5\end{smallmatrix}\right]. \tag{B.13}$$

For the second contribution (5.105), we have

$$\begin{aligned}(-)^{a_2+a_4}L_2&=\sum_{\substack{\{p_i^{(n_i)}\}\\ \mathfrak{p}_1\neq\mathfrak{p}_5}}{}^{\prime}\left(\prod\frac{1}{\mathfrak{p}^A\bar{\mathfrak{p}}^B}\right)(-)^{a_2+a_4}\mathcal{L}_2\delta_{\mathfrak{p}_1,\mathfrak{p}_3}\\
&=-\binom{a_0}{a_6}\mathcal{C}\left[\begin{smallmatrix}A_1\\ B_1\end{smallmatrix}\middle|\begin{smallmatrix}A_3\\ B_3\end{smallmatrix}\middle|\begin{smallmatrix}A_5 & a_0\\ B_5 & 0\end{smallmatrix}\right]\\
&\quad+\sum_{k=4}^{a_6}\binom{a_0-k-1}{a_6-k}\mathrm{G}_k\,\mathcal{C}\left[\begin{smallmatrix}A_1\\ B_1\end{smallmatrix}\middle|\begin{smallmatrix}A_3\\ B_3\end{smallmatrix}\middle|\begin{smallmatrix}A_5 & a_0-k\\ B_5 & 0\end{smallmatrix}\right]\\
&\quad+\sum_{k=4}^{a_2+a_4}\binom{a_0-k-1}{a_2+a_4-k}\mathrm{G}_k\,\mathcal{C}\left[\begin{smallmatrix}A_1\\ B_1\end{smallmatrix}\middle|\begin{smallmatrix}A_3\\ B_3\end{smallmatrix}\middle|\begin{smallmatrix}A_5 & a_0-k\\ B_5 & 0\end{smallmatrix}\right]\\
&\quad+\binom{a_0-2}{a_6-1}\left\{\widehat{\mathrm{G}}_2\,\mathcal{C}\left[\begin{smallmatrix}A_1\\ B_1\end{smallmatrix}\middle|\begin{smallmatrix}A_3\\ B_3\end{smallmatrix}\middle|\begin{smallmatrix}A_5 & a_0-2\\ B_5 & 0\end{smallmatrix}\right]+\frac{\pi}{\tau_2}\,\mathcal{C}\left[\begin{smallmatrix}A_1\\ B_1\end{smallmatrix}\middle|\begin{smallmatrix}A_3\\ B_3\end{smallmatrix}\middle|\begin{smallmatrix}A_5 & a_0-1\\ B_5 & -1\end{smallmatrix}\right]\right\}.\end{aligned} \tag{B.14}$$

The third contribution can be obtained from this by relabeling $a_6 \to a_4 + a_6$ and $a_2 + a_4 \to a_2$, moving the $\mathfrak{p}_{15}$-column to the first block and introducing an overall sign,

$$
\begin{aligned}
(-)^{a_4+a_6} L_3 = &-\binom{a_0}{a_2} \, \mathcal{C}\big[\begin{smallmatrix} A_1 & a_0 \\ B_1 & 0 \end{smallmatrix}\big|\begin{smallmatrix} A_3 \\ B_3 \end{smallmatrix}\big|\begin{smallmatrix} A_5 \\ B_5 \end{smallmatrix}\big] \\
&+ \sum_{k=4}^{a_4+a_6} \binom{a_0-k-1}{a_4+a_6-k} \mathrm{G}_k \, \mathcal{C}\big[\begin{smallmatrix} A_1 & a_0-k \\ B_1 & 0 \end{smallmatrix}\big|\begin{smallmatrix} A_3 \\ B_3 \end{smallmatrix}\big|\begin{smallmatrix} A_5 \\ B_5 \end{smallmatrix}\big] \\
&+ \sum_{k=4}^{a_2} \binom{a_0-k-1}{a_2-k} \mathrm{G}_k \, \mathcal{C}\big[\begin{smallmatrix} A_1 & a_0-k \\ B_1 & 0 \end{smallmatrix}\big|\begin{smallmatrix} A_3 \\ B_3 \end{smallmatrix}\big|\begin{smallmatrix} A_5 \\ B_5 \end{smallmatrix}\big] \\
&+ \binom{a_0-2}{a_4+a_6-1} \Big\{ \widehat{\mathrm{G}}_2 \, \mathcal{C}\big[\begin{smallmatrix} A_1 & a_0-2 \\ B_1 & 0 \end{smallmatrix}\big|\begin{smallmatrix} A_3 \\ B_3 \end{smallmatrix}\big|\begin{smallmatrix} A_5 \\ B_5 \end{smallmatrix}\big] + \frac{\pi}{\tau_2} \mathcal{C}\big[\begin{smallmatrix} A_1 & a_0-1 \\ B_1 & -1 \end{smallmatrix}\big|\begin{smallmatrix} A_3 \\ B_3 \end{smallmatrix}\big|\begin{smallmatrix} A_5 \\ B_5 \end{smallmatrix}\big] \Big\} .
\end{aligned}
\tag{B.15}
$$

The fourth contribution can be obtained by relabeling $a_6 \to a_2 + a_6$ and $a_2 + a_4 \to a_4$ in (B.14), moving the $\mathfrak{p}_{15}$-column to the second block and introducing an overall sign,

$$
\begin{aligned}
(-)^{a_2+a_6} L_4 = &-\binom{a_0}{a_4} \, \mathcal{C}\big[\begin{smallmatrix} A_1 \\ B_1 \end{smallmatrix}\big|\begin{smallmatrix} A_3 & a_0 \\ B_3 & 0 \end{smallmatrix}\big|\begin{smallmatrix} A_5 \\ B_5 \end{smallmatrix}\big] \\
&+ \sum_{k=4}^{a_2+a_6} \binom{a_0-k-1}{a_2+a_6-k} \mathrm{G}_k \, \mathcal{C}\big[\begin{smallmatrix} A_1 \\ B_1 \end{smallmatrix}\big|\begin{smallmatrix} A_3 & a_0-k \\ B_3 & 0 \end{smallmatrix}\big|\begin{smallmatrix} A_5 \\ B_5 \end{smallmatrix}\big] \\
&+ \sum_{k=4}^{a_4} \binom{a_0-k-1}{a_4-k} \mathrm{G}_k \, \mathcal{C}\big[\begin{smallmatrix} A_1 \\ B_1 \end{smallmatrix}\big|\begin{smallmatrix} A_3 & a_0-k \\ B_3 & 0 \end{smallmatrix}\big|\begin{smallmatrix} A_5 \\ B_5 \end{smallmatrix}\big] \\
&+ \binom{a_0-2}{a_2+a_6-1} \Big\{ \widehat{\mathrm{G}}_2 \, \mathcal{C}\big[\begin{smallmatrix} A_1 \\ B_1 \end{smallmatrix}\big|\begin{smallmatrix} A_3 & a_0-2 \\ B_3 & 0 \end{smallmatrix}\big|\begin{smallmatrix} A_5 \\ B_5 \end{smallmatrix}\big] + \frac{\pi}{\tau_2} \mathcal{C}\big[\begin{smallmatrix} A_1 \\ B_1 \end{smallmatrix}\big|\begin{smallmatrix} A_3 & a_0-1 \\ B_3 & -1 \end{smallmatrix}\big|\begin{smallmatrix} A_5 \\ B_5 \end{smallmatrix}\big] \Big\} .
\end{aligned}
\tag{B.16}
$$

Finally, we must consider the contribution due to $\mathcal{L}_5$, which is given by (5.107) summed over the remaining momenta. To simplify the result, we introduce the following shorthand notation[1]

$$
\begin{aligned}
\mathcal{C}\big[\begin{smallmatrix} m_1 \\ n_1 \end{smallmatrix}\big|\begin{smallmatrix} m_2 \\ n_2 \end{smallmatrix}\big|\,\big] = (-1)^{m_1+n_1+m_2+n_2} \, \mathcal{C}\big[\begin{smallmatrix} A_1 & m_1 \\ B_1 & n_1 \end{smallmatrix}\big|\begin{smallmatrix} A_3 & m_2 \\ B_3 & n_2 \end{smallmatrix}\big|\begin{smallmatrix} A_5 \\ B_5 \end{smallmatrix}\big] \\
- \, \mathcal{C}\big[\begin{smallmatrix} A_1 \\ B_1 \end{smallmatrix}\big|\begin{smallmatrix} A_3 \\ B_3 \end{smallmatrix}\big|\begin{smallmatrix} A_5 & m_1+m_2 \\ B_5 & n_1+n_2 \end{smallmatrix}\big]
\end{aligned}
\tag{B.17a}
$$

$$
\begin{aligned}
\mathcal{C}\big[\,\big|\begin{smallmatrix} m_1 \\ n_1 \end{smallmatrix}\big|\begin{smallmatrix} m_2 \\ n_2 \end{smallmatrix}\big] = (-1)^{m_2+n_2} \, \mathcal{C}\big[\begin{smallmatrix} A_1 \\ B_1 \end{smallmatrix}\big|\begin{smallmatrix} A_3 & m_2 \\ B_3 & n_2 \end{smallmatrix}\big|\begin{smallmatrix} A_5 & m_1 \\ B_5 & n_1 \end{smallmatrix}\big] \\
- (-1)^{m_1+n_1} \, \mathcal{C}\big[\begin{smallmatrix} A_1 & m_1+m_2 \\ B_1 & n_1+n_2 \end{smallmatrix}\big|\begin{smallmatrix} A_3 \\ B_3 \end{smallmatrix}\big|\begin{smallmatrix} A_5 \\ B_5 \end{smallmatrix}\big] .
\end{aligned}
\tag{B.17b}
$$

Using the partial-fraction identity (5.65) one final time to decompose the $(p-q)^\ell$ term in (5.108), we find the result,

$$
(-1)^{a_2+a_4} L_5 = \sum_{k=1}^{a_6} \binom{a_2+a_6-k-1}{a_6-k} X_k(0) + \sum_{k=1}^{a_2} \binom{a_2+a_6-k-1}{a_2-k} (-1)^k \tilde{X}_k(1) ,
\tag{B.18}
$$

1 This is not to be confused with the notation $\mathcal{C}\big[\begin{smallmatrix} A_1 \\ B_1 \end{smallmatrix}\big|\begin{smallmatrix} A_2 \\ B_2 \end{smallmatrix}\big|\varnothing\big]$ which we use for trihedral graphs with one empty block, cf. (5.10).

where we have defined

$$
\begin{aligned}
X_k(\epsilon) = &-\binom{a_4+k}{a_4} \mathcal{C}\left[\begin{smallmatrix} a_2+a_6-k \\ 0 \end{smallmatrix}\middle|\begin{smallmatrix} a_4+k \\ 0 \end{smallmatrix}\middle|\,\right] \\
&-\sum_{\ell=1}^{k}\binom{a_4+k-\ell-1}{k-\ell}(-)^{\epsilon\ell}\,\mathcal{C}\left[\begin{smallmatrix} a_2+a_6-k+\ell \\ 0 \end{smallmatrix}\middle|\begin{smallmatrix} a_4+k-\ell \\ 0 \end{smallmatrix}\middle|\,\right] \\
&+\sum_{\ell=4}^{k}\binom{a_4+k-\ell-1}{k-\ell}\mathrm{G}_\ell\,\mathcal{C}\left[\begin{smallmatrix} a_2+a_6-k \\ 0 \end{smallmatrix}\middle|\begin{smallmatrix} a_4+k-\ell \\ 0 \end{smallmatrix}\middle|\,\right] \\
&+\sum_{\ell=4}^{a_4}\binom{a_4+k-\ell-1}{a_4-\ell}\mathrm{G}_\ell\,\mathcal{C}\left[\begin{smallmatrix} a_2+a_6-k \\ 0 \end{smallmatrix}\middle|\begin{smallmatrix} a_4+k-\ell \\ 0 \end{smallmatrix}\middle|\,\right] \qquad \text{(B.19)} \\
&+\binom{a_4+k-2}{k-1}\left\{\widehat{\mathrm{G}}_2\,\mathcal{C}\left[\begin{smallmatrix} a_2+a_6-k \\ 0 \end{smallmatrix}\middle|\begin{smallmatrix} a_4+k-2 \\ 0 \end{smallmatrix}\middle|\,\right]+\frac{\pi}{\tau_2}\,\mathcal{C}\left[\begin{smallmatrix} a_2+a_6-k \\ 0 \end{smallmatrix}\middle|\begin{smallmatrix} a_4+k-1 \\ -1 \end{smallmatrix}\middle|\,\right]\right\} \\
&-\sum_{\ell=1}^{a_4}\binom{a_4+k-\ell-1}{a_4-\ell}(-)^{\ell} \\
&\quad\times\left\{\sum_{m=1}^{a_4+k-\ell}\binom{a_4+k-m-1}{a_4+k-\ell-m}(-)^{\epsilon(a_4+k-m)}\,\mathcal{C}\left[\begin{smallmatrix} a_0-m \\ 0 \end{smallmatrix}\middle|\begin{smallmatrix} m \\ 0 \end{smallmatrix}\middle|\,\right]\right. \\
&\qquad\left.+\sum_{m=1}^{\ell}\binom{a_4+k-m-1}{\ell-m}(-)^{m}(-)^{\epsilon(a_4+k-m)}\,\mathcal{C}\left[\,\middle|\begin{smallmatrix} a_0-m \\ 0 \end{smallmatrix}\middle|\begin{smallmatrix} m \\ 0 \end{smallmatrix}\right]\right\}
\end{aligned}
$$

and $\tilde{X}_k(\epsilon)$ is obtained from $X_k(\epsilon)$ by replacing all $\mathcal{C}\left[\begin{smallmatrix} m_1 \\ n_1 \end{smallmatrix}\middle|\begin{smallmatrix} m_2 \\ n_2 \end{smallmatrix}\middle|\,\right]$ by $\mathcal{C}\left[\,\middle|\begin{smallmatrix} m_1 \\ n_1 \end{smallmatrix}\middle|\begin{smallmatrix} m_2 \\ n_2 \end{smallmatrix}\right]$ and vice versa.

This completes the derivation of the three-point holomorphic subgraph reduction formula, which as stated in (5.109) is given by

$$
\mathcal{C}\left[\begin{smallmatrix} A_1 & a_2 \\ B_1 & 0 \end{smallmatrix}\middle|\begin{smallmatrix} A_3 & a_4 \\ B_3 & 0 \end{smallmatrix}\middle|\begin{smallmatrix} A_5 & a_6 \\ B_5 & 0 \end{smallmatrix}\right] = \sum_{i=1}^{5} L_i\,. \qquad \text{(B.20)}
$$

with the $L_i$ defined in (B.13)–(B.16) and (B.18). Although this final result is rather lengthy, it is straightforward to implement it on a computer and it provides simplifications for all trihedral graphs with three-point holomorphic subgraphs.

Since we made a choice which momentum to sum over and how to perform the partial fraction decompositions, the MGFs appearing on the RHS of (B.20) depend on the order in which the three blocks of the trihedral function are plugged into the formula. In particular, for certain choices, divergent trihedral MGFs appear in the result (for a detailed discussion of divergent MGFs, see Section 5.6). These are signaled by $\left[\begin{smallmatrix} 1 & 1 \\ -1 & 1 \end{smallmatrix}\right]$-subblocks, since using momentum conservation identities to simplify these lead to subblocks of the form $\left[\begin{smallmatrix} 1 & 1 \\ 0 & 0 \end{smallmatrix}\right]$, indicating a divergent graph. Looking at the explicit expressions for the $L_i$ above (and recalling that $a_0 \geq 3$) shows that such $\left[\begin{smallmatrix} 1 \\ -1 \end{smallmatrix}\right]$ columns can only appear in the last term in the fifth line of (B.19) if $a_4 = 1$. In this case a $\left[\begin{smallmatrix} 1 \\ -1 \end{smallmatrix}\right]$ column is introduced in the second (middle) block of the modular graph form from both $X_1$ and $\tilde{X}_1$. This means that if $a_4 = 1$ and the $(A_3, B_3)$ block of the original modular graph form contains a $\left[\begin{smallmatrix}1\\1\end{smallmatrix}\right]$ column, divergent graphs will be produced by (B.20).

There is an easy way to avoid these divergent graphs: From the original definition of trihedral MGFs, it is irrelevant in which order the three blocks of exponents are written. We may then rearrange the three blocks in such a way that the middle block does not contain a $\left[\begin{smallmatrix}1&1\\0&1\end{smallmatrix}\right]$-subblock.

Just like the divergence appearing upon partial fraction decomposition, this divergence is artificial—it cancels out if the divergent graphs is decomposed further. This decomposition can be done by means of the holomorphic subgraph reduction of $\left[\begin{smallmatrix}1&1\\0&0\end{smallmatrix}\right]$ blocks using Fay identities described in Section 5.6.4. For graphs in which each block contains a $\left[\begin{smallmatrix}1&1\\0&1\end{smallmatrix}\right]$ subblock, divergent graphs in intermediate steps are unavoidable and hence the techniques from Section 5.6.4 have to be used in this case.

## B.3 KINEMATIC POLES IN THREE-POINT KOBA–NIELSEN INTEGRALS

As explained in Section 5.6.2, a factor $|f^{(1)}_{ij}|^2$ in a Koba–Nielsen integral leads to a naive expansion of this integral terms of divergent MGFs. This signals a pole in one or more of the Mandelstam variables which can be made explicit by means of integration-by-parts manipulations. In this appendix we discuss the resulting explicit expressions for all three-point Koba–Nielsen integrals containing $|f^{(1)}_{ij}|^2$ factors using the notation (7.18).

If only one $|f^{(1)}_{ij}|^2$ is present in the integrand and the other $f^{(a)}$, $\overline{f^{(b)}}$ do not depend on $z_i$ or $z_j$, we can use the puncture only occurring in $|f^{(1)}_{ij}|^2$ to integrate by parts, obtaining one more term compared to (5.169),

$$\begin{aligned} W^\tau_{(1,a|1,b)}(2{,}3|2{,}3) = (-)^{a+1}\frac{s_{13}}{s_{12}} W^\tau_{(1,a|1,b)}(2{,}3|3{,}2) \\ - \frac{1}{s_{12}}\frac{\pi}{\tau_2} W^\tau_{(0,a|0,b)}(2{,}3|2{,}3)\,, \end{aligned} \tag{B.21}$$

with $a \neq 1$ or $b \neq 1$. Three more cases can be obtained from (B.21) by relabeling of the Mandelstam variables,

$$W^\tau_{(a,1|b,1)}(2{,}3|2{,}3) = W^\tau_{(1,a|1,b)}(2{,}3|2{,}3)\Big|_{s_{12}\leftrightarrow s_{23}} \tag{B.22}$$

$$W^\tau_{(1,a|1,b)}(3{,}2|3{,}2) = W^\tau_{(1,a|1,b)}(2{,}3|2{,}3)\Big|_{s_{12}\leftrightarrow s_{13}} \tag{B.23}$$

$$W^\tau_{(a,1|b,1)}(3{,}2|3{,}2) = W^\tau_{(1,a|1,b)}(2{,}3|2{,}3)\Big|_{\substack{s_{12}\to s_{23}\\ s_{23}\to s_{13}\\ s_{13}\to s_{12}}}\,, \tag{B.24}$$

where again $a \neq 1$ or $b \neq 1$.

If both punctures $i$ and $j$ of $|f^{(1)}_{ij}|^2$ also appear in other $f^{(a)}$, $\overline{f^{(b)}}$ factors, one obtains an additional term from the action of $\partial_{\bar{z}}$ on the corresponding $f^{(a)}$ according to (3.94). In this way, we obtain

$$W^\tau_{(a,1|b,1)}(3{,}2|2{,}3) = \Bigg\{ \frac{s_{23}}{s_{13}} W^\tau_{(1,a|b,1)}(2{,}3|3{,}2) + \frac{(-)^b}{s_{13}} \frac{\pi}{\tau_2} \Big[ W^\tau_{(0,a|b,0)}(2{,}3|2{,}3) + (-)^{a-1} W^\tau_{(1,a-1|b,0)}(2{,}3|3{,}2) \Big] \Bigg\} \Bigg|_{\substack{s_{12}\to s_{13} \\ s_{13}\to s_{23} \\ s_{23}\to s_{12}}} \tag{B.25}$$

$$W^\tau_{(a,1|b,1)}(3{,}2|2{,}3) = \Bigg\{ \frac{s_{23}}{s_{12}} W^\tau_{(a,1|1,b)}(2{,}3|3{,}2) + \frac{1}{s_{12}} \frac{\pi}{\tau_2} \Big[ W^\tau_{(a,0|0,b)}(2{,}3|3{,}2) - W^\tau_{(a,0|1,b-1)}(2{,}3|3{,}2) \Big] \Bigg\} \Bigg|_{\substack{s_{12}\to s_{23} \\ s_{13}\to s_{12} \\ s_{23}\to s_{13}}} , \tag{B.26}$$

where $a \neq 1$ in (B.25) and $b \neq 1$ in (B.26) and we set $\overline{f^{(-1)}} = 0$. With the help of the Mandelstam relabelings, we avoid the need of a Fay identity to write the RHS in terms of the integrals (7.18). One further case can be obtained by Mandelstam relabelings of (B.25) and (B.26),

$$W^\tau_{(a,1|b,1)}(2{,}3|3{,}2) = W^\tau_{(a,1|b,1)}(3{,}2|2{,}3)\Big|_{s_{12}\leftrightarrow s_{13}} , \tag{B.27}$$

where $a \neq 1$ or $b \neq 1$.

If two $|f^{(1)}_{ij}|^2$ factors are present in the integrand, we obtain (on top of the poles for each $|f^{(1)}_{ij}|^2$) a three-point kinematic pole $\sim \frac{1}{s_{123}}$, where the three-point Mandelstam variable $s_{123} = s_{12} + s_{13} + s_{23}$ is defined in (2.27). Hence, the integral

$$W^\tau_{(1,1|1,1)}(2{,}3|2{,}3) = \int \mathrm{d}\mu_2 \, \big|f^{(1)}_{12}\big|^2 \, \big|f^{(1)}_{23}\big|^2 \, \mathrm{KN}_3 \tag{B.28}$$

has pole structure $\frac{1}{s_{123}}\big(\frac{1}{s_{12}} + \frac{1}{s_{23}}\big)$. Finally, the integral

$$W^\tau_{(1,1|1,1)}(3{,}2|2{,}3) = -\int \mathrm{d}\mu_2 \, f^{(1)}_{12} \, \overline{f^{(1)}_{13}} \big|f^{(1)}_{23}\big|^2 \, \mathrm{KN}_3 \tag{B.29}$$

has pole structure $\frac{1}{s_{123}s_{23}}$. The permutations of (B.28) and (B.29) can again be obtained by relabeling the Mandelstam variables,

$$W^\tau_{(1,1|1,1)}(2{,}3|3{,}2) = W^\tau_{(1,1|1,1)}(3{,}2|2{,}3)\Big|_{s_{12}\leftrightarrow s_{13}} \tag{B.30}$$

$$W^\tau_{(1,1|1,1)}(3{,}2|3{,}2) = W^\tau_{(1,1|1,1)}(2{,}3|2{,}3)\Big|_{s_{12}\leftrightarrow s_{13}} . \tag{B.31}$$

The following discussion of the integration-by-parts rewriting of (B.28) and (B.29) closely follows Appendix D of [IV].

The first step is to rewrite the meromorphic integrand by means of the Fay identity $f_{12}^{(1)} f_{23}^{(1)} + f_{12}^{(2)} + \text{cyc}(1,2,3) = 0$, cf. (5.121),

$$f_{12}^{(1)} f_{23}^{(1)} = -\frac{s_{13}}{s_{123}}\Big(f_{12}^{(2)} + f_{23}^{(2)} + f_{31}^{(2)}\Big) + \frac{X_{12,3}}{s_{12}s_{123}} - \frac{X_{23,1}}{s_{23}s_{123}} \,. \tag{B.32}$$

The combinations on the right-hand side

$$X_{ij,k} = s_{ij} f_{ij}^{(1)}\Big(s_{ik} f_{ik}^{(1)} + s_{jk} f_{jk}^{(1)}\Big) = s_{ij} f_{ij}^{(1)} \partial_{z_k} \log \text{KN}_3 \tag{B.33}$$

are Koba–Nielsen derivatives which we integrate by parts to act on the antimeromorphic factors

$$\begin{aligned} &\int \mathrm{d}\mu_2\, f_{12}^{(1)} f_{23}^{(1)}\, \overline{f_{ij}^{(1)}}\, \overline{f_{jk}^{(1)}}\, \text{KN}_3 \\ &= \int \mathrm{d}\mu_2 \Big\{ -\frac{s_{13}}{s_{123}}\Big(f_{12}^{(2)}+f_{23}^{(2)}+f_{31}^{(2)}\Big)\, \overline{f_{ij}^{(1)}}\, \overline{f_{jk}^{(1)}} \\ &\quad - \frac{f_{12}^{(1)}}{s_{123}}\, \partial_{z_3}\Big(\overline{f_{ij}^{(1)}}\, \overline{f_{jk}^{(1)}}\Big) + \frac{f_{23}^{(1)}}{s_{123}}\, \partial_{z_1}\Big(\overline{f_{ij}^{(1)}}\, \overline{f_{jk}^{(1)}}\Big) \Big\}\, \text{KN}_3 \,. \end{aligned} \tag{B.34}$$

Given that $\partial_z \overline{f^{(1)}(z,\tau)} = -\frac{\pi}{\tau_2}$ within Koba–Nielsen integrals (cf. (3.90)), choosing $(i,j,k) = (1,2,3)$ and $(1,3,2)$ casts the integrals (B.28) and (B.29) into the form

$$\begin{aligned} W^{\tau}_{(1,1|1,1)}(2{,}3|2{,}3) = \int \mathrm{d}\mu_2 \Big\{ &-\frac{s_{13}}{s_{123}}\Big(f_{12}^{(2)}+f_{23}^{(2)}+f_{31}^{(2)}\Big)\, \overline{f_{12}^{(1)}}\, \overline{f_{23}^{(1)}} \\ &-\frac{1}{s_{123}}\frac{\pi}{\tau_2}\Big[f_{12}^{(1)}\overline{f_{12}^{(1)}} + f_{23}^{(1)}\overline{f_{23}^{(1)}}\Big] \Big\}\, \text{KN}_3 \end{aligned} \tag{B.35}$$

$$\begin{aligned} W^{\tau}_{(1,1|1,1)}(3{,}2|2{,}3) = \int \mathrm{d}\mu_2 \Big\{ &-\frac{s_{13}}{s_{123}}\Big(f_{12}^{(2)}+f_{23}^{(2)}+f_{31}^{(2)}\Big)\, \overline{f_{13}^{(1)}}\, \overline{f_{32}^{(1)}} \\ &+\frac{1}{s_{123}}\frac{\pi}{\tau_2}\Big[f_{12}^{(1)}\Big(\overline{f_{13}^{(1)}}+\overline{f_{23}^{(1)}}\Big) + f_{23}^{(1)}\overline{f_{23}^{(1)}}\Big] \Big\}\, \text{KN}_3 \,. \end{aligned} \tag{B.36}$$

The right-and sides can now be written in the form (7.18), resulting in

$$\begin{aligned} &W^{\tau}_{(1,1|1,1)}(2{,}3|2{,}3) \\ &= -\frac{s_{13}}{s_{123}}\Big[W^{\tau}_{(2,0|1,1)}(2{,}3|2{,}3) + W^{\tau}_{(0,2|1,1)}(2{,}3|2{,}3) + W^{\tau}_{(2,0|1,1)}(2{,}3|3{,}2)\Big] \\ &\quad -\frac{1}{s_{123}}\frac{\pi}{\tau_2}\Big[W^{\tau}_{(1,0|1,0)}(2{,}3|2{,}3) + W^{\tau}_{(0,1|0,1)}(2{,}3|2{,}3)\Big] \end{aligned} \tag{B.37}$$

$$\begin{aligned} &W^{\tau}_{(1,1|1,1)}(3{,}2|2{,}3) \\ &= -\frac{s_{13}}{s_{123}}\Big[W^{\tau}_{(2,0|1,1)}(3{,}2|2{,}3) + W^{\tau}_{(0,2|1,1)}(3{,}2|2{,}3) + W^{\tau}_{(2,0|1,1)}(3{,}2|3{,}2)\Big] \\ &\quad +\frac{1}{s_{123}}\frac{\pi}{\tau_2}\Big[W^{\tau}_{(1,0|1,0)}(3{,}2|2{,}3) + W^{\tau}_{(1,0|0,1)}(2{,}3|2{,}3) + W^{\tau}_{(0,1|0,1)}(2{,}3|2{,}3)\Big] \,, \end{aligned} \tag{B.38}$$

where (B.21)–(B.27) can be used for the two-particle poles on the RHS.

## B.4 SUBTRACTION SCHEME FOR TWO-PARTICLE POLES

In this appendix, we discuss an alternative way of treating kinematic poles in Koba–Nielsen integrals which is based on splitting the integral into a sum in which the divergent contributions cancel out. Using divergent MGFs, this can also be seen at the level of the $\alpha'$ expansion. This scheme has first appeared for genus-one integrals in [25]. The following discussion is largely identical to Appendix D.1 of [IV].

Consider the integral

$$W^{\tau}_{s_{12}} = \int \mathrm{d}\mu_{n-1}\, f^{(1)}_{12}\, \overline{f^{(1)}_{12}}\, \Phi(z_j, \bar{z}_j)\, \mathrm{KN}^{\tau}_n\,, \tag{B.39}$$

where the integrand $\Phi(z_j, \bar{z}_j)$ is tailored to admit no kinematic pole different from $s_{12}^{-1}$. This can be reconciled with the general form (7.11) of component integrals if

$$\Phi(z_j, \bar{z}_j) = f^{(a_3)}_{2i_3}\, f^{(a_4)}_{i_3 i_4} \dots f^{(a_n)}_{i_{n-1},i_n}\, \overline{f^{(b_3)}_{2j_3}}\, \overline{f^{(b_4)}_{j_3,j_4}} \dots \overline{f^{(b_n)}_{j_{n-1},j_n}} \tag{B.40}$$

exhibits no singularities of the form $|z_{kl}|^{-2}$ which for instance imposes $i_3 \neq j_3$.

The key idea of the subtraction scheme is to split the Koba–Nielsen factor of (B.39) into

$$\begin{aligned} W^{\tau}_{s_{12}} = &\int \mathrm{d}\mu_{n-1}\, f^{(1)}_{12}\, \overline{f^{(1)}_{12}}\, e^{s_{12}G(z_{12},\tau)}\, \Phi(z_j, \bar{z}_j) \\ &\times \Bigg( \underbrace{\prod_{j=3}^{n} e^{s_{1j}G(z_{1j},\tau)} - \prod_{j=3}^{n} e^{s_{1j}G(z_{2j},\tau)}}_{(i)} + \underbrace{\prod_{j=3}^{n} e^{s_{1j}G(z_{2j},\tau)}}_{(ii)} \Bigg) \widehat{\mathrm{KN}}_n\,, \end{aligned} \tag{B.41}$$

where

$$\widehat{\mathrm{KN}}_n = \prod_{2\leq i<j}^{n} \exp\big(s_{ij} G(z_{ij},\tau)\big)\,. \tag{B.42}$$

At the pole $f^{(1)}_{12}\, \overline{f^{(1)}_{12}} \sim |z_{12}|^{-2}$, the two terms in the integrand of (B.41) marked by $(i)$ cancel and as a result, the combined integral over the punctures $z_2, z_3, \dots, z_n$ in $W^{\tau}_{s_{12}}|_{(i)}$ will be finite for each term of the Taylor-expanded Koba–Nielsen factor.

For the leftover term in (B.41) marked by $(ii)$, the integrand only depends on $z_1$ via $f^{(1)}_{12}\overline{f^{(1)}_{12}} e^{s_{12}G(z_{12},\tau)}$ which can be conveniently integrated by parts using

$$f^{(1)}_{12}\, \overline{f^{(1)}_{12}}\, e^{s_{12}G(z_{12},\tau)} = -\frac{1}{s_{12}} \left[ \partial_{z_1} \left( \overline{f^{(1)}_{12}}\, e^{s_{12}G(z_{12},\tau)} \right) + \frac{\pi}{\tau_2} e^{s_{12}G(z_{12},\tau)} \right]. \tag{B.43}$$

The modified Koba–Nielsen factor in (B.41, $ii$) with $e^{s_{1j}G(z_{2j},\tau)}$ in the place of $e^{s_{1j}G(z_{1j},\tau)}$ does not alter the fact that the total $z_1$-derivative integrates to zero (there are by construction no poles in $z_{1j}, \bar{z}_{1j}$ with $j \geq 3$ that could contribute via Cauchy's theorem). Even if the measure $\mathrm{d}\mu_{n-1}$ is defined in (3.50) to not comprise an explicit integral over $z_1$, one can still discard total derivatives w.r.t. $z_1$ after undoing the fixing $z_1 = 0$, setting $z_n = 0$ and integrating $z_1, \ldots, z_{n-1}$ over the torus instead of $z_2, \ldots, z_n$.

Hence, the splitting (B.41) along with the integration by parts due to (B.43) allow to isolate the kinematic pole of the example (B.39),

$$W^{\tau}_{s_{12}}\Big|_{(ii)} = -\frac{1}{s_{12}}\frac{\pi}{\tau_2}\int \mathrm{d}\mu_{n-1}\,\Phi(z_j,\bar{z}_j)\,e^{s_{12}G(z_{12},\tau)}\,\widehat{\mathrm{KN}}^{\tau}_n \prod_{j=3}^{n} e^{s_{1j}G(z_{2j},\tau)}\,. \tag{B.44}$$

In order to obtain an $\alpha'$ expansion of the pole-free part ($i$) of (B.41), we have to expand both integrals in the difference separately. Since they have a $|f^{(1)}(z)|^2$ contribution in their integrand, the $\alpha'$ expansion will yield divergent MGFs. If these MGFs are decomposed into a sum of convergent MGFs and a polynomial in $\mathrm{E}_1$, this polynomial cancels between the two integrals in ($i$) of (B.41), yielding a finite result.

As an example, consider the case $\Phi = 1$ and $n = 3$. Then, the two integrals in ($i$) of (B.41) expand to

$$\begin{aligned}
&\int \mathrm{d}\mu_2\,\big|f^{(1)}_{12}\big|\,e^{s_{12}G_{12}+s_{13}G_{13}+s_{23}G_{23}}\\
&= \frac{\pi}{\tau_2}\mathrm{E}_1 - s_{12}\frac{\tau_2}{\pi}\,C\big[\begin{smallmatrix}0&1&1\\1&0&1\end{smallmatrix}\big] - s_{12}\Big(\frac{\tau_2}{\pi}\Big)^2 C\big[\begin{smallmatrix}0&1&1&1\\1&0&1&1\end{smallmatrix}\big]\\
&\qquad + \frac{1}{2}(s_{13}^2+s_{23}^2)\frac{\pi}{\tau_2}\mathrm{E}_1\mathrm{E}_2 - s_{13}s_{23}\Big(\frac{\tau_2}{\pi}\Big)^2 C\big[\begin{smallmatrix}0&1&2\\1&0&2\end{smallmatrix}\big] + O(\alpha'^3)
\end{aligned} \tag{B.45}$$

$$\begin{aligned}
&\int \mathrm{d}\mu_2\,\big|f^{(1)}_{12}\big|\,e^{s_{12}G_{12}+(s_{13}+s_{23})G_{23}}\\
&= \frac{\pi}{\tau_2}\mathrm{E}_1 - s_{12}\frac{\tau_2}{\pi}\,C\big[\begin{smallmatrix}0&1&1\\1&0&1\end{smallmatrix}\big] - s_{12}\Big(\frac{\tau_2}{\pi}\Big)^2 C\big[\begin{smallmatrix}0&1&1&1\\1&0&1&1\end{smallmatrix}\big]\\
&\qquad + \frac{1}{2}(2s_{13}s_{23}+s_{13}^2+s_{23}^2)\frac{\pi}{\tau_2}\mathrm{E}_1\mathrm{E}_2 + O(\alpha'^3)\,.
\end{aligned} \tag{B.46}$$

Using the divergent identity (5.172), it is clear that the difference is convergent and given by

$$\begin{aligned}
&\int \mathrm{d}\mu_2\,\big|f^{(1)}_{12}\big|\,e^{s_{12}G_{12}}\big(e^{s_{13}G_{13}+s_{23}G_{23}} - e^{s_{12}G_{12}+(s_{13}+s_{23})G_{23}}\big)\\
&= \frac{1}{2}s_{13}s_{23}\frac{\pi}{\tau_2}(3\mathrm{E}_3+\zeta_3) + O(\alpha'^3)\,.
\end{aligned} \tag{B.47}$$

# C

# INTEGRALS OF UNIFORM TRANSCENDENTALITY

In this appendix, we give more details on the rearrangements of the integrals in Section 6.2.3 and in particular derive the expressions for the conjecturally uniformly transcendental integrals $\widehat{\mathcal{I}}^{(a,0)}_{\dots}$ via integrations by parts. The text in this appendix has extensive overlap with Appendix D of [III].

The notion of transcendental weight for modular graph forms can be derived from the terminology for eMZVs (4.6), where $\omega(n_1, n_2, \dots, n_r)$ is said to have weight $n_1{+}n_2{+}\dots{+}n_r$. This convention implies weight one for $\pi$, weight $n_1{+}n_2{+}\dots{+}n_r$ for $\zeta_{n_1,n_2,\dots,n_r}$ and weight $r$ for iterated Eisenstein integrals $\mathcal{E}_0(k_1, k_2, \dots, k_r)$ in (4.15). This means that both types of Eisenstein series $\mathrm{G}_k$ and $\mathrm{E}_k$ have weight $k$ and that $\pi\nabla_0$ as well as $y = \pi\tau_2$ have weight one. Similarly, $(\pi\nabla_0)^p\mathrm{E}_k$ and $(\pi\nabla_0)^p\mathrm{E}_{2,2}$ are found to carry weight $k{+}p$ and $4{+}p$, respectively.

## C.1 PLANAR INTEGRALS

In this section, we provide a decomposition of $\mathcal{I}^{(4,0)}_{1234}$ defined in (6.30) into the integrals of uniform transcendentality. In particular, we derive (6.72).

### C.1.1 *Integration-by-parts manipulations*

The idea is to exploit the fact that the derivatives $\partial_i = \partial_{z_i}$ of the Koba–Nielsen factor given in (5.167) integrate to zero. In order to relate this to the constituents $f^{(a)}$ of the integrand $V_4(1,2,3,4)$ of $\mathcal{I}^{(4,0)}_{1234}$, cf. (6.38), the total $z_i$-derivatives have to furthermore act on suitably chosen functions of modular weight three. As we will see, the identities of interest involve the combination

$$\begin{aligned}\mathrm{X}^{(3)}_{1234} = {}& f^{(1)}_{12} f^{(1)}_{23} f^{(1)}_{34} + \frac{1}{6}\big(f^{(3)}_{12} + f^{(3)}_{34}\big) + \frac{2}{3} f^{(3)}_{23} + \frac{1}{3} f^{(1)}_{23}\big(f^{(2)}_{12} + f^{(2)}_{34}\big) \\ &+ \frac{2}{3} f^{(2)}_{23}\big(f^{(1)}_{12} + f^{(1)}_{34}\big) + \frac{1}{2}\big(f^{(1)}_{12} f^{(2)}_{34} + f^{(2)}_{12} f^{(1)}_{34}\big)\,,\end{aligned} \tag{C.1}$$

whose derivatives in $z_1, \ldots, z_4$ can be evaluated using

$$\begin{aligned} \partial_z f^{(1)}(z) &= 2f^{(2)}(z) - \left(f^{(1)}(z)\right)^2 - \widehat{\mathrm{G}}_2 \\ \partial_z f^{(2)}(z) &= 3f^{(3)}(z) - f^{(1)}(z)f^{(2)}(z) - \widehat{\mathrm{G}}_2 f^{(1)}(z) \\ \partial_z f^{(3)}(z) &= 4f^{(4)}(z) - f^{(1)}(z)f^{(3)}(z) - \widehat{\mathrm{G}}_2 f^{(2)}(z) - \mathrm{G}_4 \,. \end{aligned} \tag{C.2}$$

The virtue of the combination (C.1) is that it allows generating $V_4(1,2,3,4)$ and simpler elliptic functions by means of total derivatives in the punctures as detailed below. Indeed, by the symmetries $X^{(3)}_{1234} = -X^{(3)}_{4321}$ and $X^{(3)}_{1234} + X^{(3)}_{2134} + X^{(3)}_{2314} + X^{(3)}_{2341} = 0$ due to Fay identities [28, 204], one can show that

$$\partial_4 X^{(3)}_{1234} + \partial_2 X^{(3)}_{1432} - \partial_3 (X^{(3)}_{1423} + X^{(3)}_{1243}) = \mathrm{G}_4 + \widehat{\mathrm{G}}_2 V_2(1,2,3,4) - V_4(1,2,3,4) \,. \tag{C.3}$$

In order to relate this to the Koba–Nielsen integral $\mathcal{I}^{(4,0)}_{1234}$, we extend (C.3) to the following total derivative via (C.7) and $V_4(1,2,3,4) + \mathrm{cyc}(2,3,4) = 3\mathrm{G}_4$,

$$\begin{aligned} &\partial_4(X^{(3)}_{1234}\,\mathrm{KN}_4) + \partial_2(X^{(3)}_{1432}\,\mathrm{KN}_4) - \partial_3(X^{(3)}_{1243}\,\mathrm{KN}_4) - \partial_3(X^{(3)}_{1423}\,\mathrm{KN}_4) \\ &= \mathrm{KN}_4 \left[\widehat{\mathrm{G}}_2 V_2(1,2,3,4) - (1 + s_{1234})V_4(1,2,3,4) + \mathrm{G}_4 \right. \\ &\qquad \left. + 3(s_{13} + s_{24})\mathrm{G}_4 + \widehat{V}_4(1,2,3,4)\right] \,, \end{aligned} \tag{C.4}$$

using the four-particle Mandelstam variable $s_{1234} = s_{12} + s_{13} + s_{14} + s_{23} + s_{24} + s_{34}$ as defined in (2.27) and

$$\begin{aligned} \widehat{V}_4(1,2,3,4) = s_{12}R_{2|34|1} + s_{23}R_{3|41|2} + s_{34}R_{4|12|3} + s_{14}R_{1|23|4} \\ - s_{13}(R_{1|24|3} + R_{1|42|3}) - s_{24}(R_{2|13|4} + R_{2|31|4}) \end{aligned} \tag{C.5}$$

comprising several permutations of

$$R_{1|23|4} = f^{(1)}_{14} X^{(3)}_{1234} + V_4(1,2,3,4) \,. \tag{C.6}$$

In the remainder of this subsection, we elaborate on some of the intermediate steps in (C.4): When the derivatives act on the Koba–Nielsen factor $\mathrm{KN}_4$, they generate the terms

$$\begin{aligned} &X^{(3)}_{1234}\left(s_{14}f^{(1)}_{14} + s_{24}f^{(1)}_{24} + s_{34}f^{(1)}_{34}\right) + X^{(3)}_{1432}\left(s_{12}f^{(1)}_{12} - s_{23}f^{(1)}_{23} - s_{24}f^{(1)}_{24}\right) \\ &\quad - \left(X^{(3)}_{1243} + X^{(3)}_{1423}\right)\left(s_{13}f^{(1)}_{13} + s_{23}f^{(1)}_{23} - s_{34}f^{(1)}_{34}\right) \\ &= \left[s_{14}f^{(1)}_{14}X^{(3)}_{1234} + \mathrm{cyc}(1,2,3,4)\right] \\ &\quad - s_{13}f^{(1)}_{13}\left(X^{(3)}_{1243} + X^{(3)}_{1423}\right) - s_{24}f^{(1)}_{24}\left(X^{(3)}_{2134} + X^{(3)}_{2314}\right) \,, \end{aligned} \tag{C.7}$$

where the symmetries $X^{(3)}_{1234} = -X^{(3)}_{4321}$ and $X^{(3)}_{1234} + X^{(3)}_{2134} + X^{(3)}_{2314} + X^{(3)}_{2341} = 0$ have been used in passing to the last line. However, we have refrained from using momentum conservation in (C.4) or (C.7) so far. The term $s_{1234}V_4(1,2,3,4)$ in the second line of (C.4) has been generated by rewriting each term in (C.7) via permutations of (C.6). The coefficients of $s_{13}$ and $s_{24}$ in (C.7) require the additional intermediate step

$$\begin{aligned}-f_{13}^{(1)}(X_{1243}^{(3)}+X_{1423}^{(3)}) &= V_4(1,2,4,3)+V_4(1,4,2,3)-R_{1|24|3}-R_{1|42|3}\\ &= 3\mathrm{G}_4-V_4(1,2,3,4)-R_{1|24|3}-R_{1|42|3} \qquad (\text{C.8})\end{aligned}$$

in reproducing (C.4).

### C.1.2 *Uniform transcendentality decomposition*

Using the representation (C.1) and (6.38) of its constituents $X_{1234}^{(3)}$ and $V_4(1,2,3,4)$, $R_{1|23|4}$ can be expanded as

$$\begin{aligned}R_{1|23|4} &= f_{23}^{(1)}f_{34}^{(1)}f_{12}^{(2)}+f_{12}^{(1)}f_{34}^{(1)}f_{23}^{(2)}+f_{12}^{(1)}f_{23}^{(1)}f_{34}^{(2)}\\ &+f_{12}^{(1)}f_{23}^{(1)}f_{14}^{(2)}+f_{12}^{(1)}f_{34}^{(1)}f_{14}^{(2)}+f_{23}^{(1)}f_{34}^{(1)}f_{14}^{(2)}\\ &-\frac{1}{3}f_{12}^{(1)}f_{14}^{(1)}f_{23}^{(2)}-\frac{1}{3}f_{14}^{(1)}f_{34}^{(1)}f_{23}^{(2)}-\frac{2}{3}f_{14}^{(1)}f_{23}^{(1)}f_{12}^{(2)}-\frac{2}{3}f_{14}^{(1)}f_{23}^{(1)}f_{34}^{(2)}-\frac{1}{2}f_{14}^{(1)}f_{34}^{(1)}f_{12}^{(2)}\\ &-\frac{1}{2}f_{12}^{(1)}f_{14}^{(1)}f_{34}^{(2)}+f_{12}^{(2)}f_{14}^{(2)}+f_{12}^{(2)}f_{23}^{(2)}+f_{14}^{(2)}f_{23}^{(2)}+f_{12}^{(2)}f_{34}^{(2)}+f_{14}^{(2)}f_{34}^{(2)}+f_{23}^{(2)}f_{34}^{(2)}\\ &-\frac{5}{6}f_{14}^{(1)}f_{12}^{(3)}-\frac{5}{6}f_{14}^{(1)}f_{34}^{(3)}-\frac{1}{3}f_{14}^{(1)}f_{23}^{(3)}-f_{12}^{(1)}f_{14}^{(3)}-f_{23}^{(1)}f_{14}^{(3)}-f_{34}^{(1)}f_{14}^{(3)}+f_{23}^{(1)}f_{12}^{(3)}\\ &+f_{23}^{(1)}f_{34}^{(3)}+f_{34}^{(1)}f_{12}^{(3)}+f_{12}^{(1)}f_{34}^{(3)}+f_{12}^{(1)}f_{23}^{(3)}+f_{34}^{(1)}f_{23}^{(3)}+f_{12}^{(4)}+f_{14}^{(4)}+f_{23}^{(4)}+f_{34}^{(4)}\,. \qquad (\text{C.9})\end{aligned}$$

Hence, all the terms of $R_{1|23|4}$ involve at most three factors of $f_{ij}^{(n)}$, and none of them exhibits a subcycle $f_{ij}^{(m)}f_{ji}^{(n)}$ or $f_{ij}^{(m)}f_{jk}^{(n)}f_{ki}^{(p)}$. This means none of the resulting graphs in the order-by-order integration against monomials in $G_{ij}$ contains closed holomorphic subgraphs and hence the need for HSR is removed.

Apart from $R_{1|23|4} = R_{4|32|1}$, there are no further relations among the 12 reflection-independent permutations of $R_{1|23|4}$ in (C.6). The reflection property is sufficient to show that $\widehat{V}_4(1,2,3,4)+\text{cyc}(2,3,4)=0$.

In the momentum phase-space of four particles with $s_{13}=s_{24}$ and $s_{1234}=0$, one can solve (C.4) for

$$V_4(1,2,3,4) \cong \widehat{\mathrm{G}}_2V_2(1,2,3,4)+\mathrm{G}_4+6s_{13}\mathrm{G}_4+\widehat{V}_4(1,2,3,4)\,, \qquad (\text{C.10})$$

where the equivalence relation $\cong$ indicates that total derivatives $\partial_i(\ldots\mathrm{KN}_4)$ have to be discarded in equating ($\mathrm{KN}_4$ times) the two sides of (C.10). At the level of the integrals, this implies

$$\mathcal{I}_{1234}^{(4,0)}(s_{ij},\tau)=\mathrm{G}_4\mathcal{I}_{1234}^{(0,0)}(s_{ij},\tau)+\widehat{\mathrm{G}}_2\mathcal{I}_{1234}^{(2,0)}(s_{ij},\tau)+\widehat{\mathcal{I}}_{1234}^{(4,0)}(s_{ij},\tau)\,, \qquad (\text{C.11})$$

cf. (6.73), where we have introduced a new Koba–Nielsen integral

$$\widehat{\mathcal{I}}^{(4,0)}_{1234}(s_{ij},\tau) = \int \mathrm{d}\mu_3 \ \mathrm{KN}_4 \ \Big\{ 6 s_{13} \mathrm{G}_4 + s_{12} R_{2|34|1} + s_{23} R_{3|41|2} \tag{C.12}$$
$$+ s_{34} R_{4|12|3} + s_{14} R_{1|23|4} - s_{13}(R_{1|24|3} + R_{1|42|3}) - s_{24}(R_{2|13|4} + R_{2|31|4}) \Big\} \,.$$

Given that the $R_{i|jk|l}$ boil down to the $f^{(a)}$ with a lattice-sum representation (3.91), the coefficients in the $\alpha'$ expansion of (C.12) are guaranteed to be MGFs. By the absence of subcycles in the constituents (C.6) , the $\alpha'$ expansion of $\widehat{\mathcal{I}}^{(4,0)}_{1234}$ is expected to exhibit uniform transcendentality with weight $k{+}3$ at the order of $\alpha'^k$. This is confirmed by the leading orders in $\alpha'$,

$$\widehat{\mathcal{I}}^{(4,0)}_{1234}(s_{ij},\tau) = 6 s_{13} \mathrm{G}_4 + 2(s_{13}^2 + 2 s_{12} s_{23})\, \mathcal{C}\big[\begin{smallmatrix} 5 & 0 \\ 1 & 0 \end{smallmatrix}\big] + O(\alpha'^3) \,, \tag{C.13}$$

as can be seen from the form given in (6.73).

By inserting the decomposition (C.11) of the integral $\mathcal{I}^{(4,0)}_{1234}$ into (6.26), we arrive at the decomposition of the complete planar $\tau$-integrand as stated in (6.72).

### C.1.3 *Consistency check of the leading contributions to* $\widehat{\mathcal{I}}^{(4,0)}_{1234}$

The leading-order result (C.13) can not only been obtained from (C.11), but can also be checked in an independent calculation based on an $\alpha'$ expansion of (C.12) as follows. We use the notation

$$\mathcal{R}_{a|bc|d}\Big[\prod_{i<j} \mathrm{G}_{ij}^{n_{ij}}\Big] = \Big(\frac{\pi}{\tau_2}\Big)^{\sum_{i<j} n_{ij}} \int \mathrm{d}\mu_3 \, R_{a|bc|d} \prod_{i<j} \mathrm{G}_{ij}^{n_{ij}} \tag{C.14}$$

analogous to (6.33). In the absence of closed subcycles in the expression (C.9) for $R_{1|23|4}$, the leading order evidently vanishes,

$$\mathcal{R}_{a|bc|d}[\varnothing] = 0 \,. \tag{C.15}$$

At the subleading order in $\alpha'$, the same representation of $R_{1|23|4}$ yields

$$\begin{aligned}
\mathcal{R}_{1|23|4}[\mathrm{G}_{12}] &= -3\,\mathcal{C}\big[\begin{smallmatrix} 5 & 0 \\ 1 & 0 \end{smallmatrix}\big], & \mathcal{R}_{1|23|4}[\mathrm{G}_{13}] &= \frac{35}{6}\,\mathcal{C}\big[\begin{smallmatrix} 5 & 0 \\ 1 & 0 \end{smallmatrix}\big], \\
\mathcal{R}_{1|23|4}[\mathrm{G}_{23}] &= -3\,\mathcal{C}\big[\begin{smallmatrix} 5 & 0 \\ 1 & 0 \end{smallmatrix}\big], & \mathcal{R}_{1|23|4}[\mathrm{G}_{24}] &= \frac{35}{6}\,\mathcal{C}\big[\begin{smallmatrix} 5 & 0 \\ 1 & 0 \end{smallmatrix}\big], \\
\mathcal{R}_{1|23|4}[\mathrm{G}_{14}] &= -4\,\mathcal{C}\big[\begin{smallmatrix} 5 & 0 \\ 1 & 0 \end{smallmatrix}\big], & \mathcal{R}_{1|23|4}[\mathrm{G}_{34}] &= -3\,\mathcal{C}\big[\begin{smallmatrix} 5 & 0 \\ 1 & 0 \end{smallmatrix}\big],
\end{aligned} \tag{C.16}$$

and any other $\mathcal{R}_{a|bc|d}[\mathrm{G}_{ij}]$ can be obtained by relabeling. Finally, the contribution of $6 s_{13} \mathrm{G}_4$ in (C.12) integrates to $6 s_{13} \mathrm{G}_4 \mathcal{I}^{(0,0)}$ with $\mathcal{I}^{(0,0)} = 1 + O(\alpha'^2)$, completing the verification of (C.13) to the orders shown.

## C.2 NON-PLANAR INTEGRALS

The non-planar integrals $\mathcal{I}^{(2,0)}_{12|34}$ and $\mathcal{I}^{(4,0)}_{12|34}$ in (6.30) and (6.31) admit integration-by-parts manipulations analogous to (C.10) to be rewritten in terms of (conjecturally) uniformly transcendental integrals $\widehat{\mathcal{I}}^{(a,0)}_{12|34}$. In this section we derive these decompositions, given in (6.76) in the main text.

In the non-planar cases, the total derivatives are simpler and boil down to iterations of

$$\begin{aligned}\partial_2\big(f^{(1)}_{12}\,\mathrm{KN}_4\big) = \mathrm{KN}_4\,\big[\widehat{\mathrm{G}}_2 - (1+s_{12})V_2(1,2) \\ + 2s_{12}f^{(2)}_{12} - f^{(1)}_{12}\big(s_{23}f^{(1)}_{23} + s_{24}f^{(1)}_{24}\big)\big]\,,\end{aligned} \tag{C.17}$$

which can be used to solve for

$$(1+s_{12})V_2(1,2) \cong \widehat{\mathrm{G}}_2 + 2s_{12}f^{(2)}_{12} - f^{(1)}_{12}\big(s_{23}f^{(1)}_{23} + s_{24}f^{(1)}_{24}\big)\,. \tag{C.18}$$

Again, $\cong$ indicates that total derivatives $\partial_i(\ldots\mathrm{KN}_4)$ have been discarded in passing to the right-hand side. A similar identity can be derived by taking a $z_1$-derivative in (C.17), so the right-hand side of (C.18) turns out to be symmetric under the simultaneous exchange of $(z_1, s_{1j}) \leftrightarrow (z_2, s_{2j})$, at least up to total derivatives.

### C.2.1 *Rewriting the integral* $\mathcal{I}^{(2,0)}_{12|34}$

By applying (C.18) to both summands $V_2(1,2)$ and $V_2(3,4)$ of (6.31), one arrives at a decomposition

$$\mathcal{I}^{(2,0)}_{12|34}(s_{ij},\tau) = \frac{2\widehat{\mathrm{G}}_2\mathcal{I}^{(0,0)}(s_{ij},\tau) + \widehat{\mathcal{I}}^{(2,0)}_{12|34}(s_{ij},\tau)}{1+s_{12}} \tag{C.19}$$

involving a new integral that should be uniformly transcendental,

$$\begin{aligned}\widehat{\mathcal{I}}^{(2,0)}_{12|34}(s_{ij},\tau) &= \int \mathrm{d}\mu_3\;\mathrm{KN}_4\;\big[2s_{12}f^{(2)}_{12} - f^{(1)}_{12}\big(s_{23}f^{(1)}_{23} + s_{24}f^{(1)}_{24}\big) \\ &\qquad + 2s_{34}f^{(2)}_{34} - f^{(1)}_{34}\big(s_{41}f^{(1)}_{41} + s_{42}f^{(1)}_{42}\big)\big] \\ &= 2\int \mathrm{d}\mu_3\;\mathrm{KN}_4\;\big[2s_{12}f^{(2)}_{12} - f^{(1)}_{12}\big(s_{23}f^{(1)}_{23} + s_{24}f^{(1)}_{24}\big)\big]\,.\end{aligned} \tag{C.20}$$

In passing to the last line, we have used Mandelstam identities and the symmetries of the Koba–Nielsen factor to obtain identities such as

$$\int \mathrm{d}\mu_3\;\mathrm{KN}_4\,s_{23}f^{(1)}_{12}f^{(1)}_{23} = \int \mathrm{d}\mu_3\;\mathrm{KN}_4\,s_{14}f^{(1)}_{34}f^{(1)}_{41}\,. \tag{C.21}$$

The expansion of $\widehat{\mathcal{I}}^{(2,0)}_{12|34}$ up to the third order in $\alpha'$ is given in (6.77) and verified to be uniformly transcendental.

### C.2.2 *Rewriting the integral* $\mathcal{I}^{(4,0)}_{12|34}$

Repeated application of (C.17) leads to a total-derivative relation for the integrand of $\mathcal{I}^{(4,0)}_{12|34}$,

$$\begin{aligned}
&(1+s_{12})(1+s_{34})V_2(1,2)V_2(3,4) \\
&\cong \widehat{\mathrm{G}}_2^2 + s_{13}^2 f^{(1)}_{12} f^{(1)}_{24} f^{(1)}_{43} f^{(1)}_{31} + s_{14}^2 f^{(1)}_{12} f^{(1)}_{23} f^{(1)}_{34} f^{(1)}_{41} + 4 s_{12} s_{34} f^{(2)}_{12} f^{(2)}_{34} \\
&\quad + \widehat{\mathrm{G}}_2 \Big[ 2 s_{12} f^{(2)}_{12} + 2 s_{34} f^{(2)}_{34} + \frac{1}{2} f^{(1)}_{12} \big( s_{13} f^{(1)}_{13} + s_{14} f^{(1)}_{14} - s_{23} f^{(1)}_{23} - s_{24} f^{(1)}_{24} \big) \\
&\qquad\quad + \frac{1}{2} f^{(1)}_{34} \big( s_{24} f^{(1)}_{24} + s_{14} f^{(1)}_{14} - s_{23} f^{(1)}_{23} - s_{13} f^{(1)}_{13} \big) \Big] \qquad\qquad (\mathrm{C.22}) \\
&\quad + s_{34} f^{(2)}_{34} f^{(1)}_{12} \big( s_{13} f^{(1)}_{13} + s_{14} f^{(1)}_{14} - s_{23} f^{(1)}_{23} - s_{24} f^{(1)}_{24} \big) \\
&\quad + s_{12} f^{(2)}_{12} f^{(1)}_{34} \big( s_{24} f^{(1)}_{24} + s_{14} f^{(1)}_{14} - s_{23} f^{(1)}_{23} - s_{13} f^{(1)}_{13} \big) \,,
\end{aligned}$$

and therefore to a similar decomposition for the integral as in (C.19):

$$\mathcal{I}^{(4,0)}_{12|34}(s_{ij},\tau) = \frac{\widehat{\mathrm{G}}_2^2 \mathcal{I}^{(0,0)}(s_{ij},\tau) + \widehat{\mathrm{G}}_2 \widehat{\mathcal{I}}^{(2,0)}_{12|34}(s_{ij},\tau) + \check{\mathcal{I}}^{(4,0)}_{12|34}(s_{ij},\tau)}{(1+s_{12})^2} \,. \qquad (\mathrm{C.23})$$

Here, we have introduced the integral

$$\begin{aligned}
\check{\mathcal{I}}^{(4,0)}_{12|34}(s_{ij},\tau) = \int \mathrm{d}\mu_3 \, \mathrm{KN}_4 \Big[ & 4 s_{12} s_{34} f^{(2)}_{12} f^{(2)}_{34} + 4 s_{34} f^{(2)}_{34} f^{(1)}_{12} \big( s_{13} f^{(1)}_{13} + s_{14} f^{(1)}_{14} \big) \\
& + s_{13}^2 f^{(1)}_{12} f^{(1)}_{24} f^{(1)}_{43} f^{(1)}_{31} + s_{14}^2 f^{(1)}_{12} f^{(1)}_{23} f^{(1)}_{34} f^{(1)}_{41} \Big] \qquad (\mathrm{C.24})
\end{aligned}$$

over the terms without $\widehat{\mathrm{G}}_2$ on the right-hand side of (C.22), where relabeling identities of the form (C.21) were used to simplify the result. Again, we have independently verified (C.19) and (C.23) to the order of $\alpha'^3$ by expanding the integrals in (C.20) and (C.24) along the lines of Section C.1.3. Upon insertion into (6.26), we arrive at an alternative representation of the non-planar sector of the four-point amplitude,

$$\begin{aligned}
M_4(\tau)\big|_{\mathrm{Tr}(t^{a_1}t^{a_2})\mathrm{Tr}(t^{a_3}t^{a_4})} &= \frac{\mathrm{G}_4^2 \check{\mathcal{I}}^{(4,0)}_{12|34}}{(1+s_{12})^2} + \left( \frac{\mathrm{G}_4^2 \widehat{\mathrm{G}}_2}{(1+s_{12})^2} - \frac{7}{2} \frac{\mathrm{G}_4 \mathrm{G}_6}{1+s_{12}} \right) \widehat{\mathcal{I}}^{(2,0)}_{12|34} \qquad (\mathrm{C.25}) \\
&\quad + \left( \frac{\mathrm{G}_4^2 \widehat{\mathrm{G}}_2^2}{(1+s_{12})^2} - 7 \frac{\mathrm{G}_4 \mathrm{G}_6 \widehat{\mathrm{G}}_2}{1+s_{12}} + \frac{5}{3} \mathrm{G}_4^3 + \frac{49}{6} \mathrm{G}_6^2 \right) \mathcal{I}^{(0,0)} \,.
\end{aligned}$$

### C.2.3 *Towards a uniform-transcendentality basis*

However, this is not yet the desired uniform-transcendentality decomposition since the last line in the integrand (C.24) of $\check{\mathcal{I}}^{(4,0)}_{12|34}$ exhibits closed subcycles $f_{ij}^{(1)} f_{jk}^{(1)} f_{kl}^{(1)} f_{li}^{(1)}$. One can isolate a piece of uniform transcendentality from (C.24) by subtracting these subcycles via $V_4$, i.e.

$$\begin{aligned}\widehat{\mathcal{I}}^{(4,0)}_{12|34}(s_{ij},\tau) = \int \mathrm{d}\mu_3 \ \mathrm{KN}_4 \Big[ & 4s_{12}s_{34} f_{12}^{(2)} f_{34}^{(2)} + 4s_{34} f_{34}^{(2)} f_{12}^{(1)} (s_{13} f_{13}^{(1)} + s_{14} f_{14}^{(1)}) \\ & + s_{13}^2 \big( f_{12}^{(1)} f_{24}^{(1)} f_{43}^{(1)} f_{31}^{(1)} - V_4(1,2,4,3) \big) \\ & + s_{14}^2 \big( f_{12}^{(1)} f_{23}^{(1)} f_{34}^{(1)} f_{41}^{(1)} - V_4(1,2,3,4) \big) \Big]\end{aligned} \tag{C.26}$$

is claimed to be uniformly transcendental. Then, by the decomposition (C.11) of integrals over $V_4(i,j,k,l)$, we can relate this to (C.24) via

$$\begin{aligned}\check{\mathcal{I}}^{(4,0)}_{12|34} = \widehat{\mathcal{I}}^{(4,0)}_{12|34} + (s_{13}^2 + s_{23}^2)\mathrm{G}_4 \mathcal{I}^{(0,0)} + s_{13}^2 \big( \widehat{\mathcal{I}}^{(4,0)}_{1243} + \widehat{\mathrm{G}}_2 \mathcal{I}^{(2,0)}_{1243} \big) \\ + s_{23}^2 \big( \widehat{\mathcal{I}}^{(4,0)}_{1234} + \widehat{\mathrm{G}}_2 \mathcal{I}^{(2,0)}_{1234} \big) \ .\end{aligned} \tag{C.27}$$

Upon insertion into (C.25) , we obtain admixtures of the planar integrals $\widehat{\mathcal{I}}^{(4,0)}_{1234}$ and $\mathcal{I}^{(2,0)}_{1234}$ defined by (C.12) and (6.28), respectively, and arrive at (6.76). Similar to Section C.1.3, the results for $\mathcal{I}^{(a,0)}_{12|34}$ in (6.70) and (6.71) have been confirmed by performing an independent $\alpha'$ expansion of (C.20) and (C.26) and inserting into the above integration-by-parts relations.

## C.3 EFFICIENCY OF THE NEW REPRESENTATIONS FOR EXPANSIONS

Given the Mandelstam invariants in the integrands (C.12), (C.20) and (C.24) of $\widehat{\mathcal{I}}^{(4,0)}_{1234}$, $\widehat{\mathcal{I}}^{(2,0)}_{12|34}$ and $\widehat{\mathcal{I}}^{(4,0)}_{12|34}$, the $k^{\text{th}}$ order in their $\alpha'$ expansion can be computed from less than $k$ factors of $G_{ij}$ from the Koba–Nielsen factor. However, the variety of $f_{ij}^{a}$ along with the different $s_{kl}$ in the integrands increases the number of independent calculations w.r.t. relabeling the punctures and momenta at a fixed order of the Koba–Nielsen expansion. Hence, the combinatorial efficiency of the new representations (6.72) and (6.76) of $M_4(\tau)$ for higher-order $\alpha'$ expansion should be comparable to the old one in (6.26).

Instead, the main advantages of the integrals $\widehat{\mathcal{I}}^{(4,0)}_{1234}$, $\widehat{\mathcal{I}}^{(2,0)}_{12|34}$ and $\widehat{\mathcal{I}}^{(4,0)}_{12|34}$ are the following:

- The integrand of $\widehat{\mathcal{I}}^{(4,0)}_{1234}$ in (C.12) does not share the term $f_{12}^{(1)} f_{23}^{(1)} f_{34}^{(1)} f_{41}^{(1)}$ of the $V_4(1,2,3,4)$ function. Like this, expansion of $\widehat{\mathcal{I}}^{(4,0)}_{1234}$ bypasses numerous HSRs that introduced a spurious complexity into the calculations of Section 6.2.1.

- The $\alpha'$ expansions of $\mathcal{I}^{(a,0)}_{12|34}$ in Section 6.2.2 contained conditionally convergent or divergent lattice sums. Lack of absolute convergence is caused by the terms $(f_{ij}^{(1)})^2$ in the representation (6.60) of $V_2(i,j)$. Both of them are manifestly absent in the integrands (C.20) and (C.24) of $\widehat{\mathcal{I}}^{(2,0)}_{12|34}$ and $\widehat{\mathcal{I}}^{(4,0)}_{12|34}$.

# D

# FURTHER DETAILS ABOUT CHAPTER 7

This appendix contains further details about the differential equations for a generating series of Koba–Nielsen integrals discussed in Chapter 7. It has extensive text overlap with the Appendices B.5, C and E of [IV].

## D.1 COMPONENT INTEGRALS VERSUS $n$-POINT STRING AMPLITUDES

In this appendix, we specify the component integrals $W^{\tau}_{(A|B)}(\sigma|\rho)$ (7.5) that enter $n$-point one-loop closed-string amplitudes of the bosonic, heterotic and type-II theories in more detail.

Even though the zero modes of the world-sheet bosons couple the chiral halves of the closed string at genus $g > 0$, it is instructive to first review the analogous open-string correlators:

- The $n$-point correlators of massless vertex operators of the open superstring are strongly constrained by its sixteen supercharges. This can be seen from the sum over spin structures in the RNS formalism [271, 272] or from the fermionic zero modes in the pure-spinor formalism [50, 63]. As a result, these correlators comprise products $\prod_k f^{(a_k)}_{i_k j_k}$ of overall weight $\sum_k a_k = n-4$ as well as admixtures of holomorphic Eisenstein series $\mathrm{G}_w \prod_k f^{(a_k)}_{i_k j_k}$ with $w + \sum_k a_k = n-4$ and $w \geq 4$ [28, 101]. When entering heterotic-string or type-II amplitudes as a chiral half, open-superstring correlator introduce component integrals $W^{\tau}_{(A|B)}(\sigma|\rho)$ with holomorphic modular weights $|A| \leq n-4$.

- In orbifold compactifications of the open superstring that preserve four or eight supercharges, the RNS spin sums are modified by the partition function, see e.g. [273] for a review. The spin-summed $n$-point correlators of massless vertex operators may therefore depend on the punctures via $\prod_k f^{(a_k)}_{i_k j_k}$ or $\mathrm{G}_{w\geq 4} \prod_k f^{(a_k)}_{i_k j_k}$ of weight $\sum_k a_k = n-2$ or $w + \sum_k a_k = n-2$ [121, 122]. The resulting component integrals $W^{\tau}_{(A|B)}(\sigma|\rho)$ in closed-string amplitudes with such chiral halves have holomorphic modular weights $|A| \leq n-2$.

- Open bosonic strings in turn allow for combinations $\prod_k f^{(a_k)}_{i_k j_k}$ and $\widehat{\mathrm{G}}_{w=2} \prod_k f^{(a_k)}_{i_k j_k}$ or $\mathrm{G}_{w\geq 4} \prod_k f^{(a_k)}_{i_k j_k}$ of weight $\sum_k a_k = n$ and $w +$ $\sum_k a_k = n$. The same is true for the $n$-point torus correlators of Kac–Moody currents entering the gauge sector of the heterotic string [III, 174]. Accordingly, closed-string amplitudes of the heterotic and bosonic theory comprise component integrals $W^\tau_{(A|B)}(\sigma|\rho)$ with modular weights $|A| \leq n$ or $|B| \leq n$ in one or two chiral halves.

The pattern of $f^{(a_k)}_{i_k j_k}$ obtained from a direct evaluation of the correlators may not immediately line up with the integrands of $W^\tau_{(A|B)}(\sigma|\rho)$. First, contractions among the world-sheet bosons introduce spurious derivatives $\partial_z f^{(1)}(z,\tau)$. Second, one may encounter arrangements of the labels $i_j$ in the first argument in "cycles" $f^{(a_1)}_{i_1 i_2} f^{(a_2)}_{i_2 i_3} \ldots f^{(a_k)}_{i_k i_1}$ rather than the "open chains" $f^{(a_2)}_{i_1 i_2} f^{(a_3)}_{i_2 i_3} \ldots f^{(a_n)}_{i_{n-1} i_n}$ characteristic to (7.5).

In both cases, combinations of integration by parts and Fay identities (5.121) are expected to reduce any term in the above correlators to the integrands of $W^\tau_{(A|B)}(\sigma|\rho)$. For instance, the methods in Appendix D.1 of [III] reduce the cycle $f^{(1)}_{12} f^{(1)}_{23} f^{(1)}_{34} f^{(1)}_{41}$ in a four-point current correlator to G4 as well as $f^{(a_2)}_{12} f^{(a_3)}_{23} f^{(a_4)}_{34}$ with $a_2 + a_3 + a_4 = 4$ and permutations.

The above properties of chiral halves set upper bounds on the modular weights $|A|, |B|$ seen in the component integrals $W^\tau_{(A|B)}(\sigma|\rho)$ of the respective closed-string amplitudes. However, additional contributions with (non-negative) weights $(|A| - k, |B| - k)$, $k \in \mathbb{N}$ arise from interactions between left and right-movers: Both the direct contractions between left- and right-moving world-sheet bosons and the contribution of (3.94) to integrations by parts convert one unit of holomorphic and non-holomorphic modular weight into a factor of $\tau_2^{-1}$, see e.g. [122, 169, 170].

For instance, the six-point amplitude of the type-II superstring in the representation of [172] is composed of component integrals $W^\tau_{(A|B)}(\sigma|\rho)$ with $(|A|, |B|) \in \{(2,2), (1,1), (0,0)\}$. The integrands associated with $|A| = |B| = 2$ can take one of the forms

$$f^{(2)}_{ij} \, \overline{f^{(2)}_{pq}} \,, \quad f^{(2)}_{ij} \, \overline{f^{(1)}_{pq}} \, \overline{f^{(1)}_{rs}} \,, \quad f^{(1)}_{ij} \, f^{(1)}_{kl} \, \overline{f^{(2)}_{pq}} \,, \quad f^{(1)}_{ij} \, f^{(1)}_{kl} \, \overline{f^{(1)}_{pq}} \, \overline{f^{(1)}_{rs}} \,, \tag{D.1}$$

with at most one overlapping label among $f^{(1)}_{ij} f^{(1)}_{kl}$ (and separately among $\overline{f^{(1)}_{pq}} \, \overline{f^{(1)}_{rs}}$).

## D.2 VERIFYING TWO-POINT CAUCHY–RIEMANN EQUATIONS

As an example of the power of the identities discussed in Chapter 5, we will prove the two-point Cauchy–Riemann equations (7.48) for component integrals in this section. The proof relies on applying dihedral momentum conservation (5.36) and dihedral HSR (5.73) order by order in $\alpha'$.

The $\alpha'$-expansion of two-point component integrals (7.12) can be expressed in closed form (cf. (7.13))

$$W^{\tau}_{(a|b)} = (-1)^a \sum_{k=0}^{\infty} s_{12}^k \frac{1}{k!}\left(\frac{\tau_2}{\pi}\right)^k C\left[\begin{smallmatrix} a & 0 & 1_k \\ 0 & b & 1_k \end{smallmatrix}\right], \quad a,b > 0, \quad (a,b) \neq (1,1), \tag{D.2}$$

where $1_k$ denotes the row vector with $k$ entries of 1. In the case $a = 0$, $b \geq 0$, it is easy to check that (7.48) is satisfied without using any identities for MGFs. The case $a > 0$, $b = 0$ follows along the lines of the derivation below by dropping the $\left[\begin{smallmatrix} 0 \\ b \end{smallmatrix}\right]$ columns everywhere and adjusting the overall sign. The final case $a = b = 1$ can be reduced to the $a = b = 0$ case by using the integration-by-parts identity (7.15).

Focusing on the coefficient of $s_{12}^k$ in (D.2), we compute the action of the Cauchy–Riemann operator $\nabla^{(a)}$ by means of (5.53), leading to

$$\begin{aligned}
&(-1)^a k!\left(\frac{\pi}{\tau_2}\right)^k \nabla^{(a)} W^{\tau}_{(a|b)}\Big|_{s_{12}^k} = a\, C\left[\begin{smallmatrix} a+1 & 0 & 1_k \\ -1 & b & 1_k \end{smallmatrix}\right] + k\, C\left[\begin{smallmatrix} a & 0 & 2 & 1_{k-1} \\ 0 & b & 0 & 1_{k-1} \end{smallmatrix}\right] \\
&= a\Big\{-C\left[\begin{smallmatrix} a+1 & 0 & 1_k \\ 0 & b-1 & 1_k \end{smallmatrix}\right] - k\, C\left[\begin{smallmatrix} a+1 & 0 & 1 & 1_{k-1} \\ 0 & b & 0 & 1_{k-1} \end{smallmatrix}\right]\Big\} \\
&\quad + k\Big\{\mathrm{G}_{a+2}\, C\left[\begin{smallmatrix} 0 & 1_{k-1} \\ b & 1_{k-1} \end{smallmatrix}\right] - \frac{1}{2}(a+2)(a+1)\, C\left[\begin{smallmatrix} a+2 & 0 & 1_{k-1} \\ 0 & b & 1_{k-1} \end{smallmatrix}\right] \\
&\quad + \sum_{\ell=4}^{a}(a+1-\ell)\mathrm{G}_\ell\, C\left[\begin{smallmatrix} a+2-\ell & 0 & 1_{k-1} \\ 0 & b & 1_{k-1} \end{smallmatrix}\right] + a\Big(\widehat{\mathrm{G}}_2\, C\left[\begin{smallmatrix} a & 0 & 1_{k-1} \\ 0 & b & 1_{k-1} \end{smallmatrix}\right] + \frac{\pi}{\tau_2}\, C\left[\begin{smallmatrix} a+1 & 0 & 1_{k-1} \\ -1 & b & 1_{k-1} \end{smallmatrix}\right]\Big)\Big\}\,,
\end{aligned} \tag{D.3}$$

where we have used (5.36) to remove the -1 entry in the first term and (5.73) to simplify the second term. Using HSR again in $C\left[\begin{smallmatrix} a+1 & 0 & 1 & 1_{k-1} \\ 0 & b & 0 & 1_{k-1} \end{smallmatrix}\right]$ and reorganizing the terms leads to (recall that $a, b > 0$ and $(a,b) \neq (1,1)$)

$$\begin{aligned}
&(-1)^a k!\left(\frac{\pi}{\tau_2}\right)^k \nabla^{(a)} W^{\tau}_{(a|b)}\Big|_{s_{12}^k} = -a\, C\left[\begin{smallmatrix} a+1 & 0 & 1_k \\ 0 & b-1 & 1_k \end{smallmatrix}\right] \\
&\quad - k\Big(-\frac{1}{2}(a+2)(a-1)\, C\left[\begin{smallmatrix} a+2 & 0 & 1_{k-1} \\ 0 & b & 1_{k-1} \end{smallmatrix}\right] + (a+1)\mathrm{G}_{a+2}\, C\left[\begin{smallmatrix} 0 & 1_{k-1} \\ b & 1_{k-1} \end{smallmatrix}\right] \\
&\quad + \sum_{\ell=4}^{a+1}(\ell-1)\mathrm{G}_\ell\, C\left[\begin{smallmatrix} a+2-\ell & 0 & 1_{k-1} \\ 0 & b & 1_{k-1} \end{smallmatrix}\right]\Big)\,.
\end{aligned} \tag{D.4}$$

Comparing this to the coefficient of $s_{12}^k$ in (7.48) shows that they agree.

## D.3 PROOF OF $s_{ij}$-FORM OF PRODUCT OF KRONECKER–EISENSTEIN SERIES

In this appendix, we prove that the ubiquitous product of doubly-periodic Kronecker–Eisenstein series admits what we call an $s_{ij}$-form

(for any choice of $1 \leq i < j \leq n$) that is amenable to evaluating the differential operator given in (7.62). This form is given by

$$\begin{aligned}
&\prod_{p=2}^{n} \Omega(z_{p-1,p}, \xi_p) \\
&= (-1)^{j-i+1} \sum_{k=i+1}^{j} \Omega(\{1,\dots,i-1\} \sqcup \{k-1,\dots,i+1\}) \Omega(z_{ij}, \xi_k) \\
&\quad \times \Omega(\{j-1,\dots,k\} \sqcup \{j+1,\dots,n\}) \,. \qquad \text{(D.5)}
\end{aligned}$$

Note that $i < k \leq j$, so that $\{k-1,\dots,i+1\}$ and $\{j-1,\dots,k\}$ always have to be in descending order and thus are empty when $k = i+1$ or when $k = j$, respectively. Similarly, the sequences $\{1,\dots,i-1\}$ and $\{j+1,\dots,n\}$ are always in ascending and order and thus empty for $i = 1$ or $j = n$, respectively. The first factor denotes a shuffle sum of products of a total of $k-2$ Kronecker–Eisenstein series

$$\begin{aligned}
&\Omega(\{1,\dots,i-1\} \sqcup \{k-1,\dots,i+1\}) \\
&= \sum_{\substack{(a_1,\dots,a_{k-2}) \\ \in\{\{1,\dots,i-1\} \\ \sqcup\{k-1,\dots,i+1\}\}}} \left[ \prod_{p=2}^{k-2} \Omega\left( z_{a_{p-1},a_p}, \sum_{\ell=p}^{k-2} \eta_{a_\ell} + \xi_k \right) \right] \Omega(z_{a_{k-2},i}, \eta_i + \xi_k) \,, \qquad \text{(D.6)}
\end{aligned}$$

while the last factor is a shuffle sum of products of $n-k$ Kronecker–Eisenstein series:

$$\begin{aligned}
&\Omega(\{j-1,\dots,k\} \sqcup \{j+1,\dots,n\}) \\
&= \sum_{\substack{(a_{k+1},\dots,a_n) \\ \in\{\{j-1,\dots,k\} \\ \sqcup\{j+1,\dots,n\}\}}} \Omega(z_{j,a_{k+1}}, \xi_k - \eta_j) \left[ \prod_{p=k+2}^{n} \Omega\left( z_{a_{p-1},a_p}, \sum_{\ell=p}^{n} \eta_{a_\ell} \right) \right] . \qquad \text{(D.7)}
\end{aligned}$$

Note that we use here

$$\eta_1 = -\sum_{p=2}^{n} \eta_p \,. \qquad \text{(D.8)}$$

Since the notation is a bit involved, we also give a more intuitive description of (D.5). For a fixed $k$ in the range $i < k \leq j$ we have a permutation $(a_1,\dots,a_n)$ of the range $(1,\dots,n)$ with the following boundary conditions:

1. The indices $i$ and $j$ occur at positions $k-1$ and $k$: $a_{k-1} = i$ and $a_k = j$.

2. The subsequence $(a_1,\dots,a_{k-2})$ to the left of $i$ (at position $k-1$) is a rearrangement of $\{1,\dots,k-1\} \setminus \{i\}$ such that it is obtained as a shuffle in $\{1,\dots,i-1\} \sqcup \{k-1,\dots,i+1\}$.

3. The subsequence $(a_{k+1}, \dots, a_n)$ to the right of $j$ (at position $k$) is a rearrangement of $\{k, \dots, n\} \setminus \{j\}$ such that it is obtained as a shuffle in $\{j-1, \dots, k\} \sqcup \{j+1, \dots, n\}$.

The situation is also illustrated in Figure 7.1. To each such sequence $(a_1, \dots, a_n)$ there is a product in (D.5)

$$\prod_{p=2}^{n} \Omega\left(z_{a_{p-1},a_p}, \sum_{\ell=p}^{n} \eta_{a_\ell}\right) \tag{D.9}$$

and one is summing over all possible intermediate points $i < k \leq j$.

Note that we are using that

$$\xi_k = \sum_{\ell=k}^{n} \eta_\ell = \sum_{\ell=k}^{n} \eta_{a_\ell} \tag{D.10}$$

since the subsequence $(a_k, \dots, a_n)$ is a permutation of $\{k, \dots, n\}$. Therefore also

$$\sum_{\ell=p}^{k-1} \eta_{a_\ell} + \xi_k = \sum_{\ell=p}^{n} \eta_{a_\ell} \quad \text{and} \quad \eta_i + \xi_k = \sum_{\ell=k-1}^{n} \eta_{a_\ell}\,, \tag{D.11}$$

such that the arrangement of $\eta$-arguments in (D.5) is correct to represent a different sequence of points with corresponding generating arguments.

We note that the change of variables at $n$ points

$$\xi_p = \sum_{\ell=p}^{n} \eta_p \qquad \Longleftrightarrow \qquad \eta_p = \begin{cases} \xi_p - \xi_{p+1} & \text{for } 1 < p < n \\ \xi_n & \text{for } p = n \end{cases} \tag{D.12}$$

implies that for the differential operator one has

$$\sum_{k=i+1}^{j} \partial_{\xi_k} = \partial_{\eta_j} - \partial_{\eta_i} \tag{D.13}$$

for all $1 \leq i < j \leq n$ when setting $\partial_{\eta_1} = 0$. Thus the differential operator $\sum_{k=i+1}^{j} \partial_{\xi_k}$ acts only as $\partial_{\xi_k}$ on $\Omega(z_{ij}, \xi_k)$ and vanishes on all other factors in (D.5). This is true since $\xi_k$ only contains $\eta_j$ but not $\eta_i$. The term $\xi_k - \eta_{a_k} = \xi_k - \eta_j$ and all other terms to the right of $\Omega(z_{ij}, \xi_k)$ are free of $\eta_j$ while $\xi_k + \eta_{a_{k-1}} = \xi_k + \eta_i$ and all other terms to the left of $\Omega(z_{ij}, \xi_k)$ always contains $\eta_i + \eta_j$ and are therefore also annihilated by (D.13). Thus the $s_{ij}$-form (D.5) is the correct form for evaluating for $1 \leq i < j \leq n$ that

$$\left[f_{ij}^{(1)} \sum_{k=i+1}^{j} \partial_{\xi_k} - f_{ij}^{(2)}\right] \prod_{p=2}^{n} \Omega(z_{p-1,p}, \xi_p, \tau) = \frac{1}{2}(\partial_{\eta_j} - \partial_{\eta_i})^2 \prod_{p=2}^{n} \Omega(z_{p-1,p}, \xi_p, \tau)$$
$$- (-1)^{j-i+1} \sum_{k=i+1}^{j} \wp(\xi_k, \tau) \sum_{\substack{(a_1,\dots,a_n) \\ \in S_n(i,j,k)}} \prod_{p=2}^{n} \Omega\left(z_{a_{p-1},a_p}, \sum_{\ell=p}^{n} \eta_{a_\ell}\right) \tag{D.14}$$

using (7.26). Here, we set again $\partial_{\eta_1} = 0$ and have introduced the shorthand $S_n(i, j, k)$ defined in (7.61) for the sequences obtained by all possibles shuffles occurring in (D.5) and illustrated in Figure 7.1.

The proof of the $s_{ij}$-form (D.5) of the product of Kronecker–Eisenstein series proceeds by several lemmata.

### D.3.1 *$s_{1n}$-form at $n$ points*

We begin with establishing the extreme case when $i = 1$ and $j = n$ for $n$ points. The formula (D.5) then specializes to

$$\begin{aligned}
&\prod_{p=2}^{n} \Omega(z_{p-1,p}, \xi_p) \\
&= (-1)^n \sum_{k=2}^{n} \Omega(z_{k-1,k-2}, \xi_k - \xi_{k-1}) \cdots \Omega(z_{21}, \xi_k - \xi_2) \Omega(z_{1n}, \xi_k) \\
&\quad \times \Omega(z_{n,n-1}, \xi_k - \xi_n) \cdots \Omega(z_{k+1,k}, \xi_k - \xi_{k+1}) \,,
\end{aligned} \tag{D.15}$$

with only descending parts to the left and right of $\Omega(z_{1n}, \xi_k)$.

This form can be proved by induction on $n$. For $n = 2$ there is nothing to do. Assume then that (D.15) holds for $n - 1$ points. Then

$$\begin{aligned}
&\prod_{p=2}^{n} \Omega(z_{p-1,p}, \xi_p) \\
&= \left[\prod_{p=2}^{n-1} \Omega(z_{p-1,p}, \xi_p)\right] \Omega(z_{n-1,n}, \xi_n) \\
&= \Bigg[(-1)^{n-1} \sum_{k=2}^{n-1} \Omega(z_{k-1,k-2}, \xi_k - \xi_{k-1}) \cdots \Omega(z_{21}, \xi_k - \xi_2) \Omega(z_{1,n-1}, \xi_k) \\
&\quad \times \Omega(z_{n-1,n-2}, \xi_k - \xi_{n-1}) \cdots \Omega(z_{k+1,k}, \xi_k - \xi_{k+1})\Bigg] \Omega(z_{n-1,n}, \xi_n)
\end{aligned}$$

$$
\begin{aligned}
&= (-1)^n \sum_{k=2}^{n-1} \Omega(z_{k-1,k-2}, \xi_k - \xi_{k-1}) \cdots \Omega(z_{21}, \xi_k - \xi_2)\Omega(z_{1,n}, \xi_k) \\
&\quad \times \Omega(z_{n,n-1}, \xi_k - \xi_n)\Omega(z_{n-1,n-2}, \xi_k - \xi_{n-1}) \cdots \Omega(z_{k+1,k}, \xi_k - \xi_{k+1}) \\
&\quad + \Bigg[ (-1)^{n-1} \sum_{k=2}^{n-1} \Omega(z_{k-1,k-2}, \xi_k - \xi_{k-1}) \cdots \Omega(z_{21}, \xi_k - \xi_2)\Omega(z_{1,n-1}, \xi_k - \xi_n) \\
&\quad \times \Omega(z_{n-1,n-2}, \xi_k - \xi_{n-1}) \cdots \Omega(z_{k+1,k}, \xi_k - \xi_{k+1}) \Bigg] \Omega(z_{1n}, \xi_n) \\
&= (-1)^n \sum_{k=2}^{n} \Omega(z_{k-1,k-2}, \xi_k - \xi_{k-1}) \cdots \Omega(z_{21}, \xi_k - \xi_2)\Omega(z_{1,n}, \xi_k) \\
&\quad \times \Omega(z_{n,n-1}, \xi_k - \xi_n) \cdots \Omega(z_{k+1,k}, \xi_k - \xi_{k+1}) \,, \qquad \text{(D.16)}
\end{aligned}
$$

where we have used the Fay identity (5.120b) in the form

$$
\begin{aligned}
&\Omega(z_{1,n-1}, \xi_k)\Omega(z_{n-1,n}, \xi_n) \qquad \text{(D.17)} \\
&= -\Omega(z_{1n}, \xi_k)\Omega(z_{n,n-1}, \xi_k - \xi_n) + \Omega(z_{1n}, \xi_n)\Omega(z_{1,n-1}, \xi_k - \xi_n)
\end{aligned}
$$

and have used the induction hypothesis again at shifted $\xi$-values in the last step to convert the expression into the missing summand for $k = n$.

A corollary of (D.15) is obtained simply by shifting the indices

$$
\begin{aligned}
&\prod_{p=i+1}^{j} \Omega(z_{p-1,p}, \xi_p) \\
&= (-1)^{j-i+1} \sum_{k=i+1}^{j} \Omega(z_{k-1,k-2}, \xi_k - \xi_{k-1}) \cdots \Omega(z_{i+1,i}, \xi_k - \xi_{i+1})\Omega(z_{ij}, \xi_k) \\
&\quad \times \Omega(z_{j,j-1}, \xi_k - \xi_j) \cdots \Omega(z_{k+1,k}, \xi_k - \xi_{k+1}) \,. \qquad \text{(D.18)}
\end{aligned}
$$

With (D.18) we have an expression for the product of Kronecker–Eisenstein series between $i$ and $j$. In order to obtain the full product from 1 to $n$ we need to extend to the left and right. This can be done with the help of two little lemmata.

### D.3.2 *Extending left and right*

We now want to prove (D.5) by induction on $n$ which means extending the product on the left and on the right, beginning with the right.

We first consider keeping $i$ and $j$ fixed and extend (D.5) by multiplying by $\Omega(z_{n,n+1}, \xi_{n+1})$ on the right. If $k = n$ there is nothing to do since the factors just multiplies correctly at the end of the sequence: If the original sequence is $(a_1, \ldots, a_n)$ with $a_n = n$ the new sequence is $(a_1, \ldots, a_n, a_{n+1})$ with $a_{n+1} = n + 1$ and the $\xi$-factors are correct for the action of the differential operator.

If $k < n$ then there is an $m < n$ such that $a_m = n$. Moreover, $m \geq k$. This means that in the product (D.9) somewhere in the tail to the right of $\Omega(z_{ij}, \xi_k)$ there is the subproduct

$$\begin{aligned}\Omega[n, \{a_{m+1}, \ldots, a_n\}] &= \prod_{p=m+1}^{n} \Omega\left(z_{a_{p-1},a_p}, \sum_{\ell=p}^{n} \eta_{a_\ell}\right) \\ &= \Omega\left(z_{n,a_{m+1}}, \sum_{\ell=m+1}^{n} \eta_{a_\ell}\right) \Omega[a_{m+1}, \{a_{m+2}, \ldots, a_n\}]\end{aligned} \tag{D.19}$$

and the only place where the index $n$ appears is in the very first factor that has been made explicit in the second line. Multiplying such an product by $\Omega(z_{n,n+1}, \xi_{n+1})$ leads to splicing in the index $n+1$ in all possible places:

$$\begin{aligned}&\Omega[n, \{a_{m+1}, \ldots, a_n\}]\Omega(z_{n,n+1}, \xi_{n+1}) \\ &= \Omega[n, \{a_{m+1}, \ldots, a_n\} \sqcup \{n+1\}] \\ &= \Omega[n, \{n+1, a_{m+1}, \ldots, a_n\}] + \Omega[n, \{a_{m+1}, n+1, a_{m+2}, \ldots, a_n\}] \\ &\quad + \ldots + \Omega[n, \{a_{m+1}, \ldots, a_n, n+1\}].\end{aligned} \tag{D.20}$$

The assertion (D.20) is proved by induction on the length $n - m$ of the original product. For $n - m = 0$ this is trivially true as already stated above.

Now assume the formula (D.20) is correct for products of length $n - m$. Then

$$\begin{aligned}&\Omega[n, \{a_m, \ldots, a_n\}]\Omega(z_{n,n+1}, \xi_{n+1}) \\ &= \left[\Omega\left(z_{n,n+1}, \sum_{\ell=m}^{n+1} \eta_{a_\ell}\right)\Omega\left(z_{n+1,a_m}, \sum_{\ell=m}^{n} \eta_{a_\ell}\right)\right. \\ &\quad \left. + \Omega\left(z_{n,a_m}, \sum_{\ell=m}^{n+1} \eta_{a_\ell}\right)\Omega(z_{a_m,n+1}, \xi_{n+1})\right] \Omega[a_m, \{a_{m+1}, \ldots, a_n\}] \\ &= \Omega[n, \{n+1, a_{m+1}, \ldots, a_n\}] \\ &\quad + \Omega\left(z_{n,a_m}, \sum_{\ell=m}^{n+1} \eta_{a_\ell}\right)\Omega[a_m, \{a_{m+1}, \ldots, a_n\} \sqcup \{n+1\}] \\ &= \Omega[n, \{a_m, \ldots, a_n\} \sqcup \{n+1\}],\end{aligned} \tag{D.21}$$

where we have used a Fay identity in the first step and the induction hypotheses in the second step and collected terms in the last line. This proves (D.20).

Note that the index $n+1$ always ends up to the right of $n$ in this product and so the order is preserved for them. This means that (D.7) is the correct shuffle prescription for extending from $j = n$ to any $j < n$ by shuffling in the additional indices into the reversed indices to the right of $\Omega(z_{ij}, \xi_k)$ in (D.18).

By similar methods one can also show that multiplying by $\Omega(z_{01}, \xi_1)$ to extend on the left shuffles in the index 0 in all possible places to the left of the index 1. After renaming the indices, we conclude that (D.6) is the correct shuffle prescription for extending from $i = 1$ to any $i > 1$ by shuffling in the indices $\{1, \ldots, i-1\}$ into the reversed indices to the left of $\Omega(z_{ij}, \xi_k)$ in (D.18).

The sequences constructed by this inductive method constitute the set $S_n(i, j, k)$ defined in (7.61) and illustrated in Figure 7.1. This concludes the proof of (D.5).

# E

# COMPONENT INTEGRALS $Y^{\tau}_{(a|b)}$ AT LEADING ORDER

In this appendix, which has extensive text overlap with Appendix C.3 of [V], we derive both the closed depth-one formulae (8.72), (8.73) relating non-holomorphic Eisenstein series to the $\beta^{\text{sv}}$ and the reality properties (8.83) of the latter. For this purpose, we investigate the $s_{12}^0$-order of the two-point component integrals $Y^{\tau}_{(a|b)}$ in (8.22) with $a{+}b \geq 4$, where the $(s_{12} \to 0)$-limit can be performed at the level of the integrand. By the lattice-sum representations (3.91) of the $f^{(a)}$, this limit vanishes if $a = 0$ or $b = 0$ and otherwise yields MGFs (cf. (7.13) for the expansion of the $W^{\tau}_{(a|b)}$)

$$Y^{\tau}_{(a|b)} = \frac{(\tau{-}\bar{\tau})^a}{(2\pi i)^b} \, C\big[\begin{smallmatrix} a & 0 \\ b & 0 \end{smallmatrix}\big] + O(s_{12}) \,, \qquad a,b \neq 0 \,, \qquad a+b \geq 4 \,. \tag{E.1}$$

Once the MGFs are expressed in terms of non-holomorphic Eisenstein series via (5.56), the $s_{12}^0$-orders of the component integrals can be rewritten as ($k \geq 2$, $m < k$)

$$\begin{aligned} Y^{\tau}_{(k|k)} &= \mathrm{E}_k + O(s_{12}) \,, \\ Y^{\tau}_{(k+m|k-m)} &= \frac{(-4)^m (k{-}1)! (\pi \nabla_0)^m \mathrm{E}_k}{(k{+}m{-}1)!} + O(s_{12}) \,, \\ Y^{\tau}_{(k-m|k+m)} &= \frac{(k{-}1)! (\pi \overline{\nabla}_0)^m \mathrm{E}_k}{(-4)^m (k{+}m{-}1)! y^{2m}} + O(s_{12}) \,. \end{aligned} \tag{E.2}$$

These results will now be compared to the $\alpha'$-expansion (8.36) in terms of $\beta^{\text{sv}}$ and initial values. The latter can be inferred from the Laurent polynomials (8.66) by acting with the two-point derivation $R_\eta(\epsilon_0)$ in (8.14), and one obtains

$$\exp\Big(-\frac{R_{\vec{\eta}}(\epsilon_0)}{4y}\Big) \widehat{Y}^{i\infty}_{\eta} = \frac{1}{\eta\bar{\eta}} - \frac{2\pi i}{s_{12}} + 4\pi i \sum_{k=1}^{\infty} \zeta_{2k+1} \Big(\eta + \frac{i\pi\bar{\eta}}{2y}\Big)^{2k} + O(s_{12}) \,. \tag{E.3}$$

The $s_{12}^0$-order of the generating series $Y^{\tau}_{\eta}$ receives additional contributions when the $\eta$-independent kinematic pole of (E.3) is combined with one power of $s_{12}$ from the derivations $R_\eta(\epsilon_k)$. This order exclusively stems from the depth-one part of the series (8.36) in $\beta^{\rm sv}$,

$$\sum_{k=4}^{\infty}\sum_{j=0}^{k-2}\frac{(-1)^j(k-1)}{(k-j-2)!}\beta^{\rm sv}\big[\begin{smallmatrix} j \\ k\end{smallmatrix}\big]\,R_\eta\big(\mathrm{ad}_{\epsilon_0}^{k-j-2}(\epsilon_k)\big) \tag{E.4}$$
$$= s_{12}\sum_{k=4}^{\infty}\sum_{j=0}^{k-2}(2\pi i)^{k-j-2}\frac{(k-1)!}{j!(k-j-2)!}\eta^j\bar\eta^{k-j-2}\beta^{\rm sv}\big[\begin{smallmatrix} j \\ k\end{smallmatrix}\big] + O(s_{12}^2,\partial_\eta)\,,$$

where we have inserted $R_\eta(\epsilon_k) = s_{12}\eta^{k-2}$ and $R_\eta(\epsilon_0) = -2\pi i\bar\eta\partial_\eta + O(s_{12})$. In view of (E.3) and (E.4), the overall $(s_{12}\to 0)$-limit of the generating series is given by

$$Y^\tau_\eta = \frac{1}{\eta\bar\eta} - \frac{2\pi i}{s_{12}} + 4\pi i\sum_{k=1}^{\infty}\zeta_{2k+1}\Big(\eta+\frac{i\pi\bar\eta}{2y}\Big)^{2k} \tag{E.5}$$
$$-\sum_{k=4}^{\infty}\sum_{j=0}^{k-2}(2\pi i)^{k-j-1}\frac{(k-1)!}{j!(k-j-2)!}\eta^j\bar\eta^{k-j-2}\beta^{\rm sv}\big[\begin{smallmatrix} j \\ k\end{smallmatrix}\big] + O(s_{12})\,.$$

By extracting the coefficients of $\eta^{a-1}\bar\eta^{b-1}$, we arrive at the following leading orders of the component integrals (8.22)

$$Y^\tau_{(a|b)} = \frac{(a+b-2)!}{(a-1)!(b-1)!}\Big\{\frac{2\zeta_{a+b-1}}{(4y)^{b-1}} - (a+b-1)\beta^{\rm sv}\big[\begin{smallmatrix} a-1 \\ a+b\end{smallmatrix}\big]\Big\} + O(s_{12})\,, \tag{E.6}$$

where $a,b\neq 0$ and $a+b\geq 4$. Upon comparison with the earlier expression (E.2) for the $s_{12}^0$-orders in terms of non-holomorphic Eisenstein series, one can read off (8.72) by setting $(a,b)=(k,k)$ and (8.73) by setting $(a,b)=(k\pm m,k\mp m)$ with $m<k$. Moreover, irrespective of the relation (E.2) with $\mathrm{E}_k$, the reality properties (8.23) of the $Y^\tau_{(a|b)}$ enforce the $\beta^{\rm sv}$ in (E.6) to obey

$$\overline{\beta^{\rm sv}\big[\begin{smallmatrix} a-1 \\ a+b\end{smallmatrix}\big]} = (4y)^{a-b}\beta^{\rm sv}\big[\begin{smallmatrix} b-1 \\ a+b\end{smallmatrix}\big]\,. \tag{E.7}$$

This is equivalent to (8.83), i.e. we have derived the reality properties of the $\beta^{\rm sv}$ at depth one from those of $Y^\tau_{(a|b)}$ and the explicit form of their $(s_{12}\to 0)$-limits (E.6).

# INDEX